\documentclass[10pt,twoside,openright]{report} 

\title{Beyond Lorentzian Symmetry}
\author{Ross Grassie}
\date{2021}

\usepackage[phd]{edmaths}
\usepackage{lmodern}
\usepackage{bm}

\usepackage[utf8]{inputenc}
\usepackage{amsfonts,amssymb,amstext,amsthm,amsmath,eucal,bbm,mathrsfs,amscd,bbold}
\usepackage{mathtools}
\usepackage[T1]{fontenc}
\usepackage{cite}
\usepackage[hyperindex,breaklinks,unicode,hidelinks]{hyperref}
\usepackage{tensor}
\usepackage{mdframed}

\hypersetup{%
  pdftitle   = {Beyond Lorentzian Symmetry},
  pdfkeywords = {Lie superalgebra, kinematical, Aristotelian, Bargmann, superspace},
  pdfauthor  = {Ross Grassie},
  pdfcreator = {\LaTeX\ with package \flqq hyperref\frqq},
  linkcolor=Black,
  citecolor=NavyBlue,
  urlcolor=OrangeRed,
  anchorcolor=OrangeRed,
  colorlinks=true
}

\usepackage[svgnames,table]{xcolor}
\usepackage{booktabs} 
\usepackage{graphicx}
\usepackage{enumitem}
\usepackage[small]{eulervm}
\usepackage{caption}
\usepackage{enumitem}
\usepackage{array}
\usepackage{booktabs}

\usepackage{tikz,pgfplots}
\pgfplotsset{compat=1.16}
\usepgfplotslibrary{fillbetween}
\usetikzlibrary{calc,arrows,cd,shapes.geometric,positioning,through}
\tikzset{carrollian/.style={draw,regular polygon,regular polygon sides=3,fill=DarkOrange,color=DarkOrange,minimum width=1pt,scale=0.25}}
\tikzset{lorentzian/.style={draw,regular polygon,regular polygon
    sides=4,fill=red!70!black,color=red!70!black,minimum width=1pt,scale=0.3}}

\usepackage{verbatim}

\numberwithin{equation}{subsection}

\theoremstyle{plain}
\newcounter{dummy} \numberwithin{dummy}{section}
\newtheorem{lemma}[dummy]{Lemma}

\newtheorem{theorem}[dummy]{Theorem}

\theoremstyle{definition}
\newtheorem{definition}{Definition}

\newcommand{\bH}{\mathrm{H}}
\newcommand{\bZ}{\mathrm{Z}}
\newcommand{\bJ}{\mathrm{J}}

\newcommand{\bB}{\mathrm{B}}
\newcommand{\bP}{\mathrm{P}}
\newcommand{\bQ}{\mathrm{Q}}

\newcommand{\J}{\boldsymbol{\mathrm{J}}}
\newcommand{\B}{\boldsymbol{\mathrm{B}}}
\renewcommand{\P}{\boldsymbol{\mathrm{P}}}
\newcommand{\Pt}{\widetilde{\boldsymbol{\mathrm{P}}}}

\newcommand{\Q}{\boldsymbol{\mathrm{Q}}}
\newcommand{\Bbar}{\overline{\B}}
\newcommand{\Pbar}{\overline{\P}}
\newcommand{\Hbar}{\overline{\bH}}

\newcommand{\sH}{\mathsf{H}}
\newcommand{\sZ}{\mathsf{Z}}
\newcommand{\sJ}{\mathsf{J}}
\newcommand{\sB}{\mathsf{B}}
\newcommand{\sP}{\mathsf{P}}
\newcommand{\sQ}{\mathsf{Q}}
\newcommand{\sV}{\mathsf{V}}

\newcommand{\s}{\mathfrak{s}}
\renewcommand{\r}{\mathfrak{r}}
\newcommand{\e}{\mathfrak{e}}
\newcommand{\p}{\mathfrak{p}}
\newcommand{\co}{\mathfrak{co}}
\renewcommand{\k}{\mathfrak{k}}
\newcommand{\g}{\mathfrak{g}}

\newcommand{\h}{\mathfrak{h}}

\newcommand{\gl}{\mathfrak{gl}}
\newcommand{\so}{\mathfrak{so}}
\renewcommand{\sp}{\mathfrak{sp}}
\newcommand{\n}{\mathfrak{n}}
\newcommand{\osp}{\mathfrak{osp}}
\renewcommand{\a}{\mathfrak{a}}
\renewcommand{\b}{\mathfrak{b}}
\renewcommand{\c}{\mathfrak{c}}
\renewcommand{\r}{\mathfrak{r}}
\newcommand{\z}{\mathfrak{z}}

\newcommand{\sa}{\mathfrak{sa}}

\renewcommand{\v}{\mathfrak{v}}
\newcommand{\m}{\mathfrak{m}}

\newcommand{\Agr}{\mathcal{A}}
\newcommand{\Kgr}{\mathcal{K}}
\newcommand{\Hgr}{\mathcal{H}}
\newcommand{\Rgr}{\mathcal{R}}
\newcommand{\Ggr}{\mathcal{G}}
\newcommand{\Sgr}{\mathcal{S}}
\newcommand{\Bgr}{\mathcal{B}}
\newcommand{\Sgrbar}{\overline{\mathcal{S}}}
\newcommand{\Bgrbar}{\overline{\mathcal{B}}}
\newcommand{\Ghat}{\widehat{\mathcal{G}}}
\newcommand{\Khat}{\widehat{\mathcal{K}}}
\newcommand{\Hhat}{\widehat{\mathcal{H}}}
\newcommand{\eE}{\mathcal{C}^\infty}
\newcommand{\eN}{\mathcal{N}}
\newcommand{\eO}{\mathcal{O}}

\newcommand{\Aut}{\operatorname{Aut}}
\newcommand{\Mat}{\operatorname{Mat}}
\renewcommand{\G}{\operatorname{G}}
\newcommand{\GL}{\operatorname{GL}}
\newcommand{\SO}{\operatorname{SO}}

\newcommand{\SU}{\operatorname{SU}}
\newcommand{\Sp}{\operatorname{Sp}}

\newcommand{\End}{\operatorname{End}}
\newcommand{\Hom}{\operatorname{Hom}}
\newcommand{\Cl}{C\ell}

\newcommand{\cV}{\mathscr{V}}
\newcommand{\cJ}{\mathscr{J}}
\newcommand{\cS}{\mathscr{S}}

\newcommand{\M}{\mathcal{M}}

\newcommand{\Mbar}{\overline{\M}}
\renewcommand{\H}{H}
\newcommand{\x}{\boldsymbol{x}}
\newcommand{\y}{\boldsymbol{y}}
\renewcommand{\v}{\boldsymbol{v}}
\newcommand{\w}{\boldsymbol{w}}

\newcommand{\bc}{\mathbb{c}}

\newcommand{\eT}{\mathscr{T}}
\newcommand{\eX}{\mathscr{X}}
\newcommand{\bbeta}{\boldsymbol{\beta}}

\newcommand{\bpi}{\boldsymbol{\pi}}

\newcommand{\adsg}{S10}

\newcommand{\athree}{A23}

\newcommand{\zLC}{\mathsf{LC}}
\newcommand{\zAdSC}{\mathsf{AdSC}}
\newcommand{\zdSC}{\mathsf{dSC}}
\newcommand{\zdS}{\mathsf{dS}}
\newcommand{\zAdS}{\mathsf{AdS}}
\newcommand{\zC}{\mathsf{C}}
\newcommand{\zG}{\mathsf{G}}
\newcommand{\zS}{\mathsf{A}} 
\newcommand{\zTS}{\mathsf{TA}}
\newcommand{\zAdSG}{\mathsf{AdSG}}
\newcommand{\zdSG}{\mathsf{dSG}}
\newcommand{\ztAdSG}{\mathsf{AdSG}}
\newcommand{\ztdSG}{\mathsf{dSG}}
\newcommand{\EE}{\mathbb{E}}
\newcommand{\HH}{\mathbb{H}}
\newcommand{\MM}{\mathbb{M}}

\newcommand{\RP}{\mathbb{RP}}

\let\minusplus\mp

\newcommand{\mh}{H}
\newcommand{\mz}{Z}
\newcommand{\mb}{B}
\renewcommand{\mp}{P}

\newcommand{\hh}{\mathbb{h}}
\newcommand{\hhbar}{\overline{\mathbb{h}}}
\newcommand{\zz}{\mathbb{z}}
\newcommand{\bb}{\mathbb{b}}
\newcommand{\pp}{\mathbb{p}}
\newcommand{\ppbar}{\overline{\mathbb{p}}}

\newcommand{\vt}{\boldsymbol{\theta}}
\newcommand{\JJ}{\mathbb{J}}
\newcommand{\BB}{\mathbb{B}}
\newcommand{\PP}{\mathbb{P}}

\newcommand{\cc}{\mathbb{c}}
\newcommand{\dd}{\mathbb{d}}
\newcommand{\ee}{\mathbb{e}}
\newcommand{\ff}{\mathbb{f}}
\renewcommand{\gg}{\mathbb{g}}
\renewcommand{\ll}{\mathbb{l}}
\newcommand{\mm}{\mathbb{m}}
\newcommand{\nn}{\mathbb{n}}
\newcommand{\qq}{\mathbb{q}}
\newcommand{\qqbar}{\overline{\mathbb{q}}}
\newcommand{\rr}{\mathbb{r}}
\renewcommand{\ss}{\mathbb{s}}
\newcommand{\uu}{\mathbb{u}}
\newcommand{\vv}{\mathbb{v}}

\newcommand{\xx}{\mathbb{x}}

\newcommand{\sbar}{\overline{s}}
\newcommand{\ubar}{\overline{\uu}}

\newcommand{\ii}{\mathbb{i}}
\newcommand{\jj}{\mathbb{j}}
\newcommand{\kk}{\mathbb{k}}

\newcommand{\overbar}[1]{\mkern 1.5mu\overline{\mkern-1.5mu#1\mkern-1.5mu}\mkern 1.5mu}

\newcommand{\choice}[2]{\genfrac{}{}{0pt}{}{#1}{#2}}

\definecolor{gris}{rgb}{0.5,0.5,0.5}
\newcommand{\zero}{{\color{gris}0}}

\newcommand{\spn}[1]{\operatorname{span}_\RR\left\{#1\right\}}

\renewcommand{\ker}{\operatorname{ker}}
\newcommand{\im}{\operatorname{im}}

\newcommand{\Ad}{\operatorname{Ad}}
\newcommand{\ad}{\operatorname{ad}}
\renewcommand{\Re}{\operatorname{Re}}
\renewcommand{\Im}{\operatorname{Im}}
\newcommand{\csch}{\operatorname{csch}}

\newcommand{\Rad}{\operatorname{Rad}}
\newcommand{\rad}{\operatorname{rad}}
\newcommand{\id}{\mathbb{1}}
\newcommand{\1}{\mathbb{1}}
\renewcommand{\d}{\partial}
\newcommand{\pd}{\partial}

\newcommand{\Lie}{\operatorname{Lie}}
\newcommand{\Stab}{\operatorname{Stab}}

\newcommand{\RR}{\mathbb{R}}

\newcommand{\ZZ}{\mathbb{Z}}

\usepackage{pifont}

\newcommand{\N}{\mathcal{N}}

\newcommand{\eff}[1]{\colorbox{blue!30}{$#1$}}
\newcommand{\ari}[1]{\colorbox{green!30}{$#1$}}
\newcommand{\non}[1]{\colorbox{gris!20}{\textcolor{gris!50}{$#1$}}}

\begin{document}

\pagenumbering{roman}
\maketitle

\declaration

\dedication{In Memory of Winifred Whittle}

\chapter*{Acknowledgements}
\addcontentsline{toc}{chapter}{Acknowledgements}
  \begin{singlespace}
I would like to start by thanking my supervisor José Figueroa-O'Farrill.  His enthusiasm for mathematics and physics has certainly had a profound impact on my approach to research, teaching me not to be so serious all the time and remember that there is a lot of fun to be had in exploring new material.  
\\ \\
I would also like to thank my family and friends for their constant support through what has been a very difficult few years.  In particular, I would like to thank my partner Katie Meyer for always believing in me and picking me up when I felt as though the world was falling apart.  I don't think my PhD would have been half as fun as it was without my office mates, Fred Tomlinson, Graeme Auld and Ollie Burke.  Whether it was going to the pub perhaps a bit too early on Friday or chatting complete nonsense for a few hours in the office, they made the last couple of years infinitely more enjoyable.  I think a special mention should also be reserved for Sam Stern and Manya Sahni.  If it were not for them, I would never have made it through my undergraduate of masters degrees to even start my PhD!  
\\ \\
Finally, I would like to thank my gran, Winifred Whittle.  I do not think anything I can write here can do justice to how much of a positive influence she has had on my life.  It is primarily due to the work ethic she passed on to me that this thesis exists.  
  \end{singlespace}

\chapter*{Lay Summary}
\addcontentsline{toc}{chapter}{Lay Summary}
  \begin{singlespace}
The goal for any physicist is to better understand the world around them. To complete this task, we need to construct models that estimate, characterise, or otherwise describe the particular phenomenon we are interested in. This may be how an object moves through space, what happens to it over time, what happens when it is exposed to extreme temperatures or pressures, how it interacts with other things; whatever it may be, we want to describe its place in the world. The hope is that by using insightful and robust models, we will accurately predict interesting aspects of future occurrences of the phenomenon. Equipped with the ability to produce these predictions, we may say we have crucial insight into how the world works. However, our understanding of these phenomena will be heavily influenced by the model we choose. Therefore, we need to make sure we are constructing models which are fit for purpose.
\\ \\
Constructing a model for a particular phenomenon can be thought of as consisting of two parts. The first part captures the initial conditions of the system.  Using the movement of a planet as an example, we would like to know its starting position and how fast it is moving. The second part contains what we may call the ``laws of physics'' or the ``laws of nature''. These are the ideas we believe hold true irrespective of the initial conditions; for example, Newton's gravitational laws.  As the initial conditions will necessarily be different each time we conduct an experiment, the second part of our model gives us the desired predictive power to better understand the world around us.  However, building this part of our models is challenging; therefore, we would like to recognise and exploit any inherent structure and coherence in these natural laws. Since mathematics is the chosen language of physicists, we want a mathematical principle that will capture this information for us. The current principle of choice, which guides us in producing our models of the world, is \emph{symmetry}.
\\ \\
In this thesis, we will classify symmetries and construct models of space and time that employ these symmetries.  Our hope is that future researchers will use these classifications to accurately describe and better understand the world we live in. 
  \end{singlespace}

\begin{abstract}
This thesis presents a framework in which to explore kinematical symmetries beyond the standard Lorentzian case.  This framework consists of an algebraic classification, a geometric classification, and a derivation of the geometric properties required to define physical theories on the classified spacetime geometries. The work completed in substantiating this framework for kinematical, super-kinematical, and super-Bargmann symmetries constitutes the body of this thesis.  
\\ \\
To this end, the classification of kinematical Lie algebras in spatial dimension $D = 3$, as presented in~\cite{Figueroa-OFarrill:2017ycu, Figueroa-OFarrill:2017tcy}, is reviewed; as is the classification of spatially-isotropic homogeneous spacetimes of~\cite{Figueroa-OFarrill:2018ilb}.  The derivation of geometric properties such as the non-compactness of boosts, soldering forms and vielbeins, and the space of invariant affine connections is then presented.  
\\ \\
We move on to classify the $\N=1$ kinematical Lie superalgebras in three spatial dimensions, finding  $43$ isomorphism classes of Lie superalgebras.  Once these algebras are determined, we classify the corresponding simply-connected homogeneous $(4|4)$-dimensional superspaces and show how the resulting $27$ homogeneous superspaces may be related to one another via geometric limits.  
\\ \\
Finally, we turn our attention to generalised Bargmann superalgebras.  In the present work, these will be the $\N=1$ and $\N=2$ super-extensions of the Bargmann and Newton-Hooke algebras, as well as the centrally-extended static kinematical Lie algebra, of which the former three all arise as deformations.  Focussing solely on three spatial dimensions, we find $9$ isomorphism classes in the $\N=1$ case, and we identify $22$ branches of superalgebras in the $\N=2$ case.
\end{abstract}

\tableofcontents
\addcontentsline{toc}{chapter}{Contents}
\newpage
\pagenumbering{arabic}

\chapter{Introduction} \label{chap:intro}
This thesis is concerned with the classification of (super-)kinematical symmetry algebras and their corresponding spacetime (super)geometries, as well as the determination of various geometric properties of these (super)geometries. As such, this chapter will give a brief history of kinematical symmetry, introducing the key concepts underlying the research presented here and discussing why Lorentz symmetry became the primary example of kinematical symmetry. Additionally, there will be a discussion on why we may be interested in extending beyond Lorentzian symmetry which will  include the introduction of the five kinematical spacetime classes and supersymmetry.  We conclude by outlining the rest of the thesis.

\section{A Brief History of Kinematical Symmetry}
People have been using symmetry to try and describe the world around them for thousands of years. The idea of symmetry in nature can be found in the \emph{Timaeus} by Plato and Euclid's \emph{Elements}~\cite{Brading:2003zr}. However, symmetry, in the sense in which it is used in modern physics, is a relatively new concept. In 1905, Einstein published his seminal work on special relativity, introducing the profound paradigm shift that placed symmetry at the core of how we think about the world. Prior to this work, physicists such as Newton and Maxwell had built their models of the world by first seeking to write down natural laws. Therefore, the invariance of these laws under some symmetry was recognised, but it was not seen as particularly important. However, with the advent of special relativity, the roles of symmetry and natural laws were reversed; in particular, we now seek to derive laws of nature from symmetry considerations, rather than derive the symmetry of natural laws~\cite{wigner1967symmetries}.  In the special relativity paper, Einstein showed that one could derive the transformation properties of an electromagnetic field using Lorentz invariance alone, rather than deriving them from Maxwell's equations~\cite{Gross14256}. Thus he took the known Lorentz symmetry of Maxwell's equations and demonstrated how we could view it as something more fundamental; we may view it as a symmetry of the spacetime in which the electromagnetic field is propagating. 
\\ \\
From this historical perspective, if we want to understand symmetries of spacetime, also called \emph{kinematical} symmetries, we need to look at the classical laws of nature, such as Newton's laws of motion and Maxwell's electromagnetic equations. By determining what type of symmetries they are invariant under, we may better understand what kinematical symmetries we can have. The key observation here is that many of the known laws of nature take the form of differential equations~\cite{Logan1998}; therefore, to understand the allowed kinematical symmetries, we must understand the allowed symmetries of differential equations.  Sophus Lie undertook this programme of study in the early 1870s~\cite{kastrup1984contributions}. This work, which Lie saw as a direct generalisation of Galois's earlier work on applying group theory to algebraic equations, led to his theory of continuous groups, now called \emph{Lie groups}~\cite{10.2307/987310}. Thus we find that the Lie groups provide a natural setting for describing kinematical symmetries. 
\\ \\
One of Lie's early collaborators on this project was a mathematician called Felix Klein. Interested in applying group theory, not to differential equations, but differential geometry, Klein's research diverged from Lie's, and he set out his Erlanger Programme in 1872. This programme aimed to describe a manifold using its transformation group, defining any geometric objects on the manifold by the subgroup which left the object invariant. Although the Erlanger Programme was influential, forming a uniform framework to characterise classical geometries, it was only with Élie Cartan's work on generalised spaces that the programme's connection to spacetime was made manifest~\cite{HAWKINS1984442}.  In particular, when Einstein's general theory of relativity emerged in 1915, it was shown that spacetime might be viewed as a Lorentzian manifold.  Importantly, this means that, according to Einstein, the objects in spacetime, such as electromagnetic fields, must transform under the Lorentz group.  Klein noticed that the Erlanger Programme might be related to this geometric picture of space and time~\cite{kastrup1984contributions}; however, it was Cartan, building upon the Erlanger programme, making more explicit use of Lie groups due to their relation to kinematics, that sufficiently generalised Einstein's theory.  Notably, this led him to a reformulation of Newtonian gravity in the geometric language of general relativity~\cite{cartan1923varietes, cartan1924varietes}.\footnote{At this point, it may be interesting to make the following comment.  The term \emph{kinematics} was coined by Ampère in the late 1820s with the explicit intent of merging the study of mechanics with geometry~\cite{10.1007/978-94-007-4132-4_34}.  Thus, the geometric picture of spacetime began to fall under the label of kinematics.  Klein's introduction of group theory into geometry then provided the setting for Einstein's symmetry-first approach to describing the movement of objects in a spacetime geometry.  For a discussion on the historical connection between the study of kinematics and Einstein's theory of gravity, see~\cite{martinez2009kinematics}.}  
\\ \\
Due to the overwhelming success of general relativity in describing phenomena outside the reach of classical Newtonian gravity (see~\cite{Everitt1980}), Lorentzian symmetry, and its classical, Galilean limit, were the primary kinematical symmetries considered for much of the 20th century. However, in the 1960s, people began to ask whether there may be other kinematical group choices. The first paper to attempt to classify all the possible kinematical symmetries was~\cite{doi:10.1063/1.1664490}. However, this classification imposed time-reversal and parity symmetries, which are not strictly necessary from a purely algebraic perspective. On removing these conditions, the classification was completed by Bacry and Nuyts in~\cite{doi:10.1063/1.527306}. These papers show that we can split kinematical symmetries into five classes: Lorentzian, Euclidean, Galilean, Carrollian, and Aristotelian. As mentioned above, the Lorentzian and Galilean cases are the most prevalent examples of kinematical symmetries and Euclidean symmetries, owing to their close connection with Lorentzian symmetries, frequently appear in the literature. Therefore, the novel classes were the Carrollian and Aristotelian symmetries; although it was believed that they might be ``without much physical application''. As we will argue in the next section, each class of kinematical symmetry is interesting in its own right, and we will outline a few examples of how they are used in the literature.

\section{Beyond Lorentzian Symmetries}
Following Einstein's lead in placing kinematical symmetries at the forefront of our physical theories, and assuming we have defined the correct notion of kinematical symmetry, as described in \cite{Bacry:1968zf, Bacry:1986pm, Figueroa-OFarrill:2017ycu, Figueroa-OFarrill:2017tcy, Andrzejewski:2018gmz}, we can reasonably ask,
\begin{center}
\textit{(i) how would physical objects, such as particles and electromagnetic fields, act in spacetimes described by these symmetries?, \\
(ii) do we find natural systems that are described by these symmetries?, and \\
(iii) is there any way we could extend the kinematical symmetries to other physically interesting symmetries? }
\end{center}
Systematically answering the first question is the purpose of Chapter~\ref{chap:k_spaces} in this thesis; therefore, we will defer this conversation until then.  However, we can answer the second question more succinctly here; in particular, this will be the topic of Section~\ref{subsec:intro_bls_classes}.  Furthermore, the basis for the investigations of Chapters~\ref{chap:k_superspaces} and~\ref{chap:gb_superspaces} into extending the kinematical symmetries will be reviewed in Section~\ref{subsec:intro_bls_susy}.

\subsection{Classes of Kinematical Spacetime}
\label{subsec:intro_bls_classes}
There is a growing body of literature on applications of every type of kinematical symmetry.  We will now briefly review some of the research utilising these different symmetries and their associated spacetime models.
\subsubsection{Lorentzian}
Most of the progress in 20th-century physics was made with the assumption of underlying Lorentz symmetry.  It is built into gravity theories and quantum theories, and, therefore, lies at the foundation of some of the century's most famous ideas.  There are far too many research fields influenced by Lorentz symmetry to recount here.  Therefore, we will only briefly give an account of how Lorentz symmetry appears in and impacts our understanding of the two pillars of modern physics, general relativity and quantum field theory.  This review aims to highlight the parts of these theories that may be altered by imposing one of the other types of kinematical symmetry.
\\ \\
As mentioned above, Lorentz symmetry emerged as a fundamental symmetry of spacetime through Einstein's enquiries into electromagnetism~\cite{einstein1905elektrodynamik}.  In particular, it arises from a desire to have a fixed speed of light, and impose that the laws of physics look the same in all inertial reference frames. With special relativity being built into general relativity as the local description of spacetime, the general theory of relativity thus requires the spacetime geometry to admit a Lorentzian structure.\footnote{By Lorentzian structure, we mean the following.  Let $\M$ be a $(D+1)$-dimensional real smooth manifold.  The frame bundle of $\M$ is then a principle $\GL(D+1, \RR)$ bundle over $\M$.  Let $\iota: \SO(D, 1) \rightarrow \GL(D+1, \RR)$ be the Lie group monomorphism which embeds the Lorentz group inside $\GL(D+1, \RR)$.  With this data, we may construct a principle Lorentz bundle over $\M$.  We call this reduction of the frame bundle a Lorentz structure.  Using this structure, we may define a Lorentzian metric $g$ and call $(\M, g)$ a Lorentzian manifold.  This process holds for all kinematical groups, leading to Euclidean, Galilean, Carrollian, and Aristotelian structures in the other instances.
\\ \\
Notice, this requirement of a Lorentz structure arises from the equivalence principle: the fact that locally, we should recover special relativity.  However, there is an additional principle in general relativity, which leads to another form of symmetry.  The general principle of relativity states that the laws of physics must look the same in any coordinate system.  Translated into our geometric language, this statement says that the action we write down for our gravitational theory should be invariant under diffeomorphisms.  When describing general relativity as a gauge theory, it is the group of diffeomorphisms that is presented as the gauge group, \textit{not} the Lorentz group.}
This condition on the geometry has significant consequences, which were not present in the preceding Newtonian picture of gravity. One of the critical implications of Lorentzian symmetry is that there exists a restricted subspace in spacetime with which we can be in causal contact. This means that interactions, such as gravitational forces, are no longer instantaneous in Einstein's picture.  
\\ \\
Quantum field theory is the typical language used to describe the interactions between subatomic particles; therefore, it is widely used in particle physics, atomic physics, condensed matter physics, and astrophysics~\cite{peskin2018introduction, zee2010quantum}. In this theory, fields, whose excitations define particles, propagate in a fixed, Lorentzian spacetime geometry. This propagation is described by the transformation group, or \emph{relativity} group, of the underlying spacetime, which, in a flat spacetime as described by Einstein, means that the matter must live in a module of the Poincaré group.  After quantisation, the representation of the Lorentz subgroup that acts on a particular matter field will be labelled by a (half-)integer known as the \emph{spin} of the field.  The half-integer spin fields are called fermions, and the integer spin fields are called bosons~\cite{zee2010quantum}. Bosons and fermions are then our basic building blocks for any quantum field theory we wish to write down; thus, Lorentz symmetry is built into the foundation of our ideas on quantum theory.  As we will discuss in Section~\ref{subsec:intro_bls_susy}, wishing to replace the Lorentz group with the Galilean or Carrollian groups has a profound impact on our modelling of these basic constituents of matter.  
\subsubsection{Euclidean}
In modern physics, Euclidean symmetries are frequently used as a computational tool~\cite{zee2010quantum}.  Many calculations in gravitational and quantum theories are hard in the Lorentzian case, but become tractable when the time coordinate is Wick rotated, such that we arrive at a Euclidean description \cite{osterwalder1973}.  Thus, Euclidean symmetries have been indispensable in driving research due to their close connection to the computationally less friendly Lorentz symmetries. 
\subsubsection{Galilean}
Although Galilean symmetry was historically ``superseded'' by Lorentzian symmetry, there are still numerous physical systems that admit a cleaner description when described using non-relativistic symmetries.  In particular, classical Newtonian mechanics does not require the full machinery of Lorentzian symmetry; the non-relativistic limit provides a far nicer framework for these systems.  However, there are more complex systems that make use of Galilean symmetries.  In condensed matter theory, the (fractional) quantum Hall effect admits a description as an effective field theory invariant under Galilean symmetries~\cite{Banerjee:2014pya, Son:2013rqa, Geracie:2014nka, Salgado-Rebolledo:2021wtf}.  Non-relativistic spacetimes have also been useful in extending the holographic duality beyond the AdS/CFT correspondence~\cite{Christensen:2013lma, Hartong:2016yrf, Hartong:2015zia, Bergshoeff:2014uea, Hartong:2014oma}. Recently, non-relativistic spacetimes have also been incorporated into string theory~\cite{Kluson:2018egd,Harmark:2019upf}, and supersymmetric quantum field theories~\cite{Bergshoeff:2020baa, Auzzi:2019kdd}, and have been interpreted using double field theory~\cite{Blair:2019qwi,Morand:2017fnv,Ko:2015rha,Berman:2019izh}.  They have also been systematically studied in relation to JT gravity~\cite{Grumiller:2020elf, Gomis:2020wxp} in the hope they may elucidate some outstanding problems in the search for a quantum theory of gravity.  Furthermore, numerous papers have gauged known non-relativistic algebras, or extensions thereof, to arrive at novel gravity theories~\cite{Andringa:2010it,Hartong:2015zia,Bergshoeff_2019,Bergshoeff:2014gja, Bergshoeff:2017dqq, Hansen:2018ofj, Hansen:2020pqs,Harmark:2019upf, Gomis:2018xmo,Gomis:2019nih,Bergshoeff:2020fiz,deAzcarraga:2019mdn,deAzcarraga:2002xi,Matulich:2019cdo,Papageorgiou:2009zc, Gomis:2019sqv, Bergshoeff:2019ctr, Romano:2019ulw, Concha:2020sjt,Concha:2019lhn,Aviles:2018jzw, Concha:2020ebl}.
\subsubsection{Carrollian}
These ultra-relativistic spacetimes have received increased interest over recent years due to their connection with null hypersurfaces in Lorentzian spacetimes.  Namely, the restriction of the manifold's Lorentzian structure in a $D+1$-dimensional spacetime to a $D$-dimensional null hypersurface induces a Carrollian structure on the surface.  Therefore, interesting null hypersurfaces such as black hole horizons and past and future null infinity carry a Carrollian structure by construction~\cite{Bagchi:2019clu,Ciambelli:2019lap}.  This connection has lead to exciting results regarding the physics of black hole horizons~\cite{Donnay:2019jiz} and has advanced research into flat space holography~\cite{Duval:2014uva, Ciambelli:2020ftk,Ciambelli:2020eba, Herfray:2020rvq}.
\subsubsection{Aristotelian}
The spacetime models in this class are quite distinct from the four above.  In particular, these spacetimes do not allow for a change of reference frame: we are always considering the world from one fixed coordinate system.  In the physics nomenclature, this fixed reference frame corresponds to the absence of \emph{boosts}.  By having a fixed system, Aristotelian geometries are useful for describing phenomena with a preferred reference frame or phenomena where we would like to avoid thinking about which type of boosts we will allow.  This property of being ``boost agnostic'' has recently been used to consider hydrodynamic systems~\cite{deBoer:2020xlc, Armas:2020mpr}.

\subsection{Supersymmetry}
\label{subsec:intro_bls_susy}
This classification of kinematical spacetimes provides an already fertile ground for exploration into symmetries beyond the usual Lorentzian case; however, we can still think of pushing this classification a little further.  This section will introduce supersymmetry, giving some historical context to how it first appeared in physical theories and briefly reviewing some of the outstanding problems it could help to solve.  We then describe how the kinematical symmetries may be generalised to super-kinematical symmetries.  
\\ \\
Inspired by Einstein's symmetry-first approach in building general relativity, the particle physicists of the 1920s  built their models of particle interactions by introducing a new type of symmetry~\cite{10.2307/j.ctv10vm2qt}.  Equipped with the concept of \emph{gauge} symmetry, developments in the 1960s and 1970s showed that three of the four fundamental forces, the strong, electromagnetic and weak interactions, may all be described using this language.  Thus, these forces admit a unified treatment in the Standard Model.  While this model has been incredibly successful, many physicists would like to see gravitational forces included in such a unified description.  Unfortunately, gravitational interactions, as presented by general relativity, do not have an analogous description in terms of gauge symmetry~\cite{10.2307/987310}.\footnote{Perhaps I should be more explicit here.  General relativity does not admit a description as a renormalisable Yang-Mills-type gauge theory, as the strong, electromagnetic and weak interactions do.}  Attempting to circumvent the obstructions to such a unified theory, we may look to exploit new symmetries with properties which allow them to bypass any no-go theorems, such as the Coleman-Mandula theorem~\cite{wess2020supersymmetry}.\footnote{Note, the Coleman-Mandula theorem, first presented in~\cite{PhysRev.159.1251}, states that the internal symmetries describing the particle interactions and the spacetime describing particle movement cannot be combined in a non-trivial way for a Lorentzian-relativistic quantum field theory. }  One such proposal is supersymmetry. 
\\ \\
In the Lorentzian setting, supersymmetry is a spacetime symmetry that allows us to interchange fermionic and bosonic degrees of freedom~\cite{bertolini2015lectures}.  It achieves this task by introducing a new type of symmetry generator with half-integer spin. Some of the consequences of this new type of symmetry and striking and profound.  As mentioned above, the introduction of supersymmetry allows us to construct theories which may unify gravity with the other fundamental forces.  It also provides an elegant solution to the hierarchy problem and suggests a wide range of new particles which may be used to describe dark matter.\footnote{Note, the hierarchy problem is the name given to the fact the experimental value for the mass of the Higgs boson is smaller than predicted.}  Furthermore, supersymmetric theories are often easier to analyse than their traditional counterpart; thus, supersymmetry also provides researchers with a useful test-bed for exploring new physics.\footnote{There are countless papers, books, and reviews on supersymmetry and its uses; for some of the better known instances see~\cite{Martin:1997ns, Gates:1983nr}.}
\\ \\
The first paper looking to incorporate supersymmetry into the classification of kinematical spacetimes presented above was~\cite{MR769149}. However, this paper applies the same ``by no means convincing'' assumptions of parity and time-reversal symmetries as~\cite{doi:10.1063/1.1664490}.  Other papers which extend the kinematical symmetries of~\cite{doi:10.1063/1.1664490} include~\cite{Rembielinski:1984xy} and ~\cite{Kosinski:1986ts}.  Later papers such as~\cite{Hussin_1999,CampoamorStursberg:2008hm,Huang:2014ega} tackled this problem through the process of contractions; however, to the best of this author's knowledge, the first full classification of super-kinematical symmetries and superspaces was presented in~\cite{Figueroa-OFarrill:2019ucc}.  
\\ \\
It is perhaps interesting to note at this stage that supersymmetry, beyond the Lorentzian and Euclidean cases, is not a well-explored or well-defined concept.  We notice this fact instantly by considering the symmetry generators it has in addition to the classical case.  These are defined to have half-integer spin; however, if we do not have Lorentzian or Euclidean symmetry, the concept of spin is not well-defined.  Therefore, what it means to have Galilean or Carrollian supersymmetry is \emph{a priori} unclear.  In Chapter~\ref{chap:math_prelims}, we will give one possible interpretation for these generators; namely, we will use the spin associated with the subalgebra of spatial rotation.  However, it should be noted that other researchers assume the new symmetry generators to have vanishing spin~\cite{Chapman:2015wha, Arav:2019tqm}.

\section{Outline of Thesis}
Having reviewed some critical chapters in the history of symmetry and geometry and briefly summarised why they are so fundamental to our ideas of spacetime,  we can now turn our attention towards using this knowledge to explore kinematical symmetries beyond the standard Lorentzian case.  In particular, we will build our understanding of (super-)kinematical symmetries in the following framework.  First, we will consider algebraic classifications, such as those found in~\cite{doi:10.1063/1.527306}.  These investigations will furnish us with the basic building blocks from which we may construct physical theories.  Next, we will classify spacetime (super)geometries modelled on these algebras.  We may view this geometric classification as a physical realisation of Klein's Erlanger Programme, in the spirit of Cartan.  Finally, we will investigate the geometric properties required to define physical theories on each spacetime.  The three main chapters in the body of this thesis will present the work completed towards fulfilling this framework in the kinematical, super-kinematical, and super-Bargmann instances, respectively.  In more detail, this thesis will run as follows.

\subsubsection{Chapter~\ref{chap:math_prelims}}
This chapter contains a discussion on the basic mathematical objects required in this thesis.  A graduate-level knowledge of differential geometry and algebra is assumed. The chapter is divided into three sections, algebra, geometry, and geometric properties, to align with the framework we wish to substantiate.  In Section~\ref{sec:math_prelims_alg}, we first introduce the notion of a Lie (super)algebra before defining each of the classes of Lie (super)algebra we will use in later chapters.  Section~\ref{sec:math_prelims_geo} then describes how we may integrate these Lie algebras to form Lie groups and some associated geometries.  In addition, this section defines our characterisation of a supermanifold and its relation to the corresponding Lie superalgebra and underlying classical geometry.  Finally, Section~\ref{sec:math_prelims_geo_props} provides a detailed explanation of how we will define the geometric properties of kinematical spacetimes.  

\subsubsection{Chapter~\ref{chap:k_spaces}}
This chapter gives a full substantiation of the framework set out above for the case of kinematical symmetries.  In particular, we review the algebraic classification for kinematical Lie algebras from~\cite{Figueroa-OFarrill:2017ycu, Figueroa-OFarrill:2017tcy} in Section~\ref{sec:ks_klas}, before showing how these algebras were integrated into kinematical spacetimes in Section~\ref{sec:ks_kss}.  In Section~\ref{subsec:ks_kss_gls}, there is a discussion on how the kinematical spacetimes are connected via geometric limits.  Section~\ref{sec:ks_gps} then defines the geometric properties of these spacetimes.  

\subsubsection{Chapter~\ref{chap:k_superspaces}}
In this chapter, we turn our attention to the super-kinematical case.  These symmetries have not been studied as thoroughly as the classical, kinematical symmetries; therefore, we can only present an algebraic and geometric classification.  These may be found in Sections~\ref{sec:ks_superspace_ksa} and~\ref{sec:ks_superspace_kss}, respectively. Section~\ref{sec:limits-betw-supersp} then describes how the classified superspaces may be connected via limits analogously to the kinematical spacetimes of Chapter~\ref{chap:k_spaces}. The study of the geometric properties of the resulting superspaces is left to future work.

\subsubsection{Chapter~\ref{chap:gb_superspaces}}
The last symmetries we consider are the super-Bargmann symmetries.  This case is the least studied of the three, so we can only present an algebraic classification in this instance.  The direct generalisation of the super-kinematical classification in Section~\ref{sec:ks_superspace_ksa} is found in Section~\ref{sec:gb_superspace_n1}.  Section~\ref{sec:gb_superspace_n2} then extends this classification by adding additional supersymmetric generators.  

\subsubsection{Chapter~\ref{chap:conc}}
This final chapter offers some concluding remarks, noting the possible extensions of the framework set out here, areas of further study within the framework, and highlighting some exciting ways the classified symmetries, both algebras and spacetimes, may be utilised.

\chapter{Mathematical Preliminaries} \label{chap:math_prelims}
This chapter will introduce all of the mathematical objects used to construct, describe, and explore our (super-)kinematical spacetime models.  As such, this chapter is divided into three sections.  In Section~\ref{sec:math_prelims_alg}, we begin by defining Lie algebras, gradually introducing more levels of complexity to arrive at concrete definitions for kinematical Lie algebras, generalised Bargmann algebras, kinematical Lie superalgebras, and generalised Bargmann superalgebras.  In Section~\ref{sec:math_prelims_geo}, we demonstrate how to integrate these Lie algebras to arrive at spacetime models which hold the relevant symmetries.  Finally, in Section~\ref{sec:math_prelims_geo_props}, we describe how to obtain geometric properties for kinematical spacetimes, including fundamental vector fields, soldering forms, vielbeins, and invariant connections.  Note, Einstein summation will be assumed throughout.

\section{Algebra} \label{sec:math_prelims_alg}
In this section, we will first introduce the concept of a Lie (super)algebra. This brief discussion will define many ideas that will be alluded to throughout the later chapters of this thesis, including the notion of a Lie (super)algebra, Lie (super)algebra homomorphism, Lie subalgebra and ideal. Once these basics have been reviewed, we move on to discuss kinematical Lie (super)algebras and generalised Bargmann (super)algebras. First, we discuss kinematical Lie algebras, as these are the primary objects all the other types of Lie algebra will generalise. We then discuss generalised Bargmann algebras, which may be viewed as one-dimensional abelian extensions of the underlying kinematical Lie algebra.  Kinematical Lie superalgebras are then presented as the supersymmetric extensions of the kinematical algebras.  These are the Lie superalgebras $\s = \s_{\bar{0}} \oplus \s_{\bar{1}}$, which have a kinematical Lie algebra $\k$ as $\s_{\bar{0}}$. Finally, we combine the Bargmann and supersymmetric extensions into the definition of a generalised Bargmann superalgebra. Crucially, in this section, we will present the quaternionic formalism that will play such a dominant role in Chapters~\ref{chap:k_superspaces} and~\ref{chap:gb_superspaces}.

\subsection{Lie Algebras and Lie Superalgebras} \label{subsec:math_prelims_alg_laalsa}
This section will briefly summarise some of the key definitions in Lie (super)algebra theory used throughout this thesis. The material presented here is very well established and may be found in numerous places (see \cite{warner2013foundations, helgason2001differential, kirillov2008introduction} for just a few examples). Therefore, we will state definitions, leaving any discussion or elaboration to the sources mentioned above.
\begin{definition} An $n$-dimensional real \textit{Lie algebra} $\g$ consists of an $n$-dimensional real vector space $V$ equipped with an anti-symmetric, $\mathbb{R}$-bilinear bracket
\begin{equation} \label{eq:lie_bracket_defn}
	\begin{split}
		[-, -]: V \times V &\rightarrow V \\
		(a, b) &\mapsto [a, b],
	\end{split} 
\end{equation}
which satisfies the Jacobi identity:
\begin{equation}
	[a, [b, c]] =  [[a, b], c] + [b, [a, c]] \qquad \forall a, b, c \in V.
\end{equation}
\end{definition}
Using the typical overloading of notation, we will write both the Lie algebra $\g = (V, [-, -])$ and its underlying vector space as $\g$ from now on.  Also, $\g$ and $\h$ should always be taken as Lie algebras, and we will assume we are working over $\mathbb{R}$.
\begin{definition} \label{def:la_hom}
A \textit{Lie algebra homomorphism} is an $\mathbb{R}$-linear map $f: \g \rightarrow \h$ which preserves the Lie bracket:
\begin{equation}
	f([a, b]) = [f(a), f(b)] \qquad \forall a, b, \in \g.
\end{equation}
\end{definition}
With this definition of a Lie algebra homomorphism, we note that a \textit{Lie algebra isomorphism} $i:\g \rightarrow \h$ is an injective, $\ker\, i = 0$, surjective, $\im\, i = \h$, Lie algebra homomorphism.
\begin{definition} \label{def:la_sub}
A \textit{Lie subalgebra} $\h$ is an $\mathbb{R}$-linear subspace $\h \subset \g$, such that $[\h, \h] \subset \h$.
\end{definition}
\begin{definition} \label{def:la_ideal}
An \textit{ideal} $\h \subset \g$ is an $\mathbb{R}$-linear subspace such that $[\g, \h] \subset \h$.  
\end{definition}
Let $\h \subset \g$ be an ideal and let $\g \rightarrow \g/\h$ be the canonical projection.  The vector space $\g/\h$ has a unique Lie algebra structure which makes the canonical projection a Lie algebra homomorphism. 
\\ \\
Having established some basic definitions for Lie algebras, we now turn to their supersymmetric generalisation.  
\begin{definition}
An $(m|n)$-dimensional real \textit{Lie superalgebra} $\s$ consists of an $(m|n)$-dimensional $\mathbb{Z}_2$-graded real vector space $V = V_{\bar{0}} \oplus V_{\bar{1}}$, equipped with an $\mathbb{R}$-bilinear bracket, which preserves the $\mathbb{Z}_2$ grading:
\begin{equation}
	\begin{split}
		[-, -]: V_i \times V_j &\rightarrow V_{i+j} \\
		(a, b) &\mapsto [a, b].
	\end{split} 
\end{equation}
This bracket is anti-symmetric in the super sense,
\begin{equation}
	[\lambda, \mu] = -(-1)^{ij} [\mu, \lambda],
\end{equation}
where $\lambda \in \s_i$ and $\mu \in \s_j$, and it obeys the super-Jacobi identity:
\begin{equation}
	[\lambda, [\mu, \nu]] = [[\lambda, \mu], \nu] + (-1)^{ij} [\mu, [\lambda, \nu]],
\end{equation}
where $\lambda \in \s_i$ and $\mu \in \s_j$.  
\end{definition}
As in the Lie algebra case, we will use the standard overloading of notation, calling both the Lie superalgebra $\s = (V, [-, -])$ and its underlying vector space $\s$.  For our purposes, we will usually state the dimension of the Lie superalgebra as $d = m+n$.  The definitions of Lie superalgebra homomorphism, Lie subalgebra, and ideal (in the super sense) follow \textit{mutatis mutandis} from definitions \ref{def:la_hom}, \ref{def:la_sub}, and \ref{def:la_ideal}; therefore, we will omit them here. 
\\ \\
The $\mathbb{Z}_2$ grading of the Lie superalgebra has some profound consequences, which we will exploit in Chapters~\ref{chap:k_superspaces} and \ref{chap:gb_superspaces}.  In particular, it states that $\s_{\bar{0}}$ must be a Lie subalgebra of $\s$, and $\s_1$ must be an $\s_{\bar{0}}$ module under the adjoint action.

\subsection{Kinematical Lie Algebras} \label{subsec:math_prelims_alg_kla}
Having established some basic definitions, we now define the primary object that our later investigations will centre on.
\begin{definition}
A \textit{kinematical Lie algebra} (KLA) $\k$ in $D$ spatial dimensions is a $\tfrac12 (D+1)(D+2)$-dimensional real Lie algebra containing a rotational subalgebra $\r$ isomorphic to $\so(D)$ such that, under the adjoint action of $\r$, it decomposes as
\begin{equation}
\k = \r \,\oplus\, 2V \,\oplus\, \mathbb{R},
\end{equation}
where $V$ is a $D$-dimensional $\so(D)$ vector module and $\mathbb{R}$ is a one-dimensional $\so(D)$ scalar module.
\end{definition}
We will denote the real basis for these Lie algebras as $\{\bJ_{ij}, \bB_i, \bP_i, \bH\}$, where $\bJ_{ij}$ are the generators for the subalgebra $\r$, $\bB_i$ and $\bP_i$ span our two copies of $V$, and $\bH$ spans the $\so(D)$ scalar module.  Implicit in this characterisation of kinematical Lie algebras is the assumption of space isotropy, which implies that all the generators transform as expected under the spatial rotations:
\begin{equation}\label{eq:kinematical_brackets_general}
	\begin{split}
	[\bJ_{ij}, \bJ_{kl}] &= \delta_{jk} \bJ_{il} - \delta_{ik} \bJ_{jl} - \delta_{jl} \bJ_{ik} + \delta_{il} \bJ_{jk}, \\ 			[\bJ_{ij}, \bB_k] &= \delta_{jk} \bB_{i} - \delta_{ik} \bB_j, \\
	[\bJ_{ij}, \bP_k] &= \delta_{jk} \bP_{i} - \delta_{ik} \bP_j, \\ 
	[\bJ_{ij}, \bH] &= 0.
	\end{split}
\end{equation}
Note that in $D$ spatial dimensions, we always have the $\so(D)$ invariant tensors $\delta_{ij}$ and $\epsilon_{i_1 i_2 \ldots i_D}$ with which we can define our structure constants.  Therefore, when $D=3$, we may write the above expressions more concisely.  Namely, we may define an $\so(3)$ vector module $\bJ_i$ through $\bJ_{jk} = - \epsilon_{ijk} \bJ_i$, and write
\begin{equation} \label{eq:kinematical_brackets_D3}
	[\bJ_i, \bJ_j] = \epsilon_{ijk} \bJ_k, \quad [\bJ_i, \bB_j] = \epsilon_{ijk} \bB_k, \quad [\bJ_i, \bP_j] = \epsilon_{ijk} \bP_k,
	\quad [\bJ_i, \bH] = 0.
\end{equation}
Throughout this thesis, we will frequently use the following abbreviated notation.
\begin{gather}
		[\bJ_{ij}, \bB_k] = \delta_{jk} \bB_{i} - \delta_{ik} \bB_j \quad \text{is equivalent to} \quad  [\J, \B] = \B, \nonumber\\
		[\bH, \bB_i] = \bP_i \quad \text{is equivalent to} \quad [\bH, \B] = \P, \\
		[\bB_i, \bP_j] = \delta_{ij} \bH + \bJ_{ij} \quad \text{is equivalent to} \quad [\B, \P] = \bH + \J \nonumber ,
\end{gather}
\textit{et cetera}. 

\subsection{Aristotelian Lie Algebras} \label{subsec:math_prelims_alg_ala}
An Aristotelian Lie algebra (ALA) $\a$ in $D$ spatial dimensions may be thought of as a kinematical Lie algebra $\k$ with only one copy of the $\so(D)$ vector module $V$.  More explicitly, we have the following definition.
\begin{definition}
An \textit{Aristotelian Lie algebra} (ALA) $\a$ in $D$ spatial dimensions is a $(\tfrac12 D (D-1) + D + 1)$-dimensional real Lie algebra containing a rotational subalgebra $\r$ isomorphic to $\so(D)$ such that, under the adjoint action of $\r$, it decomposes as 
\begin{equation}
\a = \r \,\oplus\, V \,\oplus\, \mathbb{R}, 
\end{equation}
where $V$ is a $D$-dimensional $\so(D)$ vector module and $\mathbb{R}$ is a one-dimensional $\so(D)$ scalar module.
\end{definition}
We will denote the real basis for these Lie algebras as $\{\bJ_{ij}, \bP_i, \bH\}$, where $\bJ_{ij}$ are the generators for the subalgebra $\r$, $\bP_i$ span our copy of $V$, and $\bH$ spans the $\so(D)$ scalar module.  Implicit in this characterisation of Aristotelian Lie algebras is the assumption of space isotropy, which implies that all the generators transform as expected under the spatial rotations:
\begin{equation}\label{eq:aristotelian_brackets_general}
	\begin{split}
	[\bJ_{ij}, \bJ_{kl}] &= \delta_{jk} \bJ_{il} - \delta_{ik} \bJ_{jl} - \delta_{jl} \bJ_{ik} + \delta_{il} \bJ_{jk}, \\ 	
	[\bJ_{ij}, \bP_k] &= \delta_{jk} \bP_{i} - \delta_{ik} \bP_j, \\ 
	[\bJ_{ij}, \bH] &= 0.
	\end{split}
\end{equation}
Since Aristotelian Lie algebras are so similar to kinematical Lie algebras, we will frequently use the term kinematical Lie algebra when referring to both.

\subsection{Generalised Bargmann Algebras} \label{subsec:math_prelims_alg_gba}
A generalised Bargmann algebra (GBA) $\hat{\k}$ in $D$ spatial dimensions may be thought of as a real one-dimensional abelian extension of a kinematical Lie algebra $\k$.  Therefore, we may think of the generalised Bargmann algebras as sitting in short exact sequences
\begin{equation}
	0 \rightarrow \mathbb{R} \rightarrow \hat{\k} \rightarrow \k \rightarrow 0.
\end{equation}
More explicitly, we have the following definition.
\begin{definition}
A \textit{generalised Bargmann algebra} (GBA) $\hat{\k}$ in $D$ spatial dimensions is a $(\tfrac12 (D+1)(D+2) + 1)$-dimensional real Lie algebra containing a rotational subalgebra $\r$ isomorphic to $\so(D)$ such that, under the adjoint action of $\r$, it decomposes as 
\begin{equation}
\hat{\k} = \k \oplus \mathbb{R} = \r \,\oplus\, 2V \,\oplus\, 2\mathbb{R}, 
\end{equation}
where $V$ is a $D$-dimensional $\so(D)$ vector module and $\mathbb{R}$ is a one-dimensional $\so(D)$ scalar module.  Additionally, we require the GBA to have the following bracket.  Denote the basis for the vector modules as $X_i$ and $Y_i$, where $1 \leq i \leq D$, and let one of the scalar modules have a basis element $Z$.  The required bracket is then
\begin{equation}
	[X_i, Y_j] = \delta_{ij} Z.
\end{equation}
\end{definition}
As in Section~\ref{subsec:math_prelims_alg_kla}, we will denote the real basis for the kinematical Lie algebra $\k$ as $\{\bJ_{ij}, \bB_i, \bP_i, \bH\}$, where $\bJ_{ij}$ are the generators for the subalgebra $\r$, $\bB_i$ and $\bP_i$ span our two copies of $V$, and $\bH$ spans the $\so(D)$ scalar module.  We will choose to denote the generator of the one-dimensional extension as $\bZ$.  Given this definition, we inherit the kinematical brackets of \eqref{eq:kinematical_brackets_general}; however, for completeness, we will restate them here alongside the new brackets brought about by the introduction of $\bZ$.
\begin{equation}\label{eq:bargmann_brackets_general}
	\begin{split}
	[\bJ_{ij}, \bJ_{kl}] &= \delta_{jk} \bJ_{il} - \delta_{ik} \bJ_{jl} - \delta_{jl} \bJ_{ik} + \delta_{il} \bJ_{jk}, \\ 
	[\bJ_{ij}, \bB_k] &= \delta_{jk} \bB_{i} - \delta_{ik} \bB_j, \\ 
	[\bJ_{ij}, \bP_k] &= \delta_{jk} \bP_{i} - \delta_{ik} \bP_j, \\ 
	[\bJ_{ij}, \bH] &= 0,
	\end{split} \quad
	\begin{split}
	[\bJ_{ij}, \bZ] &= 0, \\
	[\bB_i, \bP_j] &= \delta_{ij} \bZ.
	\end{split}
\end{equation}
The brackets of these algebras may also be summarised using the abbreviated notation introduced in Section~\ref{subsec:math_prelims_alg_kla}:
\begin{equation}
	[\bB_i, \bP_j] = \delta_{ij} \bZ \quad \text{is equivalent to} \quad [\B, \P] = \bZ.
\end{equation}

\subsection{Kinematical Lie Superalgebras} \label{subsec:math_prelims_alg_klsa}
An $\N$-extended kinematical Lie superalgebra (KLSA) $\s$ in three spatial dimensions is a real Lie superalgebra $\s = \s_{\bar{0}} \oplus \s_{\bar{1}}$, such that $\s_{\bar{0}} = \k$ is a kinematical Lie algebra for which $D=3$, and $\s_{\bar{1}}$ consists of $\N$ copies of $S$, the real four-dimensional spinor module of the rotational subalgebra $\r \cong \so(3)$.  More explicitly, we have the following definition.
\begin{definition}
An \textit{$\N$-extended kinematical Lie superalgebra} (KLSA) $\s$ in three spatial dimensions is a $(10 + 4\N)$-dimensional real Lie superalgebra containing a rotational subalgebra $\r$ isomorphic to $\so(3)$ such that, under the adjoint action of $\r$, it decomposes as
\begin{equation}
\s = \s_{\bar{0}} \oplus \s_{\bar{1}} = \r \,\oplus\, 2V \,\oplus\, \mathbb{R} \,\oplus\, \N S,
\end{equation}
where $V$ is the three-dimensional $\so(3)$ vector module, $\mathbb{R}$ is a one-dimensional $\so(3)$ scalar module, and $S$ is the four-dimensional $\so(3)$ spinor module. 
\end{definition}
As in Section~\ref{subsec:math_prelims_alg_kla}, we will denote the real basis for the kinematical Lie algebra $\s_{\bar{0}} = \k$ as $\{\bJ_i, \bB_i, \bP_i, \bH\}$, where $\bJ_i$ are the generators for the subalgebra $\r$, $\bB_i$ and $\bP_i$ span our two copies of $V$, and $\bH$ spans the $\so(D)$ scalar module.\footnote{Since we are in the special case of $D=3$, we will adopt $\bJ_i$, defined by $\bJ_{jk} = - \epsilon_{ijk} \bJ_i$, from the outset.}  We will choose to denote generators of $\s_{\bar{1}}$ as $\{ \bQ^A_a \}$, where $1 \leq A \leq \N$ and $1 \leq a \leq 4$.  Given this definition, we inherit the kinematical brackets of \eqref{eq:kinematical_brackets_D3}; however, for completeness, we will restate them here alongside the new brackets brought about by the introduction of $\{ \bQ^A_a \}$.
\begin{equation}\label{eq:superkinematical_brackets_general}
	\begin{split}
		[\bJ_i, \bJ_j] &= \epsilon_{ijk} \bJ_k, \\
		[\bJ_i, \bB_j] &= \epsilon_{ijk} \bB_k, \\
		[\bJ_i, \bP_j] &= \epsilon_{ijk} \bP_k, \\
		[\bJ_i, \bH] &= 0,
	\end{split} \quad \quad 
	\begin{split}
	[\bJ_i, \bQ^A_a] &= -\tfrac12 \delta^A_B \bQ^B_b \tensor{\Gamma}{_i^b_a},
	\end{split}
\end{equation}
where we can define $\Gamma_i$ using the Pauli matrices $\sigma_i$ as
\begin{equation}
	\Gamma_1 = \begin{pmatrix}
		i\sigma_2 & \zero \\
		\zero & i\sigma_2
	\end{pmatrix}, \quad \Gamma_2 = \begin{pmatrix}
		\zero & \sigma_3 \\
		-\sigma_3 & \zero
	\end{pmatrix}, \quad \text{and} \quad 
	\Gamma_3 = \begin{pmatrix}
		\zero & \sigma_1 \\
		-\sigma_1 & 0
	\end{pmatrix}.
\end{equation}
For now, the only other requirement we will state is that we want to focus on the instances where $[\Q, \Q] \neq 0$. 
The brackets of these algebras may also be summarised using the abbreviated notation introduced in Section~\ref{subsec:math_prelims_alg_kla}:
\begin{equation}
	[\bJ_i, \bQ^A_a] = \delta^A_B \bQ^B_b \tensor{\Gamma}{_i^b_a} \quad \text{is equivalent to} \quad [\J, \Q] = \Q.
\end{equation}
\paragraph{Quaternionic Notation}  In Chapter~\ref{chap:k_superspaces}, when discussing kinematical Lie superalgebras, we will make extensive use of the following quaternionic formalism.  Notice that $\r \cong \so(3) \cong \sp(1) \cong  \Im\,\mathbb{H}$, where $\mathbb{H}$ denotes the quaternions.  We let $\ii, \jj,$ and $\kk$ be the quaternion units such that $\ii \jj = \kk$ and $\jj \ii = -\kk$.  Under the isomorphism $\r \cong \sp(1)$, $V$ may be described as a copy of $\Im\,\mathbb{H}$ and $S$ may be described as a copy of $\mathbb{H}$ where, in both instances, $\r$ acts via left quaternion multiplication.  Using this isomorphism, we may rewrite the brackets in \eqref{eq:superkinematical_brackets_general} by invoking the following injective $\mathbb{R}$-linear maps:\footnote{It may be important to note at this stage that $\theta = \theta_4 + \theta_1 \ii + \theta_2 \jj + \theta_3 \kk$ is just a quaternion with real components $\theta_i$, there are no Grassmann variables.  We have used the $\N=1$ case as an example here, and, as written in the text, we've introduced a new index $\alpha = (A, a)$ to capture the basis for the individual quaternions $a$ and the number of quaternionic directions $A$.  We will never use these indices explicitly, so perhaps these comments are not important, but introducing new indices, even for one line, without explaining them seems impolite.  }
\begin{equation} \label{eq:quat-basis-s}
	\begin{split}
	\sJ: \Im(\mathbb{H}) \rightarrow \s_{\bar{0}} \quad &\text{such that} \quad \sJ(\omega) = \omega_i \bJ_i 
		\quad \text{where} \quad \omega = \omega_1 \ii + \omega_2 \jj + \omega_3 \kk \in \Im(\mathbb{H}), \\
	\sB: \Im(\mathbb{H}) \rightarrow \s_{\bar{0}} \quad &\text{such that} \quad \sB(\beta) = \beta_i \bB_i 
		\quad \text{where} \quad \beta = \beta_1 \ii + \beta_2 \jj + \beta_3 \kk \in \Im(\mathbb{H}), \\
	\sP: \Im(\mathbb{H}) \rightarrow \s_{\bar{0}} \quad &\text{such that} \quad \sP(\pi) = \pi_i \bP_i 
		\quad \text{where} \quad \pi = \pi_1 \ii + \pi_2 \jj + \pi_3 \kk \in \Im(\mathbb{H}), \\	
	\sQ: \mathbb{H}^{\N} \rightarrow \s_{\bar{1}} \quad &\text{such that} \quad \sQ(\theta) = \theta_{\alpha} \bQ_{\alpha}
	 	\quad \text{where} \quad \theta  \in \mathbb{H}^{\N}.
	\end{split}			
\end{equation}
With these maps defined, the brackets of \eqref{eq:superkinematical_brackets_general} become\footnote{Solely to keep the notation consistent, when referring to the $\so(3)$ scalar module basis element $\bH$ in this formalism, we will use $\sH$.  Therefore, if we consider $\sJ(\omega) = \omega_i \bJ_i$ to be the map between $\sJ$ and $\bJ$, the map between $\sH$ and $\bH$ is $\sH = \bH$. }
\begin{equation}
	\begin{split}
	[\J, \J] = \J \quad &\implies \quad [\sJ(\omega), \sJ(\omega')] = \tfrac12 \sJ([\omega, \omega']), \\
	[\J, \B] = \B \quad &\implies \quad [\sJ(\omega), \sB(\beta)] = \tfrac12 \sB([\omega, \beta]), \\
	[\J, \P] = \P \quad &\implies \quad [\sJ(\omega), \sP(\pi)] = \tfrac12 \sP([\omega, \pi]), \\
	[\J, \bH] = 0 \quad &\implies \quad [\sJ(\omega), \sH] = 0, 
	\end{split} \quad
	\begin{split}
	[\J, \Q] = \Q \quad &\implies \quad [\sJ(\omega), \sQ(\theta)] = \tfrac12 \sQ(\omega\theta),
	\end{split}
\end{equation}
where $\omega, \beta, \pi \in \Im(\mathbb{H})$, $\theta \in \mathbb{H}^{\N}$, $[\omega, \beta] := \omega\beta - \beta\omega$, and $\omega\beta$ is given by quaternion multiplication.

\subsection{Generalised Bargmann Superalgebras} \label{subsec:math_prelims_alg_gbsa}
An $\N$-extended generalised Bargmann superalgebra (GBSA) $\hat{s}$ in three spatial dimensions is a real Lie superalgebra $\s = \s_{\bar{0}} \oplus \s_{\bar{1}}$, such that $\s_{\bar{0}} = \hat{\k}$ is a generalised Bargmann algebra for which $D=3$, and $\s_{\bar{1}}$ consists of $\N$ copies of $S$, the real four-dimensional spinor module of the rotational subalgebra $\r \cong \so(3)$.  More explicitly, we have the following definition.
\begin{definition}
An \textit{$\N$-extended generalised Bargmann superalgebra} (GBSA) $\hat{s}$ in three spatial dimensions is a $(11 + 4\N)$-dimensional real Lie algebra containing a rotational subalgebra $\r$ isomorphic to $\so(3)$ such that, under the adjoint action of $\r$, it decomposes as 
\begin{equation}
	\s = \s_{\bar{0}} \oplus \s_{\bar{1}} = \r \,\oplus\, 2V \,\oplus\, 2 \mathbb{R} \,\oplus\, \N S,
\end{equation}
where $V$ is the three-dimensional $\so(3)$ vector module, $\mathbb{R}$ is a one-dimensional $\so(3)$ scalar module, and $S$ is the four-dimensional $\so(3)$ spinor module.
\end{definition}
As these algebras combine the kinematical Lie algebras' two previous generalisations, we will inherit all the notation we have already established..  In particular, we will denote the real basis for the underlying kinematical Lie algebra $\k$ as $\{\bJ_i, \bB_i, \bP_i, \bH\}$, where $\bJ_i$ are the generators for the subalgebra $\r$, $\bB_i$ and $\bP_i$ span our two copies of $V$, and $\bH$ spans the $\so(D)$ scalar module.\footnote{As in Section~\ref{subsec:math_prelims_alg_klsa}, because we are already in the special case of $D=3$, we will use the generator $\bJ_i$, defined by $\bJ_{jk} = - \epsilon_{ijk} \bJ_i$, from the outset.}  We will choose to denote the generator of the one-dimensional, Bargmann extension as $\bZ$, and we label the generators of $\s_{\bar{1}}$ as $\{ \bQ^A_a \}$.  The brackets characterising our generalised Bargmann superalgebra $\hat{\s}$ are then 
\begin{equation}\label{eq:bargmann_brackets_general}
	\begin{split}
		[\bJ_i, \bJ_j] &= \epsilon_{ijk} \bJ_k, \\
		[\bJ_i, \bB_j] &= \epsilon_{ijk} \bB_k, \\
		[\bJ_i, \bP_j] &= \epsilon_{ijk} \bP_k, \\
		[\bJ_i, \bH] &= 0,
	\end{split} \quad \quad 
	\begin{split}
	[\bJ_i, \bQ^A_a] &= \delta^A_B \bQ^B_b \tensor{\Gamma}{_i^b_a}, \\
	[\bJ_i, \bZ] &= 0, \\
	[\bB_i, \bP_j] &= \delta_{ij} \bZ,
	\end{split}
\end{equation}
where, in addition, we want to impose the condition $[\Q, \Q] \neq 0$.  We will also use the abbreviated notation set up in Sections~\ref{subsec:math_prelims_alg_kla},~\ref{subsec:math_prelims_alg_gba}, and ~\ref{subsec:math_prelims_alg_klsa}.
\paragraph{$D=3$ Quaternionic Formalism} In Chapter~\ref{chap:gb_superspaces}, when we discuss generalised Bargmann superalgebras in more detail, we will make extensive use of the quaternionic formalism first considered in Section~\ref{subsec:math_prelims_alg_klsa}.  In particular, we would like to extend this formalism to the case of generalised Bargmann algebras.   This is a straightforward procedure.  Since $\bZ$ is an $\so(3)$ scalar, the brackets of the generalised Bargmann superalgebras become\footnote{As in the kinematical Lie algebra case, to keep the notation consistent in this formalism the $\so(3)$ scalars $\bH$ and $\bZ$ will be denoted $\sH$ and $\sZ$, respectively.}
\begin{equation}
	\begin{split}
	[\J, \J] = \J \quad &\implies \quad [\sJ(\omega), \sJ(\omega')] = \tfrac12 \sJ([\omega, \omega']), \\
	[\J, \B] = \B \quad &\implies \quad [\sJ(\omega), \sB(\beta)] = \tfrac12 \sB([\omega, \beta]), \\
	[\J, \P] = \P \quad &\implies \quad [\sJ(\omega), \sP(\pi)] = \tfrac12 \sP([\omega, \pi]) ,\\
	[\J, \bH] = 0 \quad &\implies \quad [\sJ(\omega), \sH] = 0, \\
	\end{split} \quad 
	\begin{split}
	[\J, \Q] = \Q \quad &\implies \quad [\sJ(\omega), \sQ(\theta)] = \tfrac12 \sQ(\omega\theta), \\
	[\J, \bZ] = 0 \quad &\implies \quad [\sJ(\omega), \sZ] = 0,  \\
	[\B, \P] = \bZ \quad &\implies \quad [\sB(\beta), \sP(\pi)] = \Re(\bar{\beta}\pi) \sZ.
	\end{split}
\end{equation}

\section{Geometry} \label{sec:math_prelims_geo}
This section will provide the necessary definitions and discussions to understand how the different types of Lie algebra, introduced in Section~\ref{sec:math_prelims_alg}, can be integrated to describe spacetime geometries.  In particular, we will arrive at a homogeneous space description for the kinematical spacetimes and a homogeneous supermanifold description for the kinematical superspaces.  The fact we arrive at such a description is an artefact of our choice of approach.  Klein's Erlanger Programme uses a homogeneous space description of geometry; therefore, we arrive at homogeneous spacetimes by applying this programme to the kinematical story.  
\\ \\
A useful result, which will be utilised when defining and exploring the kinematical spacetimes, is an association between homogeneous spaces with respect to a Lie group $\Ggr$ and coset spaces $\Ggr/\Hgr$, where $\Hgr \subset \Ggr$ is a Lie subgroup.  To arrive at the desired definition of a kinematical spacetime and define its association to a corresponding kinematical Lie algebra, we will take the following path.  First, we will cover some standard material concerning Lie groups, coset spaces, and homogeneous spaces and how they relate to Lie algebras.  Although we may find this content in numerous places (see \cite{warner2013foundations, helgason2001differential, kirillov2008introduction, sharpe2000differential} for just a few examples), we include it here for completeness.  Additionally, this material allows us to establish the notation we will use throughout the rest of this thesis.  Anticipating the requirement to specify geometric objects from algebraic objects, we will also define exponential coordinates, which will prove useful for this purpose.  Finally, we introduce supermanifolds and Lie supergroups, and show how we may construct superisations of our kinematical spacetimes.

\subsection{Lie Groups and their Lie Algebras} \label{subsec:math_prelims_geo_lgahs}
This section will briefly summarise some key definitions in Lie (super)group theory used throughout this thesis. The material presented here is very well established and may be found in numerous places (see \cite{warner2013foundations, helgason2001differential, kirillov2008introduction} for just a few examples). Therefore, we will state definitions, leaving any discussion or elaboration to the sources mentioned above.
\\ \\
\begin{definition} An $n$-dimensional real \textit{Lie group} $\Ggr$ is a group endowed with the structure of a real $n$-dimensional $\eE$ (smooth) manifold, such that both the inversion map 
\begin{equation}
	\begin{split}
		s: \Ggr &\rightarrow \Ggr \\
		g &\mapsto g^{-1},
	\end{split}
\end{equation}
and multiplication map
\begin{equation}
	\begin{split}
		m: \Ggr \times \Ggr &\rightarrow \Ggr \\
		(g, h) &\mapsto gh,
	\end{split} 
\end{equation}
are smooth.
\end{definition}
From now on, $\Ggr$ and $\Hgr$ should always be taken as Lie groups, and we will assume we are working over $\mathbb{R}$.
\begin{definition}
A \textit{Lie group homomorphism} $f: \Ggr \rightarrow \Hgr$ is a smooth map that is also a group homomorphism.
\end{definition}
We can now demonstrate the relationship between a Lie group $\Ggr$ and its associated Lie algebra. In particular, we will see that the left-invariant vector fields of $\Ggr$ form a Lie algebra, which we will call $\g$.  Before proceeding to define the left-invariant vector fields of $\Ggr$, we first require the following definition.
\begin{definition}
Let $\varphi: \Ggr \rightarrow \Hgr$ be a Lie group homomorphism and $X \in \eX(\Ggr)$.  We will denote the tangent vector produced by $X$ acting on $p \in \Ggr$ as $X_p$. We define
\begin{equation}
	\begin{split}
		d\varphi : T\Ggr &\rightarrow T\Hgr \\
			(p, X_p) &\mapsto d\varphi (X_p).
	\end{split}	
\end{equation}
The vector field $X$ is then \textit{$\varphi$-related} to $Y \in \eX(\Hgr)$ if\footnote{This expression may also be written as $d\varphi\circ X = Y\circ\varphi$ if we do not want to make explicit reference to $p \in \Ggr$. }
\begin{equation}
	d\varphi (X_p) = Y_{\varphi(p)} \quad \quad \forall p \in \Ggr.
\end{equation}
For our purposes, it is worth noting that when acting on a smooth function $f \in C^{\infty}(\Hgr)$, the above expression becomes\footnote{As above, it is also useful to note that this expression may be written as $d\varphi\circ X (f) = Y (f)\circ\varphi$ if we do not want to make explicit use of $p \in \Ggr$.}
\begin{equation}
	d\varphi (X_p) (f) = X_p (f\circ\varphi) = Y_{\varphi(p)} (f).
\end{equation}
\end{definition}
Using the multiplication map on $\Ggr$, we can define a left translation by $g \in \Ggr$ as 
\begin{equation}
	\begin{split}
	L_g : \Ggr &\rightarrow \Ggr \\
		h &\mapsto m(g, h) = g h.
	\end{split}
\end{equation}
\begin{definition}
Let $X \in \eX(\Ggr)$.  Then $X$ is called a \textit{left-invariant vector field} if it is $L_g$-related to itself for all $g \in \Ggr$:
\begin{equation}
	dL_{g}\circ X = X\circ L_{g}.
\end{equation}
\end{definition}
We will denote the space of left-invariant vector fields on $\Ggr$ as $\eX(\Ggr)^L$.  With this definition, we notice that each left-invariant vector field $X$ is uniquely defined by its value at the identity $e \in \Ggr$.  Therefore, we have an isomorphism $T_e\Ggr \cong \eX(\Ggr)^L$.  An important consequence of this isomorphism is that $\eX(\Ggr)^L$ takes the form of a real vector space with the same dimension as the Lie group, $\Ggr$.
\\ \\
For any smooth manifold $\M$, there exists a commutator defined on the space of vector fields.  This object may be defined as follows.
\begin{definition}
Let $\M$ be a $\eE$ manifold and $\eX(\M)$ be its set of vector fields.  The \textit{commutator} is an anti-symmetric $\mathbb{R}$-bilinear map, defined
\begin{equation}
	\begin{split}
		[ - , - ] : \eX(\M) \times \eX(\M) &\rightarrow \eX(\M) \\
		(X, Y) &\mapsto [X, Y],
	\end{split}
\end{equation}
which satisfies the Jacobi identity
\begin{equation}
	[X, [Y, Z]] = [[X, Y], Z] + [Y, [X, Z]] \quad \quad \forall X, Y, Z \in \eX(\M),
\end{equation}
and acts on smooth functions of the manifold as
\begin{equation}
	[X, Y] (f) = X(Y(f)) - Y(X(f)) \quad \quad \forall f \in \eE(\M).
\end{equation}
\end{definition}
Notice, this definition is almost identical to that of a Lie bracket, given in \eqref{eq:lie_bracket_defn}.  Indeed, we will now show that the commutator of two left-invariant vector fields on $\Ggr$ is itself a left-invariant vector field.  Thus, the commutator restricts from $\eX(\Ggr)$ to $\eX(\Ggr)^L$.  Combining this result with the fact $\eX(\Ggr)^L$ forms a $\dim(\Ggr)$-dimensional real vector space, we notice that $(\eX(\Ggr)^L, [-, -])$ forms a Lie algebra associated to $\Ggr$, which we call $\g$.
\\ \\
Proving that the commutator restricts to $\eX(\Ggr)^L$ is perhaps best seen by first considering the more general setting of $\varphi$-related vector fields.  Explicitly, let $X$ and $X'$ be $\varphi$-related to $Y$ and $Y'$, and $\varphi: \Ggr \rightarrow \Hgr$ be a Lie group homomorphism.  If we can show that $[X, X']$ is $\varphi$-related to $[Y, Y']$, then we arrive at the desired result by setting $\varphi = L_g$,  $Y = X$, and $Y' = X'$.  Therefore, we want to show
\begin{equation}
	d\varphi ([X, X']_p) (f) = [Y, Y']_{\varphi(p)} (f),
\end{equation}
where $p \in \Ggr$ and $f \in \eE(\Hgr)$.  Taking the left-hand side (L.H.S), we have
\begin{equation}
	\begin{split}
		d\varphi ([X, X']_p) (f) &= [X, X']_p (f\circ\varphi) \\
		&= X_p (X'(f\circ\varphi)) - X'_p (X(f\circ\varphi)) \\
		&= X_p (d\varphi\circ X' (f)) - X'_p(d\varphi\circ X (f)) \\
		&= X_p (Y' (f)\circ\varphi) - X'_p (Y(f)\circ\varphi) \\
		&= d\varphi (X_p) (Y' (f)) - d\varphi (X'_p) (Y(f)) \\
		&=  Y_{\varphi(p)} (Y'(f)) - Y'_{\varphi(p)}(Y(f)) \\
		&= [Y, Y']_{\varphi(p)} (f).
	\end{split}
\end{equation} 
Thus, we find the desired result.  Setting $\varphi = L_g$, $Y = X$, and $Y' = X'$,  the commutator restricts to the Lie bracket on the $\dim(\Ggr)$-dimensional real vector space of left-invariant vector fields, giving the Lie algebra $\g$.

\subsection{Exponential Coordinates on a Lie Group} \label{math_prelims_exp_coords}
The above story demonstrates how we may recover a Lie algebra from a Lie group; however, our interests will lie in determining Lie groups' geometric properties based on their Lie algebras.  Therefore, we need to identify a method of utilising our Lie algebra knowledge to explore an associated Lie group.  As there may be several Lie groups with the same Lie algebra, we want to clarify from the outset which Lie groups we will be discussing.  It is a well-known result that their exists a unique (up to isomorphism) connected, simply-connected Lie group $\Ggr$ such that $\Lie(\Ggr) =\g$, and that all other connected Lie groups $\Ggr'$ with $\Lie(\Ggr')=\g$ are discrete quotients $\Ggr' = \Ggr/Z$, where $Z \subset \Ggr$ is a discrete, central subgroup \cite{kirillov2008introduction}.  Therefore, to make the mapping between $\g$ and $\Ggr$ unique, we will always consider the simply-connected Lie groups.   
\\ \\
Anticipating the need to identify geometric objects on $\Ggr$ from algebraic objects on $\Lie(\Ggr) = \g$, we begin by showing how we will construct a local chart $(U, \sigma^{-1})$ around any point $o \in \Ggr$.  This task may be achieved using the exponential map $\exp:\g \rightarrow \Ggr$; however, to understand how this map is being used, it helps to first see the following.  
\\ \\
Consider a curve $\gamma: \mathbb{R} \rightarrow \Ggr$ such that $\gamma(0) = e$, where $e \in \Ggr$ is the identity element.  In particular, we will define this curve such that it is a Lie group homomorphism: $\gamma(s+t) = \gamma(s)\gamma(t)$.  Taking the derivative of this curve, we acquire $d\gamma: T\mathbb{R} \rightarrow T\Ggr$.  Notice that, at the identity, the derivative gives us a map $\mathbb{R} \rightarrow \g$.  Letting the sole basis element of $\mathbb{R}$ map to one of the basis elements of $\g$, let us call it $X$, it can show that $\gamma$ is the unique integral curve for the chosen left-invariant vector field \cite{helgason2001differential}.  Since the curve is a Lie group homomorphism, and is unique to a particular Lie algebra element $X$, the image of $\gamma$ is called the \textit{one-parameter subgroup generated by $X$}.
\\ \\
The above story allowed us to take a single Lie algebra element and uniquely determine a Lie group structure associated with it.  Using this knowledge, we can determine our map $\g \rightarrow \Ggr$.
\begin{definition}
Let $X \in \g$ and $\gamma_X$ be its unique integral curve. The \textit{exponential map} is a smooth map defined
\begin{equation}
	\begin{split}
		\exp: \g &\rightarrow \Ggr \\
			X &\mapsto \exp(X) = \gamma_X(1).
	\end{split}
\end{equation}
\end{definition}
With this definition, the one-parameter subgroup generated by $X$ can be understood as having the group multiplication
\begin{equation}
	\exp( (s+t) X) = \exp( sX) \exp(tX) \quad \quad \forall s, t, \in \mathbb{R}.
\end{equation}
Now, choosing a point $o \in \Ggr$, we can establish a local chart using the exponential map.  Let $\exp_o: \g \rightarrow \Ggr$, such that
\begin{equation}
	\exp_o(X) = \exp(X) \, o \quad \quad \forall X \in \g.
\end{equation}
This map defines a local diffeomorphism from a neighbourhood $V$ of $(0,0,\ldots,0) \in \g$ and a neighbourhood $U$ of $o \in \Ggr$. Choosing a basis $\{e_i\}$ for $\g$, where $1 \leq i \leq n$, we can define $\sigma: \mathbb{R}^{n} \rightarrow \Ggr$, where $\sigma(\boldsymbol{c}) = \exp_o( c_i e_i )$.  We now have a local coordinate chart $(U, \sigma^{-1})$, such that $o \in U \subset \Ggr$ maps to $(0, 0, \ldots, 0) \in V \subset \g$.  We can move this chart around $\Ggr$ by using the group multiplication to change the origin, giving us an atlas for $\Ggr$. Later in this section, it will be shown how we  may utilise this method of producing coordinates on a Lie group to give us spacetime coordinates. 

\subsection{Coset Spaces} \label{subsec:math_prelims_geo_cs}
Having established the definition of a Lie group $\Ggr$, shown its connection to an associated Lie algebra $\g$, and taken some first steps towards defining geometric objects on $\Ggr$ from algebraic objects on $\g$,  we now turn to the important topic of defining coset spaces. However, before defining a coset space, we need the concept of a Lie subgroup. 
\begin{definition} A Lie subgroup $\Hgr \subset \Ggr$ is a submanifold such that $\Hgr$ is also a subgroup.
\end{definition}
For our purposes, we will take submanifold to mean that there exists a closed embedding $\phi: \Hgr \rightarrow \Ggr$.  That is
\begin{enumerate}
	\item $d \phi: T\Hgr \rightarrow T\Ggr$ is globally injective,
	\item $\phi$ is a homeomorphism, and
	\item $\phi(\Hgr) \subset \Ggr$ is closed.
\end{enumerate}
Wanting to connect our ideas of Lie subgroups with Lie subalgebras, we remark that there is a one-to-one correspondence between Lie subalgebras $\h \subset \g$ and connected immersed subgroups $\Hgr \subset \Ggr$: connected subgroups $\Hgr \subset \Ggr$ which satisfy condition 1 above, but not 2 or 3 \cite{kirillov2008introduction}.  Therefore, although knowledge of Lie subalgebras is useful in finding the possible Lie subgroups, additional work is required if we are to determine which subalgebras integrate to subgroups.
\begin{definition} Let $\Hgr \subset \Ggr$ be a Lie subgroup.  The \textit{coset space} $\Ggr/\Hgr$ is a smooth manifold equipped with the quotient topology and the group action inherited from the canonical projection $\varpi: \Ggr \rightarrow \Ggr/\Hgr$.
\end{definition}
For the coset space $\Ggr/\Hgr$ to be a smooth manifold, we require the $\Hgr$-action on $\Ggr$ to be free and proper.  From a purely group-theoretic perspective, we know that $\Hgr$ is a Lie subgroup of $\Ggr$; therefore, the $\Hgr$-action on $\Ggr$ will be free as the action of $\Ggr$ on itself is free.  Additionally, we note that a closed embedding is equivalent to a proper injective immersion.  Therefore, using our definition of a submanifold, we know that $\phi: \Hgr \rightarrow \Ggr$ must be a proper map.  This map being proper means that $\Ggr/\Hgr$ is a Hausdorff space.  Combining this result with the fact the $\Hgr$-action is free, we find that $\Ggr/\Hgr$ is indeed a smooth manifold.
\\ \\
It is interesting to consider how the exponential coordinate construction, defined earlier for $\Ggr$, may be adapted to the coset space $\Ggr/\Hgr$.  Let $\g$ and $\h$ be the Lie algebras of $\Ggr$ and $\Hgr$, respectively, and consider exponential coordinates around the identity element $e \in \Ggr$, such that $\exp_o = \exp$.  From the above discussion, we know that $\h$ must be a Lie subalgebra, so we can think of writing $\g = \m \oplus \h$, where $\m$ is some vector space complement to $\h$. We can now use the fact that there exists a local diffeomorphism $\widetilde{\exp}$, which is a slight modification of the exponential map
\begin{equation}
	\begin{split}
		\widetilde{\exp}: V_{\m} \times V_{\h} &\rightarrow U \\
		(X, Y) &\mapsto \exp(X)\exp(Y),
	\end{split}
\end{equation}
from a neighbourhood $V_{\m}$ of $(0, 0, \ldots, 0) \in \m$ and neighbourhood $V_{\h}$ of $(0, 0, \ldots, 0) \in \h$ to a neighbourhood $U$ of $e \in \Ggr$ \cite{helgason2001differential}.  Letting $\varpi(e) = eH \in \Ggr/\Hgr$ be the identity coset, notice $\varpi\circ\widetilde{\exp} (X, Y) = \widetilde{\exp}|_{\m}(X) = \exp(X)$.  Choosing a basis $\{e_i\}$ for $\m$, where $1 \leq i \leq n$, we can define $\sigma: \mathbb{R}^{n} \rightarrow \G/\H$, where $\sigma(\boldsymbol{c}) = \exp( c_i e_i )$.  We now have a local coordinate chart $(U, \sigma^{-1})$, such that $eH \in \Ggr/\Hgr$ maps to $(0, 0, \ldots, 0) \in \m$.  
\subsection{Homogeneous Spaces} \label{subsec:math_prelims_geo_hs}
Above this point, everything is solely about Lie groups and pertains directly to them.  Now we shift gear to come in direct contact with Klein's Erlanger Programme, and discuss the necessary language for our spacetime models.
\begin{definition} A Lie group $\Ggr$ is called the \textit{Lie transformation group} of a smooth manifold $\M$ if there exists a smooth $\Ggr$-action 
\begin{equation}
	\begin{split}
	\Ggr \times \M &\rightarrow \M \\
	(g, m) &\mapsto g\cdot m,
	\end{split} \qquad
\end{equation} 
such that $(g_1 g_2)\cdot m = g_1\cdot (g_2\cdot m)$ for all $g_1, g_2 \in \Ggr$ and $m \in \M$.  In particular, if this $\Ggr$-action is \textit{transitive}, that is, for all $m, n \in \M$, there exists a $g \in \Ggr$ such that $g\cdot m = n$, then $\M$ is called a \textit{homogeneous space} with respect to $\Ggr$.  Furthermore, the $\Ggr$-action is said to be \textit{effective}, if the kernel of the action $\N = \{ g \in \Ggr \, | \, g\cdot m = m, \, \forall m \in \M \}$ is the identity element, $\N = \{e\}$, or \textit{locally effective} if $\N \subset \Ggr$ is a discrete subgroup.
\end{definition}
Choosing a distinguished point $p$ in a homogeneous space $\M$, we may define the stabiliser subgroup which fixes the point as
\begin{equation}
	\Stab_{\Ggr}(p) = \{ g \in \Ggr \, | \, g\cdot p = p \}.
\end{equation}
This group is sometimes called the \textit{isotropy group} at $p$.  An important theorem states that not only is $\Hgr = \Stab_{\Ggr}(p)$ a Lie subgroup of $\Ggr$, but there exists a local diffeomorphism at $p$ such that $\M \cong \Ggr/\Hgr$~\cite{helgason2001differential}.  Given this mapping, we may seek to describe the homogeneous space $\M$ using the Lie algebras $\g$ and $\h$.  We can achieve such a description using the following definition.
\begin{definition}
A \textit{Klein pair}, or \textit{Lie pair}, $(\g, \h)$ consists of a Lie algebra $\g$ and a Lie subalgebra $\h$, such that the pair is \textit{(geometrically) realisable}.  The pair is geometrically realisable if there exists a Lie group $\Ggr'$ with Lie algebra $\g'$ and Lie subgroup $\Hgr' \subset \Ggr'$ with Lie algebra $\h'$ such that $\Ggr'/\Hgr'$ describes a homogeneous space, and there exists a Lie algebra isomorphism $\varphi: \g' \rightarrow \g$ such that $\varphi(\h') = \h$.  The homogeneous space $\Ggr'/\Hgr'$ is called a \textit{(geometric) realisation} of $(\g, \h)$.  
\end{definition}
Given a Lie pair $(\g, \h)$, we say that it is \textit{effective} if $\h$ does not contain any non-zero ideals of $\g$.  Putting all these definitions together, we have the following important result.  
There is a one-to-one correspondence between effective Lie pairs $(\g, \h)$ and homogeneous spaces $\M \cong \Ggr/\Hgr$, when we take $\Ggr$ to be the unique connected, simply-connected Lie group associated with $\g$ acting effectively on $\M$. The conditions of geometric realisability and effectiveness are required for this mapping's existence and uniqueness, respectively. 
\\ \\
At this stage, we may introduce some useful definitions which will be alluded to throughout the rest of the thesis.  Assuming we have a Lie pair $(\k, \h)$ with a vector space decomposition $\k = \m \oplus \h$, we call the pair \textit{reductive} if $[\h, \h] \subset \h$ and $[\h, \m]  \subset \m$.  A reductive pair may, in addition, be \textit{symmetric} if $[\m, \m] \subset \h$.  Alternatively, if $[\m, \m] \subset \m$, then the corresponding homogeneous space is called a \textit{principal homogeneous space}.  The intersection of these two instances then defines an \textit{affine} pair.  Explicitly, an affine Lie pair has $[\m, \m] = 0$.  
\\ \\
The above mapping between Lie pairs and homogeneous spaces tells us that if we want to classify certain types of homogeneous space, we can do so purely at the Lie algebraic level.  What we require is a classification of Lie algebras $\g$ followed by consistency checks, which make sure we have a suitable Lie subalgebra $\h$, with which we can form an effective Lie pair $(\g, \h)$.  It is this procedure that will be utilised in Section~\ref{sec:ks_kss} to find the possible kinematical spacetimes.
\\ \\
Before proceeding to apply this framework to the kinematical case explicitly, we show how the exponential coordinates for a coset space may be thought of through the lens of homogeneous spaces.  Choose a point $o \in \M$ and let $\Hgr$ be the isotropy group at $o$, such that the homogeneous space may be described locally as $\M = \Ggr/\Hgr$.  Let $\g$ and $\h$ be the Lie algebras of $\Ggr$ and $\Hgr$, respectively; and write $\g = \m \oplus \h$, where $\m$ is some vector space complement to $\h$.  Using the restricted exponential map $\widetilde{\exp}|_{\m}: \m \rightarrow \M$ identified in Section~\ref{subsec:math_prelims_geo_cs}, we may write $\exp_o:\m \rightarrow \M$, defined such that 
\begin{equation}
	\exp_o (X) = \widetilde{exp}|_{\m}(X)\, o = \exp(X)\, o \quad \quad \forall X \in \m.
\end{equation}
This map defines a local diffeomorphism from a neighbourhood $V$ of $(0,0,\ldots,0) \in \m$ and a neighbourhood $U$ of $o \in \M = \Ggr/\Hgr$. Choosing a basis $\{e_i\}$ for $\m$, where $1 \leq i \leq n$, we can define $\sigma: \mathbb{R}^{n} \rightarrow \M$, where $\sigma(\boldsymbol{c}) = \exp_o( c_i e_i )$.  We now have a local coordinate chart $(U, \sigma^{-1})$, such that $o \in U \subset \M$ maps to $(0, 0, \ldots, 0) \in V \subset \M$.  We can move this chart around $\M$ by using the group action to change the origin, giving us an atlas for $\M$.
\\ \\
There are some natural questions one can ask about the local
diffeomorphism $\exp_o : \m \to \M$ or, equivalently, the local
diffeomorphism $\sigma: \RR^{n} \to \M$.  One can ask how much of $\M$
is covered by the image of $\exp_o$.  We say that $\M$ is
\textit{exponential} if $\M = \exp_o(\m)$ and \textit{weakly
  exponential} if $\M = \overline{\exp_o(\m)}$, where the bar denotes
topological closure.  Similarly, we can ask about the domain of
validity of exponential coordinates: namely, the subspace of
$\RR^{n}$ where $\sigma$ remains injective.  In particular, if
$\sigma$ is everywhere injective, does it follow that $\sigma$ is also
surjective?  We know very little about these
questions for general homogeneous spaces, even in the reductive case.
However, there are some general theorems for the case of $\M$ a
symmetric space.

\begin{theorem}[Voglaire~\cite{MR3273068}]\label{thm:voglaire}
  Let $\M = \Ggr/\Hgr$ be a connected symmetric space with symmetric
  decomposition $\g = \m \oplus \h$ and define $\exp_o: \m \to \M$.  
  Then the following are equivalent:
  \begin{enumerate}
  \item $\exp_o: \m \to \M$ is injective
  \item $\exp_o: \m \to \M$ is a global diffeomorphism
  \item $\M$ is simply-connected and for no $X \in \m$, does $\ad_X :
    \g \to \g$ have purely imaginary eigenvalues.
  \end{enumerate}
\end{theorem}

Since our homogeneous spaces are by assumption simply-connected, the
last criterion in the theorem is infinitesimal and, therefore, easily
checked from the Lie algebra. This result makes it a relatively simple
task to determine for which of
the symmetric spaces the last criterion holds.
\\ \\
Concerning the (weak) exponentiality of symmetric spaces, we will make
use of the following result.

\begin{theorem}[Rozanov~\cite{MR2503866}]\label{thm:rozanov}
  Let $\M = \Ggr/\Hgr$ be a symmetric space with $\Ggr$ connected.
  Then
  \begin{enumerate}
  \item If $\Ggr$ is solvable, then $\M$ is weakly exponential.
  \item $\M$ is weakly exponential if and only if $\widehat{\M} =
    \Ghat/\Hhat$ is weakly exponential, where $\Ghat =
    \Ggr/\Rad(\Ggr)$ and similarly for $\Hhat$, where the
    \textit{radical} $\Rad(\Ggr)$ is the maximal connected solvable
    normal subgroup of $\Ggr$.
  \end{enumerate}
\end{theorem}

The Lie algebra of $\Rad(\Ggr)$ is the radical of the Lie algebra
$\g$, which is the maximal solvable ideal, and can be calculated
efficiently via the identification $\rad\g = [\g,\g]^\perp$, namely,
the radical is the perpendicular subspace (relative to the Killing
form, which may be degenerate) of the first derived ideal.
\\ \\
These two theorems will be used when demonstrating that the action of the
boosts are non-compact for our symmetric kinematical spacetimes. 
Note, this
is a very desirable property: if the boosts were compact, they would be
more suitably interpreted as additional rotations. 
We first find those spacetimes which satisfy the third criterion of
theorem \ref{thm:voglaire}, determining the instances for which the exponential
coordinates define a global chart.  It will be shown that, in these cases,
showing the non-compactness of the boosts only requires solving a
linear ODE.  We then find the symmetric kinematical spacetimes which
satisfy criterion 2 of theorem \ref{thm:rozanov}.  The weak exponentiality of these
spacetimes is then exploited to determine the non-compactness of
their boosts.  The remaining spacetimes require a variety of arguments
to demonstrate the non-compactness of their boosts; however, the
majority of cases are covered by these two theorems. 

\subsection{Kinematical Spacetimes} \label{subsec:math_prelims_geo_ks}
Now that we have seen how we may describe homogeneous spaces $\M \cong \Ggr/\Hgr$ in terms of the Lie algebras $\Lie(\Ggr) = \g$ and $\Lie(\Hgr) = \h$, we may apply this story to the kinematical case.  In particular, we wish to use our knowledge of the possible kinematical Lie algebras to define homogeneous spacetime geometries which hold these symmetries.  The first step in this procedure is to define, explicitly, what we mean by spacetime geometry.
\begin{definition}
A \textit{(homogeneous) kinematical spacetime} $\M$ is a homogeneous space with respect to a kinematical group $\Kgr$, such that
\begin{itemize}
	\item $\M$ is a connected, smooth manifold,
	\item $\Kgr$ acts transitively and locally effectively on $\M$ with a stabiliser subgroup $\Hgr$, and 
	\item $\Hgr \subset \Kgr$ is a Lie subgroup whose Lie algebra $\h$ contains a rotational subalgebra $\r \cong \so(D)$ and decomposes as $\h = \r \oplus V$ under the adjoint action of $\r$, where $V$ is a $D$-dimensional $\so(D)$ vector module. 
\end{itemize}
\end{definition}
Notice that not all Lie subgroups $\Hgr \subset \Kgr$ may be used to describe a kinematical spacetime.  To distinguish the Lie subalgebras $\h \subset \k$ and the Lie subgroups $\Hgr \subset \Kgr$ which may be used to describe a kinematical spacetime, we will call these subalgebras and subgroups \textit{admissible}. More explicitly, a Lie subalgebra  $\h \subset \k$, and its corresponding subgroup $\Hgr \subset \Kgr$, will be called admissible if
\begin{enumerate}
	\item $\Hgr \subset \Kgr$ is a Lie subgroup, and
	\item $\h$ decomposes under the adjoint action of $\r$ as $\h = \r \oplus V$.
\end{enumerate}
The first condition ensures that $(\k, \h)$ defines a Lie pair, and the second condition ensures that $\h$ is of the correct form.  Thus, we have the following definition.
\begin{definition}
A \textit{kinematical Lie pair} $(\k, \h)$ is a Lie pair consisting of a kinematical Lie algebra $\k$ and an admissible Lie subalgbera $\h$.
\end{definition}
From our previous discussions, we know that the connected, simply-connected kinematical spacetimes  $\tilde{\M} = \tilde{\Kgr}/\Hgr$ will be in one-to-one correspondence with kinematical Lie pairs $(\k, \h)$. 
\\ \\
With this prescription for kinematical spacetimes, we may employ the exponential coordinates defined for homogeneous spaces to provide a uniform foundation from which to investigate the differences in spacetime geometry.  Explicitly, let $\Hgr \subset \Kgr$ be the isotropy group at the point $o \in \M$. The group $\Hgr$ will always be taken as the Lie subgroup generated by the spatial rotations $\bJ_{ij}$ and the boosts $\bB_i$; therefore, $\m = \spn{\bH, \bP_i}$ for a kinematical Lie algebra.  Our coordinates are then defined by the map $\sigma: \mathbb{R}^{D+1} \rightarrow \M$, such that $\sigma(t, \x) = \exp_p(t\bH + \x\cdot\bP)$.  This map gives us a local chart $(U, \sigma^{-1})$ centred on $o \in \M$.

\subsubsection{Aristotelian Spacetimes}

Although the above story holds for the majority of the spacetime classes, Aristotelian spacetimes requires a slightly different treatment.  In particular, owing to the absence of the generator $\B$, we need to amend what we mean by an admissible subalgebra in this instance.
\begin{definition}
A \textit{(homogeneous) Aristotelian spacetime} $\M$ is a homogeneous space with respect to a Aristotelian group $\Agr$, such that
\begin{itemize}
	\item $\M$ is a connected, smooth manifold,
	\item $\Agr$ acts transitively and locally effectively on $\M$ with a stabiliser subgroup $\Rgr$, and 
	\item $\Rgr \subset \Agr$ is a Lie subgroup whose Lie algebra is the rotational subalgebra $\r \cong \so(D)$.
\end{itemize}
\end{definition}
Note, only the Lie subalgebras and Lie subgroups which satisfy the above conditions will be deemed \textit{admissible}, in the Aristotelian sense.  With this definition of an Aristotelian spacetime $\M = \Agr/\Rgr$, we may specify the form of a Lie pair $(\a, \r)$ which will be associated to $\M$.
\begin{definition}
An \textit{Aristotelian Lie pair} $(\a, \r)$ is a Lie pair consisting of an Aristotelian Lie algebra $\a$ and an admissible Lie subalgbera $\r$.
\end{definition}
Although these Lie pairs are generally distinct from the kinematical cases, there are a subset of Aristotelian Lie pairs which can arise from kinematical Lie pairs through the following procedure.  Let $(\k, \h)$ be a Lie pair containing a kinematical Lie algebra $\k$ and a Lie subalgebra $\h$.  If this pair is not effective, $\h$ must contain an ideal of $\k$; in particular, since $\h = \r \oplus V$, the only possible ideal is $\b = \spn{V}$.  To make this Lie pair effective, we may take the quotient with respect to $\b$ to arrive at the pair $(\k/\b, \h/\b)$.  This effective pair then describes an Aristotelian Lie pair.

\subsection{Lie Supergroups and Homogeneous Superspaces} \label{subsec:math_prelims_geo_lsgahss}
A Lie supergroup is defined with respect to a supermanifold in an analogous manner to how a Lie group is defined with respect to a classical manifold.  With this characterisation, the relationship between Lie superalgebras and Lie supergroups is analogous to the classical setting. Due to the increased complexity of the objects involved, this correspondence is highly non-trivial, and a full treatment of this correspondence is not necessary our current purposes.  Therefore, we note that a good introduction to this topic is found in \cite{yagi1993super} and leave this version of the story to the interested reader.  To proceed, we still need the notion of a supermanifold and Lie supergroup together with an idea of how we may tie these objects to an associated Lie superalgebra.  The method presented here will have a stronger focus on the underlying classical manifold than the one demonstrated in \cite{yagi1993super}.  Such a method is preferable for the current story since the underlying manifold describes our spacetime, which is the primary object of interest. 
\\ \\
The rest of this section is written as follows.  First, we will introduce our definition of a supermanifold before introducing Harish-Chandra pairs, which are equivalent to Lie supergroups.  We then define the superisation of a homogeneous space, which will be the geometric object describing the supersymmetric generalisations of the kinematical spacetimes of Chapter~\ref{chap:k_spaces}.  Finally, we introduce the idea of a super Lie pair, which will be the key object in the classification of kinematical superspaces in Chapter~\ref{chap:k_superspaces}.
\\ \\
Before defining our notion of a Lie supergroup, we first need to introduce supermanifolds.  We will take our definition of a supermanfiold from \cite{MR0580292}, such that we arrive at the following.
\begin{definition}
An $(m|n)$-dimensional real \textit{supermanifold} is a pair $(\M, \eO)$, where the \textit{body} $\M$ is a smooth $m$-dimensional real manifold, and the \textit{structure sheaf} $\eO$ is a sheaf of supercommutative superalgebras, extending the sheaf of smooth functions $\eE$ by the subalgebra of nilpotent elements $\eN$; that is, we have an exact sequence of sheaves of supercommutative superalgebras:
\begin{equation}
	0 \rightarrow \eN \rightarrow \eO \rightarrow \eE \rightarrow 0,
\end{equation}
where, for every point $p \in \M$, there is a neighbourhood $p \in U \subset \M$ such that
\begin{equation}
	\eO(U) \cong \eE(U) \otimes \Lambda[\theta^1, \theta^2, \ldots, \theta^n].
\end{equation}
\end{definition}
In the physics literature, superspace is typically referred to through superfields, which are functions on superspace understood in terms of their expansion as a power series in Grassmann coordinates.  These Grassmann coordinates are precisely the nilpotent basis elements $\{\theta^i\}$.  Thus, in the physics nomenclature, the structure sheaf defined above is simply the space of superfields.
\\ \\
A Lie supergroup may be defined as a group object in the category of supermanifolds; however, for our purposes, we will use the following characterisation of Lie supergroups.  The category of Lie supergroups is equivalent to the category of \textit{Harish-Chandra pairs} $(\Ggr, \s)$, where $\Ggr$ is a Lie group and $\s$ is a Lie superalgebra such that $\s_{\bar{0}} \cong \Lie(\Ggr) = \g$ and the action of $\g$ on $\s_{\bar{1}}$ lifts to an action of $\Ggr$ on $\s_{\bar{1}}$ by automorphisms \cite{MR0580292, MR760837}.  The structure sheaf of the Lie supergroup corresponding to $(\Ggr, \s)$ is then the sheaf of smooth functions $\Ggr \rightarrow \Lambda^{\bullet}\s_{\bar{1}}$, which may be interpreted as the the smooth sections of a trivial vector bundle $\Ggr \times \Lambda^{\bullet} \s_{\bar{1}}$ over $\Ggr$ \cite{MR760837}.
\\ \\
We can now consider the case where the Lie group $\Kgr$ in our Harish-Chandra pair $(\Kgr, \s)$ has an associated homogeneous space $\M = \Kgr/\Hgr$ described by the Lie pair $(\k, \h)$.  Recall that for this mapping between homogeneous space and Lie pair to be unique, $\Kgr$ must be connected and simply-connected, with $\Hgr \subset \Kgr$ closed.  We know that $\s = \s_{\bar{0}} \oplus \s_{\bar{1}}$ where $\s_{\bar{0}} = \k$ and $\s_{\bar{1}}$ must be an $\s_{\bar{0}}$-module; therefore, since $\Kgr$ is simply-connected, $\s_{\bar{1}}$ is also a $\Kgr$-module and, by restriction, a $\Hgr$-module.  This knowledge allows us to construct the homogeneous vector bundle $E = \Kgr \times_{\Hgr} \s_{\bar{1}}$.  Notice, we may now define a supermanifold $(\M, \eO)$, where the body is $\M = \Kgr/\Hgr$ and the structure sheaf $\eO$ is the smooth sections of $\Lambda^{\bullet} E$.  We will call this supermanifold the \textit{superisation} of the homogeneous space $\M$ defined by the Lie superalgebra $\s$  \cite{MR2640006}.
\\ \\
It is perhaps interesting to note that the superisations presented above all have the form of a \textit{split} supermanifold; that is, the structure sheaf $\eO$ is isomorphic to the sheaf of sections of the exterior algebra bundle of a homogeneous vector bundle $E \rightarrow \M$.  Letting $U \subset \M$ be an open subset, we have
\begin{equation}
	\eE(U) = \Gamma(U, \bigoplus_{ \geq 0} \Lambda^p E) \quad \text{and} \quad \eN(U) = \Gamma(U, \bigoplus_{ \geq 1} \Lambda^p E).
\end{equation}
This result may not be surprising given a theorem by Batchelor, stating that any smooth supermanifold admits a splitting; although the splitting may not be canonical \cite{MR536951}.
\\ \\
It is also interesting to note that any homogeneous supermanifold must be of this form.  Indeed, it was shown in \cite{MR2640006} that the homogeneous superisation of $\Kgr/\Hgr$ has the $\Hgr$-equivariant smooth functions $\Kgr \rightarrow \Lambda^{\bullet} \s_{\bar{1}}$ as structure sheaf, which are precisely the smooth section of the homogeneous vector bundle $\Lambda^{\bullet} E$ over $\M = \Kgr/\Hgr$, where $E = \Kgr \times_{\Hgr} \s_{\bar{1}}$.
\\ \\
Since all homogeneous superisations of $\Kgr/\Hgr$ are of this form, we may think of associating a unique pair $(\s, \h)$ to each homogeneous superisation.  Restricting ourselves to think solely of kinematical spacetimes $\Kgr/\Hgr$, we arrive at the following definition.
\begin{definition}
A \textit{super Lie pair} $(\s, \h)$ consists of a kinematical Lie superalgebra $\s$ and an admissible Lie subalgebra $\h$, such that the Lie pair $(\s_{\bar{0}}, \h)$ is geometrically realisable; that is, $\h \subset \s_{\bar{0}}$ contains the rotational subalgebra $\r$, decomposes as $\h = \r \oplus V$ under the adjoint action of $\r$, where $V \subset \s_{\bar{0}}$ is a vector $\r$ module, and $\h$ integrates to a Lie subgroup $\Hgr \subset \Kgr$. 
\end{definition}
As in the non-supersymmetric case, a super Lie pair $(\s, \h)$ is called \textit{effective} if $\h$ does not contain any non-zero ideals of $\s$.  We observe that the condition of being geometrically realisable is not associated with supersymmetry, whereas the condition for being effective does take $\s_{\bar{1}}$ into account.  We can, therefore, have effective super Lie pairs $(\s, \h)$ where the underlying Lie pair $(\s_{\bar{0}}, \h)$ is not effective.  In these cases, the copy of $V$ in $\h$ acts trivially on the body $\M$ of the supermanifold, but acts non-trivially on the odd coordinates.  Using physics nomenclature, $V$ generates R-symmetries in these instances. 
\\ \\
As in the classical case, there is a one-to-one correspondence between effective super Lie pairs and homogeneous superisations of homogeneous manifolds. To the best of our knowledge, this result is part of the mathematical folklore and we are not aware of any reference where this result is proved or even stated as such.
\\ \\
Just as the one-to-one correspondence between effective kinematical Lie pairs and kinematical spacetimes lifts to the supersymmetric case, the correspondence between non-effective kinematical Lie pairs and Aristotelian spacetimes also lifts to the supersymmetric case.  Explicitly, we define an \emph{Aristotelian super Lie pair} $(\sa, \r)$ as consisting of an Aristotelian Lie superalgebra $\sa$, where $\sa_{\bar{0}} = \a$ is an Aristotelian Lie algebra, and a Lie subalgebra $\r$, which is admissible in the Aristotelian sense.  We may then form an Aristotelian super Lie pair $(\sa, \r)$ from a non-effective super Lie pair $(\s, \h)$ by taking the quotient with respect to the ideal $\b = \spn{V}$, where $V$ is the vector module in the Lie subalgebra $\h = \r \oplus V$. 

\section{Geometric Properties} \label{sec:math_prelims_geo_props}
In this final section, we will introduce the geometric properties of homogeneous spaces necessary for beginning to explore the physics of each kinematical spacetime.  In particular, we will introduce fundamental vector fields, soldering forms, vielbeins, invariant connections and canonical connections.  We will see that the fundamental vector fields tell us how the rotations, boosts and spacetime translations act on our spacetime manifold; the soldering forms and vielbeins will allow us to translate between the Lie algebra and geometry; and, the various connections will tell us how to move from one point in spacetime to another.  Note, this section deals exclusively with the geometric properties in the non-supersymmetric case.  Although a supersymmetric generalisation is possible, it is lies beyond the scope of this thesis.  

\subsection{The Group Action and the Fundamental Vector Fields}
The action of the group $\Kgr$ on $\M$ is induced by left
multiplication on the group.  Indeed, we have a commuting square
\begin{equation}
  \begin{tikzcd}
    \Kgr \arrow[d, "\varpi"'] \arrow[r, "L_g"] & \Kgr \arrow[d,
    "\varpi"] \\
    \M \arrow[r, "\tau_g"] & \M
  \end{tikzcd} \qquad\qquad
  \tau_g \circ \varpi = \varpi \circ L_g,
\end{equation}
where $L_g$ is the 
diffeomorphism of $\Kgr$ given by left multiplication by $g \in \Kgr$
and $\tau_g$ is the diffeomorphism of $\M$ given by acting with $g$.
In terms of exponential coordinates, we have $g \cdot (t,\x) =
(t',\x')$ where
\begin{equation}
  g \exp(t \bH + \x \cdot \P) = \exp(t'\bH + \x'\cdot \P) h,
\end{equation}
for some $h \in \Hgr$, which typically depends on $g$, $t$, and $\x$.\footnote{Note, as stated, the exponential coordinates here are defined
with respect to the Lie group, $\exp:\k \rightarrow \Kgr$.}
\\ \\
If $g = \exp(X)$ with $X \in \h$ and if $A = t \bH + \x \cdot \P \in
\m$, the following identity will be useful:
\begin{equation}
  \label{eq:Hact}
  \exp(X)\exp(A)= \exp\left(\exp(\ad_X)A\right) \exp(X).
\end{equation}
If $\M$ is reductive, so that $[\h,\m] \subset \m$, then $\ad_X A \in
\m$ and, since $\m$ is a finite-dimensional vector space and hence
topologically complete, $\exp(\ad_X) A \in \m$ as well.  In this case,
we may act on the origin $o \in \M$, which is stabilised by $\Hgr$, to
rewrite equation~\eqref{eq:Hact} as
\begin{equation}
  \exp(X)\exp_o(A) = \exp_o\left(\exp(\ad_X)A\right),
\end{equation}
or, in terms of $\sigma$,
\begin{equation}
  \exp(X) \sigma(t,\x) = \sigma(\exp(\ad_X)(t \bH + \x \cdot \P)) = \sigma(t',\x').
\end{equation}
This latter way of writing the equation shows the action of $\exp(X)$
on the exponential coordinates $(t,\x)$, namely
\begin{equation}\label{eq:exp-coord-action}
  (t,\x) \mapsto (t',\x') \quad\text{where}\quad t' \bH + \x' \cdot \P
  := \exp(\ad_X) (t \bH + \x \cdot \P).
\end{equation}
As we will show in Section~\ref{sec:rotations}, the rotations act in the usual way: they leave
$t$ invariant and rotate $\x$, so we will normally concentrate on the
action of the boosts and translations. This requires calculating, for
example,
\begin{equation}
  \exp(v^i\bP_i) \sigma(t,\x) = \sigma(t',\x') h.
\end{equation}
In some cases, this calculation is not
practical and instead we may take $\v$ to be very small and work out
$t'$ and $\x'$ to first order in $\v$. This approximation then gives
the vector field $\xi_{\bP_i}$ generating the infinitesimal action of
$\bP_i$. To be more concrete, let $X \in \k$ and consider
\begin{equation}
  \exp(s X) \sigma(t,\x) = \sigma(t',\x') h
\end{equation}
for $s$ small.  Since for $s=0$, $t'=t$, $\x'=\x$, and $h = e$, we may
write (up to $O(s^2)$)
\begin{equation}
  \exp(s X) \sigma(t,\x) = \sigma(t + s \tau, \x + s \y) \exp(Y(s)),
\end{equation}
for some $Y(s) \in \h$ with $Y(0) = 0$, and where $\tau$ and $\y$ do
not depend on $s$.  Equivalently,
\begin{equation}
  \label{eq:inf-action}
  \exp(s X) \sigma(t,\x) \exp(-Y(s)) = \sigma(t + s \tau, \x + s \y),
\end{equation}
again up to terms in $O(s^2)$.  We now differentiate this equation
with respect to $s$ at $s=0$.  Since the equation holds up to
$O(s^2)$, the differentiated equation is exact.
\\ \\
To calculate the derivative, we recall the expression for
the differential of the exponential map (see, e.g., \cite{MR1889121})
\begin{equation}
\label{eq:d-exp}
  \left.\frac{d}{ds} \exp(X(s))\right|_{s=0} = \exp(X(0))
  D(\ad_{X(0)}) X'(0)~,
\end{equation}
where $D$ is the Maclaurin series corresponding to the analytic function
\begin{equation}
  \label{eq:function-d}
  D(z) = \frac{1-e^{-z}}{z} = 1 - \tfrac12 z + O(z^2).
\end{equation}
(We have abused notation slightly and written equations as if we were
working in a matrix group.  This is only for clarity of exposition:
the results are general.)
\\ \\
Let $A = t \bH + \x \cdot \P$. Differentiating
equation~\eqref{eq:inf-action}, we find
\begin{equation}
  X \exp(A) - \exp(A) Y'(0) = \exp(A) D(\ad_A) (\tau \bH + \y \cdot \P),
\end{equation}
and multiplying through by $\exp(-A)$ and using that $D(z)$ is
invertible as a power series with inverse the Maclaurin series
corresponding to the analytic function $F(z) = z/(1-e^{-z})$, we find
\begin{equation}
  \label{eq:master}
  G(\ad_A) X - F(\ad_A) Y'(0) = \tau \bH + \y \cdot \P,
\end{equation}
where we have introduced $G(z) = e^{-z} F(z) = z/(e^z-1)$.  It is a
useful observation that the analytic functions $F$ and $G$ satisfy the
following relations:
\begin{equation}
  \label{eq:f-and-g}
  F(z) = K(z^2) + \frac{z}{2} \quad\text{and}\quad G(z) = K(z^2) - \frac{z}{2}~,
\end{equation}
for some analytic function $K(\zeta) = 1 + \tfrac1{12}\zeta +
O(\zeta^2)$.  To see this, simply notice that $F(z) - G(z) = z$ and
that the analytic function $F(z) + G(z)$ is invariant under $z \mapsto
-z$.
\\ \\
Equation~\eqref{eq:master} can now be solved for $\tau$ and $\y$ on a
case-by-case basis.  To do this, we need to compute $G(\ad_A)$ and $F(\ad_A)$
on Lie algebra elements.  Often a pattern emerges which allows us to
write down the result.  If this fails, one can bring $\ad_A$ into
Jordan normal form and then apply the usual techniques from operator
calculus.  A good check of our calculations is that the linear map
$\k \to \eX(\M)$, sending $X$ to the vector field
\begin{equation}
  \xi_X = \tau \frac{\d}{\d t} + y^i \frac{\d}{\d x^i},
\end{equation}
should be a Lie algebra \emph{anti}-homomorphism: namely,
\begin{equation}
  [\xi_X, \xi_Y] = - \xi_{[X,Y]}.
\end{equation}
We have an anti-homomorphism since the action of $\k$ on $\M$ is
induced from the vector fields which generate left translations on
$\Kgr$ and these are right-invariant, hence obeying the opposite Lie
algebra.

\subsection{Invariant Connections}
\label{sec:invar-conn}

Let $(\k,\h)$ be a Lie pair associated to a reductive homogeneous
space. We assume that $(\k,\h)$ is effective so that $\h$ does not
contain any non-zero ideals of $\k$. We let $\k = \h \oplus \m$ denote
a reductive split, where $[\h,\m] \subset \m$. This split makes $\m$ into an
$\h$-module relative to the \textit{linear isotropy representation}
$\lambda: \h \to \gl(\m)$, where
\begin{equation}
  \lambda_X Y = [X,Y] \quad \quad \forall X\in\h \quad \text{and} \quad  Y\in\m.
\end{equation}
As shown in \cite{10.2307/2372398}, one can uniquely characterise 
the invariant affine connections on $(\k,\h)$ by their \textit{Nomizu map}
$\alpha: \m \times \m \to \m$, an $\h$-equivariant bilinear map; that
is, such that for all $X \in \h$ and $Y,Z \in \m$,
\begin{equation} \label{eq:Nomizuinv}
  [X,\alpha(Y,Z)] = \alpha([X,Y],Z) + \alpha(Y,[X,Z]).
\end{equation}
The torsion and curvature of an invariant affine connection with
Nomizu map $\alpha$ are given, respectively, by the following
expressions for all $X,Y,Z \in \m$,
\begin{equation} \label{eq:reductive_tor_and_curv}
  \begin{split}
    \Theta(X,Y) &= \alpha(X,Y) - \alpha(Y,X) - [X,Y]_\m,\\
    \Omega(X,Y) Z &= \alpha(X,\alpha(Y,Z)) - \alpha(Y,\alpha(X,Z)) -
    \alpha([X,Y]_\m,Z) - [[X,Y]_\h, Z],
  \end{split}
\end{equation}
where $[X,Y] = [X,Y]_\h + [X,Y]_\m$ is the decomposition of $[X,Y] \in
\k = \h \oplus \m$.  In particular, for the canonical invariant
connection with zero Nomizu map, we have
\begin{equation} \label{eq:tor_and_curv_can}
  \Theta(X,Y) = - [X,Y]_\m \qquad\text{and}\qquad
  \Omega(X,Y) Z = - \lambda_{[X,Y]_\h} Z.
\end{equation}
For kinematical spacetimes, we can determine the possible
Nomizu maps in a rather uniform way. Rotational invariance determines
the form of the Nomizu map up to a few parameters and then we need
only study the action of the boosts. We will see that the action of the boosts is common to all spacetimes
within a given class: Lorentzian, Riemannian, Galilean, and
Carrollian; although the curvature and torsion of the invariant
connections of course do depend on the spacetime in question.

\subsection{The Soldering Form and the Canonical Connection}
\label{sec:sold-invar-conn}

On the Lie group $\Kgr$ there is a left-invariant
$\k$-valued one-form $\vartheta$: the (left-invariant)
Maurer--Cartan one-form \cite{warner2013foundations, helgason2001differential}.
This connection is the canonical invariant connection on $\Kgr$ and it obeys the structure equation
\begin{equation}
  \label{eq:MCSE}
  d\vartheta = -\tfrac12 [\vartheta,\vartheta],
\end{equation}
where the notation hides the wedge product in the right-hand side (R.H.S).
Using exponential coordinates, we can pull back $\vartheta$ to a
neighbourhood of the origin on $\M$. The following formula, which
follows from equation \eqref{eq:d-exp}, shows how to calculate it:
\begin{equation}
  \label{eq:MC-pullback}
  \sigma^* \vartheta = D(\ad_A) (dt \bH + d\x \cdot \P),
\end{equation}
where, as before, $A = t\bH + \x\cdot \P$ and $D$ is the Maclaurin series
corresponding to the analytic function in \eqref{eq:function-d}.
\\ \\
The pull-back $\sigma^* \vartheta$ is a one-form defined near the
origin on $\M$ with values in the Lie algebra $\k$. Since $\vartheta$
is the canonical invariant connection on $\Kgr$, we can ask whether
we can construct a canonical invariant connection on $\M$ using this pull-back.
Note, such a one-form must be $\Hgr$-invariant and should take values in $\k/\h$.
Let $\m$ be a vector space complement to $\h$ in $\k$ so that as a vector
space $\k = \h \oplus \m$.  This split allows us to write
\begin{equation}
  \sigma^* \vartheta = \theta + \omega~,
\end{equation}
where $\theta$ is $\m$-valued and and $\omega$ is $\h$-valued. 
Notice that our requirement of $\Hgr$ invariance tells us that not only
are $\m$ and $\k/\h$ isomorphic as vector spaces, but they are also
isomorphic as $\h$ modules.  Notice, this is precisely the reductive
condition.  Additionally, since the Maurer-Cartan one-form defines the canonical invariant connection, with vanishing Nomizu map, on $\Kgr$, the one-form we recover from the pullback, corresponds to the canonical invariant connection on $\M$. Therefore,  if the
Lie pair $(\k,\h)$ is reductive then $\omega$ is the one-form corresponding to
the \textit{canonical invariant connection} on $\M$. The
\textit{soldering form} is then given by $\theta$.
\\ \\
The torsion and curvature of $\omega$ are easy to calculate using the
fact that $\vartheta$ obeys the Maurer--Cartan structure
equation~\eqref{eq:MCSE}.\footnote{%
  Let us emphasise that in this work, curvature always refers to the
  curvature of an invariant affine connection and hence should not be
  confused with the curvature of the associated Cartan connection,
  which is always flat for the homogeneous spaces.}
Indeed, the torsion two-form $\Theta$ is given by
\begin{equation}
  \label{eq:torsion}
  \Theta = d\theta + [\omega,\theta] = - \tfrac12 [\theta,\theta]_\m
\end{equation}
and the curvature two-form $\Omega$ by
\begin{equation}
  \label{eq:curvature}
  \Omega = d\omega + \tfrac12 [\omega,\omega] = - \tfrac12 [\theta,\theta]_\h,
\end{equation}
which agree with the expressions in
equation~\eqref{eq:tor_and_curv_can}.
\\ \\
In the non-reductive case, $\omega$ does not define a connection, but
we may still project the locally defined $\k$-valued one-form
$\sigma^*\vartheta$ to $\k/\h$. The resulting local one-form $\theta$
with values in $\k/\h$ is a soldering form which defines an
isomorphism $T_o\M \to \k/\h$ for every $o \in \M$ near the
origin. Wherever $\theta$ is invertible, the exponential coordinates
define an immersion, which may however fail to be an embedding or
indeed even injective.  In practice, it is not easy to determine
injectivity, but it is easy to determine where $\theta$ is invertible
by calculating the top exterior power of $\theta$ and checking that it
is non-zero. Provided that $\theta$ is invertible, the inverse
isomorphism is the vielbein $E$, where $E(o): \k/\h \to T_o\M$ for
every $o \in \M$ near the origin. The vielbein allows us to transport
tensors on $\k/\h$ to tensor fields on $\M$ and, as we now recall, it
takes $\Hgr$-invariant tensors on $\k/\h$ to $\Kgr$-invariant tensor
fields on $\M$.

\subsection{Invariant Tensors}
\label{sec:invariant-tensors}

It is well-known that $\Kgr$-invariant tensor fields on $\M=\Kgr/\Hgr$
are in one-to-one correspondence with $\Hgr$-invariant tensors on
$\k/\h$, and, if $\Hgr$ is connected, with $\h$-invariant tensors on
$\k/\h$.  We may assume that $\Hgr$ is indeed connected, passing 
to the universal cover of $\M$, if necessary.  In practice, given an
$(r,s)$-tensor $T$ on $\k/\h$---that is, an element of
$(\k/\h)^{\otimes r} \otimes ((\k/\h)^*)^{\otimes s}$---we can turn it
into an $(r,s)$-tensor field $\eT $ on $\M$ by contracting with
soldering forms and vielbeins as appropriate to arrive, for every $o
\in \M$, at $\eT(o) \in (T_o\M)^{\otimes r} \otimes (T^*_o\M)^{\otimes
  s}$. Moreover, if $T$ is $\Hgr$-invariant, $\eT$ is
$\Kgr$-invariant.
\\ \\
Our choice of basis for $\k$ is such that $\J$ and $\B$ span $\h$ 
and therefore $\Pbar := \P \mod \h$ and $\Hbar := \bH \mod \h$ span
$\k/\h$.  In the reductive case, $\k = \h \oplus \m$ and
$\m \cong \k/\h$ as $\h$-modules.  We will let $\eta$ and $\pi^a$
denote the canonical dual basis for $(\k/\h)^*$.
\\ \\
Invariant non-degenerate metrics are in one-to-one correspondence with
$\h$-invariant non-degenerate symmetric bilinear forms on $\k/\h$ and
characterise, depending on their signature, \textit{Lorentzian} or
\textit{Riemannian} spacetimes. On the other hand, invariant
\textit{Galilean} structures\footnote{%
  We will not distinguish notationally the $\Hgr$-invariant tensor
  from the $\Kgr$-invariant tensor field.}  consist of a pair
$(\tau,h)$, where $\tau \in (\k/\h)^*$ and $h \in S^2(\k/\h)$ are
$\h$-invariant, $h$ has co-rank $1$ and $h(\tau,-) = 0$, if we think
of $h$ as a symmetric bilinear form on $(\k/\h)^*$. On $\M$, $\tau$
gives rise to an invariant clock one-form and $h$ to an invariant
spatial metric on one-forms. \textit{Carrollian} structures are dual
to Galilean structures and consist of a pair $(\kappa,b)$, where
$\kappa \in \k/\h$ defines an invariant vector field and
$b \in S^2(\k/\h)^*$ is an invariant symmetric bilinear form of
co-rank $1$ and such that $b(\kappa,-) = 0$.  Homogeneous
\textit{Aristotelian} spacetimes admit an invariant Galilean structure
and an invariant Carrollian structure simultaneously.
\\ \\
Invariance under $\h$ implies, in particular, invariance under the
rotational subalgebra, which is non-trivial for $D \geq 2$.  Assuming
that $D\geq 2$, it is easy to write down the possible
rotationally invariant tensors, and, therefore, we need only check
invariance under $\B$.  The action of $\B$ is induced by duality from
the action on $\k/\h$ which is given by
\begin{equation}
  \lambda_{\bB_a} (\Hbar) = \overline{[\bB_a,\bH]} \quad\text{and}\quad
  \lambda_{\bB_a} (\overline{\bP}_b) = \overline{[\bB_a, \bP_b]},
\end{equation}
with the brackets being those of $\k$. In practice, we can determine
this from the explicit expression of the Lie brackets by computing the
brackets in $\k$ and simply dropping any $\B$ or $\J$ from the
right-hand side. The only possible invariants in $\k/\h$ are
proportional to $\bH$, which is invariant provided that
$[\B,\bH] = 0 \mod \h$. Dually, the only possible invariants in
$(\k/\h)^*$ are proportional to $\eta$, which is invariant provided
that there is no $X \in \k$ such that $\bH$ appears in $[\B,X]$.
Omitting the tensor product symbol, the only rotational invariants in
$S^2(\k/\h)$ are linear combinations of $\bH^2$ and $\P^2 := \delta^{ab}
\bP_a \bP_b$, whereas in $S^2(\k/\h)^*$ are $\eta^2$ and $\pi^2 =
\delta_{ab}\pi^a\pi^b$.


\chapter{Kinematical Spacetimes} \label{chap:k_spaces}
In this chapter, we consider the first of our three types of symmetry, kinematical symmetry.  These symmetries are the best-studied of the three; therefore, we can present the algebra classification, spacetime classification and explore the spacetimes' geometric properties.  Each of the following two chapters will show the progress made towards recovering a similar description in the super-kinematical and super-Bargmann cases, respectively; however, it is the kinematical case with the fullest picture to-date.  Therefore, we may view this chapter as showing the direction in which we hope to take the other two types of symmetry.  The kinematical classifications have been derived for dimensions $D \geq 1$; however, the classifications of the supersymmetric algebras and spacetimes are limited to $D=3$.  To keep dimension consistent throughout the thesis, we will focus solely on the $D = 3$ kinematical spacetimes.
\\ \\
We begin in Section~\ref{sec:ks_klas} by reviewing the classification of kinematical Lie algebras, as presented in~\cite{Figueroa-OFarrill:2017ycu, Figueroa-OFarrill:2017tcy}.  Section~\ref{sec:ks_kss} shows how Figueroa-O'Farrill and Prohazka generated a spacetime classification from the preceding Lie algebra classification in their paper~\cite{Figueroa-OFarrill:2018ilb}.  Section~\ref{subsec:ks_kss_gls} then demonstrates how the classified kinematical spacetimes are connected via geometric limits.  Finally, in Section~\ref{sec:ks_gps}, we go through each of the spacetime geometries identified in Section~\ref{sec:ks_kss} and determine some of their geometric properties.

\section{Classification of Kinematical Lie Algebras} \label{sec:ks_klas}
This section reviews the classification of kinematical Lie algebras, as presented in the papers of Figueroa-O'Farrill~\cite{Figueroa-OFarrill:2017ycu, Figueroa-OFarrill:2017tcy}.  These papers aimed to find a methodology for deriving this classification that would allow us to extend beyond the $D=3$ case, which had been completed previously by Bacry and Nuyts in~\cite{Bacry:1986pm}.  The methodology that allowed for this generalisation was to consider taking deformations of the static kinematical Lie algebra $\a$; that is, the kinematical Lie algebra which has only the kinematical brackets,
\begin{equation}
	\begin{split}
	[\bJ_{ij}, \bJ_{kl}] &= \delta_{jk} \bJ_{il} - \delta_{ik} \bJ_{jl} - \delta_{jl} \bJ_{ik} + \delta_{il} \bJ_{jk}, \\ 			
	[\bJ_{ij}, \bB_k] &= \delta_{jk} \bB_{i} - \delta_{ik} \bB_j, \\
	[\bJ_{ij}, \bP_k] &= \delta_{jk} \bP_{i} - \delta_{ik} \bP_j, \\ 
	[\bJ_{ij}, \bH] &= 0:
	\end{split}
\end{equation} 
all other brackets vanish.  We will not go into the details of this classification as they are beyond the scope of our discussion; we will only briefly note that each non-vanishing deformation gives rise to a new non-vanishing bracket.\footnote{The Lie algebra one reaches when we can no longer add any more brackets is sometimes called a \textit{rigid} Lie algebra in the literature~\cite{Parsa:2018kys, Safari:2019zmc, Safari:2020pje}.}  
\\ \\
Using this deformation theoretic method, Figueroa-O'Farrill arrived at the classification of kinematical Lie algebras in spatial dimension $D \geq 3$, presented in Table~\ref{tab:klas_Dgeq3}, and the classification of kinematical Lie algebras unique to spatial dimension $D=3$, presented in Table~\ref{tab:klas_D3}.  \\ \\
\begin{table}[h!]
  \centering
  \caption{Kinematical Lie Algebras for $D \geq 3$}
  \label{tab:klas_Dgeq3}
  \rowcolors{2}{blue!10}{white}
  \resizebox{\textwidth}{!}{
    \begin{tabular}{l|*{5}{>{$}l<{$}}|l}\toprule
      \multicolumn{1}{c|}{Label} & \multicolumn{5}{c|}{Non-zero Lie brackets in addition to $[\J,\J] = \J$, $[\J, \B] = \B$, $[\J,\P] = \P$} & \multicolumn{1}{c}{Comments}\\\midrule
	  	K1 & & & & & & $\a$ \\
      	K2 & [\bH ,\B] = \P & & & & & $\g$ \\
      	K3 & [\bH , \B] = \gamma \B & [\bH , \P] = \P & & & & $\gamma \in (-1, 1)$ \\
      	K4 & [\bH , \B] = \B & [\bH , \P] = \P & & & & \\
      	K5 & [\bH, \B] = - \B & [\bH, \P] = \P & & & & $\n_-$ \\
      	K6 & [\bH, \B] = \B + \P & [\bH, \P] = \P & & & & \\
      	K7$_\chi$ & [\bH, \B] = \chi \B + \P & [\bH, \P] = \chi \P - \B & & & & $\chi > 0$ \\
      	K8 & [\bH, \B] = \P & [\bH, \P] = -\B & & & & $\n_+$ \\
      	K9 & & & & [\B, \P] = \bH & & $\c$ \\
      	K10 & [\bH, \B] = - \varepsilon \P & & [\B, \B] = \varepsilon \J & [\B, \P] = \bH & & $\varepsilon = \pm 1 \quad  \choice{ \p}{ \e}$ \\
      	K11 & [\bH, \B] = \B & [\bH, \P] = -\P & & [\B, \P] = \bH + \J & & $\so(D+1, 1)$ \\
      	K12 & [\bH, \B] = -\varepsilon \P & [\bH, \P] = \varepsilon \B & [\B, \B] = \varepsilon \J & [\B, \P] = \bH & [\P, \P] = \varepsilon \J & $\varepsilon = \pm 1 \quad \choice{\so(D, 2) }{ \so(D+2)}$ \\ \bottomrule
    \end{tabular}
  }
  \caption*{}
\end{table}
To make sense of these tables, a few comments are required.  First, these tables are utilising the abbreviated notation first presented in Section~\ref{subsec:math_prelims_alg_kla}.  In the final column of each table, we state the names of the known Lie algebras found in the classification.  See Table~\ref{tab:notation} for a key.  Finally, the kinematical Lie algebras of Table~\ref{tab:klas_D3} all contain brackets not possible when $D\neq 3$; in particular, they have either $[\B, \B] = \P$ or $[\B, \B] = \P$.  These brackets are exclusive to $D=3$ since, in this dimension, we have the $\so(3)$-invariant vector product $\varepsilon_{i_1 i_2 i_3}$.  Explicitly, this lets us write
\begin{equation}
	[\B, \B] = \B \quad \text{which is equivalent to} \quad [\bB_i, \bB_j] = \varepsilon_{ijk} \bB_k.
\end{equation}

\begin{table}[h!]
  \centering
  \caption{Notation Summary}
  \label{tab:notation}
  \resizebox{\textwidth}{!}{
    \setlength{\extrarowheight}{2pt}
    \begin{tabular}{l|l}\toprule
      \multicolumn{1}{c|}{Notation} & \multicolumn{1}{c|}{Name} \\
      \toprule
		$\p$ & Poincaré\\
		$\g$ & Galilean  \\
		$\n_-$ & (Elliptic) Newton-Hooke \\
		$\n_+$ & (Hyperbolic) Newton-Hooke \\
      \bottomrule
    \end{tabular}
    \hspace{2cm}
    \begin{tabular}{l|l}\toprule
      \multicolumn{1}{c|}{Notation} & \multicolumn{1}{c|}{Name} \\
      \toprule
		$\e$ & Euclidean  \\
		$\c$ & Carroll\\
		$\a$ & Static \\
		$\so$ & Special Orthogonal \\
      \bottomrule
    \end{tabular}
    }
  \caption*{}
\end{table}
\begin{table}[h!]
  \centering
  \caption{Kinematical Lie Algebras Unique to $D=3$}
  \label{tab:klas_D3}
  \rowcolors{2}{blue!10}{white}
  \resizebox{\textwidth}{!}{
    \begin{tabular}{l|*{5}{>{$}l<{$}}|l}\toprule
      \multicolumn{1}{c|}{Label} & \multicolumn{5}{c|}{Non-zero Lie brackets in addition to $[\J,\J] = \J$, $[\J, \B] = \B$, $[\J,\P] = \P$} & \multicolumn{1}{c}{Comments}\\\midrule
	  	K13$_{\varepsilon}$ & & & [\B, \B] = \B &  & [\P, \P] = \varepsilon (\B - \J) & $\varepsilon = \pm 1$ \\
      	K14 & & & [\B, \B] = \B & & & \\
      	K15 & & & [\B, \B] = \P & & & \\
      	K16 & & [\bH , \P] = \P & [\B, \B] = \B & & & \\
      	K17 & [\bH, \B] = - \P & & [\B, \B] = \P & & & \\
      	K18 & [\bH, \B] = \B & [\bH, \P] = 2\P & [\B, \B] = \P & & &  \\ \bottomrule
    \end{tabular}
  }
  \caption*{}
\end{table}

\subsection{Classification of Aristotelian Lie Algebras}
\label{subsec:ks_ala_class}

The discussion in this section follows appendix A in~\cite{Figueroa-OFarrill:2018ilb}.  Before moving on to the classification of the spacetime models associated with the kinematical Lie algebras in Tables~\ref{tab:klas_Dgeq3} and~\ref{tab:klas_D3}, we will pause here for a short discussion on the classification of Aristotelian Lie algebras.  This classification is separated from the above story due to the Aristotelian algebras having a different underlying vector space from the other four kinematical symmetry classes; namely, they contain only one copy of the $\so(D)$ vector module $V$.  Owing to the relative simplicity of the Aristotelian Lie algebras, their classification only requires the Jacobi identity.  In particular, given the fixed Aristotelian brackets
\begin{equation}
	\begin{split}
	[\bJ_{ij}, \bJ_{kl}] &= \delta_{jk} \bJ_{il} - \delta_{ik} \bJ_{jl} - \delta_{jl} \bJ_{ik} + \delta_{il} \bJ_{jk}, \\ 	
	[\bJ_{ij}, \bP_k] &= \delta_{jk} \bP_{i} - \delta_{ik} \bP_j, \\ 
	[\bJ_{ij}, \bH] &= 0,
	\end{split}
\end{equation}
we may write down the most general form for the possible remaining brackets $[\bH, \P]$ and $[\P, \P]$, to obtain a two-dimensional $\mathbb{R}$ vector space, $\cV$.  We then impose the Jacobi identity to cut out an algebraic variety $\cJ \subset \cV$.  The possible brackets in $D = 3$ are 
\begin{equation}
		[\bH, \bP_i] = \alpha \bP_i \quad \text{and} \quad 
		[\bP_i, \bP_j] = \beta \bJ_{ij} + \gamma \epsilon_{ijk} \bP_k,
\end{equation}
where $\alpha, \beta, \gamma \in \mathbb{R}$.  Note, we may consistently set $\gamma$ to zero under a suitable choice of basis; thus, $\cV$ is a two-dimensional vector space.  Imposing the Jacobi identities, we arrive at the classification shown in Table~\ref{tab:alas}. \\ \\

\begin{table}[h!]
  \centering
  \caption{Aristotelian Lie Algebras for $D = 3$}
  \label{tab:alas}
  \rowcolors{2}{blue!10}{white}
  \resizebox{\textwidth}{!}{
    \begin{tabular}{l|*{2}{>{$}l<{$}}|l}\toprule
      \multicolumn{1}{c|}{Label} & \multicolumn{2}{c|}{Non-zero Lie brackets in addition to $[\J,\J] = \J$, $[\J,\P] = \P$} & \multicolumn{1}{c}{Comments}\\\midrule
	  	A1 & & & $\a$ \\
      	A2 & [\bH, \P] = \P & & \\
      	A3 & & [\P, \P] =  \varepsilon \J & $\varepsilon = \pm 1$ \\ \bottomrule
    \end{tabular}
  }
  \caption*{}
\end{table}
This method of forming an $\mathbb{R}$ vector space $\cV$ representing the remaining possible brackets and utilising the Jacobi identity to cut out an algebraic variety $\cJ \subset \cV$ will be generalised and employed in the supersymmetric classifications of Chapters~\ref{chap:k_superspaces} and~\ref{chap:gb_superspaces}.

\section{Classification of Kinematical Spacetimes} \label{sec:ks_kss}
Now that we have seen the classification of the kinematical Lie algebras, we may review how Figueroa-O'Farrill and Prohazka took these algebras and produced a classification of spacetime geometries in~\cite{Figueroa-OFarrill:2018ilb}.  Unlike the Lie algebra classification, we will present a (nearly) complete discussion on the method utilised in acquiring this classification since its direct generalisation will be employed in the supersymmetric case in Chapter~\ref{chap:k_superspaces}.
\\ \\
As stated in Section~\ref{subsec:math_prelims_geo_ks}, not all kinematical Lie algebras $\k$ necessarily produce a kinematical spacetime.  Recall, we are choosing to model our spacetime geometries as homogeneous spaces $\M$ with respect to the kinematical Lie group $\Kgr$ with $\Lie(\Kgr) = \k$.  Our homogeneous space description then relies on the choice of a point $o \in \M$ with isotropy group $\Hgr$, such that, locally, we may use the diffeomorphism $\M = \Kgr/\Hgr$.  We then have a unique algebraic description of the connected, simply-connected manifold $\tilde{\M} = \tilde{\Ggr}/\Hgr$ in terms of the effective kinematical Lie pair $(\k, \h)$, where $\h \cong \Lie(\Hgr)$.
\\ \\
The criteria which must be met for the pair $(\k, \h)$ to form a kinematical spacetime are as follows.
\begin{enumerate}
	\item \textit{(admissibility)} The Lie subalgebra $\h \subset \k$ must contain the rotational Lie subalgebra $\r \cong \so(D)$, such that, under the adjoint action of $\r$, it decomposes as $\h = \r \oplus V$, where $V$ is an $\so(D)$ vector module.
	\item \textit{(effectivity)} The Lie subalgebra $\h \subset \k$ must not contain any non-zero ideals of $\k$.
	\item \textit{(geometric realisability)} The Lie subalgebra $\h \subset \k$ must integrate to a connected Lie subgroup $\Hgr \subset \tilde{\Kgr}$.
\end{enumerate}
Therefore, the task in this section will be to take each of the kinematical Lie algebras $\k$ presented in Section~\ref{sec:ks_klas} and check for suitable Lie subalgebras $\h \subset \k$.  We will now go through each of the above criteria, discussing how we can check for Lie algebras which satisfy them.
\subsubsection{A Note of Aristotelian Spacetimes}
A quick inspection of the above criteria highlights that each of the Aristotelian Lie algebras in Table~\ref{tab:alas} will give rise to a unique connected, simply-connected Aristotelian spacetime.  First, since all Aristotelian Lie algebras necessarily contain a unique copy of the rotational subalgebra $\r \cong \so(D)$, there is only one choice of admissible subalgebra in each instance.  Second, the rotational subalgebra is semi-simple, therefore, cannot contain any non-zero ideals of $\a$, making the subalgebra trivially effective.  Finally, $\r \cong \so(D)$ integrates to a compact subgroup, and compact subgroups are always closed; therefore, $(\a, \r)$ is always geometrically realisable.  Thus, the classification of Aristotelian Lie algebras in Table~\ref{tab:alas} is also a classification of the connected, simply-connected Aristotelian spacetimes up to isomorphism.  With this in mind, the rest of the section will concentrate solely on the other classes of kinematical spacetime. 
\subsection{Admissibility} \label{subsec:ks_admissibility} Determining the number of admissible Lie subalgebras $\h \subset \k$ corresponding to a particular kinematical Lie algebra $\k$ amounts to determining the number of $\Aut(\k)$-orbits in the space of $\so(D)$ vector modules, spanned by $\B$ and $\P$.  To see why this is the case, we need to be more specific about which automorphisms we are considering and we need to revisit the definition of an admissible Lie subalgebra.  For our current purposes, we will focus exclusively on the automorphisms which fix the rotational subalgebra $\r$; that is, the automorphisms that send the generator $\J$ back to itself.  By fixing the rotational subalgebra $\r$, the only choice we have in determining our admissible subalgebra  $\h = \r \oplus V$ is in our choice of $V$.  Therefore, the number of admissible Lie subalgebras $\h \subset \k$ corresponding to a particular kinematical Lie algebra $\k$ will be the number of ways we can choose $V$.  We will now show that the number of choices we have for $V$ is exactly the number of $\Aut(\k)$-orbits on the space of $\so(D)$ vector modules in $\k$. 
\\ \\
Since $\k$ contains two copies of the $\so(D)$ vector module $V$, we have a two-dimensional real vector space of $\so(D)$ vector modules with a basis $(\B, \P)$.  The group $\Aut(\k)$ then acts on this space as
\begin{equation}
	(\B, \P) \mapsto (\B, \P) \begin{pmatrix}
		a & b \\ c & d
	\end{pmatrix} = (a\B + c\P, b\B +  d\P),
\end{equation}
such that $\Aut(\k) \subset \GL(2, \mathbb{R})$.  With this prescription, it transpires that there are three possible forms for the above matrix, see~\cite{Figueroa-OFarrill:2018ilb}.  These are
\begin{equation}
	\text{case 1} = \begin{pmatrix}
		a & b \\ c & d
	\end{pmatrix}, \quad \text{case 2} = \begin{pmatrix}
		a & 0 \\ c & d
	\end{pmatrix}, \quad \text{and} \quad \text{case 3} = \begin{pmatrix}
		a & 0 \\ 0 & d
	\end{pmatrix}.
\end{equation}
We will now see that each of these cases gives rise to a different number of $\Aut(\k)$-orbits, and, therefore, a different number of admissible Lie subalgebras $\h \subset \k$.  To show this result, we introduce an arbitrary vector module $V = \alpha \B + \beta \P = (\alpha, \beta)$, where $\alpha, \beta \in \mathbb{R}$.  In case 1, we have
\begin{equation}
	\alpha \B + \beta \P \mapsto (a\alpha + b\beta) \B + (c\alpha + d \beta) \P.
\end{equation}
Notice that for a suitable choice of $a, b, c$, and $d$, we can bring the transformed vector into any form we choose. Let us send it to $(1, 0)$.  Thus, any kinematical Lie algebra $\k$ with an automorphism group of this form will only have a single admissible Lie subalgebra associated with it.  Explicitly, it will have a Lie pair with admissible Lie subalgebra $\h = \r \oplus V$, where $V = \spn{\B}$.  In contrast, case 2 gives the transformation
\begin{equation}
	\alpha \B + \beta \P \mapsto a\alpha \B + (c\alpha + d\beta) \P.
\end{equation}
Notice that if $\alpha \neq 0$, we can always choose to bring the vector into the form $(1, 0)$, as in case 1; however, if $\alpha = 0$, we can only produce a vector module of the form $(0, 1)$.  Thus, any kinematical Lie algebra $\k$ with an automorphism group of this form will have two admissible Lie pairs associated with it.  Explicitly,  it will have a Lie pair with admissible Lie subalgebra $\h = \r \oplus V$, where $V = \spn{\B}$, and a Lie pair with admissible Lie subalgebra $\h = \r \oplus V$, where $V = \spn{\P}$.  Finally, in case 3, we have 
\begin{equation}
	\alpha \B + \beta \P \mapsto a\alpha \B + d \beta \P.
\end{equation}
Notice that if $\alpha \neq 0, \beta = 0$, we can, again, choose the vector $(1, 0)$.  Alternatively, if $\alpha = 0, \beta \neq 0$, we can choose the vector $(0, 1)$.  This covers the same two instances as case 2.  However, here, we have a third choice.  Letting $\alpha \beta \neq 0$, we can bring our vector into the form $(1, 1)$, which, in this case, is a representative of a different $\Aut(\k)$-orbit than $(1, 0)$ and $(0, 1)$.  Explicitly, in case 3, we will have a Lie pair with admissible Lie subalgebra $\h = \r \oplus V$, where $V = \spn{\B}$, a Lie pair with admissible Lie subalgebra $\h = \r \oplus V$, where $V = \spn{\P}$, and a Lie pair with admissible Lie subalgebra $\h = \r \oplus V$, where $V = \spn{\B +\P}$.
\\ \\
Anticipating the final geometric interpretation of the admissible subalgebra's basis elements, we will want to write $\h$ as the span of the spatial rotations $\J$ and the boosts $\B$.  Therefore, after determining the possible admissible Lie subalgebras, the basis may be relabelled such that $\h$ always has the desired form.  This procedure for finding the admissible subalgebras will be carried out explicitly in the super-kinematical case in Section~\ref{sec:slie-pairs}.
\subsection{Effectivity} Having determined the Lie pairs $(\k , \h)$ which contain admissible Lie subalgebras $\h  \subset \k$, finding the effective Lie pairs is relatively straightforward.  Since $\h = \spn{\J, \B}$, the only possible non-zero ideal of $\k$ in $\h$ is $\b = \spn{\B}$.  Therefore, since $[\J, \B] = \B$, we only need to check whether $[\b, X] \subset \b$ for $X \in \{\bH, \B, \P\}$.  If this property holds, the pair will not be effective; if the property does not hold, the pair will be effective.   For those Lie pairs that are not effective, we can obtain an effective Lie pair by taking the quotient with respect to the ideal, $(\k/\b, \h/\b)$.\footnote{Notice, however, that this Lie pair will no longer be admissible in the kinematical sense.  But it may still form an Aristotelian Lie pair.}  Determining the effective Lie pairs is then a matter of inspecting the Lie brackets for each Lie pair containing an admissible Lie subalgebra. 
\subsection{Geometric Realisability}  Unfortunately, unlike the admissibility and effectivity criteria above, there is no ``one-size-fits-all'' method of determining a Lie pair's geometric realisability.  For this reason, we will not labour over the details and instead refer the reader to section 4.2 in~\cite{Figueroa-OFarrill:2018ilb} for the full discussion on this criterion.  
\subsection{Classification}
Having gone through each of the criteria, the remaining Lie pairs $(\k, \h)$ may be called kinematical Lie pairs, and each corresponds to a unique connected, simply-connected homogeneous space, which is taken as the geometry for the associated kinematical spacetime.  The spacetimes which arise in this manner are listed in Table~\ref{tab:spacetimes}, taken from~\cite{Figueroa-OFarrill:2018ilb}.
\begin{table}[h!]
  \centering
  \caption{Simply-Connected Spatially-Isotropic Homogeneous Spacetimes}
  \label{tab:spacetimes}
  \rowcolors{2}{blue!10}{white}
  \resizebox{\textwidth}{!}{
    \begin{tabular}{l|*{5}{>{$}l<{$}}|l}\toprule
      \multicolumn{1}{c|}{Label} & \multicolumn{5}{c|}{Nonzero Lie brackets in addition to $[\J,\J] = \J$, $[\J, \B] = \B$, $[\J,\P] = \P$} & \multicolumn{1}{c}{Comments}\\\midrule
      \hypertarget{S1}{$\MM^4$} & [H,\B] = -\P & & [\B,\B] = -\J & [\B,\P] = H & & Minkowski\\
      \hypertarget{S2}{$\zdS_4$} & [H,\B] = -\P & [H,\P] = -\B & [\B,\B]= -\J & [\B,\P] = H & [\P,\P]= \J & de~Sitter\\
      \hypertarget{S3}{$\zAdS_4$} & [H,\B] = -\P & [H,\P] = \B & [\B,\B]= -\J & [\B,\P] = H & [\P,\P] = -\J & Anti-de~Sitter\\
      \midrule
      \hypertarget{S4}{$\EE^4$} &  [H,\B] = \P & & [\B,\B] = \J & [\B,\P] = H & & Euclidean\\
      \hypertarget{S5}{$S^4$} &  [H,\B] = \P & [H,\P] = -\B & [\B,\B]= \J & [\B,\P] = H & [\P,\P]= \J & Sphere\\
      \hypertarget{S6}{$H^4$} &  [H,\B] = \P & [H,\P] = \B & [\B,\B]= \J & [\B,\P] = H & [\P,\P] = -\J & Hyperbolic Space\\
      \midrule
      \hypertarget{S7}{$\zG$} & [H,\B] = -\P & & & & & Galilean spacetime\\
      \hypertarget{S8}{$\zdSG$} & [H,\B] = -\P & [H,\P] = -\B & & & & Galilean de~Sitter ($\zdSG= \ztdSG_{\gamma=-1}$)\\
      \hypertarget{S9}{$\ztdSG_\gamma$} & [H,\B] = -\P & [H,\P] = \gamma\B + (1+\gamma)\P & & & & Torsional Galilean de~Sitter ($\gamma\in (-1,1]$) \\
      \hypertarget{S10}{$\zAdSG$} & [H,\B] =  -\P & [H,\P] = \B & & & & Galilean Anti-de~Sitter  ($\zAdSG = \ztAdSG_{\chi=0}$)\\
      \hypertarget{S11}{$\zAdSG_\chi$} & [H,\B] = -\P & [H,\P] = (1+\chi^2) \B + 2\chi \P & & & & Torsional Galilean Anti-de~Sitter ($\chi>0$) \\
      \midrule
      \hypertarget{S13}{$\zC$} & & & & [\B,\P] = H & & Carrollian Spacetime\\
      \hypertarget{S14}{$\zdSC$} & & [H,\P] = -\B & & [\B,\P] = H & [\P,\P] = \J & Carrollian de~Sitter\\
      \hypertarget{S15}{$\zAdSC$} & & [H,\P] = \B & & [\B,\P] = H & [\P,\P] = -\J & Carrollian Anti-de~Sitter\\
      \hypertarget{S16}{$\zLC$} & [H,\B] = \B & [H,\P] = -\P & & [\B,\P] = H - \J & & Carrollian Light Cone\\
      \midrule
      \hypertarget{A21}{$\zS$} & & & & & & Aristotelian Static \\
      \hypertarget{A22}{$\zTS$} & & [H,\P] = \P & & & & Torsional Aristotelian Static\\
      \hypertarget{A23p}{$\RR\times S^3$} & & & & & [\P,\P] = \J & Einstein Static Universe\\
      \hypertarget{A23m}{$\RR\times H^3$} & & & & & [\P,\P] = - \J & Hyperbolic Einstein Static Universe\\
      \bottomrule
    \end{tabular}
  }
\end{table}
Notice, this table has been divided into the five  kinematical symmetry classes; namely, from top to bottom, Lorentzian, Riemannian, Galilean, Carrollian, and Aristotelian.  Now that we have all these spacetime models collected in one place, we can highlight the distinguishing algebraic features of each class.  In particular, we note that it is possible to determine the class of spacetime by inspecting only the $[\bH, \B]$ and $[\B, \P]$ brackets.  
\\ \\
Starting from the bottom, we can see that all Aristotelian algebras are without either of these brackets.  This fact is not surprising given that Aristotelian algebras do not have the $\B$ generators; still, when analysing kinematical algebras, the absence of $[\bH, \B]$ and $[\B, \P]$ tells us we have an Aristotelian spacetime.  Next, we have the Carrollian spacetimes, which are distinguished by the brackets $[\bH, \B] = 0$ and $[\B, \P] = \bH$.  Thus, we may think of the presence of the bracket $[\B, \P] = \bH$ as telling us that the kinematical spacetime has an associated Carrollian structure.  Conversely, for the Galilean spacetimes, this bracket vanishes; we have $[\bH, \B] = -\P$ and $[\B, \P] = 0$.  In an analogous manner, we may say that the presence of the bracket $[\bH, \B] = -\P$ tells us that the kinematical spacetime has an associated Galilean structure.  Putting the Galilean and Carrollian brackets together, such that we have $[\bH, \B] = -\P$ and $[\B, \P] = \bH$, gives us a Lorentzian spacetime.  The fact a Lorentzian spacetime combines Galilean and Carrollian kinematics has a nice interpretation in terms of limits.  In brief, when we let the speed of light $c$ go to zero, we recover Carrollian kinematics, and when we take the limit $c \rightarrow \infty$, we recover Galilean kinematics.  This story will be reviewed in the next section.  Finally, Riemannian spacetimes are characterised by the brackets $[\bH, \B] = \P$ and $[\B, \P] = \bH$.  Notice that the only change from the Lorentzian case is the change in sign for $[\bH, \B]$.  This sign change in the Lie brackets changes the signature of the $\h$-invariant non-degenerate symmetric bilinear form in $\k$, such that it integrates to a Riemannian as opposed to a Lorentzian metric.  This discussion on the characteristic Lie brackets of the various kinematical symmetry classes is summarised in Table~\ref{tab:categories}.

\begin{table}[h!]
  \centering
  \caption{The Characteristic Lie Brackets of the Kinematical Symmetry Classes}
  \label{tab:categories}
  \begin{tabular}{l| l l }\toprule
    Class & $[\bH, \B]$ & $[\B, \P]$ \\ \midrule
    Lorentzian  & $-\P$ & $\bH$ \\
    Riemannian & $\P$ & $\bH$ \\
    Galilean  & $-\P$ & 0 \\
    Carrollian & 0 & $\bH$ \\
    Aristotelian & - & - \\\bottomrule
  \end{tabular}
\end{table}

\section{Limits Between Spacetimes}
\label{subsec:ks_kss_gls}

Now that we have seen the kinematical spacetimes we can construct
from the possible kinematical Lie algebras, we can investigate how the
different models are related to one another.  In this section, we will see
that the spacetimes in Table~\ref{tab:spacetimes} are related via
geometric limits. 
It will also be shown that most of these limits
may be viewed as geometric interpretations of contractions of the
underlying kinematical Lie algebras. 
\\ \\
To set up this discussion on limits, we will first introduce the idea of a
Lie algebra contraction.  Equipped with this method of relating the
kinematical Lie algebras, we begin with the (semi-)simple Lie algebras
$\so(D + 1, 1)$, $\so(D, 2)$, and $\so(D+2)$, for the de~Sitter,
anti-de~Sitter, and round sphere, respectively.  We then show how
each of the other kinematical Lie algebras arises as a contraction
from one of these starting points.  We end this section by explaining
an additional limit which does not arise from a contraction.

\subsubsection{Contractions}
To define a Lie algebra contraction, we first choose to interpret the
Jacobi identity as cutting out an algebraic variety$\cJ$ in the real vector
space $\Lambda^2 V^* \otimes V$ of all possible anti-symmetric
$\mathbb{R}$-bilinear maps.  Notice, each point in this variety will
represent a different Lie algebra structure on the underlying vector space $V$.\footnote{This interpretation of the Jacobi identities has already been alluded to in Section~\ref{subsec:ks_ala_class}.  Additionally, it will be generalised to the supersymmetric case in Chapters~\ref{chap:k_superspaces} and~\ref{chap:gb_superspaces}, playing a crucial role in our classification of the kinematical Lie superalgebras
and generalised Bargmann superalgebras.}  However, many of these
Lie algebra structures may be equivalent.  In particular, basis changes
in the underlying vector space, given by $\GL(V)$, will give rise to
isomorphic Lie algebra structures.  Therefore, we need to investigate the
$\GL(V)$-orbits in $\cJ$.  We will see that Lie algebra contractions arise
quite naturally from these investigations.
\\ \\
Recall, an $n$-dimensional real Lie algebra consists of an 
$n$-dimensional real vector space $V$ equipped with an
anti-symmetric, $\mathbb{R}$-bilinear bracket $\varphi: \Lambda^2 V
\rightarrow V$, which satisfies the Jacobi identity.  As stated above, 
we may think of the Jacobi identity as cutting out an algebraic variety 
$\cJ \subset \Lambda^2 V^* \otimes V$.  Notice, the basis changes
of the underlying vector space $\GL(V)$ will have an induced action
on $\Lambda^2 V^* \otimes V$; in particular, since the action is tensorial,
it will preserve $\cJ$.  Each $\GL(V)$-orbits in $\cJ$ then represents a
unique (up to isomorphism) Lie algebra structure $\k$ on $V$. Now, these
orbits may, or may not, be closed with respect to the induced topology
on $\cJ$.  If the closure of the orbit contains Lie algebra structures which
are not isomorphic to the original structure $\k$, then these non-isomorphic
structures are called ``degenerations''.  Lie algebra contractions are then
a specific form of Lie algebra degeneration; in particular, they are limits
of curves in the $\GL(V)$-orbit.  Let, $g: (0, 1] \to GL(V)$, mapping $t \mapsto g_t$,
be a continuous curve with $g_1 = \id_V$.  We can then define a curve of
isomorphic Lie algebras $(V,\phi_t)$, where
\begin{equation}
  \phi_t(X,Y) := \left(g^{-1}_t\cdot\phi \right)(X,Y) = g^{-1}_t \left(\phi(g_t X,
  g_t Y)\right).
\end{equation}
If the limit $\phi_0 = \lim_{t\to 0} \phi$ exists, it defines a Lie algebra
$\g_0 = (V, \phi_0)$ which is then a contraction of $\g=(V,\phi_1)$.

\subsection{Contraction Limits}
\label{subsec:contraction_limits}

Now that we know how to formulate a Lie algebra contraction, we may see
how various contractions can implement geometric limits between our
spacetimes.  
Recall, we wished to start our exploration
of the various contraction limits with the Lie algebras $\so(D + 1, 1)$, $\so(D, 2)$,
and $\so(D+2)$.  To avoid any repetition, consider $\mathbb{R}^{D+2}$, and
define a basis $e_M = (e_i, e_{\tilde{0}}, e_{\tilde{1}})$, where $1 \leq i \leq D$.  On
this space, we define the inner product such that $\eta(e_i, e_j) = \delta_{ij}$, and,
$\eta(e_{\tilde{0}}, e_{\tilde{0}}) =: \varsigma$ and $\eta(e_{\tilde{1}}, e_{\tilde{1}}) =: \varkappa$
are the only other non-vanishing entries.  Using this basis, we can view the
typical kinematical Lie algebra generators as being embedded into the higher-dimensional
generators $\{J_{MN}\}$, for $1 \leq M, N, \leq D+2$, as
\begin{equation}
	\bJ_{ij} := J_{ij}, \quad \bB_i := J_{\tilde{0}i}, \quad \bP_i := J_{i\tilde{1}}, \quad
	\bH := J_{\tilde{0}\tilde{1}}.
\end{equation}
The Lie brackets of our $(D+2)$-dimensional Lie algebra then produce the following
kinematical Lie algebra $\k$,
\begin{equation} \label{eq:general_brackets}
	\begin{split}
	[\bJ_{ij}, \bJ_{kl}] &= \delta_{jk} \bJ_{il} - \delta_{ik} \bJ_{jl} - \delta_{jl} \bJ_{ik} + \delta_{il} \bJ_{jk}, \\ 			[\bJ_{ij}, \bB_k] &= \delta_{jk} \bB_{i} - \delta_{ik} \bB_j, \\
	[\bJ_{ij}, \bP_k] &= \delta_{jk} \bP_{i} - \delta_{ik} \bP_j, \\ 
	[\bH, \bB_i] &= \varsigma \bP_i 
	\end{split} \quad 
	\begin{split}
		[\bH, \bP_i] &= - \varkappa \bB_i \\
		[\bB_i, \bP_j] &= \delta_{ij} \bH \\
		[\bB_i, \bB_j] &= - \varsigma \bJ_{ij} \\
		[\bP_i, \bP_j] &= - \varkappa \bJ_{ij}. 
	\end{split}
\end{equation}
Given these brackets, we can now investigate how our choice of $\varsigma$ and $\varkappa$
affect our possible spacetime models.  This picture is perhaps clearest when looking
at
\begin{equation} \label{eq:sigma_kappa_exp}
	[\bB_i, \bB_j] = - \varsigma \bJ_{ij} \quad \text{and} \quad [\bP_i, \bP_j] = - \varkappa \bJ_{ij}.
\end{equation}
The first bracket above tells us that, for $\varsigma \neq 0$, the $\J$ and $\B$ generators close
to form a Lie subalgebra $\h \subset \k$, which will be isomorphic to either $\so(D, 1)$ or
$\so(D+1)$ depending on the sign of $\varsigma$.  In the geometric picture, $\h$ is our admissible
Lie subalgebra in the kinematical Lie pair $(\k, \h)$. From the discussion in Section~\ref{sec:invariant-tensors},
we know that the $\h$-invariant tensors on $\k/\h$ integrate to $\Kgr$-invariant tensor fields
on $\M \cong \Kgr/\Hgr$; therefore, the change in the signature of this algebra will directly
impact the signature of our metric on the spacetime model.  In particular, if $\varsigma = -1$, we
will induce a Lorentzian spacetime model, and, if $\varsigma = +1$, we will induce a Riemannian spacetime model.  This connection between $\varsigma$ and the invariant structure on the spacetime will be made more
precise in Section~\ref{sec:invariant-structures}.
\\ \\
The second bracket in~\eqref{eq:sigma_kappa_exp} gives us an analogous algebraic
perspective of $\varkappa$; in particular, for $\varkappa \neq 0$, the $\J$ and $\P$ generators
close to form a Lie subalgebra $\h \subset \k$, which is isomorphic to either  $\so(D, 1)$ or
$\so(D+1)$.  However, since $\P$ is not a generator of the admissible Lie subalgebra for
a kinematical Lie pair, we arrive at a different geometric perspective.  The fundamental
vector fields associated with $\P$ will move us around the spacetime manifold $\M$, whereas, since
$\J \in \h$, the fundamental vector fields associated with $\J$ will implement
transformations at a point; see Section~\ref{sec:ks_gps}.  Thus, this bracket is capturing the curvature of $\M$.
In particular, if $\varkappa = -1$, we induce a hyperbolic model, such as anti-de~Sitter
spacetime, and, if $\varkappa = +1$, we induce an elliptic model, such as de~Sitter
spacetime. 
\\ \\
Now that we have a kinematical Lie algebra $\k$ which can capture all three starting points,
$\so(D + 1, 1)$, $\so(D, 2)$, and $\so(D+2)$,
we can introduce the three-parameter family of linear transformations $g_{\kappa, c, \tau}$,
which will allow us to take the desired Lie algebra contractions.  These transformations
act on the generators by
\begin{equation}
g_{\kappa, c, \tau}\cdot \J = \J, \quad g_{\kappa, c, \tau}
\cdot\B = \tfrac{\tau}{c} \B, \quad g_{\kappa, c, \tau}\cdot
\P = \tfrac{\kappa}{c} \P, \quad g_{\kappa, c, \tau}\cdot \bH =
\tau\kappa \bH. 
\end{equation}
Putting these definitions into the brackets of $\k$, we have the transformed brackets
\begin{equation} \label{eq:transformed_brackets_1}
	\begin{split}
	[\bJ_{ij}, \bJ_{kl}] &= \delta_{jk} \bJ_{il} - \delta_{ik} \bJ_{jl} - \delta_{jl} \bJ_{ik} + \delta_{il} \bJ_{jk}, \\ 			[\bJ_{ij}, \bB_k] &= \delta_{jk} \bB_{i} - \delta_{ik} \bB_j, \\
	[\bJ_{ij}, \bP_k] &= \delta_{jk} \bP_{i} - \delta_{ik} \bP_j, \\ 
	[\bH, \bB_i] &= \tau^2 \varsigma \bP_i 
	\end{split} \quad 
	\begin{split}
		[\bH, \bP_i] &= - \kappa^2 \varkappa \bB_i \\
		[\bB_i, \bP_j] &= \tfrac{1}{c^2} \delta_{ij} \bH \\
		[\bB_i, \bB_j] &= - (\tfrac{\tau}{c})^2 \varsigma \bJ_{ij} \\
		[\bP_i, \bP_j] &= - (\tfrac{\kappa}{c})^2 \varkappa \bJ_{ij}. 
	\end{split}
\end{equation}
We now want to take the limits $\kappa \to 0$,
$c \to \infty$, and $\tau \to 0$ in turn,
corresponding to the flat, non-relativistic, and ultra-relativistic
limits, respectively.  We will see that by taking multiple limits, we arrive
at a ``web'' of relations, tying all of the kinematical spacetimes together.
Since the kinematical brackets are unaffected by
these limits, we will omit them from the following discussion, but it
should be clear that they still hold for all contractions.
\\ \\
Taking the first limit, $\kappa \rightarrow 0$, we reduce the set of non-vanishing
brackets to
\begin{equation} \label{eq:flat_limit_brackets}
	\begin{split}
		[\bH, \bB_i] &= \tau^2 \varsigma \bP_i \\
		[\bB_i, \bP_j] &= \tfrac{1}{c^2} \delta_{ij} \bH \\
		[\bB_i, \bB_j] &= - (\tfrac{\tau}{c})^2 \varsigma \bJ_{ij}. \\
	\end{split}
\end{equation}
This contraction explains multiple geometric limits depending on our
choices for $\varkappa$ and $\varsigma$.  These limits are summarised in
Table~\ref{tab:flat_limits}.
\begin{table}[h!]
  \centering
  \caption{Flat Limits}
  \label{tab:flat_limits}
  \begin{tabular}{*{2}{>{$}c<{$}}|l}\toprule
    \varsigma & \varkappa & Geometric Limit \\ \midrule
    -1 & 1 & $\hyperlink{S2}{\zdS} \to \hyperlink{S1}{\MM}$\\
    -1 & -1 & $\hyperlink{S3}{\zAdS} \to \hyperlink{S1}{\MM}$ \\\midrule
    1 & 1 & $\hyperlink{S5}{S} \to \hyperlink{S4}{\EE}$\\
    1 & -1 & $\hyperlink{S6}{H} \to \hyperlink{S4}{\EE}$\\\bottomrule
  \end{tabular}
\end{table}
From here, we can now take either the non-relativistic or ultra-relativistic limits.  First taking
$c \to \infty$, we are left with only
\begin{equation}
	[\bH, \bB_i] = \tau^2 \varsigma \bP_i.
\end{equation}
Letting $\tau$ be finite and non-vanishing, we thus arrive at the 
kinematical Lie algebra for the Galilean spacetime $\hyperlink{S7}{\zG}$.  Notice that
the sign of this bracket will depend on our starting point, whether it was
a Lorentzian or Riemannian spacetime.  Irrespective of the sign, both are
equally fair descriptions of the spacetime $\hyperlink{S7}{\zG}$; however, as we shall
see, this sign does impact the non-relativistic limits of curved spacetimes.
\\ \\
Returning to the kinematical Lie algebra given by the brackets in~\eqref{eq:flat_limit_brackets},
we can take the ultra-relativistic limit, $\tau \to 0$, giving
\begin{equation}
	[\bB_i, \bP_j] = \tfrac{1}{c^2} \delta_{ij} \bH.
\end{equation}
Letting $c$ be finite and non-vanishing, we thus arrive at the 
kinematical Lie algebra for the Carrollian spacetime $\hyperlink{S13}{\zC}$.
\\ \\
Notice that taking both the non-relativistic and ultra-relativistic limits causes all the brackets 
in~\eqref{eq:flat_limit_brackets} to vanish.  This produces a non-effective 
Lie pair, which may be used to describe an Aristotelian Lie pair; in particular, it describes
the Aristotelian Lie pair associated with the static spacetime, $\hyperlink{A21}{\zS}$.
\\ \\
Now taking the non-relativistic limit of the brackets in~\eqref{eq:transformed_brackets_1},
we find
\begin{equation}
		[\bH, \bB_i] = \tau^2 \varsigma \bP_i  \quad \text{and} \quad 
		[\bH, \bP_i] = - \kappa^2 \varkappa \bB_i. 
\end{equation}
This contraction explains multiple geometric limits depending on our
choices for $\varkappa$ and $\varsigma$.  These limits are summarised in
Table~\ref{tab:NR_limits}.
\begin{table}[h!]
  \centering
  \caption{Non-Relativistic Limits}
  \label{tab:NR_limits}
  \begin{tabular}{*{2}{>{$}c<{$}}|l}\toprule
    \varsigma & \varkappa & Geometric Limit \\ \midrule
    -1 & 1 & $\hyperlink{S2}{\zdS} \to \hyperlink{S8}{\zdSG}$\\
    -1 & -1 & $\hyperlink{S3}{\zAdS} \to \hyperlink{S10}{\zAdSG}$ \\\midrule
    1 & 1 & $\hyperlink{S5}{S} \to \hyperlink{S10}{\zAdSG}$\\
    1 & -1 & $\hyperlink{S6}{H} \to \hyperlink{S8}{\zdSG}$\\\bottomrule
  \end{tabular}
\end{table}
Notice that the non-relativistic spacetimes we contract to switch depending
on whether we begin with a Lorentzian or Riemannian spacetime.  This result
is an artefact of the sign change in $[\bH, \B]$ bracket discussed earlier.  Since
contraction limits commute, we already know that taking the flat limit will take
us to $\hyperlink{S7}{\zG}$, thus we find the limits $\hyperlink{S8}{\zdSG} \to \hyperlink{S7}{\zG}$,
 and $\hyperlink{S10}{\zAdSG} \to \hyperlink{S7}{\zG}$.  The ultra-relativistic
 limit then takes us back to $\hyperlink{A21}{\zS}$, as described previously.  
\\ \\
Finally, we consider the ultra-relativistic limit of the brackets in~\eqref{eq:transformed_brackets_1}
and arrive at
\begin{equation}
	\begin{split}
		[\bH, \bP_i] &= - \kappa^2 \varkappa \bB_i \\
		[\bB_i, \bP_j] &= \tfrac{1}{c^2} \delta_{ij} \bH \\
		[\bP_i, \bP_j] &= - (\tfrac{\kappa}{c})^2 \varkappa \bJ_{ij}. 
	\end{split}
\end{equation}
This contraction explains multiple geometric limits depending on our
choices for $\varkappa$ and $\varsigma$.  These limits are summarised in
Table~\ref{tab:UR_limits}.
\begin{table}[h!]
  \centering
  \caption{Ultra-Relativistic Limits}
  \label{tab:UR_limits}
  \begin{tabular}{*{2}{>{$}c<{$}}|l}\toprule
    \varsigma & \varkappa & Geometric Limit \\ \midrule
    -1 & 1 & $\hyperlink{S2}{\zdS} \to \hyperlink{S14}{\zdSC}$\\
    -1 & -1 & $\hyperlink{S3}{\zAdS} \to \hyperlink{S15}{\zAdSC}$ \\\midrule
    1 & 1 & $\hyperlink{S5}{S} \to \hyperlink{S14}{\zdSC}$\\
    1 & -1 & $\hyperlink{S6}{H} \to \hyperlink{S15}{\zAdSC}$\\\bottomrule
  \end{tabular}
\end{table}
Since the limits commute, we know that taking a subsequent flat limit 
will give us the spacetime $\hyperlink{S13}{\zC}$; thus, we find the limits
$ \hyperlink{S14}{\zdSC} \to \hyperlink{S13}{\zC}$ and
$\hyperlink{S15}{\zAdSC} \to \hyperlink{S13}{\zC}$.  Taking the
non-relativistic limit from $\hyperlink{S13}{\zC}$ then takes us to $\hyperlink{A21}{\zS}$,
as discussed previously. 
\\ \\
Table~\ref{tab:symmetric} summarises the spacetimes discussed in this
section so far.  They
can be characterised as those homogeneous kinematical spacetimes which
are symmetric, in the sense described in Section~\ref{subsec:math_prelims_geo_hs}.
The table divides into four sections
corresponding, from top to bottom, to Lorentzian, Riemannian, Galilean
and Carrollian symmetric spacetimes.\footnote{Note, the Arisotelian spacetimes
will not fit into this picture as they do not have the generator $\B$.  Additionally, we
have taken both the Lorentzian and Riemannian contractions into account when
describing the parameters for the Galilean spacetimes. }
\begin{table}[h!]
  \centering
  \caption{Symmetric Spacetimes}
  \label{tab:symmetric}
  \begin{tabular}{*{3}{>{$}c<{$}}|l}\toprule
    \varsigma & \varkappa & c^{-1} & Spacetime\\ \midrule
    -1 & 0 & 1 & Minkowski ($\hyperlink{S1}{\MM}$)\\
    -1 & 1 & 1 & de~Sitter ($\hyperlink{S2}{\zdS}$)\\
    -1 & -1 & 1 & anti-de~Sitter ($\hyperlink{S3}{\zAdS}$)\\\midrule
    1 & 0 & 1 & euclidean ($\hyperlink{S4}{\EE}$)\\
    1 & 1 & 1 & sphere ($\hyperlink{S5}{S}$)\\
    1 & -1 & 1 & hyperbolic ($\hyperlink{S6}{H}$)\\\midrule
    \minusplus 1 & 0 & 0 & Galilean ($\hyperlink{S7}{\zG}$)\\
    \minusplus 1 & \pm 1 & 0 & Galilean de~Sitter ($\hyperlink{S8}{\zdSG}$)\\
    \minusplus 1 &  \minusplus 1 & 0 & Galilean anti-de~Sitter ($\hyperlink{S10}{\zAdSG}$)\\\midrule
    0 & 0 & 1 & Carrollian ($\hyperlink{S13}{\zC}$)\\
    0 & 1 & 1 & Carrollian de~Sitter ($\hyperlink{S14}{\zdSC}$)\\
    0 & -1 & 1 & Carrollian anti-de~Sitter ($\hyperlink{S15}{\zAdSC}$)\\\bottomrule
  \end{tabular}
\end{table}
\\  \\
This discussion accounts for the majority of the limits between the classified
spacetimes in Table~\ref{tab:spacetimes}; however, it does not include the torsional Galilean
spacetimes $\hyperlink{S9}{\ztdSG_\gamma}$ and $\hyperlink{S11}{\zAdSG_\chi}$,
the Carrollian light cone $\hyperlink{S16}{\zLC}$, or the Aristotelian spacetimes
$\hyperlink{A22}{\zTS}$, $\hyperlink{A23p}{\RR\times S^3}$, and 
$\hyperlink{A23m}{\RR\times H^3}$.\footnote{It should be noted that here the use 
of the term ``torsional'' is not related to its use in the physics literature, where it 
states that the clock one-form of a Newton-Cartan structure satisfies $d\tau \neq 0$.}  
We will now consider each of the remaining
classes of limit in turn.
\subsection{Remaining Galilean Spacetimes}
In this short section, we will demonstrate how to recover the Galilean spacetime
$\hyperlink{S7}{\zG}$ as a geometric limit from the torsional Galilean spacetimes 
$\hyperlink{S9}{\ztdSG_\gamma}$ and $\hyperlink{S11}{\zAdSG_\chi}$.  In
particular, we will see that this geometric limit is induced by a contraction of
the underlying Lie algebras. 
\\ \\
Consider the following linear transformations
\begin{equation}
g_{t}\cdot \J = \J, \quad g_{t}\cdot\B = \B, \quad g_{t}\cdot
\P = t \P, \quad g_{t}\cdot \bH = t \bH. 
\end{equation}
Under these maps, the kinematical Lie algebras for $\hyperlink{S9}{\ztdSG_\gamma}$, 
 and $\hyperlink{S11}{\zAdSG_\chi}$ are transformed
to 
\begin{equation}
	\begin{split}
		[\bH, \B] &= - \P \\
		[\bH, \P] &= t^2 \gamma \B + t(1 + \gamma) \P
	\end{split} \quad \text{and} \quad 
	\begin{split}
		[\bH, \B] &= - \P \\
		[\bH, \P] &= t^2(1 + \chi^2) \B + 2t\chi \P,
	\end{split}
\end{equation}
respectively.  Notice that taking $t \to 0$ produces the kinematical Lie algebra
for the spacetime $\hyperlink{S7}{\zG}$ in both instances.  Thus, we arrive at
the geometric limits $\hyperlink{S9}{\ztdSG_\gamma} \to \hyperlink{S7}{\zG}$
and $\hyperlink{S11}{\zAdSG_\chi} \to \hyperlink{S7}{\zG}$.

\subsection{Limit of the Carrollian Light Cone}
In this brief section, we will demonstrate how to recover the Carrollian spacetime
$\hyperlink{S13}{\zC}$ and torsional static spacetime $\hyperlink{A22}{\zTS}$ 
as a geometric limit from the Carrollian light cone $\hyperlink{S16}{\zLC}$.  In
particular, we will see that these geometric limits are induced by a contraction of
the underlying Lie algebras. 
\\ \\
For the first limit, $\hyperlink{S16}{\zLC} \to \hyperlink{S13}{\zC}$, consider
the following linear transformations
\begin{equation}
g_{t}\cdot \J = \J, \quad g_{t}\cdot\B = \B, \quad g_{t}\cdot
\P = t \P, \quad g_{t}\cdot \bH = t \bH. 
\end{equation}
Under these maps, the kinematical Lie algebra for $\hyperlink{S16}{\zLC}$ is transformed
to 
\begin{equation}
		[\bH, \B] = t\B \quad
		[\bH, \P] = t\P \quad
		[\B, \P] = \bH + t\J.
\end{equation}
Notice that taking $t \to 0$ produces the kinematical Lie algebra
for the spacetime $\hyperlink{S13}{\zC}$.  Thus, we arrive at the geometric
limit $\hyperlink{S16}{\zLC} \to \hyperlink{S13}{\zC}$ from a contraction of the
underlying kinematical Lie algebra. 
\\ \\
The final limit for the Carrollian light cone is 
$\hyperlink{S16}{\zLC} \to \hyperlink{A22}{\zTS}$.  This will involve taking a quotient
after the appropriate contraction.  In particular, transform the basis as
\begin{equation}
g_{t}\cdot \J = \J, \quad g_{t}\cdot\B = \B, \quad g_{t}\cdot
\P = t \P, \quad g_{t}\cdot \bH =  \bH. 
\end{equation}
The transformed brackets for $\hyperlink{S16}{\zLC}$ are written
\begin{equation}
		[\bH, \B] = \B, \quad [\bH, \P] =  \P, \quad [\B, \P] = t\bH + t\J.
\end{equation}
Taking $t \to 0$ leaves only the first two brackets.  Notice, that
$[\bH, \B] = \B$ means that the span of the generators $\B$ form an ideal
in the Lie subalgebra $\h$; thus, we have a non-effective Lie pair.  Quotienting
by this ideal, we arrive at an Aristotelian Lie pair isomorphic to the one associated with
$\hyperlink{A22}{\zTS}$. 

\subsection{Aristotelian Limits}
The only contraction limits remaining are those associated with the Aristotelian
spacetimes $\hyperlink{A21}{\zS}$, $\hyperlink{A22}{\zTS}$,
$\hyperlink{A23p}{\RR\times S^3}$, and $\hyperlink{A23m}{\RR\times H^3}$.
Here we will show that the Aristotelian Lie
algebras associated with the latter three
spacetimes all have the Lie algebra corresponding to $\hyperlink{A21}{\zS}$ as a 
contraction.
\\ \\
First, let us demonstrate the limit $\hyperlink{A22}{\zTS} \to \hyperlink{A21}{\zS}$.
Consider the linear transformations 
\begin{equation}
g_{t}\cdot \J = \J, \quad g_{t}\cdot
\P = \P, \quad g_{t}\cdot \bH = t \bH. 
\end{equation}
The transformed bracket for $\hyperlink{A22}{\zTS}$ is written
\begin{equation}
		[\bH, \P] = t \P.
\end{equation}
Taking the limit $t \to 0$, this final bracket
vanishes, leaving the Aristotelian Lie algebra for $\hyperlink{A21}{\zS}$.
\\ \\
The last limits to describe are $\hyperlink{A23p}{\RR\times S^3} \to
\hyperlink{A21}{\zS}$ and $\hyperlink{A23m}{\RR\times H^3} \to 
\hyperlink{A21}{\zS}$.  Notice that the Aristotelian analogue of the brackets
in~\eqref{eq:general_brackets} are given by
\begin{equation}
	\begin{split}
		[\bJ_{ij}, \bJ_{kl}] &= \delta_{jk} \bJ_{il} - \delta_{ik} \bJ_{jl} - \delta_{jl} \bJ_{ik} + \delta_{il} \bJ_{jk}, \\ 		
		[\bJ_{ij}, \bP_k] &= \delta_{jk} \bP_{i} - \delta_{ik} \bP_j, \\ 
		[\bP_i, \bP_j] &= - \varkappa \bJ_{ij},
	\end{split}
\end{equation}
where $\varkappa = +1$ describes $\hyperlink{A23p}{\RR\times S^3}$ and
$\varkappa = -1$ describes $\hyperlink{A23m}{\RR\times H^3}$.
Now, transform these brackets using the linear transformations $g_{t}$, which act on the basis as
\begin{equation}
	g_t\cdot \J = \J, \quad g_t\cdot\P = t\P, \quad g_t\cdot \bH = \bH.
\end{equation}
The transformed brackets are then
\begin{equation}
	\begin{split}
		[\bJ_{ij}, \bJ_{kl}] &= \delta_{jk} \bJ_{il} - \delta_{ik} \bJ_{jl} - \delta_{jl} \bJ_{ik} + \delta_{il} \bJ_{jk}, \\ 		
		[\bJ_{ij}, \bP_k] &= \delta_{jk} \bP_{i} - \delta_{ik} \bP_j, \\ 
		[\bP_i, \bP_j] &= - t^2 \varkappa \bJ_{ij}. 
	\end{split}	
\end{equation}
Letting $t \to 0$, we have the contraction giving rise to the geometric limits
$\hyperlink{A23p}{\RR\times S^3} \to
\hyperlink{A21}{\zS}$ and $\hyperlink{A23m}{\RR\times H^3} \to 
\hyperlink{A21}{\zS}$.

\subsection{A Non-Contracting Limit}

The discussion in the previous sections covers the majority of the geometric limits
between the spacetimes of Table~\ref{tab:spacetimes}; however, there is one final limit which does
not arise as a contraction.  This limit is found by considering the limit $\chi \to \infty$
for the torsional Galilean algebra $\hyperlink{S11}{\zAdSG_\chi}$
\begin{equation}
		[\bH, \B] = - \P \quad [\bH, \P] = (1 + \chi^2) \B + 2\chi \P.
\end{equation}
To take this limit, we must transform the basis as follows
\begin{equation}
	\tilde{\bH} = \chi^{-1} \bH, \quad \tilde{\B} = \B, \quad \tilde{\P} = \chi^{-1} \P.
\end{equation}
The new brackets take the form
\begin{equation}
	[\tilde{\bH}, \tilde{\B}] = - \tilde{\P}, \quad [\tilde{\bH}, \tilde{\P}] = (1 + \chi^{-2}) \tilde{\B}
	+ 2\tilde{\P}.
\end{equation}
Letting $\chi \to \infty$, we arrive at the Lie algebra corresponding to the
spacetime $\hyperlink{S9}{\ztdSG_1}$.

\subsection{Summary}

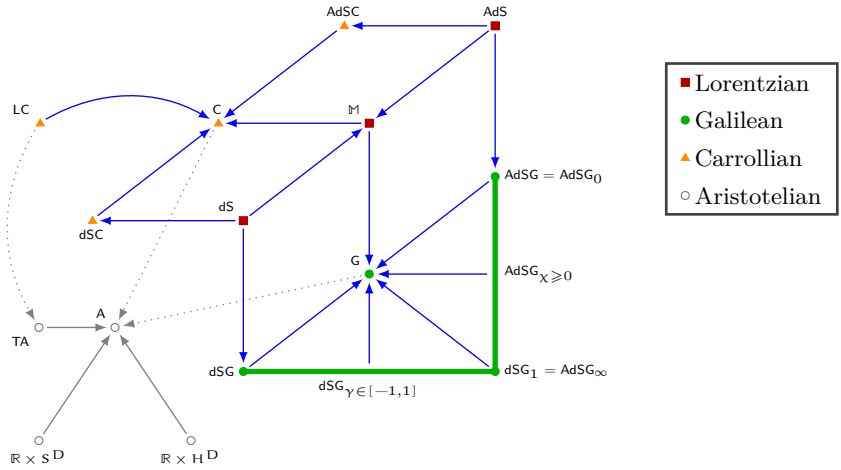
\begin{figure}[h!]
  \centering
  \begin{tikzpicture}[scale=1,>=latex, shorten >=3pt, shorten <=3pt, x=1.0cm,y=1.0cm]
    %
    %
    %
    %
    \coordinate [label=above left:{\hyperlink{S2}{\tiny $\zdS$}}] (ds) at (5.688048519056286,-0.5838170592960186);
    \coordinate [label=left:{\hyperlink{S8}{\tiny $\zdSG$}}] (dsg) at (5.688048519056286, -2.5838170592960186);
    \coordinate [label=right:{\hyperlink{S9}{\tiny $\ztdSG_1 =\ztAdSG_\infty$}}] (dsgone) at (9, -2.5838170592960186);
    \coordinate [label=below:{\hyperlink{S14}{\tiny $\zdSC$}}] (dsc) at  (3.688048519056286,-0.5838170592960186);
    \coordinate [label=above:{\hyperlink{S13}{\tiny $\zC$}}] (c) at (5.344024259528143, 0.7080914703519907);
    \coordinate [label=above left:{\hyperlink{S1}{\tiny $\MM$}}] (m) at (7.344024259528143,0.7080914703519907);
    \coordinate [label=above left:{\hyperlink{S7}{\tiny $\zG$}}] (g) at  (7.344024259528143, -1.2919085296480093);
    \coordinate [label=above:{\hyperlink{S3}{\tiny $\zAdS$}}] (ads) at (9,2);
    \coordinate [label=right:{\hyperlink{S10}{\tiny $\zAdSG =\ztAdSG_0$}}] (adsg) at (9,0);
    \coordinate [label=above:{\hyperlink{S15}{\tiny $\zAdSC$}}]  (adsc) at (7,2);
    \coordinate [label=above left:{\hyperlink{A21}{\tiny $\zS$}}] (s) at (4, -2);
    \coordinate [label=above left:{\hyperlink{S16}{\tiny $\zLC$}}]  (flc) at (3, 0.7080914703519907);
    \coordinate [label=below left:{\hyperlink{A22}{\tiny $\zTS$}}]  (ts) at (3, -2); 
    \coordinate [label=below:{\hyperlink{A23p}{\tiny $\RR\times S^D$}}]  (esu) at (3, -3.5); 
    \coordinate [label=below:{\hyperlink{A23m}{\tiny $\RR\times H^D$}}]  (hesu) at (5, -3.5); 
    %
    %
    \coordinate [label=below:{\hyperlink{S9}{\tiny $\ztdSG_{\gamma\in[-1,1]}$}}] (tdsg) at (7.344024259528143, -2.5838170592960186);
    \coordinate [label=right:{\hyperlink{S11}{\tiny $\ztAdSG_{\chi\geq0}$}}] (tadsg) at (9, -1.2919085296480093);
    %
    %
    \draw [->,line width=0.5pt,dotted,color=gray] (c) -- (s);
    \draw [->,line width=0.5pt,dotted,color=gray] (g) -- (s);
    \draw [->,line width=0.5pt,color=blue] (dsgone) -- (g);
    \draw [->,line width=0.5pt,color=blue] (7.344024259528143,-2.5838170592960186) -- (g);
    \draw [->,line width=0.5pt,color=blue] (9, -1.2919085296480093) -- (g);
    %
    %
    \draw [->,line width=0.5pt,color=blue] (adsc) -- (c);
    \draw [->,line width=0.5pt,color=blue] (dsc) -- (c);
    \draw [->,line width=0.5pt,color=blue] (ads) -- (m);
    \draw [->,line width=0.5pt,color=blue] (adsg) -- (g);
    \draw [->,line width=0.5pt,color=blue] (dsg) -- (g);
    \draw [->,line width=0.5pt,color=blue] (ds) -- (m);
    \draw [->,line width=0.5pt,color=blue] (ds) -- (dsc);
    \draw [->,line width=0.5pt,color=blue] (ds) -- (dsg);
    \draw [->,line width=0.5pt,color=blue] (m) -- (c);
    \draw [->,line width=0.5pt,color=blue] (m) -- (g);
    \draw [->,line width=0.5pt,color=blue] (ads) -- (adsc);
    \draw [->,line width=0.5pt,color=blue] (ads) -- (adsg);
    %
    %
    \draw [->,line width=0.5pt,color=blue] (flc) to [out=30,in=150] (c);
    \draw [->,line width=0.5pt,dotted,color=gray] (flc) to [out=240,in=120] (ts);
    \draw [->,line width=0.5pt,color=gray] (ts) to (s); 
    \draw [->,line width=0.5pt,color=gray] (esu) to (s); 
    \draw [->,line width=0.5pt,color=gray] (hesu) to (s); 
    %
    %
    \begin{scope}[>=latex, shorten >=0pt, shorten <=0pt, line width=2pt, color=green!70!black]
      \draw (adsg) --(dsgone);
      \draw (dsg) -- (dsgone);
    \end{scope}
    \foreach \point in {g,adsg,dsg,dsgone}
    \filldraw [color=green!70!black,fill=green!70!black] (\point) circle (1.5pt);
    \foreach \point in {ads,ds,m}
    \filldraw [color=red!70!black,fill=red!70!black] (\point) ++(-1.5pt,-1.5pt) rectangle ++(3pt,3pt);
    \foreach \point in {adsc,dsc,flc,c}
    \filldraw [color=DarkOrange,fill=DarkOrange] (\point) ++(-1pt,-1pt) -- ++(3pt,0pt) -- ++(-1.5pt,2.6pt) -- cycle;
    \foreach \point in {s,ts,esu,hesu}
    \draw [color=gray!90!black] (\point) circle (1.5pt);
    %
    %
    \begin{scope}[xshift=0.5cm]
    \draw [line width=1pt,color=gray!50!black] (10.75,-0.5) rectangle (13,1.5);
    \filldraw [color=red!70!black,fill=red!70!black] (11,1.25) ++(-1.5pt,-1.5pt) rectangle ++(3pt,3pt) ; 
    \draw (11,1.25) node[color=black,anchor=west] {\small Lorentzian}; 
    \filldraw [color=green!70!black,fill=green!70!black] (11,0.75) circle (1.5pt) node[color=black,anchor=west] {\small Galilean};
    \filldraw [color=DarkOrange,fill=DarkOrange] (11,0.25) ++(-1.5pt,-1pt) -- ++(3pt,0pt) -- ++(-1.5pt,2.6pt) -- cycle;
    \draw (11,0.25) node[color=black,anchor=west] {\small Carrollian};
    \draw [color=gray!90!black] (11,-0.25) circle (1.5pt) node[color=black,anchor=west] {\small Aristotelian};       
    \end{scope}
  \end{tikzpicture}
  \caption{Homogeneous Spacetimes in Dimension $D \geq 3$ and Their Limits.}
  \label{fig:generic-d-graph}
\end{figure}

The picture resulting from the above discussion is given in
Figure~\ref{fig:generic-d-graph}.
There are several types of limits displayed in
Figure~\ref{fig:generic-d-graph}:
\begin{itemize}
\item \emph{flat limits} in which the curvature of the canonical
  connection goes to zero: $\zAdS \to \MM$, $\zdS \to \MM$, $\zAdSC \to
  \zC$, $\zdSC \to \zC$, $\zAdSG \to \zG$ and $\zdSG \to \zG$;
\item \emph{non-relativistic limits} in which the speed of light goes
  to infinity: $\MM \to \zG$, $\zAdS \to \zAdSG$
  and $\zdS \to \zdSG$;

  In this limit there is still the notion of relativity,
  it just differs from the standard Lorentzian one.  Therefore,
   it might be more appropriate to call it the ``Galilean
  limit''.
\item \emph{ultra-relativistic limits} in which the speed of light
  goes to zero: $\MM \to \zC$, $\zAdS \to
  \zAdSC$ and $\zdS \to \zdSC$.
\item limits to non-effective Lie pairs which, after quotienting by the
  ideal generated by the boosts, result in an Aristotelian spacetime:
  the dotted arrows $\zLC \to  \zTS$, $\zC \to \zS$ and $\zG \to \zS$;
\item $\zLC \to \zC$, which is a contraction of $\so(D+1,1)$;
\item $\ztdSG_\gamma \to \zG$ and $\ztAdSG_\gamma \to \zG$, which are
  contractions of the corresponding kinematical Lie algebras;
\item limits between Aristotelian spacetimes $\zTS \to \zS$, $\RR
  \times S^D \to \zS$ and $\RR \times H^D \to \zS$; and
\item a limit $\lim_{\chi \to \infty} \zAdSG_\chi = \zdSG_1$, which is
  not due to a contraction of the kinematical Lie algebras.
\end{itemize}
Note, we can compose these limits like arrows in a commutative diagram.

\section{Geometric Properties of Kinematical Spacetimes} \label{sec:ks_gps}
The two previous sections reviewed not only the classification of the kinematical
spacetimes, but also described a method of relating the spacetimes through
geometric limits.  From our perspective, the crucial aspect of this discussion
was that both the classification and the limits arose from underlying algebraic 
procedures.  We will extend this connection between the algebra and geometric
aspects of these spacetimes in this section; in particular, we will now derive
invariant connections, invariant structures, fundamental vector fields, soldering
forms and vielbeins using techniques which make extensive use of the
underlying kinematical Lie algebra. 
\\ \\
This section is organised as follows.  In Section~\ref{sec:rotations}, we begin by
showing that the rotations $\J$ act as expected on every kinematical spacetime. 
This uniform treatment of fundamental vector fields is only possible for the rotations 
since, by definition, the rotations act the same in every kinematical spacetime; 
namely, their action is given by the kinematical brackets in~\eqref{eq:kinematical_brackets_general}.  In 
Section~\ref{sec:boosts}, we turn our attention to the action of the boosts $\B$.  
Here, we will argue that the boosts have non-compact orbits in all Lorentzian, Galilean and Carrollian
spacetimes, although, for some instances, complete proofs are postponed until 
later in the chapter. In Section~\ref{sec:nomizu},
we derive and discuss the space of invariant affine connections for each kinematical spacetime.  It will be shown that this space is heavily dependent on the
class of the kinematical spacetime being considered, therefore, this discussion
is split to describe each class separately.  In Section~\ref{sec:metric}, we derive the
geometric properties of the symmetric spacetimes highlighted in Table~\ref{tab:symmetric}.
As shown in Section~\ref{subsec:ks_kss_gls}, these spacetimes may be
considered together, with each spacetime being recovered by taking
relevant limits.  This method of taking limits will be utilised in deriving the
invariant structures, fundamental vector fields, soldering
forms and vielbeins for these spacetimes.  Unfortunately, this method
is not available when consider the torsional Galilean spacetimes, Aristotelian
spacetimes, or the Carroll light cone; therefore, the geometric properties of
these spacetimes are derived on a case-by-case basis in Sections~\ref{sec:galilean},
\ref{sec:aristotelian}, and~\ref{sec:spacetime-flc}, respectively. Note, we will use the
exponential coordinates described in Section~\ref{math_prelims_exp_coords}
throughout.  

\subsection{The Action of the Rotations}
\label{sec:rotations}

In this short section, we will derive the fundamental vector field $\xi_{\bJ_{ij}}$, 
corresponding to the action of the rotations on our kinematical spacetimes.   
We will see rotations act in
the way we may naively expect on the exponential coordinates: namely,
$t$ is a scalar and $x^i$ is a vector.
\\ \\
The infinitesimal action of the rotational generators $\bJ_{ij}$ on the
exponential coordinates can be deduced from
\begin{equation}
  [\bJ_{ij}, \bH ] = 0 \quad\text{and}\quad [\bJ_{ij}, \bP_k] = \delta_{jk}
  \bP_i - \delta_{ik} \bP_j.
\end{equation}
To be concrete, consider $\bJ_{12}$, which rotates $\bP_1$ and $\bP_2$ into
each other:
\begin{equation}
  [\bJ_{12}, \bP_1] = - \bP_2 \quad\text{and}\quad [\bJ_{12},\bP_2] = \bP_1,
\end{equation}
but leaves $\bH$ and $\bP_3,\cdots,\bP_D$ inert.  We see that
$\ad^2_{\bJ_{12}} \bP_i = - P_i$ for $i = 1,2$, so that exponentiating,
\begin{equation}
  \begin{split}
    \exp(\theta \ad_{\bJ_{12}}) (t \bH + \x \cdot \P) &= t \bH + x^1 (\cos\theta
  \bP_1 - \sin\theta \bP_2) + x^2 (\cos\theta \bP_2 + \sin\theta \bP_1) +
  x^3\bP_3 + \cdots x^D \bP_D\\
  &= t \bH + (x^1 \cos\theta + x^2 \sin\theta) \bP_1 + (x^2\cos\theta -
  x^1\sin\theta) \bP_2 + x^3 \bP_3 + \cdots + x^D \bP_D.
  \end{split}
\end{equation}
Restricting attention to the $(x^1,x^2)$ plane, we see that the orbit
of $(x^1_0, x^2_0)$ under the one-parameter subgroup
$\exp(\theta \bJ_{12})$ of rotations is
\begin{equation}
  \label{eq:rot}
    \begin{pmatrix}
    x^1(\theta) \\
    x^2(\theta)
  \end{pmatrix} = 
  \begin{pmatrix}
    \cos\theta &  \sin\theta \\
    - \sin \theta & \cos\theta 
  \end{pmatrix} \cdot
  \begin{pmatrix}
    x^1_0\\
    x^2_0
  \end{pmatrix}.
\end{equation}
Differentiating $(x^1(\theta),x^2(\theta))$ with respect to
$\theta$ yields
\begin{equation}
  \frac{d x^1}{d \theta} = x^2 \quad\text{and}\quad   \frac{d x^2}{d \theta} = -x^1,
\end{equation}
so that
\begin{equation}
  \xi_{\bJ_{12}} = x^2 \frac{\d}{\d x^1} - x^1 \frac{\d}{\d x^2}.
\end{equation}
In the general case, and in the same way, we find
\begin{equation}\label{eq:fvf-rot}
 \xi_{\bJ_{ij}} = x^j \frac{\d}{\d x^i} - x^i \frac{\d}{\d x^j},
\end{equation}
which can be checked to obey the opposite Lie algebra
\begin{equation}
  [\xi_{\bJ_{ij}}, \xi_{\bJ_{kl}}] = - \delta_{jk} \xi_{\bJ_{il}} +
  \delta_{jl} \xi_{\bJ_{ik}} + \delta_{ik} \xi_{\bJ_{jl}} - \delta_{il}
  \xi_{\bJ_{jk}} = -\xi_{[\bJ_{ij},\bJ_{kl}]}.
\end{equation}

\subsection{The Action of the Boosts}
\label{sec:boosts}

For a homogeneous space $\M=\Kgr/\Hgr$ of a kinematical Lie group
$\Kgr$ to admit a physical interpretation as a genuine spacetime, one
would seem to require that the boosts act with non-compact
orbits~\cite{Bacry:1968zf}. Otherwise, it would be more suitable to
interpret them as (additional) rotations. In other words, if $(\k,\h)$
is the Lie pair describing the homogeneous spacetime, with $\h$ the
subalgebra spanned by the rotations and the boosts, then a desirable
geometrical property of $\M$ is that for all $X = w^i \bB_i \in \h$ the
orbit of the one-parameter subgroup $\Bgr_X \subset \Hgr$ generated by
$X$ should be homeomorphic to the real line. Of course, this
requirement is strictly speaking never satisfied: the ``origin'' of
$\M$ is fixed by $\Hgr$ and, in particular, by any one-parameter
subgroup of $\Hgr$, so its orbit under any $\Bgr_X$ consists of just
one point. Therefore the correct requirement is that the
\emph{generic} orbits be non-compact. It is interesting to note that
we impose no such requirements on the space and time translations.\footnote{We could wonder whether compact orbits for space and time translations might impose that the boost orbits should be compact; however, given that our choice of coordinates is not adapted to the group action, we can say very little about how the orbits of the different group elements interact. This point may be interesting to investigate in future studies.}
\\ \\
With the exception of the Carrollian light cone
$\hyperlink{S16}{\zLC}$, which will have to be studied separately, the
action of the boosts are uniform in each class of spacetimes:
Lorentzian, Riemannian, Galilean and Carrollian. (There are no boosts
in Aristotelian spacetimes.) We can read the action of the boosts
(infinitesimally) from the Lie brackets:
\begin{itemize}
\item \emph{Lorentzian}:
  \begin{equation}
    [\B, \bH] = \P, \quad [\B, \P] = \bH \quad\text{and}\qquad [\B,\B] = \J;
  \end{equation}
\item \emph{Riemannian}:
  \begin{equation}
    [\B, \bH] = -\P, \qquad [\B, \P] = \bH \quad\text{and}\quad [\B,\B] = -\J;
  \end{equation}
\item \emph{Galilean}:
  \begin{equation}
    [\B, \bH] = \P;
  \end{equation}
\item \emph{(Reductive) Carrollian}:
  \begin{equation}
    [\B, \P] = \bH;
  \end{equation}
\item and \emph{Carrollian Light Cone ($\hyperlink{S16}{\zLC}$)}:
  \begin{equation}
    [\B, \bH] = -\B \quad\text{and}\quad [\B,\P] = \bH + \J.
  \end{equation}
\end{itemize}
Below we will calculate the action of the boosts for all spacetimes
except for the Carrollian light cone which will
be studied separately.
\\ \\
In order to simplify the calculation, it is convenient to introduce
the parameters $\varsigma$ and $c$ from Section~\ref{subsec:contraction_limits}, 
and write the infinitesimal action of the boosts as
\begin{equation}\label{eq:inf-boosts}
  [\bB_i,\bH] = -\varsigma \bP_i \quad\text{and}\quad [\bB_i,\bP_j] =
  \frac{1}{c^2} \delta_{ij} \bH.
\end{equation}
Then $(\varsigma,c^{-1}) = (-1,1)$ for Lorentzian,
$(\varsigma,c^{-1}) = (1,1)$ for Riemannian,
$(\varsigma,c^{-1}) = (-1,0)$ for Galilean and
$(\varsigma,c^{-1}) = (0,1)$ for (reductive) Carrollian spacetimes.
\\ \\
The action of the boosts on the exponential coordinates is given by
equation~\eqref{eq:exp-coord-action}, which in this case becomes
\begin{equation}
  t \bH + \x \cdot \P \mapsto \exp(\ad_{\w \cdot \B}) ( t \bH + \x \cdot \P ).
\end{equation}
From \eqref{eq:inf-boosts}, we see that
\begin{equation}
  \begin{aligned}[m]
    \ad_{\w\cdot \B} \bH &= -\varsigma \w \cdot \P \\
    \ad^2_{\w\cdot \B} \bH &= -\frac{1}{c^2} \varsigma w^2 \bH, \\
  \end{aligned}
  \quad\text{and}\quad
  \begin{aligned}[m]
    \ad_{\w\cdot \B} \P &= \frac{1}{c^2} \w \bH \\
    \ad^2_{\w\cdot \B} \P &= -\frac1{c^2} \varsigma \w (\w\cdot\P),\\
  \end{aligned}
\end{equation}
so that in all cases $\ad^3_{\w\cdot \B} = -\frac1{c^2} \varsigma w^2
\ad_{\w \cdot \B}$.  This allows us to exponentiate $\ad_{\w\cdot\B}$ easily:
\begin{equation}
  \exp(\ad_{\w\cdot\B}) = 1 + \frac{\sinh z}{z} \ad_{\w \cdot \B}
  + \frac{\cosh z - 1}{z^2} \ad^2_{\w \cdot \B},
\end{equation}
where $z^2 = -\frac1{c^2} \varsigma w^2$, and hence
\begin{equation}
  \begin{split}
    \exp(\ad_{\w\cdot\B}) t \bH &= t \cosh z \bH - \varsigma t \frac{\sinh z}{z} \w \cdot \P,\\
    \exp(\ad_{\w\cdot\B}) \x \cdot \P &= \x \cdot \P +  \frac1{c^2}
    \frac{\sinh z}{z} \x \cdot \w \bH + \frac{\cosh z - 1}{w^2}
    (\x \cdot \w) \w \cdot \P.
  \end{split}
\end{equation}
Therefore, the orbit of $(t_0,\x_0)$ under $\exp(s\w\cdot \B)$ is given
by
\begin{equation}\label{eq:boost-orbit}
  \begin{split}
    t(s) &= t_0 \cosh(s z) + \frac1{c^2} \frac{\sinh(s z)}{z}\x_0 \cdot \w,\\
    \x(s) &= \x_0^\perp - \varsigma t_0 \frac{\sinh(s z)}{z} \w +
    \frac{\cosh(s z)}{w^2} (\x_0 \cdot \w) \w,
  \end{split}
\end{equation}
where we have introduced
$\x_0^\perp := \x_0 - \frac{\x_0 \cdot \w}{w^2}\w$ to be the component
of $\x_0$ perpendicular to $\w$. It follows from this expression that
$\x^\perp(s) = \x_0^\perp$, so that the orbit lies in a plane spanned
by $\w$ and the time direction.
\\ \\
Differentiating these expressions with respect to $s$, we arrive at
the fundamental vector field $\xi_{\bB_i}$.  Indeed, differentiating
$(t(s),\x(s))$ with respect to $s$ at $s=0$, we obtain the value of
$\xi_{\w \cdot \B}$ at the point $(t_0,x_0)$.  Letting $(t_0,x_0)$
vary we obtain that
\begin{equation}\label{eq:fvf-boost}
  \xi_{\bB_i} = \frac{1}{c^2} x^i \frac{\d}{\d t} - \varsigma t \frac{\d}{\d x^i}.
\end{equation}
In particular, notice that one of the virtues of the exponential
coordinates, is that the fundamental vector fields of the stabiliser
$\h$ -- that is, of the rotations and the boosts -- are linear and, in
particular, they are complete. This will be useful in determining
whether or not the generic orbits of one-parameter subgroup of boosts
are compact.
\\ \\
Let $\exp(s \w\cdot\B)$, $s \in \RR$, be a one-parameter subgroup
consisting of boosts.  Given any $o \in \M$, its orbit under this
subgroup is the image of the map $c: \RR \to \M$, where
$c(s) := \exp(s \w\cdot \B)  o$.  As we just saw, in the
reductive examples (all but $\hyperlink{S16}{\zLC}$) the fundamental vector field
$\xi_{\w\cdot\B}$ is linear in the exponential coordinates, and hence
it is complete.  Therefore, its integral curves are one-dimensional
connected submanifolds of $\M$ and hence either homeomorphic to the
real line (if non compact) or to the circle (if compact).  The compact
case occurs if and only if the map $c$ is periodic.
\\ \\
If the exponential coordinates define a global coordinate chart (which
means, in particular, that the homogeneous space is diffeomorphic to
$\RR^{D+1}$), then it is only a matter of solving a linear ODE to
determine whether or not $c$ is periodic. In any case, we can
determine whether or not this is the case in the exponential
coordinate chart centred at the origin. For the special case of
symmetric spaces, which are the spaces obtained via limits from the
Riemannian and Lorentzian maximally symmetric spaces, we may use 
Theorem~\ref{thm:voglaire}, which gives an infinitesimal criterion for
when the exponential coordinates define a global chart. 
In particular, by inspecting Table~\ref{tab:spacetimes} and studying the
eigenvalues of $\ad_{\bH}$ and $\ad_{\bP_i}$ on $\k$, we may easily
determine the spacetimes which satisfy criterion (3) of Theorem~\ref{thm:voglaire},
and hence the spacetimes for which the exponential coordinates defines a 
diffeomorphism to $\RR^{D+1}$. These spacetimes are
$\hyperlink{S1}{\MM}$, $\hyperlink{S4}{\EE}$, $\hyperlink{S6}{H}$,
$\hyperlink{S7}{\zG}$, $\hyperlink{S8}{\zdSG}$,
$\hyperlink{S13}{\zC}$, and $\hyperlink{S15}{\zAdSC}$.
Using exponential coordinates, we will see that the orbits of boosts
in $\hyperlink{S4}{\EE}$ and $\hyperlink{S6}{H}$ are compact,
whereas the generic orbits of boosts in the other cases are
non-compact.
\\ \\
The remaining symmetric spacetimes $\hyperlink{S2}{\zdS}$,
$\hyperlink{S3}{\zAdS}$, $\hyperlink{S5}{S}$,
$\hyperlink{S10}{\zAdSG}$, and $\hyperlink{S14}{\zdSC}$ do not satisfy
the infinitesimal criterion (3) in Theorem~\ref{thm:voglaire}, and
hence the exponential coordinates are not a global chart. It may
nevertheless still be the case that the image of $\exp_o$ covers the
homogeneous spacetime (or a dense subset). It turns out that
$\hyperlink{S5}{S}$ is exponential and $\hyperlink{S10}{\zAdSG}$ is
weakly exponential. The result for $\hyperlink{S5}{S}$ is classical,
since the sphere is a compact Riemannian symmetric space, and the case
of $\hyperlink{S10}{\zAdSG}$ follows from Theorem~\ref{thm:rozanov}.
For $\hyperlink{S10}{\zAdSG}$, we have
the radicals $\rad\k = \spn{\B,\P,\bH}$ and $\rad\h = \spn{\B}$.
Therefore, $\k/\rad\k \cong \so(D) \cong \h/\rad\h$. Therefore, with
$\Khat := \Kgr/\Rad(\Kgr)$ and similarly for $\Hhat$, $\Khat/\Hhat$ is
trivially weakly exponential and hence, by Theorem~\ref{thm:rozanov},
so is $\Kgr/\Hgr$. We will see that boosts act with compact orbits in
$\hyperlink{S5}{S}$, but with non-compact orbits in
$\hyperlink{S10}{\zAdSG}$.
\\ \\
Among the symmetric spaces in Table~\ref{tab:spacetimes}, this leaves
$\hyperlink{S2}{\zdS}$, $\hyperlink{S3}{\zAdS}$, and
$\hyperlink{S14}{\zdSC}$. We treat those cases using the same
technique, which will also work for the non-symmetric
$\hyperlink{S16}{\zLC}$. Let $\M$ be a simply-connected homogeneous
spacetime and $q: \M \to \Mbar$ a covering map which is equivariant
under the action of (the universal covering group of) $\Kgr$. By
equivariance,
$q(\exp(s\w \cdot \B) o) =\exp(s\w\cdot\B)  q(o)$, so the
orbit of $o \in \M$ under the boost is sent by $q$ to the orbit of
$q(o) \in \Mbar$. Since $q$ is continuous it sends compact sets to
compact sets, so if the orbit of $q(o) \in \Mbar$ is \emph{not}
compact then neither is the orbit of $o \in \M$. For $\M$ one of
$\hyperlink{S2}{\zdS}$, $\hyperlink{S3}{\zAdS}$,
$\hyperlink{S14}{\zdSC}$, or $\hyperlink{S16}{\zLC}$, there is some
covering $q : \M \to \Mbar$ such that we can equivariantly embed
$\Mbar$ as a hypersurface in some pseudo-Riemannian space where $\Kgr$
acts linearly. It is a simple matter to work out the nature of the
orbits of the boosts in the ambient pseudo-Riemannian space (and hence
on $\Mbar$), with the caveat that what is a boost in $\Mbar$ need not
be a boost in the ambient space. Having shown that the boost orbit is
non-compact on $\Mbar$, we deduce that the orbit is non-compact on $\M$.
In this way, we will show that the generic boost orbits are non-compact
for $\hyperlink{S2}{\zdS}$, $\hyperlink{S3}{\zAdS}$,
$\hyperlink{S14}{\zdSC}$, and $\hyperlink{S16}{\zLC}$.
\\ \\
Finally, this still leaves the torsional Galilean spacetimes
$\hyperlink{S9}{\ztdSG_\gamma}$ and $\hyperlink{S11}{\ztAdSG_\chi}$, which require a
different argument to be explained when we discuss these spacetimes in
Section~\ref{sec:action-boosts-2}.

\subsection{Invariant Connections, Curvature, and Torsion for Reductive Spacetimes}
\label{sec:nomizu}

In this section, we determine the invariant affine connections for the
reductive spacetimes in Tables~\ref{tab:spacetimes}. This is equivalent to determining the space of
Nomizu maps which, can be done uniformly, a class
at a time.  We also calculate the curvature and torsion of the
invariant connections.
\\ \\
For reductive homogeneous spaces there always exists, besides the
canonical connection with vanishing Nomizu map, another interesting
connection.  It is given by the torsion-free connection
defined\footnote{%
  It is the unique Nomizu map with $\alpha(X,X)=0$ for all $X \in \m$
  and vanishing torsion and called ``canonical affine connection of
  the first kind'' in~\cite{MR0059050}.}  by
$\alpha(X,Y)= \frac{1}{2} [X,Y]_{\m}$.  The canonical and the natural
torsion-free connections coincide for symmetric spaces.
\\ \\
For any spacetime the Nomizu maps need to be rotationally invariant,
which, when $D=3$, gives us
\begin{equation}\label{eq:nomizu-rotations}
  \begin{aligned}[m]
    \alpha(\bH,\bH) &= \mu \bH \\
    \alpha(\bP_i,\bP_j) &=  \zeta \delta_{ij} \bH + \zeta' \epsilon_{ijk} \bP_k
  \end{aligned}
  \quad\quad
  \begin{aligned}[m]
  \alpha(\bH,\bP_i) &= \nu \bP_i \\
  \alpha(\bP_i,H) &= \xi \bP_i \\
  \end{aligned}
\end{equation}
for some real parameters $\mu,\nu,\zeta,\zeta',\xi$.
Now we simply impose invariance under $\bB_i$.

\subsubsection{Nomizu Maps for Lorentzian Spacetimes}
\label{sec:nomizu-maps-lorentz}

The Lorentzian spacetimes in Table~\ref{tab:spacetimes} all share the
same action of the boosts:
\begin{equation}
  \lambda_{\bB_i} \bH = \bP_i \quad\text{and}\quad \lambda_{\bB_i} \bP_i =
  \delta_{ij} \bH.
\end{equation}
  We will impose invariance explicitly in this case to illustrate the
calculation and only state the results in all other cases.
\\ \\
We calculate (remember~\eqref{eq:Nomizuinv})
\begin{equation}
  (\lambda_{\bB_k}\alpha)(\bP_i,\bP_j) = \zeta \delta_{ij} \bP_k + \zeta'
  \epsilon_{ijk} \bH - \nu \delta_{ik} \bP_j - \xi \delta_{jk} \bP_i,
\end{equation}
whose vanishing requires $\zeta = \zeta'= \nu = \xi = 0$,
as can be seen by considering $i=j\neq k$, $i=k \neq j$, and $j=k\neq i$ in turn.  
Finally,
\begin{equation}
  (\lambda_{\bB_k}\alpha)(\bH,\bH) = \mu \bP_k,
\end{equation}
whose vanishing imposes $\mu = 0$ and hence the only invariant Nomizu
map is the zero map.

\subsubsection{Nomizu Maps for Riemannian Spacetimes}
\label{sec:nomizu-maps-riemann}

The situation here is very similar to the Lorentzian case.  Now the
boosts act as
\begin{equation}
  \lambda_{\bB_i} \bH = - \bP_i \quad\text{and}\quad \lambda_{\bB_i} \bP_j =
  \delta_{ij} \bH.
\end{equation}
The results are as in the Lorentzian case: the only invariant
connection is the canonical connection.

\subsubsection{Nomizu Maps for Galilean Spacetimes}
\label{sec:nomizu-maps-galileo}

On a Galilean spacetime, the boosts act as
\begin{equation}
  \lambda_{\bB_i} \bH = \bP_i,
\end{equation}
and the $\bP_i$ are invariant.  This results in the following invariant
Nomizu maps:
\begin{equation}\label{eq:nomizu-galilean}
  \begin{aligned}[m]
    \alpha(\bH,\bH) &= (\nu + \xi) \bH \\
    \alpha(\bP_i,\bP_j) &= 0 
  \end{aligned}
  \quad\quad
  \begin{aligned}[m]
    \alpha(\bH,\bP_i) &= \nu \bP_i \\
    \alpha(\bP_i,\bH) &= \xi \bP_i \\
  \end{aligned}
\end{equation}
We will now analyse the curvature and torsion for these Nomizu maps
for each Galilean spacetime.

\paragraph{Galilean Spacetime ($\zG$)}

The torsion and curvature of the resulting connection
have the following non-zero components:
\begin{equation}
  \Theta(\bH,\bP_i) = (\nu - \xi) \bP_i \quad\text{and}\quad \Omega(\bH,\bP_i)\bH = -
  \xi^2 \bP_i.
\end{equation}
There is a unique torsion-free, flat invariant connection
corresponding to the canonical connection with $\nu = \xi = 0$.

\paragraph{Galilean de~Sitter Spacetime ($\zdSG$)}
\label{sec:galilean-de}

The torsion and curvature, given by
equation~\eqref{eq:reductive_tor_and_curv}, have the following non-vanishing
components:
\begin{equation}
  \Theta(\bH, \bP_i) = (\nu-\xi)\bP_i \quad \text{and} \quad \Omega(\bH, \bP_{i})\bH = (1-\xi^2)\bP_i.
\end{equation}
Therefore, there are two torsion-free, flat invariant connections
corresponding to $\nu=\xi=\pm 1$.  The Nomizu maps for these two
connections are
\begin{equation}
  \begin{aligned}[m]
    \alpha (\bH, \bH) &= 2 \bH \\
    \alpha (\bH, \bP_{i}) &= \bP_{i} \\
    \alpha(\bP_{i}, \bH) &= \bP_i
  \end{aligned} \quad \text{and} \quad
  \begin{aligned}[m]
    \alpha (\bH, \bH) &= -2 \bH \\
    \alpha (\bH, \bP_i) &= - \bP_i \\
    \alpha (\bP_i, \bH) &= -\bP_i.
  \end{aligned}
\end{equation}

\paragraph{Galilean anti-de~Sitter Spacetime ($\zAdSG$)}
\label{sec:galilean-anti-de}

The torsion and curvature have have the following non-zero components:
\begin{equation}
  \Theta(\bH,\bP_i) = (\nu-\xi) \bP_i \quad \text{and} \quad 
  \Omega(\bH, \bP_i) \bH = 
    - (1+\xi^{2})\bP_{i} 
\end{equation}
There are torsion-free connections, but none are flat.

\paragraph{Torsional Galilean de~Sitter Spacetime ($\ztdSG_{\gamma=1}$)}
\label{sec:tdsg1}

The torsion has the following non-zero components
\begin{equation}
  \Theta(\bH,\bP_i) = (\nu - \xi - 2 )\bP_i,
\end{equation}
whereas the only non-zero component of the curvature is
\begin{equation}
  \Omega(\bH,\bP_i)\bH = -(1 + \xi)^2\bP_i.
\end{equation}
Therefore, there exists a unique invariant connection with zero torsion
and curvature corresponding to $\nu=1$ and $\xi = -1$:
\begin{equation}
  \alpha(\bH,\bP_i) = \bP_i \quad\text{and}\quad \alpha(\bP_i,\bH) = - \bP_i.
\end{equation}

\paragraph{Torsional Galilean de~Sitter Spacetime ($\ztdSG_{\gamma\neq 1}$)}
\label{sec:other-invar-conn-tdsg}

In this instance, the torsion is given by
\begin{equation}
  \Theta(\bH,\bP_i) = (\nu - \xi - (1 + \gamma)) \bP_i
\end{equation}
and the curvature by
\begin{equation}
  \Omega(\bH,\bP_i)\bH = -(\xi + 1) (\xi + \gamma) \bP_i.
\end{equation}
Therefore, there are precisely two torsion-free, flat invariant
connections, with Nomizu maps
\begin{equation}
  \label{eq:tff-2-gamma-hd}
  \begin{aligned}[m]
    \alpha(\bH,\bH) &= (\gamma - 1) \bH\\
    \alpha(\bH,\bP_i) &= \gamma \bP_i\\
    \alpha(\bP_i,\bH) &= -\bP_i
  \end{aligned}
  \quad\quad\text{and}\quad\quad
  \begin{aligned}[m]
    \alpha(\bH,\bH) &= (1-\gamma) \bH\\
    \alpha(\bH,\bP_i) &= \bP_i\\
    \alpha(\bP_i,\bH) &= -\gamma \bP_i.
  \end{aligned}
\end{equation}

\paragraph{Torsional Galilean anti-de~Sitter Spacetime ($\ztAdSG_\chi$)}
\label{sec:other-invar-conn-tadsg}

The torsion and curvature of the connection corresponding to this
Nomizu map are given by the following non-zero components:
\begin{equation}
  \Theta(\bH,\bP_i) = (\nu - \xi - 2\chi) \bP_i \quad\text{and}\quad \Omega(\bH,\bP_i)\bH
  = - (1 + (\xi + \chi)^2) \bP_i.
\end{equation}
Therefore, we see that there are no flat invariant connections;
although there is a one-parameter family of torsion-free invariant
connections.

\subsubsection{Nomizu Maps for Carrollian Spacetimes}
\label{sec:nomizu-maps-carroll}

On a Carrollian spacetime, the boosts act as
\begin{equation}
  \lambda_{\bB_i} \bP_j = \delta_{ij} \bH,
\end{equation}
and $\bH$ is invariant.  This results in the following invariant
Nomizu maps:
\begin{equation}\label{eq:nomizu-carrollian}
  \begin{aligned}[m]
    \alpha(\bH,\bH) &= 0\\
    \alpha(\bP_i,\bP_j) &= \zeta \delta_{ij} \bH \\
  \end{aligned}
  \quad\quad
  \begin{aligned}[m]
    \alpha(\bH,\bP_i) &= 0 \\
    \alpha(\bP_i,\bH) &= 0. \\
  \end{aligned}
\end{equation}

\paragraph{Carrollian Spacetimes ($\zC$)}

In this case, the corresponding invariant connections are flat and torsion-free for
all values of $\zeta$.  

\paragraph{(Anti-)de~Sitter Carrollian Spacetimes ($\zdSC$ and $\zAdSC$)}

We will treat these two spacetimes together by introducing
$\varkappa=\pm 1$. Carrollian de~Sitter spacetime ($\hyperlink{S14}{\zdSC}$)
corresponds to $\varkappa=1$ and Carrollian anti-de~Sitter spacetime
($\hyperlink{S15}{\zAdSC}$) to $\varkappa =-1$.
\\ \\
The torsion vanishes and the curvature has the following
non-zero components:
\begin{equation}
  \Omega(\bH,\bP_i) \bP_j = \varkappa \delta_{ij} \bH \quad\text{and}\quad
  \Omega(\bP_i, \bP_j) \bP_k = \varkappa (\delta_{jk} \bP_i - \delta_{ik} \bP_j),
\end{equation}
which is never flat. Both of these results are independent of the
Nomizu map.

\paragraph{Carrollian Light Cone ($\zLC$)}
\label{sec:other-invar-conn-flc}

As show in~\cite{Figueroa-OFarrill:2018ilb}, this homogeneous
spacetime does not admit any invariant connections for $D\geq 3$. 

\subsubsection{Nomizu Maps for Aristotelian Spacetimes}
\label{sec:nomizu-maps-aristotle}

In this section, we study the space of invariant affine connections for the
Aristotelian spacetimes of Table~\ref{tab:spacetimes}. They are all
reductive, so there is a canonical invariant connection, and any other
invariant connection is determined uniquely by its Nomizu map. The
Nomizu maps $\alpha : \m \times \m \to \m$ are only subject to
equivariance under rotations and are given
by~\eqref{eq:nomizu-rotations}. They depend only on the dimension $D$
and not on the precise Aristotelian spacetime; although, of course,
the precise expression for the torsion and curvature tensors does
depend on the spacetime. We will calculate the torsion and curvature
for each spacetime below.

\paragraph{Static spacetime ($\zS$)}
\label{sec:staticinv}

The torsion and curvature of the most general invariant
connection have the following non-zero
components:
\begin{equation}
  \begin{split}
    \Theta(\bH,\bP_i) &= (\nu-\xi) \bP_i,\\
    \Theta(\bP_i,\bP_j) &= 2 \zeta' \epsilon_{ijk} \bP_k,\\
    \Omega(\bH,\bP_i) \bH &= \xi(\nu-\mu) \bP_i,\\
    \Omega(\bH,\bP_i) \bP_j &= \zeta(\mu-\nu)\delta_{ij} \bH, \\
    \Omega(\bP_i,\bP_j) \bH &= 2 \xi \zeta' \epsilon_{ijk} \bP_k, \text{and}\\
    \Omega(\bP_i,\bP_j) \bP_k &= (\zeta\xi - \zeta'^2)(\delta_{jk} \bP_i - \delta_{ik} \bP_j) + 2 \zeta \zeta' \epsilon_{ijk} \bH.
  \end{split}
\end{equation}
The torsion-free condition implies that $\zeta' = 0$.  With this value
of $\zeta'$, the above components reduce
to 
\begin{equation}
  \begin{split}
    \Theta(\bH,\bP_i) &= (\nu - \xi) \bP_i,\\
    \Omega(\bH,\bP_i)\bH &= \xi (\nu-\mu) \bP_i,\\
    \Omega(\bH,\bP_i)\bP_j &= \zeta (\mu-\nu) \delta_{ij} \bH, \text{and}\\
    \Omega(\bP_i,\bP_j) \bP_k &= \zeta \xi (\delta_{jk} \bP_i - \delta_{ik} \bP_j).
  \end{split}
\end{equation}
There are then three classes of torsion-free, flat invariant connections in addition
to the canonical connection:
\begin{enumerate}
\item $\zeta = 0$ and $\mu=\nu=\xi \neq 0$,
\item $\nu=\xi=\zeta = 0$ and $\mu \neq 0$, and
\item $\mu=\nu=\xi = 0$ and $\zeta \neq 0$.
\end{enumerate}
Since the remaining Aristotelian spacetimes all have the same Nomizu
maps as this static case, all of them will have the above torsion and
curvature components as a base, with a few additional terms included
due to the additional non-vanishing brackets of the specific
spacetime.

\paragraph{Torsional static spacetime ($\zTS$)}
\label{sec:other-invar-conn-tst}

In this instance, we get the following non-vanishing torsion and curvature components:
\begin{equation}
\begin{split}
\Theta(\bH, \bP_i) &= (\nu - \xi - 1) \bP_i, \\
\Theta(\bP_i, \bP_j) &= 2 \zeta ' \epsilon_{ijk} \bP_k, \\
\Omega(\bH, \bP_i) \bH &= \xi(\nu - \mu - 1) \bP_i, \\
\Omega(\bH, \bP_i) \bP_j &= \zeta (\mu - \nu -1) \delta_{ij} \bH - \zeta ' \epsilon_{ijk} \bP_k, \\
\Omega(\bP_i, \bP_j) \bH &= 2\xi\zeta ' \epsilon_{ijk}\bP_k, \text{and} \\
\Omega(\bP_i, \bP_j) \bP_k &= (\zeta\xi - \zeta '^2) (\delta_{jk} \bP_i - \delta_{ik} \bP_j) + 2\zeta\zeta ' \epsilon_{ijk} \bH.
\end{split}
\end{equation}
Imposing the torsion-free condition makes $\zeta'$ vanish such that we
arrive at
\begin{equation}
\begin{split}
\Theta(\bH, \bP_i) &= (\nu - \xi - 1) \bP_i, \\
\Omega(\bH, \bP_i) \bH &= \xi(\nu - \mu - 1) \bP_i, \\
\Omega(\bH, \bP_i) \bP_j &= \zeta(\mu - \nu -1) \delta_{ij} \bH, \text{and} \\
\Omega(\bP_i, \bP_j) \bP_k &= \zeta\xi (\delta_{jk} \bP_i - \delta_{ik} \bP_j).
\end{split}
\end{equation}
As in the static case, we again find three classes of torsion-free, flat invariant connection:
\begin{enumerate}
	\item $\xi=\zeta=0,$ and $\nu=1$,
	\item $\mu=\xi=\nu -1,$ and $\zeta = 0,$ and,
	\item $\xi=0,\; \nu=1,$ and $\mu=2$.
\end{enumerate}

\paragraph{Aristotelian spacetime ($\athree_\varepsilon$)}
\label{sec:other-invar-conn-a3}

The non-vanishing torsion and curvature components are 
\begin{equation}
\begin{split}
\Theta(\bH, \bP_i) &= (\nu - \xi) \bP_i, \\
\Theta(\bP_i, \bP_j) &= 2 \zeta ' \epsilon_{ijk} \bP_k, \\
\Omega(\bH, \bP_i) \bH &= \xi(\nu - \mu) \bP_i, \\
\Omega(\bH, \bP_i) \bP_j &= \zeta (\mu - \nu) \delta_{ij} \bH, \\
\Omega(\bP_i, \bP_j) \bH &= 2\xi\zeta ' \epsilon_{ijk}\bP_k, \text{and} \\
\Omega(\bP_i, \bP_j) \bP_k &= (\zeta\xi + \varepsilon - \zeta '^2) (\delta_{jk} \bP_i - \delta_{ik} \bP_j) + 2\zeta\zeta ' \epsilon_{ijk} \bH.
\end{split}
\end{equation}
As in the static and torsional static cases, imposing the torsion-free
condition sets $\zeta'=0$. This means the above components become
\begin{equation}
\begin{split}
\Theta(\bH, \bP_i) &= (\nu - \xi) \bP_i, \\
\Omega(\bH, \bP_i) \bH &= \xi(\nu - \mu) \bP_i, \\
\Omega(\bH, \bP_i) \bP_j &= \zeta(\mu - \nu) \delta_{ij} \bH, \text{and} \\
\Omega(\bP_i, \bP_j) \bP_k &= (\zeta\xi + \varepsilon) (\delta_{jk} \bP_i - \delta_{ik} \bP_j).
\end{split}
\end{equation}
Imposing flatness, we find that this requires $\varepsilon$ to vanish;
therefore, since $\varepsilon = \pm 1$, we find no torsion-free, flat
invariant connections.

\subsection{Pseudo-Riemannian Spacetimes and their Limits}
\label{sec:metric}

In this section, we wish to use the geometric limits that arose from contractions,
discussed in Section~\ref{subsec:ks_kss_gls}, to give us a unified treatment of
the geometric properties of the spacetimes admitting such a description.  
Recall, we had the following brackets in addition to the standard kinematical
brackets of~\eqref{eq:kinematical_brackets_general}
\begin{equation} \label{eq:transformed_brackets}
	[\bH, \B] = \tau^2 \varsigma \P \quad
		[\bH, \P] = - \kappa^2 \varkappa \B \quad
		[\B, \P] = \tfrac{1}{c^2} \bH \quad
		[\B, \B] = - (\tfrac{\tau}{c})^2 \varsigma \J\quad
		[\P, \P] = - (\tfrac{\kappa}{c})^2 \varkappa \J. 
\end{equation}
We will choose to absorb $\tau$ and $\kappa$ into our definition of 
$\varsigma$ and $\varkappa$, such that we have
\begin{equation}
  \label{eq:Liewithlim}
  [\bH,\B] = \varsigma  \P, \quad
  [\bH,\P] = - \varkappa  \B, \quad 
  [\B,\P] = \frac1{c^2} \bH, \quad 
  [\B,\B] = -\frac{\varsigma}{c^2} \J, \quad 
  \text{and}\quad
  [\P,\P] = -\frac{\varkappa}{c^2} \J.
\end{equation}
As before, the parameter $\varsigma$ corresponds to the signature:
$\varsigma = 1$ for Riemannian and $\varsigma=-1$ for Lorentzian. 
The parameter $\varkappa$ corresponds to
the curvature, so $\varkappa=1,0,-1$ for positive, zero and negative
curvature, respectively.\footnote{This definition is tentative due to
the possibility of having the curved Galilean spacetimes defined 
with either $\varkappa=1$ or $\varkappa=-1$.}  The limit
$c \to \infty$ corresponds to the non-relativistic limit. In the
computations below we will work with unspecified values of
$\varsigma,\varkappa,c$ and only at the end will we set them to
appropriate values to recover the results for particular spacetimes.
Some of the expressions will have (removable) singularities whenever
$\varsigma$ or $\varkappa$ vanish, so will have to think of those
cases as limits: the ultra-relativistic limit $\varsigma\to 0$ and the
flat limit $\varkappa\to 0$.

\subsubsection{Invariant Structures}
\label{sec:invariant-structures}

We will determine the form of the invariant tensors of small rank.
If $\k = \h \oplus \m$ is a reductive split then, as explained in
Section~\ref{sec:invariant-tensors}, invariant tensor
fields on a simply-connected homogeneous space $\M=\Kgr/\Hgr$ are in
bijective correspondence with $\Hgr$-invariant tensors on $\m$, and
since $\Hgr$ is connected, these are in bijective correspondence with
$\h$-invariant tensors on $\m$.
\\ \\
The action of $\h$ on $\m$ is the linear isotropy representation,
which is the restriction to $\h$ of the adjoint action:
\begin{equation}\label{eq:LIR}
  \begin{aligned}[m]
    \bJ_{ij} \cdot \bH &= 0\\
    \bJ_{ij} \cdot \bP_k &= \delta_{jk} \bP_i - \delta_{ik} \bP_j
  \end{aligned}
  \quad\text{and}\quad
  \begin{aligned}[m]
    \bB_i \cdot \bH &= - \varsigma \bP_i\\
    \bB_i \cdot \bP_j &= \frac1{c^2} \delta_{ij} \bH.
  \end{aligned}
\end{equation}
With respect to the canonical dual basis $\eta$, $\pi_i$ for $\m^*$,
the dual linear isotropy representation is the restriction of the
coadjoint action:
\begin{equation}\label{eq:LIR-dual}
  \begin{aligned}[m]
    \bJ_{ij} \cdot \eta &= 0\\
    \bJ_{ij} \cdot \pi^k &= -\delta_i^k \pi_j + \delta_j^k \pi_i
  \end{aligned}
  \quad\text{and}\quad
  \begin{aligned}[m]
    \bB_i \cdot \eta &= - \frac1{c^2} \pi_i\\
    \bB_i \cdot \pi^j &= \varsigma \delta_i^j \eta.
  \end{aligned}
\end{equation}
It follows that $\bH$ is invariant in the $\varsigma \to 0$ limit,
whereas $\eta$ is invariant in the $c\to\infty$ limit.
\\ \\
Concerning the rotationally invariant tensors of second rank, let us
observe that
\begin{equation}
  \alpha_1 \bH^2 + \beta_1 \P^2 \quad\text{is invariant}\quad \iff \quad
  \sigma \alpha_1  = \frac1{c^2} \beta_1
\end{equation}
and
\begin{equation}
  \alpha_2 \eta^2 + \beta_2 \bpi^2 \quad\text{is invariant}\quad \iff \quad
  \frac1{c^2} \alpha_2 = \sigma \beta_2.
\end{equation}
It is interesting to note that the sign $\varkappa$ of the curvature
has played no role thus far.
\\ \\
We shall now specialise to the different classes of spacetimes and
determine whether and how the structures are induced in the limit.

\paragraph{Lorentzian and Riemannian Case}
\label{sec:lorentz-riem-case}

It is clear that for the (pseudo\nobreakdash-)Riemannian case, where
$\varsigma\neq 0 \neq \frac{1}{c^{2}}$, only the metric and its
co-metric are invariant.  Keeping in mind that we wish the limit in
which the parameters $\varsigma$ and $c$ tend to zero to exist, we set
$\alpha_{1}=\frac{1}{c^{2}}$ and $\beta_{1}=\varsigma$ and similarly
for the co-metric, which leads to the invariants
\begin{equation}
  \label{eq:metrics}
  \frac{1}{c^{2}} \bH^{2} + \varsigma \P^{2} \quad \text{and} \quad \varsigma \eta^{2} + \frac{1}{c^{2}} \mathbold{\pi}^{2}  \,.
\end{equation}
For negative (positive) $\varsigma$ this is the invariant Lorentzian
(Riemannian) structure. The metric and the co-metric are not per se
the inverse of each other, although using definite values for the
limiting parameters they can be made to be.

\paragraph{Non- and Ultra-Relativistic Limits}
\label{sec:non-ultra-relat}

Let us now investigate the limits. Taking the non-relativistic limit
($c \to \infty$) of the metrics leads to the invariants
\begin{equation}
  \label{eq:gallim}
  \varsigma \P^{2} \quad \text{and} \quad \varsigma \eta^{2},
\end{equation}
which can be interpreted as the invariants that properly arise from
the Lorentzian structure. However, as~\eqref{eq:LIR-dual} shows also
$\eta$ itself is an invariant in this limit. This does not follow from the
contractions, but can be anticipated from the metrics. We could now
take the ultra-relativistic limit ($\varsigma \to 0$) of
\eqref{eq:gallim} leading to no invariant tensor. Of course, this
spacetime has the invariants $\bH, \P^{2}, \eta, \mathbold{\pi}^{2}$,
but none of these arise from the limit of the original Lorentzian and
Riemannian metrics. For the ultra-relativistic limit, we may apply 
the same logic.
\\ \\
Concluding, we have the Galilean structure $\eta, \varsigma \P^2$ and the
Carrollian structure $\bH, \frac{1}{c^{2}} \mathbold{\pi}^{2}$, where we
have left the contraction parameters for the invariants that arise
from a limit.

\subsubsection{Action of the Boosts}
\label{sec:action-boosts}

The actions of the boosts for all the Lorentzian, Riemannian, Galilean,
and reductive Carrollian spacetimes were determined in
Section~\ref{sec:boosts}, where we arrived at
equation~\eqref{eq:boost-orbit} for the orbit of $(t_0,\x_0)$ under
the one-parameter family of boosts generated by $\w \cdot \B$, which
we rewrite here as follows:
\begin{equation}
    \begin{split}
      t(s) &= t_0 \cosh(s z) + \frac1{c^2} \frac{\sinh(s z)}{z}\x_0 \cdot \w\\
      \x(s) &= \x_0^\perp - \varsigma t_0 \frac{\sinh(s z)}{z} \w +
      \cosh(s z) \frac{(\x_0 \cdot \w)}{w^2} \w ,
  \end{split}
\end{equation}
where $\x_0^\perp := \x_0 - \frac{\x_0 \cdot \w}{w^2}\w$ and $z^2 :=
-\frac1{c^2} \varsigma w^2$.  Notice that the orbits of $(0,\x_0)$
with $\x_0 \cdot \w = 0$ are point-like. To understand the nature of
the other (generic) orbits, we choose values for the
parameters. Notice that in our coset parametrisation the boosts do
not depend on $\varkappa$, but only on $\varsigma$ and $c$.  Therefore,
we shall be able to treat each class of spacetime uniformly.

\paragraph{Lorentzian boosts}
\label{sec:Lorentzian-boosts}
Here we take $\varsigma = -1$ and keep $c^{-1}$ non-zero. Then
$z^2 = \frac{w^2}{c^2}$, so $z = \left|\frac{\w}{c}\right|$, and the
orbits of the boosts are
\begin{equation}\label{eq:lor-boost-orbit}
  \begin{split}
    t(s) &= t_0 \cosh(s \left|\tfrac{\w}{c}\right|) + \frac1{c^2} \frac{\sinh(s \left|\tfrac{\w}{c}\right|)}{\left|\tfrac{\w}{c}\right|}\x_0 \cdot \w\\
    \x(s) &= \x^\perp_0 + t_0 \frac{\sinh(s \left|\tfrac{\w}{c}\right|)}{\left|\tfrac{\w}{c}\right|} \w +
    \cosh(s \left|\tfrac{\w}{c}\right|) \frac{(\x_0 \cdot \w)}{w^2} \w.
  \end{split}
\end{equation}
Let $\x = \x^\perp + y \w$, where $\x^\perp \cdot \w = 0$.  Then
$\x^\perp(s) = \x^\perp_0$ for all $s$ and the orbit takes place in
the $(t,y)$ plane.  Letting $|\w| = 1$ and $c=1$, we find
\begin{equation}\label{eq:lor-boost-orbit-too}
  t(s) = t_0 \cosh(s) + \sinh(s) y_0 \quad\text{and}\quad y(s) = t_0 \sinh(s) + \cosh(s) y_0,
\end{equation}
which is either a point (if $t_0 = y_0 = 0$), a straight line (if
$t_0 = \pm y_0 \neq 0$), or a hyperbola (otherwise). The nature of the
orbits in the exponential coordinates is clear, but only in the case
of Minkowski spacetime do the exponential coordinates provide a global
chart and hence only in that case can we deduce from this calculation
that the generic orbits are not compact. For (anti\nobreakdash-)de~Sitter
spacetime, we must argue in a different way.
\\ \\
Let $\overline{\zdS}$ denote the quotient of $\hyperlink{S2}{\zdS}$
which embeds as a quadric hypersurface in Minkowski spacetime. The
covering map $\zdS \to \overline{\zdS}$ relates the orbits of the
boosts on $\hyperlink{S2}{\zdS}$ and in the quotient $\overline{\zdS}$
and since continuous maps send compact sets to compact sets, it is
enough to show the non-compactness of the orbits in $\overline{\zdS}$.
The embedding $\overline{\zdS} \subset \RR^{D+1,1}$ is given by the
quadric
\begin{equation}
  x_1^2 + \cdots + x_D^2 + x_{D+1}^2 - x_{D+2}^2 = R^2,
\end{equation}
which is acted on transitively by $\SO(D+1,1)$. The stabiliser Lie
algebra of the point $(0,\cdots,0,R,0)$ is spanned by the $\so(D+1,1)$
generators $J_{ij}$ and $J_{i,D+2}$, so that $\bB_i = J_{i,D+2}$, which
is a boost in $\RR^{D+1,1}$. We have just shown that boosts in
Minkowski spacetime have non-compact orbits; therefore, this is the
case in $\overline{\zdS}$ and hence also in $\hyperlink{S2}{\zdS}$.
\\ \\
Similarly, let $\overline{\zAdS}$ denote the quotient of
$\hyperlink{S3}{\zAdS}$ which embeds in $\RR^{D,2}$ as the quadric
\begin{equation}
  x_1^2 + \cdots + x_D^2 - x_{D+1}^2 - x_{D+2}^2 = -R^2.
\end{equation}
The Lie algebra $\so(D,2)$ acts transitively on this quadric and the
stabiliser Lie algebra at the point $(0,\cdots,0,0,R)$ is spanned by
the $\so(D,2)$ generators $J_{ij}$ and $J_{i,D+1}$, so that
$\bB_i = J_{i,D+1}$ which is a ``boost'' in $\RR^{D,2}$. The calculation
of the orbit, in this case, is formally identical to the one for
Minkowski spacetime (in fact, it takes place in the Lorentzian plane
with coordinates $(x_i, x_{D+2})$) and we see that they are
non-compact, so the same holds in $\overline{\zAdS}$ and thus also in
$\hyperlink{S3}{\zAdS}$.

\paragraph{Euclidean ``boosts''}
\label{sec:euclidean-boosts}

Here we take $\varsigma = 1$ and keep $c^{-1}$ non-zero.  Then $z^2 =
-\frac{w^2}{c^2}$, so $z = i \left|\frac{\w}{c}\right|$, and the orbits
of the boosts are
\begin{equation}\label{eq:riem-boost-orbit}
  \begin{split}
    t(s) &= t_0 \cos(s \left|\tfrac{\w}{c}\right|) + \frac1{c^2}
    \frac{\sin(s
      \left|\tfrac{\w}{c}\right|)}{\left|\tfrac{\w}{c}\right|}\x_0
    \cdot \w\\
    \x(s) &= \x^\perp_0 - t_0 \frac{\sin(s
      \left|\tfrac{\w}{c}\right|)}{\left|\tfrac{\w}{c}\right|} \w +
    \cos(s \left|\tfrac{\w}{c}\right|) \frac{(\x_0 \cdot \w)}{w^2}
    \w.
  \end{split}
\end{equation}
As before, letting $\x = \x^\perp + y \w$, and choosing $|\w|=1$ and
$c=1$, we find that the orbit is such that $\x^\perp$ is constant and
$(t,y)$ evolve as
\begin{equation}\label{eq:riem-boost-orbit-too}
  t(s) = t_0 \cos(s) + \sin(s) y_0 \quad\text{and}\quad y(s) = -t_0 \sin(s) + \cos(s) y_0,
\end{equation}
which is either a point (if $t_0 = y_0 = 0$) or a circle (otherwise)
and in any case compact. This suffices for $\hyperlink{S4}{\EE}$ and
$\hyperlink{S6}{H}$ since the exponential coordinates give a global
chart. For $\hyperlink{S5}{S}$ it is clear that the boosts act with
compact orbits because the kinematical Lie group $\SO(D+2)$ is itself
compact, therefore, so are the one-parameter subgroups.

\paragraph{Galilean boosts}
\label{sec:galilean-boosts}

Here we take the limit $c \to \infty$ and, for definiteness,
$\varsigma = -1$.  The orbits of the boosts are then the limit $c \to
\infty$ of equation~\eqref{eq:lor-boost-orbit}:
\begin{equation}\label{eq:gal-boost-orbit}
  \begin{split}
    t(s) &= t_0 \\
    \x(s) &= \x_0 + s t_0 \w.
  \end{split}
\end{equation}
Here the orbits of $(0,\x_0)$ are point-like. The generic orbit
($t_0\neq 0$) is not periodic and hence not compact. This suffices for
$\hyperlink{S7}{\zG}$ and $\hyperlink{S8}{\zdSG}$, since the
exponential coordinates define a global chart. For
$\hyperlink{S10}{\zAdSG}$ we need to argue differently and this is
done later in this section.

\paragraph{Carrollian boosts}
\label{sec:carrollian-boosts}

Here we keep $c^{-1}$ non-zero, but take the limit $\varsigma \to 0$
in equation~\eqref{eq:boost-orbit}:
\begin{equation}\label{eq:car-boost-orbit}
  \begin{split}
    t(s) &= t_0 + s \frac1{c^2} \x_0 \cdot \w\\
    \x(s) &= \x_0.
  \end{split}
\end{equation}
Here the orbits $(t_0,\x_0)$ with $\x_0 \cdot \w =0$ are point-like,
but the other orbits are not periodic, hence not compact. This settles
it for $\hyperlink{S15}{\zAdSC}$, since the exponential coordinates
give a global chart. For the other Carrollian spacetimes we can argue
in a different way.
\\ \\
As shown in~\cite{Duval:2014uoa}, a Carrollian spacetime admits an
embedding as a null hypersurface in a Lorentzian spacetime. For the
homogeneous examples in this thesis, this was done in
\cite{Figueroa-OFarrill:2018ilb} following the embeddings of the
Carrollian spacetimes $\hyperlink{S13}{\zC}$ and
$\hyperlink{S16}{\zLC}$ as null hypersurfaces of Minkowski spacetime
described already in~\cite{Duval:2014uoa}.
\\ \\
As explained in Section~\ref{sec:boosts}, for $\hyperlink{S14}{\zdSC}$
it is enough to work with the discrete quotient $\overline{\zdSC}$,
which embeds as a null hypersurface in the hyperboloid model
$\overline{\zdS}$ of de~Sitter spacetime, which itself is a quadric
hypersurface in Minkowski spacetime. In
\cite{Figueroa-OFarrill:2018ilb}, it was shown that the boosts in
$\overline{\zdSC}$ can be interpreted as null rotations in the
(higher-dimensional) pseudo-orthogonal Lie group and the orbits of
null rotations are never compact. This is done in detail in
Section~\ref{sec:action-boosts-1} for $\hyperlink{S16}{\zLC}$.

\subsubsection{Fundamental Vector Fields}
\label{sec:fund-vect-fields}

The fundamental vector fields for rotations and boosts are linear in
exponential coordinates and given by equations~\eqref{eq:fvf-rot} and
\eqref{eq:fvf-boost}, respectively. To determine the fundamental
vector fields for the translations we must work harder.
\\ \\
Now let $A = t \bH + \x \cdot \P$.  Then we have that
\begin{equation}\label{eq:adA-on-gens}
  \begin{aligned}[m]
    \ad_A \bH &= \varkappa \x \cdot \B\\
    \ad_A \bB_i &= \varsigma t \bP_i - \frac{1}{c^2} x_i \bH\\
    \ad_A \bP_i &= \frac{\varkappa}{c^2} \bJ_{ij} x^j - \varkappa t \bB_i\\
    \ad_A \bJ_{ij} &= x_i \bP_j - x_j \bP_i
  \end{aligned}
  \quad\text{and}\quad
  \begin{aligned}[m]
    \ad^2_A \bH &= \varkappa\varsigma t \x \cdot \P - \frac{\varkappa}{c^2} x^2 \bH\\
    \ad^2_A \bB_i &= \frac{\varkappa}{c^2} \varsigma t x^j \bJ_{ij} -
    \varkappa \varsigma t^2 \bB_i - \frac{\varkappa}{c^2} x_i \x \cdot \B\\
    \ad^2_A \bP_i &= -\varkappa (\frac1{c^2} x^2 + \varsigma t^2) \bP_i + \frac{\varkappa}{c^2}
    x_i \x \cdot \P + \frac{\varkappa}{c^2} t x_i \bH\\
    \ad^2_A \bJ_{ij} &= -\varkappa t(x_i \bB_j - x_j \bB_i) + \frac{\varkappa}{c^2}
    x^k(x_i \bJ_{jk} - x_j \bJ_{ik}),
  \end{aligned}
\end{equation}
so that in general we have
\begin{equation}
  \ad_A^3 = -\varkappa (\frac1{c^2} x^2 + \varsigma t^2)\ad_A.
\end{equation}
Letting $x_\pm$ denote the two complex square roots of
$-\varkappa (\frac1{c^2} x^2 + \varsigma t^2)$, with $x_- = - x_+$, we
can rewrite this equation as $\ad_A^3 = x_+^2 \ad_A$.
\\ \\
Now, if $f(z)$ is analytic in $z$ and admits a power series expansion
$f(z) = \sum_{n=0}^\infty c_n z^n$, then
\begin{equation}
  f(\ad_A) = f(0) + \frac1{x_+}\sum_{k=0}^\infty c_{2k+1} x_+^{2k+1}
  \ad_A + \frac1{x_+^2} \sum_{k=1}^\infty c_{2k} x_+^{2k} \ad_A^2.
\end{equation}
Observing that
\begin{equation}
  \sum_{k=0}^\infty c_{2k+1} x_+^{2k+1} = \tfrac12 ( f(x_+) - f(x_-) )
  \quad\text{and}\quad
  \sum_{k=1}^\infty c_{2k} x_+^{2k} = \tfrac12 ( f(x_+) + f(x_-) -
  2 f(0) ),
\end{equation}
we arrive finally at
\begin{equation}
  \label{eq:fad}
  f(\ad_A) =f(0) + \frac1{2x_+} (f(x_+)-f(x_-)) \ad_A +
  \frac1{2x_+^2} (f(x_+) + f(x_-) - 2 f(0)) \ad_A^2.
\end{equation}
Introducing the shorthand notation:
\begin{equation}\label{eq:shorthand}
  f^+ := \tfrac12 (f(x_+) + f(x_-))\qquad\text{and}\quad
  f^- := \frac1{2x_+}(f(x_+) - f(x_-)),
\end{equation}
equation~\eqref{eq:fad} becomes
\begin{equation}
  f(\ad_A) =f(0) + f^- \ad_A + \frac1{x_+^2} (f^+ - f(0)) \ad_A^2.
\end{equation}
It follows from the above equation and
equation~\eqref{eq:adA-on-gens}, that for $f(z)$ analytic in $z$,
\begin{equation}\label{eq:fadA-on-gens}
  \begin{split}
    f(\ad_A) \bH &= f(0) \bH + f^- \varkappa \x \cdot \B +
    \tfrac1{x_+^2}(f^+-f(0))\left(\varkappa\varsigma t \x \cdot \P -
      \tfrac{\varkappa}{c^2} x^2 \bH\right)\\
    f(\ad_A) \bB_i &= f(0) \bB_i + f^-(\varsigma t \bP_i - \tfrac1{c^2} x_i
    \bH) + \tfrac1{x_+^2} (f^+-f(0))\left(- \varkappa\varsigma t^2 \bB_i -
    \tfrac{\varkappa}{c^2} x_i \x \cdot \B + \tfrac{\varkappa}{c^2}
    \varsigma t \bJ_{ij} x^j\right)\\
  f(\ad_A) \bP_i &= f^+ \bP_i + f^- (-\varkappa t \bB_i +
  \tfrac{\varkappa}{c^2} \bJ_{ij} x^j) + \tfrac1{x_+^2}(f^+-f(0))
  \tfrac{\varkappa}{c^2} x_a ( t \bH + \x \cdot \P)\\
  f(\ad_A) \bJ_{ij} &= f(0) \bJ_{ij} + f^-(x_i P_j - x_j P_i) +
  \tfrac1{x_+^2}(f^+ - f(0)) \varkappa\left(- t (x_i B_j - x_j B_i) +
    \tfrac1{c^2} x^k(x_i J_{jk} - x_j J_{ik})\right).
  \end{split}
\end{equation}
Let us calculate $\xi_{\bH} = \tau \frac{\d}{\d t} + y^i \frac{\d}{\d x^i}$, where by equation~\eqref{eq:master}
\begin{equation}
  \tau \bH + \y \cdot \P = G(\ad_A) \bH - F(\ad_A)\bbeta \cdot \B,
\end{equation}
for some $\bbeta$.  From equation~\eqref{eq:fadA-on-gens}, we have
\begin{multline}
  \tau \bH + \y \cdot \P = \bH + G^- \varkappa \x \cdot \B + \frac1{x_+^2}
  (G^+-1) \left(\varkappa\varsigma t \x \cdot \P -
    \tfrac{\varkappa}{c^2} x^2 \bH\right)\\
  -\left(\bbeta\cdot \B + F^-(\varsigma t\bbeta \cdot \P - \tfrac1{c^2} \x
    \cdot\bbeta \bH ) + \frac1{x_+^2}(F^+-1)\left( -\varkappa\varsigma t^2\bbeta \cdot\B - \tfrac{\varkappa}{c^2} \x \cdot\bbeta \x \cdot \B +
      \tfrac{\varkappa}{c^2} \varsigma t \bJ_{ij} \beta^i x^j\right) \right).
\end{multline}
By $\so(D)$-covariance, $\bbeta$ has to be proportional to $\x$, since
that is the only other vector appearing in the $\B$ terms, which means
that the $\bJ_{ij}$ term above vanishes. This leaves terms in $\B$, $\bH$,
and $\P$, which allow us to solve for $\bbeta$, $\tau$, and $\y$,
respectively. The $\B$ terms cancel if and only if
\begin{equation}
  \bbeta = \frac{G^-}{F^+} \varkappa \x,
\end{equation}
which we can re-insert into the equation to solve for $\tau$ and $\y$.
Doing so we find
\begin{equation}
    \tau = 1 - \left(\frac{x_+ \coth x_+ - 1}{x_+^2}\right)
    \frac{\varkappa}{c^2} x^2 \quad\text{and}\quad
    y^a =\left(\frac{x_+ \coth x_+ - 1}{x_+^2}\right) \varkappa
    \varsigma t x^a,
\end{equation}
so that
\begin{equation}
\xi_{\bH} = \frac{\d}{\d t} + \left(\frac{x_+ \coth x_+ -
      1}{x_+^2}\right) \varkappa \left( \varsigma t x^a \frac{\d}{\d
      x^a} - \frac{1}{c^2} x^2 \frac{\d}{\d t} \right).
\end{equation}
To calculate $\xi_{\v \cdot \P} = \tau \frac{\d}{\d t} + y^i
\frac{\d}{\d x^i}$, equation~\eqref{eq:master} says we must solve
\begin{equation}
  \tau \bH + \y \cdot \P = G(\ad_A) \v \cdot \P - F(\ad_A) \left(\bbeta \cdot \B
  + \tfrac12 \lambda^{ij} \bJ_{ij}\right),
\end{equation}
for $\lambda^{ij}$, $\bbeta$, $\tau$, and $\y$ from the components along
$\bJ_{ij}$, $\B$, $\bH$, and $\P$, respectively.  The details of the
calculation are not particularly illuminating.  Let us simply remark
that we find
\begin{equation}
  \lambda^{ij} = h_1 (v^i x^j - v^j x^i) + h_2 (\beta^i x^j - \beta^j x^i)
\end{equation}
for
\begin{equation}
  h_1 = \frac{G^-\frac{\varkappa}{c^2}}{1 -
    \frac1{x_+^2}(F^+-1)\frac{\varkappa}{c^2} x^2}
  \quad\text{and}\quad
  h_2 = \frac{-\frac1{x_+^2} (F^+-1)\frac{\varkappa}{c^2}\varsigma t}{1 -
    \frac1{x_+^2}(F^+-1)\frac{\varkappa}{c^2} x^2},
\end{equation}
and
\begin{equation}
  \bbeta = -\frac{G^-}{F^+} \varkappa t \v,
\end{equation}
so that
\begin{equation}
  \lambda^{ij} = -\frac{\varkappa}{c^2} \frac{\tanh(x_+/2)}{x_+} (v^i
  x^j - v^j x^i).
\end{equation}
Re-inserting these expressions into the equation we solve for $\tau$
and $\y$, resulting in
\begin{equation}
  \tau = \frac{x_+\coth x_+ - 1}{x_+^2} \frac{\varkappa}{c^2} t \x
  \cdot \v
\end{equation}
and
\begin{equation}
  y^i = x_+ \coth(x_+) v^i + \frac{x_+\coth x_+ - 1}{x_+^2}
  \frac{\varkappa}{c^2}  \x \cdot \v x^i.
\end{equation}
Finally, we have that
\begin{equation}
  \xi_{\bP_i} = \frac{x_+\coth x_+ - 1}{x_+^2}
    \frac{\varkappa}{c^2} x_i \left(t \frac{\d}{\d t} + x^j
      \frac{\d}{\d x^j}\right) + x_+ \coth x_+ \frac{\d}{\d x^i}.
\end{equation}
Let us summarise all the fundamental vector fields and remember that
$x_{+}= \sqrt{-\varkappa(\frac1{c^2} x^2 + \varsigma t^2)}$
\begin{equation}
  \label{eq:fundvec}
  \begin{split}
    \xi_{\bJ_{ij}} &= x^j \frac{\d}{\d x^i} - x^i \frac{\d}{\d x^j}
    \\
    \xi_{\bB_i} &= \frac{1}{c^2} x^i \frac{\d}{\d t} - \varsigma t \frac{\d}{\d x^i}
    \\
    \xi_{\bH} &= \frac{\d}{\d t}
    + \left(
      \frac{x_+ \coth x_+ - 1}{x_+^2}
    \right)
    \varkappa \left(
      \varsigma t x^i \frac{\d}{\d x^i} - \frac{1}{c^2} x^2 \frac{\d}{\d t}
    \right)
    \\
    \xi_{\bP_i} &= \frac{x_+\coth x_+ - 1}{x_+^2}
    \frac{\varkappa}{c^2} x_i
    \left(
      t \frac{\d}{\d t} + x^j \frac{\d}{\d x^j}
    \right)
    + x_+ \coth x_+ \frac{\d}{\d x^i}.
\end{split}
\end{equation}
We can now calculate the Lie brackets of the vector fields which indeed
shows the anti-homomorphism with respect to~\eqref{eq:Liewithlim}
\begin{equation}
  [\xi_{\bH},\xi_{\B}] = - \varsigma  \xi_{\P}, \quad
  [\xi_{\bH},\xi_{\P}] =  \varkappa  \xi_{\B}, \quad 
  [\xi_{\B},\xi_{\P}] =- \frac1{c^2} \xi_{\bH}, \quad 
  [\xi_{\B},\xi_{\B}] = \frac{\varsigma}{c^2} \xi_{\J}, \quad 
  \text{and}\quad
  [\xi_{\P},\xi_{\P}] = \frac{\varkappa}{c^2} \xi_{\J}.
\end{equation}
Let us emphasise that taking the limit of the vector fields and then
calculating their Lie bracket leads to the same result as just taking
just the limit of the Lie brackets, i.e., these operations commute.

\subsubsection{Soldering Form and Connection One-Form}
\label{sec:sold-form-conn}

The soldering form and the connection one-form are the two components
of the pull-back of the left-invariant Maurer--Cartan form on $\Kgr$.
We will calculate it first for all the (pseudo\nobreakdash-)Riemannian
cases and then take the flat, non-relativistic and ultra-relativistic
limit. As we will see, the exponential coordinates are well adapted
for that purpose, and the limits can then be systematically studied.
That the limits are well-defined follows from our construction since
the quantities we calculate are a power series of the contraction
parameters, $\epsilon=c^{-1}, \varkappa, \tau$ in the $\epsilon \to 0$
limit and not of their inverse.
\\ \\
For the non-flat (pseudo\nobreakdash-)Riemannian geometries
our exponential coordinates are, except for the hyperbolic case,
neither globally valid nor are quantities like the curvature very
compact. Since coordinate systems for these cases are well studied, we
will focus in the following mainly on the remaining cases. It is
useful to derive the soldering form, the invariant connection and the
vielbein in full generality since we take the limit and use them to
calculate the remaining quantities of interest.
\\ \\
We start by calculating the Maurer--Cartan form via
equation~\eqref{eq:MC-pullback} for which we again use
equation~\eqref{eq:fadA-on-gens}. We find that
\begin{multline}
  \theta + \omega
  =
  dt \bH
  + D^- \varkappa  dt \x \cdot \B
  + \frac{1}{x_+^2}(D^+-1)
  \left(
    \varkappa \varsigma t  dt \x \cdot \P
    - \tfrac{\varkappa}{c^2} x^2 dt \bH
  \right)
  \\
  + D^+ d\x \cdot \P + D^-
  (
  -\varkappa t d\x \cdot \B
  + \tfrac{\varkappa}{c^2} dx^i x^j \bJ_{ij}
  )
  + \frac1{x_+^2}(D^+-1) \tfrac{\varkappa}{c^2} \x \cdot d\x (t \bH + \x \cdot \P),
\end{multline}
which, using that
\begin{equation}
  D^- = \frac{1 - \cosh x_+}{x_+^2}, \quad D^+ = \frac{\sinh
    x_+}{x_+} \quad\text{and hence}\quad \frac1{x_+^2}(D^+-1) =
  \frac{\sinh x_+ - x_+}{x_+^3},
\end{equation}
gives the following expressions:
\begin{equation}
  \begin{split}
    \theta &= dt \bH
    + \frac{\sinh x_+}{x_+} d\x \cdot \P
    + \frac{\sinh x_+ - x_+}{x_+^3} \varkappa
    \left(
      \varsigma t dt \x\cdot\P
      + \tfrac{1}{c^2}
      (t \, \x \cdot d\x  \bH - x^2 dt \bH + \x\cdot d\x\, \x \cdot \P)
     \right)\\
    \omega &= \frac{1-\cosh x_+}{x_+^2} \varkappa
    \left(
      dt \x \cdot\B
      - t d\x \cdot \B
      - \tfrac1{c^2}  x^i dx^j \bJ_{ij}
    \right).
  \end{split}
\end{equation}
We can also evaluate the vielbein $E=E_{\bH} \eta + E_{\P} \cdot \bpi$
which leads us to
\begin{align}
  E_{\bH} &=
          \frac{\varkappa}{x_{+}^{2}}
          \left[
          \left(
          - \varsigma t^{2} - \frac{x^{2}}{c^{2}} x_{+} \csch x_{+}
          \right)
          \frac{\pd}{\pd t}
          +
          \varsigma 
          \left(
          -1+ x_{+} \csch x_{+}
          \right)
          t x^{i}
          \frac{\pd}{\pd x^{i}}
          \right]
  \\
  E_{\bP_{i}}
        &=
          \frac{\varkappa x^{i}}{c^{2} x_{+}^{2}}
          \left(
          -1 + x_{+} \csch x_{+}
          \right)
          \left(
          t \frac{\pd}{\pd t} + x^{j}\frac{\pd}{\pd x^{j}}
          \right)
          +
           x_{+} \csch x_{+} \frac{\pd}{\pd x^{i}} .
\end{align}

\subsubsection{Flat Limit, Minkowski ($\MM$) and Euclidean Spacetime ($\EE$)}
\label{sec:flat-limit}

In the flat limit $\varkappa \to 0$ the soldering form and connection
one-form are given by
\begin{equation}
    \theta = dt \bH + d\x \cdot \P \quad \text{and} \quad \omega = 0 ,
\end{equation}
respectively, where $(t,\x)$ are global coordinates. The vielbein is
given by
\begin{align}
      E&= \frac{\partial}{\partial t} \eta +  \frac{\partial}{\partial \x} \cdot \mathbold{\pi}
\end{align}
and the fundamental vector fields, taking the limit of
\eqref{eq:fundvec}, by
\begin{equation}
    \xi_{\bB_i} = \frac{1}{c^2} x^i \frac{\d}{\d t} - \varsigma t \frac{\d}{\d x^i}, \quad
    \xi_{\bH} = \frac{\d}{\d t}, \quad \text{and} \quad
    \xi_{\bP_i} =  \frac{\d}{\d x^i}.
\end{equation}
Using the soldering form and the vielbein we can now write the metric
and co-metric, given in equation~\eqref{eq:metrics}, in coordinates
\begin{align}
 g &= \varsigma dt^{2} + \frac{1}{c^{2}} d \x \cdot d \x & \tilde g &=  \frac{1}{c^{2}}  \frac{\pd}{\pd t}\otimes \frac{\pd}{\pd t} + \varsigma \delta^{ij} \frac{1}{\pd x^{i}} \otimes  \frac{1}{\pd x^{j}}.
\end{align}
Since the connection one-form vanishes the torsion and curvature evaluate to
\begin{align}
  \Omega&=0 & \Theta&=0 .
\end{align}
We can now set $\varsigma$ and $c$ to definite values to obtain the
Minkowski spacetime ($\varsigma=-1$, $c=1$), Euclidean space
($\sigma=-1$, $c=1$), Galilean spacetime ($\varsigma=1$, $c^{-1}=0$), and
Carrollian spacetime ($\varsigma=0$, $c=1$). This is obvious enough for
the first two cases so that we go straight to the Galilean spacetime.

\subsubsection{Galilean Spacetime ($\zG$)}
\label{sec:galilean-spacetime}

For Galilean spacetimes we have the fundamental vector fields
\begin{align}
    \xi_{\bB_i} &=  t \frac{\d}{\d x^i} &
    \xi_{\bH} &= \frac{\d}{\d t} &
    \xi_{\bP_i} &=  \frac{\d}{\d x^i},
\end{align}
and the invariant Galilean structure  which is characterised by the clock one-form
$\tau = dt$ and the spatial metric on one-forms
$h = \delta^{ij} \frac{\d}{\d x^i} \otimes \frac{\d}{\d x^j}$.

\subsubsection{Carrollian Spacetime ($\zC$)}
\label{sec:carrollian-spacetime}

The fundamental vector fields for the Carrollian spacetime are
\begin{align}
  \label{eq:carrvec}
    \xi_{\bB_i} &=  x^i \frac{\d}{\d t}  &
    \xi_{\bH} &= \frac{\d}{\d t} &
    \xi_{\bP_i} &=  \frac{\d}{\d x^i},
\end{align}
and the invariant Carrollian structure is given by
$\kappa = \frac{\d}{\d t}$ and $b = \delta_{ij} dx^i dx^j$. 

\subsubsection{Non-Relativistic Limit}
\label{sec:non-relat-limit}
In the non-relativistic limit $c \to \infty$ we get
$x_{+}= \sqrt{-\varkappa \varsigma t^2}$ and the soldering form and
connection one-form are given by
\begin{equation}
  \begin{split}
    \theta &= dt \bH + \frac{\sinh x_+}{x_+} d\x \cdot \P
    + \frac{\sinh  x_+ - x_+}{x_+^3} \varkappa \varsigma t dt \x\cdot\P \\
    \omega &= \frac{1-\cosh x_+}{x_+^2} \varkappa
    \left(
      dt \x \cdot \B- t d\x \cdot \B 
    \right)
  \end{split}.
\end{equation}
We take the non-relativistic limit of the vielbein and obtain
\begin{equation}
  \begin{split}
  E_{\bH} &=
          \frac{\pd}{\pd t}
          +
          \left(
          1 - x_{+} \csch x_{+}
          \right)
          \frac{x^{i}}{t}
          \frac{\pd}{\pd x^{i}}
  \\
  E_{\bP_{i}}
        &=
           x_{+} \csch x_{+} \frac{\pd}{\pd x^{i}} .
 \end{split}
\end{equation}
We can now calculate the invariant Galilean structure which is given
by the clock one-form and the spatial co-metric ($h=\varsigma \P^2$):
\begin{align}
  \tau &= \eta(\theta)= \varsigma dt
  &
    h &=  x_{+}^{2} \csch^{2} x_{+}  \delta^{ij}\frac{\pd}{\pd x^{i}} \otimes \frac{\pd}{\pd x^{j}} .
\end{align}
The fundamental vector fields are given by
\begin{equation}
  \begin{split}
    \xi_{\bB_i} &= - \varsigma t \frac{\d}{\d x^i}
    \\
    \xi_{\bH} &= \frac{\d}{\d t}
    + \left(
      \frac{x_+ \coth x_+ - 1}{x_+^2}
    \right)
    \varkappa
      \varsigma t x^i \frac{\d}{\d x^i}
    \\
    \xi_{\bP_i} &= 
    x_+ \coth x_+ \frac{\d}{\d x^i} .
\end{split}
\end{equation}

\subsubsection{Galilean de~Sitter Spacetime ($\zdSG$)}
\label{sec:dsg}

We start be setting $\varsigma=-1$ and $\varkappa=1$ so that $x_{+}=t$ and see that
\begin{equation}
  \begin{split}
    \theta &= dt \left(\bH + \frac{t - \sinh (t)}{t^2}\x\cdot\P\right) + \frac{\sinh (t)}{t} d\x\cdot\P \\
    \omega &= \frac{1 - \cosh (t)}{t^2} \left( dt \x\cdot\B - t
      d\x\cdot\B \right).
  \end{split}
\end{equation}
The soldering form is invertible for all $(t,\x)$, since
$\sinh(t)/t \neq 0$ for all $t \in \RR$. From the above soldering
form, it is easily seen that the torsion two-form vanishes and the
curvature two-form is given by
\begin{equation}
  \Omega = \frac{1}{t}\sinh (t) \bB_{i} (dt \wedge dx^{i}).
\end{equation}
The vielbein is given by
\begin{equation}
  E_{\bH} = \frac{\d}{\d t} + \left(1 - t \csch t\right)\frac{x^i}{t} \frac{\d}{\d
    x^i}\quad\text{and}\quad
  E_{\bP_i} = t\csch t \frac{\d}{\d x^i}.
\end{equation}
We can thus find the invariant Galilean structure: the clock
one-form is given by $\tau = \eta(\theta) = dt$ and the spatial co-metric
is given by
\begin{equation}
  h = t^2 \csch^2t \delta^{ij} \frac{\d}{\d x^{i}} \otimes
  \frac{\d}{\d x^{j}}.
\end{equation}
Finally, the fundamental vector fields are
\begin{align}
  \xi_{\bB_i} &= t \frac{\d}{\d x^{i}}
              \\
\xi_{\bH} &= \frac{\d}{\d t} + \left(\frac{1}{t} -
          \coth(t)\right)x^{i}\frac{\d}{\d x^{i}}
          \\
  \xi_{\bP_i} &= t \coth(t) \frac{\d}{dx^{i}} .
\end{align}
We can change coordinates to bring the Galilean structure into a nicer form;
in particular, let
\begin{equation}
	t' = t \quad \text{and} \quad x'^i = \frac{\sinh(t)}{t} x^i.
\end{equation}
In these new coordinates we find $\tau = dt'$ and
\begin{equation}
	h = \delta^{ij} \frac{\d}{\d x'^{i}} \otimes
  \frac{\d}{\d x'^{j}}.
\end{equation}

\subsubsection{Galilean Anti-de~Sitter Spacetime ($\zAdSG$)}
\label{sec:adsg}

For $\varsigma=-1$ and $\varkappa=1$ the soldering form and connection
one-form for the canonical invariant connection are
\begin{equation}\label{eq:sold-conn-adsg}
\begin{split}
  \theta &= dt \left(\bH + \frac{t - \sin t}{t^2}\x\cdot\P\right) + \frac{\sin
    t}{t} d\x\cdot\P \\
  \omega &= \frac{1 - \cos t}{t^2}\left( t \x\cdot\B - t d\x\cdot\B
  \right).
\end{split}
\end{equation}
Because of the zero of $\sin(t)/t$ at $t = \pm \pi$, the soldering
form is an isomorphism for all $\x$ and for $t \in (-\pi,\pi)$, so
that the exponential coordinates are invalid outside of that region.
Let $t_0 \in (-\pi,\pi)$ and $\x_0 \in \RR^D$.  The orbit of the point
$(t_0,\x_0)$ under the one-parameter subgroup of boosts generated by
$\w \cdot \B$ is
\begin{equation}
  t(s) = t_0 \quad\text{and}\quad \x(s) = \x_0 + s t_0 \w.
\end{equation}
The orbits are point-like for $t_0 = 0$ and straight lines for
$t_0 \neq 0$. These orbits remain inside the domain of validity of the
exponential coordinates. The generic orbits are, therefore,
non-compact.
\\ \\
The torsion two-form again vanishes and the curvature form is
\begin{equation}\label{eq:curv-adsg}
  \Omega = \frac{1}{t}\sin t \bB_{i} (dt \wedge dx^{i}).
\end{equation}
The vielbein is given by
\begin{equation}\label{eq:viel-adsg}
  E_{\bH} = \frac{\d}{\d t} + \left(1 - \frac{t}{\sin t} \right) \frac{x^i}{t} \frac{\d}{\d
    x^i}\quad\text{and}\quad
  E_{\bP_i} = \frac{t}{\sin t} \frac{\d}{\d x^i},
\end{equation}
so that the invariant Galilean structure has a clock one-form
$\tau = \eta(\theta) = dt$ and a spatial co-metric
\begin{equation}\label{eq:nc-adsg}
  h = \left(\frac{t}{\sin t}\right)^2 \delta^{ij} \frac{\d}{\d x^{i}} \otimes
  \frac{\d}{\d x^{j}}.
\end{equation}
The fundamental vector fields for Galilean AdS are
\begin{equation}\label{eq:fvf-adsg}
\begin{split}
      \xi_{\bB_{i}} &= t \frac{\d}{\d x^{i}} \\
      \xi_{\bP_{i}} &= t \cot t \frac{\d}{\d x^{i}} \\
      \xi_{\bH} &= \frac{\d}{\d t} + \left(\tfrac{1}{t} - \cot t \right)
      x^{i}\frac{\d}{\d x^{i}} .
    \end{split}
\end{equation}
As in the case of $\zdSG$, we can bring the Galilean structure into a nicer form using
a change of coordinates:
\begin{equation}
	t' = t \quad \text{and} \quad  x'^i =\frac{\sin(t)}{t}x^i.
\end{equation}
With this change of coordinates, we have $\tau = dt'$ and
\begin{equation}
	h = \delta^{ij} \frac{\d}{\d x'^{i}} \otimes
  \frac{\d}{\d x'^{j}}.
\end{equation}

\subsubsection{Ultra-Relativistic Limit}
\label{sec:ultra-relat-limit}

In the ultra-relativistic limit $\varsigma \to 0$ to the Carrollian (anti\nobreakdash-)de~Sitter 
spacetimes we get
$x_{+}= \sqrt{-\frac{\varkappa}{c^2} x^2 }$ and the soldering form and
invariant connection are
\begin{equation}
  \begin{split}
    \theta &= 
    \frac{\sinh x_+}{x_+}
    \left(
      dt \bH +  d\x \cdot \P
    \right)
    + 
    \left(
      1 - \frac{\sinh x_+}{x_+}
    \right) \frac{\x \cdot d\x}{x^{2}}
    \left(
      t \bH  +  \x \cdot \P
     \right)\\
    \omega &= \frac{\cosh x_+-1}{x^2} c^{2}
    \left(
      dt \x \cdot\B
      - t d\x \cdot \B
      - \tfrac1{c^2} \bJ_{ij} x^i dx^j
    \right) .
  \end{split}
\end{equation}
The vielbein in the ultra-relativistic limit has the following form
\begin{equation}
  \begin{split}
  E_{\bH} &=
          x_{+} \csch x_{+}
          \frac{\pd}{\pd t}
  \\
  E_{\bP_{i}}
        &=
          \frac{x^{i}}{x^{2}}
          \left(
          1 - x_{+} \csch x_{+}
          \right)
          \left(
          t \frac{\pd}{\pd t} + x^{j}\frac{\pd}{\pd x^{j}}
          \right)
          +
           x_{+} \csch x_{+} \frac{\pd}{\pd x^{i}} .
 \end{split}
\end{equation}
The ultra-relativistic limit leads to Carrollian structure consisting
of $\kappa=E_{\bH}$ and the spatial metric
$b=\frac{1}{c^{2}} \mathbold{\pi}^{2}$ given by
\begin{align}
  b=
  \frac{1}{c^{2}}
  \left(
  \frac{\sinh x_+}{x_+}
  \right)^{2} d \x \cdot d \x
  +
  \frac{1}{c^{2}}
  \left(
  1 - 
  \left(
  \frac{\sinh x_+}{x_+}
  \right)^{2}
  \right)
  \frac{(\x \cdot d\x)^{2}}{x^{2}}
  .
\end{align}
The fundamental vector fields are
\begin{equation}
  \begin{split}
    \xi_{\bB_i} &= \frac{1}{c^2} x^i \frac{\d}{\d t} 
    \\
    \xi_{\bH} &=
      x_+ \coth x_+ 
       \frac{\d}{\d t}
    \\
    \xi_{\bP_i} &=
     \frac{x^{i}}{x^{2}}
    \left(
      1 - x_+\coth x_+ 
    \right)
    \left(
      t \frac{\d}{\d t} + x^j \frac{\d}{\d x^j}
    \right)
    + x_+ \coth x_+ \frac{\d}{\d x^i} .
\end{split}
\end{equation}

\subsubsection{(Anti\nobreakdash-)de~Sitter Carrollian Spacetimes ($\zdSC$ and $\zAdSC$)}
\label{sec:de-sitter-carroll}

We will treat these two spacetimes together, such that $\varkappa=1$
corresponds to Carrollian de~Sitter ($\hyperlink{S14}{\zdSC}$) and
$\varkappa =-1$ to Carrollian anti-de~Sitter
($\hyperlink{S15}{\zAdSC}$) spacetimes. Furthermore we set $c=1$.
\\ \\
We find that the soldering form is given by
\begin{equation}
  \begin{split}
    \theta^{(\varkappa=1)} &= \frac{\sin|\x|}{|\x|} (dt \bH + d\x \cdot
    \P) + \left( 1 - \frac{\sin|\x|}{|\x|}\right) \frac{\x \cdot
      d\x}{x^2} (t \bH + \x \cdot \P)\\
    \theta^{(\varkappa=-1)} &= \frac{\sinh|\x|}{|\x|}(dt \bH + d\x \cdot
    \P) + \left( 1 - \frac{\sinh|\x|}{|\x|}\right) \frac{\x \cdot
      d\x}{x^2} (t \bH + \x \cdot \P).
  \end{split}
\end{equation}
These soldering forms are invertible whenever the functions
$\frac{\sin|\x|}{|\x|}$ (for $\varkappa=1$) or
$\frac{\sinh|\x|}{|\x|}$ (for $\varkappa=-1$) are invertible. The
latter function is invertible for all $\x$, whereas the former
function is invertible in the open ball $|\x|<\pi$.
\\ \\
The connection one-form is given by
\begin{equation}
  \begin{split}
    \omega^{(\varkappa=1)} &= \frac{\cos|\x| -1}{x^2}( dt \x \cdot
    \B - t d\x \cdot \B + dx^i x^j \bJ_{ij})\\
    \omega^{(\varkappa=-1)} &= \frac{\cosh|\x| -1}{x^2}( dt \x \cdot
    \B - t d\x \cdot \B + dx^i x^j \bJ_{ij}).
  \end{split}
\end{equation}
The canonical connection is torsion-free, since $\mathsf{(A)dSC}$ is
symmetric, but it is not flat. The curvature is given by
\begin{equation}
\begin{split}
  \Omega^{(\varkappa=1)} = & \left(\frac{\sin\,|\x|}{|\x|}\right)^2 \, dt\wedge d\x\cdot\B - \frac{\sin\,|\x|}{|\x|}\left( \frac{\sin\,|\x|}{|\x|} - 1\right) \frac{\x\cdot\B}{\x\cdot\x} dt \wedge d\x\cdot\x + \\
  & \left(\frac{\sin\,|\x|}{|\x|}\right)^2 \bJ_{ij} dx^{i}\wedge dx^{j} + \frac{2 \sin\,|\x|}{|\x|}\left( \frac{\sin\,|\x|}{|\x|} - 1\right) (x^{k}x^{j} \bJ_{ik} - t x^{j}\bB_{i})dx^{i} \wedge dx^{j}, \\
  \Omega^{(\varkappa=-1)} = & -\left(\frac{\sinh\,|\x|}{|\x|}\right)^2  dt\wedge d\x\cdot\B + \frac{\sinh\,|\x|}{|\x|}\left( \frac{\sinh\,|\x|}{|\x|} - 1\right) \frac{\x\cdot\B}{\x\cdot\x} dt \wedge d\x\cdot\x - \\
  & \left(\frac{\sinh\,|\x|}{|\x|}\right)^2 \bJ_{ij} dx^{i}\wedge dx^{j} -
  \frac{2 \sinh\,|\x|}{|\x|}\left( \frac{\sinh\,|\x|}{|\x|} - 1\right)
  (x^{k}x^{j} \bJ_{ik} - t x^{j}\bB_{i})dx^{i} \wedge dx^{j}.
\end{split}
\end{equation}
Using the soldering form, we find the vielbein $E$ to have components
\begin{equation}
\begin{split}
E_{\bH}^{(\varkappa=1)} = |\x|\csc|\x| \frac{\d}{\d t} \quad &\text{and} \quad E_{\bP_i}^{(\varkappa=1)} = \frac{x^i}{x^2} \left(1-|\x|\csc|\x|\right) \left( t \frac{\d}{\d t} + x^j \frac{\d}{\d x^j}\right) + |\x|\csc|\x|\frac{\d}{\d x^i}, \\
E_{\bH}^{(\varkappa=-1)} = |\x|\csch|\x| \frac{\d}{\d t} \quad &\text{and} \quad E_{\bP_i}^{(\varkappa=-1)} = \frac{x^i}{x^2} \left(1-|\x|\csch|\x|\right) \left( t \frac{\d}{\d t} + x^j \frac{\d}{\d x^j}\right) + |\x|\csch|\x|\frac{\d}{\d x^i}.
\end{split}
\end{equation}
The invariant Carrollian structure is given by $\kappa=E_{\bH}$ and the spatial metric
\begin{equation}
\begin{split}
b^{(\varkappa=1)} &= \left( \frac{\sin\, |\x|}{|\x|}\right)^2 d\x\cdot d\x + \left( 1 - \left( \frac{\sin\, |\x|}{|\x|}\right)^2\right)   \frac{(\x \cdot d\x)^{2}}{x^{2}}
 \\
b^{(\varkappa=-1)} &= \left( \frac{\sinh\, |\x|}{|\x|}\right)^2 d\x\cdot d\x + \left( 1 - \left( \frac{\sinh\, |\x|}{|\x|}\right)^2\right)   \frac{(\x \cdot d\x)^{2}}{x^{2}}
.
\end{split}
\end{equation}
Finally, the fundamental vector field of our curved ultra-relativistic algebras are
\begin{equation}
\begin{split}
  \xi_{\bB_i} &= x^i \frac{\d}{\d t} \\
  \xi_{\bH}^{(\varkappa=1)} &= |\x|\cot|\x| \frac{\d}{\d t}\\
  \xi_{\bH}^{(\varkappa=-1)} &= |\x|\coth|\x| \frac{\d}{\d t} \\
  \xi^{(\varkappa=1)}_{\bP_i} &= \frac{x^i}{x^2}(1-|\x|\cot|\x|)\left(t
    \frac{\d}{\d t} + x^j \frac{\d}{\d x^j}\right) + |\x|\cot|\x|
  \frac{\d}{\d x^i}\\
  \xi^{(\varkappa=-1)}_{\bP_i} &= \frac{x^i}{x^2}(1-|\x|\coth|\x|)\left(t
    \frac{\d}{\d t} + x^j \frac{\d}{\d x^j}\right) + |\x|\coth|\x|
  \frac{\d}{\d x^i}.
  \end{split}
\end{equation}

\subsection{Torsional Galilean Spacetimes}
\label{sec:galilean}

Unlike the Galilean symmetric spacetimes discussed in section
\ref{sec:metric}, some Galilean spacetimes do not arise as limits from
the (pseudo\nobreakdash-)Riemannian spacetimes: namely, the torsional
Galilean de~Sitter ($\hyperlink{S9}{\ztdSG_\gamma}$) and anti-de~Sitter
($\hyperlink{S11}{\ztAdSG_\chi}$) spacetimes, which are the
subject of this section. Galilean spacetimes can be seen as null
reductions of Lorentzian spacetimes one dimension higher and it would
be interesting to exhibit these Galilean spacetimes as null
reductions. We hope to return to this question in the future.

\subsubsection{Torsional Galilean de~Sitter Spacetime ($\ztdSG_{\gamma\neq 1}$)}
\label{sec:tdsg}

The additional brackets not involving $\J$ for $\hyperlink{S9}{\ztdSG_\gamma}$ are
$[\bH,\B] = - \P$ and $[\bH,\P] = \gamma \B + (1+\gamma) \P$, where
$\gamma \in (-1,1)$.

\paragraph{Fundamental vector fields}
\label{sec:fund-vect-fields-tdsg}

We start by determining the expressions for the fundamental vector
fields $\xi_{\bB_i}$, $\xi_{\bP_i}$, and $\xi_{\bH}$ relative to the
exponential coordinates.  The boosts are Galilean and hence act in the
usual way, with fundamental vector field
\begin{equation}
  \xi_{\bB_i} = t \frac{\d}{\d x^i}.
\end{equation}
To determine the other fundamental vector fields we must work harder.
The matrix $\ad_A$ in this basis is given by
\begin{equation}
  \ad_A  = t
  \begin{pmatrix}
    \zero  &  \gamma \\
    -1 & 1+\gamma
  \end{pmatrix},
\end{equation}
which is diagonalisable (since $\gamma \neq 1$) with eigenvalues $1$
and $\gamma$, so that $\ad_A = S \Delta S^{-1}$, with
\begin{equation}
  \Delta =
  \begin{pmatrix}
    t & \zero \\
    \zero & t \gamma
  \end{pmatrix} \quad\text{and}\quad S
  =
  \begin{pmatrix} \gamma & 1 \\
    1 & 1
  \end{pmatrix}.
\end{equation}
Therefore if $f(z)$ is analytic,
\begin{equation}
  f(\ad_A)
  = S
  \begin{pmatrix}
    f(t) & 0 \\
    0 & f(\gamma t)
  \end{pmatrix}
  S^{-1},
\end{equation}
so that
\begin{equation}
  \begin{split}
    f(\ad_A) \B &= \frac{f(\gamma t) -\gamma f(t)}{1-\gamma} \B +
    \frac{f(\gamma t) - f(t)}{1-\gamma} \P\\
    f(\ad_A) \P &= \frac{\gamma(f(\gamma t) - f(t))}{\gamma-1} \B +
    \frac{\gamma f(\gamma t) - f(t)}{\gamma-1} \P.
  \end{split}
\end{equation}
On the other hand, $\ad_A \bH = -\gamma \x \cdot \B - (1+\gamma) \x
\cdot\P$, so if $f(z) = 1 + z \widetilde{f}(z)$, then
\begin{equation}
  \begin{split}
      f(\ad_A) \bH &= \bH - \gamma \widetilde{f}(\ad_A) \x \cdot \B - (1+\gamma)
      \widetilde{f}(\ad_A) \x \cdot \P\\
      &= \bH + \frac{\gamma}{1-\gamma} \left(\gamma \widetilde{f}(\gamma t) - \widetilde{f}(t) \right)
      \x \cdot \B + \frac{1}{1-\gamma}\left( \gamma^2 \widetilde{f}(\gamma t) - \widetilde{f}(t) \right) \x \cdot \P,
  \end{split}
\end{equation}
where $\widetilde{f}(t) = (f(t)-1)/t$. With these expressions we can
now use equation~\eqref{eq:master} to solve for the fundamental vector
fields.
\\ \\
Put $X = \v\cdot \P$ and $Y'(0) = \bbeta \cdot \B$ in
equation~\eqref{eq:master} to obtain that $\tau = 0$ and
\begin{equation}
  \begin{split}
    \y \cdot \P &= \tfrac1{\gamma-1}
    \left[
      \gamma
      \left(G(\gamma t) -
        \gamma G(t)
      \right) \v \cdot \B
      + \left(
        \gamma G(\gamma t) -  G(t)
      \right) \v \cdot \P
    \right] \\
    & \quad
    - \tfrac1{1-\gamma}
    \left[
      \left(
        F(\gamma t) - \gamma F(t)
      \right)
      \bbeta \cdot \B
      + \left(
        F(\gamma t) - F(t)
      \right) \bbeta \cdot \P
    \right].
\end{split}
\end{equation}
This requires
\begin{equation}
  \bbeta = - \gamma \frac{G(\gamma t) - G(t)}{F(\gamma t) - \gamma F(t)} \v,
\end{equation}
and hence, substituting back into the equation for $\y$ and
simplifying, we obtain
\begin{equation}
  \y = t \left( -1 + \frac{(\gamma-1)e^t}{e^{\gamma t} - e^t} \right) \v,
\end{equation}
so that
\begin{equation}\label{eq:xi-P-tdsg}
  \xi_{\bP_i} = t \left( -1 + \frac{(\gamma-1)e^t}{e^{\gamma t} - e^t} \right)  \frac{\d}{\d x^i}.
\end{equation}
Finally, let $X = \bH$ and $Y'(0) = \bbeta \cdot \B$ in
equation~\eqref{eq:master} to obtain that $\tau = 1$ and
\begin{equation}
  \begin{split}
  \y \cdot \P &= \tfrac{\gamma}{1-\gamma} \left(\gamma h(\gamma t) - h(t) \right) 
      \x \cdot \B + \tfrac{1}{1-\gamma}\left( \gamma^2 h(\gamma t) -
        h(t) \right) \x \cdot \P\\
      & \quad {} - \tfrac1{1-\gamma} \left( F(\gamma t) - \gamma F(t) \right) \bbeta \cdot \B  - \tfrac1{1-\gamma} \left(F(\gamma t) -
      F(t) \right) \bbeta \cdot \P,
  \end{split}
\end{equation}
where $h(t) = (G(t)-1)/t$.  This requires
\begin{equation}
  \bbeta = \gamma \frac{\gamma h(\gamma t) - h(t)}{F(\gamma t) - \gamma,
    F(t)} \x
\end{equation}
so that
\begin{equation}
  \y = \left( 1 + \frac1t + \frac{(1-\gamma) e^t}{e^{\gamma t} - e^t}\right)\x.
\end{equation}
This means that
\begin{equation}\label{eq:xi-H-tdsg}
  \xi_{\bH} = \frac{\d}{\d t} + \left( 1 + \frac1t + \frac{(1-\gamma) e^t}{e^{\gamma t} - e^t}\right) x^i \frac{\d}{\d x^i}.
\end{equation}
We can easily check that $[\xi_{\bH}, \xi_{\bB_i}] = \xi_{\bP_i}$ and
$[\xi_{\bH},\xi_{\bP_i}] = -\gamma \xi_{\bB_i} - (1 + \gamma) \xi_{\bP_i}$.

\paragraph{Soldering form and canonical connection}
\label{sec:sold-form-canon-tdsg}

This homogeneous spacetime is reductive, so we have not just a
soldering form, but also a canonical invariant connection, which can
be determined via equation~\eqref{eq:MC-pullback}:
\begin{equation}
  \begin{split}
    \theta + \omega &= D(\ad_A) (dt \bH + d\x \cdot \P)\\
    &= dt (\bH + \tfrac\gamma{1-\gamma}(\gamma \widetilde{D}(\gamma t) - \widetilde{D}(t)) \x
    \cdot \B + \tfrac1{1-\gamma}(\gamma^2 \widetilde{D}(\gamma t) - \widetilde{D}(t)) \x \cdot\P\\
    & \quad {} + \tfrac\gamma{\gamma-1}(D(\gamma t) - D(t)) d\x\cdot\B
    + \tfrac1{\gamma-1}(\gamma D(\gamma t) - D(t)) d\x \cdot \P,
  \end{split}
\end{equation}
where now $\widetilde{D}(z) = (D(z)-1)/z$.  Substituting $D(z) = (1-e^{-z})/z$, we
find that the soldering form is given by
\begin{equation}\label{eq:theta-tdsg}
  \theta = dt \left( \bH + \frac1t \x \cdot \P \right) + \frac{e^{-t}-e^{-\gamma
        t}}{t^2(1-\gamma)} \left( dt \x - t d\x \right) \cdot \P,
\end{equation}
from where it follows that $\theta$ is invertible for all $(t,\x)$.
The canonical invariant connection is given by
\begin{equation}\label{eq:omega-tsdg}
  \omega = \left( \frac1{t^2} + \frac{\gamma e^{-t} - e^{-\gamma
        t}}{t^2(1-\gamma)} \right) (dt \x  - t d \x) \cdot \B.
\end{equation}
The torsion and curvature of the canonical invariant connection are easily
determined from equations~\eqref{eq:torsion} and~\eqref{eq:curvature},
respectively:
\begin{equation}
  \Theta =\left(\frac{1+\gamma}{1-\gamma}\right) \frac{e^{-t} - e^{-\gamma t}}{t}
  dt \wedge d\x \cdot \P \quad\text{and}\quad
  \Omega = \left(\frac{\gamma}{1-\gamma}\right) \frac{e^{-t} - e^{-\gamma t}}{t}
  dt \wedge d\x \cdot \B.
\end{equation}
This spacetime admits an invariant Galilean structure with clock form
$\tau = \eta(\theta) = dt$ and spatial metric on one-forms
$h = \delta^{ij} E_{\bP_i} \otimes E_{\bP_j}$, where $E$ is the vielbein
obtained by inverting the soldering form:
\begin{equation}
  E_{\bH} = \frac{\d}{\d t} + \left(\frac1t - \frac{\gamma-1}{e^{-t} -
      e^{-t\gamma}}\right) x^i \frac{\d}{\d x^i}\quad\text{and}\quad
  E_{\bP_i} = \frac{t(\gamma-1)}{e^{-t}-e^{-\gamma t}} \frac{\d}{\d x^i}.
\end{equation}
Therefore, the spatial co-metric of the Galilean structure is given by
\begin{equation}
  h = \frac{t^2(\gamma-1)^2}{(e^{-t} - e^{-\gamma t})^2}
  \delta^{ij}\frac{\d}{\d x^i} \otimes \frac{\d}{\d x^j}.
\end{equation}
We can simplify this expression using the following coordinate
transformation.
\begin{equation}
	t' = t \quad \text{and} \quad x'^i = \frac{e^{-t} - e^{-\gamma t}}{t(\gamma -1)} x^i.
\end{equation}
In these new coordinates, we have $\tau = dt'$ and
\begin{equation}
	h = \delta^{ij}\frac{\d}{\d x'^i} \otimes \frac{\d}{\d x'^j}.
\end{equation}

\subsubsection{Torsional Galilean de~Sitter Spacetime ($\ztdSG_{\gamma=1}$)}
\label{sec:tdsg-1}

This is $\hyperlink{S9}{\ztdSG_1}$, which is the $\gamma\to1$ limit of
the previous example. Some of the expressions in the previous section
have removable singularities at $\gamma = 1$, so it seems that
treating that case in a separate section leads to a more transparent
exposition.
\\ \\
The additional brackets not involving $\J$ are now $[\bH,\B] = - \P$ and
$[\bH,\P] = 2 \P + \B$. We start by determining the expressions for the
fundamental vector fields $\xi_{\bB_i}$, $\xi_{\bP_i}$, and $\xi_{\bH}$
relative to the exponential coordinates $(t,\x)$, where
$\sigma(t,\x) = \exp(t \bH + \x \cdot \P)$.

\paragraph{Fundamental vector fields}
\label{sec:fund-vect-fields-}

The bracket $[\bH,\B] = -\P$ shows that $\B$ acts as a Galilean
boost.  We can, therefore, immediately write down
\begin{equation}
  \xi_{\bB_i} = t \frac{\d}{\d x^i}.
\end{equation}
To find the other fundamental vector fields requires solving
equation~\eqref{eq:master} with $A = t\bH + \x \cdot \P$ and
$Y'(0)= \bbeta \cdot \B$ (for this Lie algebra) for $X = \bP_i$ and $X = \bH$.
To apply equation~\eqref{eq:master} we must first determine how to act
with $f(\ad_A)$ on the generators, where $f(z)$ is analytic in $z$.
\\ \\
We start from
\begin{equation}
  \begin{split}
    \ad_A \bH &= - \x \cdot \B - 2 \x \cdot \P\\
    \ad_A \P &= 2 t \P + t \B\\
    \ad_A \B &= -t \P.
  \end{split}
\end{equation}
It follows from the last two expressions that
\begin{equation}
  \ad_A \begin{pmatrix} \B & \P \end{pmatrix} = 
  \begin{pmatrix} \B & \P \end{pmatrix}
  \begin{pmatrix}
    \zero & t \\ -t & 2 t
  \end{pmatrix},
\end{equation}
where the matrix
\begin{equation}
  M =
  \begin{pmatrix}
    \zero & 1 \\ -1 & 2 
  \end{pmatrix}
\end{equation}
is not diagonalisable, but may be brought to Jordan normal form
$M = S J S^{-1}$, where
\begin{equation}
  J = 
  \begin{pmatrix}
    1 & \zero \\ 1 & 1
  \end{pmatrix}
  \quad\text{and}\quad
  S = S^{-1} =
  \begin{pmatrix}
    1 & -1 \\ \zero & -1
  \end{pmatrix}.
\end{equation}
It follows that for $f(z)$ analytic in $z$,
\begin{equation}
  f(\ad_A) \begin{pmatrix} \B & \P \end{pmatrix} = 
  \begin{pmatrix} \B & \P \end{pmatrix} S f(tJ) S.
\end{equation}
If $f(z) = \sum_{n=0}^\infty c_n z^n$,
\begin{equation}
  f(t J) = \sum_{n=0}^\infty c_n t^n
  \begin{pmatrix}
    1 & \zero \\ n & 1
  \end{pmatrix} =
  \begin{pmatrix}
    f(t) & \zero \\ t f'(t) & f(t)
  \end{pmatrix}.
\end{equation}
Performing the matrix multiplication, we arrive at
\begin{equation}
  \begin{split}
    f(\ad_A) \B & = (f(t) - t f'(t)) \B - t f'(t)) \P\\
    f(\ad_A) \P & = t f'(t)\B + (f(t) + t f'(t)) \P.
  \end{split}
\end{equation}
Similarly,
\begin{equation}
  f(\ad_A) \bH = f(0) \bH - 2 \x \cdot \widetilde{f}(\ad_A) \P - \x \cdot \widetilde{f}(\ad_A) \B,
\end{equation}
where $\widetilde{f}(z) = (f(z)-f(0))/z$.
\\ \\
We are now ready to apply equation~\eqref{eq:master}. Let
$X = \v \cdot \P$. Then equation~\eqref{eq:master} becomes
\begin{equation}
  \begin{split}
    \tau \bH + \y \cdot \P &= G(\ad_A) \v \cdot \P - F(\ad_A) \bbeta \cdot \B\\
    &= (G(t) + t G'(t)) \v \cdot \P + t G'(t) \v \cdot \B - (F(t) - t
    F'(t))\bbeta \cdot \B + t F'(t) \bbeta \cdot \P,
  \end{split}
\end{equation}
from where we find that $\tau = 0$,
\begin{equation}
  \bbeta = \frac{t G'(t)}{F(t) - tF'(t)} \v \quad\text{and hence}\quad
  \y = \frac{F(t)G(t) + t (F(t) G'(t)-F'(t)G(t))}{F(t) - t F'(t)} \v =
  (1-t) \v,
\end{equation}
so that
\begin{equation}
  \xi_{\bP_i} = (1-t) \frac{\d}{\d x^i},
\end{equation}
which is indeed the limit $\gamma \to 1$ of
equation~\eqref{eq:xi-P-tdsg}.
\\ \\
Now let $X = \bH$, so that equation~\eqref{eq:master} becomes
\begin{equation}
  \begin{split}
    \tau \bH + \y \cdot \P &= G(\ad_A) \bH - \bbeta \cdot F(\ad_A) \B \\
    &= \bH - 2 \x \cdot \widetilde{G}(\ad_A) \P - \x \cdot
    \widetilde{G}  (\ad_A) \B -  \bbeta \cdot  F(\ad_A)\B\\
    &= \bH - (\widetilde{G}(t) + t \widetilde{G}'(t)) \x \cdot \B -
    (F(t) -t F'(t)) \bbeta \cdot \B - (2 \widetilde{G}(t) + t
    \widetilde{G}'(t))\x \cdot \P + t F'(t) \bbeta\cdot \P, 
  \end{split}
\end{equation}
from where $\tau = 1$,
\begin{equation}
  \bbeta = \frac{\widetilde{G}(t) + t \widetilde{G}'(t)}{t F'(t) - F(t)}
  \x \quad\text{and hence}\quad
  \y = \frac{t(F'(t) \widetilde{G}(t) - F(t) \widetilde{G}'(t)) - 2
    F(t) \widetilde{G}(t)}{F(t) - t F'(t)} \x = \x.
\end{equation}
In summary,
\begin{equation}
  \xi_{\bH} = \frac{\d}{\d t} + x^i \frac{\d}{\d x^i},
\end{equation}
which is indeed the $\gamma\to 1$ limit of
equation~\eqref{eq:xi-H-tdsg}.  

\paragraph{Soldering form and canonical connection}
\label{sec:sold-form-canon-tdsg1}

To calculate the soldering form and the connection one-form for the
canonical invariant connection, we apply equation
\eqref{eq:MC-pullback}:
\begin{equation}
  \begin{split}
    \sigma^*\vartheta &= D(\ad_A) (dt \bH + d\x \cdot \P)\\
    &= dt \left( \bH - 2 \x \cdot \widetilde{D}(\ad_A) \P - \x \cdot
      \widetilde{D}(\ad_A)\B\right) + d\x \cdot D(\ad_A) \P\\
    &= dt \left( \bH - ( \widetilde{D}(t) + t  \widetilde{D}'(t))
      \x\cdot\B - (2  \widetilde{D}(t) + t  \widetilde{D}'(t)) \x
      \cdot \P \right) + (D(t) + t D'(t)) d\x \cdot \P + t D'(t) d\x
    \cdot \B.
  \end{split}
\end{equation}
Performing the calculation,
\begin{equation}
  \begin{split}
    \theta &= dt \left( \bH + \frac{1-e^{-t}}{t} \x \cdot \P \right) + e^{-t} d\x
    \cdot \P\\
    \omega &= \frac1t \left(\frac{1-e^{-t}}{t} - e^{-t}\right) (\x
    \cdot \B dt - t d\x \cdot \B),
  \end{split}
\end{equation}
which are equations~\eqref{eq:theta-tdsg} and~\eqref{eq:omega-tsdg} in
the limit $\gamma\to 1$. Notice that $\theta$ is an isomorphism for
all $(t,\x)$.
\\ \\
The torsion and curvature two-forms for the canonical
invariant connection are given by
\begin{equation}
  \Theta = -2 e^{-t} dt \wedge d \x \cdot \P \quad\text{and}\quad
  \Omega = - e^{-t} dt \wedge d \x \cdot \B.
\end{equation}
The vielbein $E$ has components
\begin{equation}
  E_{\bH} = \frac{\d}{\d t} + \frac{1-e^{t}}{t} x^i \frac{\d}{\d x^i}
  \quad\text{and}\quad
  E_{P_a} = e^t \frac{\d}{\d x^a}.
\end{equation}
The invariant Galilean structure has clock form $\tau
=\eta(\theta) = dt$ and inverse spatial co-metric
\begin{equation}
  h = \delta^{ij} E_{\bP_i}\otimes E_{\bP_j} = e^{2t}  \delta^{ij}
  \frac{\d}{\d x^i} \otimes \frac{\d}{\d x^j}.
\end{equation}
This expression can be simplified using the coordinate transformation
\begin{equation}
	t' = t \quad \text{and} \quad x'^i = e^{-t} x^i.
\end{equation}
In these coordinates, we find $\tau = dt'$ and 
\begin{equation}
h = \delta^{ij}
  \frac{\d}{\d x'^i} \otimes \frac{\d}{\d x'^j}.
\end{equation}

\subsubsection{Torsional Galilean Anti-de~Sitter Spacetime ($\ztAdSG_\chi$)}
\label{sec:spacetime-tadsg}

In this instance, the additional non-vanishing brackets are $[\bH,\B] = - \P$ and $[\bH,\P] = (1+\chi^2)\B + 2\chi \P$.

\paragraph{Fundamental vector fields}
\label{sec:fund-vect-fields-tadsg}

Since $\B$ acts via Galilean boosts we can immediately write down
\begin{equation}
  \xi_{\bB_i} = t \frac{\d}{\d x^i}.
\end{equation}
To calculate the other fundamental vector fields we employ
equation~\eqref{eq:master}. The adjoint action of
$A = t \bH + \x \cdot \P$ is given by
\begin{equation}
  \begin{split}
    \ad_A \bH &= -(1+\chi^2) \x \cdot \B - 2 \chi \x \cdot \P\\
    \ad_A \B &= - t \P\\
    \ad_A \P &= t (1+\chi^2) \B + 2 t \chi \P.
  \end{split}
\end{equation}
In matrix form,
\begin{equation}
  \ad_A \begin{pmatrix} \B & \P \end{pmatrix}= \begin{pmatrix} \B & \P \end{pmatrix}
  \begin{pmatrix}
    \zero & (1+\chi^2)t \\ -t & 2 t \chi
  \end{pmatrix}.
\end{equation}
We notice that this matrix is diagonalisable:
\begin{equation}
  \begin{pmatrix}
    \zero & (1+\chi^2) \\ -1 & 2 \chi
  \end{pmatrix} = S \Delta S^{-1}, \quad\text{where}\quad S :=
  \begin{pmatrix}
    \chi+i & \chi - i\\ 1 & 1
  \end{pmatrix}
  \quad\text{and}\quad
  \Delta :=
  \begin{pmatrix}
    \chi -i & \zero \\ \zero & \chi + i
  \end{pmatrix}.
\end{equation}
So if $f(z)$ is analytic in $z$,
\begin{equation}
  f(\ad_A) \begin{pmatrix} \B & \P \end{pmatrix} = \begin{pmatrix} \B
    & \P \end{pmatrix} S f(t\Delta) S^{-1}~,
\end{equation}
or letting $t_\pm := t(\chi \pm i)$,
\begin{equation}
  \begin{split}
    f(\ad_A) \B &= \tfrac{i}2 (f(t_+) - f(t_-)) (\P + \chi \B) + \tfrac12 (f(t_+) + f(t_-))\B\\
    f(\ad_A) \P &=-\tfrac{i}2 (f(t_+) - f(t_-))(\chi \P +
    (1+\chi^2) \B)  + \tfrac12 (f(t_+) + f(t_-))\P.
  \end{split}
\end{equation}
Similarly,
\begin{equation}
  \begin{split}
    f(\ad_A) \bH &= f(0) \bH + \frac1{\ad_A}(f(\ad_A)-f(0)) \ad_A \bH\\
    &= f(0) \bH - (1+\chi^2) \x \cdot \widetilde{f}(\ad_A) \B - 2 \chi \x \cdot \widetilde{f}(\ad_A) \P,
  \end{split}
\end{equation}
where $\widetilde{f}(z) := (f(z)-f(0))/z$. With these formulae we can
now use equation~\eqref{eq:master} to find out the expressions for the
fundamental vector fields $\xi_{\bH}$ and $\xi_{bP_i}$. Putting
$X = \v \cdot \P$ and $Y'(0) = \bbeta \cdot \B$ in
equation~\eqref{eq:master} we arrive at
\begin{equation}
  \bbeta = \frac{-i(1+\chi^2)(G(t_+)-G(t_-))}{F(t_+)+F(t_-) + i
    \chi(F(t_+)-F(t_-))} \v
\end{equation}
and hence
\begin{equation}\label{eq:xi-P-tadsg}
  \xi_{\bP_i} = t (\cot t - \chi) \frac{\d}{\d x^i}.
\end{equation}
Similarly, putting $X = \bH$ and $Y'(0) = \bbeta \cdot \B$ in 
equation~\eqref{eq:master} we find
\begin{equation}
  \bbeta = \frac{i\chi(\widetilde{G}(t_+) - \widetilde{G}(t_-)) -
    (\widetilde{G}(t_+)+\widetilde{G}(t_-))}{F(t_+) + F(t_-) + i
    \chi (F(t_+) - F(t_-))} \x
\end{equation}
and hence
\begin{equation}\label{eq:xi-H-tadsg}
  \xi_{\bH} = \frac{\d}{\d t} + \left(\tfrac1t + \chi - \cot
      t\right) x^i \frac{\d}{\d x^i}.
\end{equation}
We check that $[\xi_{\bH}, \xi_{\bB_i}] = \xi_{\bP_i}$ and
$[\xi_{\bH}, \xi_{\bP_i}] = -(1+\chi^2) \xi_{\bB_i} -2 \chi \xi_{\bP_i}$, as
expected. Another check is that taking $\chi \to 0$, we recover the
fundamental vector fields for Galilean anti-de~Sitter spacetime given
by equation~\eqref{eq:fvf-adsg}.

\paragraph{Soldering form and canonical connection}
\label{sec:sold-form-canon-tadsg}

Let us now use equation~\eqref{eq:MC-pullback} to calculate the
soldering form $\theta$ and the connection one-form $\omega$ for the
canonical invariant connection:
\begin{equation}
  \begin{split}
    \theta + \omega &= D(\ad_A) (dt \bH + d\x \cdot \P)\\
    &= dt \left(\bH - (1+\chi^2) \x \cdot \widetilde{D}(\ad_A) \B - 2
      \chi \x \cdot \widetilde{D}(\ad_A) \P\right) + d\x \cdot
    D(\ad_A) \P, \\
  \end{split}
\end{equation}
where $\widetilde{D}(z) = (D(z) -1)/z$.  Evaluating these expressions,
we find
\begin{equation}
  \theta = dt \left(\bH + \frac{(t-e^{\chi t}\sin t)}{t^2} \x \cdot
    \P\right) + \frac1{t} e^{-\chi t}\sin t d\x \cdot \P
\end{equation}
and
\begin{equation}
  \omega = \frac{1 - e^{-\chi t}(\cos t + \chi \sin t)}{t^2} (dt
  \x \cdot \B - t d\x \cdot \B).
\end{equation}
Again, the zeros of $\frac{e^{-\chi t}\sin t}{t}$ at $t = \pm \pi$
invalidate the exponential coordinates for $t \not\in
(-\pi,\pi)$.
\\ \\
The torsion and curvature of the canonical invariant connection are
easily calculated to be
\begin{equation}
  \begin{split}
    \Theta &=  -\frac{2\chi}{t} e^{-\chi t} \sin t dt \wedge d\x
    \cdot \P\\
    \Omega &= - \frac{(1+\chi^2)}{t}  e^{-\chi t} \sin t dt \wedge d\x
    \cdot \B.
  \end{split}
\end{equation}
As $\chi \to 0$, the torsion vanishes and the curvature agrees with
that of the Galilean anti-de~Sitter spacetime (\adsg) in
equation~\eqref{eq:curv-adsg}.
\\ \\
The vielbein $E$ has components
\begin{equation}
  \begin{split}
    E_{\bH} &= \frac{\d}{\d t} + \left(\frac1t - e^{\chi t}\csc t\right)x^i\frac{\d}{\d x^i}\\
    E_{\bP_i} &= t e^{\chi t}\csc t \frac{\d}{\d x^i},
  \end{split}
\end{equation}
whose $\chi \to 0$ limit agrees with equation~\eqref{eq:viel-adsg}.
The invariant Galilean structure has clock form $\tau =
\eta(\theta) = dt$ and inverse spatial metric
\begin{equation}
  h = t^2 e^{2 \chi t} \csc^2t \delta^{ij}  \frac{\d}{\d x^i}
  \otimes  \frac{\d}{\d x^j},
\end{equation}
which again agrees with equation~\eqref{eq:nc-adsg} in the limit $\chi
\to 0$.  These expressions for the Galilean structure can be simplified by changing
coordinates.  In particular, let
\begin{equation}
	t' = t \quad \text{and} \quad x'i = \frac{e^{-\chi t}}{t} \sin(t) x^i.
\end{equation}
In our new coordinates, the clock one-form is $\tau = dt'$ and the co-metric becomes
\begin{equation}
	h = \delta^{ij}  \frac{\d}{\d x'^i}
  \otimes  \frac{\d}{\d x'^j}.
\end{equation}

\subsubsection{The Action of the Boosts}
\label{sec:action-boosts-2}

In this section, we show that the generic orbits of boosts are not
compact in the torsional Galilean spacetimes discussed above. This
requires a different argument to the ones we used for the symmetric
spaces.
\\ \\
Let $\M$ be one of the torsional Galilean spacetimes discussed in this
section; that is, $\hyperlink{S9}{\ztdSG_\gamma}$ or
$\hyperlink{S11}{\ztAdSG_\chi}$, for the relevant
ranges of their parameters. The following discussion applies verbatim
to the torsional Galilean (anti\nobreakdash-)de~Sitter.
\\ \\
Our default description of $\M$ is as a simply-connected kinematical
homogeneous spacetime $\Kgr/\Hgr$, where $\Kgr$ is a simply-connected
kinematical Lie group and $\Hgr$ is the connected subgroup generated
by the boots and rotations. Our first observation is that we may
dispense with the rotations and also describe $\M$ as $\Sgr/\Bgr$,
where $\Sgr$ is the simply-connected solvable Lie group generated by
the boosts and spatio-temporal translations and $\Bgr$ is the
connected abelian subgroup generated by the boosts. This restriction from 
$\Kgr/\Hgr$ to $\Sgr/\Bgr$ can be understood as follows.  By definition, 
we know that we have a transitive $\Kgr$-action on $\M$.  Since $\Sgr$ 
is a subgroup of $\Kgr$, we find that we also have a transitive 
$\Sgr$-action on $\M$.  The typical stabiliser subgroup for this new 
action is not $\Hgr$ but $\Bgr$.  This statement tells us that $\M \cong \Sgr/\Bgr$. 
By construction, the action of the boosts will be the same on both 
$\Kgr/\Hgr$ and $\Sgr/\Bgr$, so although we started with the Klein 
pair $(\k, \h)$, we may have equally started with $(\s, \b)$ to get an 
equivalent geometric realisation of $\M$, where  the Lie algebra
$\s$ of $\Sgr$ is spanned by $\bH,\bB_i,\bP_i$ and the Lie algebra $\b$ of
$\Bgr$ is spanned by $\bB_i$ with non-zero brackets
\begin{equation}
  [\bH,\bB_i] = -\bP_i \quad\text{and}\quad [\bH,\bP_i] = \alpha \bB_i + \beta \bP_i~,
\end{equation}
for some real numbers $\alpha,\beta$ depending on the parameters
$\gamma$, $\chi$.  We may identify $\s$ with the Lie subalgebra of
$\gl(2D+1,\RR)$ given by
\begin{equation}
  \s = \left\{
    \begin{pmatrix}
      \zero & t \alpha \1 & \y \\
      -t \1 & t \beta \1 & \x \\
      \zero & \zero & \zero      
    \end{pmatrix}
\middle | (t,\x,\y) \in \RR^{2D+1}\right\},
\end{equation}
where $\1$ is the $D\times D$ identity matrix and $\b$ with the Lie
subalgebra
\begin{equation}
  \b = \left\{
    \begin{pmatrix}
      \zero & \zero & \y \\
      \zero & \zero & \zero \\
      \zero & \zero & \zero      
    \end{pmatrix}
\middle | \y \in \RR^D\right\}.
\end{equation}
The Lie algebras $\b \subset \s \subset \gl(2D+1,\RR)$ are the Lie
algebras of the subgroups
$\Bgrbar \subset \Sgrbar \subset \GL(2D+1,\RR)$ given by
\begin{equation}
  \Sgrbar = \left\{
    \begin{pmatrix}
      a(t) \1 & b(t) \1 & \y \\
      c(t) \1 & d(t) \1 & \x \\
      \zero & \zero & 1      
    \end{pmatrix}
    \middle | (t,\x,\y) \in \RR^{2D+1}\right\} \quad\text{and}\quad
  \Bgrbar = \left\{
    \begin{pmatrix}
      \1 & \zero & \y \\
      \zero & \1 & \zero \\
      \zero & \zero & 1      
    \end{pmatrix}
    \middle | \y \in \RR^D\right\},
\end{equation}
for some functions $a(t),b(t),c(t),d(t)$, which are given explicitly by
\begin{equation}\label{eq:abcd-tdsg}
  \begin{pmatrix}
    a(t) & b(t) \\ c(t) & d(t) 
  \end{pmatrix} = \frac{1}{\gamma -1}
  \begin{pmatrix}
    \gamma e^t - e^{\gamma t} & \gamma \left( e^{\gamma t} - e^t\right)\\
    e^t - e^{\gamma t} & \gamma e^{t\gamma} - e^t
  \end{pmatrix}
\end{equation}
for $\hyperlink{S9}{\ztdSG_\gamma}$ with $\gamma \in (-1,1)$,
\begin{equation}\label{eq:abcd-tdsg1}
  \begin{pmatrix}
    a(t) & b(t) \\ c(t) & d(t) 
  \end{pmatrix} =
  \begin{pmatrix}
    e^t (1-t) & e^t t \\
    -e^t t & e^t (1+t)
  \end{pmatrix}
\end{equation}
for $\hyperlink{S9}{\ztdSG_1}$, and
\begin{equation}\label{eq:abcd-tadsg}
  \begin{pmatrix}
    a(t) & b(t) \\ c(t) & d(t) 
  \end{pmatrix} = 
  \begin{pmatrix}
    e^{t\chi} (\cos t - \chi \sin t) & e^{t\chi} (1+\chi^2) \sin t\\
    -e^{t\chi} \sin t & e^{t\chi} (\cos t + \chi \sin t)
  \end{pmatrix}
\end{equation}
for $\hyperlink{S11}{\ztAdSG_\chi}$ with $\chi > 0$. The homogeneous
space $\Mbar = \Sgrbar/\Bgrbar$, if not simply-connected, is
nevertheless a discrete quotient of the simply-connected $\M$ and, as
argued at the end of Section~\ref{sec:boosts}, it is enough to show
that the orbits of boosts in $\Mbar$ are generically non-compact to
deduce that the same holds for $\M$.
\\ \\
Let us denote by $g(t,\x,\y) \in \Sgrbar$ the generic group element
\begin{equation}
  g(t,\x,\y) = \begin{pmatrix}
    a(t) \1 & b(t) \1 & \y \\
    c(t) \1 & d(t) \1 & \x \\
    \zero & \zero & 1      
  \end{pmatrix}\in \Sgrbar,
\end{equation}
so that the generic boost is given by
\begin{equation}
  g(0,0,\y) = \begin{pmatrix}
    \1 & \zero & \y \\
    \zero & \1 & \zero \\
    \zero & \zero & 1      
  \end{pmatrix}\in \Bgrbar.
\end{equation}
Parenthetically, let us remark that while it might be tempting to
identify $\Mbar$ with the submanifold of $\Sgrbar$ consisting of
matrices of the form $g(t,\x,0)$, this would not be correct. For this
to hold true, it would have to be the case that given $g(t,\x,\y)$,
there is some $g(0,0,\w)$ such that
$g(t,\x,\y) g(0,0,\w) = g(t',\x',0)$ for some $t',\x'$. As we now
show, this is only ever the case provided that $a(t) \neq 0$. Indeed,
\begin{equation}
  g(t,\x,\y)g(0,0,\w) = g(t, c(t) \w + \x, a(t)\w + \y),
\end{equation}
and hence this is of the form $g(t',\x',0)$ if and only if we can
solve $a(t) \w + \y = 0$ for $\w$. Clearly this cannot be done if
$a(t) = 0$, which may happen for
$\hyperlink{S9}{\ztdSG_{\gamma\in(0,1)}}$ at
$t = \frac{\log\gamma}{\gamma -1}$ and for
$\hyperlink{S11}{\ztAdSG_{\chi>0}}$ at
$\cos t = \pm \frac{\chi}{\sqrt{1+\chi^2}}$.
\\ \\
The action of the boosts on $\Mbar$ is induced by left multiplication
on $\Sgrbar$:
\begin{equation}
  g(0,0,\v)  g(t,\x,\y)= g(t,\x,\y + \v)
\end{equation}
which simply becomes a translation $\y \mapsto \y + \v$ in $\RR^D$.
This is non-compact in $\Sgrbar$, but we need to show that it is
non-compact in $\Mbar$.
\\ \\
The right action of $\Bgrbar$ is given by
\begin{equation}
  g(t,\x,\y) g(0,0,\w) = g(t,\x + c(t)\w, \y + a(t) \w),
\end{equation}
which is again a translation
$(\x,\y) \mapsto (\x + c(t)\w, \y + a(t)\w)$ in $\RR^{2D}$. The
quotient $\RR^{2D}/\Bgrbar$ is the quotient vector space
$\RR^{2D}/\mathbb{B}$, where $\mathbb{B}\subset \RR^{2D}$ is the image
of the linear map $\RR^D \to \RR^{2D}$ sending
$\w \to (c(t) \w, a(t) \w)$. Notice that $(a(t),c(t)) \neq (0,0)$ for
all $t$, since the matrices in $\Sgrbar$ are invertible, hence
$\BB \cong \RR^D$ and hence the quotient vector space
$\RR^{2D}/\mathbb{B} \cong \RR^{D}$. By the Heine--Borel theorem, it
suffices to show that the orbit is unbounded to conclude that it is
not compact. Let $[(\x,\y)] \in \RR^{2D}/\mathbb{B}$ denote the
equivalence class modulo $\mathbb{B}$ of $(\x,\y) \in \RR^{2D}$. The
distance $d$ between $[(\x,\y)]$ and the boosted $[(\x, \y + \v)]$ is
the minimum of the distance between $(\x,\y)$ and any point on the
coset $[(\x, \y + \v)]$; that is,
\begin{equation}
  d = \min_{\w} \|(\x + c(t)\w, \y + \v + a(t) \w) - (\x,\y)\| =
  \min_{\w} \|(c(t)\w, \v + a(t)\w)\|.
\end{equation}
Completing the square, we find
\begin{equation}
    \|(c\w, \v + a\w)\|^2 = (a^2 + c^2) \left\| \w  +
      \frac{a}{a^2+c^2} \v\right\|^2 + \frac{c^2}{a^2+c^2} \|\v\|^2,
\end{equation}
whose minimum occurs when $\w = - \frac{a}{a^2+c^2} \v$, resulting in
\begin{equation}
  d = \frac{|c(t)|}{\sqrt{a(t)^2+c(t)^2}} \|\v\|.
\end{equation}
As we rescale $\v \mapsto s \v$, this is unbounded provided that
$c(t) \neq 0$. From equations~\eqref{eq:abcd-tdsg},
\eqref{eq:abcd-tdsg1} and~\eqref{eq:abcd-tadsg}, we see that for
$\hyperlink{S9}{\ztdSG_{\gamma\in(-1,1]}}$, $c(t)=0$ if and only if
$t=0$, whereas for $\hyperlink{S11}{\ztAdSG_{\chi>0}}$, $c(t) = 0$ if
and only if $t = n \pi$ for $n \in \ZZ$, and hence, in summary, the
generic orbits are non compact.
\\ \\
Let us remark that for $\hyperlink{S11}{\ztAdSG_{\chi>0}}$, if
$t = n \pi$ for $n\neq 0$ then the exponential coordinate system
breaks down, so that we should restrict to $t \in (-\pi,\pi)$. Indeed,
using the explicit matrix representation, one can determine when the
exponential coordinates on $\Mbar$ stop being injective; that is, when
there are $(t,\x)$ and $(t',\x')$ such that
$\exp(t \bH + \x\cdot \P) = \exp(t'\bH + \x'\cdot \P) B$ for some
$B \in \Bgrbar$. In $\hyperlink{S9}{\ztdSG_{\gamma\in(-1,1]}}$ this
only happens when $t=t'$ and $\x = \x'$, but in
$\hyperlink{S11}{\ztAdSG_{\chi>0}}$ it happens whenever $t=t'= n \pi$
($n\neq 0$) and, if so, for all $\x$, $\x'$.

\subsection{Aristotelian Spacetimes}
\label{sec:aristotelian}

In this section, we study the Aristotelian
spacetimes of Table~\ref{tab:spacetimes}. In particular, we derive their fundamental
vector fields, vielbeins, soldering forms, and canonical connections.

\subsubsection{Static Spacetime ($\zS$)}
\label{sec:static}

This is an affine space and the exponential coordinates $(t,\x)$ are
affine, so that
\begin{equation}
  \xi_{\bH}= \frac{\d}{\d t} \quad\text{and}\quad \xi_{\bP_i} =
  \frac{\d}{\d x^i}.
\end{equation}
Similarly, the soldering form is $\theta = dt \bH + d\x \cdot \P$,
the canonical invariant connection vanishes, and so does the torsion.
The vielbein is
\begin{equation}
  E_{\bH} =   \xi_{\bH} \quad\text{and}\quad E_{\bP_i} = \xi_{\bP_i}.
\end{equation}
We now have both Galilean $(\tau, h)$ and Carrollian $(\kappa, b)$ structures
to define on this spacetime.  First, we find that the clock one-form is 
given by $\tau = \eta(\theta) = dt$ and the spatial co-metric of the Galilean 
structure is written
\begin{equation}
	h = \delta^{ij} \frac{\d}{\d x^i} \otimes \frac{\d}{\d x^j}.
\end{equation}
The Carrollian structure then consists of $\kappa = E_{\bH} = \tfrac{\d}{\d t}$ and
\begin{equation}
	b = d\x \cdot d\x.
\end{equation}

\subsubsection{Torsional Static Spacetime ($\zTS$)}
\label{sec:tst}

In this case, the additional non-vanishing brackets are $[\bH,\P] = \P$.

\paragraph{Fundamental vector fields}
\label{sec:fund-vect-fields-tst}

Letting $A= t\bH + \x \cdot \P$, we find $\ad_A \bH =
- \x \cdot \P$ and $\ad_A \P = t \P$.  Therefore, for any analytic
function $f$, we conclude that
\begin{equation}
  f(\ad_A) \P = f(t) \P \quad\text{and}\quad f(\ad_A) \bH = f(0) \bH -
  \frac1t (f(t) - f(0)) \x \cdot \P.
\end{equation}
Applying this to equation~\eqref{eq:master}, we find
\begin{equation}
  \begin{split}
    \xi_{\bH} &= \frac{\d}{\d t} + \left(\frac1t - \frac1{e^t-1} -1\right)
    x^i \frac{\d}{\d x^i}\\
    \xi_{\bP_i} &= \frac{t}{1-e^{-t}} \frac{\d}{\d x^i},
  \end{split}
\end{equation}
which one can check obey $[\xi_{\bH}, \xi_{\bP_i}] = - \xi_{\bP_i}$, as
expected.

\paragraph{Soldering form and canonical connection}
\label{sec:sold-form-canon-tst}

Applying the same formula to equation~\eqref{eq:MC-pullback}, we find
that the canonical invariant connection one-form vanishes in this
basis and that the soldering form is given by
\begin{equation}
  \theta = dt \left(\bH + \frac1t \left(1-\frac{1-e^{-t}}{t}\right)\x
    \cdot \P\right) + \frac{1-e^{-t}}{t} d\x \cdot \P,
\end{equation}
so that the corresponding vielbein is
\begin{equation}
  E_{\bH} = \frac{\d}{\d t} + \left(\frac1t - \frac1{1-e^{-t}} \right) x^i
  \frac{\d}{\d x^i} \quad\text{and}\quad E_{\bP_i} =
  \frac{t}{1-e^{-t}} \frac{\d}{\d x^i}.
\end{equation}
It is clear from the fact that the function $\frac{1-e^{-t}}{t}$ is
never zero that $\theta$ is invertible for all $(t,\x)$.
\\ \\
Although the canonical connection is flat, its torsion 2-form does not
vanish:
\begin{equation}
  \Theta = \frac{e^{-t}-1}{t} dt \wedge d\x \cdot \P.
\end{equation}
Choosing to change coordinates such that
\begin{equation}
	t' = t \quad \text{and} \quad x'^i = \frac{1-e^{-t}}{t} x^i
\end{equation}
and letting
\begin{equation}
	f(t') = \frac{1}{1-e^{-t'}} - \frac{1}{t'},
\end{equation}
we find that the Galilean structure has clock one-form $\tau = \eta(\theta) = dt'$ and
spatial co-metric
\begin{equation}
	h = \delta^{ij} \frac{\d}{\d x'^i} \otimes \frac{\d}{\d x'^j}.
\end{equation}
Additionally,  the Carrollian structure consists of
\begin{equation}
	\kappa = \frac{\d}{\d t'} - f(t') x'^i \frac{\d}{\d x'^i} \quad \text{and}\quad
	b = f(t')^2 x'^2 dt'^2 + 2 f(t') \x'\cdot d\x' dt' + d\x'\cdot d\x'.
\end{equation}

\subsubsection{Aristotelian Spacetime \athree$_\varepsilon$}
\label{sec:spacetime-a3}

In this instance, the additional non-vanishing brackets are $[\bP_i, \bP_j] = -\varepsilon \bJ_{ij}$.

\paragraph{Fundamental vector fields}
\label{sec:fund-vect-fields-a3}

Let $A = t \bH + \x \cdot \P$.  Then $\ad_A \bH = 0$ and
$\ad_A \bP_i = \varepsilon x^j \bJ_{ij}$.  Continuing, we find
\begin{equation}
  \ad_A^2 \bP_i = \varepsilon x^i \x \cdot \P - \varepsilon x^2 \bP_i
  \quad\text{and}\quad \ad_A^3 \bP_i = (-\varepsilon x^2) \ad_A \bP_i.
\end{equation}
Therefore, an induction argument shows that
\begin{equation}
  \ad_A^n \bP_i = (-\varepsilon x^2) \ad_A^{n-2} \bP_i \quad \forall n
  \geq 3.
\end{equation}
If $f(z)$ is analytic in $z$, then $f(\ad_A) \bH = f(0) \bH$ and
\begin{equation}
  f(\ad_A) \bP_i = \tfrac12 \left( f(x_+) + f(x_-) \right) \bP_i -
    \tfrac12 \left( f(x_+) + f(x_-) - 2 f(0) \right) \frac{x^i
      \x\cdot\P}{x^2} - \frac{\varepsilon}{2 x_+} \left(f(x_+) +
      f(x_-)\right) x^j J_{ij},
\end{equation} 
where
\begin{equation}
  x_\pm = \pm \sqrt{-\varepsilon x^2} =
  \begin{cases}
    \pm |\x| & \varepsilon = -1\\
    \pm i |\x| & \varepsilon = 1.
  \end{cases}
\end{equation}
Similarly, $\ad_A \bJ_{ij} = x^i P_j - x^j P_i$, so that
\begin{equation}
  f(\ad_A) \bJ_{ij} = f(0) \bJ_{ij} + \tfrac12
  \left(\widetilde{f}(x_+)+\widetilde{f}(x_-)\right)(x^i \bP_j - x^j \bP_i) - \frac{\varepsilon}{2
    x_+} \left(\widetilde{f}(x_+) - \widetilde{f}(x_-)\right) x^k (x^i \bJ_{kj} - x^j \bJ_{ki}),
\end{equation}
where $\widetilde{f}(z) = (f(z) - f(0))/z$.
\\ \\
Inserting these formulae in equation~\eqref{eq:master} with $X=\bH$ and
$Y'(0)=0$, we see that
\begin{equation}
  \xi_{\bH} = \frac{\d}{\d t}.
\end{equation}
If instead $X = \v \cdot \P$ and $Y'(0) = \tfrac12
\lambda^{ij}\bJ_{ij}$, we see first of all that $\tau = 0$ and that
demanding that the $\bJ_{ij}$ terms cancel,
\begin{equation}
  \lambda^{ij} = \frac{-\varepsilon \left(G(x_+)-G(x_-)\right)}{x_+
    \left(F(x_+) + F(x_-)\right)}  (x^i v^j - x^j v^i),
\end{equation}
and reinserting into equation~\eqref{eq:master}, we find that
\begin{multline}
  y^i =\tfrac12 \left(G(x_+) + G(x_-) -
    \frac{\left(G(x_+)-G(x_-)\right) \left( F(x_+)-F(x_-) \right)}{F(x_+) +
        F(x_-)}\right) v^i \\
    - \tfrac12 \left(G(x_+) + G(x_-) - 2 - \frac{\left(G(x_+)-G(x_-)\right) \left( F(x_+)-F(x_-) \right)}{F(x_+) +
        F(x_-)}\right) \frac{\v \cdot \x}{x^2} x^i.
\end{multline}
From this we read off the expression for $\xi_{\bP_i}$:
\begin{equation}
  \xi_{\bP_i} = \frac{F(x_+) G(x_-) + F(x_-)G(x_+)}{F(x_+) + F(x_-)}\frac{\d}{\d x^i} 
  + \left(1 - \frac{F(x_+) G(x_-) + F(x_-)G(x_+)}{F(x_+) + F(x_-)}\right) \frac{x^i x^j}{x^2} \frac{\d}{\d x^j},
\end{equation}
which simplifies to
\begin{equation}
  \begin{split}
    \xi_{\bP_i}^{(\varepsilon=1)} &= |\x| \cot|\x| \frac{\d}{\d x^i} + (1- |\x| \cot|\x|) \frac{x^i x^j}{x^2} \frac{\d}{\d x^j}\\
    \xi_{\bP_i}^{(\varepsilon=-1)} &= |\x| \coth|\x| \frac{\d}{\d x^i} + (1- |\x| \coth|\x|) \frac{x^i x^j}{x^2} \frac{\d}{\d x^j}.
  \end{split}
\end{equation}

\paragraph{Soldering form and canonical connection}
\label{sec:sold-form-canon-a3}

The soldering form and connection one-form for the canonical
connection are obtained from equation~\eqref{eq:MC-pullback}, which
says that
\begin{equation}
  \begin{split}
    \theta + \omega &= dt \bH + dx^i D(\ad_A) \bP_i\\
    &=  dt \bH + \tfrac12 (D(x_+)+D(x_-)) d\x \cdot \P  \\
    & \quad  - \tfrac12 (D(x_+)D(x_-) - 2) \frac{\x \cdot d\x}{x^2} \x\cdot \P-
    \frac{\varepsilon}{2 x_+}(D(x_+)-D(x_-)) x^i dx^j \bJ_{ij},
  \end{split}
\end{equation}
such that
\begin{equation}
  \begin{split}
    \theta^{(\varepsilon=1)} &= dt \bH + \frac{\sin|\x|}{|\x|} d\x \cdot \P + \left(1 - \frac{\sin|\x|}{|\x|}\right) \frac{\x \cdot d\x}{x^2} \x\cdot \P\\
    \theta^{(\varepsilon=-1)} &= dt \bH + \frac{\sinh|\x|}{|\x|} d\x \cdot \P + \left(1 - \frac{\sinh|\x|}{|\x|}\right) \frac{\x \cdot d\x}{x^2} \x\cdot \P
  \end{split}
\end{equation}
and
\begin{equation}
  \begin{split}
    \omega^{(\varepsilon=1)} &=   \frac{1 - \cos|\x|}{x^2} x^i dx^j \bJ_{ij}\\
    \omega^{(\varepsilon=-1)} &=   \frac{1 - \cosh|\x|}{x^2} x^i dx^j \bJ_{ij}.
  \end{split}
\end{equation}
It follows that if $\varepsilon=-1$ the soldering form is invertible
for all $(t,\x)$, whereas if $\varepsilon=1$ then it is invertible for
all $t$ but inside the open ball $|\x|<\pi$.  With these caveats in mind,
we find that the vielbeins are
\begin{equation}
\begin{split}
E_{\bH}^{(\varkappa=1)} = \frac{\d}{\d t} \quad &\text{and} \quad E_{\bP_i}^{(\varkappa=1)} = \frac{x^i}{x^2} \left(1-|\x|\csc|\x|\right) x^j \frac{\d}{\d x^j} + |\x|\csc|\x|\frac{\d}{\d x^i}, \\
E_{\bH}^{(\varkappa=-1)} = \frac{\d}{\d t} \quad &\text{and} \quad E_{\bP_i}^{(\varkappa=-1)} = \frac{x^i}{x^2} \left(1-|\x|\csch|\x|\right)  x^j \frac{\d}{\d x^j} + |\x|\csch|\x|\frac{\d}{\d x^i}.
\end{split}
\end{equation}
The torsion of the canonical connection vanishes, since
$[\theta,\theta]_\m = 0$.  The curvature is given by
\begin{equation}
  \begin{split}
    \Omega^{(\varepsilon=1)} &= \tfrac12 \frac{\sin^2|\x|}{x^2} dx^i
    \wedge dx^j \bJ_{ij} + \frac{\sin|\x|}{|\x|}\left(1 -
      \frac{\sin|\x|}{|\x|}\right) \frac{x^j x^k}{x^2} dx^i \wedge dx^k \bJ_{ij}\\
    \Omega^{(\varepsilon=-1)} &= -\tfrac12 \frac{\sinh^2|\x|}{x^2} dx^i
    \wedge dx^j \bJ_{ij} - \frac{\sinh|\x|}{|\x|}\left(1 -
      \frac{\sinh|\x|}{|\x|}\right) \frac{x^j x^k}{x^2} dx^i \wedge dx^k \bJ_{ij}.
  \end{split}
\end{equation}
The Galilean structure in this instance has clock one-form $\tau = dt$ and
spatial co-metric
\begin{equation}
	\begin{split}
	h^{(\varkappa =1)} &= (|\x|\csc|\x|)^2 \delta^{ij} \frac{\d}{\d x^i} \otimes \frac{\d}{\d x^j}
	+ (1- (|\x|\csc|\x|)^2) x^i x^j \frac{\d}{\d x^i} \otimes \frac{\d}{\d x^j} \\
	h^{(\varkappa =-1)} &= (|\x|\csch|\x|)^2 \delta^{ij} \frac{\d}{\d x^i} \otimes \frac{\d}{\d x^j}
	+ (1- (|\x|\csch|\x|)^2) x^i x^j \frac{\d}{\d x^i} \otimes \frac{\d}{\d x^j}.
	\end{split}
\end{equation}
The Carrollian structure is then $\kappa = \tfrac{\d}{\d t}$ and 
\begin{equation}
	\begin{split}
	b^{(\varkappa=1)} &= \left(\frac{\sin|\x|}{|\x|}\right)^2 d\x\cdot d\x + \left(1-\left(\frac{\sin|\x|}{|\x|}\right)^2 \right)
	\frac{(\x\cdot d\x)^2}{|\x|^2} \\
	b^{(\varkappa=-1)} &= \left(\frac{\sinh|\x|}{|\x|}\right)^2 d\x\cdot d\x + \left(1-\left(\frac{\sinh|\x|}{|\x|}\right)^2 \right)
	\frac{(\x\cdot d\x)^2}{|\x|^2} 
	\end{split}
\end{equation}

\subsection{Carrollian Light Cone ($\zLC$)}
\label{sec:spacetime-flc}

The Carrollian light cone $\hyperlink{S16}{\zLC}$ is a hypersurface in
Minkowski spacetime, identifiable with the future light cone. It does
not arise as a limit and has additional brackets $[\bH,\B] = \B$,
$[\bH,\P] = -\P$ and $[\B,\P] = \bH + \J$, which shows that it is a
non-reductive homogeneous spacetime.

\subsubsection{Action of the Boosts}
\label{sec:action-boosts-1}

Although it might be tempting to use that the boosts in Minkowski
spacetime act with generic non-compact orbits to deduce the same about
the boosts in $\hyperlink{S16}{\zLC}$, one has to be careful because
what we call boosts in $\hyperlink{S16}{\zLC}$ might not be
interpretable as boosts in the ambient Minkowski spacetime. Indeed, as
we will now see, boosts in $\hyperlink{S16}{\zLC}$ are actually null
rotations in the ambient Minkowski spacetime.
\\ \\
We first exhibit the isomorphism between the $\hyperlink{S16}{\zLC}$ Lie algebra and
$\so(D+1,1)$.  In the $\hyperlink{S16}{\zLC}$ Lie algebra, the boosts
and translations obey the following brackets:
\begin{equation}
  [\bH,\B] = \B, \quad [\bH,\P] = -\P, \quad\text{and}\quad [\B,\P] = \bH
  + \J.
\end{equation}
If we let $J_{MN}$ be the standard generators of $\so(D+1,1)$ with
$M = (i, \tilde{0},\tilde{1})$, $1 \leq i \leq D$, and with Lie brackets
\begin{equation}
  [J_{MN}, J_{PQ}] = \eta_{NP} J_{MQ} - \eta_{MP} J_{NQ} -  \eta_{NQ} J_{MP} + \eta_{MQ} J_{NP},
\end{equation}
where $\eta_{ij} = \delta_{ij}$, $\eta_{\tilde{0}\tilde{0}} = -1$, and
$\eta_{\tilde{1}\tilde{1}} = 1$, then the correspondence is:
\begin{equation}
  \bJ_{ij} = J_{ij}, \quad \bB_i = \tfrac1{\sqrt2}
  (J_{\tilde{0}i} + J_{i\tilde{1}}), \quad \bP_i = \tfrac1{\sqrt2} (J_{\tilde{0}i} -
  J_{i\tilde{1}}), \quad\text{and}\quad \bH = - J_{\tilde{0}\tilde{1}}.
\end{equation}
We see that, as advertised, the boosts $\bB_i$ are indeed null
rotations.
\\ \\
The boosts act linearly on the ambient coordinates $X^M$ in
Minkowski spacetime, with fundamental vector fields
\begin{equation}
  \zeta_{\bB_i} = \frac1{\sqrt2}\left( -X^{\tilde{0}}\frac{\d}{\d X^i} - X^i
    \frac{\d}{\d X^{\tilde{0}}} + X^i \frac{\d}{\d X^{\tilde{1}}} - X^{\tilde{1}}
    \frac{\d}{\d X^i}\right).
\end{equation}
Consider a linear combination $\bB = w^i \bB_i$ and let $T := X^{\tilde{0}}$, $X :=
w^i X^i$, and $Y := X^{\tilde{1}}$, so that in terms of these coordinates
and dropping the factor of $\frac1{\sqrt2}$,
\begin{equation}
  \zeta_{\bB} =  -T \frac{\d}{\d X} - X \frac{\d}{\d T} + X \frac{\d}{\d Y} - Y \frac{\d}{\d X}.
\end{equation}
This allows us to examine the orbit of this vector field while
focussing on the three-dimensional space with coordinates $T,X,Y$.  The vector field is
linear, so there is a matrix $A$ such that
\begin{equation}
  \zeta_{\bB} = \begin{pmatrix} T & X & Y \end{pmatrix} A
  \begin{pmatrix}
    \frac{\d}{\d T} \\ \frac{\d}{\d X} \\  \frac{\d}{\d Y}
  \end{pmatrix}
  \implies
  A =
  \begin{pmatrix}
    \zero & -1 & \zero\\
    -1 & \zero & 1 \\
    \zero & -1 & \zero
  \end{pmatrix}.
\end{equation}
The matrix $A$ obeys $A^3 =0$, so its exponential is
\begin{equation}
  \exp(s A) =
  \begin{pmatrix}
    1 + \tfrac12 s^2 & -s & -\tfrac12 s^2\\
    -s & 1 & s \\
    \tfrac12 s^2 & -s & 1 - \tfrac12 s^2
  \end{pmatrix}
\end{equation}
and hence the orbit of $(T_0,X_0,Y_0,\dots)$ is given by
\begin{equation}
  \begin{split}
    T(s) &= (1 + \tfrac12 s^2) T_0 - s X_0 - \tfrac{1}{2} s^2 Y_0\\
    X(s) &= -s T_0 + X_0 +s Y_0\\
    Y(s) &= \tfrac{1}{2} s^2 T_0 - s X_0 + (1-\tfrac12 s^2) Y_0,
  \end{split}
\end{equation}
with all other coordinates inert, which is clearly non-compact in the
Minkowski spacetime. But of course, this orbit lies on the future
light cone (indeed, notice that
$-T(s)^2 + X(s)^2 + Y(s)^2 = - T_0^2 + X_0^2 + Y_0^2$), which is a
submanifold, and hence the orbit is also non-compact on
$\hyperlink{S16}{\zLC}$, provided with the subspace topology.

\subsubsection{Fundamental Vector Fields}
\label{sec:fund-vect-fields-flc}

Let $A = t \bH + \x \cdot \P$ and let us calculate the action of $\ad_A$
on the generators, this time with the indices written explicitly:
\begin{equation}
  \begin{split}
    \ad_A \bB_i &= t \bB_i - x^i \bH - x^j \bJ_{ij}\\
    \ad_A \bP_i &= -t \bP_i \\
    \ad_A \bH &= x^i \bP_i\\
    \ad_A \bJ_{ij} &= x^i \bP_j - x^j \bP_i.
  \end{split}
\end{equation}
In order to compute the fundamental vector fields using
equation~\eqref{eq:master} and the soldering form using equation
\eqref{eq:MC-pullback}, we need to calculate the action of certain
universal power series on $\ad_A$ on the generators. To this end, let
us derive formulae for the action of $f(\ad_A)$, for $f(z)$ an
analytic function of $z$, on the generators. We will do this by first
calculating powers of $\ad_A$ on generators. It is clear, first of
all, that on $\P$,
\begin{equation}
  f(\ad_A) \P = f(-t) \P.
\end{equation}
On $\bH$ and $\J$ we just need to treat the constant term
separately:
\begin{equation}
  \begin{split}
    f(\ad_A) \bH &= f(0) \bH - \frac1t \left(f(-t) - f(0)\right) \x \cdot\P\\
    f(\ad_A) \bJ_{ij} &= f(0) \bJ_{ij} - \frac1t \left(f(-t) - f(0)\right)
    (x^i \bP_j - x^j \bP_i).
  \end{split}
\end{equation}
On $\B$ it is a little bit more complicated.  Notice first of all that whereas
\begin{equation}
  \ad_A^2 \bB_i = t \ad_A \bB_i - 2 x^i x^j \bP_j + x^2 \bP_i,
\end{equation}
$\ad_A^3 \bB_i = t^2 \ad_A \bB_i$.  Therefore, by induction, for all $n
\geq 1$,
\begin{equation}
  \ad_A^n \bB_i =
  \begin{cases}
    t^{n-1}\ad_A \bB_i & n~\text{odd}\\
    t^{n-1}\ad_A \bB_i + t^{n-2} (x^2 \bP_i - 2 x^i \x \cdot \P) & n~\text{even},
  \end{cases}
\end{equation}
and therefore
\begin{equation}
  f(\ad_A) \bB_i = f(t) \bB_i - \tfrac1t (f(t)-f(0))(x^i \bH + x^j \bJ_{ij}) +
  \tfrac1{t^2}(\tfrac12(f(t)+f(-t)) - f(0))(x^2 \bP_i - 2 x^i\x \cdot \P).
\end{equation}
Using these formulae, we can now apply equation~\eqref{eq:master} in
order to determine the expression of the fundamental vector fields in
terms of exponential coordinates.
\\ \\
Let us take $X = \v \cdot \P$ in equation~\eqref{eq:master}.  We must
take $Y'(0)=0$ here and find that
\begin{equation}
  \y \cdot \P = G(\ad_A) \v \cdot \P = G(-t) \v \cdot \P  \implies \y
  = \frac{t}{1-e^{-t}}\v,
\end{equation}
resulting in
\begin{equation}
  \xi_{\bP_i} = \frac{t}{1-e^{-t}} \frac{\d}{\d x^i}.
\end{equation}
Taking $X=\bH$ in equation~\eqref{eq:master}, we again must take
$Y'(0)=0$.  Doing so, we arrive at
\begin{equation}
  \tau \bH + \y \cdot \P = G(\ad_A) \bH = \bH - \frac1t \left(G(-t) - 1\right) \x
  \cdot \P \implies \tau = 1 \quad\text{and}\quad \y =
 \left(\frac{1}{t}-1-\frac{1}{e^t-1}\right)\x,
\end{equation}
resulting in
\begin{equation}
  \xi_{\bH} = \frac{\d}{\d t} +
     \left(\frac{1}{t}-1-\frac{1}{e^t-1}\right) x^i \frac{\d}{\d x^i}.
\end{equation}
One checks already that $[\xi_{\bH},\xi_{\bP_i}] = \xi_{\bP_i}$, as expected.
\\ \\
Finally, put $X=\v \cdot \B$ in equation~\eqref{eq:master} and hence
now $Y'(0)= \bbeta \cdot \B + \tfrac12 \lambda^{ij} \bJ_{ij}$.  Substituting
this in equation~\eqref{eq:master} and requiring that the $\h$-terms
vanish, we find
\begin{equation}
  \bbeta = \frac{G(t)}{F(t)} \v = e^{-t} \v \quad\text{and}\quad
  \lambda^{ij} = \frac{1-e^{-t}}{t} (v^i x^j - v^j x^i).
\end{equation}
Comparing the $\bH$ terms, we see that
\begin{equation}
  \tau =\frac{1-e^{-t}}{t} \x \cdot \v,
\end{equation}
whereas the $\P$ terms give
\begin{equation}
  \y =\frac{1-e^{-t}}{2t} x^2 \v + \frac{1-t-e^{-t}}{t^2} \x\cdot\v \x,
\end{equation}
resulting in
\begin{equation}
  \xi_{\bB_i} = \frac{1-e^{-t}}{t} x^i \frac{\d}{\d t} +
    \frac{1-e^{-t}}{2t} x^2 \frac{\d}{\d x^i} + \frac{1-t-e^{-t}}{t^2}
    x^i x^j \frac{\d}{\d x^j}.
\end{equation}
One checks that, as expected, $[\xi_{\bH}, \xi_{\bB_i}] = - \xi_{\bB_i}$ and
that $[\xi_{\bB_i},\xi_{\bP_j}] = - \delta_{ij} \xi_{\bH} - \xi_{\bJ_{ij}}$,
where $\xi_{J_{ij}} = x^j \frac{\d}{\d x^i} - x^i \frac{\d}{\d x^j}$.

\subsubsection{Soldering Form and Canonical Connection}
\label{sec:sold-form-canon-flc}

The soldering form can be calculated from
equation~\eqref{eq:MC-pullback} and projecting the result to $\k/\h$:
\begin{equation}
  \begin{split}
    \theta &= D(\ad_A)(dt \Hbar + d\x\cdot \Pbar) = dt\left(\Hbar -
    \frac{D(-t)-1}{t} \x \cdot \Pbar\right) + D(-t) d\x \cdot \Pbar \\
    &= dt \Hbar + \frac{1+t-e^t}{t^2} \x \cdot \Pbar dt + \frac{e^t-1}{t} d
    \x \cdot \Pbar.
  \end{split}
\end{equation}
It follows from the expression of $\theta$ that it is invertible for
all $(t,\x)$, since $\frac{e^t-1}{t} \neq 0$ for all $t \in \RR$.  Its
inverse, the vielbein $E$, has components
\begin{equation}
  E_{\Hbar} = \frac{\d}{\d t} + \left( \frac1t - \frac1{e^t-1}\right) x^a
  \frac{\d}{\d x^a} \quad\text{and}\quad
  E_{\overline{P}_a} = \frac{t}{e^t-1} \frac{\d}{\d x^a}.
\end{equation}
The invariant Carrollian structure is given by $\kappa = E_{\Hbar}$ and spatial
metric $b = \pi^2(\theta,\theta)$, given by
\begin{equation}
  b = \frac{(1+t-e^t)^2}{t^4} x^2 dt^2 + \frac{(e^t-1)^2}{t^2} d\x
  \cdot d\x + 2 \frac{(e^t-1)(1+t-e^t)}{t^3} \x \cdot d\x dt.
\end{equation}
This metric can be simplified using the following change of coordinates:
\begin{equation}
	t' = t \quad x'^i = \frac{e^t - 1}{t} x^i.
\end{equation}
In these coordinates, we find
\begin{equation}
	b = d\x'\cdot d\x' - 2 \x'\cdot d\x' dt' + \x'\cdot \x' dt'^2.
\end{equation}
One final change of coordinates,
\begin{equation}
	\hat{t} = t' \quad \text{and} \quad \hat{x}^i = e^{-t'} x'^i,
\end{equation}
brings the metric into the form
\begin{equation}
	b = e^{2\hat{t}} d\hat{\x} \cdot d\hat{\x}.
\end{equation}
With all of these changes to the coordinate system, the vielbein is also altered
such that
\begin{equation}
	\kappa = \frac{\d}{\d \hat{t}}.
\end{equation}

\section{Conclusion} \label{sec:ks_conc}
This chapter discussed how we might arrive at geometric properties for various spacetime models from the classification of their underlying kinematical Lie algebras.  In Section~\ref{sec:ks_klas}, we reviewed the classification of kinematical Lie algebras in spatial dimension $D = 3$ due to Figueroa-O'Farrill, before discussing the subsequent classification of (spatially-isotropic) simply-connected homogeneous spacetimes in Section~\ref{sec:ks_kss}.  We gave a more detailed review of the latter classification as direct generalisations of the methods utilised here will be at the heart of the classification of kinematical superspaces in Chapter~\ref{chap:k_superspaces}.  Finally, in Section~\ref{sec:ks_gps}, we derived numerous geometric properties for each spacetime model by employing our knowledge of the underlying Lie algebra and its use in constructing the spacetime geometry.  This procedure of algebraic classification, geometric classification, and geometric property derivation, gives us a rigorous framework in which to explore spacetime symmetries beyond the Lorentzian case.  Indeed, Chapters~\ref{chap:k_superspaces} and~\ref{chap:gb_superspaces} describe the progress towards substantiating this framework in the super-kinematical and super-Bargmann instances, respectively.  Since both of these instances are generalisations of the kinematical case, we will see many of the foundational results presented here, including those regarding geometric realisability and geometric limits, will prove invaluable in deriving similar results in these supersymmetric cases.  


\chapter{Kinematical Superspaces} \label{chap:k_superspaces}
In the previous chapter, we saw how we might arrive at a systematic study of kinematical spacetimes and their geometric properties, starting from the classification of the spacetimes' underlying Lie algebras.  This chapter will highlight the progress made towards this goal in the super-kinematical case.  In particular, we will give complete classifications of kinematical Lie superalgebras and their corresponding kinematical superspaces.
\\ \\
Recall, an $\N$-extended kinematical Lie superalgebra (KLSA) $\s$ in three spatial dimensions is a real Lie superalgebra $\s = \s_{\bar{0}} \oplus \s_{\bar{1}}$, such that $\s_{\bar{0}} = \k$ is a kinematical Lie algebra for which $D=3$, and $\s_{\bar{1}}$ consists of $\N$ copies of $S$, the real four-dimensional spinor module of the rotational subalgebra $\r \cong \so(3)$.  Here, we will focus solely on the $\N=1$ case. 
\\ \\
This chapter is organised as follows.  In Section~\ref{sec:ks_superspace_ksa}, we present the classification of $\N=1$ kinematical Lie superalgebras in $D=3$.  As part of this classification, we will demonstrate how to unpack our quaternionic formalism, using the $\N=1$ Poincaré superalgebra as our example.  Additionally, Section~\ref{sec:ks_superspace_ksa} will contain discussions on the $\N=1$ Aristotelian Lie superalgebras in $D=3$, as well as the central extensions and automorphisms of the classified kinematical and Aristotelian Lie superalgebras.  In Section~\ref{sec:ks_superspace_kss}, we use the Lie superalgebras' automorphisms to classify the possible super Lie pairs, and, thus, the possible kinematical superspaces.  Finally, in Section~\ref{sec:limits-betw-supersp}, we demonstrate how the kinematical superspaces are related to one another via geometric limits. 

\section{Classification of Kinematical Superalgebras} \label{sec:ks_superspace_ksa}
In this section, we begin by setting up our classification problem.  In particular, Section~\ref{subsec:kss_alg_setup} starts by defining which kinematical Lie algebras we shall be extending and discussing how we will derive the supersymmetric brackets. Next, it gives some preliminary results, which will be useful for limiting repetition in our calculations, and describes the Lie superalgebras' basis transformations, allowing us to identify isomorphic kinematical Lie superalgebras.  With the setup established, we classify the kinematical Lie superalgebras in Section~\ref{sec:class-kinem-lie}, summarising our findings and unpacking the quaternionic formalism in Section~\ref{sec:summary}.  Sections~\ref{sec:class-arist-lie} and~\ref{sec:central-extensions} classify the Aristotelian Lie superalgebras and the central extensions of the kinematical Lie superalgebras, respectively. Section~\ref{sec:autom-kinem-lie} then determines the automorphisms of the kinematical Lie superalgebras, which we will use to classify the kinematical superspaces later.  

\subsection{Setup for the Classification}
\label{subsec:kss_alg_setup}

For this chapter, we will combine the classifications in Tables~\ref{tab:klas_Dgeq3} and~\ref{tab:klas_D3}, so we deal with all $D=3$ kinematical Lie algebras at once.  These algebras are summarised in Table~\ref{tab:kla}.  We will now outline how we aim to build $\N=1$ supersymmetric extensions of these kinematical Lie algebras.
\begin{table}[h!]
  \centering
  \caption{Kinematical Lie Algebras in $D=3$}
  \label{tab:kla}
  \setlength{\extrarowheight}{2pt}
  \rowcolors{2}{blue!10}{white}
  \begin{tabular}{l|*{5}{>{$}l<{$}}|l}\toprule
    \multicolumn{1}{c|}{K\#} & \multicolumn{5}{c|}{Non-zero Lie brackets (besides $[\J,-]$ brackets)} & \multicolumn{1}{c}{Comment}\\\toprule
    \hypertarget{KLA1}{1} & & & & & & static \\
    \hypertarget{KLA2}{2} & [\bH ,\B] = -\P & & & & & Galilean \\
    \hypertarget{KLA3}{3$_{\gamma\in[-1,1]}$} & [\bH ,\B] = \gamma \B & [\bH ,\P] = \P & & & & \\
    \hypertarget{KLA4}{4$_{\chi\geq0}$} & [\bH ,\B] = \chi \B + \P & [\bH ,\P] = \chi \P - \B & & & & \\
    \hypertarget{KLA5}{5} & [\bH ,\B] = \B + \P & [\bH , \P] = \P & & & &  \\
    \hypertarget{KLA6}{6} & & & [\B,\P] = \bH  & & & Carroll \\
    \hypertarget{KLA7}{7} & [\bH ,\B] = \P & & [\B,\P] = \bH  & [\B,\B] = \J & & Euclidean \\
    \hypertarget{KLA8}{8}& [\bH ,\B] = - \P & & [\B,\P] = \bH  & [\B,\B] = - \J & & Poincaré \\
    \hypertarget{KLA9}{9}& [\bH ,\B] = \B & [\bH ,\P] = -\P &  [\B,\P] = \bH  - \J & & & $\so(4,1)$ \\
    \hypertarget{KLA10}{10}& [\bH ,\B] = \P & [\bH ,\P] = -\B & [\B,\P] = \bH  &  [\B,\B]= \J &  [\P,\P] = \J & $\so(5)$ \\
    \hypertarget{KLA11}{11}& [\bH ,\B] = -\P & [\bH ,\P] = \B & [\B,\P] = \bH  &  [\B,\B]= -\J &  [\P,\P] = -\J & $\so(3,2)$ \\
    \hypertarget{KLA12}{12}& & & & [\B,\B]= \B &  [\P,\P] = \B-\J & \\
    \hypertarget{KLA13}{13}& & & & [\B,\B]= \B & [\P,\P] = \J-\B & \\
    \hypertarget{KLA14}{14}& & & & [\B,\B] = \B & &  \\
    \hypertarget{KLA15}{15}& & & & [\B, \B] = \P & &  \\
    \hypertarget{KLA16}{16}& & [\bH ,\P] = \P & & [\B,\B] = \B & &  \\
    \hypertarget{KLA17}{17}& [\bH ,\B] = -\P & & & [\B,\B] = \P & &  \\
    \hypertarget{KLA18}{18}& [\bH ,\B] = \B & [\bH ,\P] = 2\P & & [\B,\B] = \P & & \\\bottomrule
  \end{tabular}
\end{table}
\\ \\
Let $\s$ be a kinematical Lie superalgebra where $\s_{\bar 0} = \k$ is
a kinematical Lie algebra from Table~\ref{tab:kla}.  To determine $\s$,
we need to specify the additional Lie brackets: $[\bH,\Q]$, $[\B,\Q]$,
$[\P,\Q]$ and $[\Q,\Q]$, subject to the super-Jacobi identity.  There are
four components to the super-Jacobi identity in a Lie superalgebra $\s =
\s_{\bar 0} \oplus \s_{\bar 1}$:
\begin{enumerate}
\item The $(\s_{\bar 0}, \s_{\bar 0},\s_{\bar 0})$ super-Jacobi identity simply
  says that $\s_{\bar 0}$ is a Lie algebra, which in our case is one
  of the kinematical Lie algebras $\k$ in Table~\ref{tab:kla}.
\item The $(\s_{\bar 0}, \s_{\bar 0},\s_{\bar 1})$ super-Jacobi identity says that
  $\s_{\bar 1}$ is a representation of $\s_{\bar 0}$ and, by
  restriction, also a representation of any Lie subalgebra of
  $\s_{\bar 0}$: for example, $\r$ in our case.
\item The $(\s_{\bar 0}, \s_{\bar 1},\s_{\bar 1})$ super-Jacobi identity says that
  the component of the Lie bracket $\bigodot^2 \s_{\bar 1} \to
  \s_{\bar 0}$ is $\s_{\bar 0}$-equivariant.  In particular, in our
  case, it is $\r$-equivariant.
\item The $(\s_{\bar 1}, \s_{\bar 1},\s_{\bar 1})$ component does not
  seem to have any representation-theoretic reformulation and needs to
  be checked explicitly.
\end{enumerate}
Our strategy will be the following.  We shall first determine the
space of $\r$-equivariant brackets $[\bH,\Q]$, $[\B,\Q]$, $[\P,\Q]$ and
$[\Q,\Q]$, which will turn out to be a $22$-dimensional real vector
space $\cV$. For each kinematical Lie algebra $\k = \s_{\bar 0}$ in
Table~\ref{tab:kla}, we then determine the algebraic variety
$\cJ \subset \cV$ cut out by the super-Jacobi identity.  We are eventually
interested in \emph{supersymmetry} algebras and hence we will restrict
attention to Lie superalgebras $\s$ for which $[\Q,\Q] \neq 0$, which
define a sub-variety $\cS \subset \cJ$.\footnote{Note, we restrict ourselves 
to the cases where $[\Q, \Q] \neq 0$ as our interests lie in spacetime
supersymmetry: we would like supersymmetry transformations to
generate geometric transformations of the spacetime.}  The isomorphism classes of
kinematical Lie superalgebras (with $[\Q,\Q]\neq 0$) are in one-to-one
correspondence with the orbits of $\cS$ under the subgroup
$\G \subset \GL(\s_{\bar 0}) \times \GL(\s_{\bar 1})$ which acts by
automorphisms of $\k = \s_{\bar 0}$, since we have fixed $\k$ from the
start.  The group $\G$ contains not just the automorphisms of the
kinematical Lie algebra $\k$ which act trivially on $\r$, but also
transformations which are induced by automorphisms of the quaternion
algebra.  We shall return to an explicit description of such
transformations below.\footnote{Notice, this strategy is a direct
generalisation of the one outlined in Section~\ref{subsec:ks_ala_class},
used to classify the Aristotelian Lie algebras. }
\\ \\
Let us start by determining the $\r$-equivariant brackets: $[\bH,\Q]$,
$[\B,\Q]$, $[\P,\Q]$ and $[\Q,\Q]$.  The bracket $[\bH,\Q]$ is an
$\r$-equivariant endomorphism of the spinor module $\Q$.  If we
identify $\r$ with the imaginary quaternions and $\Q$ with the
quaternions, the action of $\r$ on $\Q$ is via left quaternion
multiplication.  The endomorphisms of the representation $S$, which
commute with the action of $\r$, consist of left multiplication by
reals and right multiplication by quaternions, but for real numbers,
left and right multiplications agree, since the reals are central in
the quaternion algebra.  Hence the most general $\r$-equivariant
$[\bH,\Q]$ bracket takes the form\footnote{See Section~\ref{subsec:math_prelims_alg_klsa}
for a discussion on the quaternionic formalism employed here.  In this chapter, we will use $s$ as opposed to $\theta$ to denote the quaternion parameterising our supercharges $\sQ$.  There is nothing sinister in this change of notation; only $s$ was required for different purposes in Chapter~\ref{chap:gb_superspaces}, and I latterly preferred using $\theta$ as opposed to $s$ for parameterising the supercharges, hence its use in setting up the quaternionic formalism in Sections~\ref{subsec:math_prelims_alg_klsa} and~\ref{subsec:math_prelims_alg_gbsa}.}
\begin{equation}\label{eq:hq-bracket}
  [\sH,\sQ(s)] = \sQ(s \hh) \qquad\text{for some}\qquad \hh = h_1 \ii + h_2
  \jj + h_3 \kk + h_4 \in \HH.
\end{equation}
The brackets $[\B,\Q]$ and $[\P,\Q]$ are $\r$-equivariant
homomorphisms $V \otimes S \to S$, where $V$ and $S$ are the vector
and spinor modules of $\so(3)$.  There is an $\r$-equivariant map
$V \otimes S \to S$ given by the ``Clifford action'', which in this
language is left multiplication by $\Im \HH$ on $\HH$.  Its kernel is
the 8-dimensional real representation $W$ of $\r$ with spin
$\frac32$.
Therefore, the space of
$\r$-equivariant homomorphisms $V \otimes S \to S$ is isomorphic to
the space of $\r$-equivariant endomorphisms of $S$, which, as we saw
before, is a copy of the quaternions.  In summary, the $[\B,\Q]$ and
$[\P,\Q]$ brackets take the form
\begin{equation}\label{eq:bq-pq-brackets}
  \begin{split}
    [\sB(\beta), \sQ(s)] &= \sQ(\beta s \bb) \qquad\text{for some}\qquad \bb = b_1 \ii + b_2
    \jj + b_3 \kk + b_4 \in \HH\\
    [\sP(\pi), \sQ(s)] &= \sQ(\pi s \pp) \qquad\text{for some}\qquad \pp = p_1 \ii + p_2
    \jj + p_3 \kk + p_4 \in \HH,
  \end{split}
\end{equation}
for all $\beta,\pi \in \Im\HH$ and $s \in \HH$.
\\ \\
Finally, we look at the $[\Q,\Q]$ bracket, which is an
$\r$-equivariant linear map $\bigodot^2 S \to \k = \RR \oplus 3 V$.
The symmetric square $\bigodot^2S$ is a 10-dimensional $\r$-module
which decomposes as $\RR \oplus 3 V$.  Indeed, on $S$, we have
an $\r$-invariant inner product given by
\begin{equation}
  \left<s_1,s_2\right> = \Re (\sbar_1 s_2)\qquad\text{where}\qquad
  s_1,s_2 \in \HH.
\end{equation}
It is clearly invariant under left multiplication by unit quaternions:
$\left<\uu s_1, \uu s_2\right> = \left<s_1,s_2\right>$ for all $\uu \in
\Sp(1)$.  We can use this inner product to identify $\bigodot^2 S$ with the
symmetric endomorphisms of $S$: linear maps $\lambda : S \to S$ such
that $\left<\lambda(s_1), s_2\right> = \left<s_1, \lambda
  (s_2)\right>$.  Letting $L_\qq$ and $R_\qq$ denote left and
right quaternion multiplication by $\qq \in \HH$, the space of symmetric
endomorphisms of $S$ is spanned by the identity endomorphism and
$L_\ii R_\ii$, $L_\ii R_\jj$, $L_\ii R_\kk$, $L_\jj R_\ii$, $L_\jj
R_\jj$, $L_\jj R_\kk$, $L_\kk R_\ii$, $L_\kk R_\jj$ and $L_\kk
R_\kk$.  The nine non-identity symmetric endomorphisms transform
under $\r$ according to three copies of $V$.  Since $\r$ acts on $S$
via left multiplication, it commutes with the $R_\qq$ and hence the
three copies of $V$ are
\begin{equation} \label{eq:vector_mods}
  \spn{L_\ii R_\ii, L_\jj R_\ii, L_\kk R_\ii} \oplus \spn{L_\ii R_\jj,
    L_\jj R_\jj, L_\kk R_\jj} \oplus \spn{L_\ii R_\kk, L_\jj R_\kk,
    L_\kk R_\kk}.
\end{equation}
The space of $\r$-equivariant linear maps $\bigodot^2 S \to 3 V
\oplus \RR$ is thus isomorphic to the space of $\r$-equivariant
endomorphisms of $\RR \oplus 3 V = \RR \oplus (\RR^3 \otimes V)$,
which is given by
\begin{equation}
  \End_\r\left(\RR \oplus (\RR^3 \otimes V) \right) \cong \End(\RR) \oplus
  \left(\End(\RR^3) \otimes \id_V\right).
\end{equation}
The second component of this isomorphism simply states that the $\r$-equivariant endomorphisms 
do not act on the $\so(3)$ vector indices of the vector modules and rotate 
the three vector modules into one another. In particular, since $\r$ acts via 
left quaternion multiplication, the $\r$-equivariant maps act via right 
quaternion multiplication. 
In summary, the $\r$-equivariant $[\Q,\Q]$ bracket is given by
polarisation from the following
\begin{equation}
  [\sQ(s), \sQ(s)] = c_0 |s|^2 \sH + \Re(\sbar \JJ s \bc_1)
  + \Re(\sbar \BB s  \bc_2) + \Re(\sbar \PP s
  \bc_3),
\end{equation}
where $c_0 \in \RR$,
$\bc_1, \bc_2, \bc_3 \in \Im\HH$ and
where we have introduced the shorthands
\begin{equation}
  \JJ = \bJ_1 \ii + \bJ_2 \jj + \bJ_3 \kk, \quad   \BB = \bB_1 \ii + \bB_2 \jj +
  \bB_3 \kk, \quad\text{and}\quad \PP = \bP_1 \ii + \bP_2 \jj + \bP_3 \kk.
\end{equation}
Notice that if $\omega \in \Im\HH$, then $\sJ(\omega) = \Re(\bar\omega
\JJ)$, and similarly $\sB(\beta) = \Re(\bar\beta \BB)$ and $\sP(\pi) =
\Re(\bar\pi \PP)$, for $\beta,\pi \in \Im \HH$, so that we can rewrite
the $[\Q,\Q]$ bracket as
\begin{equation}\label{eq:QQdiagonal}
  [\sQ(s), \sQ(s)] =c_0 |s|^2 \sH - \sJ(s \bc_1 \sbar) - \sB(s \bc_2
  \sbar) - \sP(s \bc_3 \sbar),
\end{equation}
which polarises to give
\begin{equation}\label{eq:QQ}
  [\sQ(s), \sQ(s')] = c_0 \Re(\sbar s') \sH - \tfrac12 \sJ(s'
  \bc_1 \sbar + s \bc_1 \sbar') - \tfrac12
  \sB(s' \bc_2 \sbar + s \bc_2 \sbar') -
  \tfrac12 \sP(s' \bc_3 \sbar + s \bc_3 \sbar').
\end{equation}
In summary, we have that the $\r$-equivariant brackets by which we
extend the kinematical Lie algebra $\k$ live in a 22-dimensional real
vector space of parameters $\hh,\bb,\pp \in \HH$, $\bc_1,
\bc_2, \bc_3 \in \Im \HH$ and $c_0 \in \RR$.

\subsubsection{Preliminary Results}
\label{sec:some-preliminary-results}

In this brief section, we will go through each of the super-Jacobi identity components, determining any possible universal conditions that may aid our classification.  Note, since we are using a kinematical Lie algebra $\k$ as $\s_{\bar{0}}$, we do not need to consider the $(\s_{\bar{0}}, \s_{\bar{0}}, \s_{\bar{0}})$ component as this will automatically be satisfied.   

\paragraph{$(\s_{\bar{0}}, \s_{\bar{0}}, \s_{\bar{1}})$}  ~\\ \\
As mentioned above, this component of the super-Jacobi identity says that $\s_{\bar 1}$ is an $\s_{\bar 0}$-module,
where $\s_{\bar 0}=\k$ is the underlying kinematical Lie algebra.  The
super-Jacobi identity
\begin{equation}
  [X,[Y,\sQ(s)]] - [Y,[X,\sQ(s)]] = [[X,Y],\sQ(s)] \qquad\text{for all
    $X,Y\in\k$}
\end{equation}
gives relations between the parameters $\hh,\bb,\pp \in \HH$ appearing in
the Lie brackets.

\begin{lemma}\label{lem:kmod}
  The following relations between $\hh,\bb,\pp \in \HH$ are implied by the
  corresponding $\k$-brackets:
  \begin{equation}
    \begin{split}
      [\bH,\B] = \lambda \B + \mu \P & \implies [\bb,\hh] = \lambda \bb + \mu \pp\\
      [\bH,\P] = \lambda \B + \mu \P & \implies [\pp,\hh] = \lambda \bb + \mu \pp\\
      [\B,\B] = \lambda \B + \mu \P + \nu \J & \implies \bb^2 = \tfrac12 \lambda \bb + \tfrac12 \mu \pp + \tfrac14 \nu\\
      [\P,\P] = \lambda \B + \mu \P + \nu \J & \implies \pp^2 = \tfrac12 \lambda \bb + \tfrac12 \mu \pp + \tfrac14 \nu\\
      [\B,\P] = \lambda \bH & \implies \bb \pp + \pp \bb = 0\quad\text{and}\quad [\bb,\pp] = \lambda \hh.
    \end{split}
  \end{equation}
\end{lemma}

\begin{proof}
  The $[\bH,\B,\Q]$ super-Jacobi identity says for all $\beta \in \Im\HH$ and
  $s \in\HH$,
  \begin{equation}
    [[\sH,\sB(\beta)],\sQ(s)] = [\sH, [\sB(\beta),\sQ(s)]] - [\sB(\beta),
    [\sH, \sQ(s)]],
  \end{equation}
  which becomes
  \begin{equation}
    \lambda \sQ(\beta s \bb) + \mu \sQ(\beta s \pp) = \sQ(\beta s \bb \hh) -
    \sQ(\beta s \hh \bb).
  \end{equation}
  Since $\sQ$ is real linear and injective, it follows that
  \begin{equation}
    \lambda \beta s \bb + \mu \beta s \pp = \beta s [\bb,\hh],
  \end{equation}
  which, since it must hold for all $\beta \in \Im\HH$ and $s \in
  \HH$, becomes
  \begin{equation}
    [\bb,\hh] = \lambda \bb + \mu \pp,
  \end{equation}
  as desired.  Similarly, the $[\bH,\P,\Q]$ super-Jacobi identity gives the
  second equation in the lemma.  The third equation follows from the
  $[\B,\B,\Q]$ super-Jacobi identity, which says that for all
  $\beta,\beta'\in\Im\HH$ and $s\in\HH$,
  \begin{equation}
    [[\sB(\beta),\sB(\beta')],\sQ(s)] =
    [\sB(\beta),[\sB(\beta'),\sQ(s)]] -  [\sB(\beta'),[\sB(\beta),\sQ(s)]],
  \end{equation}
  which becomes
  \begin{equation}
    \tfrac12 \lambda \sQ([\beta,\beta'] s \bb) + \tfrac12 \mu
    \sQ([\beta,\beta'] s \pp) + \tfrac14 \nu \sQ([\beta,\beta']s) =
    \sQ(\beta\beta's \bb^2) - \sQ(\beta'\beta s \bb^2).
  \end{equation}
  Again by linearity and injectivity of $\sQ$, this is equivalent to
  \begin{equation}
    \tfrac12 \lambda [\beta,\beta'] s \bb + \tfrac12 \mu
    [\beta,\beta'] s \pp + \tfrac14 \nu [\beta,\beta']s =
    [\beta,\beta']s \bb^2,
  \end{equation}
  which, being true for all $\beta,\beta'\in\Im\HH$ and $s \in \HH$, gives
  \begin{equation}
    \tfrac12 \lambda \bb + \tfrac12 \mu \pp + \tfrac14 \nu = \bb^2,
  \end{equation}
  as desired.  The fourth identity in the lemma follows similarly from
  the $[\P,\P,\Q]$ super-Jacobi identity.  Finally, we consider the
  $[\B,\P,\Q]$ super-Jacobi identity, which says that for all
  $\beta,\pi\in\Im\HH$ and $s \in\HH$,
  \begin{equation}
    [[\sB(\beta),\sP(\pi)],\sQ(s)] = [\sB(\beta),[\sP(\pi),\sQ(s)]] -
    [\sP(\pi),[\sB(\beta),\sQ(s)]],
  \end{equation}
  which expands to
  \begin{equation}
    -\lambda \Re(\beta\pi) \sQ(s\hh) = \sQ(\beta\pi s \pp \bb) - \sQ(\pi\beta s \bb \pp)
  \end{equation}
  or, equivalently,
  \begin{equation}\label{eq:BPQ-aux}
    -\lambda \Re(\beta\pi) s \hh = \beta\pi s \pp \bb - \pi \beta s \bb \pp,
  \end{equation}
  for all $\beta,\pi \in \Im\HH$ and $s \in \HH$.  For any two
  imaginary quaternions $\beta,\pi$, we have that
  \begin{equation}
    \beta\pi = \tfrac12 [\beta,\pi] + \Re(\beta\pi),
  \end{equation}
  which allows us to rewrite equation~\eqref{eq:BPQ-aux} as
  \begin{equation}
    \Re(\beta\pi) s (\lambda \hh - \bb \pp + \pp \bb) + \tfrac12 [\beta,\pi] s (
    \pp \bb + \bb \pp) = 0.
  \end{equation}
  Taking $\beta = \pi$ and $s=1$ we see that $\lambda \hh = [\bb,\pp]$ and taking
  $\beta$ and $\pi$ to be orthogonal and $s=1$, that $\pp \bb + \bb \pp = 0$, as
  desired.
\end{proof}

\paragraph{$(\s_{\bar{0}}, \s_{\bar{1}}, \s_{\bar{1}})$} ~\\ \\
Due to the large number of parameters in the $[\s_{\bar{0}}, \s_{\bar{1}}]$
and $[\s_{\bar{1}}, \s_{\bar{1}}]$ brackets, the components 
$[\bH,\Q,\Q]$, $[\B,\Q,\Q]$ and $[\P,\Q,\Q]$ of the
super-Jacobi identity are best studied on a case-by-case basis.

\paragraph{$(\s_{\bar{1}}, \s_{\bar{1}}, \s_{\bar{1}})$} ~\\ \\
The last component of the super-Jacobi identity to consider is the $(\s_{\bar{1}}, \s_{\bar{1}}, \s_{\bar{1}})$
case $[\Q,\Q,\Q]$, which gives the following universal condition.

\begin{lemma}\label{lem:qqq}
  The $[\Q,\Q,\Q]$ component of the super-Jacobi identity implies
  \begin{equation}
    c_0 \hh = \tfrac12 \bc_1  + \bc_2 \bb + \bc_3 \pp.
  \end{equation}
\end{lemma}

\begin{proof}
  The $[\Q,\Q,\Q]$ component of the super-Jacobi identity is totally
  symmetric and hence, by polarisation, it is uniquely determined by
  its value on the diagonal.  In other words, it is equivalent to
  \begin{equation}
    [[\sQ(s),\sQ(s)], \sQ(s)] \stackrel{!}{=} 0 \qquad\text{for all $s \in \HH$.}
  \end{equation}
  Using equation~\eqref{eq:QQdiagonal}, this becomes
  \begin{equation}
    [c_0 |s|^2 \sH - \sJ(s \bc_1 \sbar) - \sB(s
    \bc_2 \sbar) - \sP(s \bc_3 \sbar), \sQ(s)]
    \stackrel{!}{=} 0,
  \end{equation}
  which expands to
  \begin{equation}
    c_0 |s|^2 \sQ(s\hh) - \tfrac12 \sQ(s \bc_1 \sbar s) - \sQ(s
    \bc_2 \sbar s \bb) - \sQ(s
    \bc_3 \sbar s \pp) \stackrel{!}{=} 0.
  \end{equation}
  Since $\sQ$ is injective, this becomes
  \begin{equation*}
    |s|^2 s (c_0 \hh - \tfrac12 \bc_1 -  \bc_2 \bb -
    \bc_3 \pp)  \stackrel{!}{=} 0.
  \end{equation*}
  This must hold for all $s \in \HH$, so in particular for $s=1$,
  proving the lemma.
\end{proof}

\subsubsection{Basis Transformations}
\label{sec:automorphisms}

As mentioned above, once we determine the sub-variety $\cS$ cut out by
the super-Jacobi identity, we need to quotient by the action of the subgroup
$\G \subset \GL(\s_{\bar 0}) \times \GL(\s_{\bar 1})$, which acts by
automorphisms of $\s_{\bar 0} = \k$ in order to arrive at the
isomorphism classes of Lie superalgebras.  In this section, we describe
the subgroup $\G$ in more detail.  There are two kinds of elements of
$\G$, those which act trivially on the rotational subalgebra $\r$ and
those which do not.  The latter consist of inner automorphisms of
$\k$, which are generated infinitesimally by the adjoint action of
$\J$, $\B$ and $\P$.  The ones generated by $\J$ are particularly easy
to describe in the quaternionic formulation, and we shall do so now in
more detail.
\\ \\
Let $\uu \in \Sp(1)$ be a unit norm quaternion.  Conjugation by $\uu$
defines a homomorphism $\Ad : \Sp(1) \to \Aut(\HH)$ whose kernel is
the central subgroup of $\Sp(1)$ consisting of $\pm 1$.  It is a
classical result that these are all the automorphisms of $\HH$.  Hence
$\Aut(\HH) \cong \SO(3)$, acting trivially on the real quaternions and
rotating the imaginary quaternions.  The action of $\Aut(\HH)$ on $\s$
leaves $\bH$ invariant and acts on the remaining generators by
pre-composing the linear maps $\sJ$, $\sB$, $\sP$ and $\sQ$ with
$\Ad_{\uu}$.  More precisely, let $\widetilde \sH = \sH$,
$\widetilde\sJ = \sJ \circ \Ad_{\uu}$, $\widetilde\sB = \sB \circ \Ad_{\uu}$,
$\widetilde\sP = \sP \circ \Ad_{\uu}$ and $\widetilde\sQ = \sQ \circ \Ad_{\uu}$.
Since the Lie brackets of $\k$ are given in terms of quaternion
multiplication, this transformation is an automorphism of $\k$, and we
have a group homomorphism $\Aut(\HH) \to \Aut(\k)$.  The action on the
remaining brackets (those involving $\Q$) is as follows.
The Lie brackets of $\s$ which involve $\Q$ are given by
\begin{equation}
  \begin{split}
    [\sH, \sQ(s)] &= \sQ(s\hh)\\
    [\sJ(\omega), \sQ(s)] &= \tfrac12 \sQ(\omega s)\\
    [\sB(\beta), \sQ(s)] &= \sQ(\beta s \bb)\\
    [\sP(\pi), \sQ(s)] &= \sQ(\pi s \pp)\\
    [\sQ(s), \sQ(s)] &= c_0 |s|^2 \sH - \sJ(s \bc_1 \sbar) -
    \sB(s \bc_2 \sbar) - \sP(s \bc_3 \sbar),
  \end{split}
\end{equation}
and hence under conjugation by $\uu \in \Sp(1)$,
\begin{equation}
  \begin{split}
    [\widetilde \sH, \widetilde \sQ(s)] &= \widetilde\sQ(s\widetilde \hh)\\
    [\widetilde \sJ(\omega), \widetilde\sQ(s)] &= \tfrac12 \widetilde\sQ(\omega s)\\
    [\widetilde \sB(\beta), \widetilde\sQ(s)] &= \widetilde\sQ(\beta s \widetilde \bb)\\
    [\widetilde\sP(\pi), \widetilde\sQ(s)] &= \widetilde\sQ(\pi s \widetilde \pp)\\
    [\widetilde\sQ(s), \widetilde\sQ(s)] &= c_0 |s|^2 \widetilde \sH - \widetilde\sJ(s \widetilde{\bc}_1 \sbar) -
    \widetilde\sB(s \widetilde{\bc}_2 \sbar) - \widetilde\sP(s \widetilde{\bc}_3 \sbar),
  \end{split}
\end{equation}
where $\widetilde \hh = \bar \uu \hh \uu$, $\widetilde \bb = \bar \uu \bb
\uu$, $\widetilde \pp = \bar \uu \pp \uu$, and $\widetilde{\bc}_i = \bar \uu
\bc_i \uu$ for $i=1,2,3$.  In other words, the scalar parameters $c_0$,
$\Re \hh$, $\Re \bb$ and $\Re \pp$ remain inert, but the imaginary
quaternion parameters $\Im \hh, \Im \bb, \Im \pp, \bc_{1,2,3}$ are
simultaneously rotated.  
\\ \\
There are other automorphisms of $\k$ which do transform $\r$: those
are the inner automorphisms generated by $\B$ and $\P$. Their
description depends on the precise form of $\k$ but they will not play
a rôle in our discussion.
\\ \\
In addition to these, $\G$ also consists of automorphisms of $\k$
which leave $\r$ intact.  If a linear map $\Phi: \s \to \s$ restricts to an
automorphism of $\k$, then it is in particular $\r$-equivariant.  The
most general $\r$-equivariant linear map $\Phi : \s \to \s$ sends
$(\J,\bH,\B,\P,\Q) \mapsto (\J, \widetilde \bH, \widetilde \B, \widetilde
\P, \widetilde \Q)$, where
\begin{equation}
  \begin{split}
    \widetilde \sH &= \mu \sH\\
    \widetilde \sB(\beta) &= a\sB(\beta) + c\sP(\beta) + e \sJ(\beta)\\
    \widetilde \sB(\beta) &= b\sB(\beta) + d\sP(\beta) + f \sJ(\beta)\\
    \widetilde \sQ(s) &= \sQ(s\qq)
  \end{split}
\end{equation}
where $\mu \in \GL(1,\RR) = \RR^\times$, $\qq \in \GL(1,\HH) =
\HH^\times$ and $\begin{pmatrix} \zero & a & b \\ \zero & c & d \\ 1 &
  e & f \end{pmatrix} \in \GL(3,\RR)$.  The automorphisms (which fix $\r$) of the kinematical Lie algebras  \hyperlink{KLA1}{$\mathsf{K1}$}-\hyperlink{KLA11}{$\mathsf{K11}$} 
in Table~\ref{tab:kla} were derived in \cite[§§3.1]{Figueroa-OFarrill:2018ilb}.  The
automorphisms of the remaining kinematical Lie algebras in the table
are listed below (see Table~\ref{tab:aut-kla}).  In particular, we
find that, although the precise form of the automorphisms depends on $\k$, a
common feature is that the coefficients $e,f$ are always zero, so we
will set them to zero from now on without loss of generality.
\\ \\
Assuming that the pair $(A=\begin{pmatrix}a & b \\ c & d
\end{pmatrix}, \mu) \in \GL(2,\RR) \times \RR^\times$ is an
automorphism of $\k = \s_{\bar 0}$, the brackets involving $\Q$ change
as follows:
\begin{equation}
  \begin{split}
    [\widetilde \sH, \widetilde\sQ(s)] &= \widetilde\sQ(s\widetilde\hh)\\
    [\widetilde \sB(\beta), \widetilde\sQ(s)] &= \widetilde\sQ(\beta s \widetilde\bb)\\
    [\widetilde \sP(\pi), \widetilde\sQ(s)] &= \widetilde\sQ(\pi s \widetilde\pp)\\
    [\widetilde\sQ(s), \widetilde\sQ(s)] &= \widetilde c_0 |s|^2
    \widetilde \sH - \widetilde\sJ(s \widetilde\bc_1 \sbar) - \widetilde
    \sB(s \widetilde \bc_2 \sbar) - \widetilde \sP(s \widetilde
    \bc_3\sbar),
  \end{split}
\end{equation}
where $\widetilde \sJ(\omega) = \sJ(\omega)$ and
\begin{equation}\label{eq:autk-on-params}
  \begin{aligned}[m]
    \widetilde\hh &= \mu \qq \hh \qq^{-1}\\
    \widetilde\bb &= \qq (a \bb + c \pp) \qq^{-1}\\
    \widetilde\pp &=  \qq (b \bb + d \pp) \qq^{-1}\\
    \widetilde c_0 &= c_0 \frac{|\qq|^2}{\mu}
  \end{aligned}
  \qquad\qquad
  \begin{aligned}[m]
    \widetilde \bc_1 &= \qq \bc_1 \qqbar\\
    \widetilde \bc_2 &= \frac{1}{ad - bc} \qq (d \bc_2 - b \bc_3) \qqbar\\
    \widetilde \bc_3 &= \frac{1}{ad - bc} \qq (a \bc_3 - c \bc_2) \qqbar.\\
  \end{aligned}
\end{equation}
In summary, the group $\G$ by which we must quotient the sub-variety
$\cS$, cut out by the super-Jacobi identity (and $[\Q,\Q]\neq 0$), acts as
follows on the generators:
\begin{equation}
  \begin{split}
    \sJ &\mapsto \sJ \circ \Ad_{\uu}\\
    \sB &\mapsto a \sB \circ \Ad_{\uu} + c \sP \circ \Ad_{\uu}\\
    \sP &\mapsto b \sB \circ \Ad_{\uu} + d \sP \circ \Ad_{\uu}\\
    \sH &\mapsto \mu \sH\\
    \sQ &\mapsto \sQ \circ \Ad_{\uu} \circ R_\qq
  \end{split}
\end{equation}
where $\mu\in \RR$ and $\qq \in \HH$ are non-zero, $\uu \in \Sp(1)$  and
$A:=\begin{pmatrix}a & b \\ c & d \end{pmatrix} \in \GL(2,\RR)$ with
$(A,\mu)$ an automorphism of $\k$.
\\ \\
Let $\Aut_\r(\k)$ denote the subgroup of $\GL(2,\RR) \times \RR^\times$
consisting of such $(A,\mu)$.  These subgroups are listed in
\cite[§3.1]{Figueroa-OFarrill:2018ilb} for the kinematical Lie
algebras \hyperlink{KLA1}{$\mathsf{K1}$}-\hyperlink{KLA11}{$\mathsf{K11}$} 
in Table~\ref{tab:kla}.  We will collect them in
Table~\ref{tab:aut-kla} for convenience and in addition also record
them for the remaining kinematical Lie algebras 
\hyperlink{KLA12}{$\mathsf{K12}$}-\hyperlink{KLA18}{$\mathsf{K18}$} in
Table~\ref{tab:kla}.

\begin{table}[h!]
  \centering
  \caption{Automorphisms of Kinematical Lie Algebras (Acting Trivially
  on $\r$)}
  \label{tab:aut-kla}
  \begin{tabular}{l|>{$}l<{$}}\toprule
    \multicolumn{1}{c|}{K\#} & \multicolumn{1}{c}{Typical $(A,\mu) \in \GL(2,\RR) \times \RR^\times$}\\
    \toprule
    \hyperlink{KLA1}{1} & \left(\begin{pmatrix} a & b \\ c & d \end{pmatrix}, \mu\right) \\[10pt]
    \hyperlink{KLA2}{2} & \left(\begin{pmatrix} a & \zero \\ c & d \end{pmatrix}, \frac{d}{a}\right) \\[10pt]
    \hyperlink{KLA3}{3$_{\gamma\in(-1,1)}$} & \left(\begin{pmatrix} a & \zero \\ \zero &
        d \end{pmatrix}, 1\right) \\[10pt]
    \hyperlink{KLA3}{3$_{-1}$} & \left(\begin{pmatrix} a & \zero \\ \zero & d \end{pmatrix}, 1\right), \left(\begin{pmatrix} \zero & b \\ c & \zero \end{pmatrix}, -1\right) \\[10pt]
    \hyperlink{KLA3}{3$_{1}$} & \left(\begin{pmatrix} a & b \\ c & d \end{pmatrix}, 1\right) \\[10pt]
    \hyperlink{KLA4}{4$_{\chi>0}$} & \left(\begin{pmatrix} a & b \\ -b & a \end{pmatrix}, 1\right) \\[10pt]
    \hyperlink{KLA4}{4$_0$} & \left(\begin{pmatrix} a & b \\ -b & a \end{pmatrix}, 1\right), \left(\begin{pmatrix} a & b \\ b & -a \end{pmatrix}, -1\right) \\[10pt]
    \hyperlink{KLA5}{5} & \left(\begin{pmatrix} a & \zero \\ c & a \end{pmatrix}, 1\right)  \\[10pt]
    \hyperlink{KLA6}{6} & \left(\begin{pmatrix} a & b \\ c & d \end{pmatrix}, ad-bc\right) \\[10pt]
    \hyperlink{KLA7}{7},\hyperlink{KLA8}{8} & \left(\begin{pmatrix} 1 & \zero \\ c & d \end{pmatrix}, d \right), \left(\begin{pmatrix} -1 & \zero \\ c & d \end{pmatrix}, -d\right)  \\[10pt]
    \hyperlink{KLA9}{9} & \left(\begin{pmatrix} a & \zero \\ \zero & a^{-1} \end{pmatrix}, 1\right),  \left(\begin{pmatrix} \zero & b \\ b^{-1} & \zero \end{pmatrix}, -1\right) \\[10pt]
    \hyperlink{KLA10}{10},\hyperlink{KLA11}{11} & \left(\begin{pmatrix} a & b \\ -b & a \end{pmatrix}, 1\right), \left(\begin{pmatrix} a & b \\ b & -a \end{pmatrix}, -1\right),\quad a^2 + b^2 = 1\\[10pt]
    \hyperlink{KLA12}{12},\hyperlink{KLA13}{13} & \left(\begin{pmatrix} 1 & \zero \\ \zero & 1 \end{pmatrix}, \mu\right) , \left(\begin{pmatrix} 1 & \zero \\ \zero & -1 \end{pmatrix}, \mu\right) \\[10pt]
    \hyperlink{KLA14}{14} & \left(\begin{pmatrix} 1 & \zero \\ \zero & d \end{pmatrix}, \mu\right) \\[10pt]
    \hyperlink{KLA15}{15} & \left(\begin{pmatrix} a & \zero \\ c & a^2 \end{pmatrix}, \mu\right) \\[10pt]
    \hyperlink{KLA16}{16} & \left(\begin{pmatrix} 1 & \zero \\ \zero & d \end{pmatrix}, 1\right) \\[10pt]
    \hyperlink{KLA17}{17} & \left(\begin{pmatrix} a & \zero \\ c & a^2 \end{pmatrix}, a\right) \\[10pt]
    \hyperlink{KLA18}{18} & \left(\begin{pmatrix} a & \zero \\ \zero & a^2 \end{pmatrix}, 1\right) \\[10pt]
    \bottomrule
    \end{tabular}
\end{table}

\subsection{Classification of Kinematical Lie Superalgebras}
\label{sec:class-kinem-lie}

We now proceed to analyse each kinematical Lie algebra
$\k$ in Table~\ref{tab:kla} in turn and impose the super-Jacobi identity for
the corresponding Lie superalgebras extending $\k$.  We recall that we
are only interested in those Lie superalgebras where $[\Q,\Q] \neq 0$,
so $c_0,  \bc_1, \bc_2, \bc_3$ cannot all simultaneously vanish.

\subsubsection{Kinematical Lie Algebras Without Supersymmetric Extensions}
\label{sec:kinem-lie-algebr-1}

There are three kinematical Lie algebras which cannot be extended to a
kinematical superalgebra: \hyperlink{KLA9}{$\so(4,1)$},
\hyperlink{KLA10}{$\so(5)$} and the Euclidean algebra
(\hyperlink{KLA7}{$\mathsf{K7}$} in Table~\ref{tab:kla}).

\subsubsection*{The Euclidean Algebra}
\label{sec:euclidean-algebra}

From Lemma~\ref{lem:kmod}, we find that $\pp = \hh = 0$ and $\bb^2 =
\frac14$, so, in particular, $\bb \in \RR$, and, from Lemma~\ref{lem:qqq},
we find that $\bc_2 \bb + \frac12 \bc_1 = 0$.  The
$[\bH,\Q,\Q]$ component of the super-Jacobi identity shows that
$\bc_2 = 0$, so that also $\bc_1 = 0$.  The
$[\P,\Q,\Q]$ component of the super-Jacobi identity is trivially satisfied,
whereas the $[\B,\Q,\Q]$ component shows that $\bc_3 = 0$,
and also that $c_0 = 0$.  In summary, there is no kinematical
superalgebra extending the Euclidean algebra for which $[\Q,\Q] \neq
0$; although there is a kinematical superalgebra where $[\sB(\beta),
\sQ(s)] = \pm \frac12 \sQ(\beta s)$, where both choices of sign
are related by an automorphism of $\k$: e.g., time reversal
$(\J,\B,\P,\bH) \mapsto (\J, -\B, \P, -\bH)$ or parity $(\J,\B,\P,\bH)
\mapsto (\J, -\B, -\P, \bH)$.

\subsubsection*{$\so(4,1)$}
\label{sec:so4-1}

In this case, Lemma~\ref{lem:kmod} gives that $\pp = \bb = 0$, but then
the $[\B,\P,\Q]$ component of the super-Jacobi identity cannot be satisfied,
showing that the $\so(3)$ representation on the spinor module $S$
cannot be extended to a module of $\so(4,1)$.  The result
would be different for $\N=2$ extensions, since $\so(4,1) \cong
\sp(1,1)$ does have an irreducible spinor module of
quaternionic dimension $2$.

\subsubsection*{$\so(5)$}
\label{sec:so5}

From Lemma~\ref{lem:kmod}, we find that $\pp = [\bb,\hh]$  from $[\bH,\B]=\P$,
and, in particular, $\pp \in \Im \HH$.  But then $[\P,\P] = \J$ says
that $\pp^2 = \frac14$, so that in particular $\pp \in \RR$ and non-zero,
which is a contradiction.  Again, this shows that the spinor
module $S$ of $\so(3)$ does not extend to a module of
$\so(5)$, and, again, the conclusion would be different for $\N=2$
extensions, since $\so(5) \cong \sp(2)$ does admit a quaternionic
module of quaternionic dimension $2$.

\subsubsection{Lorentzian Kinematical Superalgebras}
\label{sec:lorentz-kinem-super}

The Poincaré Lie algebra (\hyperlink{KLA8}{$\mathsf{K8}$}) and
\hyperlink{KLA11}{$\so(3,2)$} are the Lorentzian isometry Lie algebras of
the Minkowski and anti-de~Sitter spacetimes, respectively.  It is of
course well known that such spacetimes admit $\N=1$ superalgebras of
maximal dimension.  We treat them in this section for completeness.

\subsubsection*{The Poincaré Superalgebra}
\label{sec:poinc-super}

From Lemma~\ref{lem:kmod}, we see that $\pp = \hh = 0$ and that $\bb^2 =
-\tfrac14$, so that in particular $\bb \in \Im \HH$.  From
Lemma~\ref{lem:qqq}, we see that $\tfrac12 \bc_1 + \bc_2 \bb = 0$.  The
$[\P,\Q,\Q]$ component of the super-Jacobi identity is trivially satisfied,
whereas the $[\bH,\Q,\Q]$ component forces $\bc_1 = \bc_2 = 0$ and the
$[\B,\Q,\Q]$ component says $\bc_3 = 2 c_0 \bb$.  Demanding $[\Q,\Q]
\neq 0$ requires $c_0 \neq 0$.
\\ \\
Using the quaternion automorphism, we can rotate $\bb$ so that $\bb =
\frac12 \kk$ and via the automorphism of the Poincaré Lie algebra,
which rescales $\bH$ and $\P$ by the same amount, we can bring $c_0 =
1$.  In summary, we have a unique isomorphism class of kinematical Lie
superalgebras extending the Poincaré Lie algebra which has the
additional Lie brackets
\begin{equation} \label{eq:poincare}
  [\sB(\beta), \sQ(s)] = \tfrac12 \sQ(\beta s \kk)
  \qquad\text{and}\qquad [\sQ(s), \sQ(s)] = |s|^2 \sH - \sP(s\kk \sbar).
\end{equation}

\subsubsection*{The AdS Superalgebra}
\label{sec:ads-superalgebra}

Here, Lemma~\ref{lem:kmod} and Lemma~\ref{lem:qqq} give the following
relations:
\begin{equation}
  \pp = [\hh,\bb],\quad \bb=[\pp,\hh],\quad \hh = [\bb,\pp],\quad \bb^2 = -\tfrac14,\quad
  \pp^2 = -\tfrac14\quad\text{and}\quad c_0 \hh = \tfrac12 \bc_1 + \bc_2 \bb
  + \bc_3 \pp,
\end{equation}
and in addition $\bb \pp + \pp \bb = 0$, which simply states that $\bb \perp
\pp$.  These relations imply that $\bb,\pp,\hh \in \Im\HH$ and that $(2\bb, 2\pp,
2\hh)$ is an oriented orthonormal basis for $\Im\HH$.  The remaining
super-Jacobi identities give
\begin{equation}
  \bc_2 = - 2 c_0 \pp, \quad \bc_3 = 2 c_0 \bb \implies \bc_1 = -2 c_0 \hh,
\end{equation}
and some other relations which are identically satisfied.  If $c_0=0$
then $[\Q,\Q]=0$, so we requires $c_0\neq 0$.  Hence
$(\frac{\bc_1}{c_0}, \frac{\bc_2}{c_0}, \frac{\bc_3}{c_0})$ defines a
negatively oriented, orthonormal basis for $\Im\HH$.  The automorphism
group of $\HH$ acts transitively on the space of orthonormal oriented
bases, so we can choose $(2\bb, 2\pp, 2\hh) = (\ii,\jj,\kk)$ without loss of
generality.
\\ \\
The resulting Lie superalgebra becomes
\begin{equation}
  \begin{split}
    [\sH,\sQ(s)] &= \tfrac12 \sQ(s\kk)\\
    [\sB(\beta),\sQ(s)] &= \tfrac12 \sQ(\beta s\ii)\\
    [\sP(\pi),\sQ(s)] &= \tfrac12 \sQ(\pi s\jj)\\
    [\sQ(s),\sQ(s)] &= c_0 \left(|s|^2 \sH + \sJ(s\kk \sbar) + \sB(s\jj\bar
    s) - \sP(s\ii\sbar)\right).
  \end{split}
\end{equation}
We may rescale $\Q$ to bring $c_0$ to a sign, but we can then change
the sign via the automorphism of $\k$ which sends $(\J,\B,\P,\bH)
\mapsto (\J,\P,\B,-\bH)$ and the inner automorphism induced by the
automorphism of $\HH$ which sends $(\ii,\jj,\kk) \mapsto
(\jj,\ii,-\kk)$.  In summary, there is a unique kinematical Lie
superalgebra with $[\Q,\Q]\neq 0$ extending $\k = \so(3,2)$: namely,
\begin{equation}
  \begin{split}
    [\sH,\sQ(s)] &= \tfrac12 \sQ(s\kk)\\
    [\sB(\beta),\sQ(s)] &= \tfrac12 \sQ(\beta s\ii)\\
    [\sP(\pi),\sQ(s)] &= \tfrac12 \sQ(\pi s\jj)\\
    [\sQ(s),\sQ(s)] &= |s|^2 \sH + \sJ(s\kk \sbar) + \sB(s\jj\sbar) -
    \sP(s\ii\sbar).
  \end{split}
\end{equation}
To show that this Lie superalgebra is isomorphic to $\osp(1|4)$, we may
argue as follows.  We first prove that $\s_{\bar 0}$ leaves
invariant a symplectic form on $\s_{\bar 1}$.  The most general
rotationally invariant bilinear form on $\s_{\bar 1}$ is given by
\begin{equation}
  \omega(\sQ(s_1), \sQ(s_2)) := \Re (s_1 \qq \sbar_2) \qquad\text{for
    some $\qq \in \HH$.}
\end{equation}
Indeed, if $\uu \in \Sp(1)$ then
\begin{equation}
  \begin{split}
    (\uu \cdot \omega)(\sQ(s_1), \sQ(s_2)) &= \omega(\uu^{-1} \cdot
    \sQ(s_1), \uu^{-1} \cdot \sQ(s_2))\\
    &= \omega (\sQ(\ubar s_1),  \sQ(\ubar s_2))\\
    &= \Re(\ubar s_1 \qq \sbar_2 \uu)\\
    &= \Re(s_1 \qq \sbar_2)\\
    &= \omega(\sQ(s_1), \sQ(s_2)).
  \end{split}
\end{equation}
Demanding that $\omega$ be invariant under the other generators $\bH,
\B, \P$, we find that $\qq = \mu \kk$ for some $\mu \in \RR$.  
Acting infinitesimally now,
\begin{equation}
  \begin{split}
    (\sH \cdot \omega)(\sQ(s_1), \sQ(s_2)) &= -\omega([\sH, \sQ(s_1)], \sQ(s_2)) - \omega(\sQ(s_1), [\sH, \sQ(s_2)])\\
    &= - \tfrac12 \omega(\sQ(s_1\kk), \sQ(s_2)) - \tfrac12\omega(\sQ(s_1), \sQ(s_2\kk))\\
    &= -\tfrac12 \Re(s_1\kk \qq \sbar_2) + \tfrac12 \Re(s_1 \qq \kk \sbar_2)\\
    &= \tfrac12 \Re(s_1 [\qq,\kk] \sbar_2),
  \end{split}
\end{equation}
which must vanish for all $s_1,s_2 \in S$, so that $[\qq,\kk] = 0$ and
hence $\qq = \lambda \id + \mu\kk$ for some $\lambda,\mu \in \RR$. A
similar calculation with $\B$ and $\P$ shows that $\qq$ must
anti-commute with $\ii$ and $\jj$ and thus $\qq = \mu \kk$. So the
action of $\s_{\bar 0} \cong \so(3,2)$ on $\s_{\bar 1}$ defines a Lie
algebra homomorphism $\so(3,2) \to \sp(4,\RR)$, which is clearly
non-trivial. Since $\so(3,2)$ is simple, it is injective and a
dimension count shows that this is an isomorphism. But as
representations of $\so(3,2)$, $\odot^2\s_{\bar 1} \cong \wedge^2V$,
where $V$ is the 5-dimensional vector representation of $\s_{\bar 0}$,
and, since $\wedge^2 V \cong \so(V) \cong \s_{\bar 0}$, we have that
there is a one-dimensional space of $\s_{\bar 0}$-equivariant maps
$\odot^2 \s_{\bar 1} \to \s_{\bar 0}$. Since $[\Q,\Q] \neq 0$, the
bracket $\odot^2\s_{\bar 1} \to \s_{\bar 0}$ is an isomorphism. Thus
$\s$ is, by definition, isomorphic to $\osp(1|4)$.

\subsubsection{The Carroll Superalgebra}
\label{sec:carroll-superalgebra}

For $\k$ the Carroll Lie algebra (\hyperlink{KLA6}{$\mathsf{K6}$} in
Table~\ref{tab:kla}), Lemma~\ref{lem:kmod} implies that
$\pp = \bb = \hh = 0$, and then Lemma~\ref{lem:qqq} says that
$\bc_1 = 0$.  The $[\B,\Q,\Q]$ super-Jacobi says that $\bc_3 =0$, and the
$[\P,\Q,\Q]$ super-Jacobi says that $\bc_2 = 0$.  The only non-zero bracket
involving $\Q$ is
\begin{equation}
  [\sQ(s), \sQ(s)] = c_0 |s|^2 \sH,
\end{equation}
which is non-zero for $c_0 \neq 0$.  If so, we can set $c_0 = 1$ via an
automorphism of $\k$ which rescales $\bH$ and $\P$, say, by $c_0$.  In
summary, there is a unique Carroll superalgebra with brackets
\begin{equation}
  [\sQ(s), \sQ(s)] = |s|^2 \sH,
\end{equation}
in addition to those of the Carroll Lie algebra itself.  This Lie superalgebra
is a contraction of the Poincaré superalgebra.  We will show this explicitly
in Section~\ref{sec:limits-betw-supersp}.

\subsubsection{The Galilean Superalgebras}
\label{sec:galil-super}

For $\k$ the Galilean Lie algebra (\hyperlink{KLA2}{$\mathsf{K2}$} in
Table~\ref{tab:kla}), Lemma~\ref{lem:kmod} says that $\bb=\pp = 0$, and
Lemma~\ref{lem:qqq} says that $\bc_1 = 2 c_0 \hh$.  The $[\B,\Q,\Q]$
super-Jacobi identity says that $\bc_1 = 0$ and $c_0 = 0$.  The $[\P,\Q,\Q]$
super-Jacobi identity is now identically satisfied, whereas the $[\bH,\Q,\Q]$
super-Jacobi identity gives
\begin{equation}
  \hh \bc_2 + \bc_2 \bar \hh = 0 \qquad\text{and}\qquad \bc_2 + \hh \bc_3 +
  \bc_3 \bar \hh = 0.
\end{equation}
Since $\bc_2$ and $\bc_3$ cannot both vanish, we see that this is only
possible if $\hh \in \Im\HH$; therefore, these equations become $[\hh,\bc_2]
=0$ and $\bc_2 = [\bc_3,\hh]$.  There are two cases to consider,
depending on whether or not $\hh$ vanishes.  If $\hh=0$, then $\bc_2 = 0$
and $\bc_3$ is arbitrary.  If $\hh\neq 0$, then on the one hand $\bc_2$
is collinear with $\hh$, but also $\bc_2 = [\bc_3,\hh]$, which means that
$\bc_2=0$ so that $\bc_3 \neq 0$ is collinear with $\hh$.  In either
case, $\bc_3 \neq 0$ and $\hh = \psi \bc_3$, where $\psi \in \RR$
can be zero.
\\ \\
This gives rise to the following additional brackets
\begin{equation}
  [\sH, \sQ(s)] = \psi \sQ(s\bc_3) \qquad\text{and}\qquad [\sQ(s),
  \sQ(s)] = - \sP(s\bc_3 \sbar).
\end{equation}
We may use the automorphisms of $\HH$ to bring $\bc_3 = \phi \kk$, for
some non-zero $\phi \in \RR$.  We can set $\phi = 1$ by an automorphism
of $\k$ which rescales $\P$ and also $\B$ and $\bH$ suitably.  This
still leaves the freedom to set $\psi = 1$ if $\psi \neq 0$.
In summary, we have two Galilean superalgebras:
\begin{equation}
  [\sH, \sQ(s)] =
  \begin{cases}
    0\\
    \sQ(s \kk)
  \end{cases}
  \quad\text{and}\quad [\sQ(s),\sQ(s)] = - \sP(s\kk \sbar).
\end{equation}
The first one (where $[\bH,\Q] = 0$) is a contraction of the Poincaré
superalgebra, whereas the second (where $[\bH,\Q] \neq 0$) is not.

\subsubsection{Lie Superalgebras Associated with the Static Kinematical Lie Algebra}
\label{sec:lie-super-assoc-1}

The static kinematical Lie algebra is \hyperlink{KLA1}{$\mathsf{K1}$} in Table~\ref{tab:kla}.  In this
case, Lemma~\ref{lem:kmod} says that $\bb=\pp=0$ and
Lemma~\ref{lem:qqq} says that $\bc_1 = 2 c_0 \hh$.  The $[\bH,\Q,\Q]$
super-Jacobi identity says that $\hh \in \Im \HH$ and that $[\hh,\bc_i] = 0$
for $i=1,2,3$. Finally, either the $[\B,\Q,\Q]$ or $[\P,\Q,\Q]$ super-Jacobi
identities say that $\bc_1 = 0$, so that $\hh c_0 = 0$.  This means
that either $\hh=0$ or else $c_0 = 0$ (or both).
\\ \\
There are several branches:
\begin{enumerate}
\item If $c_0 = 0$ and $\hh \neq 0$, $\bc_2$ and $\bc_3$ are collinear
  with $\hh$, but cannot both be zero.  Using automorphisms of the
  static kinematical Lie algebra, and the ability to rotate vectors, we
  can bring $\hh = \tfrac12 \kk$, $\bc_2 = 0$ and $\bc_3 = \kk$, so that
  we have a unique Lie superalgebra in this case, with additional
  brackets
  \begin{equation}
    [\sH,\sQ(s)] = \tfrac12 \sQ(s\kk) \quad\text{and}\quad
    [\sQ(s),\sQ(s)] = - \sP(s\kk\sbar).
  \end{equation}
\item If $c_0 = 0$ and $\hh=0$, $\bc_2$ and $\bc_3$ are unconstrained,
  but not both zero.  We distinguish two cases, depending on whether
  or not they are linearly independent:
  \begin{enumerate}
  \item If they are linearly dependent, so that they are collinear,
    then we can use automorphisms to set $\bc_2$, say, to zero and
    $\bc_3 = \kk$.  This results in the Lie superalgebra
    \begin{equation}
      [\sQ(s),\sQ(s)] = - \sP(s\kk\sbar).
    \end{equation}
  \item If they are linearly independent, we can bring them to $\bc_2
    = \jj$ and $\bc_3 = \kk$, resulting in the Lie superalgebra
    \begin{equation}
      [\sQ(s),\sQ(s)] = - \sB(s\jj\sbar) - \sP(s\kk\sbar).
    \end{equation}
  \end{enumerate}
\item Finally, if $c_0\neq 0$, then $\hh=0$ and, again, $\bc_2$ and
  $\bc_3$ are unconstrained, but can now be zero.  Moreover we can
  rescale $\bH$ so that $c_0 = 1$.  We have three cases to consider,
  depending on whether they span a zero-, one- or two-dimensional real
  subspace of $\Im \HH$:
  \begin{enumerate}
  \item If $\bc_2 = \bc_3 = 0$, we have the Lie superalgebra
    \begin{equation}
      [\sQ(s),\sQ(s)] = |s|^2 \sH.
    \end{equation}
  \item If $\bc_2$ and $\bc_3$ span a line, then we may use the
    automorphisms to set $\bc_2 = 0$ and $\bc_3 = \kk$, resulting in
    the Lie superalgebra
    \begin{equation}
      [\sQ(s),\sQ(s)] = |s|^2 \sH - \sP(s\kk\sbar).
    \end{equation}
  \item Finally, if $\bc_2$ and $\bc_3$ are linearly independent, we
    may use the automorphisms to set $\bc_2 = \jj$ and $\bc_3 = \kk$,
    resulting in the Lie superalgebra
    \begin{equation}
      [\sQ(s),\sQ(s)] = |s|^2 \sH - \sB(s\jj\sbar) - \sP(s\kk\sbar).
    \end{equation}
  \end{enumerate}
\end{enumerate}

\subsubsection{Lie Superalgebras Associated with Kinematical Lie Algebra $\mathsf{K3}_\gamma$}
\label{sec:lie-super-assoc-3}

Here, Lemma~\ref{lem:kmod} says that $\bb = \pp = 0$ and
Lemma~\ref{lem:qqq} says that $\bc_1 = 2 c_0 \hh$.  The $[\B,\Q,\Q]$
super-Jacobi identity says that $\bc_1 = 0$ and $c_0 = 0$, whereas the
$[\P,\Q,\Q]$ super-Jacobi identity offers no further conditions.  Finally,
the $[\bH,\Q,\Q]$ super-Jacobi identity gives two conditions
\begin{equation}
  \gamma \bc_2 = \hh \bc_2 + \bc_2 \bar \hh \quad\text{and}\quad
  \bc_3 = \hh \bc_3 + \bc_3 \bar \hh,
\end{equation}
which are equivalent to
\begin{equation}
  (\gamma - 2 \Re \hh) \bc_2 = [\Im \hh, \bc_2] \quad\text{and}\quad
  (1 - 2 \Re \hh) \bc_3 = [\Im \hh, \bc_3].
\end{equation}
We see that we must distinguish two cases: $\gamma = 1$ and $\gamma
\in [-1,1)$.
\\ \\
If $\gamma \neq 1$, then we have two cases, depending on whether $\Re
\hh = \frac12$ or $\Re \hh = \frac12 \gamma$.  In the former case, $\bc_2
= 0$ and $\Im \hh$ is collinear with $\bc_3 \neq 0$, whereas, in the
latter, $\bc_3 = 0$ and $\Im \hh$ is collinear with $\bc_2 \neq 0$.
\\ \\
If $\gamma = 1$, then $\Re \hh = \frac12$ and $\bc_2$, $\Im \hh$ and
$\bc_3$ are all collinear, with at least one of $\bc_2$ and $\bc_3$
non-zero.  When $\gamma =1$, the automorphisms of $\kk$ include the
general linear group $\GL(2,\RR)$ acting on the two copies of the
vector representation.  Using this fact, we can always assume that $\bc_2 =
0$ and $\bc_3 \neq 0$.
\\ \\
In either case, all non-zero vectors are collinear and we can rotate
them to lie along the $\kk$ axis.  In the case $\gamma = 1$, we have
a one-parameter family of Lie superalgebras:
\begin{equation}
  [\sH,\sQ(s)] = \tfrac12 \sQ(s(1+\lambda \kk)) \quad\text{and}\quad
  [\sQ(s),\sQ(s)] = - \sP(s\kk \sbar),
\end{equation}
where we have used the freedom to rescale $\P$ in order to set
$\bc_3 = \kk$.  This is also a Lie superalgebra for $\gamma \neq 1$.
\\ \\
If $\gamma \neq 1$, we have an additional one-parameter family of Lie
superalgebras:
\begin{equation}
  [\sH,\sQ(s)] = \tfrac12 \sQ(s(\gamma +\lambda \kk)) \quad\text{and}\quad
 [\sQ(s),\sQ(s)] = - \sB(s\kk \sbar).
\end{equation}
The parameter $\lambda$ is essential; that is, Lie superalgebras with
different values of $\lambda$ are not isomorphic.  One way to test
this is the following.  Let $[-,-]_\lambda$ denote the above Lie
bracket.  This satisfies the super-Jacobi identity for all $\lambda \in
\RR$.  Write it as $[-,-]_\lambda = (1-\lambda) [-,-]_0 + \lambda
[-,-]_1$.  The difference $[-,-]_1 - [-,-]_0$ is a cocycle of the Lie
superalgebra with $\lambda = 0$.  The parameter would be inessential
if and only if it is a coboundary.  One can check that this is not the
case.  This same argument shows that the parameters appearing in other
Lie superalgebras are essential as well.

\subsubsection{Lie Superalgebras Associated with Kinematical Lie Algebra $\mathsf{K4}_\chi$}
\label{sec:lie-super-assoc-4}

Here, Lemma~\ref{lem:kmod} says $\bb = \pp = 0$ and Lemma~\ref{lem:qqq}
says that $\bc_1 = 2 c_0 \hh$.  Then either the $[\B,\Q,\Q]$ or
$[\P,\Q,\Q]$ super-Jacobi identities force $\bc_1=0$ and $c_0= 0$.  The
$[\bH,\Q,\Q]$ super-Jacobi identity results in the following two equations:
\begin{equation}
  \chi \bc_2 - \bc_3 = \hh \bc_2 + \bc_2 \bar \hh \quad\text{and}\quad
  \chi \bc_3 + \bc_2 = \hh \bc_3 + \bc_3 \bar \hh,
\end{equation}
or equivalently,
\begin{equation}
  (\chi - 2 \Re \hh) \bc_2 - \bc_3 = [\Im \hh, \bc_2] \quad\text{and}\quad
  (\chi - 2 \Re \hh) \bc_3 + \bc_2 = [\Im \hh, \bc_3].
\end{equation}
Taking the inner product of the first equation with $\bc_2$ and of the
second equation with $\bc_3$ and adding, we find
\begin{equation}
  (\chi - 2 \Re \hh) (|\bc_2|^2 + |\bc_3|^2) = 0, 
\end{equation}
and since $\bc_2$ and $\bc_3$ cannot both be zero, we see that $\Re \hh
= \frac\chi2$, and hence that
\begin{equation}
  [\Im \hh, \bc_2] = - \bc_3 \quad\text{and}\quad [\Im \hh, \bc_3] = \bc_2,
\end{equation}
so that $\bc_3 \perp \bc_2$.  This shows that $(\Im \hh, \bc_3, \bc_2)$
is an oriented orthogonal (but not necessarily orthonormal) basis.  We
can rotate them so that $\Im \hh = \phi \jj$, $\bc_3 = \psi \kk$ and
$\bc_2 = 2\phi\psi \ii$, but then we see that $\phi^2 =
\tfrac14$.  Using the automorphism of $\k$ which rescales $\B$ and
$\P$ simultaneously by the same amount, we can assume that $\bc_3 =
\kk$; then, if $\Im \hh = \pm \tfrac12 \jj$, we find $\bc_2 = \pm \ii$.
But the two signs are related by the automorphism of $\HH$ which sends
$(\ii,\jj,\kk) \mapsto (-\ii,-\jj,\kk)$.  In summary, we have a
unique Lie superalgebra associated with this kinematical Lie algebra:
\begin{equation}
  [\sH,\sQ(s)] = \tfrac12 \sQ(s(\chi + \jj)) \quad\text{and}\quad
  [\sQ(s), \sQ(s)] = - \sB(s\ii\sbar) - \sP(s\kk\sbar).
\end{equation}

\subsubsection{Lie Superalgebras Associated with Kinematical Lie Algebra $\mathsf{K5}$}
\label{sec:lie-super-assoc}

Here, Lemma~\ref{lem:kmod} says that $\bb=\pp=0$ and Lemma~\ref{lem:qqq}
says that $\bc_1 = 2 c_0 \hh$.  The $[\B,\Q,\Q]$ super-Jacobi identity forces
$c_0 = \bc_1 = 0$, which then satisfies the $[\P,\Q,\Q]$ super-Jacobi identity
identically.  The $[\bH,\Q,\Q]$ super-Jacobi identity gives two
further equations
\begin{equation}
  \bc_2 = \hh \bc_2 + \bc_2 \bar \hh \quad\text{and}\quad \bc_2 + \bc_3
  = \hh \bc_3 + \bc_3 \bar \hh.
\end{equation}
The first equation is equivalent to
\begin{equation}
  (1 - 2 \Re(\hh)) \bc_2 = [\Im \hh, \bc_2].
\end{equation}
If $\bc_2 \neq 0$, then $\Re \hh = \frac12$ and $\Im \hh$ is collinear
with $\bc_2$. But then the second equation says that $\bc_2 = [\Im \hh,
\bc_3]$, which is incompatible with $\bc_2$ and $\Im \hh$ being
collinear.  Therefore, $\bc_2 = 0$ and the second equation then says
that $\Re \hh = \frac12$  and that $\Im \hh$ is collinear with $\bc_3 \neq 0$.
In this instance, we have the following additional brackets
\begin{equation}
  [\sH, \sQ(s)] = \tfrac12 \sQ(s (1 + \lambda \bc_3))
  \quad\text{and}\quad [\sQ(s),\sQ(s)] = - \sP(s \bc_3 \sbar),
\end{equation}
where $\lambda \in \RR$.  We may rotate $\bc_3$ to $\psi \kk$, for some
non-zero $\psi \in \RR$.  We can then rescale $\P$ and $\B$
simultaneously by the same amount to set $\psi = 1$.  In summary, we
are left with the following one-parameter family of Lie superalgebras:
\begin{equation}
  [\sH, \sQ(s)] = \tfrac12 \sQ(s (1 + \lambda \kk))
  \quad\text{and}\quad [\sQ(s),\sQ(s)] = - \sP(s \kk \sbar).
\end{equation}
As in the case of the Lie superalgebras associated with Lie algebra
\hyperlink{KLA3}{$\mathsf{K3}_\gamma$}, the parameter $\lambda$ is essential and Lie
superalgebras with different values of $\lambda$ are not isomorphic.

\subsubsection{Lie Superalgebras Associated with Kinematical Lie Algebra $\mathsf{K12}$}
\label{sec:lie-super-assoc-12}

Lemma~\ref{lem:kmod} says that $\bb^2 = \frac12 \bb$, so that
$\bb\in \RR$, $[\hh,\pp]=0$ and $\pp^2= \frac12 (\bb-\frac12)$, so
that $\pp \in \Im\HH$.  (In particular, $\bb\pp= 0$.)
Lemma~\ref{lem:qqq} does not simplify at this stage.  The $[\bH,\Q,\Q]$
super-Jacobi identity says that $c_0 \Re \hh = 0$ and that
$\hh \bc_i + \bc_i \bar \hh = 0$ for $i=1,2,3$.  The $[\B,\Q,\Q]$
super-Jacobi identity says that $\bb\bc_1 = 0$, $\bb\bc_3 = 0$ and
$\bc_1 = (2\bb -1) \bc_2$.  Finally, the $[\P,\Q,\Q]$ super-Jacobi identity
says that $c_0 \pp = 0$, among other conditions that will turn out not
to play a rôle.
\\ \\
We have two branches depending on the value of $\bb$:
\begin{enumerate}
\item If $\bb=0$, $\pp^2= -\frac14$, so that $c_0 = 0$.  This means $\bc_1
  + \bc_2 =0$ and $\bc_3 = 2 \bc_1 \pp$ and none of $\bc_{1,2,3}$ can
  vanish.  This means that $\Re \hh = 0$ and that $\hh$ and $\bc_i$ are
  collinear for all $i=1,2,3$.  Also, $\hh$ and $\pp$ are collinear, which
  is inconsistent, unless $\hh = 0$: indeed, if $\pp$ and $\bc_i$ are
  collinear with $\hh \neq 0$, then $\bc_3 = 2\bc_1 \pp$ cannot be
  satisfied, since the L.H.S. is imaginary but the R.H.S. is real and both
  are non-zero.  Therefore, we conclude that $\hh=0$.  The condition
  $\bc_3 = 2 \bc_1 \pp$ says that there exists $\psi > 0$ such that
  $(\psi^{-1} \bc_1, 2\pp, \psi^{-1} \bc_3)$ is an oriented
  orthonormal basis, which can be rotated to $(\ii,\jj,\kk)$.  In
  other words, we can write $\bc_1 = \psi \ii$, $\pp = \frac12\jj$
  and $\bc_3 = \psi \kk$, so that $\bc_2 = - \psi \ii$.  We may
  rescale $\Q$ to bring $\psi =1$ and we may rotate $(\ii,\jj,\kk)
  \mapsto (-\ii,\jj,-\kk)$ to arrive at the following Lie
  superalgebra:
  \begin{equation}
    [\sP(\pi), \sQ(s)] = \tfrac12 \sQ(s\jj) \quad\text{and}\quad
    [\sQ(s),\sQ(s)] = \sJ(s\ii\sbar) - \sB(s\ii\sbar) + \sP(s\kk\bar
    s).
  \end{equation}
\item If $\bb=\frac12$, then $\pp=0$, $\bc_1= \bc_3 = 0$, and
  $\bc_2 = 2 c_0 \hh$ with $c_0 \neq 0$.  We have two sub-branches,
  depending on whether or not $\hh=0$.
  \begin{enumerate}
  \item If $\hh=0$, we have the following Lie superalgebra, after
    rescaling $\bH$ to set $c_0 = 1$:
    \begin{equation}
      [\sB(\beta),\sQ(s)] = \tfrac12 \sQ(\beta s) \quad\text{and}\quad
      [\sQ(s), \sQ(s)] = |s|^2 \sH.
    \end{equation}
  \item On the other hand, if $\hh \neq 0$, we may rotate it so that
    $2h = \psi \kk$ for some $\psi$ such that $\psi c_0 > 0$.  Then we
    may rescale $\bH$ and $\Q$ in such that a way that we bring
    $\psi c_0 =1$, thus arriving at the following Lie superalgebra:
    \begin{equation}
      [\sB(\beta),\sQ(s)] = \tfrac12 \sQ(\beta s), \quad [\sH,\sQ(s)] =
      \tfrac12 \sQ(s\kk) \quad\text{and}\quad
      [\sQ(s), \sQ(s)] = |s|^2 \sH - \sB(s\kk\sbar).
    \end{equation}
  \end{enumerate}
\end{enumerate}

\subsubsection{Lie Superalgebras Associated with Kinematical Lie Algebra $\mathsf{K13}$}
\label{sec:lie-super-assoc-13}

Here, Lemma~\ref{lem:kmod} says that $\bb^2= \frac12 \bb$, so that $\bb \in
\RR$ and $\pp^2=-\tfrac12(\bb-\frac12) \in \RR$.  Lemma~\ref{lem:qqq} does
not simplify further at this stage.  The $[\bH,\Q,\Q]$ super-Jacobi identity
says that $c_0 \Re \hh = 0$ and $\hh\bc_i + \bc_i \bar \hh = 0$ for
$i=1,2,3$.  The $[\B,\Q,\Q]$ super-Jacobi identity says that $\bb\bc_1 =
\bb\bc_3 = 0$, whereas $(\bb-\frac12)\bc_2 = \frac12 \bc_1$.  Finally, the
$[\P,\Q,\Q]$ super-Jacobi identity says that $\bc_1 = 2\pp \bc_3$, $\bc_3 = -2
\pp \bc_2$ and $\bc_3 = 2 \pp \bc_1$.
\\ \\
As usual we have two branches depending on the value of $\bb$:
\begin{enumerate}
\item If $\bb=0$, then $\pp^2= \frac14$.  Due to the automorphism of $\k$ which
  changes the sign of $\P$, we may assume $\pp = \frac12$ without loss
  of generality.  It follows that $\bc_1 = c_0 \hh$ and that $\bc_2 = -
  \bc_1 = - c_0 \hh$ and that $\bc_3 = \bc_1 = c_0 \hh$.  If $c_0 = 0$,
  then $\bc_i = 0$ for all $i$, so we must have $c_0 \neq 0$.  In that
  case, $\hh \in \Im \HH$ and $\hh$ is collinear with all $\bc_i$ for
  $i=1,2,3$.  We distinguish two cases, depending on whether or not
  $\hh=0$:
  \begin{enumerate}
  \item If $\hh\neq 0$, we may rotate it so that $\hh = \psi \kk$, where
    $\psi c_0 > 0$.  We may rescale $\bH \mapsto \psi^{-1} \bH$
    (which is an automorphism of $\k$) and rescale $\Q$ to bring
    $\psi c_0 = 1$.  In summary, we arrive at the following Lie
    superalgebra:
    \begin{equation}
      [\sH,\sQ(s)] = \sQ(s\kk),\quad [\sP(\pi), \sQ(s)] = \tfrac12
      \sQ(\pi s) \qquad\text{and}\quad [\sQ(s),\sQ(s)] = |s|^2 \sH -
      \sJ(s\kk\sbar) + \sB(s\kk\sbar) - \sP(s\kk\sbar).
    \end{equation}
  \item If $\hh = 0$, then we have the Lie superalgebra
    \begin{equation}
      [\sP(\pi), \sQ(s)] = \tfrac12
      \sQ(\pi s) \quad\text{and}\quad [\sQ(s),\sQ(s)] = |s|^2 \sH.
    \end{equation}
  \end{enumerate}
  
\item If $\bb=\frac12$, then $\pp=0$ and $\bc_1 = \bc_3 = 0$ with $\bc_2 =
  2 c_0 \hh$ with $c_0 \neq 0$ and $\hh \in \Im\HH$.  Again, we distinguish
  between vanishing and non-vanishing $\hh$:
  \begin{enumerate}
  \item If $\hh \neq 0$, we may rotate it so that $2h = \psi \kk$
    with $\psi c_0 > 0$.  We apply the $\k$-automorphism $\bH \mapsto
    \psi^{-1} \bH$ and rescale $\Q$ to bring $\psi c_0 = 1$, thus
    resulting in the Lie superalgebra
    \begin{equation}
      [\sH,\sQ(s)] = \tfrac12 \sQ(s\kk),\quad [\sB(\beta), \sQ(s)] = \tfrac12
      \sQ(\beta s) \quad\text{and}\quad [\sQ(s),\sQ(s)] = |s|^2 \sH -
      \sB(s\kk\sbar).      
    \end{equation}
  \item If $\hh = 0$, we arrive at the Lie superalgebra
    \begin{equation}
      [\sB(\beta), \sQ(s)] = \tfrac12
      \sQ(\beta s) \quad\text{and}\quad [\sQ(s),\sQ(s)] = |s|^2 \sH.
    \end{equation}
  \end{enumerate}
\end{enumerate}

\subsubsection{Lie Superalgebras Associated with Kinematical Lie Algebra $\mathsf{K14}$}
\label{sec:lie-super-assoc-14}

Here, Lemma~\ref{lem:kmod} says that $\pp=0$ and $2 \bb^2 = \bb$, so that $\bb
\in \RR$.  Lemma~\ref{lem:qqq} says that $\tfrac12 \bc_1 + \bc_2 \bb = c_0 \hh$.  The
$[\P,\Q,\Q]$ super-Jacobi identity says that $\bc_1=0$, so that $c_0 \hh =
\bc_2 \bb$.  The $[\B,\Q,\Q]$ super-Jacobi identity says that $(2\bb-1) \bc_2 = 0$ and
$\bb\bc_3=0$, whereas the $[\B,\Q,\Q]$ super-Jacobi identity says that $\hh
\bc_i + \bc_i \bar \hh = 0$ for $i=2,3$.
\\ \\
We have two branches, depending on the value of $\bb$:
\begin{enumerate}
\item If $\bb=0$, then $\bc_2 = 0$, and we have two sub-branches depending
  on whether or not $c_0 = 0$:
  \begin{enumerate}
  \item If $c_0 = 0$, then $\bc_3 \neq 0$, so that $\Re \hh = 0$ and $\hh$
    is collinear with $\bc_3$.  We may rotate $\bc_3$ to lie along
    $\kk$, say, and then use automorphisms of $\k$ to set $\bc_3 =
    \kk$.  If $\hh \neq 0$, we may also set it equal to $\kk$.  In
    summary, we have two isomorphism classes of Lie superalgebras
    here:
    \begin{equation}
      [\sH,\sQ(s)] =
      \begin{cases}
        0 \\
        \sQ(s\kk)
      \end{cases}
      \quad\text{and}\quad
      [\sQ(s),\sQ(s)] = - \sP(s\kk\sbar).
    \end{equation}
  \item If $c_0 \neq 0$, then $\hh=0$ and $\bc_3$ is free: if non-zero, we
    may rotate it to $\kk$ and, rescaling $\P$ with the 
    automorphisms of $\k$, we can bring it to $\kk$.  Rescaling $\bH$ we
    can bring $c_0 = 1$.  This gives two isomorphism classes of Lie
    superalgebras:
    \begin{equation}
      [\sQ(s),\sQ(s)] = |s|^2 \sH \quad\text{and}\quad
      [\sQ(s),\sQ(s)] = |s|^2 \sH - \sP(s\kk\sbar).
    \end{equation}
  \end{enumerate}
\item If $\bb=\tfrac12$, then $\bc_3 = 0$ and $\bc_2 = 2 c_0 \hh$, and we
  have two cases, depending on whether or not $\hh=0$.
  \begin{enumerate}
  \item If $\hh=0$, then $\bc_2 = 0$, and then $c_0 \neq 0$.  Rescaling
    $\bH$, we can set $c_0=1$ and we arrive at the Lie superalgebra
    \begin{equation}
      [\sB(\beta),\sQ(s)] = \tfrac12 \sQ(\beta s)
      \quad\text{and}\quad
      [\sQ(s),\sQ(s)] = |s|^2 \sH.
    \end{equation}
  \item If $\hh\neq 0$, we can rotate and rescale $\Q$ such that $\bc_2 =
    2 c_0 \hh = \kk$ and then we can rescale $\bH$ so that $c_0 = 1$.
    The resulting Lie superalgebra is now
    \begin{equation}
      [\sH,\sQ(s)] = \tfrac12 \sQ(s\kk),\quad [\sB(\beta),\sQ(s)] =
      \tfrac12 \sQ(\beta s) \quad\text{and}\quad [\sQ(s),\sQ(s)] =
      |s|^2 \sH - \sB(s\kk\sbar).
    \end{equation}
  \end{enumerate}
\end{enumerate}

\subsubsection{Lie Superalgebras Associated with Kinematical Lie Algebra $\mathsf{K15}$}
\label{sec:lie-super-assoc-15}

Here, Lemma~\ref{lem:kmod} says that $\bb = \pp = 0$, whereas
Lemma~\ref{lem:qqq} says that $\bc_1 = 2 c_0 \hh$.  The $[\B,\Q,\Q]$
super-Jacobi identity says that $\bc_1 = \bc_2 = 0$, and hence the
$[\P,\Q,\Q]$ component is identically satisfied.  Finally, the
$[\bH,\Q,\Q]$ super-Jacobi identity says that $\hh \bc_3 + \bc_3 \bar \hh = 0$,
which expands to
\begin{equation}
  2 \Re(\hh) \bc_3 + [\Im \hh, \bc_3] = 0.
\end{equation}
We have two branches of solutions:
\begin{enumerate}
\item If $c_0 = 0$, then $\bc_3 \neq 0$, $\Re \hh = 0$ and $\hh$
  is collinear with $\bc_3$.  We may rotate $\bc_3$ to lie along $\kk$
  and then rescale $\Q$ so that $\bc_3 = \kk$.  If $\hh \neq 0$, we may
  use automorphisms of $\k$ to set $\hh = \kk$ as well.  In summary, we
  have two isomorphism classes of Lie superalgebras:
  \begin{equation}
    [\sH, \sQ(s)] =
    \begin{cases}
      \sQ(s\kk)\\
      0
    \end{cases} \quad\text{and}\quad
    [\sQ(s), \sQ(s)] = - \sP(s\kk\sbar).
  \end{equation}
\item If $c_0 \neq 0$, then $\hh = 0$ and $\bc_3$ is unconstrained.  If
  non-zero, we may rotate it to lie along $\kk$, rescale $\Q$ so that
  $\bc_3 = \kk$ and then use automorphisms of $\k$ to set $c_0 = 1$.
  In summary, we have two isomorphism classes of Lie superalgebras:
  \begin{equation}
    [\sQ(s), \sQ(s)] = |s|^2 \sH \quad\text{or}\quad
    [\sQ(s), \sQ(s)] = |s|^2 \sH - \sP(s\kk\sbar).
  \end{equation}
\end{enumerate}

\subsubsection{Lie Superalgebras Associated with Kinematical Lie Algebra $\mathsf{K16}$}
\label{sec:lie-super-assoc-16}

Here, Lemma~\ref{lem:kmod} says that $\pp=0$ and $\bb(\bb-\frac12) =0$, so
that $\bb \in \RR$.  Lemma~\ref{lem:qqq} then says that $c_0 \hh = \frac12
\bc_1 + \bc_2 \bb$.  Now the $[\P,\Q,\Q]$ super-Jacobi identity says that
$c_0=0$ and $\bc_1= 0$, so that $\bc_2 \bb = 0$.  The $[\bH,\Q,\Q]$ super-Jacobi
identity says that $\hh\bc_2 + \bc_2 \bar \hh = 0$ and $\hh\bc_3 + \bc_3
\bar \hh = \bc_3$.  Finally the $[\B,\Q,\Q]$ super-Jacobi identity says that
$\bb\bc_3 = 0$ and $(\bb-\frac12)\bc_2= 0$.
\\ \\
Notice that if $\bb=\frac12$ then $\bc_3=0$ and $\bc_2 = 0$,
contradicting $[\Q,\Q]\neq 0$, so we must have $\bb=0$.  Now $\bc_2=0$
and hence $\bc_3 \neq 0$.  It then follows that $\Re \hh = \frac12$ and
$\Im \hh$ is collinear with $\bc_3$.  We can rescale $\P$ (which is an
automorphism of $\k$) and rotate so that $\bc_3 = \kk$, so that $\hh =
\frac12 (1 + \lambda \kk)$ for $\lambda \in \RR$.  The resulting
one-parameter family of Lie superalgebras is then
\begin{equation}
  [\sH,\sQ(s)] = \tfrac12 \sQ(s(1+\lambda \kk)) \quad\text{and}\quad
  [\sQ(s), \sQ(s)] = - \sP(s\kk\sbar).
\end{equation}
As in the case of the Lie superalgebras associated with Lie algebras
\hyperlink{KLA3}{$\mathsf{K3}_\gamma$} and \hyperlink{KLA5}{$\mathsf{K5}$}, the
parameter $\lambda$ is essential and Lie superalgebras with different
values of $\lambda$ are not isomorphic.

\subsubsection{Lie Superalgebras Associated with Kinematical Lie Algebra $\mathsf{K17}$}
\label{sec:lie-super-assoc-17}

Here, Lemma~\ref{lem:kmod} simply sets $\bb = \pp = 0$ and
Lemma~\ref{lem:qqq} says $\bc_1 = 2 c_0 \hh$.  The $[\P,\Q,\Q]$ super-Jacobi
identity sets $\bc_1 = 0$ and hence $c_0 \hh = 0$.  The $[\B,\Q,\Q]$
super-Jacobi identity sets $c_0 = 0$ and $\bc_2 = 0$, whereas the
$[\bH,\Q,\Q]$ super-Jacobi identity says that $\hh$ is collinear with $\bc_3
\neq 0$.  We can rotate $\bc_3$ to lie along $\kk$ and rescale $\Q$
to effectively set it to $\kk$.  Then $\hh = \frac\psi2 \kk$ for some
$\psi$ and rescaling $\bH$ allows us to set $\psi =1$.  In
summary, we have a unique Lie superalgebra associated with this
kinematical Lie algebra; namely,
\begin{equation}
  [\sH,\sQ(s)] = \tfrac12 \sQ(s\kk) \quad\text{and}\quad
  [\sQ(s), \sQ(s)] = - \sP(s\kk\sbar).
\end{equation}

\subsubsection{Lie Superalgebras Associated with Kinematical Lie Algebra $\mathsf{K18}$}
\label{sec:lie-super-assoc-18}

Here, Lemma~\ref{lem:kmod} simply sets $\bb = \pp = 0$, and
Lemma~\ref{lem:qqq} says $\bc_1 = 2 c_0 \hh$.  The $[\P,\Q,\Q]$ super-Jacobi
identity sets $\bc_1 = 0$ and $c_0 = 0$, whereas the $[\B,\Q,\Q]$
super-Jacobi identity sets $\bc_2 = 0$.  Finally, the $[\bH,\Q,\Q]$ super-Jacobi
identity says that $\Re \hh = 1$ and $\Im \hh = \lambda \bc_3$ for some
$\lambda \in \RR$.  We can rotate $\bc_3$ to lie along $\kk$ and rescale $\Q$
to effectively set it to $\kk$.  Then $\hh = 1 + \lambda \kk$. In
summary, we have a one-parameter family of Lie superalgebras
associated with this kinematical Lie algebra; namely,
\begin{equation}
  [\sH,\sQ(s)] = \sQ(s(1+\lambda \kk)) \quad\text{and}\quad
  [\sQ(s), \sQ(s)] = - \sP(s\kk\sbar).
\end{equation}
As in the case of the Lie superalgebras associated with Lie algebras
\hyperlink{KLA3}{$\mathsf{K3}_\gamma$}, \hyperlink{KLA5}{$\mathsf{K5}$} and
\hyperlink{KLA16}{$\mathsf{K16}$}, the parameter $\lambda$ is essential and
Lie superalgebras with different values of $\lambda$ are not
isomorphic.

\subsection{Summary}
\label{sec:summary}

Table~\ref{tab:klsa} summarises the results.  In that table, we list
the isomorphism classes of kinematical Lie superalgebras (with
$[\Q,\Q]\neq 0$).  Recall that the Lie brackets involving $\Q$ are the
$[\Q,\Q]$ bracket and also
\begin{equation}
  [\sH, \sQ(s)] = \sQ(s\hh), \quad [\sB(\beta),\sQ(s)] = \sQ(\beta s \bb),
  \quad [\sP(\pi), \sQ(s) ] = \sQ(\pi s \pp),
\end{equation}
for some $\hh,\bb,\pp \in \HH$.  In Table~\ref{tab:klsa}, we list any non-zero
values of $\hh,\bb,\pp$ and the $[\Q,\Q]$ bracket.  The first column is
simply the label for the Lie superalgebra, the second column is the
corresponding kinematical Lie algebra, the next columns are
$\hh,\bb,\pp$ and $[\Q,\Q]$.  The next four columns are the possible
$\so(3)$-equivariant $\ZZ$-gradings (with $\J$ of degree $0$)
compatible with the $\ZZ_2$-grading; that is, such that the parity is
the reduction modulo $2$ of the degree.  This requires, in particular,
that $q$ be an odd integer, which we can take to be $-1$ by
convention, if so desired.
\\ \\
Despite all of the Lie superalgebras in Table~\ref{tab:klsa} producing 
spacetime supersymmetry, there are some important qualitative differences 
between those Lie superalgebras that have the time translation generator 
$\bH$ in the $[\Q, \Q]$ bracket and those that do not. In particular, 
theories invariant under a supersymmetry algebra for which $[\Q, \Q] = \bH$ 
are guaranteed to have a non-negative energy spectrum~\cite{Arav:2019tqm}. 
Therefore, for the construction of phenomenological models, these 
Lie superalgebras may be of more interest.

\begin{table}[h!]
  \centering
  \caption{Kinematical Lie Superalgebras (with $[\Q,\Q]\neq 0$)}
  \label{tab:klsa}
  \setlength{\extrarowheight}{2pt}
  \rowcolors{2}{blue!10}{white}
  \resizebox{\textwidth}{!}{
    \begin{tabular}{l|l*{4}{|>{$}c<{$}}*{4}{|>{$}c<{$}}}\toprule
      \multicolumn{1}{c|}{S\#} & \multicolumn{1}{c|}{$\k$} & \multicolumn{1}{c|}{$\hh$}& \multicolumn{1}{c|}{$\bb$} & \multicolumn{1}{c|}{$\pp$} & \multicolumn{1}{c|}{$[\sQ(s),\sQ(s)]$} & w_ {\bH} & w_{\B} & w_{\P} & w_{\Q}\\
      \toprule
      \hypertarget{KLSA1}{1} & \hyperlink{KLA1}{$\mathsf{K1}$}& \tfrac12 \kk & & & -\sP(s\kk\sbar) & 0 & 2m & 2q & q \\
      \hypertarget{KLSA2}{2} & \hyperlink{KLA1}{$\mathsf{K1}$} & & & & |s|^2 \sH - \sB(s\jj\sbar) - \sP(s\kk\sbar)& 2q & 2q & 2q & q \\
      \hypertarget{KLSA3}{3} & \hyperlink{KLA1}{$\mathsf{K1}$} & & & & |s|^2 \sH - \sP(s\kk\sbar) & 2q & 2m & 2 q & q \\
      \hypertarget{KLSA4}{4} & \hyperlink{KLA1}{$\mathsf{K1}$} & & & & |s|^2 \sH & 2q & 2m & 2p & q \\
      \hypertarget{KLSA5}{5} & \hyperlink{KLA1}{$\mathsf{K1}$} & & & & - \sB(s\jj\sbar) - \sP(s\kk\sbar) & 2n & 2q & 2 q & q \\
      \hypertarget{KLSA6}{6} & \hyperlink{KLA1}{$\mathsf{K1}$} & & & & -\sP(s\kk\sbar) & 2n & 2m & 2q & q \\
      \hypertarget{KLSA7}{7} & \hyperlink{KLA2}{$\mathsf{K2}$} & \kk & & & -\sP(s\kk\sbar) & 0 & 2q & 2 q & q \\
      \hypertarget{KLSA8}{8} & \hyperlink{KLA2}{$\mathsf{K2}$} & & & & -\sP(s\kk\sbar) & 2n & 2(q-n) & 2q & q \\
      \hypertarget{KLSA9}{9$_{\gamma\in[-1,1],\lambda\in\RR}$} & \hyperlink{KLA3}{$\mathsf{K3}_\gamma$} & \tfrac12 (1 + \lambda \kk) & & & -\sP(s\kk\sbar) & 0 & 2m & 2q & q \\
      \hypertarget{KLSA10}{10$_{\gamma\in[-1,1),\lambda\in\RR}$} & \hyperlink{KLA3}{$\mathsf{K3}_\gamma$} & \tfrac12 (\gamma + \lambda \kk) & & & -\sB(s\kk\sbar) & 0 & 2q & 2p & q \\
      \hypertarget{KLSA11}{11$_{\chi\geq0}$} & \hyperlink{KLA4}{$\mathsf{K4}_\chi$} & \tfrac12 (\chi + \jj) & & & -\sB(s\ii\sbar) - \sP(s\kk\sbar) & 0 & 2q & 2q & q \\
      \hypertarget{KLSA12}{12$_{\lambda\in\RR}$} & \hyperlink{KLA5}{$\mathsf{K5}$} & \tfrac12 (1 + \lambda \kk) & & & -\sP(s\kk\sbar) & 0 & 2q & 2q & q \\
      \hypertarget{KLSA13}{13} & \hyperlink{KLA6}{$\mathsf{K6}$} & & & & |s|^2 \sH & 2q & 2m & 2(q-m) & q \\
      \hypertarget{KLSA14}{14} & \hyperlink{KLA8}{$\mathsf{K8}$} & & \tfrac12 \kk & & |s|^2 \sH - \sP(s\kk\sbar) & 2q & 0 & 2q & q \\
      \hypertarget{KLSA15}{15} & \hyperlink{KLA11}{$\mathsf{K11}$} & \tfrac12 \kk & \tfrac12 \ii & \tfrac12 \jj & |s|^2 \sH + \sJ(s\kk\sbar) + \sB(s\jj\sbar) - \sP(s\ii\sbar) & - & - & - & - \\
      \hypertarget{KLSA16}{16} & \hyperlink{KLA12}{$\mathsf{K12}$} & & & \tfrac12 \jj & \sJ(s\ii\sbar) - \sB(s\ii\sbar) + \sP(s\kk\sbar) & - & - & - & - \\
      \hypertarget{KLSA17}{17} & \hyperlink{KLA12}{$\mathsf{K12}$} & & \tfrac12 & & |s|^2 \sH & 2q & 0 & 0 & q \\
      \hypertarget{KLSA18}{18} & \hyperlink{KLA12}{$\mathsf{K12}$} & \tfrac12 \kk & \tfrac12 & & |s|^2 \sH - \sB(s \kk \sbar) & - & - & - & - \\
      \hypertarget{KLSA19}{19} & \hyperlink{KLA13}{$\mathsf{K13}$} & \kk & & \tfrac12 & |s|^2 \sH - \sJ(s\kk\sbar) + \sB(s\kk\sbar) - \sP(s\kk\sbar) & - & - & - & - \\
      \hypertarget{KLSA20}{20} & \hyperlink{KLA13}{$\mathsf{K13}$} & & & \tfrac12 & |s|^2 \sH & 2q & 0 & 0 & q \\
      \hypertarget{KLSA21}{21} & \hyperlink{KLA13}{$\mathsf{K13}$} & & \tfrac12 & & |s|^2 \sH &  2q & 0 & 0 & q \\
      \hypertarget{KLSA22}{22} & \hyperlink{KLA13}{$\mathsf{K13}$} & \tfrac12 \kk & \tfrac12 & & |s|^2 \sH - \sB(s\kk\sbar) & - & - & - & - \\
      \hypertarget{KLSA23}{23} & \hyperlink{KLA14}{$\mathsf{K14}$} & \kk & & & - \sP(s\kk\sbar) & 0 & 0 & 2q & q \\
      \hypertarget{KLSA24}{24} & \hyperlink{KLA14}{$\mathsf{K14}$} & & & & - \sP(s\kk\sbar) & 2n & 0 & 2q & q \\
      \hypertarget{KLSA25}{25} & \hyperlink{KLA14}{$\mathsf{K14}$} & & & & |s|^2 \sH & 2q & 0 & 2p & q \\
      \hypertarget{KLSA26}{26} & \hyperlink{KLA14}{$\mathsf{K14}$} & & & & |s|^2 \sH - \sP(s\kk\sbar) & 2q & 0 & 2q & q \\
      \hypertarget{KLSA27}{27} & \hyperlink{KLA14}{$\mathsf{K14}$} & & \tfrac12 & & |s|^2 \sH & 2q & 0 & 2p & q \\
      \hypertarget{KLSA28}{28} & \hyperlink{KLA14}{$\mathsf{K14}$} & \tfrac12 \kk & \tfrac12 & & |s|^2 \sH - \sB(s\kk\sbar) & - & - & - & - \\
      \hypertarget{KLSA29}{29} & \hyperlink{KLA15}{$\mathsf{K15}$} & \kk & & & -\sP(s\kk\sbar) & - & - & - & - \\
      \hypertarget{KLSA30}{30} & \hyperlink{KLA15}{$\mathsf{K15}$} & & & & -\sP(s\kk\sbar) & - & - & - & - \\
      \hypertarget{KLSA31}{31} & \hyperlink{KLA15}{$\mathsf{K15}$} & & & & |s|^2 \sH & 2q & 2m & 4m & q \\
      \hypertarget{KLSA32}{32} & \hyperlink{KLA15}{$\mathsf{K15}$} & & & & |s|^2 \sH -\sP(s\kk\sbar) & - & - & - & - \\
      \hypertarget{KLSA33}{33$_{\lambda\in\RR}$} & \hyperlink{KLA16}{$\mathsf{K16}$} & \tfrac12 (1 + \lambda \kk) & & & -\sP(s\kk\sbar) & 0 & 0 & 2q & q \\
      \hypertarget{KLSA34}{34} & \hyperlink{KLA17}{$\mathsf{K17}$} & \tfrac12 \kk & & & -\sP(s\kk\sbar) & - & - & - & - \\
      \hypertarget{KLSA35}{35$_{\lambda\in\RR}$} & \hyperlink{KLA18}{$\mathsf{K18}$} & 1 + \lambda \kk & & & -\sP(s\kk\sbar) & - & - & - & - \\
      \bottomrule
    \end{tabular}
  }
  \caption*{The first column is our identifier for $\s$, whereas the
    second column is the kinematical Lie algebra $\k = \s_{\bar 0}$ in
    Table~\ref{tab:kla}.  The next four columns specify the brackets
    of $\s$ not of the form $[\J,-]$.  Supercharges $\sQ(s)$ are
    parametrised by $s \in \HH$, whereas $\sJ(\omega)$, $\sB(\beta)$
    and $\sP(\pi)$ are parametrised by $\omega,\beta,\pi \in
    \Im\HH$. The brackets are given by $[\sH,\sQ(s)] = \sQ(s\hh)$,
    $[\sB(\beta),\sQ(s)]=\sQ(\beta s \bb)$ and
    $[\sP(\pi),\sQ(s)] = \sQ(\pi s \pp)$, for some
    $\hh,\bb,\pp\in\HH$.  (This formalism is explained in
    Section~\ref{subsec:math_prelims_alg_klsa}.)  The final four columns specify
    compatible gradings of $\s$, with $m,n,p,q \in \ZZ$ and $q$ odd.}
\end{table}

\subsubsection{Unpacking the Quaternionic Notation}
\label{sec:unpack-quat-notat}

The quaternionic formalism we have employed in the classification of kinematical Lie superalgebras, which has the virtue of uniformity and ease in computation, results in expressions that are perhaps unfamiliar and, therefore, might hinder comparison with other formulations. In this section, we will go through an example illustrating how to unpack the notation.
\\ \\
The non-zero, supersymmetric brackets of the Poincaré superalgebra
\hyperlink{KLSA14}{$\mathsf{S14}$} are given by
\begin{equation}
  [\sB(\beta), \sQ(s)] = \sQ(\tfrac12 \beta s \kk) \quad\text{and}\quad
  [\sQ(s),\sQ(s)] = |s|^2 \sH - \sP(s\kk\sbar),
\end{equation}
where
\begin{equation}
  \sB(\beta) = \sum_{i=1}^3 \beta_i \bB_i \quad\text{and}\quad \sQ(s) = \sum_{a=1}^4
  s_a \bQ_a,
\end{equation}
and where
\begin{equation}
  \beta = \beta_1 \ii + \beta_2 \jj + \beta_3 \kk
  \quad\text{and}\quad s = s_1 \ii + s_2 \jj + s_3 \kk + s_4.
\end{equation}
This allows us to simply unpack the brackets into the following expressions
\begin{equation}
  [\bB_i, \bQ_a] = \tfrac12 \sum_{b=1}^4 \bQ_b \beta_i{}^b{}_a
  \quad\text{and}\quad
  [\bQ_a, \bQ_b] = \sum_{\mu=0}^3 \bP_\mu \gamma^\mu_{ab}.
\end{equation}
Here, we have introduced $\bP_0 = \bH$, and the matrices
$\boldsymbol{\beta}_i := [\beta_i{}^b{}_a]$ are given by
\begin{equation}
  \boldsymbol{\beta}_1 =
  \begin{pmatrix}
    \zero & -\id \\
    -\id & \zero
  \end{pmatrix},
  \quad
  \boldsymbol{\beta}_2 =
  \begin{pmatrix}
    \zero & i\sigma_2\\
    -i\sigma_2 & \zero
  \end{pmatrix}
  \quad\text{and}\quad
  \boldsymbol{\beta}_3 =
  \begin{pmatrix}
    \id & \zero\\
    \zero & -\id
  \end{pmatrix}.
\end{equation}
Additionally, the symmetric matrices $\boldsymbol{\gamma}^\mu :=
[\gamma^\mu_{ab}]$ are given by
\begin{equation}
  \boldsymbol{\gamma}^0 =
  \begin{pmatrix}
    \id & \zero \\ \zero & \id
  \end{pmatrix},
  \quad
  \boldsymbol{\gamma}^1 =
  \begin{pmatrix}
     \zero & \id \\ \id & \zero
  \end{pmatrix},
  \quad
  \boldsymbol{\gamma}^2 =
  \begin{pmatrix}
    \zero & -i\sigma_2\\ i\sigma_2 & \zero
  \end{pmatrix}
  \quad\text{and}\quad
  \boldsymbol{\gamma}^3 =
  \begin{pmatrix}
    -\id & \zero \\ \zero & \id
  \end{pmatrix}.
\end{equation}
As will be shown in Section~\ref{sec:low-rank-invariants}, there is a
two-parameter family of symplectic forms on the
spinor module $S$ which are invariant under the action of
$\bB_i$ and $\bJ_i$.  They are given by
\begin{equation}
  \omega(s_1,s_2) := \Re(s_1 (\alpha \ii + \beta \jj) \sbar_2),
\end{equation}
for $\alpha,\beta \in \RR$ not both zero.  We may normalise $\omega$
such that $\alpha^2 + \beta^2 = 1$, resulting in a circle of
symplectic structures.  Relative to the standard real basis
$(\ii,\jj,\kk,1)$ for $\HH$, the matrix $\Omega$ of $\omega$ is given
by $\Omega = i \sigma_2 \otimes (-\alpha \sigma_1 + \beta \sigma_3)$,
whose inverse is $\Omega^{-1} = - \Omega$, due to the chosen
normalisation.  Let us define endomorphisms $\gamma^\mu$ of $S$ such
that $(\gamma^\mu)^a{}_b = (\Omega^{-1})^{ac} \gamma^\mu_{cb}$.
Explicitly, they are given by
\begin{equation}
  \begin{aligned}[m]
    \gamma^0 &= i\sigma_2 \otimes (\alpha \sigma_1 - \beta\sigma_3)\\
    \gamma^1 &= \sigma_3 \otimes (\alpha \sigma_1 - \beta\sigma_3)
  \end{aligned}
  \quad\quad
  \begin{aligned}[m]
    \gamma^2 &= -\mathbb{1} \otimes (\alpha \sigma_3 + \beta\sigma_1)\\
    \gamma^3 &= \sigma_1 \otimes (\alpha \sigma_1 - \beta\sigma_3).
  \end{aligned}
\end{equation}
It then follows that these endomorphisms represent the Clifford
algebra $\Cl(1,3)$:
\begin{equation}
  \gamma^\mu \gamma^\nu + \gamma^\nu \gamma^\mu = -2 \eta^{\mu\nu}
  \mathbb{1}.
\end{equation}

\subsection{Classification of Aristotelian Lie Superalgebras}
\label{sec:class-arist-lie}

Table~\ref{tab:ALAs} lists the Aristotelian Lie algebras (with
three-dimensional space isotropy), classified in
\cite[App.~A]{Figueroa-OFarrill:2018ilb}.  In this section, we classify
the $\N=1$ supersymmetric extensions of the Aristotelian Lie algebras
(with $[\Q,\Q] \neq 0$).

\begin{table}[h!]
  \centering
  \caption{Aristotelian Lie Algebras and their Spacetimes}
  \label{tab:ALAs}
  \rowcolors{2}{blue!10}{white}
  \begin{tabular}{l|*{2}{>{$}l<{$}}|l}\toprule
    \multicolumn{1}{c|}{A\#} & \multicolumn{2}{c|}{Non-zero Lie brackets}& \multicolumn{1}{c}{Spacetime}\\\midrule
    \hypertarget{ALA1}{1} & & & \hyperlink{A21}{$\zS$} \\
    \hypertarget{ALA2}{2} & [\bH,\P] = \P & & \hyperlink{A22}{$\zTS$}\\
    \hypertarget{ALA3p}{3$_+$} & & [\P,\P] =  \J & \hyperlink{A23p}{$\RR\times S^3$}\\
    \hypertarget{ALA3m}{3$_-$} & & [\P,\P] = - \J & \hyperlink{A23m}{$\RR\times H^3$} \\
    \bottomrule
  \end{tabular}
\end{table}

\subsubsection{Lie Superalgebras Associated with Aristotelian Lie
  Algebra $\mathsf{A1}$}
\label{sec:lie-super-ALA1}

We start with the static Aristotelian Lie algebra
\hyperlink{ALA1}{$\mathsf{A1}$}, whose only non-zero brackets are
$[\J,\J] = \J$ and $[\J,\P] = \P$.  Any supersymmetric extension $\s$
has possible brackets
\begin{equation}
  [\sH,\sQ(s)] = \sQ(s\hh), \quad [\sP(\pi),\sQ(s)] = \sQ(\pi s \pp)
  \quad\text{and}\quad [\sQ(s), \sQ(s)] = c_0 |s|^2 \sH - \sJ(s \bc_1
  \sbar) - \sP(s \bc_3 \sbar),
\end{equation}
for some $\hh,\pp \in \HH$, $c_0 \in \RR$ and $\bc_1, \bc_3 \in \Im
\HH$, using the same notation as in
Section~\ref{sec:class-kinem-lie}.  We can re-use
Lemmas~\ref{lem:kmod} and \ref{lem:qqq}, by setting $\bb = 0$ and
$\bc_2 = 0$ and ignoring $\B$.  Doing so, we find that $\pp = 0$ and
that $\bc_1 = 2 c_0 \hh$.  The $[\bH,\Q,\Q]$ component of the super-Jacobi
identity gives $c_0 \Re\hh = 0$ (which already follows from
Lemma~\ref{lem:qqq}), $\bc_1 \hhbar + \hh \bc_1 = 0$ and $\bc_3 \hhbar
+ \hh \bc_3 = 0$.  The $[\P,\Q,\Q]$ component of the super-Jacobi identity
says that $[s\bc_1 \sbar, \pi] = 0$ for all $\pi \in \Im\HH$ and $s
\in \HH$, which says $\bc_1 =0$ and hence $c_0 \hh = 0$.  This gives
rise to two branches:
\begin{enumerate}
\item If $c_0 = 0$, then $\bc_3 \neq 0$ and the condition $\bc_3 \hhbar
  + \hh \bc_3 = 0$ is equivalent to $[\Im\hh,\bc_3] = - 2 \bc_3 \Re\hh$,
  which says $\Re\hh = 0$, and, therefore, that $\hh$ and $\bc_3$ are
  collinear.  We can change basis so that $\bc_3 = \kk$ and $\hh =
  \kk$ if non-zero.  This leaves two possible Lie superalgebras
  depending on whether or not $\hh = 0$:
  \begin{equation}
    [\sH, \sQ(s)] =
    \begin{cases}
      \sQ(s\kk)\\
      0
    \end{cases}
\quad\text{and}\quad [\sQ(s), \sQ(s)] =
    - \sP(s\kk\sbar).
  \end{equation}
\item If $c_0 \neq 0$, then $\hh = 0$ and $\bc_3$ is free.  We can set
  $c_0 = 1$ and, if non-zero, we can also set $\bc_3 = \kk$.  This
  gives two possible Lie superalgebras:
  \begin{equation}
    [\sQ(s), \sQ(s)] =
    \begin{cases}
      |s|^2 \sH \\
      |s|^2 \sH - \sP(s\kk\sbar).
    \end{cases}
  \end{equation}
\end{enumerate}

\subsubsection{Lie Superalgebras Associated with Aristotelian Lie
  Algebra $\mathsf{A2}$}
\label{sec:lie-super-ALA2}

Let us now consider the Aristotelian Lie algebra
\hyperlink{ALA2}{$\mathsf{A2}$}, with additional bracket
$[\bH,\P] = \P$.  Lemma~\ref{lem:kmod} says $\pp = 0$ and
Lemma~\ref{lem:qqq} says that $\bc_1 = 2 c_0 \hh$.  The
$[\bH,\Q,\Q]$ component of the super-Jacobi identity implies that
$c_0 \Re\hh = 0$ (which is redundant),
$\bc_1 \hhbar + \hh \bc_1 = 0$ and $\bc_3 \hhbar + \hh \bc_3 = \bc_3$,
whereas the $[\P,\Q,\Q]$ component results in
$[s\bc_1 \sbar, \pi] = 2 c_0 |s|^2\pi$ for all $\pi \in \Im\HH$ and
$s \in \HH$.  This can only be the case if $c_0 = 0$; therefore, 
$\bc_1=0$, which then forces $\bc_3 \neq 0$.  The equation
$\bc_3 \hhbar + \hh \bc_3 = \bc_3$ results in
$[\Im\hh, \bc_3] = (1-2\Re\hh) \bc_3$, which implies
$\Re\hh = \tfrac12$ and $\Im\hh$ collinear with $\bc_3$.  We can
change basis so that $\bc_3 = \kk$ and we end up with a one-parameter
family of Lie superalgebras with brackets
\begin{equation}
  [\sH,\sQ(s)] = \sQ(\tfrac12 s (1 + \lambda \kk)), \quad [\sQ(s),
  \sQ(s)] = - \sP(s\kk\sbar)
\end{equation}
for $\lambda \in \RR$, in addition to $[\sH,\sP(\pi)] = \sP(\pi)$.

\subsubsection{Lie Superalgebras Associated with Aristotelian Lie
  Algebras $\mathsf{A3}_{\pm}$}
\label{sec:lie-super-ALA3}

Finally, we consider the Aristotelian Lie algebras \hyperlink{ALA3p}{$\mathsf{A3}_\pm$}
with bracket $[\P,\P] = \pm \J$.  Lemma~\ref{lem:kmod} says
that $[\hh,\pp] =0$ and $\pp^2 = \pm \tfrac14$, whereas
Lemma~\ref{lem:qqq} says that $c_0 \hh = \tfrac12\bc_1 + \bc_3 \pp$.
The $[\bH,\Q,\Q]$ super-Jacobi says $c_0\Re\hh = 0$, $\bc_1 \hhbar + \hh \bc_1
= 0$ and $\bc_3 \hhbar + \hh \bc_3 = 0$, whereas the $[\P,\Q,\Q]$
super-Jacobi gives the following relations:
\begin{equation}
  c_0 \Re(\sbar\pi s \pp) = 0, \quad \pi s \pp \bc_3 \sbar - s \bc_3
  \ppbar \sbar \pi = \tfrac12 [\pi, s\bc_1
  \sbar]\quad\text{and}\quad
  \pi s \pp \bc_1 \sbar - s \bc_1
  \ppbar \sbar \pi = \pm \tfrac12 [\pi, s\bc_3 \sbar].
\end{equation}
We must distinguish two cases depending on the choice of signs.
\begin{enumerate}
\item Let's take the $+$ sign.  Then $\pp^2 = \tfrac14 \in \RR$.
  Without loss of generality, we can take $\pp = \tfrac12$ by changing
  the sign of $\P$ if necessary.  Then the $[\P,\Q,\Q]$ super-Jacobi
  equations say that $\bc_1 = \bc_3$ and hence $c_0 \hh = \bc_1$.  If
  $c_0 = 0$, then $\bc_1 = \bc_3 = 0$, hence we take $c_0 \neq 0$ and
  thus $\Re \hh = 0$.  We can change basis so that $c_0 = 1$ and hence
  $\hh = \bc_1 = \bc_3$.  If non-zero, we can take them all equal to
  $\kk$.  In summary, we have two possible Aristotelian Lie
  superalgebras extending \hyperlink{ALA3p}{$\mathsf{A3}_+$}, with
  brackets $[\sP(\pi), \sP(\pi')] = \tfrac12 \sJ([\pi,\pi'])$ and, in
  addition, either
  \begin{equation}
    [\sP(\pi), \sQ(s)] = \sQ(\tfrac12 \pi s), \quad\text{and}\quad
    [\sQ(s), \sQ(s) ] = |s|^2 \sH
  \end{equation}
  or
  \begin{equation}
    [\sH,\sQ(s)] = \sQ(s\kk), \quad [\sP(\pi), \sQ(s)] = \sQ(\tfrac12
    \pi s) \quad\text{and}\quad
    [\sQ(s), \sQ(s) ] = |s|^2 \sH - \sJ(s\kk\sbar) - \sP(s\kk\sbar).
  \end{equation}
  
\item Let us now take the $-$ sign.  Here $\pp^2 = -\frac14$, so that
  $\pp \in \Im\HH$ (and $\pp \neq 0$) and $\Im\hh$ is collinear
  with $\pp$.  The $[\bH,\Q,\Q]$ super-Jacobi equations force $\hh = 0$ and
  the $[\P,\Q,\Q]$ super-Jacobi equations force $c_0 = 0$ and
  $\bc_3 \pp = -\tfrac12 \bc_1$. 
  This means that $(\bc_1, 2\pp, \bc_3)$ is an oriented orthonormal
  frame for $\Im\HH$ and hence we can rotate them so that
  $(\bc_1, 2\pp, \bc_3) = (-\jj,\ii,\kk)$, for later uniformity.  This
  results in the Aristotelian Lie superalgebra extending
  \hyperlink{ALA3m}{$\mathsf{A3}_-$} by the following brackets in
  addition to $[\sP(\pi), \sP(\pi')] = -\tfrac12 \sJ([\pi,\pi'])$:
  \begin{equation}
    [\sP(\pi), \sQ(s)] = \sQ(\tfrac12 \pi s \ii) \quad\text{and}\quad
    [\sQ(s), \sQ(s)] = \sJ(s\jj\sbar) - \sP(s\kk\sbar).
  \end{equation}
\end{enumerate}
These results are summarised in Table~\ref{tab:alsa} below, together
with the possible compatible $\ZZ$-gradings.  This table also classifies the homogeneous Aristotelian superspaces for similar reasons to the non-supersymmetric case; see Section~\ref{sec:ks_kss} for a definite explanation. 

\begin{table}[h!]
  \centering
  \caption{Aristotelian Lie Superalgebras (with $[\Q,\Q]\neq 0$)}
  \label{tab:alsa}
  \setlength{\extrarowheight}{2pt}
  \rowcolors{2}{blue!10}{white}
  \begin{tabular}{l|l*{3}{|>{$}c<{$}}*{3}{|>{$}c<{$}}}\toprule
    \multicolumn{1}{c|}{S\#} & \multicolumn{1}{c|}{$\a$} & \multicolumn{1}{c|}{$\hh$}& \multicolumn{1}{c|}{$\pp$} & \multicolumn{1}{c|}{$[\sQ(s),\sQ(s)]$} & w_ {\bH} & w_{\P} & w_{\Q}\\
    \toprule
    \hypertarget{ALSA36}{36} &  \hyperlink{ALA1}{$\mathsf{A1}$} & \kk & & - \sP(s\kk\sbar) & 0 & 2q & q \\
    \hypertarget{ALSA37}{37} &  \hyperlink{ALA1}{$\mathsf{A1}$} & & & - \sP(s\kk\sbar) & 2n & 2q & q \\
    \hypertarget{ALSA38}{38} &  \hyperlink{ALA1}{$\mathsf{A1}$} & & & |s|^2 \sH & 2q & 2p & q \\
    \hypertarget{ALSA39}{39} &  \hyperlink{ALA1}{$\mathsf{A1}$} & & & |s|^2 \sH - \sP(s\kk\sbar) & 2q & 2q & q \\
    \hypertarget{ALSA40}{40$_{\lambda\in\RR}$} &  \hyperlink{ALA2}{$\mathsf{A2}$} & \tfrac12(1 + \lambda \kk) & & -\sP(s\kk\sbar) & 0 & 2q & q\\
    \hypertarget{ALSA41}{41} & \hyperlink{ALA3p}{$\mathsf{A3}_+$} & & \tfrac12 & |s|^2 \sH & 2q & 0 & q \\
    \hypertarget{ALSA42}{42} & \hyperlink{ALA3p}{$\mathsf{A3}_+$} & \kk & \tfrac12 & |s|^2 \sH - \sJ(s\kk\sbar) - \sP(s\kk\sbar) & - & - & - \\
    \hypertarget{ALSA43}{43} & \hyperlink{ALA3m}{$\mathsf{A3}_-$} & & \tfrac12\ii &  \sJ(s\jj\sbar) - \sP(s\kk\sbar) & - & - & - \\
    \bottomrule
  \end{tabular}
  \caption*{The first column is our identifier for $\s$, whereas the
    second column is the Aristotelian Lie algebra $\a = \s_{\bar 0}$
    in Table~\ref{tab:ALAs}.  The next three columns specify the
    brackets of $\s$ not of the form $[\J,-]$.  Supercharges $\sQ(s)$
    are parametrised by $s \in \HH$, whereas $\sJ(\omega)$ and
    $\sP(\pi)$ are parametrised by $\omega,\pi \in \Im\HH$. The
    brackets are given by $[\sH,\sQ(s)] = \sQ(s\hh)$ and
    $[\sP(\pi),\sQ(s)] = \sQ(\pi s \pp)$, for some $\hh,\pp\in\HH$.
    (The formalism is explained in Section~\ref{subsec:math_prelims_alg_klsa}.)
    The final three columns are compatible gradings of $\s$, with
    $n,p,q \in \ZZ$ and $q$ odd.}
\end{table}

\subsection{Central Extensions}
\label{sec:central-extensions}

In this section, we determine the possible central extensions of the
kinematical and Aristotelian Lie superalgebras.
\\ \\
We start with the kinematical Lie superalgebras. Let
$\s = \s_{\bar 0} \oplus \s_{\bar 1}$ be one of the Lie superalgebras
in Table~\ref{tab:klsa}. By a \textit{central extension} of $\s$, we
mean a short exact sequence of Lie superalgebras \begin{equation}
  \begin{tikzcd}
    0 \arrow[r] & \z \arrow[r] & \widehat\s \arrow[r] & \s \arrow[r] & 0,
  \end{tikzcd}
\end{equation}
where $\z$ is central in $\widehat\s$.  We may choose a vector space
splitting and view (as a vector space) $\widehat\s = \s \oplus \z$ and
the Lie bracket is given, for $(X,z), (Y,z') \in \s \oplus \z$, by
\begin{equation}
  [(X,z), (Y,z')]_{\widehat\s} = \left( [X,Y]_{\s}, \omega(X,Y) \right),
\end{equation}
where $\omega : \wedge^2\s \to \z$ is a cocycle.  (Here $\wedge$ is
taken in the super sense, so that it is symmetric on odd elements.)
Central extensions of $\s$ are classified up to isomorphism by the
Chevalley--Eilenberg cohomology group $H^2(\s)$, which, by
Hochschild--Serre, can be computed from the subcomplex relative to the
rotational subalgebra $\r \subset \s_{\bar 0}$.  Indeed, we have the
isomorphism \cite{MR0054581}
\begin{equation}
  H^2(\s) \cong H^2(s,\r).
\end{equation}
Let $W = \spn{\bH,\B,\P,\Q}$.  Then the cochains in $C^2(s,\r)$ are
$\r$-equivariant maps $\wedge^2W \to \RR$, or, equivalently,
$\r$-invariant vectors in $\wedge^2 W^*$.  This is a two-dimensional
real vector space which, in quaternionic language, is given for $x,y \in \RR$ by
\begin{equation}
  \omega(\sB(\beta), \sP(\pi)) = x \Re(\beta \pi) = - \omega(\sP(\pi),
  \sB(\beta)) \quad\text{and}\quad \omega(\sQ(s_1), \sQ(s_2)) = y
  \Re(s_1\sbar_2).
\end{equation}
The cocycle conditions (i.e., the Jacobi identities of the central
extension $\widehat\s$) have several components.  Letting $\sV$ stand
for either $\sB$ or $\sP$, the cocycle conditions are given by
\begin{equation}
  \begin{split}
    \omega([\sH,\sV(\alpha)], \sV(\beta)) + \omega(\sV(\alpha), [\sH,\sV(\beta)]) &= 0,\\
    \omega([\sV(\alpha),\sV(\beta)], \sV(\gamma)) + \text{cyclic} &= 0,\\
    \omega([\sH,\sQ(s)],\sQ(s)) &= 0,\\
    2 \omega([\sV(\alpha),\sQ(s)], \sQ(s)) + \omega([\sQ(s),\sQ(s)],\sV(\alpha)) &= 0.
  \end{split}
\end{equation}
The first two of the above equations only involve the even
generators and hence depend only on the underlying kinematical Lie
algebra, whereas the last two equations do depend on the precise
superalgebra we are dealing with.  In the case of Aristotelian Lie
superalgebras, there is no $\B$ and hence $\sV = \sP$ in the above
equations and, of course, the cocycle can only modify the $[\Q,\Q]$
bracket and hence the cocycle conditions are simply
\begin{equation}
  \omega([\sH,\sQ(s)],\sQ(s)) = 0 \quad\text{and}\quad
    \omega([\sP(\alpha),\sQ(s)], \sQ(s)) = 0.
\end{equation}
The calculations are routine, and we will not give any details, but
simply collect the results in Table~\ref{tab:central-ext_1}, where $\bZ$
is the basis for the one-dimensional central ideal $\z = \spn{\bZ}$, and
where we list only the brackets which are liable to change under
central extension.

\begin{table}[h!]
  \centering
  \caption{Central Extensions of Kinematical and Aristotelian Lie Superalgebras}
  \label{tab:central-ext_1}
  \setlength{\extrarowheight}{2pt}
  \rowcolors{2}{blue!10}{white}
  \begin{tabular}{l*{2}{|>{$}c<{$}}}\toprule
    \multicolumn{1}{c|}{S\#} & \multicolumn{1}{c|}{$[\sB(\beta),\sP(\pi)]$} & \multicolumn{1}{c}{$[\sQ(s),\sQ(s)]$} \\
    \toprule
    \hyperlink{KLSA1}{1} & & |s|^2 \bZ - \sP(s\kk\sbar) \\
    \hyperlink{KLSA4}{4} & -\Re(\beta\pi) \sZ & |s|^2 \sH \\
    \hyperlink{KLSA5}{5} & & |s|^2 \sZ - \sB(s\jj\sbar) - \sP(s\kk\sbar) \\
    \hyperlink{KLSA6}{6} & & |s|^2 \sZ -\sP(s\kk\sbar) \\
    \hyperlink{KLSA7}{7} & & |s|^2 \sZ -\sP(s\kk\sbar) \\
    \hyperlink{KLSA8}{8} & & |s|^2 \sZ -\sP(s\kk\sbar) \\
    \hyperlink{KLSA10}{10$_{\gamma=0,\lambda\in\RR}$} & & |s|^2 \sZ -\sB(s\kk\sbar) \\
    \hyperlink{KLSA11}{11$_{\chi=0}$} & & |s|^2 \sZ - \sB(s\ii\sbar) - \sP(s\kk\sbar) \\
    \hyperlink{KLSA13}{13} & -\Re(\beta\pi) (\sH + \sZ) & |s|^2 \sH \\
    \hyperlink{KLSA23}{23} & & |s|^2 \sZ - \sP(s\kk\sbar) \\
    \hyperlink{KLSA24}{24} & & |s|^2 \sZ - \sP(s\kk\sbar) \\
    \hyperlink{KLSA29}{29} & & |s|^2 \sZ -\sP(s\kk\sbar) \\
    \hyperlink{KLSA30}{30} & & |s|^2 \sZ -\sP(s\kk\sbar) \\
    \hyperlink{KLSA34}{34} & & |s|^2 \sZ -\sP(s\kk\sbar) \\
    \midrule
    \hyperlink{ALSA36}{36} & - & |s|^2 \sZ -\sP(s\kk\sbar) \\
    \hyperlink{ALSA37}{37} & - & |s|^2 \sZ -\sP(s\kk\sbar) \\
    \bottomrule
  \end{tabular}
  \caption*{The first column is our identifier for $\s$, whereas the
    other two columns are the possible central terms in the central
    extension $\widehat\s$.  Here $\beta,\pi \in\Im\HH$ and $s \in
    \HH$ are (some of) the parameters defining the Lie brackets in the
    quaternionic formalism explained in Section~\ref{subsec:math_prelims_alg_klsa}.}
\end{table}

\subsection{Automorphisms of Kinematical Lie Superalgebras}
\label{sec:autom-kinem-lie}

In the next section, we will classify the homogeneous superspaces
associated to the kinematical Lie superalgebras.  As we will explain
below, the first stage is to classify ``super Lie pairs'' up to
isomorphism.  To that end, it behoves us to determine the group of
automorphisms of the Lie superalgebras in Table~\ref{tab:klsa}, to
which we now turn.
\\ \\
Without loss of generality, we can restrict to automorphisms which are
the identity when restricted to $\r$: we call them $\r$-fixing
automorphisms.  Following from our discussion in
Section~\ref{sec:automorphisms}, these are parametrised by triples
\begin{equation}
  \left(A := \begin{pmatrix}a & b\\ c & d\end{pmatrix}, \mu, 
  \qq\right) \in \GL(2,\RR) \times \RR^\times \times \HH^\times
\end{equation}
subject to the condition that the associated linear transformations
leave the Lie brackets in $\s$ unchanged.
\\ \\
It is easy to read off from equation~\eqref{eq:autk-on-params} what
$(A, \mu,\qq)$ must satisfy for the $\r$-equivariant linear
transformation $\Phi : \s \to \s$ defined by them to be an
automorphism of $\s$, namely:
\begin{equation}\label{eq:aut-s}
  \begin{aligned}[m]
    \hh\qq &= \mu \qq \hh \\
    \bb\qq &= \qq (a \bb + c \pp) \\
    \pp\qq &=  \qq (b \bb + d \pp) \\
    \mu c_0 &= |\qq|^2 c_0
  \end{aligned}
  \quad\quad
  \begin{aligned}[m]
    \qq\bc_1\qqbar &= \bc_1 \\
    \qq\bc_2\qqbar &= a \bc_2 + b \bc_3\\
    \qq\bc_3 \qqbar &= c \bc_2 + d \bc_3.\\
  \end{aligned}
\end{equation}
It is then a straightforward -- albeit lengthy -- process to go
through each Lie superalgebra in Table~\ref{tab:klsa} and solve
equations \eqref{eq:aut-s} for $(A, \mu,\qq)$. In particular, $(A,\mu)
\in \Aut_\r(\k)$ and they are given in Table~\ref{tab:aut-kla}.  The
results of this section are summarised in Tables~\ref{tab:aut-klsa}
and \ref{tab:aut-klsa-extra}, which list the $\r$-fixing automorphisms
for the Lie superalgebras \hyperlink{KLSA1}{$\mathsf{S1}$}-
\hyperlink{KLSA15}{$\mathsf{S15}$}
and \hyperlink{KLSA16}{$\mathsf{S16}$}-\hyperlink{KLSA35}{$\mathsf{S35}$}, 
respectively, in Table~\ref{tab:klsa}.
\\ \\
The first six Lie superalgebras in Table~\ref{tab:klsa} are
supersymmetric extensions of the static kinematical Lie algebra for
which $(A,\mu)$ can be any element in $\GL(2,\RR) \times \RR^\times$.

\subsubsection{Automorphisms of Lie Superalgebra $\mathsf{S1}$}
\label{sec:autom-kinem-lie-1}

Here $\hh = \frac12\kk$, $\bb=\pp=0$, $c_0 = 0$, $\bc_1 = \bc_2 = 0$
and $\bc_3 = \kk$.  The invariance conditions~\eqref{eq:aut-s} give
\begin{equation}
  \mu \qq \kk = \kk\qq, \quad b \kk = 0 \quad\text{and}\quad d \kk
  = \qq \kk \qqbar.
\end{equation}
The second equation requires $b=0$.  The third equation says that the
real linear map $\alpha_\qq: \HH \to \HH$ defined by $\alpha_\qq (\xx)
= \qq \xx \qqbar$ preserves the $\kk$-axis in $\Im \HH$.

\begin{lemma}\label{lem:dk=qkqbar}
  Let $\qq\kk\qqbar = d \qq$ for some $d\in \RR$.  Then either $d =
  |\qq|^2$ and $\qq \in \spn{1,\kk}$ or $d = - |\qq|^2$ and $\qq \in
  \spn{\ii,\jj}$.
\end{lemma}

\begin{proof}
  Taking the quaternion norm of both sides of the equation $\qq \kk
  \qqbar = d \qq$, and using that $\qq \neq 0$, we see that $d =
  \pm |\qq|^2$ and hence right multiplying by $\qq$, the equation
  becomes $\pm \kk\qq = \qq \kk$.  If $\kk \qq = \qq\kk$,
  then $\qq \in \spn{1,\kk}$ and $d= |\qq|^2$, whereas if $-\kk\qq =
  \qq\kk$, then $\qq \in \spn{\ii,\jj}$ and $d = -|\qq|^2$.
\end{proof}
Taking the quaternion norm of the first equation shows that $\mu =
\pm 1$ and hence that $d = \mu |\qq|^2$.  In summary, we have that the
typical automorphism $(A,\mu,\qq)$ takes one of two possible forms:
\begin{equation}
  \begin{split}
    &A = \begin{pmatrix} a & \zero \\ c & |\qq|^2
    \end{pmatrix},\quad \mu= 1 \quad\text{and}\quad \qq = q_4 + q_3 \kk\\
   \text{or}\quad &A =  \begin{pmatrix}
      a & \zero \\ c & - |\qq|^2
    \end{pmatrix}, \quad \mu = - 1 \quad\text{and}\quad \qq = q_1 \ii
    + q_2 \jj.
  \end{split}
\end{equation}

\subsubsection{Automorphisms of Lie Superalgebra $\mathsf{S2}$}
\label{sec:autom-kinem-lie-2}

Here $\hh = \bb = \pp = 0$, $c_0=1$, $\bc_1 = 0$, $\bc_2 = \jj$ and
$\bc_3 = \kk$.  The invariance conditions~\eqref{eq:aut-s} give
\begin{equation}
  \mu = |\qq|^2, \quad a \jj + b \kk = \qq \jj \qqbar
  \quad\text{and}\quad c \jj + d \kk = \qq \kk \qqbar.
\end{equation}
The last two equations say that the real linear map $\alpha_\qq : \HH
\to \HH$ defined earlier preserves the $(\jj,\kk)$-plane in $\Im \HH$.

\begin{lemma}\label{lem:conj-plane}
  The map $\alpha_\qq : \HH \to \HH$ preserves the $(\jj,\kk)$-plane
  in $\Im \HH$ if and only if $\qq \in \spn{1,\ii} \cup
  \spn{\jj,\kk}$.
\end{lemma}

\begin{proof}
  Since $\qq \neq 0$, we can write it as $\qq = |\qq| \uu$, for some
  unique $\uu \in \Sp(1)$ and $\alpha_\qq = |\qq|^2 \alpha_\uu$.
  The map $\alpha_\qq$ preserves separately the real and imaginary
  subspaces of $\HH$, and $\alpha_\qq$ preserves the $(\jj,\kk)$-plane
  if and only if $\alpha_\uu$ does.  But, for $\uu \in \Sp(1)$,
  $\alpha_\uu$ acts on $\Im \HH$ by rotations and hence if $\alpha_\uu$
  preserves $(\jj,\kk)$-plane, it also preserves the perpendicular
  line, which, in this case, is the $\ii$-axis.  Additionally, since it must preserve length,
  $\alpha_\uu (\ii) = \pm \ii$.  It follows that $\alpha_\qq(\ii) = \pm |\qq|^2
  \ii$, so that $\alpha_\qq$ too preserves the $\ii$-axis.  By an
  argument similar to that of Lemma~\ref{lem:dk=qkqbar} it follows
  that $\qq$ belongs either to the complex line in $\HH$
  generated by $\ii$ or to its perpendicular complement.
\end{proof}
From the lemma, we have two cases to consider: $\qq = q_4 + q_1 \ii$ or
$\qq = q_2 \jj + q_3 \kk$.  In each case, we can use the last two
equations to solve for $a,b,c,d$ in terms of the components of $\qq$.
Summarising, we have that the typical automorphism $(A,\mu,\qq)$ takes
one of two possible forms:
\begin{equation}
  \begin{split}
    &A = \begin{pmatrix} q_4^2-q_1^2 & 2 q_1 q_4 \\ -2 q_1 q_4 & q_4^2 - q_1^2
    \end{pmatrix},\quad \mu= q_1^2+q_4^2 \quad\text{and}\quad \qq = q_4 + q_1 \ii\\
   \text{or}\quad &A =  \begin{pmatrix}
     q_2^2-q_3^2 & 2 q_2 q_3 \\ 2 q_2 q_3 & q_3^2 - q_2^2
    \end{pmatrix}, \quad \mu = q_2^2+q_3^2 \quad\text{and}\quad \qq = q_2 \jj
    + q_3 \kk.
  \end{split}
\end{equation}

\subsubsection{Automorphisms of Lie Superalgebra $\mathsf{S3}$}
\label{sec:autom-kinem-lie-3}

Here $\hh=\bb=\pp=0$, $c_0=1$, $\bc_1 = \bc_2 = 0$ and $\bc_3 = \kk$.
The invariance conditions~\eqref{eq:aut-s} give
\begin{equation}
  \mu = |\qq|^2, \quad b\kk = 0 \quad\text{and}\quad d\kk = \qq \kk \qqbar.
\end{equation}
This is very similar to the case of the Lie superalgebra \hyperlink{KLSA1}{$\mathsf{S1}$}; in
particular, Lemma~\ref{lem:dk=qkqbar} applies.  The typical
automorphism $(A,\mu,\qq)$ takes one of two possible forms:
\begin{equation}
  \begin{split}
    &A = \begin{pmatrix} a & \zero \\ c & |\qq|^2
    \end{pmatrix},\quad \mu= |\qq|^2 \quad\text{and}\quad \qq = q_4 + q_3 \kk\\
   \text{or}\quad &A =  \begin{pmatrix}
      a & \zero \\ c & - |\qq|^2
    \end{pmatrix}, \quad \mu = |\qq|^2 \quad\text{and}\quad \qq = q_1 \ii
    + q_2 \jj.
  \end{split}
\end{equation}

\subsubsection{Automorphisms of Lie Superalgebra $\mathsf{S4}$}
\label{sec:autom-kinem-lie-4}

Here, $\hh=\bb=\pp=0$, $c_0=1$ and $\bc_1 = \bc_2 = \bc_3 = 0$.  The
only condition is $\mu = |\qq|^2$.  Hence the typical automorphism
$(A,\mu,\qq)$ takes the form
\begin{equation}
  A =
  \begin{pmatrix}
    a & b \\ c & d
  \end{pmatrix},
  \quad \mu = |\qq|^2 \quad\text{and}\quad \qq \in \HH^\times.
\end{equation}

\subsubsection{Automorphisms of Lie Superalgebra $\mathsf{S5}$}
\label{sec:autom-kinem-lie-5}

Here, $\hh= \bb = \pp = 0$, $c_0 = 0$, $\bc_1 = 0$, $\bc_2 = \jj$ and
$\bc_3 = \kk$.  The invariance conditions~\eqref{eq:aut-s} are here as for the
Lie superalgebra \hyperlink{KLSA2}{$\mathsf{S2}$}, except that $\mu$ is
unconstrained.  In other words, the typical automorphism $(A,\mu,\qq)$
takes one of two possible forms:
\begin{equation}
  \begin{split}
    &A = \begin{pmatrix} q_4^2-q_1^2 & 2 q_1 q_4 \\ -2 q_1 q_4 & q_4^2 - q_1^2
    \end{pmatrix},\quad \mu \quad\text{and}\quad \qq = q_4 + q_1 \ii\\
   \text{or}\quad &A =  \begin{pmatrix}
     q_2^2-q_3^2 & 2 q_2 q_3 \\ 2 q_2 q_3 & q_3^2 - q_2^2
    \end{pmatrix}, \quad \mu \quad\text{and}\quad \qq = q_2 \jj
    + q_3 \kk.
  \end{split}
\end{equation}

\subsubsection{Automorphisms of Lie Superalgebra $\mathsf{S6}$}
\label{sec:autom-kinem-lie-6}

Here, $\hh=\bb=\pp=0$, $c_0= 0$, $\bc_1 = \bc_2 = 0$ and $\bc_3= \kk$.
This is similar to Lie superalgebra \hyperlink{KLSA3}{$\mathsf{S3}$}, except
that $\mu$ remains unconstrained.  In summary, the typical
automorphisms $(A,\mu,\qq)$ takes one of two possible forms:
\begin{equation}
  \begin{split}
    &A = \begin{pmatrix} a & \zero \\ c & |\qq|^2
    \end{pmatrix},\quad \mu \quad\text{and}\quad \qq = q_4 + q_3 \kk\\
   \text{or}\quad &A =  \begin{pmatrix}
      a & \zero \\ c & - |\qq|^2
    \end{pmatrix}, \quad \mu \quad\text{and}\quad \qq = q_1 \ii + q_2 \jj.
  \end{split}
\end{equation}
The next two Lie superalgebras (\hyperlink{KLSA7}{$\mathsf{S7}$} and
\hyperlink{KLSA8}{$\mathsf{S8}$}) are supersymmetric extensions of the
Galilean Lie algebra, where $(A,\mu)$ take the form
\begin{equation}
  A =
  \begin{pmatrix}
    a & b \\ c & d 
  \end{pmatrix} \quad\text{and}\quad \mu = \frac{d}{a}.
\end{equation}

\subsubsection{Automorphisms of Lie Superalgebra $\mathsf{S7}$}
\label{sec:autom-kinem-lie-7}

Here, $\hh=\kk$, $\bb = \pp = 0$, $c_0 = 0$, $\bc_1 = \bc_2 = 0$ and
$\bc_3= \kk$.  The invariance conditions~\eqref{eq:aut-s} are
\begin{equation}
  d \qq\kk = a \kk \qq \quad\text{and}\quad d\kk = \qq \kk \qqbar.
\end{equation}
Multiplying the second equation on the right by $\qq$, using the
first equation and the fact that $\qq \neq 0$, results in $a =
d^2/|\qq|^2$, so that $a > 0$.  Taking the quaternion norm of the
first equation shows that $a = |d|$, so that $a = |\qq|^2$.  The first
equation now follows from the second, and that is solved by
Lemma~\ref{lem:dk=qkqbar}.
\\ \\
In summary the typical automorphism $(A,\mu,\qq)$ takes one of two
possible forms:
\begin{equation}
  \begin{split}
    &A = \begin{pmatrix} |\qq|^2 & \zero \\ c & |\qq|^2
    \end{pmatrix},\quad \mu = 1 \quad\text{and}\quad \qq = q_4 + q_3 \kk\\
   \text{or}\quad &A =  \begin{pmatrix}
      |\qq|^2 & \zero \\ c & - |\qq|^2
    \end{pmatrix}, \quad \mu=-1 \quad\text{and}\quad \qq = q_1 \ii + q_2 \jj.
  \end{split}
\end{equation}

\subsubsection{Automorphisms of Lie Superalgebra $\mathsf{S8}$}
\label{sec:autom-kinem-lie-8}

Here, $\hh=\bb=\pp=0$, $c_0=0$, $\bc_1=\bc_2 =0$ and $\bc_3 = \kk$.
The invariance conditions~\eqref{eq:aut-s} reduce to just $d\kk =
\qq\kk\qqbar$, which we solve by Lemma~\ref{lem:dk=qkqbar}.  In
summary, the typical automorphism $(A,\mu,\qq)$ is the same here as in the
previous Lie superalgebra, except that $a$ is unconstrained (but
non-zero).  It can thus take one of two possible forms:
\begin{equation}
  \begin{split}
    &A = \begin{pmatrix} a & \zero \\ c & |\qq|^2
    \end{pmatrix},\quad \mu =\frac{|\qq|^2}{a} \quad\text{and}\quad \qq = q_4 + q_3 \kk\\
   \text{or}\quad &A = \begin{pmatrix}
      a & \zero \\ c & - |\qq|^2
    \end{pmatrix}, \quad \mu=-\frac{|\qq|^2}{a} \quad\text{and}\quad \qq = q_1 \ii + q_2 \jj.
  \end{split}
\end{equation}
The next two classes of Lie superalgebras are associated with the
one-parameter family of kinematical Lie algebras
\hyperlink{KLA3}{$\mathsf{K3}_\gamma$}, whose typical automorphisms
$(A,\mu)$ depend on the value of $\gamma \in [-1,1]$.  In the interior
of the interval, it takes the form
\begin{equation}
  A =
  \begin{pmatrix}
    a & \zero \\ \zero & d
  \end{pmatrix} \quad\text{and}\quad \mu = 1.
\end{equation}
At the boundaries, this is enhanced: at $\gamma = -1$, one can also
have automorphisms of the form
\begin{equation}
  A =
  \begin{pmatrix}
    \zero & b \\ c & \zero
  \end{pmatrix} \quad\text{and}\quad \mu = -1,
\end{equation}
whereas, at $\gamma = 1$, the typical automorphism takes the form
\begin{equation}
  A =
  \begin{pmatrix}
    a & b \\ c & d
  \end{pmatrix} \quad\text{and}\quad \mu = 1.
\end{equation}

\subsubsection{Automorphisms of Lie Superalgebra $\mathsf{S9}_{\gamma,\lambda}$}
\label{sec:autom-kinem-lie-9}

Here, $\hh= \tfrac12 (1 + \lambda \kk)$, $\bb=\pp=0$, $c_0=0$, $\bc_1 =
\bc_2 = 0$ and $\bc_3 = \kk$.  The invariance
conditions~\eqref{eq:aut-s} reduce to $b=0$ and, in addition, 
\begin{equation}
  \mu \qq (1 + \lambda \kk) = (1+\lambda \kk) \qq
  \quad\text{and}\quad d\kk = \qq\kk\qqbar.
\end{equation}
Taking the norm of the first equation, we find that $\mu = \pm 1$.  If
$\mu =1$, then $\lambda[\kk,\qq] =0$ so that either $\lambda\neq 0$,
in which case $\qq \in \spn{1,\kk}$, or $\lambda = 0$, and $\qq$ is not
constrained by this equation.  The second equation is dealt with by
Lemma~\ref{lem:dk=qkqbar}, which implies that $d =
\pm |\qq|^2$, and since $\qq \neq 0$, $d \neq 0$.  This precludes the
case $\mu = -1$ by inspecting the possible automorphisms $(A,\mu)$  of
$\k$.  In summary, for generic $\gamma$ and $\lambda$, the typical
automorphism $(A,\mu,\qq)$ takes the form
\begin{equation}
  A =
  \begin{pmatrix}
    a & \zero \\ \zero & |\qq|^2
  \end{pmatrix}, \quad \mu = 1 \quad\text{and}\quad \qq = q_4 + q_3
  \kk,
\end{equation}
which is enhanced for $\gamma = 1$ (but $\lambda$ still generic) to
\begin{equation}
  A =
  \begin{pmatrix}
    a & \zero \\ c & |\qq|^2
  \end{pmatrix}, \quad \mu = 1 \quad\text{and}\quad \qq = q_4 + q_3
  \kk.
\end{equation}
If $\lambda = 0$, then the automorphisms are enhanced by the addition
of $(A,\mu,\qq)$ of the form
\begin{equation}
  A =
  \begin{pmatrix}
    a & \zero \\ \zero & -|\qq|^2
  \end{pmatrix}, \quad \mu = 1 \quad\text{and}\quad \qq = q_1\ii + q_2 \jj,
\end{equation}
for generic $\gamma$ or, for $\gamma = 1$ only,  also
\begin{equation}
  A =
  \begin{pmatrix}
    a & \zero \\ c & -|\qq|^2
  \end{pmatrix}, \quad \mu = 1 \quad\text{and}\quad \qq = q_1\ii +
  q_2 \jj.
\end{equation}

\subsubsection{Automorphisms of Lie Superalgebra $\mathsf{S10}_{\gamma,\lambda}$}
\label{sec:autom-kinem-lie-10}

Here, $\hh= \tfrac12(\gamma + \lambda\kk)$, $\bb=\pp=0$, $c_0=0$,
$\bc_1=\bc_3=0$ and $\bc_2 = \kk$.  The invariance
conditions~\eqref{eq:aut-s} imply that $c=0$ and also
\begin{equation}
  \mu \qq (\gamma + \lambda\kk) = (\gamma + \lambda\kk) \qq
  \quad\text{and}\quad a\kk = q\kk\qqbar.
\end{equation}
It is very similar to the previous Lie superalgebra, except here
$\gamma \neq 1$.  Lemma~\ref{lem:dk=qkqbar} now says that either $a
= |\qq|^2$ and $\qq = q_4 + q_3 \kk$ or $a = - |\qq|^2$ and $\qq = q_1
\ii + q_2 \jj$. In particular, since $\qq \neq 0$, $a\neq 0$.  From
the expressions for the automorphisms $(A,\mu)$ of $\k$, we see that
$\mu = 1$.  This means that the first equation says $\qq$ commutes
with $\gamma + \lambda \kk$.   If $\lambda = 0$, this condition is
vacuous, but if $\lambda \neq 0$, then it forces $\qq = q_4 + q_3\kk$
and hence $a = |\qq|^2$.
\\ \\
In summary, for $\lambda \neq 0$, we have that
$(A,\mu,\qq)$ takes the form
\begin{equation}
    A =
    \begin{pmatrix}
      |q|^2 & \zero \\ \zero & d \end{pmatrix}, \quad \mu = 1
    \quad\text{and}\quad \qq = q_4 +  q_3\kk,
\end{equation}
whereas, if $\lambda = 0$, it can also take the form
\begin{equation}
    A =
    \begin{pmatrix}
      -|q|^2 & \zero \\ \zero & d \end{pmatrix}, \quad \mu = 1
    \quad\text{and}\quad \qq = q_1\ii +  q_2\jj.
\end{equation}
The next Lie superalgebra is based on the kinematical Lie algebra
\hyperlink{KLA4}{$\mathsf{K4}_\chi$}, whose automorphisms $(A,\mu)$ take the form
\begin{equation}
  A =
  \begin{pmatrix}
    a & b \\ -b & a
  \end{pmatrix} \quad\text{and}\quad \mu = 1
\end{equation}
for generic $\chi$, whereas, if $\chi = 0$, then they can also be of
the form
\begin{equation}
  A =
  \begin{pmatrix}
    a & b \\ b & -a
  \end{pmatrix} \quad\text{and}\quad \mu = -1.
\end{equation}

\subsubsection{Automorphisms of Lie Superalgebra $\mathsf{S11}_\chi$}
\label{sec:autom-kinem-lie-11}

Here, $\hh = \tfrac12 (\chi + \jj)$, $\bb=\pp=0$, $c_0=0$, $\bc_1 = 0$,
$\bc_2 = \ii$ and $\bc_3 = \kk$.  The invariance
conditions~\eqref{eq:aut-s} reduce to
\begin{equation}
  \mu \qq (\chi + \jj) = (\chi + \jj)\qq, \quad \qq\ii\qqbar = a\ii +
  b\kk \quad\text{and}\quad \qq\kk\qqbar = c\ii + d\kk.
\end{equation}
The last two equations are solved via Lemma~\ref{lem:conj-plane}:
either $\qq = q_4 + q_2\jj$ or else $\qq = q_1\ii + q_3\kk$.  This
latter case can only happen when $\chi = 0$.  Substituting these
possible expressions for $\qq$ in the last two equations, we determine
the entries of the matrix $A$.
\\ \\
In summary, $(A,\mu,\qq)$ takes the form
\begin{equation}
  A =
  \begin{pmatrix}
    q_4^2 - q_2^2 & - 2 q_2 q_4 \\ 2 q_2 q_4 & q_4^2 - q_2^2
  \end{pmatrix}, \quad \mu = 1 \quad\text{and}\quad \qq = q_4 + q_2\jj,
\end{equation}
and, (only) if $\chi = 0$, it can also take the form
\begin{equation}
  A =
  \begin{pmatrix}
    q_1^2 - q_3^2 & 2 q_1 q_3 \\ 2 q_1 q_3 & q_3^2 - q_1^2
  \end{pmatrix}, \quad \mu = -1 \quad\text{and}\quad \qq = q_1\ii + q_3\kk.
\end{equation}
The next Lie superalgebra is the supersymmetric extension of the
kinematical Lie algebra \hyperlink{KLA5}{$\mathsf{K5}$}, whose automorphisms
$(A,\mu)$ are of the form
\begin{equation}
  A =
  \begin{pmatrix}
    a & \zero \\ c & a
  \end{pmatrix} \quad\text{and}\quad \mu = 1.
\end{equation}

\subsubsection{Automorphisms of Lie Superalgebra $\mathsf{S12}_\lambda$}
\label{sec:autom-kinem-lie-12}

Here, $\hh = \tfrac12 (1 + \lambda \kk)$, $\bb=\pp=0$, $c_0 =0$, $\bc_1
= \bc_2 = 0$ and $\bc_3 = \kk$.  The invariance
conditions~\eqref{eq:aut-s} reduce to
\begin{equation}
  \qq\hh = \hh\qq \quad\text{and}\quad a\kk  = \qq \kk \qqbar.
\end{equation}
The second equation is solved via Lemma~\ref{lem:dk=qkqbar}, which
says that either $a = |\qq|^2$ and $\qq = q_4 + q_3 \kk$ or $a =
- |\qq|^2$ and $\qq = q_1\ii + q_2\jj$.  The first equation is
identically satisfied  if $\lambda =0$, but otherwise it forces
$\qq = q_4 + q_3 \kk$ and hence $a= |\qq|^2$.  In summary, for general
$\lambda$, an automorphism $(A,\mu,\qq)$ takes the form
\begin{equation}
  A =
  \begin{pmatrix}
    |\qq|^2 & \zero \\ c & |\qq|^2
  \end{pmatrix}, \quad \mu = 1 \quad\text{and}\quad \qq = q_4 + q_3\kk,
\end{equation}
whereas if $\lambda = 0$, it can also take the form
\begin{equation}
  A =
  \begin{pmatrix}
    -|\qq|^2 & \zero \\ c & -|\qq|^2
  \end{pmatrix}, \quad \mu = 1 \quad\text{and}\quad \qq = q_1\ii + q_2\jj.
\end{equation}
The next Lie superalgebra is the supersymmetric extension of the
Carroll algebra, whose automorphisms $(A,\mu)$ take the form
\begin{equation}
  A =
  \begin{pmatrix}
    a & b \\ c & d
  \end{pmatrix} \quad\text{and}\quad \mu = ad - bc.
\end{equation}

\subsubsection{Automorphisms of Lie Superalgebra $\mathsf{S13}$}
\label{sec:autom-kinem-lie-13}

Here $\hh = \bb = \pp = 0$, $c_0 =1$ and $\bc_1 = \bc_2 = \bc_3 = 0$.
The invariance conditions \eqref{eq:aut-s} reduce to a single
condition: $ad - bc = |\qq|^2$.  The automorphisms $(A,\mu,\qq)$ are of
the form
\begin{equation}
  A  =   \begin{pmatrix}
    a & b \\ c & d
  \end{pmatrix}, \quad \mu = ad - bc = |\qq|^2 \quad\text{and}\quad
  \qq \in\HH^\times.
\end{equation}
The next Lie superalgebra is the Poincaré superalgebra whose
($\r$-fixing) automorphisms $(A,\mu)$ can take one of two possible
forms:
\begin{equation}
  \begin{split}
    &A = \begin{pmatrix} 1 & \zero \\ c & d
    \end{pmatrix} \quad\text{and}\quad \mu = d\\
   \text{or}\quad &A = \begin{pmatrix}
      -1 & \zero \\ c & d
    \end{pmatrix}\quad\text{and}\quad \mu = -d.
  \end{split}
\end{equation}

\subsubsection{Automorphisms of Lie Superalgebra $\mathsf{S14}$}
\label{sec:autom-kinem-lie-14}

Here $\hh=\pp=0$, $\bb= \tfrac12 \kk$, $c_0=1$, $\bc_1 = \bc_2 = 0$
and $\bc_3 = \kk$.  The invariance conditions~\eqref{eq:aut-s}
translate into
\begin{equation}
  \pm \qq \kk = \kk\qq, \quad d = \pm |\qq|^2  \quad\text{and}\quad
  d\kk = \qq\kk\qqbar,
\end{equation}
where the signs are correlated and the last equation follows from the first two.
\\ \\
Choosing the plus sign, $\qq\kk = \kk\qq$, so that $\qq =
q_4 + q_3 \kk$ and $d = |\qq|^2$, whereas choosing the minus sign,
$\qq\kk = - \kk\qq$, so that $\qq = q_1 \ii + q_2 \jj$ and $d =
-|\qq|^2$.
\\ \\
In summary, automorphisms $(A,\mu,\qq)$ of the Poincaré superalgebra
take the form
\begin{equation}
  \begin{split}
    &A = \begin{pmatrix} 1 & \zero \\ c & |\qq|^2
    \end{pmatrix}, \quad \mu = |\qq|^2 \quad\text{and}\quad
    \qq = q_4 + q_3 \kk\\
    \text{or}\quad
    &A = \begin{pmatrix} -1 & \zero \\ c & -|\qq|^2
    \end{pmatrix}, \quad \mu = |\qq|^2 \quad\text{and}\quad
    \qq = q_1\ii + q_2 \jj.
  \end{split}
\end{equation}
The next Lie superalgebra is the AdS superalgebra, whose ($\r$-fixing)
automorphisms $(A,\mu)$ are of the form
\begin{equation}
  A =
  \begin{pmatrix}
    a & b \\ \mp b & \pm a
  \end{pmatrix} \quad\text{and}\quad \mu = \pm 1,
\end{equation}
where $a^2 + b^2 = 1$.

\subsubsection{Automorphisms of Lie Superalgebra $\mathsf{S15}$}
\label{sec:autom-kinem-lie-15}

Here $\hh=\tfrac12\kk$, $\bb=\tfrac12\ii$, $\pp=\tfrac12\jj$, $c_0 =
1$, $\bc_1 = \kk$, $\bc_2 = \jj$ and $\bc_3= \ii$.  The invariance
conditions~\eqref{eq:aut-s} include $\mu = |\qq|^2$, which forces $\mu
= 1$.  Taking this into account, another of the invariance
conditions~\eqref{eq:aut-s} is $\qq\kk = \kk \qq$, which together with
$|\qq|=1$, forces $\qq = e^{\theta\kk}$.  The remaining invariance
conditions are
\begin{equation}
  a \qq \ii - b \qq \jj = \ii\qq, \quad
  b \qq \ii + a \qq \jj = \jj\qq, \quad
  a\jj\qq + b \ii\qq = \qq\jj \quad\text{and}\quad
  a\ii\qq - b\jj\qq = \qq\ii.
\end{equation}
Given the expression for $\qq$, these are solved by $a = \cos2\theta$
and $b = \sin2\theta$.  In summary, the ($\r$-fixing) automorphisms
$(A,\mu,\qq)$ of the AdS superalgebra are of the form
\begin{equation}
  A =
  \begin{pmatrix}
    \cos2\theta & \sin2\theta \\ -\sin2\theta & \cos2\theta
  \end{pmatrix}, \quad \mu = 1 \quad\text{and}\quad \qq = e^{\theta\kk}.
\end{equation}
The next three Lie superalgebras in Table~\ref{tab:klsa} are
supersymmetric extensions of the kinematical Lie algebra
\hyperlink{KLA12}{$\mathsf{K12}$} in Table~\ref{tab:kla}, whose $\r$-fixing
automorphisms $(A,\mu)$ take the following form:
\begin{equation}
  A =
  \begin{pmatrix}
    1 & \zero \\ \zero & \pm 1
  \end{pmatrix} \quad\text{and}\quad \mu \in \RR^\times.
\end{equation}

\subsubsection{Automorphisms of Lie Superalgebra $\mathsf{S16}$}
\label{sec:autom-kinem-lie-16}

Here $\hh=\bb=0$, $\pp= \tfrac12\jj$, $c_0= 0$, $\bc_1 = -\ii$, $\bc_2
= \ii$ and $\bc_3 = -\kk$.  The invariance conditions~\eqref{eq:aut-s}
reduce to
\begin{equation}
  \pm \qq \jj = \jj\qq, \quad \pm \qq\kk = \kk \qq
  \quad\text{and}\quad \qq\ii\qqbar = \ii.
\end{equation}
It follows from the last equation that $|\qq|=1$ and hence that
$\qq\ii = \ii\qq$.  Depending on the (correlated) signs of the first
two equations, we find that, for the plus sign, $\qq$ commutes with
$\ii$, $\jj$ and $\kk$ and hence $\qq \in \RR$, but since $|\qq|=1$,
we must have $\qq = \pm 1$.  For the minus sign, we find that $\qq$
commutes with $\ii$ but anticommutes with $\jj$ and $\kk$, so that
$\qq = \pm \ii$, after taking into account that $|\qq|=1$.  In
summary, the automorphisms $(A,\mu,\qq)$ of this Lie
superalgebra take one of two possible forms:
\begin{equation}
  \begin{split}
    &A = \begin{pmatrix} 1 & \zero \\ \zero & 1
    \end{pmatrix},\quad \mu \in \RR^\times \quad\text{and}\quad \qq
    = \pm 1\\
   \text{or}\quad &A = \begin{pmatrix}
      1 & \zero \\ \zero & -1
    \end{pmatrix},\quad \mu \in \RR^\times \quad\text{and}\quad \qq
    = \pm \ii.
  \end{split}
\end{equation}

\subsubsection{Automorphisms of Lie Superalgebra $\mathsf{S17}$}
\label{sec:autom-kinem-lie-17}

Here $\hh = \pp = 0$, $\bb = \tfrac12$, $c_0 =1$ and $\bc_1 = \bc_2 =
\bc_3 = 0$.  There is only one invariance condition: namely, $\mu
= |\qq|^2$, and hence the automorphisms $(A,\mu,\qq)$ take the form
\begin{equation}
  A =
  \begin{pmatrix}
    1 & \zero \\ \zero & \pm 1
  \end{pmatrix}, \quad \mu = |\qq|^2 \quad\text{and}\quad \qq \in \HH^\times.
\end{equation}

\subsubsection{Automorphisms of Lie Superalgebra $\mathsf{S18}$}
\label{sec:autom-kinem-lie-18}

Here $\hh=\tfrac12\kk$, $\bb = \tfrac12$, $\pp = 0$, $c_0=1$, $\bc_1 =
\bc_3 = 0$ and $\bc_2=\kk$.  The invariance
conditions~\eqref{eq:aut-s} reduce to
\begin{equation}
  \mu \qq\kk = \kk\qq, \quad \mu = |\qq|^2 \quad\text{and}\quad \kk
  = \qq\kk\qqbar.
\end{equation}
From the first equation we see that $\mu = \pm 1$, but from the
second it must be positive, so $\mu = 1$, which says implies that
$|\qq| = 1$ and hence that $\qq$ commutes with $\kk$.  In summary, the
typical automorphism $(A,\mu,\qq)$ takes the form
\begin{equation}
  A =
  \begin{pmatrix}
    1 & \zero \\ \zero & \pm 1
  \end{pmatrix}, \quad \mu = 1 \quad\text{and}\quad \qq = e^{\theta \kk}.
\end{equation}
The next four Lie superalgebras in Table~\ref{tab:klsa} are
supersymmetric extensions of the kinematical Lie algebra \hyperlink{KLA13}{$\mathsf{K13}$} in
Table~\ref{tab:kla}, whose typical $\r$-fixing automorphisms $(A,\mu)$
take the form
\begin{equation}
  A =
  \begin{pmatrix}
    1 & \zero \\ \zero & \pm 1
  \end{pmatrix} \quad\text{and}\quad \mu \in \RR^\times.
\end{equation}

\subsubsection{Automorphisms of Lie Superalgebra $\mathsf{S19}$}
\label{sec:autom-kinem-lie-19}

Here $\hh=\kk$, $\bb =0$, $\pp = \tfrac12$, $c_0=1$, $\bc_1 = \bc_3 =
\kk$ and $\bc_2= -\kk$.  The invariance conditions~\eqref{eq:aut-s}
are given by
\begin{equation}
  \mu \qq \kk = \kk\qq, \quad \mu = |\qq|^2 \quad\text{and}\quad
  d\qq = \qq.
\end{equation}
The last equation says that $d=1$, whereas the first says that $\mu =
\pm 1$, but from the second equation it is positive and thus $\mu =
1$.  This also means $|\qq|=1$ and that $\qq\kk=\kk\qq$.  In summary,
the typical automorphism $(A,\mu,\qq)$ of $\s$ takes the form
\begin{equation}
  A =
  \begin{pmatrix}
    1 & \zero \\ \zero & 1
  \end{pmatrix}, \quad \mu = 1 \quad\text{and}\quad \qq = e^{\theta\kk}.
\end{equation}

\subsubsection{Automorphisms of Lie Superalgebra $\mathsf{S20}$}
\label{sec:autom-kinem-lie-20}

Here $\hh=\bb =0$, $\pp = \tfrac12$, $c_0=1$, $\bc_1 = \bc_2 = \bc_3=
0$.  The invariance conditions~\eqref{eq:aut-s} are given by
\begin{equation}
  d\qq = \qq \quad\text{and}\quad \mu = |\qq|^2.
\end{equation}
The first equation simply sets $d= 1$ and, in summary, the typical
automorphism of $\s$ is takes the form
\begin{equation}
  A =
  \begin{pmatrix}
    1 & \zero \\ \zero & 1
  \end{pmatrix}, \quad \mu = |\qq|^2 \quad\text{and}\quad \qq \in \HH^\times.
\end{equation}

\subsubsection{Automorphisms of Lie Superalgebra $\mathsf{S21}$}
\label{sec:autom-kinem-lie-22}

Here $\hh=\pp=0$, $\bb=\tfrac12$, $c_0=1$ and $\bc_1 = \bc_2 = \bc_3 =
0$.  The only invariance condition is $\mu = |\qq|^2$, so that the
typical automorphism $(A,\mu,\qq)$ takes the form
\begin{equation}
  A =
  \begin{pmatrix}
    1 & \zero \\ \zero & \pm 1
  \end{pmatrix}, \quad \mu = |\qq|^2 \quad\text{and}\quad \qq \in \HH^\times.
\end{equation}

\subsubsection{Automorphisms of Lie Superalgebra $\mathsf{S22}$}
\label{sec:autom-kinem-lie-21}

Here $\hh = \tfrac12\kk$, $\bb=\tfrac12$, $\pp=0$, $c_0=1$,
$\bc_1=\bc_3=0$ and $\bc_2=\kk$.  The invariance
conditions~\eqref{eq:aut-s} reduce to
\begin{equation}
  \mu \qq \kk = \kk\qq \quad\text{and}\quad \mu = |\qq|^2.
\end{equation}
The first equation says that $\mu = \pm 1$, but the second equation
says it is positive, so that $\mu = 1$ and $|\qq|=1$.  Furthermore,
$\qq$ commutes with $\kk$, so that $\qq = e^{\theta\kk}$.  In summary,
the typical automorphism $(A,\mu,\qq)$ takes the form
\begin{equation}
  A =
  \begin{pmatrix}
    1 & \zero \\ \zero & \pm 1
  \end{pmatrix}, \quad \mu = 1 \quad\text{and}\quad \qq = e^{\theta\kk}.
\end{equation}
The next six Lie superalgebras in Table~\ref{tab:klsa} are
supersymmetric extensions of the kinematical Lie algebra
\hyperlink{KLA14}{$\mathsf{K14}$} in Table~\ref{tab:kla}, whose $\r$-fixing
automorphisms $(A,\mu)$ take the form
\begin{equation}
  A =
  \begin{pmatrix}
    1 & \zero \\ \zero & d
  \end{pmatrix} \quad\text{and}\quad \mu \in \RR^\times.
\end{equation}

\subsubsection{Automorphisms of Lie Superalgebra $\mathsf{S23}$}
\label{sec:autom-kinem-lie-23}

Here $\hh= \kk$, $\bb=\pp=0$, $c_0=0$, $\bc_1 = \bc_2 = 0$ and $\bc_3
= \kk$.  The invariance conditions~\eqref{eq:aut-s} reduce to
\begin{equation}
  \mu \qq \kk = \kk \qq \quad\text{and}\quad d \kk = \qq \kk \qqbar.
\end{equation}
The first equation says that $\mu = \pm 1$, so that $\pm \qq \kk = \kk
\qq$.  The second equation follows from Lemma~\ref{lem:dk=qkqbar}:
either $d = |\qq|^2$ and hence $\qq = q_4 + q_3 \kk$ or $d = -|\qq|^2$
and hence $\qq = q_1 \ii + q_2 \jj$.  In summary, the typical
automorphism $(A,\mu,\qq)$ takes one of two possible forms:
\begin{equation}
  \begin{split}
    &A = \begin{pmatrix} 1 & \zero \\ \zero & |\qq|^2
    \end{pmatrix},\quad \mu =1 \quad\text{and}\quad \qq
    = q_4 + q_3\kk\\
   \text{or}\quad &A = \begin{pmatrix}
      1 & \zero \\ \zero & -|\qq|^2
    \end{pmatrix},\quad \mu = -1 \quad\text{and}\quad \qq = q_1 \ii
    + q_2 \jj.
  \end{split}
\end{equation}

\subsubsection{Automorphisms of Lie Superalgebra $\mathsf{S24}$}
\label{sec:autom-kinem-lie-24}

Here $\hh=\bb=\pp=0$, $c_0=0$, $\bc_1 = \bc_2 = 0$ and $\bc_3 = \kk$.
Hence the only invariance condition is $d\kk = \qq\kk\qqbar$.
Lemma~\ref{lem:dk=qkqbar} says that either $d = |\qq|^2$ and hence
$\qq = q_4 + q_3\kk$ or else $d = - |\qq|^2$ and hence $\qq = q_1 \ii
+ q_2 \jj$.  In summary, the typical automorphism $(A,\mu,\qq)$ takes
one of two possible forms:
\begin{equation}
  \begin{split}
    &A = \begin{pmatrix} 1 & \zero \\ \zero & |\qq|^2
    \end{pmatrix},\quad \mu \in\RR^\times \quad\text{and}\quad \qq
    = q_4 + q_3\kk\\
   \text{or}\quad &A = \begin{pmatrix}
      1 & \zero \\ \zero & -|\qq|^2
    \end{pmatrix},\quad \mu \in\RR^\times \quad\text{and}\quad \qq = q_1 \ii
    + q_2 \jj.
  \end{split}
\end{equation}

\subsubsection{Automorphisms of Lie Superalgebra $\mathsf{S25}$}
\label{sec:autom-kinem-lie-25}

Here $\hh=\bb=\pp=0$, $c_0=1$ and $\bc_1 = \bc_2 = \bc_3 = 0$, so that
the only invariance condition is $\mu = |\qq|^2$.  In summary, the
typical automorphism $(A,\mu,\qq)$ takes the form
\begin{equation}
  A =
  \begin{pmatrix}
    1 & \zero \\ \zero & d
  \end{pmatrix}, \quad \mu = |\qq|^2 \quad\text{and}\quad \qq \in \HH^\times.
\end{equation}

\subsubsection{Automorphisms of Lie Superalgebra $\mathsf{S26}$}
\label{sec:autom-kinem-lie-26}

Here $\hh=\bb=\pp=0$, $c_0=1$, $\bc_1 = \bc_2 = 0$ and $\bc_3 = \kk$,
so that there are two conditions in \eqref{eq:aut-s}:
\begin{equation}
  \mu = |\qq|^2 \quad\text{and}\quad d \kk = \qq \kk \qqbar.
\end{equation}
The second equation can be solved via Lemma~\ref{lem:dk=qkqbar}:
either $d = |\qq|^2$ and $\qq = q_4 + q_3\kk$ or $d= - |\qq|^2$ and
$\qq = q_1 \ii + q_2 \jj$.  In summary, the automorphisms
$(A,\mu,\qq)$ take one of two possible forms:
\begin{equation}
  \begin{split}
    &A = \begin{pmatrix} 1 & \zero \\ \zero & |\qq|^2
    \end{pmatrix},\quad \mu =|\qq|^2 \quad\text{and}\quad \qq
    = q_4 + q_3\kk\\
   \text{or}\quad &A = \begin{pmatrix}
      1 & \zero \\ \zero & -|\qq|^2
    \end{pmatrix},\quad \mu =|\qq|^2 \quad\text{and}\quad \qq = q_1 \ii
    + q_2 \jj.
  \end{split}
\end{equation}

\subsubsection{Automorphisms of Lie Superalgebra $\mathsf{S27}$}
\label{sec:autom-kinem-lie-27}

Here $\hh=\pp=0$, $\bb=\frac12$, $c_0= 1$ and $\bc_1 = \bc_2 = \bc_3 =
0$, so that the only invariance condition is $\mu = |\qq|^2$.
Therefore the typical automorphism $(A,\mu,\qq)$ takes the form
\begin{equation}
  A =
  \begin{pmatrix}
    1 & \zero \\ \zero & d
  \end{pmatrix}, \quad \mu = |\qq|^2 \quad\text{and}\quad \qq \in \HH^\times.
\end{equation}

\subsubsection{Automorphisms of Lie Superalgebra $\mathsf{S28}$}
\label{sec:autom-kinem-lie-28}

Here $\hh=\bb=\frac12$, $\pp=0$, $c_0=1$, $\bc_1 = \bc_3 = 0$ and
$\bc_2 = \kk$.  The invariance conditions~\eqref{eq:aut-s} reduce to
the following:
\begin{equation}
  \mu = |\qq|^2, \quad \mu \qq = \qq \quad\text{and}\quad \kk = \qq
  \kk \qqbar.
\end{equation}
From the second equation we see that $\mu = 1$, so that from the first
$|\qq| = 1$ and hence $\kk \qq = \qq \kk$, so that $\qq =
e^{\theta\kk}$.  In summary, the typical automorphism $(A,\mu,\pp)$
takes the form
\begin{equation}
  A =
  \begin{pmatrix}
    1 & \zero \\ \zero & d
  \end{pmatrix}, \quad \mu = 1 \quad\text{and}\quad \qq = e^{\theta\kk}.
\end{equation}
The next four Lie superalgebras in Table~\ref{tab:klsa} are
supersymmetric extensions of the kinematical Lie algebra
\hyperlink{KLA15}{$\mathsf{K15}$} in Table~\ref{tab:kla}, whose $\r$-fixing
automorphisms $(A,\mu)$ take the form
\begin{equation}
  A =
  \begin{pmatrix}
    a & \zero \\ c & a^2
  \end{pmatrix} \quad\text{and}\quad \mu \in \RR^\times.
\end{equation}

\subsubsection{Automorphisms of Lie Superalgebra $\mathsf{S29}$}
\label{sec:autom-kinem-lie-29}

Here $\bb=\pp=0$, $\hh= \kk$, $c_0 = 0$, $\bc_1 = \bc_2 = 0$ and
$\bc_3 = \kk$.  The invariance conditions~\eqref{eq:aut-s} result in
\begin{equation}
  \mu\qq\kk = \kk\qq \quad\text{and}\quad a^2 \kk = \qq\kk\qqbar.
\end{equation}
Taking the norm of the first equation, we see that $\mu = \pm 1$, and
of the second equation, $a^2=|\qq|^2$.  This then says that $\qq$
commutes with $\kk$, so that $\mu = 1$ and $\qq = q_4 + q_3\kk$.  In
summary, the typical automorphism $(A,\mu,\qq)$ takes the form
\begin{equation}
  A =
  \begin{pmatrix}
    \pm |\qq| & \zero \\ c & |\qq|^2
  \end{pmatrix}, \quad \mu = 1 \quad\text{and}\quad \qq = q_4 +
  q_3\kk.
\end{equation}

\subsubsection{Automorphisms of Lie Superalgebra $\mathsf{S30}$}
\label{sec:autom-kinem-lie-30}

Here $\hh=\bb=\pp =0$, $c_0 = 0$, $\bc_1 = \bc_2 = 0$ and $\bc_3 =
\kk$.  The only invariance condition is $a^2\kk = \qq\kk\qqbar$.
Taking the norm, $a^2 = |\qq|^2$ and hence $\kk\qq = \qq\kk$ and thus
$\qq = q_4 + q_3 \kk$.  Hence the typical automorphism $(A,\mu,\qq)$
takes the form
\begin{equation}
  A =
  \begin{pmatrix}
    \pm |\qq| & \zero \\ c & |\qq|^2
  \end{pmatrix}, \quad \mu \in \RR^\times \quad\text{and}\quad \qq
  = q_4 + q_3 \kk.
\end{equation}

\subsubsection{Automorphisms of Lie Superalgebra $\mathsf{S31}$}
\label{sec:autom-kinem-lie-31}

Here $\hh=\bb=\pp=0$, $c_0=1$ and $\bc_1 = \bc_2 = \bc_3 = 0$, so that
the only invariance condition is $\mu = |\qq|^2$.  In summary, the
typical automorphism $(A,\mu,\qq)$ takes the form
\begin{equation}
  A =
  \begin{pmatrix}
    a & \zero \\ c & a^2
  \end{pmatrix}, \quad \mu = |\qq|^2 \quad\text{and}\quad \qq \in \HH^\times.
\end{equation}

\subsubsection{Automorphisms of Lie Superalgebra $\mathsf{S32}$}
\label{sec:autom-kinem-lie-32}

Here $\hh=\bb=\pp=0$, $c_0=1$, $\bc_1 = \bc_2 = 0$ and $\bc_3 = \kk$,
so that there are two invariance conditions:
\begin{equation}
  \mu = |\qq|^2 \quad\text{and}\quad a^2 \kk = \qq \kk \qqbar.
\end{equation}
The second shows that $a^2 = |\qq|^2$ and hence $\qq$ commutes with
$\kk$, so that $\qq = q_4 + q_3 \kk$.  In summary, the
typical automorphism $(A,\mu,\qq)$ takes the form
\begin{equation}
  A =
  \begin{pmatrix}
    \pm |\qq| & \zero \\ c & |\qq|^2
  \end{pmatrix}, \quad \mu = |\qq|^2 \quad\text{and}\quad \qq = q_4
  + q_3 \kk.
\end{equation}
The next Lie superalgebra in Table~\ref{tab:klsa} is a one-parameter
family of supersymmetric extensions of the kinematical Lie algebra
\hyperlink{KLA16}{$\mathsf{K16}$} in Table~\ref{tab:kla}, whose $\r$-fixing
automorphisms $(A,\mu)$ take the form
\begin{equation}
  A =
  \begin{pmatrix}
    1 & \zero \\ \zero & d
  \end{pmatrix} \quad\text{and}\quad \mu =1.
\end{equation}

\subsubsection{Automorphisms of Lie Superalgebra $\mathsf{S33}$}
\label{sec:autom-kinem-lie-33}

Here $\hh= \frac12 (1 + \lambda \kk)$, $\bb = \pp = 0$, $c_0 = 0$,
$\bc_1 = \bc_2 = 0$ and $\bc_3 = \kk$.  There are two invariance
conditions:
\begin{equation}
  \qq (1 + \lambda \kk) = (1 + \lambda \kk) \qq \quad\text{and}\quad
  d \kk = \qq \kk \qqbar.
\end{equation}
For the second equation we use Lemma~\ref{lem:dk=qkqbar} and for the
first equation we must distinguish between $\lambda =0$ and $\lambda
\neq 0$.  In the latter case, we have that $\qq = q_4 + q_3 \kk$ so
that only the $d = |\qq|^2$ of the lemma survives.  If $\lambda = 0$,
both branches survive.  In summary, for $\lambda \neq 0$, the typical
automorphism $(A,\mu,\qq)$ takes the form
\begin{equation}
  A =
  \begin{pmatrix}
    1 & \zero \\ \zero & |\qq|^2
  \end{pmatrix}, \quad \mu = 1 \quad\text{and}\quad \qq = q_4 + q_3\kk,
\end{equation}
whereas if $\lambda = 0$ we have additional automorphisms of the form
\begin{equation}
  A =
  \begin{pmatrix}
    1 & \zero \\ \zero & -|\qq|^2
  \end{pmatrix}, \quad \mu = 1 \quad\text{and}\quad \qq = q_1\ii + q_2\jj.
\end{equation}
The next Lie superalgebra in Table~\ref{tab:klsa} is the
supersymmetric extension of the kinematical Lie algebra
\hyperlink{KLA17}{$\mathsf{K17}$} in Table~\ref{tab:kla}, whose $\r$-fixing
automorphisms $(A,\mu)$ take the form
\begin{equation}
  A =
  \begin{pmatrix}
    a & \zero \\ c & a^2
  \end{pmatrix} \quad\text{and}\quad \mu =a.
\end{equation}

\subsubsection{Automorphisms of Lie Superalgebra $\mathsf{S34}$}
\label{sec:autom-kinem-lie-34}

Here $\hh = \frac12 \kk$, $\bb = \pp = 0$, $c_0 = 0$, $\bc_1 = \bc_2 =
0$ and $\bc_3 = \kk$.  The invariance conditions are
\begin{equation}
  a \qq \kk = \kk\qq \quad\text{and}\quad a^2\kk = \qq \kk \qqbar.
\end{equation}
Taking norms of the first equation gives $a = \pm 1$ and hence $\pm
\qq\kk = \kk \qq$ and of the second equation $a^2 = |\qq|^2$ and hence
$\qq \kk = \kk \qq$.  This shows that $a = 1$ and hence $|\qq| = 1$,
so that $\qq = e^{\theta\kk}$.  In summary, the typical automorphism
$(A,\mu,\qq)$ takes the form
\begin{equation}
  A =
  \begin{pmatrix}
    1 & \zero \\ c & 1
  \end{pmatrix}, \quad \mu = 1 \quad\text{and}\quad \qq = e^{\theta\kk}.
\end{equation}
The last Lie superalgebra in Table~\ref{tab:klsa} is a one-parameter
family of supersymmetric extensions of the kinematical Lie algebra
\hyperlink{KLA18}{$\mathsf{K18}$} in Table~\ref{tab:kla}, whose $\r$-fixing
automorphisms $(A,\mu)$ take the form
\begin{equation}
  A =
  \begin{pmatrix}
    a & \zero \\ \zero & a^2
  \end{pmatrix} \quad\text{and}\quad \mu =1.
\end{equation}

\subsubsection{Automorphisms of Lie Superalgebra $\mathsf{S35}$}
\label{sec:autom-kinem-lie-35}

Here $\hh = 1 + \lambda \kk$, $\bb = \pp = 0$, $c_0 = 0$, $\bc_1 =
\bc_2 = 0$ and $\bc_3 = \kk$.  The invariance
conditions~\eqref{eq:aut-s} reduce to
\begin{equation}
  \qq (1 + \lambda \kk) = (1 + \lambda \kk) \qq \quad\text{and}\quad
  a^2\kk = \qq\kk\qqbar.
\end{equation}
Taking the norm of the second equation, $a^2 = |\qq|^2$ so that $\qq
\kk = \kk \qq$ and hence $\qq = q_4 + q_3 \kk$.  This also solves the
first equation, independently of the value of $\lambda$.  In summary,
the typical automorphism $(A,\mu,\qq)$ takes the form
\begin{equation}
  A =
  \begin{pmatrix}
    \pm |\qq| & \zero \\ \zero & |\qq|^2
  \end{pmatrix}, \quad \mu = 1 \quad\text{and}\quad \qq = q_4 + q_3\kk.
\end{equation}

\subsubsection{Summary}
\label{sec:summary-1}

Tables~\ref{tab:aut-klsa} and \ref{tab:aut-klsa-extra} summarise the above discussion and lists the
typical automorphisms of each of the Lie superalgebras in Table~\ref{tab:klsa}.

\begin{table}[h!]
  \centering
  \caption{Automorphisms of Kinematical Lie Superalgebras}
  \label{tab:aut-klsa}
  \begin{tabular}{l|>{$}l<{$}}\toprule
    \multicolumn{1}{c|}{S\#} & \multicolumn{1}{c}{Typical $(A,\mu,\qq) \in \GL(2,\RR) \times \RR^\times \times \HH^\times$}\\
    \toprule
    \hypertarget{SAut1}{1} & \left(\begin{pmatrix} a & \zero \\ c & |\qq|^2 \end{pmatrix}, 1, q_4 + q_3 \kk\right),
   \left(\begin{pmatrix} a & \zero \\ c & -|\qq|^2 \end{pmatrix}, -1, q_1\ii + q_2 \jj\right)\\[10pt]
    \hypertarget{SAut2}{2} & \left(\begin{pmatrix} q_4^2-q_1^2 & 2 q_1 q_4 \\ -2 q_1 q_4 & q_4^2 - q_1^2
    \end{pmatrix}, q_1^2+q_4^2, q_4 + q_1 \ii\right), \left(\begin{pmatrix}
     q_2^2-q_3^2 & 2 q_2 q_3 \\ 2 q_2 q_3 & q_3^2 - q_2^2
    \end{pmatrix}, q_2^2+q_3^2, q_2 \jj + q_3 \kk\right)\\[10pt]
    \hypertarget{SAut3}{3} & \left(\begin{pmatrix} a & \zero \\ c & |\qq|^2 \end{pmatrix}, |\qq|^2, q_4 + q_3 \kk\right),
   \left(\begin{pmatrix} a & \zero \\ c & -|\qq|^2 \end{pmatrix}, |\qq|^2, q_1\ii + q_2 \jj\right)\\[10pt]
    \hypertarget{SAut4}{4} & \left(\begin{pmatrix} a & b \\ c & d \end{pmatrix}, |\qq|^2, \qq \right)\\[10pt]
    \hypertarget{SAut5}{5} & \left(\begin{pmatrix} q_4^2-q_1^2 & 2 q_1 q_4 \\ -2 q_1 q_4 & q_4^2 - q_1^2
    \end{pmatrix}, \mu, q_4 + q_1 \ii\right), \left(\begin{pmatrix}
     q_2^2-q_3^2 & 2 q_2 q_3 \\ 2 q_2 q_3 & q_3^2 - q_2^2
    \end{pmatrix}, \mu, q_2 \jj + q_3 \kk\right)\\[10pt]
    \hypertarget{SAut6}{6} & \left(\begin{pmatrix} a & \zero \\ c & |\qq|^2 \end{pmatrix}, \mu , q_4 + q_3 \kk\right),
   \left(\begin{pmatrix} a & \zero \\ c & -|\qq|^2 \end{pmatrix}, \mu , q_1\ii + q_2 \jj\right)\\[10pt]
    \hypertarget{SAut7}{7} & \left(\begin{pmatrix} |\qq|^2 & \zero \\ c & |\qq|^2 \end{pmatrix}, 1, q_4 + q_3 \kk\right),
   \left(\begin{pmatrix} |\qq|^2 & \zero \\ c & -|\qq|^2 \end{pmatrix}, -1, q_1\ii + q_2 \jj\right)\\[10pt]
    \hypertarget{SAut8}{8} & \left(\begin{pmatrix} a & \zero \\ c & |\qq|^2 \end{pmatrix}, \frac{|\qq|^2}{a}, q_4 + q_3 \kk\right),
   \left(\begin{pmatrix} a & \zero \\ c & -|\qq|^2 \end{pmatrix}, -\frac{|\qq|^2}{a}, q_1\ii + q_2 \jj\right)\\[10pt]
    \hypertarget{SAut9a}{9$_{\gamma\neq 1, \lambda \neq 0}$} & \left(\begin{pmatrix} a & \zero \\ \zero & |\qq|^2 \end{pmatrix}, 1, q_4 + q_3 \kk\right)\\[10pt]
    \hypertarget{SAut9b}{9$_{\gamma= 1, \lambda \neq 0}$} & \left(\begin{pmatrix} a & \zero \\ c & |\qq|^2 \end{pmatrix}, 1, q_4 + q_3 \kk\right)\\[10pt]
    \hypertarget{SAut9c}{9$_{\gamma\neq 1, \lambda= 0}$} & \left(\begin{pmatrix} a & \zero \\ \zero & |\qq|^2 \end{pmatrix}, 1, q_4 + q_3 \kk\right), \left(\begin{pmatrix} a & \zero \\ \zero & -|\qq|^2 \end{pmatrix}, 1, q_1\ii + q_2 \jj\right)\\[10pt]
    \hypertarget{SAut9d}{9$_{\gamma= 1, \lambda = 0}$} &  \left(\begin{pmatrix} a & \zero \\ c & |\qq|^2 \end{pmatrix}, 1, q_4 + q_3 \kk\right), \left(\begin{pmatrix} a & \zero \\ c & -|\qq|^2 \end{pmatrix}, 1, q_1\ii + q_2 \jj\right)\\[10pt]
    \hypertarget{SAut10a}{10$_{\gamma,\lambda\neq0}$} &  \left(\begin{pmatrix} |\qq|^2 & \zero \\ \zero & d \end{pmatrix}, 1, q_4 + q_3 \kk\right)\\[10pt]
    \hypertarget{SAut10b}{10$_{\gamma,\lambda=0}$} &  \left(\begin{pmatrix} |\qq|^2 & \zero \\ \zero & d \end{pmatrix}, 1, q_4 + q_3 \kk\right), \left(\begin{pmatrix} -|\qq|^2 & \zero \\ \zero & d \end{pmatrix}, 1, q_1\ii + q_2 \jj\right)\\[10pt]
    \hypertarget{SAut11a}{11$_{\chi> 0}$} & \left(\begin{pmatrix} q_4^2 - q_2^2 & - 2 q_2 q_4 \\ 2 q_2 q_4 & q_4^2 - q_2^2 \end{pmatrix}, 1, q_4 + q_2\jj\right)\\[10pt]
    \hypertarget{SAut11b}{11$_{\chi= 0}$} & \left(\begin{pmatrix} q_4^2 - q_2^2 & - 2 q_2 q_4 \\ 2 q_2 q_4 & q_4^2 - q_2^2 \end{pmatrix}, 1, q_4 + q_2\jj\right), \left(\begin{pmatrix} q_1^2 - q_3^2 & 2 q_1 q_3 \\ 2 q_1 q_3 & q_3^2 - q_1^2 \end{pmatrix}, -1, q_1\ii + q_3\kk\right)\\[10pt]
    \hypertarget{SAut12a}{12$_{\lambda\neq 0}$} & \left(\begin{pmatrix} |\qq|^2 & \zero \\ c & |\qq|^2 \end{pmatrix}, 1, q_4 + q_3\kk\right)\\[10pt]
    \hypertarget{SAut12b}{12$_{\lambda= 0}$} & \left(\begin{pmatrix} |\qq|^2 & \zero \\ c & |\qq|^2 \end{pmatrix}, 1, q_4 + q_3\kk\right), \left(\begin{pmatrix} -|\qq|^2 & \zero \\ c & -|\qq|^2 \end{pmatrix}, 1, q_1\ii + q_2\jj\right)\\[10pt]
    \hypertarget{SAut13}{13} & \left(\begin{pmatrix} a & b \\ c & d \end{pmatrix}, ad-bc = |\qq|^2, \qq\right)\\[10pt]
    \hypertarget{SAut14}{14} & \left(\begin{pmatrix} 1 & \zero \\ c & |\qq|^2 \end{pmatrix}, |\qq|^2, q_4 + q_3\kk\right), \left(\begin{pmatrix} -1 & \zero \\ c & -|\qq|^2 \end{pmatrix}, |\qq|^2 , q_1\ii + q_2\jj\right)\\[10pt]
    \hypertarget{SAut15}{15} & \left(\begin{pmatrix} \cos2\theta & -\sin2\theta \\ \sin2\theta & \cos2\theta \end{pmatrix}, 1, e^{\theta \kk}\right)\\[10pt]
    \bottomrule
  \end{tabular}
\end{table}

\begin{table}[h!]
  \centering
  \caption{Automorphisms of Kinematical Lie Superalgebras (continued)}
  \label{tab:aut-klsa-extra}
  \begin{tabular}{l|>{$}l<{$}}\toprule
    \multicolumn{1}{c|}{S\#} & \multicolumn{1}{c}{Typical $(A,\mu,\qq) \in \GL(2,\RR) \times \RR^\times \times \HH^\times$}\\
    \toprule
    \hypertarget{SAut16}{16} &\left(\begin{pmatrix} 1 & \zero \\ \zero & 1 \end{pmatrix},\mu, \pm 1 \right),
         \left(\begin{pmatrix} 1 & \zero \\ \zero & -1 \end{pmatrix}, \mu, \pm\ii \right)\\[10pt]
    \hypertarget{SAut17}{17} & \left(\begin{pmatrix} 1 & \zero \\ \zero & \pm 1 \end{pmatrix}, |\qq|^2, \qq \right)\\[10pt]
    \hypertarget{SAut18}{18} & \left(\begin{pmatrix} 1 & \zero \\ \zero & \pm 1 \end{pmatrix}, 1, e^{\theta\kk} \right)\\[10pt]
    \hypertarget{SAut19}{19} & \left(\begin{pmatrix} 1 & \zero \\ \zero & 1 \end{pmatrix}, 1, e^{\theta\kk} \right)\\[10pt]
    \hypertarget{SAut20}{20} & \left(\begin{pmatrix} 1 & \zero \\ \zero & 1 \end{pmatrix}, |\qq|^2, \qq \right)\\[10pt]
    \hypertarget{SAut21}{21} & \left(\begin{pmatrix} 1 & \zero \\ \zero & \pm 1 \end{pmatrix}, |\qq|^2, \qq \right)\\[10pt]
    \hypertarget{SAut22}{22} & \left(\begin{pmatrix} 1 & \zero \\ \zero & \pm 1 \end{pmatrix}, 1, e^{\theta\kk} \right)\\[10pt]
    \hypertarget{SAut23}{23} & \left(\begin{pmatrix} 1 & \zero \\ \zero & |\qq|^2 \end{pmatrix}, 1, q_4 + q_3\kk\right),
         \left(\begin{pmatrix} 1 & \zero \\ \zero & -|\qq|^2 \end{pmatrix}, -1, q_1\ii + q_2\jj\right)\\[10pt]
    \hypertarget{SAut24}{24} &\left(\begin{pmatrix} 1 & \zero \\ \zero & |\qq|^2 \end{pmatrix}, \mu , q_4 + q_3\kk\right),
         \left(\begin{pmatrix} 1 & \zero \\ \zero & -|\qq|^2 \end{pmatrix}, \mu , q_1\ii + q_2\jj\right)\\[10pt]
    \hypertarget{SAut25}{25} & \left(\begin{pmatrix} 1 & \zero \\ \zero & d \end{pmatrix}, |\qq|^2 , \qq \right)\\[10pt]
    \hypertarget{SAut26}{26} &\left(\begin{pmatrix} 1 & \zero \\ \zero & |\qq|^2 \end{pmatrix}, |\qq|^2 , q_4 + q_3\kk\right),
         \left(\begin{pmatrix} 1 & \zero \\ \zero & -|\qq|^2 \end{pmatrix}, |\qq|^2 , q_1\ii + q_2\jj\right)\\[10pt]
    \hypertarget{SAut27}{27} & \left(\begin{pmatrix} 1 & \zero \\ \zero & d \end{pmatrix}, |\qq|^2 , \qq \right)\\[10pt]
    \hypertarget{SAut28}{28} & \left(\begin{pmatrix} 1 & \zero \\ \zero & d \end{pmatrix}, 1 , e^{\theta\kk}\right)\\[10pt]
    \hypertarget{SAut29}{29} & \left(\begin{pmatrix} \pm|\qq| & \zero \\ c & |\qq|^2 \end{pmatrix}, 1 , q_4 + q_3\kk\right)\\[10pt]
    \hypertarget{SAut30}{30} & \left(\begin{pmatrix} \pm|\qq| & \zero \\ c & |\qq|^2 \end{pmatrix}, \mu , q_4 + q_3\kk\right)\\[10pt]
    \hypertarget{SAut31}{31} & \left(\begin{pmatrix} a & \zero \\ c & a^2 \end{pmatrix}, |\qq|^2 , \qq\right)\\[10pt]
    \hypertarget{SAut32}{32} & \left(\begin{pmatrix} \pm|\qq| & \zero \\ c & |\qq|^2 \end{pmatrix}, |\qq|^2 , q_4 + q_3\kk\right)\\[10pt]
    \hypertarget{SAut33a}{33$_{\lambda \neq 0}$} & \left(\begin{pmatrix} 1 & \zero \\ \zero & |\qq|^2 \end{pmatrix}, 1 , q_4 + q_3\kk\right)\\[10pt]
    \hypertarget{SAut33b}{33$_{\lambda = 0}$} & \left(\begin{pmatrix} 1 & \zero \\ \zero & |\qq|^2 \end{pmatrix}, 1 , q_4 + q_3\kk\right),
                       \left(\begin{pmatrix} 1 & \zero \\ \zero & -|\qq|^2 \end{pmatrix}, 1 , q_1\ii + q_2\jj\right)\\[10pt]
    \hypertarget{SAut34}{34} & \left(\begin{pmatrix} 1 & \zero \\ c & 1 \end{pmatrix}, 1, e^{\theta\kk} \right)\\[10pt]
    \hypertarget{SAut35}{35$_\lambda$} & \left(\begin{pmatrix} \pm|\qq| & \zero \\ \zero & |\qq|^2 \end{pmatrix}, 1 , q_4 + q_3\kk\right)\\[10pt]
    \bottomrule
  \end{tabular}
\end{table}

\section{Classification of Kinematical Superspaces} \label{sec:ks_superspace_kss}
Now that we have a complete classification of the possible $\N=1$ kinematical Lie superalgebras in three spatial dimensions, we can turn to the corresponding kinematical superspace classification.  As discussed in Section~\ref{subsec:math_prelims_geo_lsgahss}, each homogeneous superisation of a kinematical spacetime $\Kgr/\Hgr$, corresponds to a unique effective super Lie pair $(\s, \h)$, where $\s = \s_{\bar{0}} \oplus \s_{\bar{1}}$, with $\s_{\bar{0}}=\k$ the kinematical Lie algebra associated with $\Kgr$, and $\h \subset \s_{\bar{0}} = \k$ is the Lie subalgebra associated with $\Hgr$.  To establish the possible effective super Lie pairs, this section runs as follows. In Section~\ref{sec:slie-pairs}, we will use the automorphisms derived in Section~\ref{sec:autom-kinem-lie} to identify the admissible super Lie pairs for each KLSA.  Then, in Section~\ref{sec:effective-super-lie}, we determine which of the admissible Lie pairs are effective.  We then give a brief account of the possible Aristotelian homogeneous superspaces in Section~\ref{sec:arist-super-lie} before summarising our findings in Section~\ref{sec:summary-3}.  In Section~\ref{sec:low-rank-invariants}, we end by discussing the low-rank invariants of the kinematical superspaces.
Notice the method used in determining the kinematical superspaces is a direct generalisation of the classification method used in the kinematical spacetime case: we find the admissible Lie subalgebras and then restrict ourselves to those Lie pairs which are effective.  Geometric realisability does not need to be studied in this instance since we assume the underlying kinematical spacetime is geometrically realisable.  This property then trivially extends to the superised geometry.

\subsection{Admissible Super Lie Pairs}
\label{sec:slie-pairs}

We are now ready to classify the admissible super Lie pairs, up to
isomorphism.  We recall these are pairs $(\s,\h)$, where $\s$ is one
of the kinematical Lie superalgebras in Table~\ref{tab:klsa} and $\h$
is a Lie subalgebra $\h \subset \k = \s_{\bar 0}$ which is admissible
in the sense of Section~\ref{subsec:math_prelims_geo_ks}; that is, it contains
the rotational subalgebra $\r$ and, as an $\r$ module,
$\h = \r \oplus V$ where $V \subset \k$ is a copy of the vector
module.  Two super Lie pairs $(\s,\h)$ and $(\s,\h')$ are
isomorphic if there is an automorphism of $\s$ which maps $\h$
(isomorphically) to $\h'$.  As in
Section~\ref{sec:ks_kss}, our strategy in classifying
admissible super Lie pairs up to isomorphism will be to take each
kinematical Lie superalgebra $\s$ in Table~\ref{tab:klsa} in turn,
determine the admissible subalgebras $\h$ and study the action of the
automorphisms in Tables~\ref{tab:aut-klsa} and
\ref{tab:aut-klsa-extra} on the space of admissible subalgebras in
order to select one representative from each orbit.  In particular,
every admissible super Lie pair $(\s,\h)$ defines a unique admissible
Lie pair $(\k,\h)$ which, if effective and geometrically realisable,
is associated with a unique simply-connected kinematical homogeneous
spacetime $\Kgr/\Hgr$.  That being the case, we may think of the super
Lie pair $(\s,\h)$ as a homogeneous kinematical superspacetime which
superises $\Kgr/\Hgr$.
\\ \\
Without loss of generality -- since an admissible subalgebra $\h$
contains $\r$ -- the vectorial complement $V$ can be taken to be the
span of $\alpha \bB_i + \beta \bP_i$, $i=1,2,3$, for some
$\alpha,\beta \in \RR$ not both zero, since the spans of
$\{\bJ_i, \alpha \bB_i + \beta \bP_i\}$ and of
$\{\bJ_i, \alpha \bB_i + \beta \bP_i + \gamma \bJ_i\}$ coincide for all
$\gamma \in \RR$.  We will often use the shorthand
$V = \alpha \B + \beta \P$.  The determination of the possible
admissible subalgebras can be found in
\cite[§§3.1-2]{Figueroa-OFarrill:2018ilb}, but we cannot simply import
the results of that paper wholesale because here we are only allowed
to act with automorphisms of $\s$ and not just of $\k$.
\\ \\
As in that paper, and as discussed in Section~\ref{subsec:ks_admissibility}, we will eventually change basis in the Lie
superalgebra $\s$ so that the admissible subalgebra $\h$ is spanned by
$\J$ and $\B$.  Hence, in determining the possible super Lie pairs, we
will keep track of the required change of basis, ensuring, where
possible, that $(\s,\h)$ is reductive; that is, such that $\bH, \bP_i,
\bQ_a$ (defined by equation~\eqref{eq:quat-basis-s}) span a subspace $\m
\subset \s$ complementary to $\h$ and such that $[\h,\m]\subset \m$.
This is equivalent to requiring that the span $\m_{\bar 0}$ of $\bH,
\bP_i$ satisfies $[\h,\m_{\bar 0}] \subset \m_{\bar 0}$, since the $\bQ_a$
span $\s_{\bar 1}$ and $[\h, \s_{\bar 1} ] \subset \s_{\bar 1}$ by
virtue of $\s$ being a Lie superalgebra.
\\ \\
It follows by inspection of
\cite[§§3.1-2]{Figueroa-OFarrill:2018ilb} that the Lie superalgebras
$\s$ whose automorphisms are listed in Table~\ref{tab:aut-klsa} are
extensions of kinematical Lie algebras $\k$ for which \emph{any}
vectorial subspace $V = \alpha \B + \beta \P$ defines an admissible
subalgebra $\h = \r \oplus V \subset \k$.  It is then a simple matter
to determine the orbits of the action of the automorphisms listed in
Table~\ref{tab:aut-klsa} on the space of vectorial subspaces and hence
to arrive at a list of possible inequivalent super Lie pairs
$(\s, \h)$ for such $\s$.
\\ \\
It also follows by inspection of
\cite[§§3.1-2]{Figueroa-OFarrill:2018ilb} that, of the remaining Lie
superalgebras (i.e., those whose automorphisms are listed in
Table~\ref{tab:aut-klsa-extra}), most are extensions of kinematical Lie
algebras possessing a unique vectorial subspace $V$ for which
$\h = \r \oplus V$ is an admissible subalgebra.  The exceptions are
those Lie superalgebras
\hyperlink{KLSA23}{$\mathsf{S23}$}--\hyperlink{KLSA28}{$\mathsf{S28}$}
and \hyperlink{KLSA33}{$\mathsf{S33}_\lambda$}, which are extensions
of the kinematical Lie algebras \hyperlink{KLA14}{$\mathsf{K14}$} and
\hyperlink{KLA16}{$\mathsf{K16}$}, respectively, for which there are
precisely two vectorial subspaces leading to admissible subalgebras.
\\ \\
Let us concentrate first on the Lie superalgebras
\hyperlink{KLSA1}{$\mathsf{S1}$}--\hyperlink{KLSA15}{$\mathsf{S15}$},
whose automorphisms are listed in Table~\ref{tab:aut-klsa}.  As
mentioned above, for $V$ any vectorial subspace, $\h = \r
\oplus V$ is an admissible subalgebra.  We need to determine the
orbits of the action of the automorphisms in
Table~\ref{tab:aut-klsa}.  Since $V = \alpha \B + \beta \P$, this is
equivalent to studying the action of the matrix part $A$ of the
automorphism $(A,\mu,\qq)$ on non-zero
vectors $(\alpha,\beta) \in \RR^2$.  In fact, since $(\alpha,\beta)$
and $(\lambda\alpha,\lambda\beta)$ for $0 \neq \lambda \in \RR$ denote
the same vectorial subspace, we must study the action of the subgroup
of $\GL(2,\RR)$ defined by the matrices $A$ in the automorphism group
on the projective space $\RP^1$.  The map $(A,\mu,\qq) \mapsto A$
defines a group homomorphism from the automorphism group of a Lie
superalgebra $\s$ to $\GL(2,\RR)$.  We will let $\Agr$ denote the
image of this homomorphism: it is a subgroup of $\GL(2,\RR)$ and it is
the action of $\Agr$ on $\RP^1$ that we need to investigate.  Of
course, $\Agr$ depends on $\s$, even though we choose not to overload
the notation by making this dependence explicit.
\\ \\
It follows by inspection of Table~\ref{tab:aut-klsa}, that for $\s$
any of the Lie superalgebras \hyperlink{SAut2}{$\mathsf{S2}$},
\hyperlink{SAut4}{$\mathsf{S4}$}, \hyperlink{SAut5}{$\mathsf{S5}$},
\hyperlink{SAut11a}{$\mathsf{S11}_{\chi\geq 0}$},
\hyperlink{SAut13}{$\mathsf{S13}$} and
\hyperlink{SAut15}{$\mathsf{S15}$}, the subgroup
$\Agr \subset \GL(2,\RR)$ acts transitively on $\RP^1$ and hence for
such Lie superalgebras there is a unique admissible subalgebra spanned
by $\J$ and $\B$. 
\\ \\
In contrast, if $\s$ is any of the Lie superalgebras
\hyperlink{SAut1}{$\mathsf{S1}$}, \hyperlink{SAut3}{$\mathsf{S3}$},
\hyperlink{SAut6}{$\mathsf{S6}$}, \hyperlink{SAut7}{$\mathsf{S7}$},
\hyperlink{SAut8}{$\mathsf{S8}$},
\hyperlink{SAut9b}{$\mathsf{S9}_{\gamma=1,\lambda\in\RR}$},
\hyperlink{SAut12a}{$\mathsf{S12}_{\lambda\in\RR}$} and
\hyperlink{SAut14}{$\mathsf{S14}$}, the subgroup $\Agr \subset
\GL(2,\RR)$ acts with two orbits on $\RP^1$.  For example, consider
the Lie superalgebra \hyperlink{SAut1}{$\mathsf{S1}$}, for which any $A
\in \Agr$ takes the form
\begin{equation}
  \begin{pmatrix}
    a & \zero \\ c & d
  \end{pmatrix} \quad\text{for some $a,c,d \in \RR$ with $a,d\neq 0$,}
\end{equation}
and act as
\begin{equation}
  \begin{pmatrix}
    \alpha \\ \beta 
  \end{pmatrix}\mapsto   \begin{pmatrix}
    a & \zero \\ c & d
  \end{pmatrix}   \begin{pmatrix}
    \alpha \\ \beta 
  \end{pmatrix} =   \begin{pmatrix}
    a \alpha \\ d \beta + c \alpha
  \end{pmatrix}.
\end{equation}
If $\alpha \neq 0$, we can choose $c = -d\beta/\alpha$ to bring
$(\alpha,\beta)$ to $(a\alpha, 0)$, which is projectively equivalent to
$(1,0)$.  On the other hand, if $\alpha = 0$, then we cannot change
that via automorphisms and hence we have $(0,\beta)$, which is
projectively equivalent to $(0,1)$.  In summary, we have two
inequivalent admissible subalgebras with vectorial subspaces $V=\B$
and $V=\P$.  The same result holds for the other Lie superalgebras in
this list.
\\ \\
For the cases where $V=\P$ we change basis in the Lie superalgebra
$\s$ so that the admissible subalgebra $\h$ is spanned by $\J$ and
$\B$.  This results in different brackets, which we now proceed to
list.
\\ \\
\begin{table}[h!]
  \centering
  \caption{Super Lie pairs (with $V=\P$)}
  \label{tab:slp-vp-1}
  \setlength{\extrarowheight}{2pt}
  \rowcolors{2}{blue!10}{white}
    \begin{tabular}{l|*{3}{>{$}l<{$}}*{3}{|>{$}c<{$}}}\toprule
      \multicolumn{1}{c|}{S\#} & \multicolumn{3}{c|}{$\k$ brackets} & \multicolumn{1}{c|}{$\hh$} & \multicolumn{1}{c|}{$\pp$} & \multicolumn{1}{c}{$[\sQ(s),\sQ(s)]$}\\
      \toprule
      \hyperlink{KLSA1}{1} & & & & \tfrac12 \kk & & -\sB(s\kk\sbar) \\
      \hyperlink{KLSA3}{3} & & & & & & |s|^2 \sH - \sB(s\kk\sbar) \\
      \hyperlink{KLSA6}{6} & & & & & & -\sB(s\kk\sbar) \\
      \hyperlink{KLSA7}{7} & [\bH,\P]=-\B & & & \kk & & -\sB(s\kk\sbar) \\
      \hyperlink{KLSA8}{8} & [\bH,\P]=-\B & & & & & -\sB(s\kk\sbar) \\
      \hyperlink{KLSA9}{9$_{\gamma=1,\lambda\in\RR}$} & [\bH,\B]=\B & [\bH,\P]=\P & & \tfrac12 (1 + \lambda \kk) & & -\sB(s\kk\sbar) \\
      \hyperlink{KLSA12}{12$_{\lambda\in\RR}$} & [\bH,\B]=\B & [\bH,\P] = \B + \P & & \tfrac12 (1 + \lambda \kk) & & -\sB(s\kk\sbar) \\
      \hyperlink{KLSA14}{14} & [\bH,\P] = \B & [\B,\P] = \bH & [\P,\P] = - \J & & \tfrac12 \kk & |s|^2 \sH + \sB(s\kk\sbar) \\
      \bottomrule
    \end{tabular}
\end{table}
Finally, if $\s$ is any of the Lie superalgebras
\hyperlink{SAut9a}{$\mathsf{S9}_{\gamma\neq 1,\lambda\in\RR}$} and
\hyperlink{SAut10a}{$\mathsf{S10}_{\gamma,\lambda\in\RR}$}, the
subgroup $\Agr \subset \GL(2,\RR)$ acts with three orbits.  Indeed,
the matrices $A \in \Agr$ are now diagonal and of the form
\begin{equation}
  \begin{pmatrix}
    a & \zero \\ \zero & d
  \end{pmatrix},
\end{equation}
where at least one of $a,d$ can take \emph{any} non-zero value.  If
$(\alpha,\beta)$ is such that $\alpha = 0$ or $\beta = 0$, we cannot
alter this via automorphisms and hence projectively we have either
$(1,0)$ or $(0,1)$.  If $\alpha\beta \neq 0$, then we can always bring
it to $(1,1)$ or $(-1,-1)$ via an automorphism, but these are
projectively equivalent.  In summary, we have three orbits,
corresponding to $V=\B$, $V = \P$ and $V = \B + \P$.
\\ \\
When $V = \P$, the Lie brackets of
\hyperlink{KLSA9}{$\mathsf{S9}_{\gamma\neq 1,\lambda\in\RR}$} in the
new basis are given by
\begin{equation}
 [\bH,\B] = \B, \quad [\bH,\P]=\gamma\P,\quad [\sH,\sQ(s)]=\sQ(\tfrac12 s
 (1 + \lambda\kk)) \quad\text{and}\quad [\sQ(s),\sQ(s)] =
 -\sB(s\kk\sbar),
\end{equation}
and those of \hyperlink{KLSA10}{$\mathsf{S10}_{\gamma,\lambda\in\RR}$}
by
\begin{equation}
  [\bH,\B] = \B, \quad [\bH,\P]=\gamma\P,\quad [\sH,\sQ(s)]=\sQ(\tfrac12 s
  (\gamma + \lambda\kk)) \quad\text{and}\quad [\sQ(s),\sQ(s)] =
  -\sP(s\kk\sbar).
\end{equation}
On the other hand, when $V = \B + \P$, the Lie brackets of
\hyperlink{KLSA9}{$\mathsf{S9}_{\gamma\neq 1,\lambda\in\RR}$} in the
new basis are given by
\begin{equation}
  \begin{aligned}[m]
    [\bH,\B] &= -\P\\
    [\bH,\P] &= \gamma\B + (1+\gamma)\P\\
  \end{aligned}
  \quad\quad
  \begin{aligned}[m]
    [\sH,\sQ(s)] &= \sQ(\tfrac12 s (1 + \lambda\kk))\\
    [\sQ(s),\sQ(s)] &= \tfrac{1}{1-\gamma}(\gamma \sB(s\kk\sbar) + \sP(s\kk\sbar)),
  \end{aligned}
\end{equation}
and those of \hyperlink{KLSA10}{$\mathsf{S10}_{\gamma,\lambda\in\RR}$}
by
\begin{equation}
  \begin{aligned}[m]
    [\bH,\B] &= -\P\\
    [\bH,\P] &= \gamma\B + (1+\gamma)\P\\
  \end{aligned}
  \quad\quad
  \begin{aligned}[m]
    [\sH,\sQ(s)] &= \sQ(\tfrac12 s (\gamma + \lambda\kk))\\
    [\sQ(s),\sQ(s)] &= \tfrac{1}{\gamma-1}(\sB(s\kk\sbar) + \sP(s\kk\sbar)).
  \end{aligned}
\end{equation}
Now we turn to the Lie superalgebras whose automorphisms are listed in
Table~\ref{tab:aut-klsa-extra}.  If $\s$ is one such Lie superalgebra,
not every vectorial subspace leads to an admissible subalgebra.  From
the results in \cite[§§3.1-2]{Figueroa-OFarrill:2018ilb}, we have that
Lie superalgebras
\hyperlink{SAut16}{$\mathsf{S16}$}--\hyperlink{SAut22}{$\mathsf{S22}$}
admit a unique admissible subalgebra with $V = \B$, whereas for the
Lie superalgebras
\hyperlink{SAut29}{$\mathsf{S29}$}--\hyperlink{SAut32}{$\mathsf{S32}$},
\hyperlink{SAut34}{$\mathsf{S34}$} and
\hyperlink{SAut35}{$\mathsf{S35}_{\lambda\in\RR}$} also admit a unique
admissible subalgebra with $V = \P$.  Finally, the Lie superalgebras
\hyperlink{SAut23}{$\mathsf{S23}$}--\hyperlink{SAut28}{$\mathsf{S28}$}
and \hyperlink{SAut33a}{$\mathsf{S33}_{\lambda\in\RR}$} admit precisely
two admissible subalgebras with $V= \B$ and $V= \P$, which cannot be
related by automorphisms.
\\ \\
\begin{table}[h!]
  \centering
  \caption{More super Lie pairs (with $V=\P$)}
  \label{tab:slp-vp-2}
  \setlength{\extrarowheight}{2pt}
  \rowcolors{2}{blue!10}{white}
    \begin{tabular}{l|*{3}{>{$}l<{$}}*{3}{|>{$}c<{$}}}\toprule
      \multicolumn{1}{c|}{S\#} & \multicolumn{3}{c|}{$\k$ brackets} & \multicolumn{1}{c|}{$\hh$} & \multicolumn{1}{c|}{$\pp$} & \multicolumn{1}{c}{$[\sQ(s),\sQ(s)]$}\\
      \toprule
      \hyperlink{KLSA23}{23} & [\P,\P] = \P & & & \kk & & -\sB(s\kk\sbar) \\
      \hyperlink{KLSA24}{24} & [\P,\P] = \P & & & & & - \sB(s\kk\sbar) \\
      \hyperlink{KLSA25}{25} & [\P,\P] = \P & & & & & |s|^2 \sH \\
      \hyperlink{KLSA26}{26} & [\P,\P] = \P & & & & & |s|^2 \sH - \sB(s\kk\sbar) \\
      \hyperlink{KLSA27}{27} & [\P,\P]= \P & & & & \tfrac12  & |s|^2 \sH \\
      \hyperlink{KLSA28}{28} & [\P,\P]= \P & & & \tfrac12 \kk & \tfrac12 & |s|^2 \sH -\sP(s\kk\sbar) \\
      \hyperlink{KLSA29}{29} & [\P,\P]= \B & & & \kk & & -\sB(s\kk\sbar) \\
      \hyperlink{KLSA30}{30} & [\P,\P] = \B & & & & & -\sB(s\kk\sbar) \\
      \hyperlink{KLSA31}{31} & [\P,\P] = \B & & & & & |s|^2 \sH \\
      \hyperlink{KLSA32}{32} & [\P,\P] = \B & & & & & |s|^2 \sH - \sB(s\kk\sbar) \\
      \hyperlink{KLSA33}{33$_{\lambda\in\RR}$} & [\bH,\B] = \B & [\P,\P] = \P & & \tfrac12 (1+\lambda \kk) & & - \sB(s\kk\sbar) \\
      \hyperlink{KLSA34}{34} & [\bH,\P] = -\B & [\P,\P] = \B & & \tfrac12 \kk & & - \sB(s\kk\sbar) \\
      \hyperlink{KLSA35}{35$_{\lambda\in\RR}$} & [\bH,\P] = \P & [H,\B] =2\B & [\P,\P] = \B & 1+\lambda \kk & & - \sB(s\kk\sbar) \\
      \bottomrule
    \end{tabular}
\end{table}
Table~\ref{tab:super-lie-pairs} summarises the above results.  For
each Lie superalgebra $\s$ in Table~\ref{tab:klsa} it lists the
admissible subalgebras $\h$ and hence the possible super Lie pairs
$(\s,\h)$.  The notation for $\h$ is simply the generators of the
vectorial subspace $V \subset \h$, where the span of
$\alpha \bB_i + \beta \bP_i$ is abbreviated as $\alpha \B + \beta \P$.
The blue entries correspond to effective super Lie pairs, whereas the
green and greyed out correspond to non-effective super Lie pairs: the
green ones giving rise to Aristotelian superspaces upon quotienting by the ideal $\b = \spn{\B}$, as described in Section~\ref{subsec:math_prelims_geo_ks}.  In Section~\ref{sec:class-arist-lie}, we classified
Aristotelian Lie superspaces by classifying their corresponding
Aristotelian Lie superalgebras (see Table~\ref{tab:alsa}) and in
Section~\ref{sec:arist-super-lie} we exhibit the precise
correspondence between the Aristotelian non-effective super Lie pairs
and the Aristotelian superspaces (see
Table~\ref{tab:aristo-correspondence}).

\begin{table}[h!]
  \centering
  \caption{Summary of super Lie pairs}
  \label{tab:super-lie-pairs}
  \resizebox{\textwidth}{!}{
    \setlength{\extrarowheight}{2pt}
    \begin{tabular}{l|l*{3}{|>{$}c<{$}}}\toprule
      \multicolumn{1}{c|}{$\s$} & \multicolumn{1}{c|}{$\k$} & \multicolumn{3}{c}{$V \subset \h$}\\
      \toprule
      \hyperlink{KLSA1}{$\mathsf{S1}$} & \hyperlink{KLA1}{$\mathsf{K1}$} & \ari{\B} & \non{\P} & \\
      \hyperlink{KLSA2}{$\mathsf{S2}$} & \hyperlink{KLA1}{$\mathsf{K1}$} & \ari{\B} & & \\
      \hyperlink{KLSA3}{$\mathsf{S3}$} & \hyperlink{KLA1}{$\mathsf{K1}$} & \ari{\B} & \ari{\P} & \\
      \hyperlink{KLSA4}{$\mathsf{S4}$} & \hyperlink{KLA1}{$\mathsf{K1}$} & \ari{\B} & & \\
      \hyperlink{KLSA5}{$\mathsf{S5}$} & \hyperlink{KLA1}{$\mathsf{K1}$} & \ari{\B} & & \\
      \hyperlink{KLSA6}{$\mathsf{S6}$} & \hyperlink{KLA1}{$\mathsf{K1}$} & \ari{\B} & \non{\P} & \\
      \hyperlink{KLSA7}{$\mathsf{S7}$} & \hyperlink{KLA2}{$\mathsf{K2}$} & \eff{\B} & \non{\P} & \\
      \hyperlink{KLSA8}{$\mathsf{S8}$} & \hyperlink{KLA2}{$\mathsf{K2}$} & \eff{\B} & \non{\P} & \\
      \hyperlink{KLSA9}{$\mathsf{S9}_{\gamma\in[-1,1),\lambda\in\RR}$} & \hyperlink{KLA3}{$\mathsf{K3}_\gamma$} & \ari{\B} & \non{\P} & \eff{\B + \P}\\
      \hyperlink{KLSA9}{$\mathsf{S9}_{\gamma=1,\lambda\in\RR}$} &  \hyperlink{KLA3}{$\mathsf{K3}_{\gamma=1}$} & \ari{\B} & \non{\P} &\\
      \hyperlink{KLSA10}{$\mathsf{S10}_{\gamma\in[-1,1),\lambda\in\RR}$} & \hyperlink{KLA3}{$\mathsf{K3}_\gamma$} & \non{\B} & \ari{\P} & \eff{\B + \P} \\
      \hyperlink{KLSA11}{$\mathsf{S11}_{\chi\geq0}$} & \hyperlink{KLA4}{$\mathsf{K4}_\chi$} & \eff{\B} & & \\
      \bottomrule
    \end{tabular}
    \hspace{2cm}
    \begin{tabular}{l|l*{2}{|>{$}c<{$}}}\toprule
      \multicolumn{1}{c|}{$\s$} & \multicolumn{1}{c|}{$\k$} & \multicolumn{2}{c}{$V \subset \h$}\\
      \toprule
      \hyperlink{KLSA12}{$\mathsf{S12}_{\lambda\in\RR}$} & \hyperlink{KLA5}{$\mathsf{K5}$} & \eff{\B} & \non{\P} \\
      \hyperlink{KLSA13}{$\mathsf{S13}$} & \hyperlink{KLA6}{$\mathsf{K6}$} & \eff{\B} &  \\
      \hyperlink{KLSA14}{$\mathsf{S14}$} & \hyperlink{KLA8}{$\mathsf{K8}$} & \eff{\B} & \eff{\P } \\
      \hyperlink{KLSA15}{$\mathsf{S15}$} & \hyperlink{KLA11}{$\mathsf{K11}$} & \eff{\B} &  \\
      \hyperlink{KLSA16}{$\mathsf{S16}$} & \hyperlink{KLA12}{$\mathsf{K12}$} & \ari{\B} &  \\
      \hyperlink{KLSA17}{$\mathsf{S17}$} & \hyperlink{KLA12}{$\mathsf{K12}$} & \eff{\B} &  \\
      \hyperlink{KLSA18}{$\mathsf{S18}$} & \hyperlink{KLA12}{$\mathsf{K12}$} & \eff{\B} & \\ 
      \hyperlink{KLSA19}{$\mathsf{S19}$} & \hyperlink{KLA13}{$\mathsf{K13}$} & \ari{\B} & \\ 
      \hyperlink{KLSA20}{$\mathsf{S20}$} & \hyperlink{KLA13}{$\mathsf{K13}$} & \ari{\B} & \\ 
      \hyperlink{KLSA21}{$\mathsf{S21}$} & \hyperlink{KLA13}{$\mathsf{K13}$} & \eff{\B} & \\ 
      \hyperlink{KLSA22}{$\mathsf{S22}$} & \hyperlink{KLA13}{$\mathsf{K13}$} & \eff{\B} & \\ 
      \hyperlink{KLSA23}{$\mathsf{S23}$} & \hyperlink{KLA14}{$\mathsf{K14}$} & \ari{\B} & \non{\P} \\
      \bottomrule
    \end{tabular}
    \hspace{2cm}
    \begin{tabular}{l|l*{2}{|>{$}c<{$}}}\toprule
      \multicolumn{1}{c|}{$\s$} & \multicolumn{1}{c|}{$\k$} & \multicolumn{2}{c}{$V \subset \h$}\\
      \toprule
      \hyperlink{KLSA24}{$\mathsf{S24}$} & \hyperlink{KLA14}{$\mathsf{K14}$} & \ari{\B} & \non{\P} \\
      \hyperlink{KLSA25}{$\mathsf{S25}$} & \hyperlink{KLA14}{$\mathsf{K14}$} & \ari{\B} & \ari{\P} \\
      \hyperlink{KLSA26}{$\mathsf{S26}$} & \hyperlink{KLA14}{$\mathsf{K14}$} & \ari{\B} & \ari{\P} \\
      \hyperlink{KLSA27}{$\mathsf{S27}$} & \hyperlink{KLA14}{$\mathsf{K14}$} & \eff{\B} & \ari{\P} \\
      \hyperlink{KLSA28}{$\mathsf{S28}$} & \hyperlink{KLA14}{$\mathsf{K14}$} & \eff{\B} & \ari{\P} \\
      \hyperlink{KLSA29}{$\mathsf{S29}$} & \hyperlink{KLA15}{$\mathsf{K15}$} & \non{\P} & \\ 
      \hyperlink{KLSA30}{$\mathsf{S30}$} & \hyperlink{KLA15}{$\mathsf{K15}$} & \non{\P} & \\ 
      \hyperlink{KLSA31}{$\mathsf{S31}$} & \hyperlink{KLA15}{$\mathsf{K15}$} & \ari{\P} & \\ 
      \hyperlink{KLSA32}{$\mathsf{S32}$} & \hyperlink{KLA15}{$\mathsf{K15}$} & \ari{\P} & \\ 
      \hyperlink{KLSA33}{$\mathsf{S33}_{\lambda\in\RR}$} & \hyperlink{KLA16}{$\mathsf{K16}$} & \ari{\B} & \non{\P} \\
      \hyperlink{KLSA34}{$\mathsf{S34}$} & \hyperlink{KLA17}{$\mathsf{K17}$} & \non{\P} & \\
      \hyperlink{KLSA35}{$\mathsf{S35}_{\lambda\in\RR}$} & \hyperlink{KLA18}{$\mathsf{K18}$} & \non{\P} &\\
      \bottomrule    
    \end{tabular}
  }
  \caption*{The blue pairs (e.g., \eff{\scriptsize\B}) are effective; the
    green pairs (e.g., \ari{\scriptsize\B}) though not effective, give rise to
    Aristotelian superspaces; whereas the greyed out pairs (e.g.,
    \non{\scriptsize\B}) are not effective and will not be considered further.}
\end{table}

\subsection{Effective Super Lie Pairs}
\label{sec:effective-super-lie}

Recall that a super Lie pair $(\s,\h)$ is said to be
\emph{effective} if $\h$ does not contain an ideal of $\s$.  Since
$\h \subset \k$ and contains the rotational subalgebra, which has
non-vanishing brackets with $\Q$, the only possible ideal of $\s$
contained in $\h$ would be the vectorial subspace $V \subset \h$.  It
is then a simple matter to inspect the super Lie pairs determined in
the previous section and selecting those for which $V$ is not an ideal of
$\s$.  Those super Lie pairs have been highlighted in blue in
Table~\ref{tab:super-lie-pairs}.  Additionally, we highlight in green the 
non-effective super Lie pairs that can give rise to Aristotelian superspaces.  
Though we could ignore these cases, leaving the identification of Aristotelian 
superspaces to Section~\ref{sec:arist-super-lie}, we identify them here for completeness. 
\\ \\
We now take each effective super Lie pair
in turn, change basis if needed so that $V$ is spanned by $\B$, and
then list the resulting brackets in that basis.  Every such super Lie
pair $(\s,\h)$ determines a Lie pair $(\k,\h)$.  If the Lie pair
$(\k,\h)$ is effective (and geometrically realisable), then $(\s,\h)$
describes a homogeneous superisation of one of the spatially-isotropic
homogeneous spacetimes in Table~\ref{tab:spacetimes}.  We remark
that there are effective super Lie pairs $(\s,\h)$ for which the
underlying Lie pair $(\k,\h)$ is not effective.  In those cases, there
are no boosts on the body of the superspacetime, but instead there
are R-symmetries in the odd coordinates.
\\ \\
As usual, in writing the Lie brackets of $\s$ below, we do not include
any bracket involving $\J$, which are given in
equation~\eqref{eq:superkinematical_brackets_general}, and instead give any non-zero
additional brackets.

\subsubsection{Galilean Superspaces}
\label{sec:super-g}

Galilean spacetime is described by $(\k,\h)$, where $\k$ has the
additional bracket $[\bH,\B] = - \P$.  There are two possible
superisations $(\s,\h)$, with brackets
\begin{equation}
  [\sH, \sQ(s)] =
  \begin{cases}
    \sQ(s\kk) \\ 0
  \end{cases} \quad\text{and}\quad
  [\sQ(s), \sQ(s)] = - \sP(s\kk\sbar).
\end{equation}
These are associated with Lie superalgebras \hyperlink{KLSA7}{$\mathsf{S7}$}
and \hyperlink{KLSA8}{$\mathsf{S8}$} in Table~\ref{tab:klsa}.

\subsubsection{Galilean de~Sitter Superspace}
\label{sec:super-dsg}

Galilean de~Sitter spacetime is described by $(\k,\h)$, where $\k$ has the
additional brackets $[\bH,\B] = - \P$ and $[\bH,\P] = -\B$.  There are two
one-parameter families of superisations $(\s,\h)$, with brackets
\begin{equation}
  [\sH, \sQ(s)] =\sQ(\tfrac12 s (\pm 1+\lambda\kk)) \quad\text{and}\quad
  [\sQ(s), \sQ(s)] = - \tfrac12 (\sB(s\kk\sbar) \minusplus \sP(s\kk\sbar))
\end{equation}
for $\lambda \in \RR$.  They are associated with Lie superalgebras
\hyperlink{KLSA9}{$\mathsf{S9}_{\gamma=-1,\lambda}$} and
\hyperlink{KLSA10}{$\mathsf{S10}_{\gamma=-1,\lambda}$}, respectively.

\subsubsection{Torsional Galilean de~Sitter Superspaces}
\label{sec:super-tdsg}

Torsional Galilean de~Sitter spacetime is described by $(\k,\h)$, where
$\k$ has the additional brackets $\b[H,\B] = - \P$ and
$[\bH,\P] = \gamma \B + (1+ \gamma) \P$, where $\gamma\in(-1,1)$.  There
are two one-parameter families of superisations $(\s,\h)$, with brackets
\begin{equation}
  [\sH, \sQ(s)] =\sQ(\tfrac12 s (1+\lambda\kk)) \quad\text{and}\quad
  [\sQ(s), \sQ(s)] = \tfrac1{1-\gamma} (\gamma \sB(s\kk\sbar) + \sP(s\kk\sbar))
\end{equation}
and
\begin{equation}
  [\sH, \sQ(s)] =\sQ(\tfrac12 s (\gamma+\lambda\kk)) \quad\text{and}\quad
  [\sQ(s), \sQ(s)] = \tfrac1{\gamma-1} (\sB(s\kk\sbar) + \sP(s\kk\sbar))
\end{equation}
for $\lambda \in \RR$.  The associated Lie superalgebras are
\hyperlink{KLSA9}{$\mathsf{S9}_{\gamma,\lambda}$} and
\hyperlink{KLSA10}{$\mathsf{S10}_{\gamma,\lambda}$}, respectively.
\\ \\
For $\gamma=1$, with additional brackets $[\bH,\B] = -\P$ and $[\bH,\P] =
\B + 2 \P$, there is a one-parameter family of superisations, with brackets
\begin{equation}
  [\sH, \sQ(s)] =\sQ(\tfrac12 s (1+\lambda\kk)) \quad\text{and}\quad
  [\sQ(s), \sQ(s)] = \sB(s\kk\sbar) + \sP(s\kk\sbar).
\end{equation}
The associated Lie superalgebras are \hyperlink{KLSA12}{$\mathsf{S12}_{\lambda}$}.

\subsubsection{Galilean Anti-de~Sitter Superspace}
\label{sec:super-adsg}

Galilean anti-de~Sitter spacetime is described by $(\k,\h)$, where $\k$ has the
additional brackets $[\bH,\B] = -\P$ and $[\bH,\P] = \B$.  It admits a
superisation $(\s,\h)$, with brackets 
\begin{equation}
  [\sH, \sQ(s)] =\sQ(\tfrac12 s \jj) \quad\text{and}\quad
  [\sQ(s), \sQ(s)] = - \sB(s\ii\sbar) + \sP(s\kk\sbar),
\end{equation}
which corresponds to the Lie superalgebra \hyperlink{KLSA11}{$\mathsf{S11}_{\chi = 0}$}, after changing the sign of $\P$.

\subsubsection{Torsional Galilean Anti-de~Sitter Superspace}
\label{sec:super-tadsg}

Torsional Galilean anti-de~Sitter spacetime is described by $(\k,\h)$, where $\k$ has the
additional brackets $[\bH,\B] = \chi \B + \P$ and $[\bH,\P] = \chi \P -
\B$, where $\chi > 0$.  There is a unique superisation $(\s,\h)$, with brackets
\begin{equation}
  [\sH, \sQ(s)] =\sQ(\tfrac12 s (\chi + \jj)) \quad\text{and}\quad
  [\sQ(s), \sQ(s)] = - \sB(s\ii\sbar) - \sP(s\kk\sbar).
\end{equation}
For uniformity, we change basis so that $[\bH,\B] = -\P$ as for all
Galilean spacetimes.  Then the resulting super Lie pair $(\s,\h)$ is
determined by the brackets $[\bH,\B] = -\P$, $[\bH,\P] = (1+\chi^2)\B +
2\chi \P$ and, in addition,
\begin{equation}
  [\sH, \sQ(s)] =\sQ(\tfrac12 s (\chi + \jj)) \quad\text{and}\quad
  [\sQ(s), \sQ(s)] = \sB(s\kk(\chi + \jj)\sbar) + \sP(s\kk\sbar),
\end{equation}
corresponding to the Lie superalgebra \hyperlink{KLSA11}{$\mathsf{S11}_\chi$}.

\subsubsection{Carrollian Superspace}
\label{sec:super-c}

Carrollian spacetime is described by $(\k,\h)$, where $\k$ has the
additional brackets $[\B,\P] = \bH$.  It admits a superisation
$(\s,\h)$, with brackets
\begin{equation}
  [\sQ(s), \sQ(s)] = |s|^2 \sH,
\end{equation}
which corresponds to the Lie superalgebra \hyperlink{KLSA13}{$\mathsf{S13}$}.

\subsubsection{Minkowski Superspace}
\label{sec:super-m}

Minkowski superspace arises as a superisation of Minkowski
spacetime, described by $(\k,\h)$ with brackets $[\bH,\B] = -\P$,
$[\B,\P] = \bH$ and $[\B,\B] = -\J$ and, in addition,
\begin{equation}
  [\sB(\beta),\sQ(s)] = \sQ(\tfrac12\beta s \kk)
  \quad\text{and}\quad
  [\sQ(s),\sQ(s)] = |s|^2 \sH - \sP(s\kk\sbar).
\end{equation}
This is, of course, the Poincaré superalgebra \hyperlink{KLSA14}{$\mathsf{S14}$}.

\subsubsection{Carrollian Anti-de~Sitter Superspace}
\label{sec:super-adsc}

Carrollian anti-de~Sitter spacetime is described as $(\k,\h)$, where
the $\k$ brackets are given by $[\bH,\P] = \B$, $[\B,\P] = \bH$ and
$[\P,\P] = -\J$.  It admits a unique superisation $(\s,\h)$ with
brackets (we have rotated $\kk$ to $\ii$)
\begin{equation}
  [\sP(\pi), \sQ(s)] = \sQ(\tfrac12 \pi s\ii) \quad\text{and}\quad
  [\sQ(s),\sQ(s)] = |s|^2 \sH + \sB(s\ii\sbar).
\end{equation}
We remark that just as with Carrollian anti-de~Sitter and Minkowski
spacetimes, which are both homogeneous spacetimes of the Poincaré
group, their superisations have isomorphic supersymmetry algebras:
namely, the Poincaré superalgebra \hyperlink{KLSA14}{$\mathsf{S14}$}.

\subsubsection{Anti-de~Sitter Superspace}
\label{sec:super-ads}

Anti-de~Sitter spacetime is described kinematically as $(\k,\h)$ with
brackets
\begin{equation}
  [\bH,\B] =-\P, \quad [\bH,\P] = \B, \quad [\B,\P] = \bH, \quad [\B,\B]
  = -\J \quad\text{and}\quad [\P,\P] = -\J.
\end{equation}
It admits a unique superisation $(\s,\h)$, with additional brackets
(where we have rotated $(\ii,\jj,\kk) \mapsto (\kk,\ii,\jj)$ for uniformity)
\begin{gather}
    [\sH,\sQ(s)] = \sQ(\tfrac12 s \jj), \quad [\sB(\beta), \sQ(s)] =
    \sQ(\tfrac12 \beta s \kk), \quad [\sP(\pi),\sQ(s)] = \sQ(\tfrac12
    \pi s\ii) \nonumber \\
    \quad\text{and}\quad [\sQ(s),\sQ(s)] = |s|^2 \sH +
    \sJ(s\jj\sbar) + \sB(s\ii\sbar) - \sP(s\kk\sbar).
\end{gather}
The associated Lie superalgebra is \hyperlink{KLSA15}{$\mathsf{S15}$}, which
is isomorphic to $\osp(1|4)$.

\subsubsection{Super-Spacetimes Extending $\RR \times S^3$}
\label{sec:super-rxS3}

These correspond to the effective super Lie pairs associated with the
Lie superalgebras \hyperlink{KLSA21}{$\mathsf{S21}$} and
\hyperlink{KLSA22}{$\mathsf{S22}$}.  The super Lie pairs $(\s,\h)$ are
effective, but the underlying Lie pair $(\k,\h)$ is not.  Indeed, the
brackets of $\k$ are now $[\B,\B] = \B$ and $[\P,\P]= \J - \B$, from
where we see that $\B$ spans an ideal of $\k$; although not one of
$\s$, due to the brackets
\begin{equation}
  [\sB(\beta), \sQ(s)] = \sQ(\tfrac12\beta s) \quad\text{and}\quad
  [\sQ(s),\sQ(s)] = |s|^2 \sH,
\end{equation}
for $\s$ the Lie superalgebra \hyperlink{KLSA21}{$\mathsf{S21}$} or
\begin{equation}
  [\sH, \sQ(s)] = \sQ(\tfrac12 s \kk), \quad [\sB(\beta), \sQ(s)] =
  \sQ(\tfrac12\beta s) \quad\text{and}\quad [\sQ(s),\sQ(s)] = |s|^2
  \sH - \sB(s\kk\sbar),
\end{equation}
for $\s$ the Lie superalgebra \hyperlink{KLSA22}{$\mathsf{S22}$}.  In both
superspaces, $\B$ does not generate boosts but R-symmetries.  The
underlying spacetime in both cases is the Einstein static universe
$\RR \times S^3$.\footnote{The naming of this manifold may be a slight misnomer.  When referring to the Einstein static universe, we typically mean the Lorentzian manifold with topology $\RR \times S^3$; however, here, we refer to an Aristotelian manifold with the same topology.  This discrepancy is an artefact of how we defined the classification problem: we classify only effective (super) Lie pairs, and the Lorentz action on the Einstein static universe is not effective. Therefore, the Lorentzian description of this manifold does not appear in our classification; instead, we find an Aristotelian description of this manifold since the rotational Lie subgroup does act effectively.  In particular, there exists an $\SO(D)$-equivariant diffeomorphism between $\Kgr/\Hgr$ and $\Agr/\Rgr$, where $\Agr$ is the Lie group generated by the rotations and spatio-time translations and $\Rgr$ is the Lie subgroup generated by the rotations. A similar story holds for the $\RR \times H^3$ case in the next section. }

\subsubsection{Super-Spacetimes Extending $\RR \times H^3$}
\label{sec:super-rxH3}

These correspond to the effective super Lie pairs associated with the
Lie superalgebras \hyperlink{KLSA17}{$\mathsf{S17}$} and
\hyperlink{KLSA18}{$\mathsf{S18}$}.  The super Lie pairs $(\s,\h)$ are
effective, but the underlying Lie pair $(\k,\h)$ is not.  Indeed, the
brackets of $\k$ are $[\B,\B] = \B$ and $[\P,\P] = \B - \J$, so that
$\B$ spans an ideal $\v\subset \k$.  The resulting Aristotelian
spacetime $(\k/\v,\r)$ is the hyperbolic version of the Einstein
static universe \hyperlink{A23m}{$\RR \times H^3$}.
\\ \\
For $\s$ the Lie superalgebra \hyperlink{KLSA17}{$\mathsf{S17}$}, the brackets are
\begin{equation}
  [\sB(\beta), \sQ(s)] = \sQ(\tfrac12\beta s) \quad\text{and}\quad
  [\sQ(s),\sQ(s)] = |s|^2 \sH,
\end{equation}
so that $\B$ does not span an ideal of $\s$.  In other words, $\B$ does
not generate boosts in the underlying homogeneous spacetime, but
rather R-symmetries.
\\ \\
A similar story holds for $\s$ the Lie superalgebra
\hyperlink{KLSA18}{$\mathsf{S18}$}, with the additional brackets
\begin{equation}
  [\sH, \sQ(s)] = \sQ(\tfrac12 s \kk), \quad [\sB(\beta), \sQ(s)] =
  \sQ(\tfrac12\beta s) \quad\text{and}\quad [\sQ(s),\sQ(s)] = |s|^2
  \sH - \sB(s\kk\sbar).
\end{equation}
Again, the generator $\B$ is to be interpreted as an R-symmetry.

\subsubsection{Super-Spacetimes Extending the Static Aristotelian Spacetime}
\label{sec:super-S}

This corresponds to the Lie superalgebras \hyperlink{KLSA27}{$\mathsf{S27}$}
and \hyperlink{KLSA28}{$\mathsf{S28}$}.  In either case the resulting super
Lie pair $(\s,\h)$ is effective, but the underlying Lie pair $(\k,\h)$
is not since $[\B,\B] = \B$ spans an ideal of $\k$.  The homogeneous
spacetime associated with the non-effective $(\k,\h)$ is the
Aristotelian static spacetime \hyperlink{A21}{$\zS$}.
\\ \\
As in the previous cases, the generators $\B$ do not act as boosts but
rather as R-symmetries, as evinced by the brackets:
\begin{equation}
  [\sB(\beta), \sQ(s)] = \sQ(\tfrac12\beta s) \quad\text{and}\quad
  [\sQ(s),\sQ(s)] = |s|^2 \sH
\end{equation}
for $\s$ the Lie superalgebra \hyperlink{KLSA27}{$\mathsf{S27}$}, or
\begin{equation}
  [\sH, \sQ(s)] = \sQ(\tfrac12 s\kk), \quad
  [\sB(\beta), \sQ(s)] = \sQ(\tfrac12\beta s) \quad\text{and}\quad
  [\sQ(s),\sQ(s)] = |s|^2 \sH - \sB(s\kk\sbar)
\end{equation}
for $\s$ the Lie superalgebra \hyperlink{KLSA28}{$\mathsf{S28}$}.

\subsection{Aristotelian Homogeneous Superspaces}
\label{sec:arist-super-lie}

The super Lie pairs $(\s,\h)$ in green in
Table~\ref{tab:super-lie-pairs} are such that the vectorial subspace
$V \subset \h$ is an ideal $\v$ of $\s$.  Quotienting $\s$ by this
ideal yields a Lie superalgebra $\sa \cong \s/\v$ with
$\a = \sa_{\bar 0}$ an Aristotelian Lie algebra
(see Table~\ref{tab:alas} for a classification).
The resulting Aristotelian super Lie pair $(\sa,\r)$ is effective by
construction and geometrically realisable.  It is then a simple matter
to identify the Aristotelian Lie superalgebra to which each of those
non-effective super Lie pairs in Table~\ref{tab:super-lie-pairs}
leads.  We summarise this in Table~\ref{tab:aristo-correspondence},
which exhibits the correspondence between Aristotelian super Lie pairs
in Table~\ref{tab:super-lie-pairs} and Aristotelian Lie superalgebras
in Table~\ref{tab:alsa}.  We identify the super Lie pair $(\s,\h)$ by
the label for $\s$ as in Table~\ref{tab:klsa} and the ideal
$\v \subset \h$.

\begin{table}[h!]
  \centering
  \caption{Correspondence Between Non-Effective Super Lie Pairs and
    Aristotelian Superalgebras}
  \label{tab:aristo-correspondence}
  \resizebox{\textwidth}{!}{
    \rowcolors{2}{blue!10}{white}
    \begin{tabular}{l|>{$}l<{$}|l}\toprule
      \multicolumn{1}{c|}{$\s$} & \multicolumn{1}{c|}{$\v$} & \multicolumn{1}{c}{$\sa$}\\\midrule
      \hyperlink{KLSA1}{$\mathsf{S1}$} & \B & \hyperlink{ALSA36}{$\mathsf{S36}$} \\
      \hyperlink{KLSA2}{$\mathsf{S2}$} & \B & \hyperlink{ALSA39}{$\mathsf{S39}$} \\
      \hyperlink{KLSA3}{$\mathsf{S3}$} & \B & \hyperlink{ALSA39}{$\mathsf{S39}$} \\
      \hyperlink{KLSA3}{$\mathsf{S3}$} & \P & \hyperlink{ALSA38}{$\mathsf{S38}$} \\
      \hyperlink{KLSA4}{$\mathsf{S4}$} & \B & \hyperlink{ALSA38}{$\mathsf{S38}$} \\
      \hyperlink{KLSA5}{$\mathsf{S5}$} & \B & \hyperlink{ALSA37}{$\mathsf{S37}$} \\
      \hyperlink{KLSA6}{$\mathsf{S6}$} & \B & \hyperlink{ALSA37}{$\mathsf{S37}$}\\
      \hyperlink{KLSA9}{$\mathsf{S9}_{\gamma\in[-1,1),\lambda\in\RR}$} & \B & \hyperlink{ALSA40}{$\mathsf{S40}_\lambda$} \\
      \hyperlink{KLSA9}{$\mathsf{S9}_{\gamma=1,\lambda\in\RR}$} & \B &  \hyperlink{ALSA40}{$\mathsf{S40}_\lambda$} \\
      \bottomrule
    \end{tabular}
    \hspace{1cm}
    \begin{tabular}{l|>{$}l<{$}|l} \toprule
      \multicolumn{1}{c|}{$\s$} & \multicolumn{1}{c|}{$\v$} & \multicolumn{1}{c}{$\sa$}\\\midrule
      \hyperlink{KLSA10}{$\mathsf{S10}_{\gamma\in[-1,0)\cup(0,1),\lambda\in\RR}$} & \P &  \hyperlink{ALSA40}{$\mathsf{S40}_\lambda$} \\
      \hyperlink{KLSA10}{$\mathsf{S10}_{\gamma=0,\lambda\neq 0}$} & \P & \hyperlink{ALSA36}{$\mathsf{S36}$} \\
      \hyperlink{KLSA10}{$\mathsf{S10}_{\gamma=0,\lambda= 0}$} & \P & \hyperlink{ALSA37}{$\mathsf{S37}$} \\
      \hyperlink{KLSA16}{$\mathsf{S16}$} & \B & \hyperlink{ALSA43}{$\mathsf{S43}$} \\
      \hyperlink{KLSA19}{$\mathsf{S19}$} & \B & \hyperlink{ALSA42}{$\mathsf{S42}$} \\
      \hyperlink{KLSA20}{$\mathsf{S20}$} & \B & \hyperlink{ALSA41}{$\mathsf{S41}$} \\
      \hyperlink{KLSA23}{$\mathsf{S23}$} & \B & \hyperlink{ALSA36}{$\mathsf{S36}$} \\
      \hyperlink{KLSA24}{$\mathsf{S24}$} & \B & \hyperlink{ALSA37}{$\mathsf{S37}$} \\
      \bottomrule
    \end{tabular}
    \hspace{1cm}
    \begin{tabular}{l|>{$}l<{$}|l} \toprule
      \multicolumn{1}{c|}{$\s$} & \multicolumn{1}{c|}{$\v$} & \multicolumn{1}{c}{$\sa$}\\\midrule
      \hyperlink{KLSA25}{$\mathsf{S25}$} & \B & \hyperlink{ALSA38}{$\mathsf{S38}$} \\
      \hyperlink{KLSA25}{$\mathsf{S25}$} & \P & \hyperlink{ALSA38}{$\mathsf{S38}$} \\
      \hyperlink{KLSA26}{$\mathsf{S26}$} & \B & \hyperlink{ALSA39}{$\mathsf{S39}$} \\
      \hyperlink{KLSA26}{$\mathsf{S26}$} & \P & \hyperlink{ALSA38}{$\mathsf{S38}$} \\
      \hyperlink{KLSA27}{$\mathsf{S27}$} & \P & \hyperlink{ALSA41}{$\mathsf{S41}$} \\
      \hyperlink{KLSA28}{$\mathsf{S28}$} & \P & \hyperlink{ALSA42}{$\mathsf{S42}$} \\
      \hyperlink{KLSA31}{$\mathsf{S31}$} & \P & \hyperlink{ALSA38}{$\mathsf{S38}$} \\
      \hyperlink{KLSA32}{$\mathsf{S32}$} & \P & \hyperlink{ALSA38}{$\mathsf{S38}$} \\
      \hyperlink{KLSA33}{$\mathsf{S33}_{\lambda\in\RR}$} & \B & \hyperlink{ALSA40}{$\mathsf{S40}_\lambda$} \\
      \bottomrule
    \end{tabular}
  }
\end{table}

\subsection{Summary}
\label{sec:summary-3}

Table~\ref{tab:superspaces} lists the homogeneous superspaces we have
classified in this section. Each superspacetime is a superisation of an
underlying spatially-isotropic, homogeneous (kinematical or
Aristotelian) spacetime, which we list in Table~\ref{tab:spacetimes}. Let us recall that
Table~\ref{tab:spacetimes} is divided into five sections,
corresponding to the different invariant structures which the
homogeneous spacetimes admit, as discussed in Chapter~\ref{chap:k_spaces}.  We
have a similar division of Table~\ref{tab:superspaces}: with the
superisations of spacetimes admitting a Lorentzian, Galilean,
Carrollian, Aristotelian (with R-symmetries) and Aristotelian (without
R-symmetries) structures, respectively.  All spacetimes admit
superisations with the exception of the Riemannian spaces, de~Sitter
spacetime ($\hyperlink{S2}{\zdS}_4$) and two of the Carrollian
spacetimes: Carrollian de~Sitter ($\hyperlink{S14}{\zdSC}$) and the
Carrollian light-cone ($\hyperlink{S16}{\zLC}$).

\begin{table}[h!]
  \centering
  \caption{Simply-Connected Spatially-Isotropic Homogeneous Superspaces}
  \label{tab:superspaces}
  \setlength{\extrarowheight}{2pt}
  \rowcolors{2}{blue!10}{white}
  \begin{tabular}{l|l|l|l*{4}{|>{$}c<{$}}}\toprule
    \multicolumn{1}{c|}{SM\#} & \multicolumn{1}{c|}{$\M$} & \multicolumn{1}{c|}{$\s$} & \multicolumn{1}{c|}{$\k$ (or $\a$)} & \multicolumn{1}{c|}{$\hh$}& \multicolumn{1}{c|}{$\bb$} & \multicolumn{1}{c|}{$\pp$} & \multicolumn{1}{c}{$[\sQ(s),\sQ(s)]$} \\
    \toprule
    \hypertarget{SM1}{1} & \hyperlink{S1}{$\MM^4$} & \hyperlink{KLSA14}{$\mathsf{S14}$} & \hyperlink{KLA8}{$\mathsf{K8}$} & & \tfrac12 \kk & & |s|^2 \sH - \sP(s\kk\sbar) \\
    \hypertarget{SM2}{2} & \hyperlink{S3}{$\zAdS_4$} & \hyperlink{KLSA15}{$\mathsf{S15}$} & \hyperlink{KLA11}{$\mathsf{K11}$} & \tfrac12 \jj & \tfrac12 \kk & \tfrac12 \ii & |s|^2 \sH + \sJ(s\jj\sbar) + \sB(s\ii\sbar) - \sP(s\kk\sbar) \\
    \midrule
    \hypertarget{SM3}{3} & \hyperlink{S7}{$\zG$} & \hyperlink{KLSA7}{$\mathsf{S7}$} & \hyperlink{KLA2}{$\mathsf{K2}$} & \kk & & & -\sP(s\kk\sbar) \\
    \hypertarget{SM4}{4} & \hyperlink{S7}{$\zG$} & \hyperlink{KLSA8}{$\mathsf{S8}$} & \hyperlink{KLA2}{$\mathsf{K2}$} & & & & -\sP(s\kk\sbar)  \\
    \hypertarget{SM5}{5$_{\lambda\in\RR}$} & \hyperlink{S8}{$\zdSG$} & \hyperlink{KLSA9}{$\mathsf{S9}_{-1,\lambda}$}& \hyperlink{KLA3}{$\mathsf{K3}_{-1}$} & \tfrac12 (1 + \lambda \kk) & & & -\tfrac12 (\sB(s\kk\sbar) - \sP(s\kk\sbar)) \\      
    \hypertarget{SM6}{6$_{\lambda\in\RR}$} & \hyperlink{S8}{$\zdSG$} & \hyperlink{KLSA10}{$\mathsf{S10}_{-1,\lambda}$} & \hyperlink{KLA3}{$\mathsf{K3}_{-1}$} & \tfrac12 (-1 + \lambda \kk) & & & -\tfrac12 (\sB(s\kk\sbar) + \sP(s\kk\sbar))  \\
    \hypertarget{SM7}{7$_{\gamma\in(-1,1),\lambda\in\RR}$} & \hyperlink{S9}{$\ztdSG_\gamma$} & \hyperlink{KLSA9}{$\mathsf{S9}_{\gamma,\lambda}$} & \hyperlink{KLA3}{$\mathsf{K3}_\gamma$} & \tfrac12 (1 + \lambda \kk) & & & \tfrac{1}{1-\gamma}(\gamma\sB(s\kk\sbar) + \sP(s\kk\sbar)) \\
    \hypertarget{SM8}{8$_{\gamma\in(-1,1),\lambda\in\RR}$} & \hyperlink{S9}{$\ztdSG_\gamma$} &  \hyperlink{KLSA10}{$\mathsf{S10}_{\gamma,\lambda}$} & \hyperlink{KLA3}{$\mathsf{K3}_\gamma$} & \tfrac12 (\gamma + \lambda \kk) & & & \tfrac1{\gamma-1}(\sB(s\kk\sbar) + \sP(s\kk\sbar)) \\
    \hypertarget{SM9}{9$_{\lambda\in\RR}$} & \hyperlink{S9}{$\ztdSG_{\gamma=1}$} & \hyperlink{KLSA12}{$\mathsf{S12}_\lambda$} & \hyperlink{KLA3}{$\mathsf{K3}_1$} & \tfrac12 (1 + \lambda \kk) & & & \sB(s\kk\sbar) + \sP(s\kk\sbar) \\
    \hypertarget{SM10}{10} & \hyperlink{S10}{$\zAdSG$} & \hyperlink{KLSA11}{$\mathsf{S11}_0$} & \hyperlink{KLA4}{$\mathsf{K4}_0$} & \tfrac12 \jj & & & -\sB(s\ii\sbar) + \sP(s\kk\sbar) \\
    \hypertarget{SM11}{11$_{\chi>0}$} & \hyperlink{S11}{$\ztAdSG_\chi$} & \hyperlink{KLSA11}{$\mathsf{S11}_\chi$} & \hyperlink{KLA4}{$\mathsf{K4}_\chi$} & \tfrac12 (\chi + \jj) & & & \sB(s\kk(\chi + \jj)\sbar) + \sP(s\kk\sbar)  \\
    \midrule
    \hypertarget{SM12}{12} & \hyperlink{S13}{$\zC$} & \hyperlink{KLSA13}{$\mathsf{S13}$} & \hyperlink{KLA6}{$\mathsf{K6}$} & & & & |s|^2 \sH \\
    \hypertarget{SM13}{13} & \hyperlink{S15}{$\zAdSC$} & \hyperlink{KLSA14}{$\mathsf{S14}$} & \hyperlink{KLA8}{$\mathsf{K8}$} & & & \tfrac12 \ii & |s|^2 \sH + \sB(s\ii\sbar) \\
    \midrule
    \hypertarget{SM14}{14} & \hyperlink{A23m}{$\RR \times H^3$} & \hyperlink{KLSA17}{$\mathsf{S17}$} & \hyperlink{KLA12}{$\mathsf{K12}$} & & \tfrac12 & & |s|^2 \sH \\
    \hypertarget{SM15}{15} & \hyperlink{A23m}{$\RR \times H^3$} & \hyperlink{KLSA18}{$\mathsf{S18}$} & \hyperlink{KLA12}{$\mathsf{K12}$} & \tfrac12 \kk & \tfrac12 & & |s|^2 \sH - \sB(s \kk \sbar) \\
    \hypertarget{SM16}{16} & \hyperlink{A23p}{$\RR \times S^3$} & \hyperlink{KLSA21}{$\mathsf{S21}$} & \hyperlink{KLA13}{$\mathsf{K13}$} & & \tfrac12 & & |s|^2 \sH \\
    \hypertarget{SM17}{17} & \hyperlink{A23p}{$\RR \times S^3$} & \hyperlink{KLSA22}{$\mathsf{S22}$} & \hyperlink{KLA13}{$\mathsf{K13}$} & \tfrac12 \kk & \tfrac12 & & |s|^2 \sH - \sB(s\kk\sbar) \\
    \hypertarget{SM18}{18} & \hyperlink{A21}{$\zS$} & \hyperlink{KLSA27}{$\mathsf{S27}$} & \hyperlink{KLA14}{$\mathsf{K14}$} & & \tfrac12 & & |s|^2 \sH \\
    \hypertarget{SM19}{19} & \hyperlink{A21}{$\zS$} & \hyperlink{KLSA28}{$\mathsf{S28}$} & \hyperlink{KLA14}{$\mathsf{K14}$} & \tfrac12 \kk & \tfrac12 & & |s|^2 \sH - \sB(s\kk\sbar) \\
    \midrule
    \hypertarget{SM20}{20} &  \hyperlink{A21}{$\zS$} & \hyperlink{ALSA36}{$\mathsf{S36}$} & \hyperlink{ALA1}{$\mathsf{A1}$} & \kk & - & & - \sP(s\kk\sbar) \\
    \hypertarget{SM21}{21} &  \hyperlink{A21}{$\zS$} & \hyperlink{ALSA37}{$\mathsf{S37}$} & \hyperlink{ALA1}{$\mathsf{A1}$} & & -& & - \sP(s\kk\sbar)  \\
    \hypertarget{SM22}{22} &  \hyperlink{A21}{$\zS$} & \hyperlink{ALSA38}{$\mathsf{S38}$} & \hyperlink{ALA1}{$\mathsf{A1}$} & & - & & |s|^2 \sH  \\
    \hypertarget{SM23}{23} & \hyperlink{A21}{$\zS$} & \hyperlink{ALSA39}{$\mathsf{S39}$} & \hyperlink{ALA1}{$\mathsf{A1}$} & & - & & |s|^2 \sH - \sP(s\kk\sbar)  \\
    \hypertarget{SM24}{24$_{\lambda\in\RR}$} &  \hyperlink{A22}{$\zTS$} & \hyperlink{ALSA40}{$\mathsf{S40}_\lambda$} & \hyperlink{ALA2}{$\mathsf{A2}$} & \tfrac12(1 + \lambda \kk) & - & & -\sP(s\kk\sbar)  \\
    \hypertarget{SM25}{25} & \hyperlink{A23p}{$\RR\times S^3$} & \hyperlink{ALSA41}{$\mathsf{S41}$} & \hyperlink{ALA3p}{$\mathsf{A3}_+$} & & - & \tfrac12 & |s|^2 \sH  \\
    \hypertarget{SM26}{26} & \hyperlink{A23p}{$\RR\times S^3$} & \hyperlink{ALSA42}{$\mathsf{S42}$} & \hyperlink{ALA3p}{$\mathsf{A3}_+$} &\kk & - & \tfrac12 & |s|^2 \sH - \sJ(s\kk\sbar) - \sP(s\kk\sbar)  \\
    \hypertarget{SM27}{27} & \hyperlink{A23m}{$\RR\times H^3$} & \hyperlink{ALSA43}{$\mathsf{S43}$} & \hyperlink{ALA3m}{$\mathsf{A3}_-$} & & - & \tfrac12\ii &  \sJ(s\jj\sbar) - \sP(s\kk\sbar)  \\
    \bottomrule
  \end{tabular}
  \caption*{The first column is our identifier for the superspace,
    whereas the second column is the underlying homogeneous spacetime
    it superises.  The next two columns are the isomorphism classes of
    kinematical Lie superalgebra and kinematical Lie algebra,
    respectively.  The next columns specify the brackets
    of $\s$ not of the form $[\J,-]$ in a basis where $\h$ is spanned
    by $\J$ and $\B$.  As explained in Section~\ref{subsec:math_prelims_alg_klsa},
    supercharges $\sQ(s)$ are parametrised by $s \in \HH$, whereas
    $\sJ(\omega)$, $\sB(\beta)$ and $\sP(\pi)$ are parametrised by
    $\omega,\beta,\pi \in \Im\HH$. The brackets are given by
    $[\sH,\sQ(s)] = \sQ(s\hh)$, $[\sB(\beta),\sQ(s)]=\sQ(\beta s \bb)$
    and $[\sP(\pi),\sQ(s)] = \sQ(\pi s \pp)$, for some
    $\hh,\bb,\pp\in\HH$. The table is divided into five sections from
    top to bottom: Lorentzian, Galilean, Carrollian, Aristotelian with
    R-symmetries and Aristotelian.}
\end{table}
 
\subsection{Low-Rank Invariants}
\label{sec:low-rank-invariants}

In this section, we exhibit the low-rank invariants of the homogeneous
superspaces in Table~\ref{tab:superspaces}, all of which are
reductive.  Indeed, a homogeneous supermanifold with super Lie pair
$(\s,\h)$, where $\h \subset \k = \s_{\bar 0}$, is reductive if and
only the underlying homogeneous manifold $(\k, \h)$ is also reductive.  This is
because if $\k = \h \oplus \m$ is a reductive split, then so is
$\s = \h \oplus (\m \oplus S)$, with $S = \s_{\bar 1}$: the bracket
$[\h,\m] \subset \m$ because $(\k,\h)$ is reductive and the bracket
$[\h, S] \subset S$ because $\h \in \s_{\bar 0}$ and
$S = \s_{\bar 1}$. In \cite{Figueroa-OFarrill:2018ilb}, it is shown
that all the homogeneous spacetimes in Table~\ref{tab:spacetimes} are
reductive with the exception of the Carrollian light cone $\zLC$,
which in any case does not admit any $(4|4)$-dimensional
superisation. Hence all the superspaces in Table~\ref{tab:superspaces}
are reductive.
\\ \\
Let $(\s,\h)$ be the super Lie pair associated with one of the
homogeneous superspaces in Table~\ref{tab:superspaces}.  We will write
$\s = \h \oplus \m$, where we have promoted $\m$ to a vector
superspace $\m = \m_{\bar 0} \oplus \m_{\bar 1}$, with $\k = \h
\oplus \m_{\bar 0}$ a reductive split and $\m_{\bar 1} = \s_{\bar 1} =
S$.
\\ \\
Invariant tensors on the simply-connected superspace with super Lie
pair $(\s,\h)$ are in one-to-one correspondence with $\h$-invariant
tensors on $\m$.  Since $\h$ contains the rotational subalgebra $\r
\cong \so(3)$, $\h$-invariant tensors are in particular also
rotationally invariant.  It is not difficult to write down the
rotationally invariant tensors of low order.
\\ \\
As an $\r$ module, $\m = \RR \oplus V \oplus S$, where $\RR$ is the
trivial one-dimensional module, $V$ is the vector
three-dimensional module and $S$ is the spinor
four-dimensional module.  Under the isomorphism $\r = \sp(1) =
\Im \HH$, $\m = \RR \oplus \Im \HH \oplus 
\HH$, where the integrated action of a unit-norm quaternion $\uu \in
\Sp(1)$ on $(h, p, s) \in \m$ is given by
\begin{equation}
  \uu \cdot (h, p, s) = (h, \uu p \bar \uu, \uu s).
\end{equation}
Let $\bH, \bP_i, \bQ_a$ denote a basis for $\m$, where $\bP_i$ and $\bQ_a$ have
been defined in equation \eqref{eq:quat-basis-s}.  We let $\eta,
\pi^i, \theta^a$ denote the canonically dual basis for $\m^*$.
There is a rotationally invariant line in $\m$: namely, the span of
$\bH$, which lives in $\m_{\bar 0}$.  Dually, there is a rotationally invariant
line in $\m^*$, which is the span of $\eta$.  These are all the
rotationally invariant tensors of rank $1$.
\\ \\
Let us now consider rank $2$.  As an $\Sp(1)$ module, $\m
\otimes \m$ has the following invariants.  First of all, we have
$\bH^2$, which is the only invariant featuring $\bH$.  Another invariant
is $\P^2 := \sum_i \bP_i \otimes \bP_i$, which corresponds to the
$\Sp(1)$-invariant inner product  $\left<-,-\right> : \Im\HH \times
\Im\HH \to \RR$ given by $\left<\alpha,\beta\right> =
\Re(\alpha \bar\beta) = - \Re(\alpha\beta)$.  If $\qq \in \HH$ is any
quaternion, the real bilinear form
\begin{equation}
  \omega_{\qq} : \HH \to \HH \to \RR \quad\text{defined by}\quad
  \omega_{\qq}(s_1,s_2) = \Re(s_1 \qq \sbar_2)
\end{equation}
is $\Sp(1)$-invariant: symmetric if $\qq$ is real and symplectic if $\qq$
is imaginary (and non-zero).  This gives rise to four
$\Sp(1)$-invariants quadratic in $\Q$: $\sum_a \bQ_a \otimes \bQ_a$ and
the triplet $\sum_{a,b} I_{ab} \bQ_a \otimes \bQ_b$,
$\sum_{a,b} J_{ab} \bQ_a \otimes \bQ_b$ and
$\sum_{a,b} K_{ab} \bQ_a \otimes \bQ_b$, where $I,J,K$ are the matrices
representing right-multiplication by the quaternions $\ii$, $\jj$,
$\kk$; that is,
\begin{equation}
  \sQ(s\ii) = \sum_{a,b=1}^4 \bQ_a I_{ab} s_b, \quad   \sQ(s\jj) =
  \sum_{a,b=1}^4 \bQ_a J_{ab} s_b \quad\text{and}\quad
  \sQ(s\kk) = \sum_{a,b=1}^4 \bQ_a K_{ab} s_b.
\end{equation}
Similarly there are several rotational invariants in
$\m^* \otimes \m^*$: $\eta^2$ and, in addition, the symmetric tensors
$\pi^2$ and $\theta^2$, and the triplet of symplectic forms
$\omega_I$, $\omega_J$ and $\omega_K$, defined as follows:
\begin{equation}
  \begin{split}
    \pi^2(\sP(\alpha'),\sP(\alpha)) &= \Re(\alpha' \bar\alpha) = - \Re(\alpha'\alpha)\\
    \theta^2(\sQ(s'), \sQ(s)) &= \Re(s'\sbar)\\
    \omega_I(\sQ(s'), \sQ(s)) &= \Re(s'\ii\sbar)\\
    \omega_J(\sQ(s'), \sQ(s)) &= \Re(s'\jj\sbar)\\
    \omega_K(\sQ(s'), \sQ(s)) &= \Re(s'\kk\sbar).
  \end{split}
\end{equation}
To investigate the invariant tensors on $(\s,\h)$, we need to
investigate the action of $\B$ on the tensors.  For the classical
invariants (i.e., those not involving $\bQ_a$ or $\theta^a$), we may
consult \cite{Figueroa-OFarrill:2018ilb}: the Lorentzian metric (and
the corresponding cometric) are invariant for the Lorentzian
spacetimes, the clock one-form and spatial cometric for the Galilean
spacetimes, the Carrollian vector and the spatial metric for the
Carrollian spacetimes.  The generators $\B$ act trivially on
Aristotelian spacetimes, so the rotationally invariant tensors are the
invariant tensors.  For the invariants involving $\bQ_a$ or $\theta^a$,
we need to examine how $\B$ acts on $S$.
\\ \\
As can be gleaned from Table~\ref{tab:superspaces}, $\B$ acts
trivially on $\Q$ in most cases.  The exceptions are Minkowski and AdS
superspaces and the Aristotelian superspaces where $\B$ acts via
R-symmetries. Hence, in all other superspaces, the four rotational
invariants in $\m_{\bar 1} \otimes \m_{\bar 1}$ defined above and
$\theta^2$, $\omega_I$, $\omega_J$ and $\omega_K$ in
$\m^*_{\bar 1} \otimes \m^*_{\bar 1}$ are $\h$-invariant.  This
situation continues to hold for the Aristotelian superspaces with
R-symmetry, namely
\hyperlink{SM14}{$\mathsf{SM14}$}--\hyperlink{SM19}{$\mathsf{SM19}$}.
Indeed, one can show that all the rotational invariants which are
quadratic in $\Q$ or in the $\theta^a$ are also R-symmetry invariant.
Indeed, the R-symmetry generator $\bB_i$ acts on $\m_{\bar 1}$ in the
same way as the infinitesimal rotation generator $\bJ_i$.
\\ \\
Hence it is only for \hyperlink{SM1}{Minkowski} and
\hyperlink{SM2}{$\zAdS$} superspaces that the $\h$-invariants do not agree
with the $\r$-invariants.  For both of these superspaces, $\h \cong
\so(3,1)$, acting in the same way on the spinors:
\begin{equation}
  [\sB(\beta), \sQ(s)] = \sQ(\tfrac12 \beta s \kk).
\end{equation}
It is a simple calculation to see that the following are
$\h$-invariant: 
$\sum_{a,b} I_{ab} \bQ_a \otimes \bQ_b$,
$\sum_{a,b} J_{ab} \bQ_a \otimes \bQ_b$, $\omega_I$ and $\omega_J$.
\\ \\
Since $\h$ is isomorphic to the Lorentz subalgebra,
we recover the well-known fact that there are two independent
Lorentz-invariant symplectic structures on the Majorana spinors.  This
does not contradict the fact that the Majorana spinor representation
$S$ of $\so(3,1)$ is irreducible as a \emph{real} representation,
since its complexification (the Dirac spinor representation)
decomposes as a direct sum of the two Weyl spinor representations,
each one having a Lorentz-invariant symplectic structure.

\section{Limits Between Superspaces}
\label{sec:limits-betw-supersp}

In this section, we exhibit some limits between the superspaces in
Table~\ref{tab:superspaces} and interpret them in terms of
contractions of the underlying Lie superalgebras.
\\ \\
As we will show, a limit between two superspaces induces a limit of
the underlying homogeneous spacetimes.  These were discussed in
Section~\ref{subsec:ks_kss_gls}.  Our discussion will closely follow 
that in Section~\ref{subsec:ks_kss_gls}.  There, contractions of a
Lie algebra $\g = (V, \phi)$, where $V$ is a finite-dimensional real
vector space and $\phi: \wedge^2 V \to V$ is a linear map satisfying
the Jacobi identity, were defined as limits of curves in the space of
Lie brackets.  If $g: (0, 1] \to GL(V)$, mapping $t \mapsto g_t$, is a
continuous curve with $g_1 = \id_V$, we can define a curve of
isomorphic Lie algebras $(V,\phi_t)$, where
\begin{equation}
  \phi_t(X,Y) := \left(g^{-1}_t\cdot\phi \right)(X,Y) = g^{-1}_t \left(\phi(g_t X,
  g_t Y)\right).
\end{equation}
If the limit $\phi_0 = \lim_{t\to 0} \phi$ exists, it defines a Lie algebra
$\g_0 = (V, \phi_0)$, which is then a contraction of $\g=(V,\phi_1)$.
\\ \\
In the current case, we will contract Lie superalgebras $\s = (V,
\phi)$, where $V$ is now a real finite-dimensional super vector space
and $\phi: \wedge^2 V \to V$ is a linear map, where
$\wedge^2$ is defined in the super sense, satisfying the super-Jacobi
identity.  We will define contractions of $\s$ in a completely
analogous manner.

\subsection{Contractions of the AdS Superalgebra}
\label{sec:ads-limits}

We begin with the superalgebra for the AdS superspace
\hyperlink{SM2}{$\mathsf{SM2}$}, whose generators $\J$, $\B$, $\P$,
$\bH$ and $\Q$ satisfy the following brackets (in our abbreviated notation):
\begin{equation}
  \begin{aligned}[m]
    [\J, \J] &= \J \\
    [\J, \B] &= \B \\
    [\J, \P] &= \P \\
    [\J, \Q] &= \Q
  \end{aligned}
  \qquad\qquad
  \begin{aligned}[m]
    [\bH, \B] &= -\P \\
    [\bH, \P] &= \B \\
    [\B, \P] &= \bH \\
    [\B, \B] &= -\J \\
    [\P, \P] &= -\J
  \end{aligned} \qquad\qquad
  \begin{aligned}[m]
    [\bH, \Q] &= \Q \\
    [\B, \Q] &= \Q \\
    [\P, \Q] &= \Q \\
    [\Q, \Q] &= H + \J + \B - \P.
  \end{aligned}
\end{equation}
Consider the following three-parameter family of
linear transformations $g_{\kappa, c, \tau}$ defined by
\begin{equation}
g_{\kappa, c, \tau}\cdot \J = \J, \qquad g_{\kappa, c, \tau}
\cdot\B = \tfrac{\tau}{c} \B, \qquad g_{\kappa, c, \tau}\cdot
\P = \tfrac{\kappa}{c} \P, \qquad g_{\kappa, c, \tau}\cdot \bH =
\tau\kappa \bH, \qquad g_{\kappa, c, \tau}\cdot\Q = \tfrac{\kappa\tau}{c}\Q. 
\end{equation}
The action on the even generators is as in
Section~\ref{subsec:ks_kss_gls} and the action on $\Q$ is chosen to ensure that the
bracket $[\Q, \Q]$ has well-defined limits as $\kappa \to 0$, $c \to
\infty$ or $\tau\to 0$.
\\ \\
The brackets involving $\J$ remain unchanged for 
the above transformations and the remaining brackets
become
\begin{equation}
  \begin{aligned}[m]
    [\bH, \B] &= -\tau^2 \P \\
    [\bH, \P] &= \kappa^2 \B \\
    [\B, \P] &= \tfrac{1}{c^2} \bH
  \end{aligned} \qquad\qquad
  \begin{aligned}[m]
    [\B, \B] &= -\tfrac{\tau^2}{c^2}\J \\
    [\P, \P] &= -\tfrac{\kappa^2}{c^2}\J\\
    [\bH, \Q] &= \kappa\tau\Q
  \end{aligned} \qquad\qquad
  \begin{aligned}[m]
    [\B, \Q] &= \tfrac{\tau}{c}\Q \\
    [\P, \Q] &= \tfrac{\kappa}{c}\Q \\
    [\Q, \Q] &= \tfrac{1}{c} \bH + \tfrac{\kappa\tau}{c}\J + \kappa\B - \tau\P.
  \end{aligned}
\end{equation}
We now want to take the limits $\kappa \to 0$,
$c \to \infty$, and $\tau \to 0$ in turn,
corresponding to the flat, non-relativistic, and ultra-relativistic
limits, respectively. Notice that the limits of the brackets between
the even generators will produce the same Lie algebra contractions as
in Section~\ref{subsec:ks_kss_gls}. Thus we cannot have a limit from
one superspace to another unless there exists a limit between their
underlying homogeneous spacetimes.
\\ \\
Taking the flat limit $\kappa \to 0$, we are left with
\begin{equation}
[\bH, \B] = -\tau^2 \P, \quad [\B, \P] = \tfrac{1}{c^2} \bH, 
\quad [\B, \B] = -\tfrac{\tau^2}{c^2}\J, \quad
[\B, \Q] = \tfrac{\tau}{c}\Q \quad \text{and} \quad
 [\Q, \Q] = \tfrac{1}{c} \bH - \tau\P.
\end{equation}
For $\tfrac{\tau}{c}\neq 0$, this is the Poincaré superalgebra
(\hyperlink{KLSA14}{$\mathsf{S14}$}).  Thus, we obtain the limit
$\hyperlink{SM2}{\mathsf{SM2}} \to \hyperlink{SM1}{\mathsf{SM1}}$.
Subsequently taking the non-relativistic limit $c \to \infty$,
the brackets reduce to
\begin{equation}
[\bH, \B] = -\tau^2 \P  \qquad\text{and}\qquad [\Q, \Q] = - \tau\P.
\end{equation}
For $\tau \neq 0$, this shows us that we have the limit
$\hyperlink{SM1}{\mathsf{SM1}} \to
\hyperlink{SM4}{\mathsf{SM4}}$.
\\ \\
Alternatively, we could have taken the ultra-relativistic limit 
$\tau \to 0$, which, for $c \neq 0$, gives us the
Carroll superalgebra (\hyperlink{KLSA13}{$\mathsf{S13}$}):
\begin{equation}
[\B, \P] = \tfrac{1}{c^2} \bH  \qquad\text{and}\qquad [\Q, \Q] =
\tfrac{1}{c} H.
\end{equation}
Thus, we have $\hyperlink{SM1}{\mathsf{SM1}}\to
\hyperlink{SM12}{\mathsf{SM12}}$.
\\ \\
Returning to the AdS superalgebra
(\hyperlink{KLSA15}{$\mathsf{S15}$}) and taking the non-relativistic
limit $c \to \infty$, we find
\begin{equation}
[\bH, \B] = -\tau^2 \P, \qquad [H, \P] = \kappa^2 \B,
\qquad [\bH, \Q] = \kappa\tau\Q \quad\text{and}\quad
[\Q, \Q] = \kappa\B - \tau\P.
\end{equation}
For $\tau\kappa \neq 0$, this is $\hyperlink{KLSA11}{\mathsf{S11}_0}$
(under a suitable basis change).  Therefore, we have 
$\hyperlink{SM2}{\mathsf{SM2}} \to \hyperlink{SM10}{\mathsf{SM10}}$.
Because these limits commute, we may now take
the flat limit to arrive at \hyperlink{SM4}{$\mathsf{SM4}$}.  
\\ \\
Finally, we may take the ultra-relativistic limit of
$\zAdS$ (\hyperlink{KLSA15}{$\mathsf{S15}$}).  This limit leaves the
brackets
\begin{equation}
[\bH, \P] = \kappa^2 \B, \quad [\B, \P] = \tfrac{1}{c^2} \bH,
\quad [\P, \P] = -\tfrac{\kappa^2}{c^2}\J, \quad
[\P, \Q] = \tfrac{\kappa}{c}\Q \quad\text{and}\quad
[\Q, \Q] = \tfrac{1}{c} \bH + \kappa\B,
\end{equation}
for $\tfrac{\kappa}{c} \neq 0$. Thus, we arrive at
\hyperlink{SM13}{$\mathsf{SM13}$}. Subsequently taking the flat limit,
we find \hyperlink{SM12}{$\mathsf{SM12}$}, as expected.
\\ \\
We can also take limits from the superspaces discussed above to
non-effective super Lie pairs, which will have associated 
Aristotelian superspaces.  Since all of the above superspaces
have either \hyperlink{SM4}{$\mathsf{SM4}$} or \hyperlink{SM12}{$\mathsf{SM12}$}
as a limit, we will only show the limits to Aristotelian 
superspaces coming form these two cases.  Beginning 
with \hyperlink{SM4}{$\mathsf{SM4}$}, we can use the 
transformation
\begin{equation}
g_t\cdot\B = t\B, \quad g_t\cdot \bH = \bH, \quad g_t\cdot \P = \P
\quad\text{and}\quad g_t\cdot\Q = \Q
\end{equation}
and the limit $t\to 0$ to obtain \hyperlink{SM21}{$\mathsf{SM21}$}.
Using the same transformation and limit, we can also start with 
\hyperlink{SM12}{$\mathsf{SM12}$} and find \hyperlink{SM22}{$\mathsf{SM22}$}.

\subsection{Remaining Galilean Superspaces}
\label{sec:lim-galilean}

We have shown that we obtain the other Lorentzian and 
two Carrollian superspaces as limits of the AdS superspace
\hyperlink{SM2}{$\mathsf{SM2}$}: namely, Minkowski
(\hyperlink{SM1}{$\mathsf{SM1}$}), Carroll
(\hyperlink{SM12}{$\mathsf{SM12}$}) and Carrollian anti-de~Sitter
(\hyperlink{SM13}{$\mathsf{SM13}$}) superspaces.  In addition, we also
obtain two superisations of Galilean spacetimes: a superisation
\hyperlink{SM4}{$\mathsf{SM4}$} of the flat Galilean spacetime and the
superisation \hyperlink{SM10}{$\mathsf{SM10}$} of Galilean anti-de~Sitter spacetime.  But what about the superisations of other Galilean
spacetimes?

\subsubsection{Flat Galilean Superspaces}
\label{sec:lim-g}

From \hyperlink{SM2}{$\mathsf{SM2}$}, we obtained the Galilean
superspace \hyperlink{SM4}{$\mathsf{SM4}$}.  There is a second
superisation \hyperlink{SM3}{$\mathsf{SM3}$} of the flat Galilean
homogeneous spacetime, from which we can also reach
\hyperlink{SM4}{$\mathsf{SM4}$}.  Indeed, using the transformations
\begin{equation}
  g_t\cdot\B = t\B, \quad g_t\cdot H = t H, 
  \quad g_t\cdot \P = t\P \quad\text{and}\quad g_t\cdot \Q =
  \sqrt{t} \Q,
\end{equation}
on the Lie superalgebra for \hyperlink{SM3}{$\mathsf{SM3}$}, and taking the
limit $t\to 0$, we find the Lie superalgebra for 
\hyperlink{SM4}{$\mathsf{SM4}$}.  Thus, we have $\hyperlink{SM3}{\mathsf{SM3}}
\to \hyperlink{SM4}{\mathsf{SM4}}$.
\\ \\
Beginning with \hyperlink{SM3}{$\mathsf{SM3}$}, we may also 
consider the transformation
\begin{equation}
  g_t\cdot\B = t\B, \quad g_t\cdot \bH =  \bH, \quad g_t\cdot \P = t\P
  \quad\text{and}\quad g_t\cdot \Q = \sqrt{t} \Q,
\end{equation}
and the limit $t\to 0$.  This procedure will give us a non-effective
super Lie pair corresponding to \hyperlink{SM20}{$\mathsf{SM20}$}.

\subsubsection{Galilean de~Sitter Superspaces}
\label{sec:lim-dsg}

The superspaces \hyperlink{SM5}{$\mathsf{SM5}_\lambda$} and
\hyperlink{SM6}{$\mathsf{SM6}_\lambda$} arise as the $\gamma \to -1$
limit of \hyperlink{SM7}{$\mathsf{SM7}_{\gamma, \lambda}$} and
\hyperlink{SM8}{$\mathsf{SM8}_{\gamma, \lambda}$}, respectively.  This
fact has already been noted in Section~\ref{sec:super-dsg}.
Section~\ref{sec:super-tdsg} demonstrated that
\hyperlink{SM9}{$\mathsf{SM9}_\lambda$} is the $\gamma \to 1$ limit of
\hyperlink{SM7}{$\mathsf{SM7}_{\gamma, \lambda}$} and
\hyperlink{SM8}{$\mathsf{SM8}_{\gamma, \lambda}$}.
\\ \\
The superalgebras associated with these five superspaces take the 
general form
\begin{equation} \label{eq:galilean-de-sitter-brackets}
  \begin{aligned}[m]
    [\sH, \sB(\beta)] &= - \sP(\beta) \\
    [\sH, \sP(\pi)] &= \gamma \sB(\pi) + (1+\gamma) \sP(\pi)
  \end{aligned} \quad
  \begin{aligned}[m]
    [\sH, \sQ(s)] &= \tfrac{1}{2} \sQ(s(\eta + \lambda\kk)) \\
    [\sQ(s), \sQ(s)] &= \rho \sB(s\kk\sbar) + \sigma \sP(s\kk\sbar)
  \end{aligned} 
\end{equation}
for some $\eta, \rho, \sigma \in \mathbb{R}$, where $\gamma \in
[-1,1]$ and $\lambda \in \RR$ are the parameters of the Lie
superalgebras.  Using the transformations
\begin{equation}
g_t\cdot \B = \B, \quad g_t\cdot \bH = t\bH, \quad g_t\cdot \P = t \P
\quad\text{and}\quad g_t\cdot \Q = \sqrt{\omega t} \Q,
\end{equation}
where $\omega \in \mathbb{R}$, and taking the limit $t \to 0$, 
the above brackets become
\begin{equation}
[\sH, \sB(\beta)] = -\sP(\beta) \quad\text{and}\quad [\sQ(s), \sQ(s)]
= \omega\sigma \sP(s\kk\sbar).
\end{equation}
Therefore, by choosing $\omega = -\sigma^{-1}$, we can always
recover \hyperlink{SM4}{$\mathsf{SM4}$}.
\\ \\
There is a second superisation of the flat Galilean homogeneous
spacetime, namely \hyperlink{SM3}{$\mathsf{SM3}$}. There does not
seem to be any Lie superalgebra contraction that gives
\hyperlink{SM3}{$\mathsf{SM3}$}, but as we will see below, there are 
non-contracting limits (involving taking $\lambda \to \pm \infty$)
which take the superspaces \hyperlink{SM5}{$\mathsf{SM5}_\lambda$},
\hyperlink{SM6}{$\mathsf{SM6}_\lambda$},
\hyperlink{SM7}{$\mathsf{SM7}_{\gamma, \lambda}$},
\hyperlink{SM8}{$\mathsf{SM8}_{\gamma, \lambda}$} and
\hyperlink{SM9}{$\mathsf{SM9}_\lambda$} to
\hyperlink{SM3}{$\mathsf{SM3}$}.

\subsubsection{Galilean Anti-de~Sitter Superspaces}

The superspace \hyperlink{SM10}{$\mathsf{SM10}$} is, by definition, 
the $\chi \to 0$ limit of \hyperlink{SM11}{$\mathsf{SM11}_\chi$}.
These algebras take the form
\begin{equation}
  \begin{aligned}[m]
    [\sH, \sB(\beta)] &= - \sP(\beta) \\
    [\sH, \sP(\pi)] &= (1+\chi^2) \sB(\pi) + \chi \sP(\pi)
  \end{aligned} \quad\quad
  \begin{aligned}[m]
    [\sH, \sQ(s)] &= \tfrac{1}{2} \sQ(s(\chi + \jj)) \\
    [\sQ(s), \sQ(s)] &= - \sB(s\ii\sbar) - \sP(s\kk\sbar),
  \end{aligned}
\end{equation}
where $\chi \geq 0$ is the parameter of the Lie superalgebra.  Using
the same transformations as in the Galilean de~Sitter case, but with
$\omega = 1$, we find
\begin{equation}
  [\sH, \sB(\beta)] = \sP(\beta) \quad\text{and}\quad [\sQ(s), \sQ(s)] = -
  \sP(s\kk\sbar).
\end{equation}
Thus, we find \hyperlink{SM4}{$\mathsf{SM4}$} as a limit of both
\hyperlink{SM10}{$\mathsf{SM10}$} and \hyperlink{SM11}{$\mathsf{SM11}_\chi$}.
\\ \\
We cannot obtain \hyperlink{SM3}{$\mathsf{SM3}$} as a limit
of these superspaces as \hyperlink{SM3}{$\mathsf{SM3}$} has collinear
$\hh$ and $\bc_3$, whereas \hyperlink{SM10}{$\mathsf{SM10}$} and 
\hyperlink{SM11}{$\mathsf{SM11}_\chi$} have orthogonal $\hh$ and $\bc_3$.

\subsubsection{Non-Contracting Limits}
\label{sec:gal-non-contracting-lim}

In Section~\ref{subsec:ks_kss_gls}, it was shown that
$\lim_{\chi\to\infty} \hyperlink{S11}{\zAdSG_\chi} =
\hyperlink{S9}{\ztdSG_1}$, but this limit is not induced by a Lie algebra
contraction since the Lie algebras are non-isomorphic for different
values of $\chi$.  Does this limit extend to the superspaces?
\\ \\
Beginning with \hyperlink{SM11}{$\mathsf{SM11}_\chi$}, change basis such
that
\begin{equation}
  \bH' = \chi^{-1} \bH, \quad \B' = \B, \quad \P' = \chi^{-1} \P \quad
  \text{and} \quad
  \Q' = \chi^{-1/2} \Q,
\end{equation}
under which the brackets become
\begin{equation}
  \begin{aligned}[m]
    [\sH', \sB'(\beta)] &= -\sP'(\beta) \\
    [\sH', \sP'(\pi)] &= 2 \sP'(\pi) + (1+\chi^{-2})\sB'(\pi)
  \end{aligned} \quad\quad
  \begin{aligned}[m]
    [\sH', \sQ'(s)] &= \tfrac{1}{2\chi} \sQ'(s(\chi + \jj)) \\
    [\sQ'(s), \sQ'(s)] &= - \chi^{-1} \sB'(s\ii\sbar) + \sB'(s\kk\sbar) +
    \sP(s\kk\sbar).
  \end{aligned}
\end{equation}
Taking the limit $\chi\to\infty$, we find
\begin{equation}
  \begin{aligned}[m]
    [\sH', \sB'(\beta)] &= -\sP'(\beta) \\
    [\sH', \sP'(\pi)] &= 2 \sP'(\pi) + \sB'(\pi)
  \end{aligned} \quad\quad
  \begin{aligned}[m]
    [\sH', \sQ'(s)] &= \tfrac{1}{2} \sQ'(s) \\
    [\sQ'(s), \sQ'(s)] &= - \sB'(s\kk\sbar) +
    \sP(s\kk\sbar).
  \end{aligned}
\end{equation}
This Lie superalgebra is precisely that for \hyperlink{SM9}{$\mathsf{SM9_0}$}.
Thus, we inherit this limit from the underlying homogeneous 
spacetimes.
\\ \\
The superspaces \hyperlink{SM5}{$\mathsf{SM5}_\lambda$},
\hyperlink{SM6}{$\mathsf{SM6}_\lambda$},
\hyperlink{SM7}{$\mathsf{SM7}_{\gamma, \lambda}$},
\hyperlink{SM8}{$\mathsf{SM8}_{\gamma, \lambda}$} and
\hyperlink{SM9}{$\mathsf{SM9}_\lambda$} all have an additional
parameter $\lambda$, and we can ask what happens if we take the limit
$\lambda \to \pm \infty$ in these cases.  This is again a
non-contracting limit, since the Lie superalgebras with different
values of $\lambda \in \RR$ are not isomorphic.
\\ \\
Using the general form of the brackets stated
in~\eqref{eq:galilean-de-sitter-brackets}, consider a change of
basis
\begin{equation}
  \B'=\B, \quad \bH' = 2 \lambda^{-1} \bH, \quad \P' = 2 \lambda^{-1}
  \P \quad\text{and}\quad \Q' = \lambda^{-\tfrac{1}{2}} \Q.
\end{equation}
In our new basis, the brackets become
\begin{equation} 
  \begin{aligned}[m]
    [\sH', \sB'(\beta)] &= - \sP'(\beta) \\
    [\sH', \sP'(\pi)] &= 4\lambda^{-2}\gamma\sB'(\pi) + 2\lambda^{-1} (1+\gamma) \sP'(\pi)
  \end{aligned} \quad
  \begin{aligned}[m]
    [\sH', \sQ'(s)] &= \sQ'(s(\lambda^{-1}\eta + \kk)) \\
    [\sQ'(s), \sQ'(s)] &= \lambda^{-1} \rho \sB'(s\kk\sbar) + \tfrac{\sigma}{2} \sP'(s\kk\sbar).
  \end{aligned} 
\end{equation}
Taking either $\lambda \to \infty$ or $\lambda \to -\infty$, we find
\begin{equation}
[\sH', \sB'(\beta)] = - \sP'(\beta), \quad [\sH', \sQ'(s)] = \sQ'(s\kk), 
\quad [\sQ'(s), \sQ'(s)] = \tfrac{\sigma}{2} \sP'(s\kk\sbar).
\end{equation}
Rescaling both $\B'$ and $\P'$ by $\tfrac{\sigma}{2}$, we recover the Lie
superalgebra for \hyperlink{SM3}{$\mathsf{SM3}$}.
\\ \\
Figure~\ref{fig:super-limits} below illustrates the different
superspaces and the limits between them.  The families
\hyperlink{SM5}{$\mathsf{SM5}_\lambda$},
\hyperlink{SM6}{$\mathsf{SM6}_\lambda$},
\hyperlink{SM7}{$\mathsf{SM7}_{\gamma, \lambda}$},
\hyperlink{SM8}{$\mathsf{SM8}_{\gamma, \lambda}$} and
\hyperlink{SM9}{$\mathsf{SM9}_\lambda$} fit together into a
two-dimensional space, which also includes
\hyperlink{SM3}{$\mathsf{SM3}$} as their common limits
$\lambda \to \pm \infty$.  This space may be described as follows.  If we
fix $\lambda \in \RR$, then
\begin{equation}
  \lim_{\gamma\to 1} \hyperlink{SM7}{\mathsf{SM7}_{\gamma, \lambda}}
  =\hyperlink{SM9}{\mathsf{SM9}_\lambda} \quad\text{whereas}\quad
  \lim_{\gamma \to -1} \hyperlink{SM7}{\mathsf{SM7}_{\gamma,
      \lambda}} = \hyperlink{SM5}{\mathsf{SM5}_\lambda}. 
\end{equation}
Similarly, again fixing $\lambda \in \RR$, we have
\begin{equation}
  \lim_{\gamma\to 1} \hyperlink{SM8}{\mathsf{SM8}_{\gamma, \lambda}}
  =\hyperlink{SM9}{\mathsf{SM9}_\lambda} \quad\text{whereas}\quad
  \lim_{\gamma \to -1} \hyperlink{SM8}{\mathsf{SM8}_{\gamma,
      \lambda}} = \hyperlink{SM6}{\mathsf{SM6}_\lambda}.
\end{equation}
This gives rise to the following two-dimensional parameter spaces:
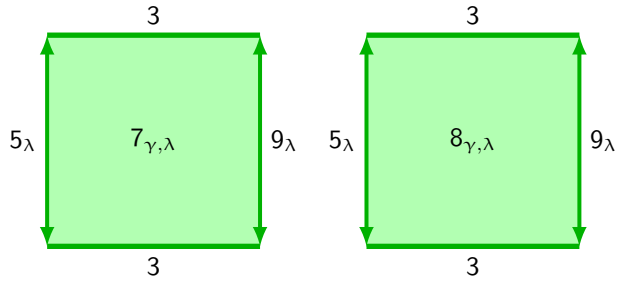
\begin{center}
  \begin{tikzpicture}[>=latex, x=1.0cm,y=1.0cm,scale=0.7]
    %
    %
    %
    %
    \coordinate (bl1) at (-2,0);
    \coordinate (br1) at (2,0);
    \coordinate (tl1) at (-2,4);
    \coordinate (tr1) at (2,4);
    \coordinate (bl2) at (4,0);
    \coordinate (br2) at (8,0);
    \coordinate (tl2) at (4,4);
    \coordinate (tr2) at (8,4);
    %
    %
    \fill [color=green!30!white] (bl1) -- (tl1) -- (tr1) -- (br1) -- (bl1);
    \fill [color=green!30!white] (bl2) -- (tl2) -- (tr2) -- (br2) -- (bl2);
    %
    %
    \node at (0,2) {$\hyperlink{SM7}{\mathsf{7}_{\gamma,\lambda}}$};
    \node at (6,2) {$\hyperlink{SM8}{\mathsf{8}_{\gamma,\lambda}}$};
    %
    %
    \draw [<->,line width=1.5pt,color=green!70!black] (bl1) -- (tl1) node [midway,left] {$\hyperlink{SM5}{\mathsf{5}_\lambda}$};
    \draw [<->,line width=1.5pt,color=green!70!black] (br1) -- (tr1) node [midway,right] {$\hyperlink{SM9}{\mathsf{9}_\lambda}$};
    \draw [<->,line width=1.5pt,color=green!70!black] (bl2) -- (tl2) node [midway,left] {$\hyperlink{SM6}{\mathsf{5}_\lambda}$};
    \draw [<->,line width=1.5pt, color=green!70!black] (br2) -- (tr2) node [midway,right] {$\hyperlink{SM9}{\mathsf{9}_\lambda}$};
    \draw [-, line width=2pt, color=green!70!black] (bl1) -- (br1) node [midway,below] {$\hyperlink{SM3}{\mathsf{3}}$}; 
    \draw [-, line width=2pt, color=green!70!black] (bl2) -- (br2) node [midway,below] {$\hyperlink{SM3}{\mathsf{3}}$}; 
    \draw [-, line width=2pt, color=green!70!black] (tl1) -- (tr1) node [midway,above] {$\hyperlink{SM3}{\mathsf{3}}$}; 
    \draw [-, line width=2pt, color=green!70!black] (tl2) -- (tr2) node [midway,above] {$\hyperlink{SM3}{\mathsf{3}}$}; 
  \end{tikzpicture}
\end{center}
We then flip the square on the right horizontally and glue the two
squares along their common $\hyperlink{SM9}{\mathsf{9}_\lambda}$ edge
to obtain the following picture
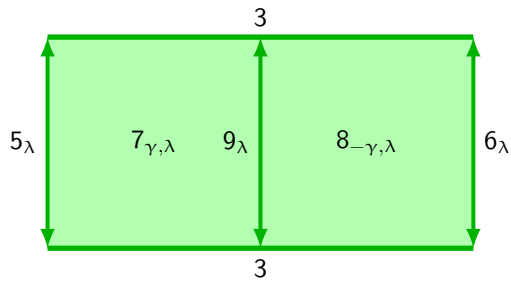
\begin{center}
  \begin{tikzpicture}[>=latex,  x=1.0cm,y=1.0cm,scale=0.7]
    %
    %
    %
    %
    \coordinate (bl) at (-2,0);
    \coordinate (br) at (6,0);
    \coordinate (bm) at (2,0);    
    \coordinate (tl) at (-2,4);
    \coordinate (tr) at (6,4);
    \coordinate (tm) at (2,4);
    %
    %
    \fill [color=green!30!white] (bl) -- (tl) -- (tr) -- (br) -- (bl);
    %
    %
    \node at (0,2) {$\hyperlink{SM7}{\mathsf{7}_{\gamma,\lambda}}$};
    \node at (4,2) {$\hyperlink{SM8}{\mathsf{8}_{-\gamma,\lambda}}$};
    %
    %
    \draw [<->,line width=1.5pt,color=green!70!black] (bl) -- (tl) node [midway,left] {$\hyperlink{SM5}{\mathsf{5}_\lambda}$};
    \draw [<->,line width=1.5pt,color=green!70!black] (br) -- (tr) node [midway,right] {$\hyperlink{SM6}{\mathsf{6}_\lambda}$};
    \draw [<->,line width=1.5pt,color=green!70!black] (bm) -- (tm) node [midway,left] {$\hyperlink{SM9}{\mathsf{9}_\lambda}$};
    \draw [-, line width=2pt, color=green!70!black] (bl) -- (br) node [midway,below] {$\hyperlink{SM3}{\mathsf{3}}$}; 
    \draw [-, line width=2pt, color=green!70!black] (tl) -- (tr) node [midway,above] {$\hyperlink{SM3}{\mathsf{3}}$}; 
  \end{tikzpicture}
\end{center}
We now glue the top and bottom edges to arrive at the following
cylinder:

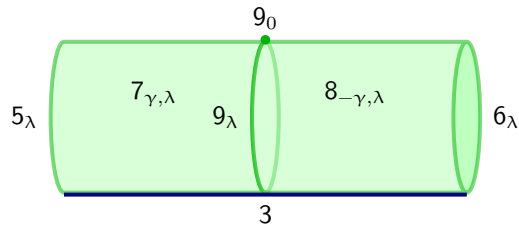
\begin{center}
  \begin{tikzpicture}[>=stealth, aspect=1.5,x=1.0cm,y=1.0cm,scale=0.7,color=green!70!black,line width=1.5pt]
    %
    %
    %
    %
    \node [name=cyl1, draw, cylinder, cylinder uses custom fill, 
    cylinder body fill=green!30!white, minimum height=3cm, minimum
    width=2cm,opacity=0.5] {};
    \node [name=cyl2, draw, cylinder, cylinder uses custom fill, cylinder end fill=green!50!white,
    cylinder body fill=green!30!white, minimum height=3cm, minimum width=2cm,above=0pt of cyl1.before top, anchor=after bottom,opacity=0.5] {};
    %
    %
    \draw [color=blue!50!black, line width=1.5pt] (cyl1.before bottom)
    -- (cyl2.after top) node [midway, below]{\hyperlink{SM3}{$\mathsf{3}$}};
    %
    %
    \coordinate [label=left:{\hyperlink{SM5}{$\mathsf{5}_\lambda$}}] (5) at (cyl1.bottom);
    \coordinate [label=right:{\hyperlink{SM6}{$\mathsf{6}_\lambda$}}] (6) at (cyl2.top);
    \coordinate [label=above:{\hyperlink{SM9}{$\mathsf{9}_0$}}] (90) at (cyl1.before top); 
    \coordinate [label=left:{\hyperlink{SM9}{$\mathsf{9}_\lambda$}}] (9) at (cyl2.bottom); 
    \coordinate [label=:{$\hyperlink{SM7}{\mathsf{7}_{\gamma,\lambda}}$}] (7) at (cyl1.center);
    \coordinate [label=:{$\hyperlink{SM8}{\mathsf{8}_{-\gamma,\lambda}}$}] (8) at (cyl2.center);    
    %
    %
    \foreach \point in {90}
    \filldraw [color=green!70!black,fill=green!70!black] (\point) circle (1.5pt);
  \end{tikzpicture}
\end{center}
Finally, we collapse the ``edge'' labelled
\hyperlink{SM3}{$\mathsf{3}$} to a point, arriving at the object in
Figure~\ref{fig:super-limits}.

\subsection{Aristotelian Limits}
\label{sec:aristo-lim}

There are two kinds of superisations of Aristotelian spacetimes: the
ones where $\B$ acts as R-symmetries and the ones where $\B$ acts
trivially.  We treat them in turn.

\subsubsection{Aristotelian Superspaces with R-Symmetry}
\label{sec:r-sym-lim}
 
The homogeneous spacetimes \hyperlink{A23m}{$\RR\times H^3$} and \hyperlink{A23p}{$\RR\times S^3$} 
underlying the homogeneous superspaces \hyperlink{SM14}{$\mathsf{SM14}$}
-  \hyperlink{SM17}{$\mathsf{SM17}$} have \hyperlink{A21}{$\zS$} as their limit.  
Therefore, we could expect \hyperlink{SM14}{$\mathsf{SM14}$} - 
\hyperlink{SM17}{$\mathsf{SM17}$} to have either \hyperlink{SM18}{$\mathsf{SM18}$} or
\hyperlink{SM19}{$\mathsf{SM19}$} as limits.  The relevant contraction uses
the transformation
\begin{equation}
g_t\cdot\B = \B, \quad g_t\cdot \bH = \bH \quad\text{and} \quad
g_t\cdot\P = t\P.
\end{equation}
Taking the limit $t \to 0$, the $[\P, \P]$ bracket vanishes
leaving all other brackets unchanged. Thus, we find
$\hyperlink{SM14}{\mathsf{SM14}}
\to\hyperlink{SM18}{\mathsf{SM18}}$,
$\hyperlink{SM16}{\mathsf{SM16}} \to
\hyperlink{SM18}{\mathsf{SM18}}$, $\hyperlink{SM15}{\mathsf{SM15}}
\to \hyperlink{SM19}{\mathsf{SM19}}$ and
$\hyperlink{SM17}{\mathsf{SM17}} \to \hyperlink{SM19}{\mathsf{SM19}}$.
\\ \\
Taking into account the form of $\hh$, and the $[\Q, \Q]$ bracket
for each of these superspaces, we notice that each homogeneous
spacetime has two superspaces associated with it.  One for which
\begin{equation}
\bb = \tfrac{1}{2} \quad\text{and}\quad [\sQ(s), \sQ(s)] = |s|^2 \sH,
\end{equation}
and one for which
\begin{equation}
\bb = \tfrac{1}{2}, \quad \hh = \tfrac{1}{2}\kk \quad\text{and}\quad
[\sQ(s), \sQ(s)] = |s|^2 \sH - \sB(s\kk\sbar).
\end{equation}
Using transformations which act as
\begin{equation}
g_t\cdot \bH = t\bH, \quad g_t\cdot\Q = \sqrt{t}\Q
\end{equation}
and trivially on $\J, \B,$ and $\P$, we find the brackets of the 
latter superspaces described by
\begin{equation}
\bb = \tfrac{1}{2}, \quad \hh = \tfrac{t}{2} \kk, \quad \text{and}
\quad [\sQ(s), \sQ(s)] = |s|^2 \sH - t\sB(s\kk\sbar).
\end{equation}
Therefore, taking the limit $t\to 0$, we find the former
superspaces.  Thus, we get the limits $\hyperlink{SM15}{\mathsf{SM15}}
\to \hyperlink{SM14}{\mathsf{SM14}}$, $\hyperlink{SM17}{\mathsf{SM17}}
\to\hyperlink{SM16}{\mathsf{SM16}}$ and $\hyperlink{SM19}{\mathsf{SM19}}
\to\hyperlink{SM18}{\mathsf{SM18}}$.
\\ \\
All of the above superspaces have \hyperlink{SM18}{$\mathsf{SM18}$} 
as a limit.  Therefore, we will only consider the limits of this 
superspace to those Aristotelian superspaces without R-symmetry.
Letting
\begin{equation}
g_t\cdot\B = t\B, \quad  g_t\cdot \bH = \bH, \quad g_t\cdot\P = \P,
\quad g_t\cdot\Q = \Q,
\end{equation}
and taking the limit $t\to 0$, we arrive at a non-effective super 
Lie pair corresponding to \hyperlink{SM22}{$\mathsf{SM22}$}.

 \subsubsection{Aristotelian Superspaces without R-Symmetry}
\label{sec:w-o-r-sym-lim}

The Aristotelian homogeneous spacetimes \hyperlink{A23p}{$\RR\times S^3$},
\hyperlink{A23m}{$\RR\times H^3$}, and \hyperlink{A22}{$\zTS$} have \hyperlink{A21}{$\zS$}
as their limit; therefore, we would expect their superisations to have 
have one or more of \hyperlink{SM20}{$\mathsf{SM20}$}-\hyperlink{SM23}{$\mathsf{SM23}$} 
as limits.  For \hyperlink{A22}{$\zTS$} to have \hyperlink{A21}{$\zS$} as its 
limit, we require the transformation
\begin{equation}
g_t\cdot \B = \B, \quad g_t\cdot \bH = t\bH \quad\text{and}\quad g_t\cdot \P = \P. 
\end{equation}
Wanting to ensure $[\Q, \Q] \neq 0$, and that the limit $t\to 0$
is well-defined, we need $g_t\cdot \Q = \sqrt{t} \Q$.  Taking this limit,
we find $\hyperlink{SM24}{\mathsf{SM24}_\lambda}\to\hyperlink{SM21}{\mathsf{SM21}}$.
\\ \\
To get \hyperlink{A21}{$\zS$}  from \hyperlink{A23p}{$\RR\times S^3$},
we need the transformation
\begin{equation}
g_t\cdot \B = \B, \quad g_t\cdot \bH = \bH \quad\text{and}\quad g_t\cdot \P = t\P. 
\end{equation}
Using this transformation and taking the limit $t\to 0$, we
find $\hyperlink{SM25}{\mathsf{SM25}}\to\hyperlink{SM22}{\mathsf{SM22}}$.
However, the limit is not well-defined for \hyperlink{SM26}{$\mathsf{SM26}$}
due to $\P$ in the expression for $[\Q, \Q]$.  In this case, we additionally
require $g_t\cdot\Q = \sqrt{t} \Q$.  Then $\hyperlink{SM26}{\mathsf{SM26}}
\to \hyperlink{SM20}{\mathsf{SM20}}$.  Another choice of 
transformation,
\begin{equation}
g_t\cdot \B = \B, \quad g_t\cdot \bH = t\bH, \quad g_t\cdot \P = t\P
\quad\text{and}\quad g_t\Q = \sqrt{t}\Q,
\end{equation}
for \hyperlink{SM26}{$\mathsf{SM26}$}, gives \hyperlink{SM23}{$\mathsf{SM23}$}  in the 
limit $t\to 0$.  Thus, we also have $\hyperlink{SM26}{\mathsf{SM26}}
\to\hyperlink{SM23}{\mathsf{SM23}}$.
\\ \\
Finally, to get \hyperlink{A21}{$\zS$}  from \hyperlink{A23m}{$\RR\times H^3$},
we use the transformation
\begin{equation}
g_t\cdot \B = \B, \quad g_t\cdot \bH = \bH, \quad g_t\cdot \P = t\P. 
\end{equation}
To ensure the limit $t\to 0$ is well-defined, we subsequently need
$g_t\cdot\Q = \sqrt{t} \Q$.  This transformation with the limit gives
$\hyperlink{SM27}{\mathsf{SM27}}\to\hyperlink{SM21}{\mathsf{SM21}}$.
\\ \\
There are only two underlying Aristotelian homogeneous spacetimes
which have more than one superisation.  These are
\hyperlink{A21}{$\zS$} and \hyperlink{A23p}{$\RR\times S^3$}.
In the latter case, we find the superisation
\hyperlink{SM25}{$\mathsf{SM25}$} as the limit of \hyperlink{SM26}{$\mathsf{SM26}$}
using the transformation
\begin{equation}
g_t\cdot \B = \B, \quad g_t\cdot \bH = t\bH, \quad g_t\cdot \P = \P
\quad\text{and}\quad g_t\cdot \Q = \sqrt{t} \Q,
\end{equation}
and taking $t \to 0$.  In the former case, the superisations
\hyperlink{SM22}{$\mathsf{SM22}$} and \hyperlink{SM21}{$\mathsf{SM21}$} can be
found as limits of \hyperlink{SM23}{$\mathsf{SM23}$} using the 
transformations
\begin{equation}
g_t\cdot \B = \B, \quad g_t\cdot \bH = t\bH, \quad g_t\cdot \P = \P 
\quad\text{and}\quad g_t\cdot \Q = \sqrt{t} \Q,
\end{equation}
and
\begin{equation}
g_t\cdot \B = \B, \quad g_t\cdot \bH = \bH, \quad g_t\cdot \P = t\P
\quad\text{and}\quad g_t\cdot \Q = \sqrt{t} \Q,
\end{equation}
respectively.  We also have 
\begin{equation}
g_t\cdot \B = \B, \quad g_t\cdot \bH = t\bH, \quad g_t\cdot \P = \P
\quad\text{and}\quad g_t\cdot \Q = \Q,
\end{equation}
giving the limit $\hyperlink{SM20}{\mathsf{SM20}}\to\hyperlink{SM21}{\mathsf{SM21}}$.

\subsection{A Non-Contracting Limit}
\label{sec:aristo-non-contracting-lim}

Use the following change of basis on the Lie superalgebra for
\hyperlink{SM24}{$\mathsf{SM24}_\lambda$},
\begin{equation}
\B'=\B, \quad \bH' = 2 \lambda^{-1} \bH, \quad \P' = \P, 
\quad \Q' = \Q.
\end{equation}
The brackets then become
\begin{equation}
[\sH', \sP(\pi)'] = 2 \lambda^{-1} \sP(\pi)', \quad 
[\sH', \sQ'(s)] = \sQ'(s(\lambda^{-1} + \kk)), \quad 
[\sQ'(s), \sQ'(s)] = -\sP'(s\kk\sbar).
\end{equation}
Taking the limits $\lambda \to \pm \infty$, we find the superspace
\hyperlink{SM20}{$\mathsf{SM20}$}.  Therefore, the line of superspaces
\hyperlink{SM24}{$\mathsf{SM24}_\lambda$} compactifies to a circle
with \hyperlink{SM20}{$\mathsf{SM20}$} as the point at infinity.

\subsection{Summary}
\label{sec:limit-summary}

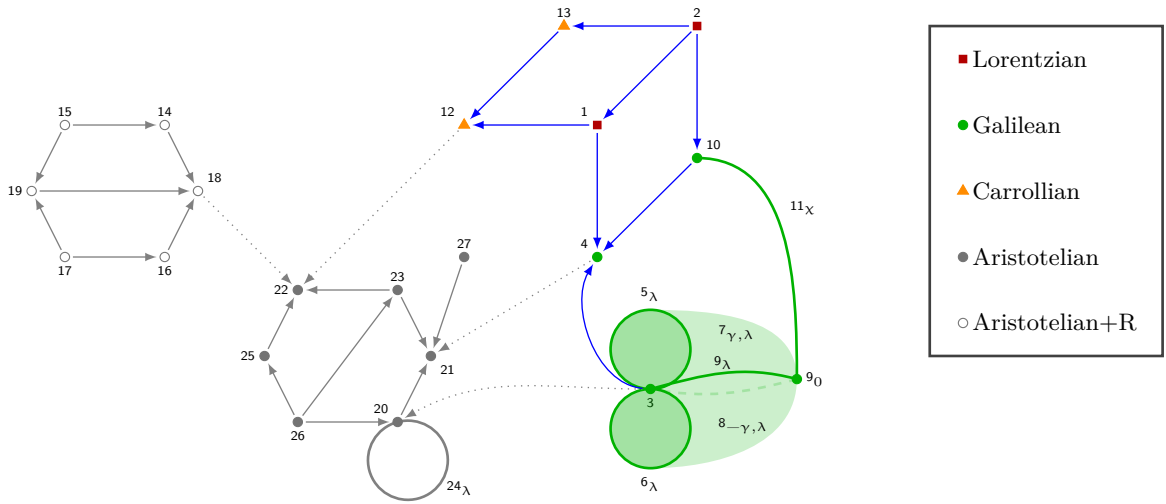
\begin{figure}[h!]
  \centering
  \begin{tikzpicture}[scale=1.75,>=latex, shorten >=3pt, shorten <=3pt, x=1.0cm,y=1.0cm]
    %
    %
    %
    %
    %
    \coordinate [label=above left:{\hyperlink{SM1}{\tiny $\mathsf{1}$}}] (1) at (6.75, 0.75);
    \coordinate [label=above:{\hyperlink{SM2}{\tiny $\mathsf{2}$}}] (2) at (7.5, 1.5);
    \coordinate [label=above left:{\hyperlink{SM4}{\tiny $\mathsf{4}$}}] (4) at  (6.75, -0.25);
    \coordinate [label=above right:{\hyperlink{SM10}{\tiny $\mathsf{10}$}}] (10) at (7.5, 0.5); 
    \coordinate [label=above left:{\hyperlink{SM12}{\tiny $\mathsf{12}$}}] (12) at (5.75,0.75);
    \coordinate [label=above:{\hyperlink{SM13}{\tiny $\mathsf{13}$}}]  (13) at (6.5,1.5);
    \coordinate [label=below:{\hyperlink{SM3}{\tiny $\mathsf{3}$}}] (3) at  (7.15, -1.25); 
    \coordinate [label=above:{\hyperlink{SM5}{\tiny $\mathsf{5}_{\lambda}$}}] (5) at (7.15, -.65);
    \coordinate [label=below:{\hyperlink{SM6}{\tiny $\mathsf{6}_{\lambda}$}}] (6) at (7.15, -1.85);
    \coordinate [label=right:{\hyperlink{SM9}{\tiny $\mathsf{9}_{0}$}}] (9) at (8.25, -1.175);
    \coordinate [label=above:{\hyperlink{SM14}{\tiny $\mathsf{14}$}}]  (14) at (3.5,0.75);
    \coordinate [label=above:{\hyperlink{SM15}{\tiny $\mathsf{15}$}}]  (15) at (2.75,0.75);    
    \coordinate [label=below:{\hyperlink{SM16}{\tiny $\mathsf{16}$}}]  (16) at (3.5,-0.25);
    \coordinate [label=below:{\hyperlink{SM17}{\tiny $\mathsf{17}$}}]  (17) at (2.75,-0.25);
    \coordinate [label=above right:{\hyperlink{SM18}{\tiny $\mathsf{18}$}}]  (18) at (3.75,.25);    
    \coordinate [label=left:{\hyperlink{SM19}{\tiny $\mathsf{19}$}}]  (19) at (2.5,.25);    
    \coordinate [label=above left:{\hyperlink{SM20}{\tiny $\mathsf{20}$}}] (20) at (5.25, -1.5);
    \coordinate [label=below right:{\hyperlink{SM21}{\tiny $\mathsf{21}$}}] (21) at (5.5, -1);
    \coordinate [label=left:{\hyperlink{SM22}{\tiny $\mathsf{22}$}}] (22) at (4.5, -.5);
    \coordinate [label=above:{\hyperlink{SM23}{\tiny $\mathsf{23}$}}] (23) at (5.25, -.5);
    \coordinate [label=left:{\hyperlink{SM25}{\tiny $\mathsf{25}$}}]  (25) at (4.25, -1); 
    \coordinate [label=below:{\hyperlink{SM26}{\tiny $\mathsf{26}$}}]  (26) at (4.5, -1.5);
    \coordinate [label=right:{\hyperlink{SM24}{\tiny $\mathsf{24}_{\lambda}$}}]  (24) at (5.55,-2); 
    \coordinate [label=above:{\hyperlink{SM27}{\tiny $\mathsf{27}$}}]  (27) at (5.75, -0.25);
    \coordinate [label=above:{\hyperlink{SM11}{\tiny $\mathsf{11}_\chi$}}] (11) at (8.3, 0); 
    %
    %
    \path [fill=green!70!black,opacity=0.2, line width=.1mm] (9) to [in=0,out=90] (5) arc (90:270:0.3) to  [opacity=0,in=165, out=15] (9);
    \path [fill=green!70!black,opacity=0.2, line width=.1mm] (9) to [in=0,out=270] (6) arc (270:90:0.3) to  [opacity=0,in=165, out=15] (9);
    \draw[>=latex, shorten >=0pt, shorten <=0pt, line width=1pt, color=green!70!black, fill=green!70!black,fill opacity=.2] (3) arc (90:450:0.3);
    \draw[>=latex, shorten >=0pt, shorten <=0pt, line width=1pt, color=green!70!black, fill=green!70!black,fill opacity=.2] (3) arc (-90:270:0.3); 
    \draw [>=latex, shorten >=0pt, shorten <=0pt, line width=1pt, color=green!70!black] (10) to [in=90,out=0] (9);
    \draw [>=latex, shorten >=0pt, shorten <=0pt, line width=1pt, color=green!70!black] (9) to [in=15,out=165] (3);
    \draw [>=latex, shorten >=0pt, shorten <=0pt, line width=1pt, dashed, opacity = 0.2, color=green!70!black] (9) to [in=350,out=195] (3);
    \draw[>=latex, shorten >=0pt, shorten <=0pt, line width=1pt, color=gray,rotate=-30] (20) arc (-225:135:0.3); 
    %
    %
    \draw [->,line width=0.5pt,dotted,color=gray] (18) -- (22);
    \draw [->,line width=0.5pt,dotted,color=gray] (12) -- (22);
    \draw [->,line width=0.5pt,dotted,color=gray] (4) -- (21);
    \draw [->,line width=0.5pt,dotted,color=gray] (3) to [in=30, out=180] (20);
    %
    %
    \draw [->,line width=0.5pt,color=blue] (2) -- (13);
    \draw [->,line width=0.5pt,color=blue] (2) -- (1);
    \draw [->,line width=0.5pt,color=blue] (2) -- (10);
    \draw [->,line width=0.5pt,color=blue] (13) -- (12);
    \draw [->,line width=0.5pt,color=blue] (1) -- (12);
    \draw [->,line width=0.5pt,color=blue] (1) -- (4);
    \draw [->,line width=0.5pt,color=blue] (10) -- (4); 
    \draw [->,line width=0.5pt,color=blue] (3) to [in=-120,out=175] (4);
    %
    %
    \draw [->,line width=0.5pt,color=gray] (15) to (19); 
    \draw [->,line width=0.5pt,color=gray] (17) to (19); 
    \draw [->,line width=0.5pt,color=gray] (14) to (18); 
    \draw [->,line width=0.5pt,color=gray] (16) to (18); 
    \draw [->,line width=0.5pt,color=gray] (15) to (14); 
    \draw [->,line width=0.5pt,color=gray] (19) to (18); 
    \draw [->,line width=0.5pt,color=gray] (17) to (16); 
    %
    %
    \draw [->,line width=0.5pt,color=gray] (27) to (21); 
    \draw [->,line width=0.5pt,color=gray] (20) to (21);    
    \draw [->,line width=0.5pt,color=gray] (26) to (20); 
    \draw [->,line width=0.5pt,color=gray] (26) to (25); 
    \draw [->,line width=0.5pt,color=gray] (25) to (22); 
    \draw [->,line width=0.5pt,color=gray] (23) to (22); 
    \draw [->,line width=0.5pt,color=gray] (23) to (21); 
    \draw [->,line width=0.5pt,color=gray] (26) to (23);  
    %
    %
    \coordinate [label=below:{\hyperlink{SM7}{\tiny $\mathsf{7}_{\gamma, \lambda}$}}] (7) at (7.8, -0.7);
    \coordinate [label=below:{\hyperlink{SM8}{\tiny $\mathsf{8}_{-\gamma, \lambda}$}}] (8) at (7.85,-1.4);
    \coordinate [label=above:{\hyperlink{SM9}{\tiny $\mathsf{9}_{\lambda}$}}] (9_2) at (7.7,-1.17);
    %
    %
    \foreach \point in {3, 4, 9, 10}
    \filldraw [color=green!70!black,fill=green!70!black] (\point) circle (1pt);
    \foreach \point in {1, 2}
   \node[lorentzian] at (\point) {};
    \foreach \point in {12,13}
    \node[carrollian] at (\point) {}; 
    \foreach \point in {14, 15, 16, 17, 18, 19}
    \draw [color=gray!90!black] (\point) circle (1pt);
    \foreach \point in {20, 21, 22, 23, 25, 26, 27}
    \draw [color=gray!90!black,fill=gray!90!black] (\point) circle (1pt);
    %
    %
    \begin{scope}[xshift=-1.5cm]
    \draw [line width=1pt,color=gray!50!black] (10.75,-1) rectangle (12.5,1.5);
    \node[lorentzian] at (11,1.25) {};
    \draw (11,1.25) node[color=black,anchor=west] {\small Lorentzian};
    \filldraw [color=green!70!black,fill=green!70!black] (11,0.75) circle (1pt) node[color=black,anchor=west] {\small Galilean};
    \node[carrollian] at (11,0.25) {};
    \draw (11,0.25) node[color=black,anchor=west] {\small Carrollian};
    \draw [color=gray!90!black,fill=gray!90!black] (11,-0.25) circle (1pt) node[color=black,anchor=west] {\small Aristotelian};       
    \draw [color=gray!90!black] (11, -0.75) circle (1pt) node[color=black,anchor=west] {\small Aristotelian+R}; 
    \end{scope}
  \end{tikzpicture}
  \caption{Homogeneous Superspaces and their Limits.\\
    (Numbers are hyperlinked to the corresponding superspaces in
    Table~\ref{tab:superspaces}.)}
  \label{fig:super-limits}
\end{figure}

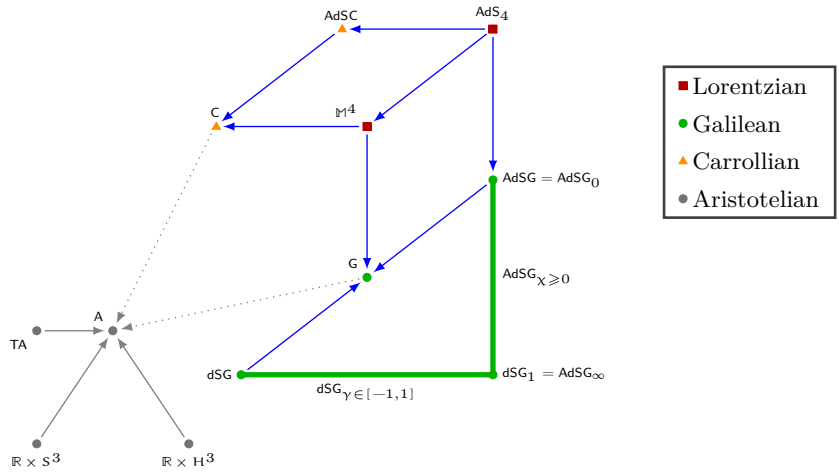
\begin{figure}[h!]
  \centering
  \begin{tikzpicture}[scale=1,>=latex, shorten >=3pt, shorten <=3pt, x=1.0cm,y=1.0cm]
    %
    %
    %
    %
    \coordinate [label=left:{\hyperlink{S8}{\tiny $\zdSG$}}] (dsg) at (5.688048519056286, -2.5838170592960186);
    \coordinate [label=right:{\hyperlink{S9}{\tiny $\ztdSG_1 =\ztAdSG_\infty$}}] (dsgone) at (9, -2.5838170592960186);
    \coordinate [label=above:{\hyperlink{S13}{\tiny $\zC$}}] (c) at (5.344024259528143, 0.7080914703519907);
    \coordinate [label=above left:{\hyperlink{S1}{\tiny $\MM^4$}}] (m) at (7.344024259528143,0.7080914703519907);
    \coordinate [label=above left:{\hyperlink{S7}{\tiny $\zG$}}] (g) at  (7.344024259528143, -1.2919085296480093);
    \coordinate [label=above:{\hyperlink{S3}{\tiny $\zAdS_4$}}] (ads) at (9,2);
    \coordinate [label=right:{\hyperlink{S10}{\tiny $\zAdSG =\ztAdSG_0$}}] (adsg) at (9,0);
    \coordinate [label=above:{\hyperlink{S15}{\tiny $\zAdSC$}}]  (adsc) at (7,2);
    \coordinate [label=above left:{\hyperlink{A21}{\tiny $\zS$}}] (s) at (4, -2);
    \coordinate [label=below left:{\hyperlink{A22}{\tiny $\zTS$}}]  (ts) at (3, -2); 
    \coordinate [label=below:{\hyperlink{A23p}{\tiny $\RR\times S^3$}}]  (esu) at (3, -3.5); 
    \coordinate [label=below:{\hyperlink{A23m}{\tiny $\RR\times H^3$}}]  (hesu) at (5, -3.5); 
    %
    %
    \coordinate [label=below:{\hyperlink{S9}{\tiny $\ztdSG_{\gamma\in[-1,1]}$}}] (tdsg) at (7.344024259528143, -2.5838170592960186);
    \coordinate [label=right:{\hyperlink{S11}{\tiny $\ztAdSG_{\chi\geq0}$}}] (tadsg) at (9, -1.2919085296480093);
    %
    %
    \draw [->,line width=0.5pt,dotted,color=gray] (c) -- (s);
    \draw [->,line width=0.5pt,dotted,color=gray] (g) -- (s);
    %
    %
    \draw [->,line width=0.5pt,color=blue] (adsc) -- (c);
    \draw [->,line width=0.5pt,color=blue] (ads) -- (m);
    \draw [->,line width=0.5pt,color=blue] (adsg) -- (g);
    \draw [->,line width=0.5pt,color=blue] (dsg) -- (g);
    \draw [->,line width=0.5pt,color=blue] (m) -- (c);
    \draw [->,line width=0.5pt,color=blue] (m) -- (g);
    \draw [->,line width=0.5pt,color=blue] (ads) -- (adsc);
    \draw [->,line width=0.5pt,color=blue] (ads) -- (adsg);
    %
    %
    \draw [->,line width=0.5pt,color=gray] (ts) to (s); 
    \draw [->,line width=0.5pt,color=gray] (esu) to (s); 
    \draw [->,line width=0.5pt,color=gray] (hesu) to (s); 
    %
    %
    \begin{scope}[>=latex, shorten >=0pt, shorten <=0pt, line width=2pt, color=green!70!black]
      \draw (adsg) --(dsgone);
      \draw (dsg) -- (dsgone);
    \end{scope}
    \foreach \point in {g,adsg,dsg,dsgone}
    \filldraw [color=green!70!black,fill=green!70!black] (\point) circle (1.5pt);
    \foreach \point in {ads,m}
    \filldraw [color=red!70!black,fill=red!70!black] (\point) ++(-1.5pt,-1.5pt) rectangle ++(3pt,3pt);
    \foreach \point in {adsc,c}
    \filldraw [color=DarkOrange,fill=DarkOrange] (\point) ++(-1pt,-1pt) -- ++(3pt,0pt) -- ++(-1.5pt,2.6pt) -- cycle;
    \foreach \point in {s,ts,esu,hesu}
    \filldraw [color=gray!90!black] (\point) circle (1.5pt);
    %
    %
    \begin{scope}[xshift=0.5cm]
    \draw [line width=1pt,color=gray!50!black] (10.75,-0.5) rectangle (13,1.5);
    \filldraw [color=red!70!black,fill=red!70!black] (11,1.25) ++(-1.5pt,-1.5pt) rectangle ++(3pt,3pt) ; 
    \draw (11,1.25) node[color=black,anchor=west] {\small Lorentzian}; 
    \filldraw [color=green!70!black,fill=green!70!black] (11,0.75) circle (1.5pt) node[color=black,anchor=west] {\small Galilean};
    \filldraw [color=DarkOrange,fill=DarkOrange] (11,0.25) ++(-1.5pt,-1pt) -- ++(3pt,0pt) -- ++(-1.5pt,2.6pt) -- cycle;
    \draw (11,0.25) node[color=black,anchor=west] {\small Carrollian};
    \filldraw [color=gray!90!black] (11,-0.25) circle (1.5pt) node[color=black,anchor=west] {\small Aristotelian};       
    \end{scope}
  \end{tikzpicture}
  \caption{Limits Between Superisable Spacetimes}
  \label{fig:sub-limits}
\end{figure}

The picture resulting from the above discussion is given in
Figure~\ref{fig:super-limits}.  Except for
$\hyperlink{SM4}{\mathsf{SM3}} \to \hyperlink{SM4}{\mathsf{SM4}}$, the
limits from the families \hyperlink{SM5}{$\mathsf{SM5}_\lambda$},
\hyperlink{SM6}{$\mathsf{SM6}_\lambda$},
\hyperlink{SM7}{$\mathsf{SM7}_{\gamma, \lambda}$},
\hyperlink{SM8}{$\mathsf{SM8}_{\gamma, \lambda}$},
\hyperlink{SM9}{$\mathsf{SM9}_\lambda$} and
\hyperlink{SM11}{$\mathsf{SM11}_\chi$} to
\hyperlink{SM4}{$\mathsf{SM4}$} are not shown explicitly in order to
improve readability.  Neither is the limit between
\hyperlink{SM24}{$\mathsf{SM24}_\lambda$} and
\hyperlink{SM21}{$\mathsf{SM21}$} shown.
\\ \\
For comparison, we extract from
Figure~\ref{fig:generic-d-graph} the subgraph corresponding to
spacetimes which admit superisations and show it in
Figure~\ref{fig:sub-limits}.  There are arrows between these two
pictures: taking a superspace to its corresponding spacetime, but
making this explicit seems beyond our artistic abilities.
\\ \\
Nevertheless, interpreting Figures~\ref{fig:super-limits} and
\ref{fig:sub-limits} as posets, with arrows defining the partial
order, the map taking a superspace to its underlying spacetime is
surjective by construction (we consider only superisable spacetimes)
and order preserving, as shown at the start of this section.  As can
be gleaned from Table~\ref{tab:superspaces}, the fibres of this map
are often quite involved, clearly showing the additional ``internal''
structure in the superspace which allows for more than one possible
superisation of a spacetime.
\\ \\
We should mention that, despite appearances, superspaces
$\hyperlink{SM3}{\mathsf{SM3}}$ and $\hyperlink{SM4}{\mathsf{SM4}}$
share the same underlying spacetime: namely, the Galilean spacetime
$\hyperlink{S7}{\zG}$.  Notice that superspaces
$\hyperlink{SM21}{\mathsf{SM21}}$ and
$\hyperlink{SM22}{\mathsf{SM22}}$, which are ``terminal'' in the
partial order, correspond to the static Aristotelian spacetime
$\hyperlink{A21}{\mathsf{\zS}}$.  With the exception of
$\lim_{\chi\to \infty} \hyperlink{SM11}{\mathsf{SM11}_\chi} =
\hyperlink{SM9}{\mathsf{SM9}_0}$, all other non-contracting limits
between superspaces induce limits between the underlying spacetimes,
which arise from contractions of the kinematical Lie algebras: the
limits $|\lambda|\to \infty$ of
\hyperlink{SM5}{$\mathsf{SM5}_\lambda$} and
\hyperlink{SM6}{$\mathsf{SM6}_\lambda$} induce the contraction
$\hyperlink{S8}{\zdSG} \to \hyperlink{S7}{\zG}$, whereas the limits
$|\lambda|\to \infty$ of
\hyperlink{SM7}{$\mathsf{SM7}_{\gamma, \lambda}$},
\hyperlink{SM8}{$\mathsf{SM8}_{\gamma, \lambda}$} and
\hyperlink{SM9}{$\mathsf{SM9}_\lambda$} induce the contractions
$\hyperlink{S9}{\zdSG_\gamma} \to \hyperlink{S7}{\zG}$, where
$\gamma = 1$ for \hyperlink{SM9}{$\mathsf{SM9}_\lambda$}.

\section{Conclusion} \label{sec:ks_superspace_conc}
In this chapter, we discussed the classification of $\N=1$ kinematical Lie superalgebras in three spatial dimensions and, subsequently, the classification of the kinematical superspaces which arise from these algebras.  
\\ \\
The Lie superalgebras were classified by solving the super-Jacobi identities
in a quaternionic reformulation, which made the computations no harder
than multiplying quaternions and paying close attention to the action
of automorphisms in order to ensure that there is no repetition in our
list.  Since we are interested in supersymmetry, we focussed on Lie
superalgebras where the supercharges were not abelian: i.e., we demand
that $[\Q,\Q] \neq 0$ and, subject to that condition, we classified
Lie superalgebras which extend either kinematical or Aristotelian Lie
algebras.  The results are contained in Tables~\ref{tab:klsa} and
\ref{tab:alsa}, respectively.
\\ \\
There are two salient features of these classifications.  Firstly, not
every kinematical Lie algebra admits a supersymmetric extension: in
some cases because of our requirement that $[\Q,\Q] \neq 0$, but in
other cases (e.g., $\so(5)$, $\so(4,1)$,...) because the
four-dimensional spinor module of $\so(3)$ does not extend to
a module of these Lie algebras.
\\ \\
Secondly, some kinematical Lie algebras admit more than one
non-isomorphic supersymmetric extension. For example, the Galilean Lie
algebra admits two supersymmetric extensions, but only one of them
($\hyperlink{KLSA8}{\mathsf{S8}}$) can be obtained as a contraction
of $\osp(1|4)$.  By far most of the Lie superalgebras in our
classification cannot be so obtained and hence are not listed in
previous classifications.  Nevertheless, our ``moduli space'' of Lie
superalgebras is connected, if not always by contractions.  For
example, the other supersymmetric extension of the Galilean algebra
($\hyperlink{KLSA7}{\mathsf{S7}}$) can be obtained as a
non-contracting limit of some of the multi-parametric families of Lie
superalgebras in the limit as one of the parameters goes to
$\pm \infty$, in effect compactifying one of the directions in the
parameter space into a circle.
\\ \\
We classified the corresponding superspaces via their super Lie pairs
$(\s,\h)$, where $\s$ is a kinematical Lie superalgebra and $\h$ an
admissible subalgebra.  Every such pair ``superises'' a pair
$(\k,\h)$, where $\k = \s_{\bar 0}$ is a kinematical Lie algebra.  As
discussed in Section~\ref{subsec:math_prelims_geo_ks}, effective and geometrically
realisable pairs $(\k,\h)$ are in bijective correspondence with
simply-connected homogeneous spacetimes, and hence the super Lie pairs 
$(\s,\h)$ are in bijective correspondence with superisations of such
spacetimes.  These are listed in Table~\ref{tab:superspaces}.
\\ \\
There are several salient features of that table.  Firstly, many
spacetimes admit more than one inequivalent superisation.  Whereas
Minkowski and AdS spacetimes admit a unique ($\N{=}1$) superisation,
and so too do the (superisable) Carrollian spacetimes.  Many of the
Galilean spacetimes admit more than one and in some cases even a
circle of superisations.
\\ \\
Secondly, there are effective super Lie pairs $(\s,\h)$ for which the
underlying pair $(\k,\h)$ is not effective.  This means that the
``boosts'' act trivially on the underlying spacetime, but non-trivially
in the superspace: in other words, the ``boosts'' are actually
R-symmetries.  Since $(\k,\h)$ is not effective, this means that it
describes an Aristotelian spacetime and this gives rise to the class
of Aristotelian superspaces with R-symmetry.
\\ \\
Thirdly, there are three superspaces in our list which also appear in
\cite{deMedeiros:2016srz}: namely, Minkowski
($\hyperlink{SM1}{\mathsf{SM1}}$) and AdS
($\hyperlink{SM2}{\mathsf{SM2}}$) superspaces, but also the
Aristotelian superspace $\hyperlink{SM26}{\mathsf{SM26}}$, whose
underlying manifold appears in \cite{deMedeiros:2016srz} as the
Lorentzian Lie group $\RR \times \SU(2)$ with a bi-invariant
metric.
\\ \\
Lastly, just like Minkowski ($\hyperlink{S1}{\MM^4}$) and Carrollian
AdS ($\hyperlink{S15}{\zAdSC}$) spacetimes are homogeneous under the
Poincaré group, their (unique) superisations
($\hyperlink{SM1}{\mathsf{SM1}}$ and
$\hyperlink{SM13}{\mathsf{SM13}}$, respectively) are homogeneous under
the Poincaré supergroup, suggesting a sort of correspondence or
duality.

\chapter{Generalised Bargmann Superspaces} \label{chap:gb_superspaces}
 In this chapter, we consider the last of our three types of symmetry, super-Bargmann symmetry. These symmetries are the least-studied of the three; therefore, we can only present the algebraic classification in this instance. 
\\ \\ 
Before introducing any supersymmetry, recall that a generalised Bargmann algebra (GBA) $\hat{\k}$ in $D$ spatial dimensions  may be thought of as a real one-dimensional abelian extension of a kinematical Lie algebra $\k$.  In particular, this enhancement of a kinematical Lie algebra requires an additional $\so(3)$ scalar module in the underlying vector space.  Let $\bZ$ span this extra copy of $\mathbb{R}$.  The classification of these extensions was presented in \cite{Figueroa-OFarrill:2017ycu} and followed a similar method to that used in the classification of kinematical Lie algebras, as discussed in Section~\ref{sec:ks_klas}.  In particular, rather than classifying the Lie algebras we can recover as deformations of the static kinematical Lie algebra, they classified the Lie algebras we can recover as deformations of the static kinematical Lie algebra's universal central extension.  The centrally-extended static kinematical Lie algebra is spanned by $\J, \B, \P, \bH$, and $\bZ$, with non-vanishing brackets 
\begin{equation}
	[\J, \J] = \J \quad [\J, \B] = \B \quad [\J, \P] = \P \quad [\B, \P] = \bZ.
\end{equation}
All other centrally-extended kinematical Lie algebras are then deformations of this algebra; therefore, these brackets are common to all such algebras. The results of this classification, for $D=3$, are shown in Table~\ref{tab:ce_algebras}, taken from \cite{Figueroa-OFarrill:2017ycu}.\footnote{As in Chapter~\ref{chap:k_superspaces}, we will only consider the generalised Bargmann algebras in $D=3$.}  The three sections of this table, starting from the top, are the non-trivial central extensions, the trivial central extensions, and, finally, the non-central extensions of kinematical Lie algebras. 
\begin{table}[h!]
  \centering
  \caption{Centrally-Extended Kinematical Lie Algebras in $D=3$}
  \label{tab:ce_algebras}
  \rowcolors{2}{blue!10}{white}
  \resizebox{\textwidth}{!}{
    \begin{tabular}{l|*{5}{>{$}l<{$}}|l}\toprule
      \multicolumn{1}{c|}{KLA} & \multicolumn{5}{c|}{Non-zero Lie brackets in addition to $[\J ,\J] = \J$, $[\J , \B] = \B$, $[\J ,\P] = \P$} & \multicolumn{1}{c}{Comments}\\\midrule
      1 & [\B ,\P] = \bZ & &  &  & & $\hat{\a}$ \\
      2 & [\B ,\P] = \bZ & [\bH,\B] = \B & [\bH,\P]= -\P &  & & $\hat{\n}_-$\\
      3 & [\B ,\P] = \bZ & [\bH,\B] = \P & [\bH,\P]= -\B & & & $\hat{\n}_+$\\
      4 & [\B ,\P] = \bZ & [\bH,\B]=-\P & & & & $\hat{\g}$\\
	  \hline
      5 & [\B ,\P] = \bH & [\bH,\B] = \P &  &  & [\B , \B] = \J & $\e \oplus \mathbb{R}Z$ \\
      6 & [\B ,\P] = \bH & [\bH,\B] = -\P &  &  & [\B , \B] = -\J & $\p \oplus \mathbb{R}Z$\\
      7 & [\B ,\P] = \bH + \J & [\bH,\B] = -\B & [\bH,\P]= \P & & & $\so(4,1) \oplus \mathbb{R}Z$\\
      8 & [\B ,\P] = \bH & [\bH,\B]= \P & [\bH , \P] = -\B & [\P, \P] = \J & [\B, \B] = \J & $\so(5) \oplus \mathbb{R}Z$\\
      9 & [\B ,\P] = \bH & [\bH,\B] = -\P &  [\bH, \P] = \B & [\P, \P] = -\J  & [\B, \B] = -\J & $\so(3, 2) \oplus \mathbb{R}Z$ \\
      \hline
      10 & [\B ,\P] = \bZ & [\bH,\B] = \B & [\bH,\P]= \P & [\bH, \bZ] = 2 \bZ & & \\
      11 & [\B ,\P] = \bZ & [\bH,\B] = \gamma \B & [\bH,\P]= \P & [\bH, \bZ] = (\gamma + 1) \bZ & & $\gamma \in (-1, 1)$\\
      12 & [\B ,\P] = \bZ & [\bH,\B]= \B+ \P & [\bH, \P] = \P & [\bH, \bZ] = 2 \bZ &  & \\
      13 & [\B ,\P] = \bZ & [\bH,\B] = \alpha \B + \P & [\bH,\P]= -\B + \alpha \P & [\bH, \bZ] = 2\alpha \bZ & & $\alpha > 0$ \\
      14 & [\B ,\P] = \bZ & [\bZ,\B] = \P & [\bH,\P]= \P & [\bH, \bZ] = \bZ & [\B, \B] = \J & $\co(4) \ltimes \mathbb{R}^4 $\\
      15 & [\B ,\P] = \bZ & [\bZ,\B]= -\P & [\bH, \P] = \P & [\bH, \bZ] = \bZ &  [\B, \B] = - \J & $\co(3,1) \ltimes \mathbb{R}^{3, 1}$ \\ 
      \bottomrule
    \end{tabular}
    }
\end{table}  \\ \\
In this chapter, we will focus solely on the first of these sections, and, from now on, it shall be exclusively the algebras of this section that we are referring to when using the term \textit{generalised Bargmann algebras}. It is useful for our later calculations to define a \textit{universal} generalised Bargmann algebra.  In addition to the standard kinematical brackets given in \eqref{eq:kinematical_brackets_D3}, this algebra has non-vanishing brackets
\begin{equation}
	[\B, \P] = \bZ \qquad [\bH, \B] = \lambda \B + \mu \P \qquad [\bH, \P] = \eta \B + \varepsilon \P,
\end{equation}
where $\lambda, \mu, \eta, \varepsilon \in \mathbb{R}$.\footnote{Note, the universal generalised Bargmann algebra is not a Lie algebra for arbitrary $\lambda, \mu, \eta, \varepsilon$.  It is used here simply as a computational tool.}  Setting these four parameters to certain values allows us to reduce to the four cases of interest.  For example, 
$\hat{\g}$ is given by setting $\lambda=\eta=\varepsilon=0$ and $\mu=-1$.  By beginning with the universal algebra, and only picking our parameters, and, thus, our algebra, when we can no longer make progress in the universal case, we reduce the amount of repetition in our calculations.
\begin{table}[h!]
  \centering
  \caption{Generalised Bargmann Algebras in $D=3$}
  \label{tab:gb_algebras}
  \rowcolors{2}{blue!10}{white}
    \begin{tabular}{l|*{5}{>{$}l<{$}}|l}\toprule
      \multicolumn{1}{c|}{GBA} & \multicolumn{5}{c|}{Non-zero Lie brackets in addition to $[\J,\J] = \J$, $[\J, \B] = \B$, $[\J,\P] = \P$} & \multicolumn{1}{c}{Comments}\\\midrule
      \hypertarget{a}{1} & [\B,\P] = \bZ & &  &  & & $\hat{\a}$ \\
      \hypertarget{n-}{2} & [\B,\P] = \bZ & [\bH,\B] = \B & [\bH,\P]= -\P &  & & $\hat{\n}_-$\\
      \hypertarget{n+}{3} & [\B,\P] = \bZ & [\bH,\B] = \P & [\bH,\P]= -\B & & & $\hat{\n}_+$\\
      \hypertarget{g}{4} & [\B,\P] = \bZ & [\bH,\B]=-\P & & & & $\hat{\g}$\\
      \bottomrule
    \end{tabular}
\end{table}
\\ \\
Our strategy for classifying generalised Bargmann superalgebras will be analogous to the strategy used in Section~\ref{sec:ks_superspace_ksa} to classify the kinematical Lie superalgebras.  In particular, a super-extension $\s$ of one of our generalised Bargmann algebras $\hat{\k}$ will be a Lie superalgebra such that $\s_{\bar{0}}= \hat{\k}$.  To determine the super-extensions of the generalised Bargmann algebras, we, therefore, begin by letting $\s_{\bar{0}}$ be our universal generalised Bargmann algebra.  We then need to specify the Lie brackets $[\bH, \Q]$, $[\bZ, \Q]$, $[\B, \Q]$, $[\P, \Q]$, and $[\Q, \Q]$.  Each of the $[\s_{\bar{0}}, \s_{\bar{1}}]$ components of the bracket must be an $\r$-equivariant endomorphism of $\s_{\bar{1}}$, while the $[\s_{\bar{1}}, \s_{\bar{1}}]$ component must be an $\r$-equivariant map $\bigodot^2 \s_{\bar{1}} \rightarrow \s_{\bar{0}}$.  The space of possible brackets will be a real vector space $\cV$.  We then use the super-Jacobi identity to cut out an algebraic variety $\cJ \subset \cV$.  Since we are exclusively interested in supersymmetric extensions, we restrict ourselves to those Lie superalgebras for which $[\Q, \Q]$ is non-vanishing, which define a sub-variety $\cS \subset \cJ$.\footnote{As in Chapter~\ref{chap:k_superspaces}, we restrict ourselves to the cases where $[Q, Q] \neq 0$ as our interests lie in spacetime supersymmetry: we would like supersymmetry transformations to generate geometric transformations of the spacetime.}  This sub-variety may be unique to each generalised Bargmann algebra; therefore, it is at this stage we start to set the parameters of the universal algebra, where applicable. The isomorphism classes of the remaining Lie superalgebras are then in one-to-one correspondence with the orbits of $\cS$ under the subgroup $\G \subset \GL(\s_{\bar{0}}) \times \GL(\s_{\bar{1}})$.  The group $\G$ contains the automorphisms of $\s_{\bar{0}} = \hat{\k}$ and additional transformations which are induced by the endomorphism ring of $\s_{\bar{1}}$. The form of this subgroup will be discussed in the $\N=1$ and $\N=2$ cases in sections \ref{subsubsec:N1_S_autos} and \ref{subsubsec:N2_S_autos}, respectively.  
\\ \\
In the $\N=1$ case, we will identify each orbit of $\cS$ explicitly, giving a full classification of the generalised Bargmann superalgebras in this instance.  However, in the $\N=2$ case, we will only identify the non-empty branches of $\cS$.  Each branch will have a unique set of $[\s_{\bar{0}}, \s_{\bar{1}}]$ and $[\s_{\bar{1}}, \s_{\bar{1}}]$ brackets for the associated generalised Bargmann algebra.  Thus, we can highlight the form of the possible super-extensions without spending too much time pinpointing exact coefficients.
\\ \\
The rest of this chapter is organised as follows.  In Section~\ref{sec:gb_superspace_n1}, we classify the $\N=1$ generalised Bargmann superalgebras in $D=3$.  We begin by generalising the setup for the kinematical Lie superalgebra classification presented in Section~\ref{subsec:kss_alg_setup} before proceeding to the classification itself in Section~\ref{subsec:N1_class} and summarising in Section~\ref{subsec:N1_summary}.  As part of our summary, we will demonstrate how to unpack our quaternionic formalism, using one of the $\N=1$ Bargmann superalgebras as our example. In Section~\ref{sec:gb_superspace_n2}, we move on to classify the $\N=2$ generalised Bargmann superalgebras in $D=3$.  This case is considerably more involved than the $\N=1$ case; therefore, after an analogous discussion on the initial setup of the classification problem, we require an intermediate step in which we define four possible branches of generalised Bargmann superalgebra in $\cS$. In Section~\ref{subsec:N2_class}, we go through the lengthy procedure of identifying the sub-branches which contain valid generalised Bargmann superalgebra structures, and we summarise our findings in Section~\ref{subsec:N2_summary}.

\section{Classification of $\N=1$ Generalised Bargmann Superalgebras} \label{sec:gb_superspace_n1}
Our investigation into generalised Bargmann superalgebras begins with the simplest
case, $\N=1$.  Following on from Section~\ref{subsec:math_prelims_alg_gbsa}, Section~\ref{subsec:N1_setup} will complete our set up for this 
case by specifying the precise form of the $[\s_{\bar{0}}, \s_{\bar{1}}]$ and $[\s_{\bar{1}}, \s_{\bar{1}}]$ brackets.  We then give some preliminary results that will be useful in the
classification of the $\N=1$ extensions, and 
define the group of basis transformations $\G \subset \GL(\s_{\bar{0}}) \times \GL(\s_{\bar{1}})$, which will allow us to pick out a single representative for each isomorphism class.  In Section~\ref{subsec:N1_class}, the classification is given before we summarise the results in Section~\ref{subsec:N1_summary}.
\subsection{Setup for the $\N=1$ Calculation} \label{subsec:N1_setup}
We note that, in addition to the standard kinematical Lie brackets, the brackets for the universal generalised Bargmann superalgebra are
\begin{equation}
	\begin{split}
		[\sB(\beta), \sP(\pi)] &= \Re(\bar{\beta}\pi) \sZ \\ 
		[\sH, \sB(\beta)] &= \lambda \sB(\beta) + \mu \sP(\beta) \\
		[\sH, \sP(\pi)] &= \eta \sB(\pi) + \varepsilon \sP(\pi) ,
	\end{split}
\end{equation}
where $\beta, \pi \in \Im(\mathbb{H})$ and $\lambda, \mu, \eta, \varepsilon \in \mathbb{R}$.  We now want to specify the possible $[\s_{\bar{0}}, \s_{\bar{1}}]$
and $[\s_{\bar{1}}, \s_{\bar{1}}]$ brackets.  From Section~\ref{subsec:kss_alg_setup},
we have
\begin{equation} \label{eq:N1_general_brackets}
	\begin{split}
		[\sJ(\omega), \sQ(\theta)] &= \tfrac12 \sQ(\omega \theta) \\
		[\sB(\beta), \sQ(\theta)] &= \sQ(\beta \theta \bb) \\
		[\sP(\pi), \sQ(\theta)] &= \sQ(\pi \theta \pp) \\
		[\sH, \sQ(\theta)] &= \sQ(\theta\hh), \\
	\end{split}
\end{equation}
where $\omega, \pi, \beta \in \Im(\mathbb{H})$, $\theta, \bb, \pp, \hh \in \mathbb{H}$.
Since $\bZ$ is just another $\so(3)$ scalar module, and, therefore, the analysis of the 
bracket $[\bZ, \Q]$ will be identical to that of $[\bH, \Q]$, we know we can write
\begin{equation} \label{eq:N1_Z_bracket}
	[\sZ, \sQ(\theta)] = \sQ(\theta\zz),
\end{equation}
where $\zz \in \mathbb{H}$.  Having added this additional generator, 
the possible $[\s_{\bar{1}}, \s_{\bar{1}}]$ brackets are now $\so(3)$-equivariant elements
of $\Hom_{\mathbb{R}}(\bigodot^2 S, \s_{\bar{0}}) = \Hom_{\mathbb{R}}(3V \oplus \mathbb{R}, 3V \oplus 2 \mathbb{R})
= 9\, \Hom_{\mathbb{R}}(V, V) \oplus 2\, \Hom_{\mathbb{R}}(\mathbb{R}, \mathbb{R})$.
As may have been expected, given that we did not alter the vectorial sector of the algebra, the number of $\Hom_{\mathbb{R}}(V, V)$ elements does not change. However, now that we have an additional $\so(3)$ scalar module, we have an additional scalar map, so
\begin{equation} \label{eq:N1_QQ}
	[\sQ(\theta), \sQ(\theta)] = n_0 |\theta|^2 \sH + n_1 |\theta|^2 \sZ - \sJ(\theta\nn_2\bar{\theta}) - \sB(\theta\nn_3\bar{\theta}) - \sP(\theta\nn_4\bar{\theta}),
\end{equation}
where $n_0, n_1 \in \mathbb{R}$ and $\nn_2, \nn_3, \nn_4 \in \Im(\mathbb{H})$.  This expression polarises to
\begin{equation}
	[\sQ(\theta), \sQ(\theta')] = n_0 \Re(\bar{\theta}\theta') \sH + n_1 \Re(\bar{\theta}\theta') \sZ - \sJ(\theta'\nn_2\bar{\theta}+ \theta \nn_2\bar{\theta'}) 
	- \sB(\theta'\nn_3\bar{\theta}+ \theta \nn_3\bar{\theta'}) -\sP(\theta'\nn_4\bar{\theta}+ \theta \nn_4\bar{\theta'}).
\end{equation}
\subsubsection{Preliminary Results} \label{subsubsec:N1_S_preliminaries}
Following the example of Section~\ref{subsec:kss_alg_setup}, we will now consider the super-Jacobi identity
and use its components to derive universal conditions that may aid our classification.  We have three components of super-Jacobi identity to consider
\begin{enumerate}
	\item $(\s_{\bar{0}}, \s_{\bar{0}}, \s_{\bar{1}})$,
	\item $(\s_{\bar{0}}, \s_{\bar{1}}, \s_{\bar{1}})$, and 
	\item $(\s_{\bar{1}}, \s_{\bar{1}}, \s_{\bar{1}})$.
\end{enumerate}
We do not need to consider the $(\s_{\bar{0}}, \s_{\bar{0}}, \s_{\bar{0}})$ case as we already know
that these are satisfied by the generalised Bargmann algebras. Equally, we do not need to include $\J$ in our investigations as the identities involving the rotational subalgebra $\r$ impose the $\so(3)$-equivariance of the brackets, which we already have by construction.  Now, let us consider each component of the identity in turn.  In the following
discussions, we will only write down explicitly those identities which are not trivially satisfied.
\\
\paragraph{$(\s_{\bar{0}}, \s_{\bar{0}}, \s_{\bar{1}})$} \label{par:N1_S_P_001} ~\\ \\
By imposing these super-Jacobi identities, we ensure that $\s_{\bar{1}}$ is an 
$\s_{\bar{0}}$ module, not just an $\so(3)$ module. 
The $(\s_{\bar{0}}, \s_{\bar{0}}, \s_{\bar{1}})$ identities can be summarised as follows.
\begin{lemma}\label{lem:N1_001}
  The following relations between $\hh, \zz,\bb,\pp \in \mathbb{H}$ are implied by the
  corresponding $\k$-brackets:
  \begin{equation}
    \begin{split}
    	  [\bH,\bZ] = \lambda \bH + \mu \bZ & \implies [\zz, \hh] = \lambda \hh + \mu \zz \\
      [\bH,\B] = \lambda \B + \mu \P & \implies [\bb,\hh] = \lambda \bb + \mu \pp\\
      [\bH,\P] = \lambda \B + \mu \P & \implies [\pp,\hh] = \lambda \bb + \mu \pp\\
      [\bZ,\B] = \lambda \B + \mu \P & \implies [\bb,\zz] = \lambda \bb + \mu \pp\\
      [\bZ,\P] = \lambda \B + \mu \P & \implies [\pp,\zz] = \lambda \bb + \mu \pp\\
      [\B,\B] = \lambda \B + \mu \P + \nu \J & \implies \bb^2 = \tfrac12 \lambda \bb + \tfrac12 \mu \pp + \tfrac14 \nu\\
      [\P,\P] = \lambda \B + \mu \P + \nu \J & \implies \pp^2 = \tfrac12 \lambda \bb + \tfrac12 \mu \pp + \tfrac14 \nu\\
      [\B,\P] = \lambda \bH + \mu \bZ & \implies \bb \pp + \pp\bb = 0\quad\text{and}\quad [\bb,\pp] = \lambda \hh  + \mu \zz.
    \end{split}
  \end{equation}
\end{lemma}
\begin{proof}
All the results excluding $\bZ$ are taken from Lemma~\ref{lem:kmod}, and 
the $[\bZ, \B]$ and $[\bZ, \P]$ results are the same \textit{mutatis mutandis} as $[\bH, \B]$ and $[\bH, \P]$.
Therefore, the only new results are those for $[\bH, \bZ]$ and $[\B, \P]$.  The $[\bH, \bZ, \Q]$ 
identity is written
\begin{equation}
	[\sH, [\sZ, \sQ(\theta)] = [[\sH, \sZ], \sQ(\theta)] + [\sZ, [\sH, \sQ(\theta)]].
\end{equation}
Substituting in the relevant brackets, we find
\begin{equation}
	\sQ(\theta\zz\hh) = \lambda \sQ(\theta\hh) + \mu \sQ(\theta\zz) + \sQ(\theta\hh\zz).
\end{equation}
Using the injectivity of $\sQ$, we arrive at
\begin{equation}
	[\zz, \hh] = \lambda \hh + \mu \zz.
\end{equation}
Finally, the $[\B, \P, \Q]$ identity is
\begin{equation}
	[\sB(\beta), [\sP(\pi), \sQ(\theta)]] = [[\sB(\beta), \sP(\pi)], \sQ(\theta)] + [ \sP(\pi), [\sB(\beta), \sQ(\theta)]].
\end{equation}
Substituting in the brackets from \eqref{eq:N1_general_brackets}, we arrive at
\begin{equation}
	\sQ(\beta\pi \theta \pp\bb) = \Re(\bar{\beta}\pi) (\lambda \sQ(\theta\hh) + \mu \sQ(\theta\zz)) + \sQ(\pi\beta \theta \bb\pp).
\end{equation}
Letting $\beta = \pi = \ii$, we find
\begin{equation}
	[\bb, \pp] = \lambda \hh + \mu \zz.
\end{equation}
Now, let $\beta = \ii$ and $\pi = \jj$.  In this case, the $\lambda$ and $\mu$ terms vanish and
we are left with
\begin{equation}
	\bb\pp + \pp\bb = 0.
\end{equation}
\end{proof}
\paragraph{$(\s_{\bar{0}}, \s_{\bar{1}}, \s_{\bar{1}})$} \label{par:N1_S_P_011} ~ \\ \\
By imposing these super-Jacobi identities, we ensure that the $[\Q, \Q]$ bracket is an $\s_{\bar{0}}$-equivariant map  $\bigodot \s_{\bar{1}} \rightarrow \s_{\bar{0}}$.  The $(\s_{\bar{0}}, \s_{\bar{1}}, \s_{\bar{1}})$ identities can be difficult to
study if we are trying to be completely general; however, we know that 
all four algebras in Table~\ref{tab:gb_algebras} can be written as specialisations
of the universal generalised Bargmann algebra:
\begin{equation}
	[\B, \P] = \bZ \qquad [\bH, \B] = \lambda \B + \mu \P \qquad [\bH, \P] = \eta \B + \varepsilon \P,
\end{equation}
where $\lambda, \mu, \eta, \varepsilon \in \mathbb{R}$.  Therefore, we may use the
brackets of this algebra to obtain the following result.
\begin{lemma} \label{lem:N2_011}
The $[\bH, \Q, \Q]$ identity produces the conditions
\begin{equation}
	\begin{split}
		0 &= n_i \Re(\hh) \quad \text{where} \quad i \in \{0, 1\} \\
		0 &= \hh \nn_2 + \nn_2 \bar{\hh} \\
		\lambda \nn_3 + \eta \nn_4 &= \hh \nn_3 + \nn_3 \bar{\hh} \\
		\mu \nn_3 + \varepsilon \nn_4 &= \hh \nn_4 + \nn_4 \bar{\hh}.
	\end{split}
\end{equation}
The $[\bZ, \Q, \Q]$ identity produces the conditions
\begin{equation}
	\begin{split}
		0 &= n_i \Re(\hh) \quad \text{where} \quad i \in \{0, 1\} \\
		0 &= \hh \nn_j + \nn_j \bar{\hh} \quad \text{where} \quad j \in \{2, 3, 4\}.
	\end{split}
\end{equation}
The $[\B, \Q, \Q]$ identity produces the conditions
\begin{equation}
	\begin{split}
		0 &= n_0 \Re(\bar{\theta}\beta \theta \bb) \\
		0 &= \Re(\bar{\beta} \theta (\nn_4 + 2 n_1 \bar{\bb}) \bar{\theta}) \\
		0 &= \theta \nn_2 \overbar{\beta \theta \bb} + \beta \theta \bb \nn_2 \bar{\theta} \\
		\lambda n_0 |\theta|^2 \beta + \tfrac12 [\beta, \theta\nn_2\bar{\theta}] &= \theta \nn_3 \overbar{\beta \theta \bb}+ 
		\beta \theta \bb \nn_3 \bar{\theta} \\
		\mu n_0 |\theta|^2 \beta &= \theta \nn_4 \overbar{\beta \theta \bb} + \beta \theta \bb \nn_4 \bar{\theta}.
	\end{split}
\end{equation}
The $[\P, \Q, \Q]$ identity produces the conditions
\begin{equation}
	\begin{split}
		0 &= n_0 \Re(\bar{\theta}\pi \theta \pp) \\
		0 &= \Re(\bar{\pi} \theta (\nn_3 - 2 n_1 \bar{\pp}) \bar{\theta}) \\
		0 &= \theta \nn_2 \overbar{\pi \theta \pp} + \pi \theta \pp \nn_2 \bar{\theta} \\
		\eta n_0 |\theta|^2 \pi &= \theta \nn_3 \overbar{\pi \theta \pp}+ \pi \theta \pp \nn_3 \bar{\theta} \\
		\varepsilon n_0 |\theta|^2 \pi + \tfrac12 [\pi, \theta\nn_2\bar{\theta}] &= \theta \nn_4 \overbar{\pi \theta \pp} + \pi \theta \pp \nn_4 \bar{\theta}.
	\end{split}
\end{equation}
\end{lemma}
\begin{proof}
The $[\bH, \Q, \Q]$ super-Jacobi identity is written
\begin{equation}
	[\sH, [\sQ(\theta), \sQ(\theta)]] = 2 [ [\sH, \sQ(\theta)], \sQ(\theta)].
\end{equation}
Using \eqref{eq:N1_general_brackets} and \eqref{eq:N1_QQ}, we find
\begin{equation}
	\begin{split}
		- \sB( \theta (\lambda \nn_3 + \eta \nn_4) \bar{\theta}) - \sP(\theta(\mu \nn_3 + \varepsilon \nn_4) \bar{\theta})
		= &\, 2 n_0 \Re(\overbar{\theta\hh}\theta) \sH + 2 n_1 \Re(\overbar{\theta\hh} \theta) \sZ\\ & - \sJ(\theta \nn_2 \overbar{\theta\hh} 
		+ \theta\hh \nn_2 \bar{\theta}) - \sB(\theta \nn_3 \overbar{\theta\hh} + \theta\hh \nn_3 \bar{\theta}) - \sP(\theta\nn_4 \overbar{\theta\hh}
		+ \theta\hh \nn_4 \bar{\theta}).
	\end{split}
\end{equation}
Comparing $\sH$, $\sZ$, $\sJ$, $\sB$, and $\sP$ coefficients, and using the injectivity and 
linearity of the maps $\sJ, \sB$, and $\sP$, we find
\begin{equation}
	\begin{split}
		0 &= n_i \Re(\hh) \quad \text{where} \quad i \in \{0, 1\} \\
		0 &= \hh \nn_2 + \nn_2 \bar{\hh} \\
		\lambda \nn_3 + \eta \nn_4 &= \hh \nn_3 + \nn_3 \bar{\hh} \\
		\mu \nn_3 + \varepsilon \nn_4 &= \hh \nn_4 + \nn_4 \bar{\hh}.	
	\end{split}
\end{equation}
The calculations for the $[\bZ, \Q, \Q]$ identity follows in an analogous manner.  The key difference
is this case is that the L.H.S.
vanishes in all instances since $\bZ$ commutes with all basis elements.  The $[\B, \Q, \Q]$ identity
is 
\begin{equation}
	[\sB(\beta), [\sQ(\theta), \sQ(\theta)]] = 2 [[\sB(\beta), \sQ(\theta)], \sQ(\theta)].
\end{equation}
Substituting in the relevant brackets, the L.H.S. becomes
\begin{equation}
	\text{L.H.S.} = -\lambda n_0 |\theta|^2 \sB(\beta) - \tfrac12 \sB([\beta, \theta \nn_2\bar{\theta}]) - \Re(\bar{\beta} \theta\nn_4 \bar{\theta}) \sZ,
\end{equation}
and the R.H.S. becomes
\begin{equation}
	\begin{split}
		\text{R.H.S.} = & \, 2 n_0 \Re(\bar{\theta}\beta \theta\bb) \sH + 2 n_1 \Re(\bar{\theta}\beta \theta\bb) \sZ \\
		& - \sJ(\theta\nn_2 \overbar{\beta \theta\bb} + \beta \theta \bb \nn_2 \bar{\theta}) - \sB(\theta\nn_3 \overbar{\beta \theta\bb} 
		+ \beta \theta \bb \nn_3 \bar{\theta}) - \sP(\theta\nn_4 \overbar{\beta \theta\bb} + \beta \theta \bb \nn_4 \bar{\theta}).
	\end{split}
\end{equation}
Again, comparing coefficients and using the injectivity and linearity of our maps, we
get
\begin{equation}
	\begin{split}
		0 &= n_0 \Re(\bar{\theta}\beta \theta \bb) \\
		0 &= \Re(\bar{\beta} \theta (\nn_4 + 2 n_1 \bar{\bb}) \bar{\theta}) \\
		0 &= \theta \nn_2 \overbar{\beta \theta \bb} + \beta \theta \bb \nn_2 \bar{\theta} \\
		\lambda n_0 |\theta|^2 \beta + \tfrac12 [\beta, \theta\nn_2\bar{\theta}] &= \theta \nn_3 \overbar{\beta \theta \bb}+ 
		\beta \theta \bb \nn_3 \bar{\theta} \\
		\mu n_0 |\theta|^2 \beta &= \theta \nn_4 \overbar{\beta \theta \bb} + \beta \theta \bb \nn_4 \bar{\theta}.
	\end{split}
\end{equation}
The $[\P, \Q, \Q]$ results follow in identical fashion by replacing $\bb$ with $\pp$ and $\beta$
with $\pi$.
\end{proof}
\paragraph{$(\s_{\bar{1}}, \s_{\bar{1}}, \s_{\bar{1}})$} \label{par:N1_S_P_111} ~\\ \\
The last super-Jacobi identity to consider is the $(\s_{\bar{1}}, \s_{\bar{1}}, \s_{\bar{1}})$ case,
$[\Q, \Q, \Q]$.
\begin{lemma}\label{lem:N1_111} 
  The $[\Q,\Q,\Q]$ identity produces the condition
  \begin{equation}
    n_0 \hh + n_1 \zz = \tfrac12 \nn_2  + \nn_3 \bb + \nn_4 \pp.
  \end{equation}
\end{lemma}
\begin{proof}
The $[\Q, \Q, \Q]$ identity is 
\begin{equation}
	0 = [[\sQ(\theta), \sQ(\theta)], \sQ(\theta)].
\end{equation}
Using \eqref{eq:N1_QQ}, and the brackets in \eqref{eq:N1_general_brackets} and \eqref{eq:N1_Z_bracket}, 
we find
\begin{equation}
	\begin{split}
		0 &= [n_0 |\theta|^2 \sH + n_1 |\theta|^2 \sZ - \sJ(\theta\nn_2\bar{\theta}) - \sB(\theta\nn_3\bar{\theta}) - \sP(\theta\nn_4
		\bar{\theta}), \sQ(\theta) ] \\
		&= n_0 |\theta|^2 \sQ(\theta\hh) + n_1 |\theta|^2 \sQ(\theta\zz) - \tfrac12 \sQ(\theta\nn_2\bar{\theta}\theta) - \sQ(\theta\nn_3\bar{\theta}\theta\bb) - \sQ(\theta\nn_4
		\bar{\theta}\theta\pp).
	\end{split}
\end{equation}
Since $\sQ$ is injective, this gives us
\begin{equation}
	n_0 \hh + n_1 \zz = \tfrac12 \nn_2  + \nn_3 \bb + \nn_4 \pp.
\end{equation}
\end{proof}
\subsubsection{Basis Transformations} \label{subsubsec:N1_S_autos}
As well as modifying the super-Jacobi identities presented in Section~\ref{subsec:kss_alg_setup}, the new $\so(3)$ scalar also impacts the subgroup $\G \subset \GL(\s_{\bar{0}}) \times \GL(\s_{\bar{1}})$ of basis transformation for kinematical Lie superalgebras.  All the automorphisms in $\G$ generated by $\so(3)$ 
remain the same for $\bb$, $\pp$, and $\hh$, but we may now add how $\zz$ transforms.  These automorphisms
act by rotating the three imaginary quaternionic bases $\ii$, $\jj$, and $\kk$ by an element
of $\SO(3)$.  In particular, we have a homomorphism $\Ad: \Sp(1) \rightarrow \Aut(\mathbb{H})$
defined such that for $\uu \in \Sp(1)$ and $\ss \in \mathbb{H}$, $\Ad_{\uu}(\ss) = \uu \ss \bar{\uu}$.  The map $\Ad_{\uu}$ then acts trivially on the real component of $\ss$ and via $\SO(3)$ rotations on
$\Im(\mathbb{H})$.  Therefore, $\tilde{\sB} = \sB\circ \Ad_{\uu}$, $\tilde{\sP} = \sP\circ \Ad_{\uu}$, $\tilde{\sH} = \sH$, $\tilde{\sQ} = \sQ\circ \Ad_{\uu}$. Since $\bZ$ is an $\so(3)$ scalar, $\tilde{\sZ} = \sZ$.  Substituting this with $\tilde{\sQ} = \sQ\circ \Ad_{\uu}$ into the
$[\bZ, \Q]$ bracket, we find that $\tilde{\zz} = \bar{\uu}\zz \uu$.  Additionally, substituting these transformations into the $[\Q, \Q]$ bracket, we see that $n_1$ remains inert.  The other type of transformations to consider
are the $\so(3)$-equivariant maps $\s \rightarrow \s$.  Since we now have two $\so(3)$ 
scalars, we can have
\begin{equation}
	\begin{split}
		\tilde{\sH} &= a \sH + b \sZ \\
		\tilde{\sZ} &= c \sH + d \sZ
	\end{split} \quad \text{where} \quad 
	\begin{pmatrix}
		a & b \\ c & d
	\end{pmatrix} \in \GL(2, \mathbb{R}).
\end{equation}
The $\so(3)$ vector and spinor maps remain unchanged from those given in Section~\ref{subsec:kss_alg_setup}.
In particular, $\tilde{\sQ}(s) = \sQ(s \qq)$ for $\qq \in \mathbb{H}^\times$.
Substituting $\tilde{\sH}$, $\tilde{\sZ}$ and $\tilde{\sQ}$ into the brackets
\begin{equation}
	\begin{split}
		[\tilde{\sH}, \tilde{\sQ}(\theta)] &= \tilde{\sQ}(\theta\tilde{\hh}) \\
		[\tilde{\sZ}, \tilde{\sQ}(\theta)] &= \tilde{\sQ}(\theta\tilde{\zz}) \\
	\end{split} \qquad 
		[\tilde{\sQ}(\theta), \tilde{\sQ}(\theta)] = \tilde{n_0} |\theta|^2 \tilde{\sH} + \tilde{n_1} |\theta|^2 \tilde{\sZ} - \tilde{\sJ}(\theta\tilde{\nn_2}\bar{\theta}) - \tilde{\sB}(\theta\tilde{\nn_3}\bar{\theta}) - \tilde{\sP}(\theta\tilde{\nn_4}\bar{\theta}),
\end{equation}
we find 
\begin{equation}
	\begin{split}
		\tilde{\hh} &= \qq (a \hh + b \zz) \qq^{-1} \\
		\tilde{\zz} &= \qq (c \hh + d \zz) \qq^{-1}
	\end{split} \quad 
	\begin{split}
		\tilde{n_0} &= \frac{|\qq|^2}{ad -bc} (d n_0 - c n_1) \\
		\tilde{n_1} &= \frac{|\qq|^2}{ad- bc} (a n_1 - b n_0).
	\end{split}
\end{equation}
These amendments mean that the transformation in $\G$ produce the following basis changes
\begin{equation}
\begin{split}
\sJ &\mapsto \sJ\circ \Ad_{\uu} \\
\sB &\mapsto e \sB\circ \Ad_{\uu} + f \sP\circ \Ad_{\uu} \\
\sP &\mapsto h \sB\circ \Ad_{\uu} + i \sP\circ \Ad_{\uu} \\
\sH &\mapsto a \sH + b \sZ \\
\sZ &\mapsto c \sH + d \sZ \\
\sQ &\mapsto \sQ\circ \Ad_{\uu} \circ R_\qq.
\end{split}
\end{equation}
These transformations may be summarised by $(A = \big( \begin{smallmatrix} a & b \\ c & d \end{smallmatrix} \big),
C = \big(\begin{smallmatrix} e & f \\ h & i \end{smallmatrix}\big), \qq, \uu) \in \GL(\mathbb{R}^2) \times \GL(\mathbb{R}^2)
\times \mathbb{H}^\times \times \mathbb{H}^\times$.
\subsection{Classification} \label{subsec:N1_class}
The calculations for classifying the super-extensions of $\hat{\n}_{\pm}$ and $\hat{\g}$ all follow identically.  It
will, therefore, only be stated once below.  However, the central extension of the static 
kinematical Lie algbera is a little different, so will be treated first.
\subsubsection{$\hat{\a}$}
Using the preliminary results from Lemma~\ref{lem:N1_001} in Section~\ref{par:N1_S_P_001}, we find
$\bb=\pp=\zz=0$.  Substituting these quaternions into the $[\B, \Q, \Q]$ and $[\P, \Q, \Q]$
identities with the relevant brackets, we get $\nn_2 = 0$, $\nn_3=0$ and  $\nn_4 = 0$ .  
Then, wanting $[\Q, \Q] \neq 0$, the $[\bH, \Q, \Q]$ conditions tells us that $\Re(\hh)=0$.  Finally,
Lemma~\ref{lem:N1_111} reduces to $n_0 \hh = 0$.  Therefore, we have two possible cases: one in 
which $n_0 = 0$ and $\hh \in \Im(\mathbb{H})$ and another in which $\hh = 0$ and
$n_0$ is unconstrained.  In the former case, the subgroup $\G \subset \GL(\s_{\bar{0}}) \times \GL(\s_{\bar{1}})$ can be used to 
set $\hh = \kk$ and $n_1=1$, such that the only non-vanishing brackets 
involving $\sQ$ are
\begin{equation}
	[\sH, \sQ(\theta)] = \sQ(\theta\kk) \quad \text{and} \quad [\sQ(\theta), \sQ(\theta)] = |\theta|^2 \sZ.
\end{equation}
Notice, however, that this case also allows for $\hh = 0$, leaving only
\begin{equation}
	[\sQ(\theta), \sQ(\theta)] = |\theta|^2 \sZ.
\end{equation}
In the latter case, we can use $\G$ to scale $n_0$ and 
$n_1$, so the non-vanishing brackets are
\begin{equation}
	[\sQ(\theta), \sQ(\theta)] = |\theta|^2 \sH + |\theta|^2 \sZ.
\end{equation}
\subsubsection{$\hat{\n}_{\pm}$ and $\hat{\g}$} \label{subsubsec:gb_superspace_n1_g}
Using the preliminary results of Lemmas~\ref{lem:N1_001} and~\ref{lem:N1_111}, we
 instantly find $\bb = \pp = \zz = 0$, and, subsequently, $n_0 \hh = \tfrac12 \nn_2$.
The super-Jacobi identity $[\B, \Q, \Q]$ then tells us that $n_0 = \nn_2 = \nn_4 = 0$ and the identity
$[\P, \Q, \Q]$ gives us $\nn_3 = 0$.  Thus, the $(\s_{\bar{1}}, \s_{\bar{1}}, \s_{\bar{1}})$ 
condition is trivially satisfied.  The only remaining condition is from $[\bH, \Q, \Q]$, which
tells us $n_1\Re(\hh)=0$.  Since we want $[\Q, \Q] \neq 0$, we must have $n_1 \neq 0$,
therefore, $\hh \in \Im(\mathbb{H})$.  Using $\G$ to set
$\hh = \kk$ and $n_1 = 1$, we have non-vanishing brackets
\begin{equation}
	[\sH, \sQ(\theta)] = \sQ(\theta\kk) \quad \text{and} \quad [\sQ(\theta), \sQ(s)] = |\theta|^2 \sZ.
\end{equation}
Similar to the $\hat{\a}$ case, the restriction $\hh \in \Im(\mathbb{H})$ does not remove the choice
$\hh = 0$.  Therefore, we may also have
\begin{equation}
	[\sQ(\theta), \sQ(s)] = |\theta|^2 \sZ
\end{equation}
as the only non-vanishing bracket.
\subsection{Summary} \label{subsec:N1_summary}
Table~\ref{tab:N1_classification} lists all the $\N=1$ generalised Bargmann superalgebras we have classified.  Each Lie superalgebra is an $\N=1$ super-extension of one of the generalised Bargmann algebras given in Table~\ref{tab:gb_algebras}, taken from \cite{Figueroa-OFarrill:2017ycu}.  It is interesting to compare this list of $\N=1$ super-extensions of centrally-extended kinematical Lie algebras to the list of centrally-extended $\N=1$ kinematical Lie superalgebras given in Table~\ref{tab:central-ext}.  This table is a reduced and adapted version of one given in Section~\ref{sec:central-extensions}, where we have only kept those extensions built upon the static, Newton-Hooke, and Galilean algebras.
\\ \\
Notice that only one of the generalised Bargmann superalgebras is present in the classification of centrally-extended kinematical Lie superalgebras, \hyperlink{S1}{S1}.  Although it does not match exactly, we can use the basis transformations in $\G \subset \GL(\s_{\bar{0}}) \times \GL(\s_{\bar{1}})$ to bring it into the same form as the second superalgebra in Table~\ref{tab:central-ext}.  
\\ \\
The fact that these tables only have one Lie superalgebra in common is, perhaps, unsurprising. The Lie superalgebras presented in Table~\ref{tab:N1_classification} almost exclusively have $[\Q, \Q] = \bZ$.  Thus, before introducing the new generator $\bZ$, these Lie superalgebras would have had $[\Q, \Q] =0$.  By construction, such Lie superalgebras were left out of the classification in Section~\ref{sec:class-kinem-lie}.  This may explain why there is so little crossover between these tables.  
\begin{table}[h!]
  \centering
  \caption{$\N=1$ Generalised Bargmann Superalgebras (with $[\Q,\Q]\neq 0$)}
  \label{tab:N1_classification}
  \setlength{\extrarowheight}{2pt}
  \rowcolors{2}{blue!10}{white}
    \begin{tabular}{l|l*{5}{|>{$}c<{$}}}\toprule
      \multicolumn{1}{c|}{S} & \multicolumn{1}{c|}{$\k$} & \multicolumn{1}{c|}{$\hh$}& \multicolumn{1}{c|}{$\zz$}& \multicolumn{1}{c|}{$\bb$} & \multicolumn{1}{c|}{$\pp$} & \multicolumn{1}{c}{$[\sQ(\theta),\sQ(\theta)]$}\\
      \toprule
      1 & \hyperlink{a}{$\hat{\a}$} & & & & & |\theta|^2 \sZ  \\
      2 & \hyperlink{a}{$\hat{\a}$} & \kk & & & & |\theta|^2 \sZ  \\ 
      \hypertarget{S3}{3} & \hyperlink{a}{$\hat{\a}$} & & &&  & |\theta|^2 \sH + |\theta|^2 \sZ \\
      4 & \hyperlink{n-}{$\hat{\n}_-$} & & & & & |\theta|^2 \sZ  \\
      5 & \hyperlink{n-}{$\hat{\n}_-$} & \kk & & & & |\theta|^2 \sZ  \\
      6 & \hyperlink{n+}{$\hat{\n}_+$} &  & & & & |\theta|^2 \sZ  \\      
      7 & \hyperlink{n+}{$\hat{\n}_+$} & \kk & & & & |\theta|^2 \sZ  \\
      8 & \hyperlink{g}{$\hat{\g}$} & & & & & |\theta|^2 \sZ  \\
      9 & \hyperlink{g}{$\hat{\g}$} & \kk & & & & |\theta|^2 \sZ  \\
      \bottomrule
    \end{tabular}
    \caption*{The first column gives each generalised Bargmann superalgebra $\s$ a unique identifier,
    and the second column tells us the underlying generalised Bargmann algebra $\k$.
    The next four columns tells us how the $\s_{\bar{0}}$ generators $\bH, \bZ, \B$, and $\P$ act on $\Q$.
    Recall, the action of $\J$ is fixed, so we do not need to state this explicitly.  The final column then specifies the $[\Q, \Q]$ bracket. 
    }
\end{table}
\begin{table}[h!]
  \centering
  \caption{Central Extensions of $\N=1$ Kinematical Lie Superalgebras}
  \label{tab:central-ext}
  \setlength{\extrarowheight}{2pt}
  \rowcolors{2}{blue!10}{white}
  \begin{tabular}{l*{6}{|>{$}c<{$}}}\toprule
    \multicolumn{1}{c|}{$\k$} & \multicolumn{1}{c|}{$[\sB(\beta),\sP(\pi)]$} & \multicolumn{1}{c|}{$\hh$} & \multicolumn{1}{c|}{$\zz$} & \multicolumn{1}{c|}{$\bb$}  & \multicolumn{1}{c|}{$\pp$} & \multicolumn{1}{c}{$[\sQ(\theta),\sQ(\theta)]$} \\
    \toprule
    $\a$ & & \tfrac12 \kk & & & & |\theta|^2 \sZ - \sP(\theta\kk\bar{\theta}) \\
    $\a$ & \Re(\bar{\beta}\pi) \sZ & & & & & |\theta|^2 \sH \\
    $\a$ & & & & & & |\theta|^2 \sZ - \sB(\theta\jj\bar{\theta}) - \sP(\theta\kk\bar{\theta}) \\
    $\a$ & & & & & & |\theta|^2 \sZ -\sP(\theta\kk\bar{\theta}) \\
    $\g$ & & \kk & & & & |\theta|^2 \sZ -\sP(\theta\kk\bar{\theta}) \\
    $\g$ & & & & & & |\theta|^2 \sZ -\sP(\theta\kk\bar{\theta}) \\
    $\n_+$ & & \tfrac12 \jj & & & & |\theta|^2 \sZ - \sB(\theta\ii\bar{\theta}) - \sP(\theta\kk\bar{\theta}) \\
    \bottomrule
  \end{tabular}
  \caption*{The first column identifies the kinematical Lie algebra $\k$ underlying the extensions.
  The second column indicates whether the central extension has been introduced in the $[\B, \P]$
  bracket.  The next four columns show the $[s_{\bar{0}}, \s_{\bar{1}}]$ brackets for the KLSA.  As 
  we can see, the only non-vanishing case is $[\sH, \sQ(\theta)] = \sQ(\theta\hh)$, where $\theta \in \mathbb{H}$ and
  $\hh \in \Im(\mathbb{H})$.  The final column then tells us whether the central extension enters
  the $[\Q, \Q]$ bracket.}
\end{table}
\subsubsection{Unpacking the Notation}
Although the quaternionic formulation of these superalgebras is convenient for our purposes, it is perhaps unfamiliar to the reader. Therefore, in this section, we will convert one of the $\N=1$ super-extension for the Bargmann algebra (\hyperlink{S9}{S9}) into a more conventional format. The brackets for this algebra, excluding the $\s_{\bar{0}}$ brackets, which are shown in Table~\ref{tab:gb_algebras}, take the form 
\begin{equation}
	[\sH, \sQ(\theta)] = \sQ(\theta\kk) \quad \text{and} \quad [\sQ(\theta), \sQ(\theta)] = |\theta|^2 \sZ.
\end{equation}
Letting $\sQ(\theta) = \sum_{a=1}^{4} \theta_a \bQ_a$, where $\theta = \theta_4 + \theta_1 \ii + \theta_2 \jj + \theta_3 \kk$, we can rewrite these brackets as
\begin{equation}
	[\bH, \bQ_a] = \sum_{a = 1}^{4} \bQ_b \tensor{\Sigma}{^b_a} \quad \text{and} \quad  [\bQ_a, \bQ_b] = \delta_{ab} \bZ,
\end{equation}
where, for $\sigma_2$ being the second Pauli matrix, 
\begin{equation}
	\Sigma = \begin{pmatrix}
		 \zero & i \sigma_2 \\ -i \sigma_2 & \zero
	\end{pmatrix}.
\end{equation}

\section{Classification of $\N=2$ Generalised Bargmann Superalgebras} \label{sec:gb_superspace_n2}
Having established the introduction of the central extension $\bZ$ into our classification problem in Section~\ref{sec:gb_superspace_n1}, we now look to introduce an additional $\so(3)$ spinor module.  Section~\ref{subsec:N2_setup} will describe the setup up for the classification of the $\N=2$ generalised Bargmann superalgebras, before extending the preliminary results from Section~\ref{subsec:N1_setup} to this case.  It is also in this section that the group of basis transformations $\G$ will be adapted for extended supersymmetry.  The number of additional parameters in this case means that there are several branches of super-extension for each generalised Bargmann algebra.  In Section~\ref{subsec:N2_branches}, we use the preliminary results from the $(\s_{\bar{0}}, \s_{\bar{0}}, \s_{\bar{1}})$ component of the super-Jacobi identity to identify four possible branches of generalised Bargmann superalgebra in the sub-variety $\cS \subset \cJ$.  Each branch is then explored in detail in Section~\ref{subsec:N2_class} where we identify the non-empty sub-branches, which are summarised in Section~\ref{subsec:N2_summary}.
\subsection{Setup for the $\N=2$ Calculation} \label{subsec:N2_setup}
Recall that, in addition to the standard kinematical Lie brackets, the brackets for the universal generalised Bargmann superalgebra are
\begin{equation}
	\begin{split}
		[\sB(\beta), \sP(\pi)] &= \Re(\bar{\beta}\pi) \sZ \\ 
		[\sH, \sB(\beta)] &= \lambda \sB(\beta) + \mu \sP(\beta) \\
		[\sH, \sP(\pi)] &= \eta \sB(\pi) + \varepsilon \sP(\pi),
	\end{split} 
\end{equation}
where $\beta, \pi \in \Im(\mathbb{H})$ and $\lambda, \mu, \eta, \varepsilon \in \mathbb{R}$.  Because we now have two spinor modules, the brackets involving
$\s_{\bar{1}}$ need to be adapted from those given in Section~\ref{subsec:N1_setup}.  We will continue to use the map $\sQ$ for the odd dimensions; however, it now acts on $\vt$, a vector in $\mathbb{H}^2$.  We will choose $\mathbb{H}^2$ to
be a left quaternionic vector space such that $\mathbb{H}$ acts linearly from the 
left and all $2 \times 2$ $\mathbb{H}$ matrices act on the right.  Therefore,
writing $\vt$ out in its components, we have
\begin{equation}
\vt= \begin{pmatrix}
\theta_1 & \theta_2
\end{pmatrix},
\end{equation}
where $\theta_1, \theta_2 \in \mathbb{H}$.
The $[\s_{\bar{0}}, \s_{\bar{1}}]$ brackets are again the $\so(3)$-equivariant 
endomorphisms of $\s_{\bar{1}}$.  Since we choose $\so(3)$ to act via
left quaternionic multiplication, the commuting endomorphisms are all
those that may act on the right.  In the present case, these are elements 
of $\Mat_{2}(\mathbb{H})$.  Thus the brackets containing the $\so(3)$ 
scalars are 
\begin{equation} \label{eq:N2_S_general_scalar_brackets}
\begin{split}
[\sH, \sQ(\vt)] &= \sQ(\vt\mh) \\
[\sZ, \sQ(\vt)] &= \sQ(\vt\mz),
\end{split}
\end{equation}
where $\mh, \mz \in \Mat_2(\mathbb{H})$.
In Section~\ref{subsec:N1_setup}, $[\J, \Q] $, $[\B, \Q]$ and $[\P, \Q]$ were described by Clifford multiplication $V \otimes S \rightarrow S$, which is given by left quaternionic multiplication by $\Im(\mathbb{H})$.  The space of such maps was isomorphic to the space of $\r$-equivariant maps of $S$, which is a copy of the quaternions.  Now with $\s_{\bar{1}} = S \oplus S$, we have four possible endomorphisms of this type.  Labelling the two spinor modules $S_1$ and $S_2$, we may use the Clifford action to map $S_1$ to $S_1$, $S_1$ to $S_2$, $S_2$ to $S_1$, or $S_2$ to $S_2$.  All of these maps may be summarised as follows
\begin{equation}  \label{eq:N2_S_general_vector_brackets}
\begin{split}
[\sJ(\omega), \sQ(\vt)] &= \tfrac12 \sQ(\omega \vt) \\
[\sB(\beta), \sQ(\vt)] &= \sQ(\beta \vt \mb) \\
[\sP(\pi), \sQ(\vt)] &= \sQ(\pi \vt \mp).
\end{split}
\end{equation}
Here, $\omega, \beta,\pi \in \Im(\mathbb{H})$ and $\mb, \mp \in \Mat_2(\mathbb{H})$.  
Finally, consider the $[\Q, \Q]$ bracket.  This will consist of the $\so(3)$-equivariant
$\mathbb{R}$-linear maps $\bigodot^2 \s_{\bar{1}} \rightarrow \s_{\bar{0}}$.  To write down these
maps, we make use of the $\so(3)$-invariant inner product on $\s_{\bar{1}}$
\begin{equation}
\langle \vt, \vt' \rangle = \Re(\vt \vt^\dagger) \quad \text{where} \quad \vt, \vt' \in \mathbb{H}^2 \quad 
\vt^\dagger = \bar{\vt}^T.
\end{equation}
This bracket's $\so(3)$-invariance is clear on considering left multiplication by
$\uu \in \sp(1)$ and noting $\sp(1) \cong \so(3)$.  We can now use this bracket 
to identify $\bigodot^2 \s_{\bar{1}}$ with the symmetric $\mathbb{R}$-linear endomorphisms of
$\s_{\bar{1}} \cong S^2 \cong \mathbb{H}^2$, i.e.\ the maps $\mu: \mathbb{H}^2 \rightarrow \mathbb{H}^2$ such
that $\langle \mu(\vt), \vt' \rangle = \langle \vt, \mu(\vt') \rangle$.  A general 
$\mathbb{R}$-linear map of $\mathbb{H}^2$ may be written
\begin{equation}
\mu(\vt) = \qq \vt M \quad \text{where} \quad \qq \in \mathbb{H} \; \;  \text{and} \; \;  M \in \Mat_2(\mathbb{H}).
\end{equation}
Now, inserting this definition into the condition for a symmetric endomorphism, we 
obtain the following two cases: 
\begin{enumerate}
	\item $\qq \in \mathbb{R}$ and $M = M^\dagger$
	\item $\qq \in \Im(\mathbb{H})$ and $M = -M^\dagger$.
\end{enumerate}
The first instance gives us our $\so(3)$ scalar modules in $\bigodot^2 \s_{\bar{1}}$; therefore, 
these will map to either $\bH$ or $\bZ$ in $\s_{\bar{0}}$ to ensure we have
$\so(3)$-equivariance.  The condition on $M$ states that it must be of the form
\begin{equation}
M = \begin{pmatrix}
a & b + \mm \\ b - \mm & c
\end{pmatrix} = 
a \begin{pmatrix}
1 & \zero \\ \zero & \zero
\end{pmatrix} + 
b \begin{pmatrix}
 \zero & 1 \\ 1 & \zero
\end{pmatrix} + 
c \begin{pmatrix}
 \zero & \zero \\ \zero & 1
\end{pmatrix} + 
\mm \begin{pmatrix}
\zero & 1 \\ -1 &  \zero
\end{pmatrix},
\end{equation}
where $a, b, c \in \mathbb{R}$ and $\mm \in \Im(\mathbb{H})$.  We can make sense of
this result using the decomposition of the odd part of the superalgebra, 
$\s_{\bar{1}} \cong S^2 \cong S \otimes \mathbb{R}^2$, and our knowledge of
the maps in  $\Hom(\bigodot^2 S, \s_{\bar{0}})$ derived from Section~\ref{subsec:kss_alg_setup}.  
Symmetrising the decomposed $\s_{\bar{1}}$, we get $\bigodot^2 S^2 \cong \bigodot^2 S \otimes \bigodot^2 \mathbb{R}^2 \oplus \bigwedge^2 S \otimes \bigwedge^2 \mathbb{R}^2$.  Recall that $\bigodot^2 S \cong \RR \oplus 3 V$, where the scalar component is equivalent to the real span of the endomorphism $L_1 R_1$, and the vector components are equivalent to the real span of the endomorphisms $L_\qq R_{\qq'}$, where $\qq, \qq' \in \Im(\HH)$.  Notice, the scalar component is the only map that requires multiplication on the left by a real scalar.  Since case 1 demands that the spinor module is multiplied on the left by a real scalar, we would expect that the terms in $M$ which are associated with the $\bigodot^2 S \otimes \bigodot^2 \RR^2$ component of $\bigodot^2 S^2$ would also multiply the spinor module on the right by a real scalar. Indeed, we find that the first three terms of $M$ correspond to maps inside $\bigodot^2 S \otimes \bigodot^2 \mathbb{R}^2$: they all take the form of the scalar component of $\bigodot^2 S$, $L_1 R_1$, multiplied by a basis element of $\bigodot^2 \mathbb{R}^2$.  
\\ \\
To complete our decomposition of $M$ for this case, note that $\Lambda^2 S \cong 3 \RR \oplus V$, where the three copies of $\RR$ correspond to the endomorphisms $L_1 R_\ii$, $L_1 R_\jj$, and $L_1 R_\kk$ and $V$ corresponds to $\spn{L_\ii R_1, L_\jj R_1, L_\kk R_1}$.  Again, case 1 imposes that we only consider the maps containing $L_1$, which can be succinctly written as an imaginary quaternion acting on the right.  Thus, the final term in our decomposition is part of a map inside $\Lambda^2 S \otimes \Lambda^2 \RR^2$ and takes the form of an imaginary quaternion multiplied by the basis element of $\Lambda^2 \RR^2$. 
\\ \\
Now, the second case gives us our $\so(3)$ vector modules in 
$\bigodot^2 S^2$; therefore, these will map to $\B$, $\P$, or $\J$ to ensure 
$\so(3)$-equivariance.  The condition on $M$ in this case produces
\begin{equation}
M = \begin{pmatrix}
\nn & d + \ll \\ -d + \ll & \rr
\end{pmatrix} = 
\nn \begin{pmatrix}
1 & \zero \\ \zero & \zero
\end{pmatrix} + 
\ll \begin{pmatrix}
 \zero & 1 \\ 1 & \zero
\end{pmatrix} + 
\rr \begin{pmatrix}
 \zero & \zero \\ \zero & 1
\end{pmatrix} + 
d \begin{pmatrix}
\zero & 1 \\ -1 &  \zero
\end{pmatrix},
\end{equation}
where $\nn, \ll, \rr \in \Im(\mathbb{H})$ and $d \in \mathbb{R}$.  Again, we can understand
this result through the decomposition of $\s_{\bar{1}}$ and its symmetrisation, $\bigodot^2 S^2 \cong \bigodot^2 S \otimes \bigodot^2 \mathbb{R}^2 \oplus \bigwedge^2 S \otimes \bigwedge^2 \mathbb{R}^2$.  From Section~\ref{subsec:kss_alg_setup},
we know that the $\so(3)$ vectors in $\bigodot^2 S$ come from 
simultaneous left and right quaternionic multiplication by $\Im(\mathbb{H})$. Since case 2 imposes
that we must consider the maps which multiply $\s_{\bar{1}}$ on the left by an imaginary
quaternion, we may expect that the $M$ associated with the $\bigodot^2 S \otimes \bigodot^2 \RR^2$ component of $\bigodot^2 S^2$ will also
multiply on the right by an imaginary quaternion.  Indeed, the first three components 
of $M$ correspond to maps inside $\bigodot^2 S \otimes \bigodot^2 \RR^2$: they consist of a map in $\bigodot^2 S$
of the form $L_\qq R_{\qq'}$, for $\qq, \qq' \in \Im(\HH)$, multiplied by a basis element
of $\bigodot^2 \RR^2$.  
\\ \\
The final term in our decomposition comes from the $\bigwedge^2 S \otimes \bigwedge^2 \mathbb{R}^2$
component of $\bigodot^2 S^2$.  Recall, the only anti-symmetric endomorphisms of $S$
which involve multiplication on the left by an imaginary quaternion are in 
$\spn{L_\ii R_1, L_\jj R_1, L_\kk R_1}$, which transforms as an $\so(3)$ vector module.
Therefore, we would anticipate that the terms in $M$ associated with $\bigwedge^2 S \otimes \bigwedge^2 \mathbb{R}^2$
would correspond to a real scalar multiplied by the $\Lambda^2 \RR^2$
basis element, which is indeed the case.
\\ \\
Putting all this together, we may write
\begin{equation}
[\sQ(\vt), \sQ(\vt)] = \langle \vt, \vt N_0 \rangle \sH + \langle \vt, \vt N_1 \rangle \sZ + \langle \vt, \JJ \vt N_2 \rangle
+ \langle \vt, \BB \vt N_3 \rangle + \langle \vt, \PP \vt N_4 \rangle,
\end{equation}
where $N_0$, $N_1$ are quaternion Hermitian, and $N_2$, $N_3$, $N_4$ are quaternion
skew-Hermitian, as stated above, and 
\begin{equation}
\JJ = \bJ_1 \ii + \bJ_2 \jj + \bJ_3 \kk \quad \BB = \bB_1 \ii + \bB_2 \jj + \bB_3 \kk \quad 
\PP = \bP_1 \ii + \bP_2 \jj + \bP_3 \kk.
\end{equation}
Using the fact that $\Re(\bar{\omega}\JJ) = \sJ(\omega)$ and $N_i = N_i^\dagger$ for 
$i \in \{0, 1\}$, we can write
\begin{equation}
[\sQ(\vt), \sQ(\vt)] = \Re(\vt N_0 \vt^\dagger) \sH + \Re(\vt N_1 \vt^\dagger) \sZ - \sJ(\vt N_2 \vt^\dagger) 
- \sB(\vt N_3 \vt^\dagger) - \sP(\vt N_4 \vt^\dagger).
\end{equation}
This polarises to
\begin{equation}
\begin{split}
[\sQ(\vt), \sQ(\vt')] = \frac{1}{2} \big( & \Re(\vt N_0 \vt'^\dagger + \vt' N_0 \vt^\dagger) \sH 
+ \Re(\vt N_1 \vt'^\dagger + \vt' N_1 \vt^\dagger) \sZ \\ & - \sJ(\vt N_2 \vt'^\dagger + \vt' N_2 \vt^\dagger) 
- \sB(\vt N_3 \vt'^\dagger + \vt' N_3 \vt^\dagger) - \sP(\vt N_4 \vt'^\dagger + \vt' N_4 \vt^\dagger) \big) .
\end{split}
\end{equation}

\subsubsection{Preliminary Results} \label{subsubsec:N2_S_preliminaries}
As in the $\N=1$ case, we can form a number of universal results that will help us when 
investigating the super-extensions of the generalised Bargmann algebras.  The following 
subsections will cover the $(\s_{\bar{0}}, \s_{\bar{0}}, \s_{\bar{1}})$, 
$(\s_{\bar{0}}, \s_{\bar{1}}, \s_{\bar{1}})$, and $(\s_{\bar{1}}, \s_{\bar{1}}, \s_{\bar{1}})$
components of the super-Jacobi identity, respectively. \\
\paragraph{$(\s_{\bar{0}}, \s_{\bar{0}}, \s_{\bar{1}})$} \label{par:N2_S_P_011} ~\\ \\
In the $\N=1$ case, $\hh, \zz, \bb, \pp \in \mathbb{H}$,
and in the $\N=2$ case $\mh, \mz, \mb, \mp \in \Mat_2(\mathbb{H})$.  Since
$\mathbb{H}$ and $\Mat_2(\mathbb{H})$ are both associative, non-commutative
algebras, the algebraic manipulations are the same in both cases.  Therefore, the 
$\N=1$ results generalise to the $\N=2$ case; the only difference being that the
variables are $2 \times 2$ $\mathbb{H}$ matrices rather than $\mathbb{H}$ elements. 
\begin{lemma}\label{lem:N2_001}
  The following relations between $\mh, \mz, \mb, \mp \in \Mat_2(\mathbb{H})$ are implied by the
  corresponding $\k$-brackets:
  \begin{equation}
    \begin{split}
    	  [\bH,\bZ] = \lambda \bH + \mu \bZ & \implies [\mz, \mh] = \lambda \mh + \mu \mz \\
      [\bH,\B] = \lambda \B + \mu \P & \implies [\mb,\mh] = \lambda \mb + \mu \mp\\
      [\bH,\P] = \lambda \B + \mu \P & \implies [\mp,\mh] = \lambda \mb + \mu \mp\\
      [\bZ,\B] = \lambda \B + \mu \P & \implies [\mb,\mz] = \lambda \mb + \mu \mp\\
      [\bZ,\P] = \lambda \B + \mu \P & \implies [\mp,\mz] = \lambda \mb + \mu \mp\\
      [\B,\B] = \lambda \B + \mu \P + \nu \J & \implies \mb^2 = \tfrac12 \lambda \mb + \tfrac12 \mu \mp + \tfrac14 \nu\\
      [\P,\P] = \lambda \B + \mu \P + \nu \J & \implies \mp^2 = \tfrac12 \lambda \mb + \tfrac12 \mu \mp + \tfrac14 \nu\\
      [\B,\P] = \lambda \bH + \mu \bZ & \implies \mb \mp + \mp\mb = 0\quad\text{and}\quad [\mb,\mp] = \lambda \mh  + \mu \mz.
    \end{split}
  \end{equation}
\end{lemma}
\begin{proof}
See the proof of Lemma~\ref{lem:N1_001} for the algebraic manipulations that produce the
above results.
\end{proof}
\paragraph{$(\s_{\bar{0}}, \s_{\bar{1}}, \s_{\bar{1}})$} \label{par:N2_S_P_011} ~\\ \\
As in the $\N=1$ case, we use the universal generalised Bargmann algebra to simplify
our analysis here.  Recall the brackets for this algebra are
\begin{equation}
	[\B, \P] = \bZ \qquad [\bH, \B] = \lambda \B + \mu \P \qquad [\bH, \P] = \eta \B + \varepsilon \P,
\end{equation}
where $\lambda, \mu, \eta, \varepsilon \in \mathbb{R}$.  Using these brackets, we obtain the 
following result.
\begin{lemma} \label{lem:N2_011}
\begin{equation}
	\begin{split}
		\text{The $[\bH, \Q, \Q]$ identity produces the conditions} \\
		0 &= \mh N_i + N_i \mh^\dagger \quad \text{where} \quad i \in \{0, 1, 2\} \\
		\lambda N_3 + \eta N_4 &= \mh N_3 + N_3 \mh^\dagger \\
		\mu N_3 + \varepsilon N_4 &= \mh N_4 + N_4 \mh^\dagger. \\
	\text{The $[\bZ, \Q, \Q]$ identity produces the conditions} \\
	0 &= \mz N_i + N_i \mz^\dagger \quad \text{where} \quad i \in \{0, 1, 2, 3, 4\}. \\
	\text{The $[\B, \Q, \Q]$ identity produces the conditions} \\
		0 &= \mb N_0 - N_0 \mb^\dagger \\
		N_4 &= \mb N_1 - N_1 \mb^\dagger \\
		0 &= \beta \vt \mb N_2 \vt^\dagger + \vt N_2 (\beta \vt \mb)^\dagger \\
		\lambda \Re(\vt N_0 \vt^\dagger) \beta + \tfrac12 [\beta, \vt N_2 \vt^\dagger] &= \beta \vt \mb N_3 \vt^\dagger
		+ \vt N_3 (\beta \vt \mb)^\dagger \\
		\mu \Re(\vt N_0\vt^\dagger) \beta &= \beta \vt \mb N_4 \vt^\dagger + \vt N_4 (\beta \vt \mb)^\dagger. \\
	\text{The $[\P, \Q, \Q]$ identity produces the conditions} \\
		0 &= \mp N_0 - N_0 \mp^\dagger \\
		-N_3 &= \mp N_1 - N_1 \mp^\dagger \\
		0 &= \pi \vt \mp N_2 \vt^\dagger + \vt N_2 (\pi \vt \mp)^\dagger \\
		\eta \Re(\vt N_0\vt^\dagger) \pi &= \pi \vt \mp N_3 \vt^\dagger + \vt N_3 (\pi \vt \mp)^\dagger \\
		\varepsilon \Re(\vt N_0 \vt^\dagger) \pi + \tfrac12 [\pi, \vt N_2 \vt^\dagger] &= \pi \vt \mp N_4 \vt^\dagger
		+ \vt N_4 (\pi \vt \mp)^\dagger , \\
	\end{split}		
\end{equation}
where $\beta, \pi \in \Im(\mathbb{H})$ and $\vt \in \mathbb{H}^2$.
\end{lemma}
\begin{proof}
Beginning with the $[\bH, \Q, \Q]$ identity, we have
\begin{equation}
	[\sH, [\sQ(s), \sQ(s)]] = 2 [[\sH, \sQ(s)], \sQ(s)].
\end{equation}
Focussing on the L.H.S., note the general form of the $[\Q, \Q]$ bracket
has components along each of the $\s_{\bar{0}}$ basis elements; however,
$\bH$ only commutes with $\B$ and $\P$.  Therefore, 
\begin{equation}
	\begin{split}
		L.H.S. &= -[\sH, \sB(\vt N_3\vt^\dagger)] - [\sH, \sP(\vt N_4\vt^\dagger)] \\
		&= - \sB( \vt (\lambda N_3 + \eta N_4) \vt^\dagger) - \sP( \vt (\mu N_3 + \varepsilon N_4) \vt^\dagger ).
	\end{split}
\end{equation}
Substituting $[\sH, \sQ(\vt)] = \sQ(\vt\mh)$ into the R.H.S. and using the polarised form of the $[\Q, \Q]$ bracket,
we find
\begin{equation}
	\begin{split}
		R.H.S. = & \Re( \vt (\mh N_0  + N_0 \mh^\dagger) \vt^\dagger) \sH + \Re( \vt (\mh N_1  + N_1 \mh^\dagger) \vt^\dagger) \sZ \\
		& - \sJ( \vt (\mh N_2 + N_2 \mh^\dagger) \vt^\dagger ) - \sB( \vt (\mh N_3 + N_3 \mh^\dagger) \vt^\dagger )
		- \sP( \vt (\mh N_4 + N_4 \mh^\dagger) \vt^\dagger ).
	\end{split}
\end{equation}
Comparing coefficients and using the injectivity and linearity of the maps $\sJ, \sB$ and $\sP$, we get the
desired conditions.  The $[\bZ, \Q, \Q]$ result follows in an analogous manner.  Consider the 
$[\B, \Q, \Q]$ Jacobi identity
\begin{equation}
	[ \sB(\beta), [\sQ(\vt), \sQ(\vt)]] = 2 [[\sB(\beta), \sQ(\vt)], \sQ(\vt)].
\end{equation}
Since $\B$ commutes with $\bZ$ and $\B$, the L.H.S. takes the following form
\begin{equation}
	\begin{split}
		L.H.S. &= [\sB(\beta), \Re(\vt N_0\vt^\dagger) \sH - \sJ(\vt N_2\vt^\dagger) - \sP(\vt N_4 \vt^\dagger)] \\
		&= - \Re(\vt N_0 \vt^\dagger) (\lambda \sB(\beta) + \mu \sP(\beta) ) +\tfrac12 \sB([\vt N_2\vt^\dagger, \beta])
		- \Re(\bar{\beta} \vt N_4 \vt^\dagger) \sZ.
	\end{split}
\end{equation}
Turning attention to the R.H.S., we find
\begin{equation}
	\begin{split}
		R.H.S. = & \Re( \beta \vt \mb N_0 \vt^\dagger + \vt N_0 \mb^\dagger \vt^\dagger \bar{\beta}) \sH
		+ \Re( \beta \vt \mb N_1 \vt^\dagger + \vt N_1 \mb^\dagger \vt^\dagger \bar{\beta}) \sZ \\
		& - \sJ(\beta \vt \mb N_2 \vt^\dagger + \vt N_2 \mb^\dagger \vt^\dagger \bar{\beta}) 
		- \sB(\beta \vt \mb N_3 \vt^\dagger + \vt N_3 \mb^\dagger \vt^\dagger \bar{\beta})
		- \sP(\beta \vt \mb N_4 \vt^\dagger + \vt N_4 \mb^\dagger \vt^\dagger \bar{\beta}).
	\end{split}
\end{equation}
Using the property $\bar{\beta} = -\beta$, since $\beta \in \Im(\mathbb{H})$, and the cyclic
property of $\Re$, the first two terms can have their coefficients written in the form
\begin{equation}
	\Re( \beta \vt \mb N_i \vt^\dagger + \vt N_i \mb^\dagger \vt^\dagger \bar{\beta})
	= \Re( \beta \vt (\mb N_i - N_i \mb^\dagger ) \vt^\dagger ),
\end{equation} 
for $i \in \{0, 1\}$.  Again, comparing coefficients we obtain the desired results.  The 
$[\P, \Q, \Q]$ case follows identically.
\end{proof}
\paragraph{$(\s_{\bar{1}}, \s_{\bar{1}}, \s_{\bar{1}})$} \label{par:N2_S_P_111} ~\\ \\
The last super-Jacobi identity component to consider is the $(\s_{\bar{1}}, \s_{\bar{1}}, \s_{\bar{1}})$ case,
$[\Q, \Q, \Q]$.
\begin{lemma}\label{lem:N2_111}
The $[\Q,\Q,\Q]$ identity produces the condition
\begin{equation}
\Re(\vt N_0 \vt^\dagger) \vt \mh + \Re(\vt N_1 \vt^\dagger) \vt \mz = \tfrac12 \vt N_2 \vt^\dagger \vt + \vt N_3 \vt^\dagger \vt \mb 
+ \vt N_4 \vt^\dagger \vt \mp.
\end{equation}
\end{lemma}
\begin{proof}
The $[\Q, \Q, \Q]$ identity is written
\begin{equation}
	0 = [[\sQ(\vt), \sQ(\vt)], \sQ(\vt)].
\end{equation}
Substituting in the $[\Q, \Q]$ bracket, this becomes
\begin{equation}
	0 = [ \Re(\vt N_0 \vt^\dagger) \sH + \Re(\vt N_1 \vt^\dagger) \sZ - \sJ(\vt N_2 \vt^\dagger) 
- \sB(\vt N_3 \vt^\dagger) - \sP(\vt N_4 \vt^\dagger), \sQ(\vt)]. 
\end{equation}
Finally, using the brackets in \eqref{eq:N2_S_general_scalar_brackets} and \eqref{eq:N2_S_general_vector_brackets} and the injectivity of $\sQ$, we obtain the desired result.
\end{proof}
\subsubsection{Basis Transformations} \label{subsubsec:N2_S_autos}
We will investigate the subgroup $\G \subset \GL(\s_{\bar{0}}) \times 
\GL(\s_{\bar{1}})$ by first looking at the transformations induced by the adjoint
action of the rotational subalgebra $\r \cong \so(3)$.  
We will then look at the $\so(3)$-equivariant maps transforming
the basis of the underlying vector space.  These will act via Lie algebra automorphisms in $\s_{\bar{0}}$
and endomorphisms of the $\so(3)$ module $S^2$ in $\s_{\bar{1}}$. 
Note, in the former case, where the automorphism is induced by
$\ad_{\bJ_i}$, each $\so(3)$ module will transform into itself, while,
in the latter case, when the transformation is some $\so(3)$-equivariant map,
the modules transform into one another.  For completeness, at the end
of the section, we determine the automorphisms of each generalised Bargmann algebra.
\\ \\
Recall that $\Sp(1)$ is the double-cover of $\Aut(\mathbb{H})$, and $\Aut(\mathbb{H}) 
\cong \SO(3)$.  We, therefore, write $\lambda \in \Aut(\mathbb{H})$ as $\lambda(\ss) =
\uu\ss\bar{\uu}$ for some $\uu \in \Sp(1)$, which will act trivially on the real component 
of $\ss$ and rotate the imaginary components.  Using this result, we can represent the action
of $\Aut(\mathbb{H})$ on the $\so(3)$ vector modules in $\s_{\bar{0}}$ by pre-composing the linear maps
$\sJ$, $\sB$, and $\sP$ with $\Ad_{\uu}$, for $\uu \in \Sp(1)$.  To preserve the kinematical
brackets in $[\s_{\bar{0}}, \s_{\bar{0}}]$, we must pre-compose with the same $\uu$
for each map. Note, $\so(3)$ acts trivially on $\bH$ and $\bZ$, so
these basis elements will be left invariant under these automorphisms. 
For $\s_{\bar{1}}$, we restrict to the individual copies of $S$ through diagonal
matrices.  To preserve the $[\J, \Q]$ bracket, we must 
pre-compose with the same $\uu$ as above.  Therefore, we write
$\tilde{\sQ}(\vt) = \sQ(\uu \vt\bar{\uu} \mathbb{1})$.  We can now investigate how these automorphisms
affect our brackets
\begin{equation}
\begin{gathered}
\begin{split}
[\sJ(\omega), \sQ(\vt)] &= \tfrac12 \sQ(\omega \vt) \\
[\sB(\beta), \sQ(\vt)] &= \sQ(\beta \vt \mb) \\
[\sP(\pi), \sQ(\vt)] &= \sQ(\pi \vt \mp)
\end{split} \qquad
\begin{split}
[\sH, \sQ(\vt)] &= \sQ(\vt\mh) \\
[\sZ, \sQ(\vt)] &= \sQ(\vt\mz)
\end{split} \\ \\
[\sQ(\vt), \sQ(\vt)] = \Re(\vt N_0 \vt^\dagger) \sH + \Re(\vt N_1 \vt^\dagger) \sZ - \sJ(\vt N_2 \vt^\dagger) 
- \sB(\vt N_3 \vt^\dagger) - \sP(\vt N_4 \vt^\dagger).
\end{gathered}
\end{equation}
Transforming the basis, we have
\begin{equation} \label{eq:transformed_brackets}
\begin{gathered}
\begin{split}
[\tilde{\sJ}(\omega), \tilde{\sQ}(\vt)] &= \tfrac12 \tilde{\sQ}(\omega \vt) \\
[\tilde{\sB}(\beta), \tilde{\sQ}(\vt)] &= \tilde{\sQ}(\beta \vt \tilde{\mb}) \\
[\tilde{\sP}(\pi), \tilde{\sQ}(\vt)] &= \tilde{\sQ}(\pi \vt \tilde{\mp})
\end{split} \qquad
\begin{split}
[\tilde{\sH}, \tilde{\sQ}(\vt)] &= \tilde{\sQ}(\vt\tilde{\mh}) \\
[\tilde{\sZ}, \tilde{\sQ}(\vt)] &= \tilde{\sQ}(\vt\tilde{\mz})
\end{split} \\ \\
[\tilde{\sQ}(\vt), \tilde{\sQ}(\vt)] = \Re(s\vt\widetilde{N_0} \vt^\dagger) \tilde{\sH} + \Re(\vt \widetilde{N_1} \vt^\dagger) \tilde{\sZ}
 - \tilde{\sJ}(\vt \widetilde{N_2} \vt^\dagger) - \tilde{\sB}(\vt \widetilde{N_3} \vt^\dagger) - \tilde{\sP}(\vt \widetilde{N_4} \vt^\dagger),
 \end{gathered}
\end{equation}
with $\tilde{\sH} = \sH$, $\tilde{\sZ} = \sZ$, $\tilde{\sJ} = \sJ\circ \Ad_{\uu}$, 
$\tilde{\sB} = \sB\circ \Ad_{\uu}$, $\tilde{\sP} = \sP\circ \Ad_{\uu}$, and $\tilde{\sQ} = \sQ\circ \Ad_{\uu}$,
where it is understood that $\Ad_{\uu}$ acts diagonally on the $\s_{\bar{1}}$ basis, $\sQ$.  The transformed
matrices are 
\begin{equation}
\begin{split}
\tilde{\mh} &= D \mh D^{-1}\\\
\tilde{\mz} &= D \mz D^{-1}
\end{split} \quad
\begin{split}
\tilde{\mb} &= D \mb D^{-1} \\
\tilde{\mp} &= D \mp D^{-1}
\end{split} \quad
\widetilde{N_i} = D N_i D^\dagger,
\end{equation}
where $D = \uu \mathbb{1}$ for $\uu \in \Sp(1)$ and $i \in \{0, 1, ..., 4\}$.  Therefore, $D^{-1} = D^\dagger = \bar{\uu}\mathbb{1}$.
These automorphisms simultaneously rotate all quaternions, all the components of the matrices $\mh, \mz, \mb, \mp$ and $N_i$, by the same $\Sp(1)$ element.
\\ \\
Next, we want to consider the $\so(3)$-equivariant linear maps 
which leave the rotational subalgebra invariant: $(\sJ, \sB, \sP, \sH, \sZ, \sQ) \rightarrow
(\sJ, \tilde{\sB}, \tilde{\sP}, \tilde{\sH}, \tilde{\sZ}, \tilde{\sQ})$.  These take the general form
\begin{equation} \label{eq:autos_2}
\begin{split}
\tilde{\sH} &= a \sH + b \sZ \\
\tilde{\sZ} &= c \sH + d \sZ \\
\tilde{\sB}(\beta) &= e \sB(\beta) + f \sP(\beta) + g \sJ(\beta) \\
\tilde{\sP}(\pi) &= h \sB(\pi) + i \sP(\pi) + j \sJ(\pi) \\
\tilde{\sQ}(\vt) &= \sQ(\vt M),
\end{split}
\end{equation}
where $a, ..., j \in \mathbb{R}$ and $M \in \GL(\mathbb{H}^2)$.  Crucially, 
\begin{equation}
	A = \begin{pmatrix}
		a & b \\ c & d
	\end{pmatrix} \in \GL(2, \mathbb{R}) \quad \text{and} \quad
	C = \begin{pmatrix}
		e & f & g \\ h & i & j \\ \zero & \zero & 1
	\end{pmatrix} \in \GL(3, \mathbb{R}),
\end{equation}
act on $(\sH, \sZ)^T$ and $(\sB, \sP, \sJ)^T$, respectively.  Each of the generalised
Bargmann algebra allows different transformations of this type; however, there are some
important general results.  Therefore, we will begin by working through the analysis of
these maps with the universal generalised Bargmann algebra before focussing on each algebra separately.
\\ \\
As in the $\N = 1$ case, the checking of brackets that include $\J$ is really verifying that
the above maps are $\so(3)$-equivariant, so this does not give us any information not already presented.  The first bracket 
we will consider is $[\B, \P] = \bZ$.  Substituting in the maps of \eqref{eq:autos_2}, we find
the following important results:
\begin{equation}
	d = ei - fh, \quad c = 0, \quad \text{and} \quad g = j = 0.
\end{equation}
The vanishing of $c$ tells us that $d \neq 0$ if we are to have $A \in \GL(2, \mathbb{R})$.  Also,
the vanishing of $g$ and $j$ shows that we can reduce $C$ to an element of $\GL(2, \mathbb{R})$,
\begin{equation}
	C = \begin{pmatrix}
		e & f \\ h & i 
	\end{pmatrix},
\end{equation}
acting on  $(\sB, \sP)^T$.  The remaining $[\s_{\bar{0}}, \s_{\bar{0}}]$ brackets are $[\bH, \B]$ and $[\bH, \P]$, which produce
\begin{equation} \label{eq:AB_constraints}
	\begin{split}
		0 &= \lambda e (a-1) + \eta af - \mu h\\
		0 &= \lambda f - \varepsilon af + \mu (i - ea) 
	\end{split} \quad \text{and} \quad
	\begin{split}
		0 &= \eta (e-ai) + \varepsilon h - \lambda ah \\
		0 &= \eta f + \varepsilon i (1-a) - \mu ah,
	\end{split}
\end{equation}
respectively.  Clearly, these conditions are dependent on the exact choice of generalised 
Bargmann algebra, so we will leave these results in this form for now.
\\ \\
Now, since the $[\s_{\bar{0}}, \s_{\bar{1}}]$ and $[\s_{\bar{1}}, \s_{\bar{1}}]$ brackets
are so far independent of the chosen algebra, the following results will hold for all the generalised Bargmann algebras.  
Reusing \eqref{eq:transformed_brackets}, in this instance we find
\begin{equation}
\tilde{\mh} = M (a \mh + b\mz) M^{-1} \quad 
\tilde{\mz} = d M \mz M^{-1} \quad
\tilde{\mb} = M (e\mb + f \mp) M^{-1} \quad 
\tilde{\mp} = M (h \mb + i\mp) M^{-1}  \nonumber
\end{equation} 
\begin{equation} \label{eq:N2_S_basis_transformations}
\begin{split}
\widetilde{N_0} &= \frac{1}{a} M N_0 M^\dagger \\
\widetilde{N_1} &= \frac{1}{ad} M (a N_1 - b N_0) M^\dagger
\end{split} \quad
\begin{split}
\widetilde{N_2} &= M N_2 M^\dagger \\
\widetilde{N_3} &= \frac{1}{ie-fh} M (i N_3 - h N_4) M^\dagger \\
\widetilde{N_4} &= \frac{1}{ie-fh} M (e N_4 - f N_3) M^\dagger.
\end{split}
\end{equation}
Putting the two types of transformation in $\G$ together, we have 
\begin{equation}
\begin{split}
\sJ &\mapsto \sJ\circ \Ad_{\uu} \\
\sB &\mapsto e \sB\circ \Ad_{\uu} + f \sP\circ \Ad_{\uu} \\
\sP &\mapsto h \sB\circ \Ad_{\uu} + i \sP\circ \Ad_{\uu} \\
\sH &\mapsto a \sH + b \sZ \\
\sZ &\mapsto d \sZ \\
\sQ &\mapsto \sQ\circ \Ad_{\uu} \circ R_M.
\end{split}
\end{equation}
These transformations may be summarised by $(A = \big( \begin{smallmatrix} a & b \\ 0 & d\end{smallmatrix} \big),
C = \big(\begin{smallmatrix} e & f \\ h & i \end{smallmatrix}\big), M, \uu) \in \GL(\mathbb{R}^2) \times \GL(\mathbb{R}^2)
\times \GL(\mathbb{H}^2) \times \mathbb{H}^\times$.  Now that we have the most general element of
the subgroup $G \subset \GL(\s_{\bar{0}}) \times \GL(\s_{\bar{1}})$ for
$\s_{\bar{0}} = \k$ the universal generalised Bargmann algebra, we can restrict ourselves to 
the automorphisms of $\s_{\bar{0}}$ and set the parameters
$\lambda, \mu, \eta, \varepsilon \in \mathbb{R}$ to determine the automorphism group
for each of the generalised Bargmann algebras.  The results of this investigation are presented
in Table~\ref{tab:GBA-auts}.
\\
\paragraph{$\hat{\a}$} ~\\ \\
In this instance, all the conditions vanish as $\lambda = \mu = \eta
= \varepsilon = 0$; therefore, the matrices $A$ and $C$ are left as stated above.
\\ 
\paragraph{$\hat{\n}_-$}  ~\\ \\
Having $\lambda = - \varepsilon = 1$ and $\mu = \eta = 0$, the
conditions in \eqref{eq:AB_constraints} become
\begin{equation}
	\begin{split}
		0 &= e (a -1) \\
		0 &= f (1 +a) 
	\end{split}  \quad \text{and} \quad
	\begin{split}
		0 &= h (1+a) \\
		0 &= i (1-a).
	\end{split}
\end{equation}
Notice, if $a \notin \{ \pm 1\}$ then $C$ must vanish, which cannot happen if we are to retain
the basis elements $\B$ and $\P$.  Therefore, we are left with two cases: $a = 1$ and $a = -1$.
In the former instance, we have automorphisms with
\begin{equation}
	A = \begin{pmatrix}
		1 & b \\ \zero & ei
	\end{pmatrix} \quad \text{and} \quad 
	C = \begin{pmatrix}
		e & \zero \\ \zero & i
	\end{pmatrix}.
\end{equation}
In the latter instance, we have
\begin{equation}
	A = \begin{pmatrix}
		-1 & b \\ \zero & -hf
	\end{pmatrix} \quad \text{and} \quad 
	C = \begin{pmatrix}
		0 & h \\ f & \zero
	\end{pmatrix}.
\end{equation}
\paragraph{$\hat{\n}_+$} ~\\ \\ In this case, $\lambda = \varepsilon = 0$ and $\mu = - \eta = 1$.
Therefore, our constraints become
\begin{equation}
	\begin{split}
		0 &= h + af \\
		0 &= i - ae
	\end{split} \quad \text{and} \quad
	\begin{split} 
		0 &= e - ai \\
		0 &= f + ah.
	\end{split}
\end{equation}
Taking the expressions for $h$ and $i$ from the conditions on the left and substituting them
into the conditions on the right, we find
\begin{equation}
	0 = (1-a^2) f \qquad 0 = (1 - a^2) e.
\end{equation}
If $a^2 \neq 1$, we would need both $f$ and $e$ to vanish, which contradicts our assumption
that $C \in \GL(2, \mathbb{R})$.  Therefore, we need $a^2 = 1$, which presents two cases:
$a = 1$ and $a=-1$.  In the former instance, we find automorphisms of the form
\begin{equation}
	A = \begin{pmatrix}
		1 & b \\ \zero & e^2 + h^2
	\end{pmatrix} \quad \text{and} \quad 
	C = \begin{pmatrix}
		e & h \\ -h & e
	\end{pmatrix}.
\end{equation}
In the latter instance, we get
\begin{equation}
	A = \begin{pmatrix}
		-1 & b \\ \zero & -e^2 -h^2
	\end{pmatrix} \quad \text{and} \quad
	C = \begin{pmatrix}
		e & h \\ h & -e
	\end{pmatrix}.
\end{equation}
\paragraph{$\hat{\g}$} ~\\ \\Finally, we have $\lambda = \eta = \varepsilon = 0$ and $\mu = -1$,
which, when substituted into \eqref{eq:AB_constraints}, produces
\begin{equation}
		0 = h \quad \text{and} \quad i = ae.
\end{equation}
Therefore, automorphisms for the Bargmann algebra take the form
\begin{equation}
	A = \begin{pmatrix}
		a & b \\ \zero & ae^2
	\end{pmatrix} \quad \text{and} \quad
	C = \begin{pmatrix}
		e & f \\ \zero & ae
	\end{pmatrix}.
\end{equation}
\begin{table}[h!]
  \centering
  \caption{Automorphisms of the Generalised Bargmann Algebras}
  \label{tab:GBA-auts}
  \setlength{\extrarowheight}{2pt}
  \begin{tabular}{l|>{$}l<{$}}\toprule
    \multicolumn{1}{c|}{$\k$} & \multicolumn{1}{c}{General $(A, C) \in \GL(\mathbb{R}^2) \times \GL(\mathbb{R}^2)$} \\
    \toprule
    $\hat{\a}$ & \left(\begin{pmatrix} a & b \\ \zero & d\end{pmatrix}, \quad  \begin{pmatrix} e & f \\ h & i \end{pmatrix}\right) \\
    $\hat{\n}_-$ & \left(\begin{pmatrix}
		1 & b \\ \zero & ei
	\end{pmatrix}, \quad 
	\begin{pmatrix}
		e & \zero \\ \zero & i
	\end{pmatrix}\right) \cup \left(\begin{pmatrix}
		-1 & b \\ \zero & -hf
	\end{pmatrix}, \quad 
	 \begin{pmatrix}
		\zero & h \\ f & \zero
	\end{pmatrix}\right) \\
    $\hat{\n}_+$ & \left(\begin{pmatrix}
		1 & b \\ \zero & e^2 + h^2
	\end{pmatrix}, \quad 
	 \begin{pmatrix}
		e & h \\ -h & e
	\end{pmatrix}\right) \cup \left(\begin{pmatrix}
		-1 & b \\ \zero & -e^2 -h^2
	\end{pmatrix}, \quad 
	 \begin{pmatrix}
		e & h \\ h & -e
	\end{pmatrix}\right) \\
    $\hat{\g}$ & \left(\begin{pmatrix}
		a & b \\ \zero & ie
	\end{pmatrix},\quad 
	\begin{pmatrix}
		e & f \\ \zero & ae
	\end{pmatrix}\right) \\
    \bottomrule
  \end{tabular}
\end{table}
\subsection{Establishing Branches} \label{subsec:N2_branches}
Before proceeding to the discussion in which the non-empty sub-branches of $\cS$ are identified, we first establish the possible $[\s_{\bar{0}}, \s_{\bar{1}}]$
brackets.  More specifically, we establish the possible forms for $\mz, \mh, \mb, \mp \in \Mat_2(\mathbb{H})$.
In this section, we focus solely on the results of Lemma~\ref{lem:N2_001}
concerning the $(\s_{\bar{0}}, \s_{\bar{0}}, \s_{\bar{1}})$ component of the super-Jacobi identity.
Using the universal generalised Bargmann algebra, we find that $\mb, \mp \in \Mat_2(\mathbb{H})$,
which encode the brackets $[\B, \Q]$ and $[\P, \Q]$, respectively, form a
double complex.  Analysing this structure, we identify four possible
cases:
\begin{enumerate}
	\item $\mb = 0$ and $\mp = 0$
	\item $\mb = 0$ and $\mp \neq 0$
	\item $\mb \neq 0$ and $\mp = 0$
	\item $\mb \neq 0$ and $\mp \neq 0$. 
\end{enumerate}
Taking each of these cases in turn, we find forms for $\mz$ and $\mh$ to establish four branches in $\cS$ which may contain generalised Bargmann superalgebras.  These branches
will form the basis for our investigations into the possible super-extensions for each
of the generalised Bargmann algebras in Section~\ref{subsec:N2_class}.
\\ \\
Using the results of Lemma~\ref{lem:N2_001}, we notice that $\mb^2 = \mp^2 = 0$ and 
$\mb\mp + \mp\mb = 0$; therefore, $\mb$ and $\mp$ are the differentials of a double complex in which
the modules are $\s_{\bar{1}}$.  What does 
this mean for the form of $\mb$ and $\mp$?  Notice that we could simply set $\mb$ and $\mp$ to zero.
However, assuming at least one component of these matrices is non-vanishing, we find the following
cases.  Take $\mp$ as our example and let
\begin{equation}
	\mp = \begin{pmatrix} \pp_1 & \pp_2 \\ \pp_3 & \pp_4 \end{pmatrix}.
\end{equation}
The fact that this squares to zero tells us
\begin{equation} \label{eq:sqrd_mats}
\pp_1^2 + \pp_2 \pp_3 = 0 \qquad \pp_1 \pp_2 + \pp_2 \pp_4 = 0 \qquad \pp_3 \pp_1 + \pp_4 \pp_3 = 0
\qquad \pp_3 \pp_2 + \pp_4^2 = 0.
\end{equation}
There are two cases, $\pp_3 = 0$ and $\pp_3 \neq 0$, which we shall now consider in turn.
\\ \\
In the $\pp_3 =0$ case, the constraints in \eqref{eq:sqrd_mats} become
\begin{equation}
	\pp_1^2 = 0 \qquad \pp_1 \pp_2 + \pp_2 \pp_4 = 0 \qquad  \pp_4^2 = 0.
\end{equation}
Therefore, $\pp_1 = \pp_4 = 0$ and $\pp_2$ is unconstrained, leaving the matrix
\begin{equation}
	\mp = \begin{pmatrix}\zero & \pp_2 \\ \zero & \zero \end{pmatrix}.
\end{equation}
In the $\pp_3 \neq 0$ case, we can use the first and third constraints of 
\eqref{eq:sqrd_mats} to get $\pp_2 = - \pp_1^2 \pp_3^{-1}$ and 
$\pp_4 = - \pp_3 \pp_1 \pp_3^{-1}$, respectively.  These choices trivially satisfy the second and 
fourth constraints such that we arrive at
\begin{equation}
	\mp = \begin{pmatrix} \pp_1 & -\pp_1^2 \pp_3^{-1} \\ \pp_3 & -\pp_3\pp_1\pp_3^{-1} \end{pmatrix}.
\end{equation}
In a completely analogous manner, we find
\begin{equation}
	\mb = \begin{pmatrix}\zero & \bb_2 \\ \zero & \zero \end{pmatrix} \quad \text{and} \quad \mb =
	\begin{pmatrix} \bb_1 & -\bb_1^2 \bb_3^{-1} \\ \bb_3 & -\bb_3\bb_1\bb_3^{-1} \end{pmatrix}.
\end{equation}
Now, what does the anti-commuting condition, $\mb\mp + \mp\mb = 0$, tell us about the non-vanishing matrices?  
Assuming, for now, that $\mb$ and $\mp$ are non-vanishing, we have four options:
\begin{enumerate}
	\item $\pp_3 \neq 0$, $\bb_3 \neq 0$, 
	\item $\pp_3 \neq 0$, $\bb_3 = 0$, 
	\item $\pp_3 = 0$, $\bb_3 \neq 0$, and
	\item $\pp_3 = 0$, $\bb_3 = 0$.
\end{enumerate}
\paragraph{\textbf{Option 1}}Here, we will find three distinct sub-options.  Interestingly, 
these three sub-options are equivalent to options 2, 3, and 4 above.  Substituting
the matrices associated with  $\pp_3 \neq 0$ and $\bb_3 \neq 0$ into 
$\mb\mp + \mp\mb = 0$ gives us
\begin{equation} \label{eq:N2_anti_commuting_conditions}
	\begin{split}
		 0 &= \bb_1\pp_1 - \bb_1^2 \bb_3^{-1}\pp_3 + \pp_1 \bb_1 - \pp_1^2 \pp_3^{-1} \bb_3 \\
		 0 &= -\bb_1 \pp_1^2 \pp_3^{-1} + \bb_1^2 \bb_3^{-1} \pp_3 \pp_1 \pp_3^{-1} - \pp_1 \bb_1^2 
		 \bb_3^{-1} + \pp_1^2 \pp_3^{-1} \bb_3 \bb_1 \bb_3^{-1} \\
		 0 &= \bb_3 \pp_1 - \bb_3 \bb_1 \bb_3^{-1} \pp_3 + \pp_3 \bb_1 - \pp_3 \pp_1 \pp_3^{-1} \bb_3 \\
		 0 &= - \bb_3 \pp_1^2 \pp_3^{-1} + \bb_3 \bb_1 \bb_3^{-1} \pp_3 \pp_1 \pp_3^{-1} - \pp_3 \bb_1^2 \bb_3^{-1}
		 + \pp_3 \pp_1 \pp_3^{-1} \bb_3 \bb_1 \bb_3^{-1} .
	\end{split}
\end{equation}
Multiplying the first of these conditions on the right by $\bb_1\bb_3^{-1}$ and adding it to the second condition,
we obtain
\begin{equation}
0 = \bb_1 (\pp_1 - \bb_1\bb_3^{-1}\pp_3)(\bb_1\bb_3^{-1} - \pp_1 \pp_3^{-1}).
\end{equation}
Since the quaternions have no zero-divisors, one of these terms must vanish.  The
vanishing of the second is equivalent to the vanishing of the third, so we have two sub-options:
\begin{enumerate}[label=1.\arabic*]
	\item $\bb_1 = 0$, and
	\item $\bb_1\bb_3^{-1} = \pp_1 \pp_3^{-1}$.
\end{enumerate}
In the latter case, the third and fourth conditions of \eqref{eq:N2_anti_commuting_conditions} 
are trivially satisfied, but in the former case,
a little more work is required.  Setting $\bb_1 = 0$, we obtain
\begin{equation}
0 = \bb_3 \pp_1^2\pp_3^{-1} \quad \text{and} \quad 0 = \bb_3 \pp_1 - \pp_3 \pp_1 \pp_3^{-1} \bb_3.
\end{equation}
Again, using the fact the quaternions have no zero-divisors, these conditions mean this sub-option 
further divides into two:
\begin{enumerate}[label=1.1.\arabic*]
	\item $\bb_3 = 0$, and
	\item $\pp_1 = 0$,
\end{enumerate}
with $\pp_3$ left free. Recall that to arrive at these options we first made a choice to multiply 
the first condition of \eqref{eq:N2_anti_commuting_conditions} by $\bb_1\bb_3^{-1}$.  We could equally have multiplied
by $\pp_1\pp_3^{-1}$ such that case 1.1 above read $\pp_1 = 0$.  (Notice, the second
case is symmetric, so would remain the same in this instance.)  Analogous subsequent
calculations would lead to sub-options $\pp_3 = 0$ and $\bb_1 = 0$.  Putting all of this together, 
we have four sub-options to consider:
\begin{equation}
	\begin{alignedat}{2}
		\text{Sub-option 1:} \quad \mb &= 0 \quad \hspace*{26.25mm} \mp = \begin{pmatrix} \pp_1 & - \pp_1^2 \pp_3^{-1} \\
		\pp_3 & -\pp_3\pp_1\pp_3^{-1} \end{pmatrix}   \\
		\text{Sub-option 2:} \quad \mb &= \begin{pmatrix} \zero & \zero \\ \bb_3 & \zero \end{pmatrix} \quad
		\hspace*{3.25mm} \quad \qquad \mp = \begin{pmatrix} \zero & \zero \\ \pp_3 & \zero \end{pmatrix}  \\
		\text{Sub-option 3:} \quad \mb &= \begin{pmatrix} \bb_1 & -\bb_1^2 \bb_3^{-1} \\ \bb_3 &
		 -\bb_3\bb_1\bb_3^{-1} \end{pmatrix} \quad \mp = 0  \\
		 \text{Sub-option 4:} \quad \mb &= \begin{pmatrix} \bb_1 & -\bb_1^2 \bb_3^{-1} \\ \bb_3 &
		 -\bb_3\bb_1\bb_3^{-1} \end{pmatrix} \quad  \mp = \begin{pmatrix} \pp_1 & - \pp_1^2 \pp_3^{-1} \\
		\pp_3 & -\pp_3\pp_1\pp_3^{-1} \end{pmatrix} \quad \text{where} \quad \bb_1\bb_3^{-1} 
		= \pp_1 \pp_3^{-1} .
	\end{alignedat}
\end{equation}
In fact, this list can be simplified further.  For all generalised Bargmann algebras,
we can choose a transformation $(\mathbb{1}, \mathbb{1},
M, 1)$, where, $M$ takes the form
\begin{equation} \label{eq:N2_diagonalising_auto}
	M = \begin{pmatrix}
		1 & -\bb_1 \bb_3^{-1} \\ \zero & 1
	\end{pmatrix},
\end{equation}
such that sub-option 4 becomes sub-option 2.  In summary, the $\pp_3 \neq 0$ and 
$\bb_3 \neq 0$ assumption lead to three separate sub-options.
\begin{equation}
	\begin{alignedat}{2}
		\text{Sub-option 1:} \quad \mb &= 0 \quad \hspace*{26.25mm} \mp = \begin{pmatrix} \pp_1 & - \pp_1^2 \pp_3^{-1} \\
		\pp_3 & -\pp_3\pp_1\pp_3^{-1} \end{pmatrix}   \\
		\text{Sub-option 2:} \quad \mb &= \begin{pmatrix} \zero & \zero \\ \bb_3 & \zero \end{pmatrix} \quad
		\hspace*{3.25mm} \quad \qquad \mp = \begin{pmatrix} \zero & \zero \\ \pp_3 & \zero \end{pmatrix}  \\
		\text{Sub-option 3:} \quad \mb &= \begin{pmatrix} \bb_1 & -\bb_1^2 \bb_3^{-1} \\ \bb_3 &
		 -\bb_3\bb_1\bb_3^{-1} \end{pmatrix} \quad \mp = 0.
	\end{alignedat}
\end{equation}
\paragraph{\textbf{Option 2}}Letting $\pp_3 \neq 0$ and $\bb_3 = 0$, the anti-commuting condition
tells us 
\begin{equation}
	0=\bb_2\pp_3 \qquad \text{and} \qquad  \pp_1\bb_2 = \bb_2 \pp_3 \pp_1 \pp_3^{-1}.
\end{equation}
Using the first condition, $\bb_2 = 0$, and, with $\bb_2 = 0$, we are left with sub-option 1 
above.  \\
\paragraph{\textbf{Option 3}}
Now, consider $\pp_3 = 0$ and $\bb_3 \neq 0$.  Substituting the relevant forms of $\mp$ and $\mb$ 
into the anti-commuting condition, $\mb\mp + \mp\mb=0$, we find
\begin{equation}
	0=\pp_2\bb_3 \qquad \text{and} \qquad \bb_1\pp_2 = \pp_2 \bb_3 \bb_1 \bb_3^{-1}.
\end{equation}
This is identical to option 2 only $\bb$ and $\pp$ have been swapped.  Therefore,
we have a similar result: $\pp = 0$ such that we have sub-option 3 above. \\
\paragraph{\textbf{Option 4}}
The final case to consider is $\pp_3 = 0$ and $\bb_3 = 0$, where 
\begin{equation}
	\mp = \begin{pmatrix}\zero & \pp_2 \\ \zero & \zero\end{pmatrix} \quad \text{and}
		\quad \mb = \begin{pmatrix}\zero & \bb_2 \\ \zero & \zero\end{pmatrix}.
\end{equation}
These strictly upper-triangular matrices are equivalent to the strictly
lower-triangular matrices of sub-option 2 above.  Thus, again, we find no new cases to
carry forward.
\\ \\
To simplify the rest of the calculations, we will choose to use the transformation in 
\eqref{eq:N2_diagonalising_auto} for all generalised Bargmann algebras and all options.
Combining the case in which both $\mb$ and $\mp$ vanish with the non-vanishing
options, we find
\begin{equation}
		\begin{alignedat}{2}
		\text{Case 1:} \quad \mb &= 0 \quad \hspace*{26.25mm}  \mp = 0 \\
		\text{Case 2:} \quad \mb &= 0 \quad \hspace*{26.25mm} \mp = \begin{pmatrix} \zero & \zero \\
		\pp_3 & \zero \end{pmatrix}   \\
		\text{Case 3:} \quad \mb &= \begin{pmatrix} \zero & \zero \\ \bb_3 & \zero \end{pmatrix} \quad
		\hspace*{3.25mm} \quad \qquad \mp = \begin{pmatrix} \zero & \zero \\ \pp_3 & \zero \end{pmatrix}  \\
		\text{Case  4:} \quad \mb &= \begin{pmatrix} \zero & \zero \\ \bb_3 & \zero \end{pmatrix} 
		\hspace*{3.25mm} \quad \quad \qquad \mp = 0.
	\end{alignedat}
\end{equation}
In all cases, it is a straight-forward computation to show that 
$[\mb, \mp] = \mz$ tells us that $\mz = 0$.  Therefore, we are left with only $\mh$ to determine.  
From the results in Lemma~\ref{lem:N2_001}, the conditions we have including $\mb$, $\mp$ and  $\mh$ are 
\begin{equation} \label{eq:general_BP}
[\mb,\mh] = \lambda \mb + \mu \mp \quad \text{and} \quad [\mp,\mh] = \eta \mb + \varepsilon \mp.
\end{equation}
\paragraph{\textbf{Case 1}} The vanishing of $\mb$ and $\mp$ in this instance, when substituted into \eqref{eq:general_BP},
means we do not obtain any conditions on $\mh$.  Thus, we find a branch with matrices
\begin{equation}
		\mb = \mp = \mz = 0 \quad \text{and} \quad \mh \quad \text{unconstrained} .
\end{equation}
\paragraph{\textbf{Case 2}}
Notice that the vanishing of $\mb$ means that the second condition in \eqref{eq:general_BP} becomes
\begin{equation}
	\varepsilon \begin{pmatrix} \zero & \zero \\ \pp_3 & \zero \end{pmatrix} = \begin{pmatrix} \zero & \zero \\ \pp_3 & \zero \end{pmatrix} 
	\begin{pmatrix} \hh_1 & \hh_2 \\ \hh_3 & \hh_4 \end{pmatrix} - \begin{pmatrix} \hh_1 & \hh_2 \\ \hh_3 & \hh_4 \end{pmatrix}
	\begin{pmatrix} \zero & \zero \\ \pp_3 & \zero \end{pmatrix}. 
\end{equation}
This gives us two constraints
\begin{equation}
	0 = \hh_2 \pp_3 \qquad \text{and} \qquad \varepsilon \pp_3 = \pp_3 \hh_1 - \hh_4 \pp_3.
\end{equation}
The first constraint here tells us that either $\hh_2$ or $\pp_3$ must vanish.  In the latter instance, we recover the matrices from case 1.  In the former instance, we can use the second constraint to write $\hh_4$ in terms of $\hh_1$.  Thus, we find a branch with matrices
\begin{equation}
	\mb = \mz = 0 \quad \mp = \begin{pmatrix} \zero & \zero \\ \pp_3 & \zero \end{pmatrix} \quad 
	\mh = \begin{pmatrix} \hh_1 & \zero \\ \hh_3 & \pp_3 \hh_1 \pp_3^{-1} - \varepsilon \end{pmatrix} .
\end{equation}
The first condition in \eqref{eq:general_BP} does not add any new branches to those already given
as, with $\mb = 0$, it reduces to $0 = \mu \mp$.  Therefore, for those generalised Bargmann algebras
with $\mu \neq 0$, it gives the branch identified in Case 1, and, for those with $\mu = 0$, it
leaves $\pp_3$ free to fix $\hh_4$ as prescribed for the branch presented in this case.
\\
\paragraph{\textbf{Case 3}}
Substituting the $\mb$ and $\mp$ associated with this case into 
\eqref{eq:general_BP}, we get the following constraints
\begin{equation} \label{eq:branch4_h_constraints}
	0 = \hh_2 \bb_3 \qquad 0 = \hh_2 \pp_3 \qquad \lambda \bb_3 + \mu \pp_3 = \bb_3 \hh_1 - \hh_4 \bb_3 \qquad
	\eta \bb_3 + \varepsilon \pp_3 = \pp_3 \hh_1 - \hh_4 \pp_3.
\end{equation}
The first two constraints above tell us that if $\hh_2 \neq 0$, then we again arrive at the branch with
$\mb= \mp = \mz = 0$ and $\mh$ unconstrained. Letting $\hh_2 = 0$, we focus on the second two constraints.
Notice, for this branch to be distinct from the other two, we require $\bb_3 \neq 0$ and $\pp_3 \neq 0$.  
These assumptions allow us to take inverses of both $\bb_3$ and $\pp_3$ in the following calculations.
Multiplying the third constraint on the right by $\bb_3^{-1}$, we can rearrange for $\hh_4$ and substitute 
this into the fourth constraint to get
\begin{equation}
	\eta \bb_3 + \varepsilon \pp_3 = \pp_3 \hh_1 - \bb_3 \hh_1 \bb_3^{-1} \pp_3 + \mu \pp_3 \bb_3^{-1} \pp_3 
	+ \lambda \pp_3.
\end{equation}
Multiplying this expression by $\bb_3^{-1}$ on the left and rearranging, we find
\begin{equation}
	[\uu, \hh_1] = -\mu \uu^2 + (\varepsilon - \lambda) \uu + \eta,
\end{equation}
where $\uu = \bb_3^{-1} \pp_3$.  Alternatively, we could have chosen to multiply the fourth condition on
the right by $\pp_3^{-1}$ to get our expression for $\hh_4$ and substituted this into the third constraint.
Multiplying this on the left by $\pp_3^{-1}$ produces the similar condition
\begin{equation}
	[\vv, \hh_1] = \eta \vv^2 + (\lambda - \varepsilon) \vv + \mu,
\end{equation}
where $\vv = \pp_3^{-1} \bb_3$.  Depending on the generalised Bargmann algebra in question, one
of these will prove more useful than the other. We will leave these constraints in this form to be analysed 
separately for each generalised Bargmann algebra.  Note, this analysis show that, depending on the
algebra in question, we may find super-extensions for which $\mb \neq 0$ and $\mp \neq 0$.  Thus,
we can think of promoting this case to a branch.
\\
\paragraph{\textbf{Case 4}}The calculations for this case are nearly identical to those for Case 2.
The vanishing of $\mp$ means that the first constraint in \eqref{eq:general_BP} produces
\begin{equation}
	0 = \hh_2 \bb_3 \qquad \text{and} \qquad \lambda \bb_3 = \bb_3 \hh_1 - \hh_4 \bb_3.
\end{equation}
From the first expression above, if $\hh_2 \neq 0$, we recover the branch presented in Case 1.
However, setting $\hh_2 = 0$, $\bb_3$
is general, and we can use the second constraint to write $\hh_4$ in terms of $\hh_1$ and $\bb_3$.
Thus, we find a branch with matrices
\begin{equation}
	\mp = \mz = 0 \quad \mb = \begin{pmatrix} \zero & \zero \\ \bb_3 & \zero \end{pmatrix} \quad 
	\mh = \begin{pmatrix} \hh_1 & \zero \\ \hh_3 & \bb_3 \hh_1 \bb_3^{-1} - \lambda \end{pmatrix}.
\end{equation}
The second constraint in \eqref{eq:general_BP} does not produce any new branches for
$\mb, \mp,$ and $\mh$.  Substituting in $\mp = 0$,  it becomes $0 = \eta \mb$.  Therefore, 
if $\eta \neq 0$, $\mb$ must vanish leaving the branch from Case 1; and, if $\eta =0$,
$\bb_3$ is left free so we can write $\hh_4$ as prescribed for the branch presented
here.
\\ \\
In summary, we have the following three branches for all generalised Bargmann algebras
\begin{enumerate}
	\item $\mb=\mp=\mz=0 \quad \text{and} \quad \mh \quad \text{unconstrained}$
	\item $\mb = \mz = 0 \quad \mp = \begin{pmatrix} \zero & \zero \\ \pp_3 & \zero \end{pmatrix} \quad 
	\mh = \begin{pmatrix} \hh_1 & \zero \\ \hh_3 & \pp_3 \hh_1 \pp_3^{-1} - \varepsilon\end{pmatrix}$
	\item  $\mp = \mz = 0 \quad \mb = \begin{pmatrix} \zero & \zero \\ \bb_3 & \zero \end{pmatrix} \quad 
	\mh = \begin{pmatrix} \hh_1 & \zero \\ \hh_3 & \bb_3 \hh_1 \bb_3^{-1} - \lambda \end{pmatrix}$.
\end{enumerate}
There is also a possible fourth branch depending on the generalised Bargmann algebra:
\begin{equation}
	\mz = 0 \quad \mh = \begin{pmatrix} \hh_1 & \zero \\ \hh_3 & \hh_4 \end{pmatrix} \quad 
	\mb = \begin{pmatrix} \zero & \zero \\ \bb_3 & \zero \end{pmatrix} \quad 
	\mp = \begin{pmatrix} \zero & \zero \\ \pp_3 & \zero \end{pmatrix},
\end{equation}
subject to
\begin{equation}
	[\uu, \hh_1] = -\mu \uu^2 + (\varepsilon - \lambda) \uu + \eta \quad 
	\text{and} \quad [\vv, \hh_1] = \eta \vv^2 + (\lambda - \varepsilon) \vv + \mu
\end{equation}
where $\uu = \bb_3^{-1} \pp_3$ and $\vv = \pp_3^{-1} \bb_3$.
\subsection{Classification} \label{subsec:N2_class}
In this section, we complete the story started in Section~\ref{subsec:N2_branches}.  Each branch we identified in Section~\ref{subsec:N2_branches} encodes the possible $[\s_{\bar{0}}, \s_{\bar{1}}]$ brackets for a generalised Bargmann superalgebra $\s$.  Here, we take each branch in turn and find corresponding $[\Q, \Q]$ brackets.  Since
our interests are in supersymmetry, we will always impose the condition that $[\Q, \Q] \neq 0$;
therefore, we are only interested in branches for which at least one of the $N_i$ matrices
does not vanish.  Note, the imposition of this condition means that the various branches 
 identified here belong to the sub-variety $\cS$ of the real
algebraic variety cut out by the super-Jacobi identity $\cJ \subset \cV$.
\\ \\
We will begin our investigation into each branch by stating the associated matrices, $\mb, \mp, \mh, \mz \in \Mat_2(\mathbb{H})$.  These matrices are then substituted into the conditions from Lemmas~\ref{lem:N2_011} and~\ref{lem:N2_111}, which use the Lie brackets of the universal generalised Bargmann algebra.  This process produces a system of equations containing $\mb, \mp, \mh, \mz$ encoding the $[\s_{\bar{0}}, \s_{\bar{1}}]$ components of the bracket, the matrices $N_i$ for $ i \in \{0, 1, ..., 4\}$ encoding the $[\s_{\bar{1}}, \s_{\bar{1}}]$ components of the bracket, and the four parameters of the universal generalised Bargmann algebra, $\lambda, \mu, \eta, \varepsilon \in \mathbb{R}$.  Any conditions which do not contain one of the parameters $\lambda, \mu, \eta, \varepsilon$ are analysed and possible dependencies among the $N_i$ matrices are found.  Once these dependencies have been established, we start setting parameters to consider the various generalised Bargmann algebras.  In branches 1 and 2, we will see that multiple generalised Bargmann algebras produce the same set of conditions.  In these instances, we will highlight the relevant algebras but only analyse the system once to avoid repetition.  
\\ \\
In branches $\mathsf{2}$, $\mathsf{3}$ and $\mathsf{4}$, we find that the vanishing of certain matrices $N_i$ imposes the vanishing of other $N_i$. Thus, we end up with a chain of dependencies, which lead to different sub-branches.  These sub-branches will be labelled such that sub-branches with a larger branch number will have more non-vanishing matrices $N_i$.  For example, sub-branch $\mathsf{2.2}$ may have non-vanishing $N_0$ and $N_1$, but sub-branch $\mathsf{2.3}$ may additionally have non-vanishing $N_3$. Within each sub-branch, we regularly find two options: one in which $N_0$ vanishes, leaving $\mh$ free, and one in which $\mh=0$ such that $N_0$ is unconstrained.  Using sub-branch $\mathsf{2.2}$ as our example, the former instance, with $N_0 = 0$, will be labelled $\mathsf{2.2.i}$, and the latter instance will be labelled $\mathsf{2.2.ii}$.  In branch 4, we will find some instances in which both $N_0$ and $\mh$ can be non-vanishing.  Using sub-branch 4.3 as an example, we will label these cases as $\mathsf{4.3.iii}$.
\\ \\
Each sub-branch is designed to have a unique set of non-vanishing matrices.  However, the components within the matrices are not completely fixed by the super-Jacobi identity.  Therefore, each sub-branch is given as a tuple $(\mathcal{M}_{\k , \, \mathsf{X} }, \mathcal{C}_{\k , \, \mathsf{X} } )$, where $\k$ labels the underlying generalised Bargmann algebra, and $\mathsf{X}$ will be the branch number.  This tuple consists of $\mathcal{M}$, the subset of non-vanishing matrices in $\{\mb, \mp, \mh, \mz, N_0, N_1, N_2, N_3, N_4\}$ describing the branch, and $\mathcal{C}$, the set of constraints on the components of the matrices.  After stating $(\mathcal{M}, \mathcal{C})$ for a given sub-branch, we proceed to a discussion on possible parameterisations of the super-extensions in the sub-branch.  In particular, the aim of these discussions is to highlight the existence of super-extensions in the sub-branch.  First we set as many of the parameters to zero as possible.  In general, this will involve setting $\mh$ to zero along with a small number of components in the matrices $N_i$.  Then, using any residual transformations in the group $\G \subset \GL(\s_{\bar{0}}) \times \GL(\s_{\bar{1}})$, we fix the remaining parameters.  Once the existence of super-extensions has been established, we introduce some other parameters to produce further examples of generalised Bargmann superalgebras contained within the sub-branch.
\\ \\
Recall, we build the $[\Q, \Q]$ bracket from the $N_i$ matrices as follows
\begin{equation}
	[\sQ(\vt), \sQ(\vt)] = \Re(\vt N_0\vt^\dagger) \sH + \Re(\vt N_1\vt^\dagger) \sZ - \sJ(\vt N_2\vt^\dagger)  - \sB(\vt N_3\vt^\dagger)  - \sP(\vt N_4\vt^\dagger),
\end{equation}
where $N_0$ and $N_1$ are quaternion Hermitian, $N_i^\dagger = N_i$, and $N_2$, $N_3$ and $N_4$ are quaternion skew-Hermitian,
$N_j^\dagger = - N_j$.  Throughout this section, we will use the following forms for the quaternion Hermitian matrices:
\begin{equation} \label{eq:N2_C_H_Herm_Mats}
	N_0 = \begin{pmatrix} a & \qq \\ \bar{\qq} & b \end{pmatrix} \quad \text{and} \quad 
	N_1 = \begin{pmatrix} c & \rr \\ \bar{\rr} & d \end{pmatrix},
\end{equation}
where $a, b, c, d \in \mathbb{R}$, and $\qq, \rr \in \mathbb{H}$.  The
quaternion skew-Hermitian matrices will be defined
\begin{equation} \label{eq:N2_C_H_SkewHerm_Mats}
	N_3 = \begin{pmatrix} \ee & \ff \\ -\bar{\ff} & \gg \end{pmatrix} \quad \text{and} \quad 
	N_4 = \begin{pmatrix} \nn & \mm \\ -\bar{\mm} & \ll \end{pmatrix},
\end{equation}
where $\ee, \gg, \nn, \ll \in \Im(\mathbb{H})$ and $\ff, \mm \in \mathbb{H}$.
We will only briefly need to consider parts of the $N_2$ matrix explicitly; therefore, we will define its components as necessary.
\subsubsection{Branch 1} \label{subsubsec:N2_C_branch1}
\begin{equation}
	\mb=\mp=\mz=0 \quad \text{and} \quad \mh \quad \text{unconstrained}.
\end{equation}
Using the remaining conditions from the $(\s_{\bar{0}}, \s_{\bar{1}}, \s_{\bar{1}})$ and $(\s_{\bar{1}},
\s_{\bar{1}}, \s_{\bar{1}})$ components of the super-Jacobi identity, we can look to find some expressions for the matrices $N_i$.  The
conditions derived from the $[\B, \Q, \Q]$ identity in Lemma~\ref{lem:N2_011} immediately give 
$N_4 = 0$ due to the vanishing of $\mb$. Similarly, the $[\P, \Q, \Q]$ conditions give us $N_3 = 0$ due to the vanishing of $\mp$.  We are thus left with
\begin{equation}
	\begin{split}
		0 &= \mh N_i + N_i \mh^\dagger \quad i \in \{0, 1, 2\} \\
		0 &= \Re(\vt N_0\vt^\dagger) \vt \mh - \tfrac12 \vt N_2 \vt^\dagger \vt
	\end{split} \qquad 
	\begin{split}
		0 &= \mu \Re(\vt N_0 \vt^\dagger) \\
		0 &= \eta \Re(\vt N_0 \vt^\dagger) \\
		0 &= \lambda \Re(\vt N_0\vt^\dagger) \beta + \tfrac12 [\beta, \vt N_2 \vt^\dagger ]  \\
		0 &= \varepsilon \Re(\vt N_0\vt^\dagger) \pi + \tfrac12 [\pi, \vt N_2 \vt^\dagger] \\
	\end{split} \qquad \forall \beta, \pi \in \Im(\mathbb{H}), \; \forall \vt \in \mathbb{H}^2.
\end{equation}
Since $[\cc, \dd]$ is perpendicular to both $\cc$ and $\dd$ for $\cc, \dd \in \mathbb{H}$, the final two
conditions can be reduced to 
\begin{equation}
	0 = \lambda \Re(\vt N_0\vt^\dagger) \qquad 0 = [\beta, \vt N_2 \vt^\dagger ]  \qquad
	0 = \varepsilon \Re(\vt N_0\vt^\dagger) \qquad 0 = [\pi, \vt N_2 \vt^\dagger] .
\end{equation}
Substituting $\vt = (1, 0)$, $\vt = (0, 1)$, and $\vt = (1, 1)$ into the $N_2$ conditions above, we find that 
\begin{equation}
N_2 = \begin{pmatrix} \zero & e \\ -e & \zero\end{pmatrix},
\end{equation}
where $e \in \mathbb{R}$.  Now substituting $\vt = (1, \ii)$ into the $N_2$ conditions, we find
\begin{equation}
0 = -2 e [\beta, \ii].
\end{equation}
We can choose any $\beta \in \Im(\mathbb{H})$; therefore, we may choose
$\beta = \jj$.  Thus we find that $e$ must vanish, making $N_2 = 0$.
This result reduces the conditions further:
\begin{equation} \label{eq:N2_branch1_general_conditions}
	\begin{split}
		0 &= \mh N_i + N_i \mh^\dagger \quad i \in \{0, 1\} \\
		0 &= \Re(\vt N_0\vt^\dagger) \vt \mh 
	\end{split} \qquad 
	\begin{split}
		0 &= \mu \Re(\vt N_0 \vt^\dagger) \\
		0 &= \eta \Re(\vt N_0 \vt^\dagger) \\
		0 &= \lambda \Re(\vt N_0\vt^\dagger) \beta \\
		0 &= \varepsilon \Re(\vt N_0\vt^\dagger) \pi \\
	\end{split} \qquad \forall \beta, \pi \in \Im(\mathbb{H}), \; \forall \vt \in \mathbb{H}^2.
\end{equation}
Focussing on the conditions common to all generalised Bargmann algebras,
i.e.\ those conditions which do not contain $\lambda$, $\mu$, $\eta$, or $\varepsilon$,
we have only
\begin{equation}
	\begin{split}
		0 &= \mh N_i + N_i \mh^\dagger \quad i \in \{0, 1\} \\
		0 &= \Re(\vt N_0\vt^\dagger) \vt \mh.
	\end{split}
\end{equation}
Since the second condition must hold for all $\vt \in \mathbb{H}^2$, we find
that either
\begin{enumerate}[label=(\roman*)]
	\item $N_0 = 0$ and $\mh \neq 0$, or
	\item $N_0 \neq 0$ and $\mh = 0$.
\end{enumerate}
We can now split this analysis in two depending on the generalised
Bargmann algebra of interest. First, we will discuss the algebras in which at least 
one of the parameters $\lambda, \mu, \eta, \varepsilon$
are non-vanishing.  Subsequently, we will consider the algebras in which 
all of these parameters vanish.  The former instance encapsulates 
$\hat{\n}_{\pm}$ and $\hat{\g}$, and the latter encapsulates $\hat{\a}$.  
\\
\paragraph{\textit{$\hat{\n}_{\pm}$ and $\hat{\g}$}}~\\ \\
All of these algebras have non-vanishing values for at least one of the parameters,
$\lambda, \mu, \eta, \varepsilon$.  Therefore, all have the conditions for Branch 1 reduce to
\begin{equation}
	\begin{split}
		0 &= \mh N_i + N_i \mh^\dagger \quad i \in \{0, 1\} \\
		0 &= \Re(\vt N_0 \vt^\dagger) \\
		0 &= \Re(\vt N_0\vt^\dagger) \vt \mh.
	\end{split}
\end{equation}
Substituting $\vt = (1, 0)$, $\vt = (0, 1)$, and $\vt = (1, 1)$ 
into the second condition above, we find that 
\begin{equation}
N_0 = \begin{pmatrix} \zero & \Im(\qq) \\ -\Im(\qq) & \zero \end{pmatrix}.
\end{equation}
Now substitute $\vt = (1, \ii)$ into this condition,
using the convention that $\qq = q_1 \ii + q_2 \jj + q_3 \kk$,
to find
\begin{equation}
 	0 = \Re(\ii \bar{\qq}) = q_1.
\end{equation}
Using $\vt = (1, \jj)$ and $\vt= (1, \kk)$, we get analogous expressions for $q_2$ and $q_3$, so $\qq = 0$.
Therefore, $N_0 = 0$, and we cannot produce a super-extension in sub-branch $\mathsf{1.ii}$ for
these generalised Bargmann algebras.
\\ \\
The only remaining matrices are $\mh$ and $N_1$, such that
\begin{equation}
0 = \mh N_i + N_i \mh^\dagger,
\end{equation}
with no constraints on $\mh$ and $N_1 = N_1^\dagger$.  So far, we have not used any basis transformations
for this branch; therefore, we can choose $N_1$ to be the canonical quaternion Hermitian form, $\mathbb{1}$.
The above condition then states that $\mh^\dagger = - \mh$.  Thus, this branch produces one non-empty sub-branch for $\hat{\n}_\pm$ and $\hat{\g}$, with the set of non-vanishing \hypertarget{N2_ng_1}{matrices} given by
\begin{equation}
 \mathcal{M}_{\hat{\n}_\pm\,  \text{and} \, \hat{\g}, \, \mathsf{1.i}} = \Big\{ \mh = \begin{pmatrix} \hh_1 & \hh_2 \\ -\bar{\hh_2} & \hh_3 \end{pmatrix}, \quad  N_1 = \begin{pmatrix}
	1 & \zero \\ \zero & 1
\end{pmatrix} \Big\} .
\end{equation}
Although already explicit in the forms of $\mh$ and $N_1$, we note that the set of constraints for this sub-branch is
\begin{equation}
	\mathcal{C}_{\hat{\n}_\pm\,  \text{and} \, \hat{\g}, \, \mathsf{1.i}} = \{ \mh^\dagger = - \mh, \quad  N_1 = N_1^\dagger \} .
\end{equation}
Our only comment on $\mh$ going into the analysis of this branch was that it was unconstrained; therefore, we may choose to have $\mh = 0$.  Thus there is certainly a super-extension in this sub-branch, one with only $N_1 = \mathbb{1}$ non-vanishing.  However, wanting to introduce some more parameters, we may let $\hh_1$, $\hh_2$ and $\hh_3$ be non-vanishing.  These quaternions can be fixed using the group of basis transformations $\G \subset \GL(\s_{\bar{0}})\times\GL(\s_{\bar{1}})$ by noticing that $\mh^\dagger = -\mh$ tells us that $\mh \in \sp(2)$.  Therefore, the residual $\Sp(2) \subset \GL(\s_{\bar{1}})$ which fixes $N_1 = \mathbb{1}$ acts on $\mh$ via the adjoint action of $\Sp(2)$ on its Lie algebra.  Thus, we can make $\mh$ diagonal and choose the two imaginary quaternions parameterising it, arriving at 
\begin{equation}
	 \mh = \begin{pmatrix} \ii & \zero \\ \zero & \jj \end{pmatrix} \quad \text{and} \quad N_1 = \begin{pmatrix}
	1 & \zero \\ \zero & 1
\end{pmatrix} .
\end{equation}
\paragraph{$\hat{\a}$} ~\\ \\
Since $\hat{\a}$ has $\lambda = \mu = \eta = \varepsilon = 0$, the conditions in \eqref{eq:N2_branch1_general_conditions} become
\begin{equation}
	\begin{split}
		0 &= \mh N_i + N_i \mh^\dagger \quad i \in \{0, 1\} \\
		0 &= \Re(\vt N_0\vt^\dagger) \vt \mh.
	\end{split}
\end{equation}
Unlike the $\hat{\n}_{\pm}$ and $\hat{\g}$ case, these conditions do not instantly
set $N_0=0$; therefore, we may have super-extensions with either $(\mathsf{i})$ $N_0 = 0$ and $\mh \neq 0$, or $(\mathsf{ii})$ $N_0 \neq 0$ and $\mh=0$.
First, setting $\mh \neq 0$, we know this imposes $N_0 = 0$, and,
as in the $\hat{\n}_\pm$ and $\hat{\g}$ case, we may use the basis transformations to set 
$N_1 = \mathbb{1}$, such that $\mh^\dagger = -\mh$.  
Therefore, one of the possible super-extensions for $\hat{\a}$ has non-vanishing \hypertarget{N2_a_1i}{matrices}
\begin{equation} 
 	\mathcal{M}_{\hat{\a}, \, \mathsf{1.i}} = \Big\{ \mh = \begin{pmatrix} \hh_1 & \hh_2 \\ -\bar{\hh_2} & \hh_3 \end{pmatrix}, \quad  N_1 = \begin{pmatrix}
	1 & \zero \\ \zero & 1
\end{pmatrix} \Big\} .
\end{equation}
As before, the set of conditions for this super-extension is
\begin{equation}
		\mathcal{C}_{\hat{\a}, \, \mathsf{1.i}} = \{ \mh^\dagger = -\mh, \quad N_1^\dagger = N_1 \} ,
\end{equation}
and we can use $\G$ to fix the quaternions in $\hh$.  Alternatively, setting $N_0 \neq 0$, we need $\mh= 0$.  
Thus the second possible super-extension in this
\hypertarget{N2_a_1ii}{branch} has
\begin{equation}
	\mathcal{M}_{\hat{\a}, \, \mathsf{1.ii}} = \Big\{ N_0 = \begin{pmatrix}
		a & \qq \\ \bar{\qq} & b
	\end{pmatrix},\quad N_1 = \begin{pmatrix}
		c & \rr \\ \bar{\rr} & d
	\end{pmatrix} \Big\} \quad \text{and} \quad 
	\mathcal{C}_{\hat{\a}, \, \mathsf{1.ii}} = \{ N_0^\dagger = N_0, \quad N_1^\dagger = N_1 \} .		
\end{equation}
Since the primary constraint on these matrices is that both be non-vanishing, we can choose to have $b$, $\qq$, $c$ and $\rr$ vanish.  Using the scaling symmetry of the $\s_{\bar{0}}$ basis elements present in $\G \subset \GL(\s_{\bar{0}}) \times \GL(\s_{\bar{1}})$, we can write down the super-extension
\begin{equation}
	N_0 = \begin{pmatrix}
		1 & \zero \\ \zero & \zero
	\end{pmatrix}\quad N_1 = \begin{pmatrix}
		\zero & \zero \\ \zero & 1
	\end{pmatrix}.
\end{equation}
Therefore, this sub-branch is not empty.  Additionally, we may choose to keep all the parameters in the matrices of $\mathcal{M}_{\hat{\a}, \, \mathsf{1.ii}}$ and use the basis transformations to fix them.  In particular, we can let $N_0 = \mathbb{1}$.  This choice leaves us with a residual $\Sp(2)$ action with which to fix the parameters of $N_1$, which may give us $N_1 = \mathbb{1}$.  
\subsubsection{Branch 2} \label{subsubsec:N2_C_branch2}
\begin{equation}
	\mb = \mz = 0 \quad \mp = \begin{pmatrix} \zero & \zero \\ \pp_3 & \zero \end{pmatrix} \quad 
	\mh = \begin{pmatrix} \hh_1 & \zero \\ \hh_3 & \pp_3 \hh_1 \pp_3^{-1} - \varepsilon\end{pmatrix}.
\end{equation}
As above, it is useful to exploit the vanishing matrices of the branch to simplify the conditions from Lemmas~\ref{lem:N2_011} and~\ref{lem:N2_111}.  In particular, the $[\B, \Q, \Q]$ super-Jacobi identity tells us $N_4 = 0$ due to the vanishing of $\mb$.  The rest of the $[\B, \Q, \Q]$ conditions tells us that
\begin{equation}
	\begin{split}
		0 &= \lambda \Re(\vt N_0 \vt^\dagger) \beta + \tfrac12 [\beta, \vt N_2 \vt^\dagger] \\
		0 &= \mu \Re(\vt N_0\vt^\dagger) \beta .			
	\end{split}
\end{equation}
The $[\P, \Q, \Q]$ conditions become
\begin{equation}
	\begin{split}
		0 &= \mp N_0 - N_0 \mp^\dagger \\	
		-N_3 &= \mp N_1 - N_1 \mp^\dagger \\
		0 &= \pi \vt \mp N_2 \vt^\dagger + \vt N_2 (\pi \vt \mp)^\dagger \\
		\eta \Re(\vt N_0\vt^\dagger) \pi &= \pi \vt \mp N_3 \vt^\dagger + \vt N_3 (\pi \vt \mp)^\dagger \\
		0 &= \varepsilon \Re(\vt N_0 \vt^\dagger) \pi + \tfrac12 [\pi, \vt N_2 \vt^\dagger]. \\
	\end{split}
\end{equation}
Since the conditions from the $[\bZ,\Q, \Q]$ identity are all satisfied due to $\mz = 0$, the final conditions are
\begin{equation}
	\begin{split}
		0 &= \mh N_i + N_i \mh^\dagger \quad \text{where} \quad i \in \{0, 1, 2\} \\
		\lambda N_3 &= \mh N_3 + N_3 \mh^\dagger \\
		0 &= \mu N_3, \\
	\end{split}
\end{equation}
from $[\bH, \Q, \Q]$.  The result from Lemma~\ref{lem:N2_111} then gives us 
\begin{equation}
	\Re(\vt N_0 \vt^\dagger) \vt \mh = \tfrac12 \vt N_2 \vt^\dagger \vt.
\end{equation}
As in Branch 1, the conditions
\begin{equation}
			0 = \lambda \Re(\vt N_0 \vt^\dagger) \beta + \tfrac12 [\beta, \vt N_2 \vt^\dagger] \quad \text{and} \quad 
			0 = \varepsilon \Re(\vt N_0 \vt^\dagger) \pi + \tfrac12 [\pi, \vt N_2 \vt^\dagger],
\end{equation}
tell us that $N_2 = 0$.  Therefore, the conditions reduce further to
\begin{equation}  \label{eq:N2_C_branch2_constraints}
	\begin{split}
		0 &= \mu N_3 \\
		0 &= \mh N_i + N_i \mh^\dagger \quad \text{where} \quad i \in \{0, 1\} \\
		0 &= \mp N_0 - N_0 \mp^\dagger \\
		0 &= \lambda \Re(\vt N_0 \vt^\dagger) \beta\\
		0 &= \mu \Re(\vt N_0\vt^\dagger) \beta \\
		0 &= \varepsilon \Re(\vt N_0 \vt^\dagger) \pi \\
		0 &= \Re(\vt N_0\vt^\dagger) \vt \mh 
	\end{split}
	\begin{split}
		\lambda N_3 &= \mh N_3 + N_3 \mh^\dagger \\
		-N_3 &= \mp N_1 - N_1 \mp^\dagger \\
		\eta \Re(\vt N_0\vt^\dagger) \pi &= \pi \vt \mp N_3 \vt^\dagger + \vt N_3 (\pi \vt \mp)^\dagger \\
	\end{split} \quad \forall \beta, \pi \in \Im(\mathbb{H}), \forall \vt \in \mathbb{H}^2.
\end{equation}
We can now use the following two conditions common to all generalised Bargmann algebras to highlight
the possible sub-branches:
\begin{equation}
		-N_3 = \mp N_1 - N_1 \mp^\dagger \quad \text{and} \quad
		0 = \Re(\vt N_0\vt^\dagger) \vt \mh.
\end{equation}
Substituting the $N_1$ from \eqref{eq:N2_C_H_Herm_Mats} and the
$N_3$ from \eqref{eq:N2_C_H_SkewHerm_Mats} into the first condition here, we can write
\begin{equation}
	N_3 = \begin{pmatrix} \zero & c\bar{\pp_3} \\ - c\pp_3 & \bar{\rr}\bar{\pp_3} - \pp_3 \rr \end{pmatrix}.
\end{equation}
This result tells us that $N_3$ is dependent on $N_1$: if $N_1 = 0$ then
$N_3=0$.  Therefore, we may organise our investigation into the possible super-extensions by
considering each of the following sub-branches in turn
\begin{enumerate}
	\item $N_1 = 0$ and $N_3 = 0$,
	\item $N_1 \neq 0$ and $N_3 = 0$,
	\item $N_1 \neq 0$ and $N_3 \neq 0$.
\end{enumerate}
Next, consider the condition from the $[\Q, \Q, \Q]$ identity:
\begin{equation}
	0 = \Re(\vt N_0 \vt^\dagger) \vt \mh.
\end{equation}
Notice, this is identical to the condition from the $[\Q, \Q, \Q]$ identity we found in \hyperlink{N2_a_1i}{Branch 1}.  Therefore, as before, we have two cases to consider in each sub-branch: 
\begin{enumerate}[label=(\roman*)]
	\item $N_0 = 0$ and $\mh \neq 0$, and
	\item  $N_0 \neq 0$ and $\mh = 0$.
\end{enumerate}
We will now consider each generalised Bargmann algebra in turn to determine
whether they have super-extensions associated to these sub-branches.
\\
\paragraph{$\hat{\a}$} ~\\ \\
In addition to the conditions already discussed in producing the possible sub-branches,
\begin{equation} \label{eq:N2_C_branch2_a_conditions1}
	-N_3 = \mp N_1 - N_1 \mp^\dagger \quad \text{and} \quad
		0 = \Re(\vt N_0\vt^\dagger) \vt \mh,
\end{equation}
substituting $\lambda = \mu = \eta = \varepsilon = 0$ into \eqref{eq:N2_C_branch2_constraints} leaves us with
\begin{equation} \label{eq:N2_C_branch2_a_conditions2}
	\begin{split}
		0 &= \mh N_i + N_i \mh^\dagger \quad \text{where} \quad i \in \{0, 1, 3\} \\
		0 &= \mp N_0 - N_0 \mp^\dagger \\
		0 &= \pi \vt \pp N_3 \vt^\dagger + \vt N_3 (\pi \vt \mp)^\dagger. \\
	\end{split}
\end{equation}
None of these conditions force the vanishing of any more $N_i$; therefore, \textit{a priori}
we may find super-extensions in each of the sub-branches.  The only restriction
to the matrices so far has been the re-writing of $N_3$:
\begin{equation}
	N_3 = \begin{pmatrix} \zero & c\bar{\pp_3} \\ - c\pp_3 & \bar{\rr}\bar{\pp_3} - \pp_3 \rr \end{pmatrix}.
\end{equation}
\paragraph{\textbf{Sub-Branch 2.1}} Setting $N_1 = N_3 = 0$, we are left with only $N_0$, subject to 
\begin{equation}
		0 = \mh N_0 + N_0 \mh^\dagger \quad \text{and} \quad
		0 = \mp N_0 - N_0 \mp^\dagger.
\end{equation}
We know that we may have two possible cases for this sub-branch: either $(\mathsf{i})$ $N_0 = 0$ and $\mh \neq 0$,
or $(\mathsf{ii})$ $N_0 \neq 0$ and $\mh = 0$.  Since we need $N_0 \neq 0$ for a supersymmetric extension,
we must have the latter case.  This leaves only the second condition above with which to 
restrict the form of $N_0$.  Since $\pp_3 \neq 0$, this tells us
\begin{equation} \label{eq:case2_N0_constraints}
	0 = a \quad \text{and} \quad 0 = \pp_3\qq - \bar{\qq} \bar{\pp_3}.
\end{equation}
Thus the \hypertarget{N2_a_21ii}{sub-branch} is given by
\begin{equation}
	 \mathcal{M}_{\hat{\a}, \, \mathsf{2.1.ii}} = \Big\{ \mp = \begin{pmatrix} \zero & \zero \\ \pp_3 & \zero \end{pmatrix}, \quad 
	N_0 = \begin{pmatrix} \zero & \qq \\ \bar{\qq} & b \end{pmatrix} \Big\} \quad \text{and} \quad 
	\mathcal{C}_{\hat{\a}, \, \mathsf{2.1.ii}} = \{ 0 =  \pp_3\qq - \bar{\qq} \bar{\pp_3} \}.
\end{equation}
This sub-branch is parameterised by two collinear quaternions $\pp_3$ and $\qq$, and a single real scalar $b$, such that it defines an 8-dimensional space in the sub-variety $\cS$.  Notice that we can choose either $\qq = 0$ or $b = 0$ and this sub-branch remains supersymmetric.  Choosing the former case, we can use the endomorphisms of $\s_{\bar{1}}$ to set $\pp_3 = \ii$ and the scaling symmetry of $\sH$ to produce
\begin{equation}
	 \mp = \begin{pmatrix} \zero & \zero \\ \ii & \zero \end{pmatrix} \quad \text{and} \quad
	N_0 = \begin{pmatrix} \zero & \zero \\ \zero & 1 \end{pmatrix}.
\end{equation}
In the latter case, we can still choose $\pp_3 = \ii$, and the condition in $\mathcal{C}_{\hat{\a}, \, \mathsf{2.1.ii}}$ will impose that $\qq$ must also lie along $\ii$.  Again using the scaling symmetry of $\sH$ in $\G \subset \GL(\s_{\bar{0}}) \times \GL(\s_{\bar{1}})$, we arrive at
\begin{equation}
	 \mp = \begin{pmatrix} \zero & \zero \\ \ii & \zero \end{pmatrix} \quad \text{and} \quad 
	N_0 = \begin{pmatrix} \zero & \ii \\ -\ii & \zero \end{pmatrix}.
\end{equation}
These two examples turn out to be the only super-extensions in this sub-branch.  Keeping both $b$ and $\qq$ at the outset, we can use the endomorphisms of $\s_{\bar{1}}$ to set $b = 0$ while imposing that $\pp_3$ and $\qq$ lie along $\ii$.  Thus, in this case, we could always retrieve the second example above.
\\
\paragraph{\textbf{Sub-Branch 2.2}} Setting $N_1 \neq 0$ but keeping $N_3 = 0$, the conditions in  \eqref{eq:N2_C_branch2_a_conditions1} and \eqref{eq:N2_C_branch2_a_conditions2} become
\begin{equation}
	\begin{split}
		0 &= \mp N_i - N_i \mp^\dagger  \\
		0 &= \mh N_i + N_i \mh^\dagger
	\end{split}  \quad \text{where} \quad i \in \{0, 1\}.
\end{equation}
Importantly, we can now have super-extensions in either of the two cases: $(\mathsf{i})$ $N_0 = 0$ and $\mh \neq 0$, or $(\mathsf{ii})$ $N_0 \neq 0$ and $\mh = 0$.  In the former case, in which $N_0 = 0$, \eqref{eq:N2_C_branch2_a_conditions1} and \eqref{eq:N2_C_branch2_a_conditions2} become
\begin{equation}
		0 = \mp N_1 - N_1 \mp^\dagger \quad \text{and} \quad 
		0 = \mh N_1 + N_1 \mh^\dagger.
\end{equation}
The first of these conditions tells us that 
\begin{equation}
	N_1 = \begin{pmatrix}
		\zero & \rr \\ \bar{\rr} & d
	\end{pmatrix},
\end{equation}
such that $0 = \pp_3\rr - \bar{\rr}\bar{\pp_3}$.  Substituting this $N_1$ into the latter condition, we find
\begin{equation}
	\begin{split}
		0 &= \hh_1 \rr + \rr \overbar{\pp_3\hh_1\pp_3^{-1}} \\
		0 &= \Re(\hh_3\rr) + d \Re(\hh_1).
	\end{split}
\end{equation}
Assuming $\rr \neq 0$ and $\hh_1 \neq 0$, take the real part of the first constraint to get $\Re(\hh_1) = 0$.  Alternatively, with $\rr = 0$, $d \neq 0$ for $N_1 \neq 0$; therefore, the second constraint would also impose $\Re(\hh_1) = 0$.  This result allows us to simply the constraints to 
\begin{equation}
		0 = \Re(\hh_1) \qquad 0 = [\hh_1, \rr\pp_3] \qquad 0 = \Re(\hh_3\rr).
\end{equation}
In fact, the second constraint above is satisfied by 
\begin{equation}
	0 = \pp_3\rr - \bar{\rr}\bar{\pp_3},
\end{equation}
so the set of constraints on this sub-branch becomes
\begin{equation}
	\mathcal{C}_{\hat{\a},\, \mathsf{2.2.i}} = \{ 0 = \Re(\hh_1), \quad 0 = \Re(\hh_3\rr), \quad 0 = \pp_3\rr - \bar{\rr}\bar{\pp_3} \}.
\end{equation}
Subject to these constraints, we have the following non-vanishing \hypertarget{N2_a_22i}{matrices}
\begin{equation}
	\mathcal{M}_{\hat{\a},\, \mathsf{2.2.i}} = \Big\{ \mp = \begin{pmatrix} \zero & \zero \\ \pp_3 & \zero \end{pmatrix}, \quad 
	\mh = \begin{pmatrix} \hh_1 & \zero \\ \hh_3 & \pp_3 \hh_1 \pp_3^{-1} \end{pmatrix}, \quad
	N_1 = \begin{pmatrix} \zero & \rr \\ \bar{\rr} & d \end{pmatrix} \Big\}.
\end{equation}
This sub-branch consists of two collinear quaternions $\pp_3$ and $\rr$, one quaternion $\hh_3$ that is perpendicular to these two in $\Im(\mathbb{H})$, and one imaginary quaternion $\hh_1$.  In addition, there is a single real scalar, $d$.  Notice that if $\mh$ vanishes, we produce a system that is equivalent to the one found in Sub-Branch \hyperlink{N2_a_21ii}{$\mathsf{2.1.ii}$}; therefore, this sub-branch is certainly non-empty.  However, to investigate the role of $\mh$, we will require at least one of its components to be non-vanishing.  To simplify $\mh$ as far as possible, let $\hh_1 = 0$.  Now we can choose either $\rr$ or $d$ to vanish while maintaining supersymmetry.  Letting $\rr = 0$, we can use the endomorphisms of $\s_{\bar{1}}$ on $\pp_3$ and $\hh_3$, and employ the scaling of $\sZ$ on $N_1$ to arrive at
\begin{equation}
	\mp = \begin{pmatrix} 0 & 0 \\ \ii& 0 \end{pmatrix}, \quad 
	\mh = \begin{pmatrix} 0 & 0 \\ \ii & 0 \end{pmatrix}, \quad
	N_1 = \begin{pmatrix} 0 & 0\\ 0 & 1 \end{pmatrix}.
\end{equation}
Thus, there exist super-extensions in this sub-branch for which $\mh \neq 0$.  Wanting to be more be a little more general, we can choose for only $\hh_3$ to vanish.  Then, using the endomorphisms in $\GL(\s_{\bar{1}})$, we can set $\rr = d \ii$ such that $\pp_3$ also lies along $\ii$.  Utilising the scaling symmetry of $\sP$ and $\sZ$ in $\GL(\s_{\bar{0}})$, we can remove the constants from the matrices $\mp$ and $N_1$ to get
\begin{equation}
	\mp = \begin{pmatrix}
		\zero & \zero \\ \ii & \zero
	\end{pmatrix} \quad \text{and} \quad
	N_1 = \begin{pmatrix}
		\zero & \ii \\ -\ii & 1
	\end{pmatrix}.
\end{equation}
Employing the residual endomorphisms of $\s_{\bar{1}}$, we can now choose $\hh_1$ to lie along $\ii$. This change allows us to use the scaling symmetry of $\sH$ in $\GL(\s_{\bar{0}})$ such that $\mh$ becomes
\begin{equation}
	\mh = \begin{pmatrix}
		\ii & \zero \\ \zero & \ii
	\end{pmatrix}.
\end{equation}
Now, returning to the latter case, in which $\mh$ vanishes and $N_0 \neq 0$, we have only
\begin{equation}
		0 = \mp N_i - N_i \mp^\dagger \quad \text{where} \quad i \in \{0, 1\},
\end{equation}
which tells us that 
\begin{equation}
	N_0 = \begin{pmatrix}
		\zero & \qq \\ \bar{\qq} & b
	\end{pmatrix} \quad \text{and} \quad N_1 = \begin{pmatrix}
		\zero & \rr \\ \bar{\rr} & d
	\end{pmatrix},
\end{equation} 
where
\begin{equation}
	0 = \pp_3\qq - \bar{\qq}\bar{\pp_3} \quad \text{and} \quad 0 = \pp_3\rr - \bar{\rr}\bar{\pp_3}.
\end{equation} 
Therefore, the set of non-vanishing \hypertarget{N2_a_22ii}{matrices} is given by
\begin{equation}
	\mathcal{M}_{\hat{\a},\,\mathsf{2.2.ii}} = \Big\{ \mp = \begin{pmatrix} \zero & \zero \\ \pp_3 & \zero \end{pmatrix}, \quad 
	N_0 = \begin{pmatrix} \zero & \qq \\ \bar{\qq} & b \end{pmatrix}, \quad 
	N_1 = \begin{pmatrix} \zero & \rr \\ \bar{\rr} & d \end{pmatrix} \Big\},
\end{equation}
subject to 
\begin{equation}
	\mathcal{C}_{\hat{\a},\,\mathsf{2.2.ii}} = \{ 0 = \pp_3\qq - \bar{\qq} \bar{\pp_3} \quad \text{and} \quad 0 = \pp_3\rr - \bar{\rr}\bar{\pp_3} \}.
\end{equation}
Notice that the matrices $N_0$ and $N_1$ and the constraints on their components take the same form as the matrix $N_0$ and its constraints in Sub-Branch \hyperlink{N2_a_21ii}{$\mathsf{2.1.ii}$}.  However, this sub-branch is distinct.  Notice that, using the endomorphisms of $\s_{\bar{1}}$ and the conditions in $\mathcal{C}_{\hat{\a},\,\mathsf{2.2.ii}}$, we can make all the quaternions parameterising this sub-branch of $\cS$ lie along $\ii$.  The scaling symmetry of $\sP$ may then be employed to set $\pp_3 = \ii$, leaving only $b$ and $d$ unfixed.  The last of the endomorphisms of $\s_{\bar{1}}$ may set one of these parameters to zero, but not both; therefore, we cannot have $N_0 = N_1$, which would be a necessary condition for this sub-branch to be equivalent to $(\mathcal{M}_{\hat{\a},\,\mathsf{2.1.ii}}, \mathcal{C}_{\hat{\a},\,\mathsf{2.1.ii}})$.  However, we can fix all the parameters of this sub-branch.  Had we chosen $\qq = b \ii$ with the initial $\s_{\bar{1}}$ endomorphism and set $d = 0$, we could scale $\sH$ and $\sZ$ to find
\begin{equation}
	\mp = \begin{pmatrix} \zero & \zero \\ \ii& \zero \end{pmatrix}, \quad 
	N_0 = \begin{pmatrix} \zero & \ii \\ -\ii & 1 \end{pmatrix}, \quad 
	N_1 = \begin{pmatrix} \zero & \ii \\ -\ii & \zero \end{pmatrix}.
\end{equation}
Thus, this sub-branch is non-empty and we can fix all parameters in each super-extension it contains. 
\\
\paragraph{\textbf{Sub-Branch 2.3}} 
Finally, with $N_1 \neq 0$ and $N_3 \neq 0$, we can substitute $\vt = (0, 1)$ into
\begin{equation}
	0 = \pi \vt \mp N_3 \vt^\dagger + \vt N_3 (\pi \vt \mp)^\dagger
\end{equation} 
to find $c = 0$.  Therefore, $N_1$ and $N_3$ are reduced to 
\begin{equation}
	N_1 = \begin{pmatrix}
	\zero & \rr \\ \bar{\rr} & d
	\end{pmatrix} \quad \text{and} \quad 
	N_3 = \begin{pmatrix}
	\zero & \zero \\ \zero & \bar{\rr}\bar{\pp_3} - \pp_3 \rr 
	\end{pmatrix}.
\end{equation}
Recall, the condition
\begin{equation}
	0 = \Re(\vt N_0 \vt^\dagger) \vt \mh
\end{equation}
tells us that either $(\mathsf{i})$ $N_0 = 0$ and $\mh \neq 0$, or $(\mathsf{ii})$ $N_0 \neq 0$ and $\mh = 0$.
Letting $N_0 = 0$, the final conditions for this sub-branch are
\begin{equation}
		0 = \mh N_i + N_i \mh^\dagger \quad \text{where} \quad i \in \{1, \, 3\}.
\end{equation}
From the discussion in Sub-Branch \hyperlink{N2_a_22i}{$\mathsf{2.2.i}$}, we know that
the $N_1$ case produces the constraints
\begin{equation} 
	0 = \Re(\hh_1), \quad 0 = [\hh_1, \rr\pp_3], \quad \text{and} \quad 0 = \Re(\hh_3 \rr) .
\end{equation}
Interestingly, the $N_3$ condition adds no new constraints to this set; therefore, we have 
\begin{equation} 
	\mathcal{C}_{\hat{\a}, \, \mathsf{2.3.i}} = \{ 0 = \Re(\hh_1), \quad 0 = [\hh_1, \rr\pp_3],
	\quad 0 = \Re(\hh_3 \rr) \}.
\end{equation}
The corresponding \hypertarget{N2_a_23i}{matrices} for this sub-branch are given by
\begin{equation}
	\mathcal{M}_{\hat{\a}, \, \mathsf{2.3.i}} = \Big\{ \mp = \begin{pmatrix} \zero & \zero \\ \pp_3 & \zero \end{pmatrix}, \quad 
	\mh = \begin{pmatrix} \hh_1 & \zero \\ \hh_3 & \pp_3 \hh_1 \pp_3^{-1} \end{pmatrix}, \quad
	N_1 = \begin{pmatrix} \zero & \rr \\ \bar{\rr} & d \end{pmatrix}, \quad
	N_3 = \begin{pmatrix} \zero & \zero \\ \zero & \bar{\rr}\bar{\pp_3} - \pp_3 \rr \end{pmatrix} \Big\}.
\end{equation}
To establish the existence of super-extensions in this sub-branch, begin by setting $\mh = 0$ and $d = 0$.  The endomorphisms of $\s_{\bar{1}}$ may be used to set $\pp_3$ to lie along $\ii$ and scale $\rr$ such that $\rr \in \Sp(1)$.  We can then utilise  the automorphisms of $\mathbb{H}$ and the scaling symmetry of $\sP$ and $\sB$ in $\GL(\s_{\bar{0}})$ to set $\rr$ and fix the parameters in $\mp$ and $N_3$.  This leaves us with a super-extension whose matrices are written
\begin{equation}
	 \mp = \begin{pmatrix} \zero & \zero \\ \ii & \zero \end{pmatrix}, \quad 
	N_1 = \begin{pmatrix} \zero & 1 + \jj \\ 1-\jj & \zero \end{pmatrix}, \quad
	N_3 = \begin{pmatrix} \zero & \zero \\ \zero & \kk \end{pmatrix}.
\end{equation}
Using this parameterisation, we can also introduce $\hh_1$.  Substituting $\pp_3 = \ii$ and $\rr = 1 + \jj$ into the constraints of $\mathcal{C}_{\hat{\a}, \, \mathsf{2.3.i}}$, we find
\begin{equation}
	\mp = \begin{pmatrix} \zero & \zero \\ \ii & \zero \end{pmatrix}, \quad 
	\mh = \begin{pmatrix} \ii - \kk & \zero \\ \zero & \ii + \kk \end{pmatrix}, \quad
	N_1 = \begin{pmatrix} \zero & 1 + \jj \\ 1-\jj & \zero \end{pmatrix}, \quad
	N_3 = \begin{pmatrix} \zero & \zero \\ \zero & \kk \end{pmatrix}.
\end{equation}
Looking to include $\hh_3$ or $d$ leads to the introduction of parameters that cannot be fixed using the basis transformations $\G \subset \GL(\s_{\bar{0}}) \times \GL(\s_{\bar{1}})$ and the constraints.
\\ \\
In the latter case, for which $\mh = 0$ and $N_0 \neq 0$, the only remaining condition is
\begin{equation}
		0 = \mp N_0 - N_0 \mp^\dagger,
\end{equation}
which we know from the previous sub-branches, tells us that $\pp_3$ and $\qq$
are collinear, and that $a = 0$.  Therefore, the non-vanishing
\hypertarget{N2_a_23ii}{matrices} for this sub-branch are
\begin{equation}
	\mathcal{M}_{\hat{\a},\, \mathsf{2.3.ii}} = \Big\{\mp = \begin{pmatrix} \zero & \zero \\ \pp_3 & \zero \end{pmatrix}, \quad 
	N_0 = \begin{pmatrix} \zero & \qq \\ \bar{\qq} & b \end{pmatrix}, \quad
	N_1 = \begin{pmatrix} \zero & \rr \\ \bar{\rr} & d \end{pmatrix}, \quad
	N_3 = \begin{pmatrix} \zero & \zero \\ \zero & \bar{\rr}\bar{\pp_3} - \pp_3 \rr \end{pmatrix} \Big\},
\end{equation}
and the constraints are given by
\begin{equation}
	\mathcal{C}_{\hat{\a}, \, \mathsf{2.3.ii}} = \{ 0 = \pp_3\qq - \bar{\qq}\bar{\pp_3} \}.
\end{equation}
This sub-branch of $\cS$ has 13 real parameters, being parameterised by two collinear quaternions $\pp_3$ and $\qq$, an additional quaternion $\rr$ and two real scalars, $b$ and $d$.  Letting $d = 0$ and $\qq = 0$, we can use the same transformations as in Sub-Branch \hyperlink{N2_a_23i}{$\mathsf{2.3.i}$} to fix 
\begin{equation}
	 \mp = \begin{pmatrix} \zero & \zero \\ \ii & \zero \end{pmatrix}, \quad 
	N_1 = \begin{pmatrix} \zero & 1 + \jj \\ 1-\jj & \zero \end{pmatrix}, \quad
	N_3 = \begin{pmatrix} \zero & \zero \\ \zero & \kk \end{pmatrix}.
\end{equation}
Subsequently employing the scaling symmetry of $\sH$ in $\GL(\s_{\bar{0}})$, we can fix $b$ such that
\begin{equation}
	N_0 = \begin{pmatrix} \zero & \zero \\ \zero & 1 \end{pmatrix}.
\end{equation}
Therefore, there are certainly super-extensions in this sub-branch.  We can introduce either $\qq$ or $d$ while continuing to fix all the parameters of the super-extension; however, attempting to include both leads to the inclusion of a parameter that we cannot fix with the constraints of $\mathcal{C}_{\hat{\a}, \, \mathsf{2.3.ii}}$ and basis transformations in $\G$.
\\
\paragraph{$\hat{\n}_-$}~\\ \\
Setting $\mu = \eta = 0$, $\lambda = 1$ and $\varepsilon = -1$, the conditions in \eqref{eq:N2_C_branch2_constraints} reduce to 
\begin{equation}
	\begin{split}
		0 &= \mh N_i + N_i \mh^\dagger \quad \text{where} \quad i \in \{0, 1\} \\
		0 &= \mp N_0 - N_0 \mp^\dagger \\
		0 &= \Re(\vt N_0 \vt^\dagger) \beta\\
		0 &= - \Re(\vt N_0 \vt^\dagger) \pi \\
		0 &= \Re(\vt N_0\vt^\dagger) \vt \mh
	\end{split} \qquad 
	\begin{split}
		0 &= \pi \vt \mp N_3 \vt^\dagger + \vt N_3 (\pi \vt \mp)^\dagger \\
		N_3 &= \mh N_3 + N_3 \mh^\dagger \\
		-N_3 &= \mp N_1 - N_1 \mp^\dagger. \\
	\end{split}
\end{equation}
From the third and fourth condition, we instantly get $N_0 = 0$.  Therefore, we cannot have any solutions along sub-branches satisfying case $(\mathsf{ii})$ for $\hat{\n}_-$.  We are left with 
\begin{equation} \label{eq:N2_n-_22_conditions}
	\begin{split}
		0 &= \mh N_1 + N_1 \mh^\dagger \\
		0 &= \pi \vt \mp N_3 \vt^\dagger + \vt N_3 (\pi \vt \mp)^\dagger
	\end{split} \qquad 
	\begin{split}
		N_3 &= \mh N_3 + N_3 \mh^\dagger \\
		-N_3 &= \mp N_1 - N_1 \mp^\dagger. \\
	\end{split}
\end{equation}
\paragraph{\textbf{Sub-Branch 2.1}} Since $N_1$ and $N_3$ are the only possible non-vanishing
matrices encoding the $[\Q, \Q]$ bracket, we cannot have a super-extension in this branch.
\\
\paragraph{\textbf{Sub-Branch 2.2}} With $N_3 = 0$, we are left with
\begin{equation} 
		0 = \mh N_1 + N_1 \mh^\dagger \quad \text{and} \quad 
		0 = \mp N_1 - N_1 \mp^\dagger.
\end{equation}
The latter condition tells us that $\pp_3$ and $\rr$ are collinear and $0 = c\pp_3$.  Since we must have $\pp_3 \neq 0$ in this branch, we have $c = 0$.  Using this result, the first condition above tells us
\begin{equation}
	\begin{split}
		0 &= \hh_1 \rr + \rr \overbar{\pp_3\hh_1\pp_3^{-1}+1} \\
		0 &= \Re(\hh_3\rr) + d ( \Re(\hh_1) +1).
	\end{split}
\end{equation}
In fact, utilising the collinearity of $\pp_3$ and $\rr$, the first of these constraints becomes
\begin{equation}
	0 = (2 \Re(\hh_1) + 1 ) \Re(\pp_3 \rr).
\end{equation}
Thus, we have
\begin{equation}
	\mathcal{C}_{\hat{\n}_-,\, \mathsf{2.2.i}} = \{ 0 = \bar{\rr}\bar{\pp_3} - \pp_3 \rr, \quad 0 = (2 \Re(\hh_1) + 1 ) \Re(\pp_3 \rr),
	\quad 0 = \Re(\hh_3\rr) + d ( \Re(\hh_1) +1) \}.
\end{equation}
The non-vanishing \hypertarget{N2_n-_22i}{matrices} in this instance are
\begin{equation}
	\mathcal{M}_{\hat{\n}_-, \, \mathsf{2.2.i}} = \Big\{\mp = \begin{pmatrix} \zero & \zero \\ \pp_3 & \zero \end{pmatrix}, \quad 
	\mh = \begin{pmatrix} \hh_1 & \zero \\ \hh_3 & \pp_3 \hh_1 \pp_3^{-1} + 1 \end{pmatrix}, \quad
	N_1 = \begin{pmatrix} \zero & \rr \\ \bar{\rr} & d \end{pmatrix} \Big\}.	
\end{equation}
Therefore, the sub-branch in $\cS$ for these super-extensions of $\hat{\n}_-$ is parameterised by two collinear quaternions $\pp_3$ and $\rr$, two quaternions encoding the action of $\sH$ on $\s_{\bar{1}}$, $\hh_1$ and $\hh_3$, and one real scalar $d$. Notice, this is the first instance in which setting some parameters to zero imposes particular values for other parameters in the extension.  
In particular, the vanishing of $\rr$ imposes $\Re(\hh_1) = -1$ by the third constraint in $\mathcal{C}_{\hat{n}_-,\, \mathsf{2.2.i}}$, since $d \neq 0$ in this instance.  However, if $\rr \neq 0$, the second constraint implies $2\Re(\hh_1) = -1$.  In the former case, we can set $\hh_3$ and the imaginary part of $\hh_1$ to zero.  Using the endomorphisms of $\s_{\bar{1}}$ to set $\pp_3 = \ii$, we can subsequently employ the scaling symmetry of $\sH$ and $\sZ$ to obtain a super-extension with matrices
\begin{equation}
	\mp = \begin{pmatrix} \zero & \zero \\ \ii & \zero \end{pmatrix}, \quad 
	\mh = \begin{pmatrix} 1 & \zero \\ \zero & \zero \end{pmatrix} \quad  \text{and} \quad 
	N_1 = \begin{pmatrix} \zero & \zero \\ \zero & 1 \end{pmatrix}.
\end{equation}
Therefore, there exist super-extensions in this sub-branch for which $\rr = 0$.  Letting $\rr \neq 0$, we may again use the endomorphisms of $\s_{\bar{1}}$ to impose that $\pp_3$ lies along $\ii$; however, due to the first constraint in $\mathcal{C}_{\hat{\n}_-, \, \mathsf{2.2.i}}$, this also means that $\rr$ lies along $\ii$. Utilising the scaling symmetry of the $\s_{\bar{0}}$ basis elements, we may write down the matrices
\begin{equation}
	\mp = \begin{pmatrix} \zero & \zero \\ \ii & \zero \end{pmatrix}, \quad 
	\mh = \begin{pmatrix} 1 & \zero \\ \zero & -1 \end{pmatrix} \quad  \text{and} \quad 
	N_1 = \begin{pmatrix} \zero & \ii \\ -\ii & \zero \end{pmatrix}.
\end{equation}
Thus, super-extensions for which $\rr \neq 0$ exist in this sub-branch.  In both cases, residual $\s_{\bar{1}}$ endomorphisms may be used to set $\hh_3$ and the imaginary part of $\hh_1$ should we choose to include them. 
\\ 
\paragraph{\textbf{Sub-Branch 2.3}} Setting $N_3 \neq 0$, we must now consider 
\begin{equation}
	0 = \pi \vt \mp N_3 \vt^\dagger + \vt N_3 (\pi \vt \mp)^\dagger,
\end{equation}
which, on substituting in $\vt = (0, 1)$, tells us that $c = 0$.  Therefore, as in sub-branch \hyperlink{N2_n-_22}{2.2}, the first condition
of \eqref{eq:N2_n-_22_conditions} tells us 
\begin{equation} \label{eq:N2_n-_23_N1_conditons}
	\begin{split}
		0 &= \hh_1 \rr + \rr \overbar{\pp_3\hh_1\pp_3^{-1}+1} \\
		0 &= \Re(\hh_3\rr) + d ( \Re(\hh_1) +1).
	\end{split}
\end{equation}
However, unlike sub-branch \hyperlink{N2_n-_22i}{2.2}, $\rr$ and $\pp_3$ are not collinear
since the imaginary part of $\pp_3\rr$ makes up the only non-vanishing component of $N_3$:
\begin{equation}
	N_3 = \begin{pmatrix}
		\zero & \zero \\ \zero & \bar{\rr}\bar{\pp_3} - \pp_3 \rr 
	\end{pmatrix}.
\end{equation}
Substituting this $N_3$ into its condition from the $[\bH, \Q, \Q]$ identity, we find
\begin{equation}
	(1 - 2 \Re(\hh_4)) \Im(\ll) = [ \Im(\hh_4), \Im(\ll)] ,
\end{equation}
where $\hh_4 = \pp_3 \hh_1 \pp_3^{-1} + 1$ and $\ll = \bar{\rr}\bar{\pp_3} - \pp_3 \rr $.  Since $\Im(\ll)$ is
perpendicular to $[ \Im(\hh_4), \Im(\ll)]$, both sides of this expression must vanish separately.  Substituting 
$\hh_4$ and $\ll$ into the above expressions, we find
\begin{equation}
	0 = (1 + 2 \Re(\hh_1) ) \Im(\pp_3 \rr) \quad \text{and} \quad 0 = [\hh_1, \rr\pp_3].
\end{equation}
As stated above, $\rr$ and $\pp_3$ are not collinear; therefore, the first constraint here tells us that
\begin{equation}
	2 \Re(\hh_1) = - 1. 
\end{equation}
Substituting this result into the second constraint in \eqref{eq:N2_n-_23_N1_conditons}, we find
\begin{equation}
	2 \Re(\hh_3\rr) = -d. 
\end{equation}
Putting all these results together, the constraints are
\begin{equation}
	\mathcal{C}_{\hat{\n}_-,\,\mathsf{2.3.i}} = \{ 2 \Re(\hh_1) = - 1, \quad 2 \Re(\hh_3 \rr) = -d, \quad 0 = [\hh_1, \rr\pp_3] \},
\end{equation}
for the non-vanishing \hypertarget{N2_n-_23i}{matrices}
\begin{equation}
	\mathcal{M}_{\hat{\n}_-,\,\mathsf{2.3.i}} = \Big\{ \mp = \begin{pmatrix} \zero & \zero \\ \pp_3 & \zero \end{pmatrix}, \quad 
	\mh = \begin{pmatrix} \hh_1 & \zero \\ \hh_3 & \pp_3 \hh_1 \pp_3^{-1} + 1 \end{pmatrix}, \quad
	N_1 = \begin{pmatrix} \zero & \rr \\ \bar{\rr} & d \end{pmatrix}, \quad
	N_3 = \begin{pmatrix} \zero & \zero \\ \zero &  \bar{\rr}\bar{\pp_3} - \pp_3 \rr \end{pmatrix}	 \Big\}.		
\end{equation}
Notice, the sub-branch in $\cS$ describing these super-extensions of $\hat{\n}_-$ is parameterised by four quaternions $\pp_3$, $\hh_1$, $\hh_3$ and $\rr$, and one real scalar $d$.  Wanting to establish the existence of super-extensions in this sub-branch, we can choose to set $\hh_3$, $d$, and the imaginary part of $\hh_1$ to zero.  Then, utilising the endomorphisms of $\s_{\bar{1}}$, we can impose that $\pp_3$ must lie along $\ii$ and that $\rr$ must have unit norm.  Subsequently employing $\Aut(\mathbb{H})$ to fix $\rr$, we can finally scale $\sH$, $\sZ$, $\sP$, and $\sB$ to get the super-extension
\begin{equation}
	\mp = \begin{pmatrix} \zero & \zero \\ \ii & \zero \end{pmatrix}, \quad 
	\mh = \begin{pmatrix} 1 & \zero \\ \zero & -1 \end{pmatrix}, \quad
	N_1 = \begin{pmatrix} \zero & 1 + \jj \\ 1 - \jj & \zero \end{pmatrix}, \quad
	N_3 = \begin{pmatrix} \zero & \zero \\ \zero &  \kk \end{pmatrix}.
\end{equation}
Having established that this sub-branch is not empty, we may look to introduce the components we have set to zero for this example.  Notably, we may introduce the imaginary part of $\hh_1$ while still fixing all parameters using the basis transformations $\G \subset \GL(\s_{\bar{0}}) \times \GL(\s_{\bar{1}})$.  However, the inclusion of either $\hh_3$ or $d$ will introduce parameters that cannot be fixed.
\\
\paragraph{$\hat{\n}_+$ and $\hat{\g}$}~\\ \\
Substituting $\lambda = \varepsilon = 0$, $\mu = \pm 1$ into the conditions of \eqref{eq:N2_C_branch2_constraints},\footnote{Whether 
we are in the $\hat{\n}_+$ or $\hat{\g}$ case makes no difference: the distinction between
the two is the value of $\eta$, which, if non-vanishing, would add the condition 
\begin{equation}
	0 = \Re(\vt N_0\vt^\dagger) \pi.
\end{equation} 
This condition sets $N_0 = 0$, but we already have this result from another condition.
Therefore, the super-extensions are the same for both of these generalised Bargmann algebras.}
we instantly
have $N_3 = 0$ and
\begin{equation}
	\begin{split}
		0 &= \mh N_i + N_i \mh^\dagger \quad \text{where} \quad i \in \{0, 1\} \\
		0 &= \mp N_i - N_i \mp^\dagger \quad \text{where} \quad i \in \{0, 1\} \\
		0 &= \Re(\vt N_0\vt^\dagger) \vt \mh \\
		0 &= \pm \Re(\vt N_0\vt^\dagger) \beta. \\
	\end{split}
\end{equation}
The final condition here states that $N_0 = 0$; therefore, $N_1$ is the only possible 
non-vanishing matrix of those encoding $[\Q, \Q]$.  This result tells us there will be no
sub-branch 2.1 or 2.3 for these algebras and no sub-branch satisfying case $(\mathsf{ii})$, in which $N_0 \neq 0$.  Therefore,
the conditions reduce to 
\begin{equation} \label{eq:N2_C_branch2_ng_conditions}
		0 = \mh N_1 + N_1\mh^\dagger \quad \text{and} \quad 
		0 = \mp N_1 - N_1 \mp^\dagger.
\end{equation}
Under the assumption that $\pp_3 \neq 0$ for this branch of super-extensions,
the latter condition tells us that $c = 0$ and that $\pp_3$ and $\rr$ are collinear:
\begin{equation}
	0 = \bar{\rr} \bar{\pp_3} - \pp_3 \rr.
\end{equation}
Substituting these results into the first condition, we find
\begin{equation}
	0 = \Re(\hh_1), \quad 0 = [\hh_1, \rr\pp_3] \quad \text{and} \quad 0 = \Re(\hh_3\rr).
\end{equation}
Notice that, since $\pp_3$ and $\rr$ are collinear, the second constraint is instantly
satisfied.  Thus, our constraints reduce to 
\begin{equation}
	\mathcal{C}_{\hat{\n}_+ \,\text{and}\,\hat{\g}, \, \mathsf{2.2.i}} = \{ 0 = \Re(\hh_1), \quad 0 = \Re(\hh_3\rr), \quad 0 = \bar{\rr} \bar{\pp_3} - \pp_3 \rr \}.
\end{equation}
The non-vanishing \hypertarget{N2_ng_22i}{matrices} in this instance are
\begin{equation}
	\mathcal{M}_{\hat{\n}_+\,\text{and}\,\hat{\g},\,\mathsf{2.2.i}} = \Big\{ \mp = \begin{pmatrix} \zero & \zero \\ \pp_3 & \zero \end{pmatrix}, \quad 
	\mh = \begin{pmatrix} \hh_1 & \zero \\ \hh_3 & \pp_3 \hh_1 \pp_3^{-1} \end{pmatrix}, \quad
	N_1 = \begin{pmatrix} \zero & \rr \\ \bar{\rr} & d \end{pmatrix} \Big\}.	
\end{equation}
This sub-branch has identical $(\mathcal{M}, \mathcal{C})$ to sub-branch \hyperlink{N2_a_22i}{$\mathsf{2.2.i}$} for $\hat{\a}$.  Therefore, for a discussion on the existence of such super-extensions, we refer the reader to the discussion found there.
\subsubsection{Branch 3} \label{subsubsec:N2_C_branch3}
\begin{equation}
	\mp = \mz = 0 \quad \mb = \begin{pmatrix} \zero & \zero \\ \bb_3 & \zero \end{pmatrix} \quad 
	\mh = \begin{pmatrix} \hh_1 & \zero \\ \hh_3 & \bb_3 \hh_1 \bb_3^{-1} - \lambda \end{pmatrix}.
\end{equation}
Exploiting the vanishing of $\mz$ and $\mp$, we can reduce the conditions from Lemmas~\ref{lem:N2_011}
and~\ref{lem:N2_111}.  In particular, the vanishing of $\mp$, when substituted into the conditions from the $[\P, \Q, \Q]$ super-Jacobi identity, tells us that $N_3 = 0$ and
\begin{equation}
	\begin{split}
		0 &= \eta \Re(\vt N_0\vt^\dagger) \pi \\
		0 &= \varepsilon \Re(\vt N_0 \vt^\dagger) \pi + \tfrac12 [\pi, \vt N_2 \vt^\dagger]. \\		
	\end{split}
\end{equation}
The $[\B, \Q, \Q]$ identity then produce
\begin{equation}
	\begin{split}
		0 &= \mb N_0 - N_0 \mb^\dagger \\	
		N_4 &= \mb N_1 - N_1 \mb^\dagger \\
		0 &= \beta \vt \mb N_2 \vt^\dagger + \vt N_2 (\beta \vt \mb)^\dagger \\
		0 &= \lambda \Re(\vt N_0 \vt^\dagger) \beta + \tfrac12 [\beta, \vt N_2 \vt^\dagger] \\
		\mu \Re(\vt N_0\vt^\dagger) \beta &= \beta \vt \mb N_4 \vt^\dagger + \vt N_4 (\beta \vt \mb)^\dagger. \\
	\end{split}
\end{equation}
The conditions from the $[\bZ, \Q, \Q]$ identity are satisfied since $\mz=0$, and, lastly, the $[\bH, \Q, \Q]$ super-Jacobi identity
produces
\begin{equation}
	\begin{split}
		0 &= \mh N_i + N_i \mh^\dagger \quad \text{where} \quad i \in \{0, 1, 2\} \\
		0 &= \eta N_4\\
		\varepsilon N_4 &= \mh N_4 + N_4 \mh^\dagger.
	\end{split}
\end{equation}	
From Lemma~\ref{lem:N2_111}, we get
\begin{equation}
		\Re(\vt N_0\vt^\dagger) \vt \mh  = \tfrac12 \vt N_2 \vt^\dagger \vt.
\end{equation}
As in both previous branches, the conditions 
\begin{equation}
	\begin{split}
		0 &= \lambda \Re(\vt N_0 \vt^\dagger) \beta + \tfrac12 [\beta, \vt N_2 \vt^\dagger] \\
		0 &= \varepsilon \Re(\vt N_0 \vt^\dagger) \pi + \tfrac12 [\pi, \vt N_2 \vt^\dagger], \\
	\end{split}
\end{equation}
tell us $N_2 = 0$, such that, putting everything together, we have
\begin{equation} \label{eq:N2_C_branch3_constraints}
	\begin{split}
		0 &= \eta N_4\\
		0 &= \mh N_i + N_i \mh^\dagger \quad \text{where} \quad i \in \{0, 1\} \\
		0 &= \mb N_0 - N_0 \mb^\dagger \\
		0 &= \eta \Re(\vt N_0\vt^\dagger) \pi \\
		0 &= \lambda \Re(\vt N_0 \vt^\dagger) \beta \\
		0 &= \varepsilon \Re(\vt N_0 \vt^\dagger) \pi \\
		0 &= \Re(\vt N_0\vt^\dagger) \vt \mh
	\end{split}
	\begin{split}
		\varepsilon N_4 &= \mh N_4 + N_4 \mh^\dagger. \\
		N_4 &= \mb N_1 - N_1 \mb^\dagger \\
		\mu \Re(\vt N_0\vt^\dagger) \beta &= \beta \vt \mb N_4 \vt^\dagger + \vt N_4 (\beta \vt \mb)^\dagger \\
	\end{split} \quad \forall \beta, \pi \in \Im(\mathbb{H}), \forall \vt \in \mathbb{H}.
\end{equation}
We can now use some of the conditions common to all generalised Bargmann algebras to identify possible sub-branches with which we can organise our investigations.  Substituting the $N_1$ from \eqref{eq:N2_C_H_Herm_Mats} and the $N_4$ from \eqref{eq:N2_C_H_SkewHerm_Mats} into the condition
\begin{equation}
	N_4 = \mb N_1 - N_1 \mb^\dagger ,
\end{equation} 
we can write $N_4$ in terms of the parameters in $N_1$ and $\mb$:
\begin{equation} \label{eq:case_3_N4}
	N_4 = \begin{pmatrix} \zero & -c\bar{\bb_3} \\ c\bb_3 & \bb_3 \rr  - \bar{\rr}\bar{\bb_3}\end{pmatrix}.
\end{equation}
Notice that this means $N_4$ is completely dependent on $N_1$: if $N_1 = 0$ then $N_4 = 0$.
Therefore, in general, we have the following sub-branches:
\begin{enumerate}
	\item $N_1 = 0$ and $N_4 = 0$,
	\item $N_1 \neq 0$ and $N_4 = 0$,
	\item $N_1 \neq 0$ and $N_4 \neq 0$.
\end{enumerate}
Also, as in Branches 1 and 2, the condition derived from the $[\Q, \Q, \Q]$ identity tells us that either $N_0$ or $\mh$ vanishes.  We will consider both of these cases within each sub-branch, identifying them as
\begin{enumerate}[label=(\roman*)]
	\item $N_0 = 0$ and $\mh \neq 0$, and
	\item  $N_0 \neq 0$ and $\mh = 0$.
\end{enumerate}
\paragraph{$\hat{\a}$}~\\ \\
Setting $\lambda = \mu = \eta =\varepsilon = 0$, the conditions in \eqref{eq:N2_C_branch3_constraints}
reduce to 
\begin{equation} \label{eq:N2_C_branch3_a_constraints}
	\begin{split}
		0 &= \mh N_i + N_i \mh^\dagger \quad \text{where} \quad i \in \{0, 1, 4 \} \\
		0 &= \mb N_0 - N_0 \mb^\dagger \\
		0 &= \beta \vt \mb N_4 \vt^\dagger + \vt N_4 (\beta \vt \mb)^\dagger \\
		0 &= \Re(\vt N_0\vt^\dagger) \vt \mh \\
		N_4 &= \mb N_1 - N_1 \mb^\dagger. \\
	\end{split}
\end{equation}
As in Branch 2, none of these conditions force the vanishing of any more $N_i$; therefore,
super-extensions may be found in each of the sub-branches.  In fact, because of the symmetry
of the generators $\sB$ and $\sP$ in this generalised Bargmann algebra, we may use
automorphisms to transform the above conditions into those
in \eqref{eq:N2_C_branch2_a_conditions1} and \eqref{eq:N2_C_branch2_a_conditions2}, which describe
the super-extensions of $\hat{\a}$ in Branch 2.  More explicitly, substitute the transformation with matrices
\begin{equation}
	A = \begin{pmatrix}
		1 & \zero \\ \zero & 1
	\end{pmatrix}, \quad 
	C = \begin{pmatrix}
		\zero & -1 \\ 1 & \zero
	\end{pmatrix}, \quad \text{and} \quad
	M = \begin{pmatrix}
		1 & \zero \\ \zero & 1
	\end{pmatrix},
\end{equation}
and the quaternion $\uu = 1$, into \eqref{eq:N2_S_basis_transformations}.  Putting the transformed matrices
into the conditions of \eqref{eq:N2_C_branch3_a_constraints}, we recover the conditions of 
\eqref{eq:N2_C_branch2_a_conditions1} and \eqref{eq:N2_C_branch2_a_conditions2}.
Therefore, all the super-extensions of $\hat{\a}$ in this branch are equivalent to the super-extensions
of Branch 2.  Thus, for this particular generalised Bargmann algebra, this branch produces no
new super-extensions.
\\
\paragraph{$\hat{\n}_-$}~\\ \\
Setting $\mu = \eta = 0$, $\lambda = 1$ and $\varepsilon = -1$, the conditions of \eqref{eq:N2_C_branch3_constraints} become
\begin{equation} \label{eq:case_3_constraints}
	\begin{split}
		0 &= \mh N_i + N_i \mh^\dagger \quad \text{where} \quad i \in \{0, 1\} \\
		0 &= \mb N_0 - N_0 \mb^\dagger \\
		0 &= \Re(\vt N_0 \vt^\dagger) \beta \\
		0 &= - \Re(\vt N_0 \vt^\dagger) \pi \\
		0 &= \Re(\vt N_0\vt^\dagger) \vt \mh
	\end{split}
	\begin{split}
		- N_4 &= \mh N_4 + N_4 \mh^\dagger. \\
		N_4 &= \mb N_1 - N_1 \mb^\dagger \\
		0 &= \beta \vt \mb N_4 \vt^\dagger + \vt N_4 (\beta \vt \mb)^\dagger. \\
	\end{split}
\end{equation}
The conditions 
\begin{equation}
		0 = \Re(\vt N_0 \vt^\dagger) \beta \quad \text{and} \quad 0 = - \Re(\vt N_0 \vt^\dagger) \pi
\end{equation}
tell us that $N_0$ must vanish, leaving only
\begin{equation} \label{eq:N2_C_n-_constraints}
	\begin{split}
		0 &= \mh N_1 + N_1 \mh^\dagger \\
		0 &= \beta \vt \mb N_4 \vt^\dagger + \vt N_4 (\beta \vt \mb)^\dagger
	\end{split}\quad 
	\begin{split}
		- N_4 &= \mh N_4 + N_4 \mh^\dagger \\
		N_4 &= \mb N_1 - N_1 \mb^\dagger. \\
	\end{split}
\end{equation}
Notice, this result tells us that we cannot have any sub-branches satisfying case $(\mathsf{ii})$; therefore, all sub-branches $(\mathcal{M}, \mathcal{C})$ discussed below will have a subscript ending in $\mathsf{i}$.  Like the $\hat{\a}$ case,
this generalised Bargmann algebra allows for an automorphism which transforms the conditions for this branch into
the conditions for Branch 2.  However, in this instance, this branch will produce some distinct super-extensions.  This result
is a consequence of the parameters $\varepsilon$ and $\lambda$ and their appearance in $\mh$.  In Branch 2, the 
matrix $\mh$ is written as
\begin{equation}
	\mh = \begin{pmatrix}
		\hh_1 & \zero \\ \hh_3 & \pp_3\hh_1\pp_3^{-1} - \varepsilon
	\end{pmatrix},
\end{equation}
and in this branch, it is written
\begin{equation}
	\mh = \begin{pmatrix}
		\hh_1 & \zero \\ \hh_3 & \pp_3\hh_1\pp_3^{-1} - \lambda
	\end{pmatrix}.
\end{equation}
Since $\hat{\n}_-$ has $\varepsilon = -1$ and $\lambda = 1$, this matrix differs in these branches, if only
be a sign.  Thus, although the investigations into the super-extensions of $\hat{\n}_-$ in this branch will be very similar to 
those in the previous branch, we will give a partial presentation of them here to demonstrate any consequences of this
change in sign.  In particular, we will omit the discussions on the existence of super-extensions and parameter fixing
as these require only trivial adjustments from the discussions found in Branch 2.
\\
\paragraph{\textbf{Sub-Branch 3.1}} As $N_0 = 0$, we cannot have both $N_1$ and $N_4$ vanish; therefore,
there is no super-extension in this sub-branch.
\\
\paragraph{\textbf{Sub-Branch 3.2}} Letting $N_1 \neq 0$ and $N_4 = 0$, we are left with only the conditions
\begin{equation}
		0 = \mh N_1 + N_1 \mh^\dagger \quad \text{and} \quad 
		0 = \mb N_1 - N_1 \mb^\dagger.
\end{equation}
The second condition above tells us that 
\begin{equation}
	0 = c\bb_3 \quad \text{and} \quad 0 = \bb_3 \rr  - \bar{\rr}\bar{\bb_3}.
\end{equation}
As $\bb_3 \neq 0$ by assumption, $c = 0$.  Substituting this result into the first
condition above, we find 
\begin{equation}
	\begin{split}
		0 &= \hh_1 \rr  + \rr \overbar{\bb_3\hh_1\bb_3^{-1} - 1} \\
		0 &= \Re(\hh_3\rr) + d ( \Re(\hh_1) - 1 ).
	\end{split}
\end{equation}
Using the collinearity of $\bb_3$ and $\rr$, the first of these constraints tells us that
\begin{equation}
	0 = (2 \Re(\hh_1) - 1) \Re(\bb_3\rr).
\end{equation}
Therefore, the constraints in this instance are given by
\begin{equation}
	\mathcal{C}_{\hat{\n}_-,\,\mathsf{3.2.i}} = \{ 0 = \bb_3 \rr  - \bar{\rr}\bar{\bb_3}, \quad
	0 = (2 \Re(\hh_1) - 1) \Re(\bb_3\rr), \quad 0 = \Re(\hh_3\rr) + d ( \Re(\hh_1) - 1 ) \}.
\end{equation}
The non-vanishing \hypertarget{N2_n-_32i}{matrices} in this instance are
\begin{equation}
	\mathcal{M}_{\hat{\n}_-,\,\mathsf{3.2.i}} = \Big\{ \mb = \begin{pmatrix} \zero & \zero \\ \bb_3 & \zero \end{pmatrix}, \quad 
	\mh = \begin{pmatrix} \hh_1 & \zero \\ \hh_3 & \bb_3\hh_1\bb_3^{-1} -1 \end{pmatrix}, \quad
	N_1 = \begin{pmatrix} \zero & \rr \\ \bar{\rr} & d \end{pmatrix} \Big\}.
\end{equation}
This sub-branch of $\cS$ is parameterised by two collinear quaternions $\bb_3$ and $\rr$, two quaternions encoding the action of $\sH$ on $\s_{\bar{1}}$, $\hh_1$ and $\hh_3$, and one real scalar $d$.  Notice that the real component of $\hh_1$ varies depending on whether $\rr$ vanishes.  Together with the super-extensions in Sub-Branch \hyperlink{N2_n-_22i}{$\mathsf{2.2.i}$} for $\hat{\n}_-$, these are the only super-extensions that demonstrate this type of dependency.   If $\rr = 0$, the first two constraints of $\mathcal{C}_{\hat{\n}_-,\,\mathsf{3.2.i}}$ are trivial, and the third condition tell us that $\Re(\hh_1) = 1$, since $d \neq 0$ for $N_1 \neq 0$.  However, if $\rr \neq 0$, the second constraint requires $2 \Re(\hh_1) = 1$.  In this instance, the third constraint then becomes $ 2 \Re(\hh_3 \rr) = d$.  As the matrices and conditions for this sub-branch are so similar to those in \hyperlink{N2_n-_22i}{$\mathsf{2.2.i}$}, we refer the reader to the discussion on existence of super-extensions and parameter fixing presented there.  
\\
\paragraph{\textbf{Sub-Branch 3.3}} Finally, let $N_1 \neq 0$ and $N_4 \neq 0$.  
The condition 
\begin{equation}
	0 = \beta \vt \mb N_4 \vt^\dagger + \vt N_4 (\beta \vt \mb)^\dagger
\end{equation}
 imposes $c = 0$, such that
\begin{equation} \label{eq:case3_n-_Ns}
	N_1 = \begin{pmatrix} \zero & \rr \\ \bar{\rr} & d \end{pmatrix} \quad \text{and} \quad
	N_4 = \begin{pmatrix} \zero & \zero \\ \zero & \bb_3\rr - \bar{\rr} \bar{\bb_3}  \end{pmatrix}.
\end{equation}
This result reduces the conditions in \eqref{eq:N2_C_n-_constraints} to
\begin{equation} \label{eq:N2_C_n-_subbranch33_constraints}
	\begin{split}
		0 &= \mh N_1 + N_1 \mh^\dagger \\
		- N_4 &= \mh N_4 + N_4 \mh^\dagger. \\
	\end{split}
\end{equation}
Substituting the $N_4$ from \eqref{eq:case3_n-_Ns} into the second condition above, we have
\begin{equation}
	- \ll = \hh_4 \ll +\ll \bar{\hh_4},
\end{equation}
where $\ll = \bb_3 \rr-\bar{\rr}\bar{\bb_3}$ and $\hh_4 = \bb_3\hh_1\bb_3^{-1} - 1$.  We can rewrite
this condition as
\begin{equation}
	(1+2 \Re(\hh_4) ) \ll = [\ll, \hh_4].
\end{equation}
Notice that the R.H.S. of this expression must lie in $\Im(\mathbb{H})$ and be orthogonal
to $\ll$, which is imaginary by construction.  Therefore, both sides of this expression must vanish independently:
\begin{equation}
	0 = (1 + 2 \Re(\hh_4)) \ll \qquad 0 = [\ll, \hh_4].
\end{equation}
Substituting $\ll$ and $\hh_4$ into these constraints, we find 
\begin{equation} \label{eq:case3_n-_N4_condition}
	0 = (2 \Re(\hh_1) - 1) (\bb_3 \rr - \bar{\rr}\bar{\bb_3}) \quad \text{and} \quad 0 = [\hh_1, \rr\bb_3],
\end{equation}
respectively.  For $N_4$ to not vanish,
we must have $\Im(\bb_3\rr) \neq 0$, so, by the first constraint above, we need $2 \Re(\hh_1) = 1$.
The first condition in \eqref{eq:N2_C_n-_subbranch33_constraints} produces the same constraints as
in Sub-Branch \hyperlink{N2_n-_32i}{$\mathsf{3.2}$}; namely,
\begin{equation}
	\begin{split}
		0 &= \hh_1 \rr  + \rr \overbar{\bb_3\hh_1\bb_3^{-1} - 1} \\
		0 &= \Re(\hh_3\rr) + d ( \Re(\hh_1) - 1 ).
	\end{split}
\end{equation}
Notice that the requirement of setting $2 \Re(\hh_1) = 1$ makes the second constraint here $2 \Re(\hh_3\rr) = d$, and
says that the first constraint is equivalent to $0 = [\hh_1, \rr\bb_3]$.  Therefore, the constraints on this sub-branch are given by
\begin{equation}
		\mathcal{C}_{\hat{\n}_-,\,\mathsf{3.3.i}} = \{ d = 2 \Re(\hh_3\rr), \quad 1 = 2 \Re(\hh_1) \quad \text{and} \quad 0 = [\hh_1, \rr\bb_3] \},
\end{equation}
and the non-vanishing \hypertarget{N2_n-_33i}{matrices} are
\begin{equation}
	\mathcal{M}_{\hat{\n}_-,\,\mathsf{3.3.i}} = \Big\{ \mb = \begin{pmatrix} \zero & \zero \\ \bb_3 & \zero \end{pmatrix}, \quad 
	\mh = \begin{pmatrix} \hh_1 & \zero \\ \hh_3 & \bb_3\hh_1\bb_3^{-1} -1 \end{pmatrix}, \quad
	N_1 = \begin{pmatrix} \zero & \rr \\ \bar{\rr} & d \end{pmatrix}, \quad
	N_4 = \begin{pmatrix} \zero & \zero \\ \zero & \bb_3\rr - \bar{\rr} \bar{\bb_3} \end{pmatrix}	\Big\}.
\end{equation}
For the discussion on existence of super-extensions and how to fix the parameters of the matrices describing this sub-branch of $\cS$, we refer the reader to Sub-Branch \hyperlink{N2_n-_23i}{$\mathsf{2.3.i}$}.  The application of the discussion to the present case requires only minor adjustments.
\\
\paragraph{$\hat{\n}_+$}~\\ \\
Substituting $\lambda = \varepsilon = 0$, $\mu = 1$, and $\eta = -1$ into the results for
Lemmas~\ref{lem:N2_011} and~\ref{lem:N2_111}, we find
\begin{equation}
	\begin{split}
		0 &= - N_4\\
		0 &= \mh N_i + N_i \mh^\dagger \quad \text{where} \quad i \in \{0, 1, 4\} \\
		0 &= \mb N_0 - N_0 \mb^\dagger \\
		0 &= - \Re(\vt N_0\vt^\dagger) \pi \\
		0 &= \Re(\vt N_0\vt^\dagger) \vt \mh
	\end{split}
	\begin{split}
		N_4 &= \mb N_1 - N_1 \mb^\dagger \\
		\Re(\vt N_0\vt^\dagger) \beta &= \beta \vt \mb N_4 \vt^\dagger + \vt N_4 (\beta \vt \mb)^\dagger. \\
	\end{split}
\end{equation}
Therefore, $N_4$ vanishes, and $N_0$ vanishes by $0 = - \Re(\vt N_0\vt^\dagger) \pi$.
This leaves us with
\begin{equation}
		0 = \mh N_1 + N_1 \mh^\dagger \quad \text{and} \quad 
		0 = \mb N_1 - N_1 \mb^\dagger.
\end{equation}
Notice that these conditions are similar to those of \eqref{eq:N2_C_branch2_ng_conditions}, which describe
the super-extensions of $\hat{\n}_+$ in Branch 2.  In fact, we can utilise the automorphisms of $\hat{\n}_+$
to transform the above conditions into those in \eqref{eq:N2_C_branch2_ng_conditions}.  Unlike the $\hat{\n}_-$
case, since $\hat{\n}_+$ has vanishing $\varepsilon$ and $\lambda$, there is no discrepancy between
the transformed matrices and those of Branch 2; therefore, the super-extensions of $\hat{\n}_+$ in Branches 2 and 3
are equivalent.  Thus, we have no new super-extensions here.
\\
\paragraph{$\hat{\g}$}~\\ \\
Substituting $\lambda = \eta = \varepsilon = 0$ and $\mu = -1$ into \eqref{eq:N2_C_branch3_constraints},
we have
\begin{equation} \label{eq:N2_C_branch3_g_constraints}
	\begin{split}
		0 &= \mh N_i + N_i \mh^\dagger \quad \text{where} \quad i \in \{0, 1, 4\} \\
		0 &= \mb N_0 - N_0 \mb^\dagger \\
		0 &= \Re(\vt N_0\vt^\dagger) \vt \mh
	\end{split}
	\begin{split}
		N_4 &= \mb N_1 - N_1 \mb^\dagger \\
		- \Re(\vt N_0\vt^\dagger) \beta &= \beta \vt \mb N_4 \vt^\dagger + \vt N_4 (\beta \vt \mb)^\dagger. \\
	\end{split}
\end{equation}
With these conditions, we can now investigate the three sub-branches.
\\
\paragraph{\textbf{Sub-Branch 3.1}} We cannot have $N_1 = N_4 = 0$, since the vanishing
of $N_4$ means $N_0 =0$ through
\begin{equation}
	- \Re(\vt N_0\vt^\dagger) \beta = \beta \vt \mb N_4 \vt^\dagger + \vt N_4 (\beta \vt \mb)^\dagger.
\end{equation}
This would cause all $N_i$ to vanish such that $[\Q, \Q] = 0$. Therefore, 
there is no super-extension in this sub-branch.
\\
\paragraph{\textbf{Sub-Branch 3.2}} With only $N_1 \neq 0$, the conditions reduce to
\begin{equation}
		0 = \mh N_1 + N_1 \mh^\dagger \quad \text{and} \quad
		0 = \mb N_1 - N_1 \mb^\dagger.
\end{equation}
Notice that this is the same set of conditions as the $\hat{\n}_+$ case above.  Therefore,
we may expect the analysis for this generalised Bargmann algebra to be analogous. However,
there is a very important distinction.  In the $\hat{\n}_+$ case, we were able to use the automorphisms
to transform the conditions into those of Sub-Branch 2.2.  This automorphism is not permitted by
the generalised Bargmann algebra $\hat{\g}$.  Therefore, although the analysis will be the same
\textit{mutatis mutandis} as that of Sub-Branch 2.2, the resulting super-extensions will be distinct.
\\ \\
Now, since $N_1$ is the only possible non-vanishing matrix in the $[\Q, \Q]$ bracket, it must have non-zero
components.  The latter condition above tells us that $c=0$ and $\bb_3$ and $\rr$
are collinear quaternions, while the former condition imposes
\begin{equation}
	\begin{split}
		0 &= \hh_1 \rr + \rr \overbar{\bb_3\hh_1\bb_3^{-1}} \\
		0 &= \Re(\hh_3\rr) + d \Re(\hh_1).
	\end{split}
\end{equation}
Notice that if $\rr = 0$, we need $d \neq 0$ for the existence of a super-extension; therefore, the
final constraint above would impose $\Re(\hh_1) = 0$.  Similarly, if $\rr \neq 0$, the first constraint
would also enforce $\Re(\hh_1) = 0$.  Thus, in all super-extensions, we require $\Re(\hh_1) = 0$.
Using this result, these two constraints simplify to 
\begin{equation}
		0 = [\hh_1, \rr\bb_3] \quad \text{and} \quad 0 = \Re(\hh_3\rr).
\end{equation}
However, since $\bb_3$ and $\rr$ are collinear and it is only the imaginary part
of $\rr\bb_3$ that will contribute to $[\hh_1, \rr\bb_3]$, the first of these constraints
is already satisfied.  Therefore, the final set of constraints on this sub-branch is 
\begin{equation}
	\mathcal{C}_{\hat{\g},\,\mathsf{3.2.i}} = \{ 0 = \bb_3\rr - \bar{\rr} \bar{\bb_3}, \quad 0 = \Re(\hh_1), 
		\quad 0 = \Re(\hh_3\rr) \}.
\end{equation}
Subject to these constraints, we have non-vanishing \hypertarget{N2_g_32i}{matrices} are 
\begin{equation}
	\mathcal{M}_{\hat{\g},\,\mathsf{3.2.i}} = \Big\{ \mb = \begin{pmatrix} \zero & \zero \\ \bb_3 & \zero \end{pmatrix}, \quad 
	\mh = \begin{pmatrix} \hh_1 & \zero \\ \hh_3 & \bb_3 \hh_1 \bb_3^{-1} \end{pmatrix}, \quad
	N_1 = \begin{pmatrix}  \zero & \rr \\ \bar{\rr} & d \end{pmatrix} \Big\}.
\end{equation}
Since this $(\mathcal{M}, \mathcal{C})$ is analogous to the one found in Branch \hyperlink{N2_ng_22i}{$\mathsf{2}$} 
for $\hat{\n}_+$ and $\hat{\g}$, we will omit the discussion on existence of super-extensions and
parameter fixing.  
\\
\paragraph{\textbf{Sub-Branch 3.3}} Finally, with $N_4 \neq 0$, we can think of setting $N_0 \neq 0$
and $\mh = 0$.  But first, try setting $N_0 = 0$ to allow $\mh \neq 0$.  The conditions in \eqref{eq:N2_C_branch3_g_constraints} become
\begin{equation}
	\begin{split}
		0 &= \mh N_i + N_i \mh^\dagger \quad \text{where} \quad i \in \{1, 4\} \\
		N_4 &= \mb N_1 - N_1 \mb^\dagger \\
		0 &= \beta \vt \mb N_4 \vt^\dagger + \vt N_4 (\beta \vt \mb)^\dagger. \\
	\end{split}
\end{equation}
Notice that the second condition above allows us to write $N_4$ in terms of $\mb$ and $N_1$:
\begin{equation}
	N_4 = \begin{pmatrix} \zero & -c\bar{\bb_3} \\ c\bb_3 & \bb_3 \rr  - \bar{\rr}\bar{\bb_3}\end{pmatrix}.
\end{equation}
The third condition then imposes $c = 0$, since $\bb_3 \neq 0$, leaving us with
\begin{equation}
	N_1 = \begin{pmatrix}
		\zero & \rr \\ \bar{\rr} & d
	\end{pmatrix} \quad \text{and} \quad 
	N_4 = \begin{pmatrix}
		\zero & \zero \\ \zero & \bb_3 \rr - \bar{\rr} \bar{\bb_3}
	\end{pmatrix}.
\end{equation}
Using these matrices in the final conditions,
\begin{equation}
			0 = \mh N_i + N_i \mh^\dagger \quad \text{where} \quad i \in \{1, 4\},
\end{equation}
produces 
\begin{equation}
 	0 = [\hh_1, \rr\bb_3],
\end{equation}
when $i =4$, and, when $i = 1$, we obtain
\begin{equation}
	\begin{split}
		0 &= \hh_1 \rr + \rr \overbar{\bb_3\hh_1\bb_3^{-1}} \\
		0 &= \Re(\hh_3\rr) + d \Re(\hh_1).
	\end{split}
\end{equation}
Since $\rr \neq 0$ for $N_4 \neq 0$, the first condition here states that $\Re(\hh_1) = 0$.
Therefore, the constraints on the parameters of this super-extension are given by
\begin{equation}
	\mathcal{C}_{\hat{\g},\,\mathsf{3.3.i}} = \{ 0 = \Re(\hh_1), \quad  0 = [\hh_1, \rr\bb_3], \quad
	 	0 = \Re(\hh_3\rr) \}.
\end{equation}
The \hypertarget{N2_g_33i}{non-vanishing} matrices associated with this sub-branch are
\begin{equation}
	\mathcal{M}_{\hat{\g},\,\mathsf{3.3.i}} = \Big\{ \mb = \begin{pmatrix}
		\zero & \zero \\ \bb_3 & \zero
	\end{pmatrix}, \quad
	\mh = \begin{pmatrix}
		\hh_1 & \zero \\ \hh_3 & \bb_3 \hh_1\bb_3^{-1}
	\end{pmatrix}, \quad
	N_1 = \begin{pmatrix}
		\zero & \rr \\ \bar{\rr} & d
	\end{pmatrix}, \quad
	N_4 = \begin{pmatrix}
		\zero & \zero \\ \zero & \bb_3 \rr - \bar{\rr} \bar{\bb_3}
	\end{pmatrix} \Big\}.
\end{equation}
The $(\mathcal{M}, \mathcal{C})$ of this sub-branch is the same \textit{mutatis mutandis} as that of Sub-Branch \hyperlink{N2_a_23i}{$\mathsf{2.3.i}$} for $\hat{\a}$; therefore, we refer the reader to the discussion found there on existence of super-extensions and parameter fixing. 
\\ \\
Finally, let $N_0 \neq 0$ such that $\mh = 0$.  The conditions remaining from \eqref{eq:N2_C_branch3_g_constraints} are
\begin{equation}
	\begin{split}
		0 &= \mb N_0 - N_0 \mb^\dagger \\
		N_4 &= \mb N_1 - N_1 \mb^\dagger \\
		- \Re(\vt N_0\vt^\dagger) \beta &= \beta \vt \mb N_4 \vt^\dagger + \vt N_4 (\beta \vt \mb)^\dagger. \\
	\end{split}
\end{equation}
We know how the second condition acts from the discussion at the beginning of this branch.  The first of these conditions tells us
\begin{equation}
	0 = a \quad \text{and} \quad 0 = \bb_3\qq - \bar{\qq}\bar{\bb_3},
\end{equation}
and the third, substituting in $\vt = (0, 1)$, produces
\begin{equation}
	- b = 2 c |\bb_3|^2.
\end{equation}
Now, substituting $\vt = (1, \ss)$ into the third condition, we find
\begin{equation}
	- 2 \Re(\ss\bar{\qq}) - b |\ss|^2 = 2c |\ss|^2 |\bb_3|^2.
\end{equation}
Therefore, using the previous result and letting $\ss = 1$, $\ss = \ii$, $\ss = \jj$
and $\ss = \kk$, we see that all components of $\qq$ must vanish.  We thus \hypertarget{N2_g_33ii}{have} non-vanishing matrices
\begin{equation}
	\mathcal{M}_{\hat{\g},\,\mathsf{3.3.ii}} = \Big\{ \mb = \begin{pmatrix}
		\zero & \zero \\ \bb_3 & \zero
	\end{pmatrix}, \quad
	N_0 = \begin{pmatrix}
		\zero & \zero \\ \zero & -2c |\bb_3|^2
	\end{pmatrix}, \quad
	N_1 = \begin{pmatrix}
		c & \rr \\ \bar{\rr} & d
	\end{pmatrix}, \quad
	N_4 = \begin{pmatrix}
		\zero & -c\bar{\bb_3} \\ c\bb_3 & \bb_3 \rr - \bar{\rr} \bar{\bb_3}
	\end{pmatrix} \Big\}.
\end{equation}
Interestingly, there are no additional constraints to the parameters of this sub-branch; therefore, $\mathcal{C}_{\hat{\g},\,\mathsf{3.3.ii}}$ is empty. Notice the sub-branch of $\cS$ for this type of super-extension is parameterised by two quaternions $\bb_3$ and $\rr$, and two real scalars $c$ and $d$.  To demonstrate that this sub-branch is not empty, we begin by setting both $\rr$ and $d$ to zero.  This choice allows us to utilise the endomorphisms of $\s_{\bar{1}}$ to set $\bb_3 = \ii$ and $c=1$.  Employing the scaling symmetry of the basis elements, we arrive at
\begin{equation} \label{eq:notation_example}
	 \mb = \begin{pmatrix}
		\zero & \zero \\ \ii & \zero
	\end{pmatrix}, \quad
	N_0 = \begin{pmatrix}
		\zero & \zero \\ \zero & 1
	\end{pmatrix}, \quad
	N_1 = \begin{pmatrix}
		1 & \zero \\ \zero & \zero
	\end{pmatrix}, \quad
	N_4 = \begin{pmatrix}
		\zero & \ii \\ \ii & \zero
	\end{pmatrix}.
\end{equation}
We may now look to introduce $\rr$ and $d$.  Again, using the endomorphisms of $\s_{\bar{1}}$, we can impose that $\bb_3$ must lie along $\ii$, set $|\rr|^2 = 1$, and choose $\sqrt{2} c =1$.  This choice for $\rr$ imposes that $\rr \in \Sp(1)$, and we may utilise $\Aut(\mathbb{H})$ to fix $\sqrt{2} \rr = 1 + \ii$.  Having chosen $\rr \neq 0$, we can always employ the residual endomorphisms of $\s_{\bar{1}}$ to set $d = 0$.  Using the only remaining symmetry, the scaling of $\sH$, $\sZ$, $\sB$, and $\sP$, we find
\begin{equation}
	 \mb = \begin{pmatrix}
		\zero & \zero \\ \sqrt{2} \ii & \zero
	\end{pmatrix}, \quad
	N_0 = \begin{pmatrix}
		\zero & \zero \\ \zero & 1
	\end{pmatrix}, \quad
	N_1 = \begin{pmatrix}
		1 & 1+\ii \\ 1-\ii & \zero
	\end{pmatrix}, \quad
	N_4 = \begin{pmatrix}
		\zero & \ii \\ \ii & 2\ii
	\end{pmatrix}.
\end{equation}
\subsubsection{Branch 4} \label{subsubsec:N2_C_branch4}
\begin{equation}
	\mz = 0 \quad \mh = \begin{pmatrix} \hh_1 & \zero \\ \hh_3 & \hh_4 \end{pmatrix} \quad 
	\mb = \begin{pmatrix} \zero & \zero \\ \bb_3 & \zero \end{pmatrix} \quad 
	\mp = \begin{pmatrix} \zero & \zero \\ \pp_3 & \zero \end{pmatrix},
\end{equation}
subject to
\begin{equation} \label{eq:N2_branch4_condition}
	[\uu, \hh_1] = - \mu \uu^2 + (\lambda - \varepsilon) \uu + \eta \quad \text{or} \quad [\vv, \hh_1] = \eta \vv^2 + (\lambda - \varepsilon) \vv + \mu,
\end{equation}
where $0 \neq \uu = \bb_3^{-1} \pp_3$ and $0 \neq \vv = \pp_3^{-1} \bb_3$.
Recall, we keep both of these constraints as, depending on the generalised Bargmann algebra under investigation,
one of them will prove more useful than the other.  
We still need to determine the generalised
Bargmann algebras for which this branch could provide a super-extension.  Therefore, we will consider
each algebra in turn, and analyse those for which the above constraints may hold.
\\
\paragraph{$\hat{\a}$}~\\ \\ Setting $\lambda = \mu = \eta = \varepsilon = 0$ in 
\eqref{eq:N2_branch4_condition}, we could still get a super-extension, as long
as we impose
\begin{equation}
	0 = [\uu, \hh_1].
\end{equation}
Throughout this section, we will choose to write parameters in terms of $\bb_3$; therefore, 
we write $\pp_3 = \bb_3\uu$ and $\hh_4 = \bb_3 \hh_1 \bb_3^{-1}$, where
$\uu \in \mathbb{H}$.  Notice that the significance of $\uu$ is only manifest when 
$\hh_1 \neq 0$: when $\hh_1$ vanishes, we are simply replacing $\pp_3$ with $\uu$.  However,
since $\uu$ will be important is several instances, we will always use this notation.
\\ \\
Since neither $\mb$ nor $\mp$ vanish, there are no immediate results as in the three previous
branches: all the conditions of Lemmas~\ref{lem:N2_011} and \ref{lem:N2_111} must be
taken into consideration.  However, as with Branches 2 and 3, we can organise our 
investigations based on dependencies.  In particular, the conditions
\begin{equation}
	\begin{split}
			N_4 &= \mb N_1 - N_1 \mb^\dagger \\
			-N_3 &= \mp N_1 - N_1 \mp^\dagger \\
			\tfrac12 [\beta, \vt N_2 \vt^\dagger] &= \beta \vt \mb N_3 \vt^\dagger
			+ \vt N_3 (\beta \vt \mb)^\dagger \\	
			\tfrac12 [\pi, \vt N_2 \vt^\dagger] &= \pi \vt \mp N_4 \vt^\dagger
			+ \vt N_4 (\pi \vt \mp)^\dagger	,
	\end{split}
\end{equation}
show us that if $N_1$ vanishes, so must $N_2, N_3$, and $N_4$.  Additionally, the vanishing of
either $N_3$ or $N_4$ means we must have $N_2 = 0$.  Therefore, we can divide
our investigations into the following sub-branches.
\begin{enumerate}
	\item $N_1 = N_2 = N_3 = N_4 = 0$
	\item $N_1 \neq 0$ and $N_2 = N_3 = N_4 = 0$
	\item $N_1 \neq 0$, $N_3 \neq 0$, and $N_2 = N_4 = 0$
	\item $N_1 \neq 0$, $N_4 \neq 0$, and $N_2 = N_3 = 0$
	\item $N_1 \neq 0$, $N_3 \neq 0$, $N_4 \neq 0$, and $N_2 = 0$
	\item $N_1 \neq 0$, $N_2 \neq 0$, $N_3 \neq 0$, and $N_4 \neq 0$.
\end{enumerate}
Unlike Branches 1, 2 and 3, the $[\Q, \Q, \Q]$ super-Jacobi identity will not always result in the cases
$(\mathsf{i})$, in which $N_0 = 0$ and $\mh \neq 0$, or $(\mathsf{ii})$, in which $N_0 \neq 0$
and $\mh = 0$.  There are instances in which both $N_0$ and $\mh$ may not vanish.  These
cases, will be labelled $(\mathsf{iii})$.  
\\
\paragraph{\textbf{Sub-Branch 4.1}} With only $N_0$ left available, it cannot vanish for a supersymmetric
extension to exist.
Therefore, the $[\Q, \Q, \Q]$ identity,
\begin{equation}
	\Re(\vt N_0 \vt^\dagger) \vt \mh = 0,
\end{equation}
tells us we must have $\mh = 0$.  The remaining conditions are then
\begin{equation}
		0 = \mb N_0 - N_0 \mb^\dagger \quad \text{and} \quad 
		0 = \mp N_0 - N_0 \mp^\dagger,
\end{equation}
which tell us
\begin{equation}
	0 = a, \quad 0 = \bb_3 \qq - \bar{\qq} \bar{\bb_3} \quad \text{and} \quad
	0 = \bb_3\uu \qq - \bar{\qq} \bar{\uu} \bar{\bb_3}.
\end{equation}
This sub-branch thus has non-vanishing \hypertarget{N2_a_41}{matrices}
\begin{equation}
	\mb = \begin{pmatrix}
		\zero & \zero \\ \bb_3 & \zero
	\end{pmatrix}, \quad
	\mp = \begin{pmatrix}
		\zero & \zero \\ \bb_3\uu & \zero
	\end{pmatrix}, \quad
	N_0 = \begin{pmatrix}
		\zero & \qq \\ \bar{\qq} & b
	\end{pmatrix},
\end{equation}
subject to the constraints 
\begin{equation}
	 0 = \bb_3 \qq - \bar{\qq} \bar{\bb_3}, \quad 0 = \bb_3\uu \qq - \bar{\qq} \bar{\uu}\bar{\bb_3}.
\end{equation}
Notice that these matrices and constraints are very similar to $(\mathcal{M}_{\hat{\a},\,\mathsf{2.1.ii}}, \mathcal{C}_{\hat{\a},\,\mathsf{2.1.ii}})$.  In fact, employing the automorphisms of $\hat{\a}$, we can show that the above system is equivalent to Sub-Branch \hyperlink{N2_a_21ii}{$\mathsf{2.1.ii}$}.  Using the endomorphisms of $\s_{\bar{1}}$ and the constraints above, we can set $\bb_3$, $\bb_3\uu$ and $\qq$ to lie along $\ii$, and set $b = 0$.  In particular, this means that $\uu \in \mathbb{R}$.  Scaling $\sB$, $\sP$, and $\sH$, we find the matrices
\begin{equation}
 	\mb = \begin{pmatrix}
		\zero & \zero \\ \ii & \zero
	\end{pmatrix}, \quad
	\mp = \begin{pmatrix}
		\zero & \zero \\ \ii & \zero
	\end{pmatrix}, \quad
	N_0 = \begin{pmatrix}
		\zero & \ii \\ -\ii & \zero
	\end{pmatrix},
\end{equation}
which under the basis transformation with 
\begin{equation}
	C = \begin{pmatrix}
		1 & -1 \\ \zero & 1
	\end{pmatrix},
\end{equation}
recovers the maximal super-extension of Sub-Branch \hyperlink{N2_a_21ii}{$\mathsf{2.1.ii}$}.  Thus, this sub-branch does not contribute any new super-extensions to $\hat{\a}$.
\\
\paragraph{\textbf{Sub-Branch 4.2}}
The $[\Q, \Q, \Q]$ identity still imposes that either $N_0$ or $\mh$ must vanish in this
sub-branch; however, we can now consider the case where $N_0 = 0$ as we have $N_1 \neq 0$.
First, consider case $(\mathsf{i})$, with $N_0 = 0$ such that $\mh \neq 0$.  The conditions remaining are
\begin{equation}
	\begin{split}
		0 &= \mh N_1 + N_1 \mh^\dagger \\
		0 &= \mb N_1 - N_1 \mb^\dagger \\
		0 &= \mp N_1 - N_1 \mp^\dagger. \\
	\end{split}
\end{equation}
The latter two conditions tell us that $c=0$ and $\bb_3$ is collinear with $\bb_3\uu$ and $\rr$.  Substituting these
results into the first condition, we find
\begin{equation}
	\begin{split}
		0 &= \hh_1 \rr + \rr \overbar{\bb_3 \hh_1 \bb_3^{-1}} \\
		0 &= \Re(\hh_3\rr) + d \Re(\hh_1). \\
	\end{split}
\end{equation}
We know from the analysis of Branch 3 that demanding $N_1 \neq 0$ under these conditions imposes $\Re(\hh_1) = 0$; and, that having the condition
\begin{equation}
		0 = \bb_3 \rr - \bar{\rr} \bar{\bb_3}
\end{equation}
means we always satisfy the imaginary part of 
\begin{equation}
	0 = \hh_1 \rr + \rr \overbar{\bb_3 \hh_1 \bb_3^{-1}}.
\end{equation}
Putting all this together, we find the constraints on this sub-branch to be
\begin{equation}
	0 = \Re(\hh_1), \quad 0 = \Re(\hh_3 \rr) , \quad 0 = \bb_3 \rr - \bar{\rr}\bar{\bb_3}, \quad
		0 = \bb_3 \uu \rr - \bar{\rr}\bar{\uu} \bar{\bb_3}, \quad 0 = [\uu, \hh_1].
\end{equation}
The \hypertarget{N2_a_42i}{non-vanishing} matrices are then 
\begin{equation}
	\mb = \begin{pmatrix}
		\zero & \zero \\ \bb_3 & \zero
	\end{pmatrix}, \quad
	\mp = \begin{pmatrix}
		\zero & \zero \\ \bb_3\uu & \zero
	\end{pmatrix}, \quad 
	\mh = \begin{pmatrix}
		\hh_1 & \zero \\ \hh_3 & \bb_3 \hh_1 \bb_3^{-1}
	\end{pmatrix} \quad \text{and} \quad 
	N_1 = \begin{pmatrix}
		\zero & \rr \\ \bar{\rr} & d
	\end{pmatrix}.
\end{equation}
Notice that $\bb_3$, $\rr$ and $\bb_3\uu$ all being collinear implies that $\uu \in \mathbb{R}$.  Thus the final constraint is satisfied, and, as in Sub-Branch 4.1, we can use the endomorphisms of $\s_{\bar{1}}$ and the automorphisms of $\hat{\a}$ to rotate $\sB$ and $\sP$ such that we only have the matrix $\mp$, in which $\pp_3 = \ii$.  The resulting matrices and constraints are then equivalent to those found in Sub-Branch \hyperlink{N2_a_22i}{$\mathsf{2.2.i}$}, and, therefore, this sub-branch does not produce any new super-extensions for $\hat{\a}$.
\\ \\
Now, considering case $(\mathsf{ii})$, let $\mh = 0$.  The remaining conditions are
\begin{equation}
	\begin{split}
		0 &= \mb N_i - N_i \mb^\dagger \quad \text{where} \quad i \in \{0, 1\} \\
		0 &= \mp N_i - N_i \mp^\dagger \quad \text{where} \quad i \in \{0, 1\}.
	\end{split}
\end{equation}
Therefore, $N_0$ and $N_1$ take the same form in this instance: both $a$ and $c$ vanish, with $\qq$ and $\rr$ being collinear to both $\bb_3$ and $\bb_3\uu$.  In summary, the constraints are
\begin{equation}
	0 = \bb_3 \qq - \bar{\qq} \bar{\bb_3}, \quad 0 = \bb_3\uu \qq - \bar{\qq} \bar{\uu} \bar{\bb_3}, \quad 0 = \bb_3 \rr - \bar{\rr} \bar{\bb_3}, \quad 0 = \bb_3 \uu \rr - \bar{\rr} \bar{\uu} \bar{\bb_3},
\end{equation}
and the non-vanishing \hypertarget{N2_a_42ii}{matrices} are
\begin{equation}
	\mb = \begin{pmatrix}
		\zero & \zero \\ \bb_3 & \zero
	\end{pmatrix}, \quad
	\mp = \begin{pmatrix}
		\zero & \zero \\ \bb_3\uu & \zero
	\end{pmatrix}, \quad
	N_0 = \begin{pmatrix}
		\zero & \qq \\ \bar{\qq} & b
	\end{pmatrix}, \quad \text{and} \quad 
	N_1 = \begin{pmatrix}
		\zero & \rr \\ \bar{\rr} & d
	\end{pmatrix}.
\end{equation}
Through the same use of the subgroup $\G \subset \GL(\s_{\bar{0}}) \times \GL(\s_{\bar{1}})$ as discussed for case $(\mathsf{i})$, we find that this sub-branch is equivalent to \hyperlink{N2_a_22ii}{$\mathsf{2.2.ii}$} for $\hat{\a}$.
\\
\paragraph{\textbf{Sub-Branch 4.3}} Now with $N_3 \neq 0$, we can use 
\begin{equation}
	-N_3 = \mp N_1 - N_1 \mp^\dagger \quad \text{and} \quad  0 = \pi \vt \mp N_3 \vt^\dagger + \vt N_3 (\pi \vt \mp)^\dagger
\end{equation}
to first write $N_3$ in terms of $\mp$ and $N_1$ before setting $c=0$ by substituting $\vt = (0,1)$ into the latter condition.
This produces the matrix
\begin{equation}
	N_3 = \begin{pmatrix}
		\zero & \zero \\ \zero &  \bar{\rr} \bar{\uu}\bar{\bb_3} - \bb_3\uu\rr
	\end{pmatrix}.
\end{equation}
Since $N_3$ and $\mb$ are non-vanishing, the condition from the $[\Q, \Q, \Q]$ identity no longer
states that we must set either $N_0$ or $\mh$ to zero.  We have
\begin{equation}
	\Re(\vt N_0 \vt^\dagger) \vt\mh = \vt N_3 \vt^\dagger \vt \mb.
\end{equation}
Substituting $\vt = (0, 1)$ into the above condition, we find
\begin{equation} \label{eq:N2_branch4_a_43_h3}
	b \hh_3 = (\bar{\rr} \bar{\uu}\bar{\bb_3} - \bb_3\uu\rr) \bb_3 \quad \text{and} \quad b \hh_1 = 0.
\end{equation}
By assumption $N_3 \neq 0$; therefore, both $b$ and $\hh_3$ cannot vanish.  Using this result, the second constraint tells us that $\hh_1 = 0$.  Thus $\mh$ is reduced to a strictly lower-diagonal matrix.  As in Sub-Branch \hyperlink{N2_a_42i}{$\mathsf{4.2}$}, we have 
\begin{equation}
	\begin{split}
		0 &= \mb N_i - N_i \mb^\dagger \quad \text{where} \quad i \in \{0, 1\} \\
		0 &= \mp N_0 - N_0 \mp^\dagger,
	\end{split}
\end{equation}
which tell us $a$ and $c$ vanish, and
\begin{equation}
	0 = \bb_3\qq - \bar{\qq}\bar{\bb_3}, \quad 0 = \bb_3\uu\qq - \bar{\qq}\bar{\uu}\bar{\bb_3}, \quad \text{and}
	\quad 0 = \bb_3 \rr - \bar{\rr}\bar{\bb_3}.
\end{equation}
Using these results and the rewriting of $\hh_3$ in \eqref{eq:N2_branch4_a_43_h3}, the conditions from the $[\bH, \Q, \Q]$ identity are instantly satisfied.  Therefore, the constraints on the parameters of this sub-branch are
\begin{equation}
	0 = \bb_3\qq - \bar{\qq}\bar{\bb_3}, \; 0 = \bb_3\uu\qq - \bar{\qq}\bar{\uu}\bar{\bb_3}, \; 0 = \bb_3 \rr - \bar{\rr}\bar{\bb_3}, \; \text{and} \; b \hh_3 = (\bar{\rr}\bar{\uu}\bar{\bb_3} - \bb_3\uu\rr) \bb_3.
\end{equation}
Notice that the first three constraints here tell us that $\bb_3$ is collinear with both $\qq$ and $\rr$ and that $\bb_3\uu$ is collinear with $\qq$.  In particular, were we to use the endomorphisms of $\s_{\bar{1}}$ to set $\qq$ to lie along $\ii$, $\bb_3$, $\bb_3\uu$ and $\rr$ would all lie along $\ii$ as well.  Thus, $\bb_3\uu\rr \in \mathbb{R}$, such that $N_3 = 0$.  Therefore, this sub-branch is empty.
\\
\paragraph{\textbf{Sub-Branch 4.4}} This sub-branch will be very similar to the one above
due to the similarity in the conditions the super-Jacobi identity imposes on $N_3$ and $N_4$.
Using 
\begin{equation}
		N_4 = \mb N_1 - N_1 \mb^\dagger \quad \text{and} \quad 0 = \beta \vt \mb N_4 \vt^\dagger + \vt N_4 (\beta \vt \mb)^\dagger,
\end{equation}
we know $N_4$ may be written
\begin{equation}
	N_4 = \begin{pmatrix}
		\zero & \zero \\ \zero & \bb_3\rr - \bar{\rr}\bar{\bb_3}
	\end{pmatrix}.
\end{equation}
Lemma~\ref{lem:N2_111} then tells us that
\begin{equation}
	\Re(\vt N_0 \vt^\dagger) \vt\mh = \vt N_4 \vt^\dagger \vt \mp.
\end{equation}
Substituting $\vt = (0, 1)$ into this condition produces
\begin{equation} \label{eq:N2_C_a_branch44_h_condition}
	b \hh_3 = (\bb_3\rr - \bar{\rr}\bar{\bb_3}) \bb_3\uu \quad \text{and} \quad b \hh_1 = 0.
\end{equation}
Since $\bb_3\uu \neq 0$ and $\bb_3\rr - \bar{\rr}\bar{\bb_3} \neq 0$ by assumption, $b$ cannot vanish; therefore,
$\hh_1 = 0$.  The conditions
\begin{equation}
	\begin{split}
		0 &= \mp N_i - N_i \mp^\dagger \quad \text{where} \quad i \in \{0, 1\} \\
		0 &= \mb N_0 - N_0 \mb^\dagger,  \\
	\end{split}
\end{equation}
tell us that both $a$ and $c$ vanish, and 
\begin{equation}
	0 = \bb_3\qq- \bar{\qq}\bar{\bb_3}, \quad 0 =\bb_3\uu\qq - \bar{\qq}\bar{\uu}\bar{\bb_3}, \quad \text{and}
	\quad 0 = \bb_3\uu \rr - \bar{\rr}\bar{\uu}\bar{\bb_3}.
\end{equation}
Finally, we have the conditions from the $[\bH, \Q, \Q]$ identity, which impose
\begin{equation}
	0 = \Re(\hh_3 \qq) \quad \text{and} \quad 0 = \Re(\hh_3 \rr) .
\end{equation}
However, using the form of $\hh_3$ in \eqref{eq:N2_C_a_branch44_h_condition} and the collinearity of $\bb_3\uu$ with $\qq$ and $\rr$, both of these constraints are already satisfied.  Therefore, the final set of constraints on this sub-branch is
\begin{equation}
	0 = \bb_3\qq- \bar{\qq}\bar{\bb_3}, \quad 0 =\bb_3\uu\qq - \bar{\qq}\bar{\uu}\bar{\bb_3}, \quad 0 = \bb_3\uu \rr - \bar{\rr}\bar{\uu}\bar{\bb_3}, \quad b \hh_3 = (\bb_3\rr - \bar{\rr}\bar{\bb_3}) \bb_3\uu.
\end{equation}
Notice that the first three constraints tell us that $\bb_3$, $\bb_3\uu$, $\qq$, and $\rr$ are collinear.  This tells us that $\bb_3\rr \in \mathbb{R}$; therefore, significantly, $N_4 = 0$.  Thus this sub-branch is empty. 
\\
\paragraph{\textbf{Sub-Branch 4.5}}  Now with non-vanishing $N_3$ and $N_4$, we can begin by using
\begin{equation}
	\begin{split}
		-N_3 &= \mp N_1 - N_1 \mp^\dagger \\
		0 &= \pi \vt \mp N_3 \vt^\dagger + \vt N_3 (\pi \vt \mp)^\dagger
	\end{split} \quad \text{and} \quad 
	\begin{split}
		N_4 &= \mb N_1 + N_1 \mb^\dagger \\
		0 &= \beta \vt \mb N_4 \vt^\dagger + \vt N_4 (\beta \vt \mb)^\dagger,
	\end{split}
\end{equation}
to write
\begin{equation}
	N_1 =  \begin{pmatrix}
		\zero & \rr \\ \bar{\rr} & d
	\end{pmatrix}, \quad 
	N_3 = \begin{pmatrix}
		\zero & \zero \\ \zero & \bar{\rr}\bar{\uu}\bar{\bb_3}-\bb_3\uu\rr
	\end{pmatrix} \quad \text{and} \quad 
	N_4 = \begin{pmatrix}
		\zero & \zero \\ \zero & \bb_3\rr - \bar{\rr}\bar{\bb_3}
	\end{pmatrix}.
\end{equation}
Using these results, substitute $\vt = (0, 1)$ into the condition from the $[\Q, \Q, \Q]$
identity to find
\begin{equation} \label{eq:h_3_prescription}
	b \hh_3 = (\bar{\rr}\bar{\uu}\bar{\bb_3} - \bb_3\uu\rr) \bb_3 + (\bb_3\rr - \bar{\rr}\bar{\bb_3})\bb_3\uu \quad \text{and} \quad b \hh_1 = 0.
\end{equation}
As in all previous sub-branches, the $[\P, \Q, \Q]$ and $[\B, \Q, \Q]$ conditions on $N_0$ tell us
\begin{equation}
	0 = a, \quad 0 = \bb_3\qq - \bar{\qq}\bar{\bb_3} \quad \text{and} \quad 0 = \bb_3\uu\qq - \bar{\qq}\bar{\uu}\bar{\bb_3}.
\end{equation}
Finally, the $[\bH, \Q, \Q]$ identities tell us
\begin{equation}
	\begin{split}
		0 &= \Re(\hh_3 \qq) + b \Re(\hh_1)\\
		0 &= \hh_1 \qq + \qq \overbar{\bb_3\hh_1\bb_3^{-1}}
	\end{split} \quad \text{and} \quad
	\begin{split} 
		0 &= \Re(\hh_3 \rr) + d \Re(\hh_1) \\
		0 &= \hh_1 \rr + \rr \overbar{\bb_3\hh_1\bb_3^{-1}}.
	\end{split}
\end{equation}
Since, by assumption, $N_1 \neq 0$, these constraints mean we must have $\Re(\hh_1) = 0$.
If $\rr = 0$, we would need $d \neq 0$, which, when substituted into $0 = d \Re(\hh_1)$, mean $\Re(\hh_1)=0$.  Alternatively, if $\rr \neq 0$, we multiply
\begin{equation}
	0 = \hh_1 \rr + \rr \overbar{\bb_3\hh_1\bb_3^{-1}} 
\end{equation}
on the right by $\rr^{-1}$ and take the real part to obtain $\Re(\hh_1) = 0$.  Knowing this, we can use the fact $\overbar{\bb_3\hh_1\bb_3^{-1}} \in \Im(\mathbb{H})$ to rewrite the remaining imaginary part of this constraint as
\begin{equation}
	0 = [\hh_1, \rr\bb_3].
\end{equation}
Additionally, since $\Re(\hh_1)=0$, we can use $ 0 = \bb_3\qq - \bar{\qq}\bar{\bb_3}$ to instantly
satisfy the condition
\begin{equation}
 0 = \hh_1 \qq + \qq \overbar{\bb_3\hh_1\bb_3^{-1}}.
\end{equation}
These results leave us with
\begin{equation}
	\begin{split}
	\mathcal{C}_{\hat{\a},\,\mathsf{4.5.iii}} = \{
		0 &= \bb_3\qq - \bar{\qq}\bar{\bb_3}, \quad
		0 = \bb_3\uu\qq - \bar{\qq}\bar{\uu}\bar{\bb_3}, \\
		0 &= \Re(\hh_3\qq), \quad
		0 = \Re(\hh_3\rr), \quad 
		0 = \Re(\hh_1), \\
		0 &= [\hh_1, \rr\bb_3], \quad 
		0 = b \hh_1, \quad
		b \hh_3 = - 2 \Im(\bb_3\uu\rr)\bb_3 + 2 \Im(\bb_3\rr)\bb_3\uu \}.
	\end{split}
\end{equation}
Subject to these constraints, the non-vanishing \hypertarget{N2_a_45}{matrices} are
\begin{equation}
	\begin{split}
	\mathcal{M}_{\hat{\a},\,\mathsf{4.5.iii}} = \Big\{ \mb &= \begin{pmatrix}
		\zero & \zero \\ \bb_3 & \zero
	\end{pmatrix}, \quad
	\mp = \begin{pmatrix}
		\zero & \zero \\ \bb_3\uu & \zero
	\end{pmatrix}, \quad
	\mh = \begin{pmatrix}
		\hh_1 & \zero \\ \hh_3 & \bb_3\hh_1\bb_3^{-1}
	\end{pmatrix}, \\
	N_0 &= \begin{pmatrix}
		\zero & \qq \\ \bar{\qq} & b
	\end{pmatrix}, \quad
	N_1 = \begin{pmatrix}
		\zero & \rr \\ \bar{\rr} & d
	\end{pmatrix}, \quad
	N_3 = \begin{pmatrix}
		\zero & \zero \\ \zero & \bar{\rr}\bar{\uu}\bar{\bb_3}-\bb_3\uu\rr
	\end{pmatrix}, \quad 
	N_4 = \begin{pmatrix}
		\zero & \zero \\ \zero & \bb_3\rr - \bar{\rr}\bar{\bb_3}
	\end{pmatrix} \Big\}.
	\end{split}
\end{equation}
The wealth of parameters describing this sub-branch mean we will only highlight one parameterisation of these super-extensions here, though many more may exist.  In particular, we will choose to set $b$, $d$ and $\hh_3$ to zero.  We will also utilise the subgroup $\G \subset \GL(\s_{\bar{0}}) \times \GL(\s_{\bar{1}})$ to impose that $\qq$, $\bb_3$ and $\bb_3\uu$, lie along $\ii$. The residual endomorphisms of $\s_{\bar{1}}$ may then scale $\rr$ such that its norm becomes $1$. Employing $\Aut(\mathbb{H})$, we can set $\hh_1$ to lie along $\ii$ as well. Having made these choices, the constraint
\begin{equation}
	0 = [\hh_1, \rr\bb_3]
\end{equation}
tells us $\rr \in \mathbb{R} \langle 1, \ii \rangle$.  Notice that for $N_3$ and $N_4$ to be non-vanishing $\rr$ must have a real component; therefore, to simplify the form of the matrices in our example, we will choose $\rr = 1$.  The remaining constraints in $\mathcal{C}_{\hat{\a},\,\mathsf{4.5.iii}}$ are then satisfied, and we can use the scaling symmetry of the $\s_{\bar{0}}$ basis elements to produce 
\begin{equation}
	\begin{split}
	 \mb &= \begin{pmatrix}
		\zero & \zero \\ \ii & \zero
	\end{pmatrix}, \quad
	\mp = \begin{pmatrix}
		\zero & \zero \\ \ii & \zero
	\end{pmatrix}, \quad
	\mh = \begin{pmatrix}
		\ii & \zero \\ \zero & \ii
	\end{pmatrix}, \\
	N_0 &= \begin{pmatrix}
		\zero & \ii \\ -\ii & \zero
	\end{pmatrix}, \quad
	N_1 = \begin{pmatrix}
		\zero & 1 \\ 1 & \zero
	\end{pmatrix}, \quad
	N_3 = \begin{pmatrix}
		\zero & \zero \\ \zero & \ii
	\end{pmatrix} \quad \text{and} \quad 
	N_4 = \begin{pmatrix}
	 \zero & \zero \\ \zero & \ii
	\end{pmatrix}.
	\end{split}
\end{equation}
\paragraph{\textbf{Sub-Branch 4.6}} We find that this sub-branch is empty using the analysis from the previous
sub-branch. Again, we use
\begin{equation}
	\begin{split}
		-N_3 &= \mp N_1 - N_1 \mp^\dagger \\
		0 &= \pi \vt \mp N_3 \vt^\dagger + \vt N_3 (\pi \vt \mp)^\dagger
	\end{split} \quad \text{and} \quad 
	\begin{split}
		N_4 &= \mb N_1 + N_1 \mb^\dagger \\
		0 &= \beta \vt \mb N_4 \vt^\dagger + \vt N_4 (\beta \vt \mb)^\dagger,
	\end{split}
\end{equation}
to write
\begin{equation}
	N_3 = \begin{pmatrix}
		\zero & \zero \\ \zero & \bar{\rr}\bar{\uu}\bar{\bb_3}- \bb_3\uu\rr
	\end{pmatrix} \quad
	N_4 = \begin{pmatrix}
		\zero & \zero \\ \zero & \bb_3\rr - \bar{\rr}\bar{\bb_3}
	\end{pmatrix}.
\end{equation}
Substituting these matrices into 
\begin{equation}
	\begin{split}
		\tfrac12 [\beta, \vt N_2 \vt^\dagger] &= \beta \vt \mb N_3 \vt^\dagger
		+ \vt N_3 (\beta \vt \mb)^\dagger \\
		\tfrac12 [\pi, \vt N_2 \vt^\dagger] &= \pi \vt \mp N_4 \vt^\dagger
		+ \vt N_4 (\pi \vt \mp)^\dagger ,
	\end{split}
\end{equation}
the R.H.S. of both of these constraints vanishes, setting $N_2 = 0$.  Therefore, this
sub-branch is empty.
\\
\paragraph{$\hat{\n}_-$}~\\ \\
Setting $\mu = \eta = 0$, $\lambda = 1$ and 
$\varepsilon = -1$, the first condition
in \eqref{eq:N2_branch4_condition} becomes
\begin{equation}
	[\uu, \hh_1] = 2 \uu.
\end{equation}
Since $[\uu, \hh_1]$ is perpendicular to $\uu$ in $\Im(\mathbb{H})$ this branch 
cannot provide a super-extension for $\hat{\n}_-$.  
\\
\paragraph{$\hat{\n}_+$}~\\ \\In this case, for which $\lambda = \varepsilon = 0$, 
$\mu = 1$, and $\eta = -1$, the first constraint in \eqref{eq:N2_branch4_condition} gives us
\begin{equation} \label{eq:N2_branch4_n-}
	[\hh_1, \uu] = \uu^2 + 1.
\end{equation}
Taking the real part of \eqref{eq:N2_branch4_n-} produces
\begin{equation}
	\Re(\uu^2) = -1,
\end{equation}
therefore, $\uu \in \Im(\mathbb{H})$, such that $|\uu|^2 = 1$, i.e. it is
a unit-norm vector quaternion, or \textit{right versor}.  The imaginary part
of  \eqref{eq:N2_branch4_n-} imposes
\begin{equation}
	[\uu, \hh_1] = 0.
\end{equation} 
Thus, we could get a super-extension of $\hat{\n}_+$ in this branch.
Wishing to write our parameters in terms of $\bb_3$, we have
$\pp_3 = \bb_3 \uu$ and $\hh_4 = \bb_3 (\hh_1 - \uu) \bb_3^{-1}$, where
$\uu \in \Im(\mathbb{H})$, such that $\uu^2 = -1$.
\\ \\
As with the $\hat{\a}$ case above, all of the conditions
of Lemmas~\ref{lem:N2_011} and~\ref{lem:N2_111} must be taken into account.  
The conditions
\begin{equation}
		N_4 = \mb N_1 - N_1 \mb^\dagger \quad \text{and} \quad -N_3 = \mp N_1 - N_1 \mp^\dagger
\end{equation}
tell us that if $N_1 = 0$, $N_3 = 0$ and $N_4 = 0$.  Substituting these results into
\begin{equation}
	\begin{split}
		- \Re(\vt N_0\vt^\dagger) \pi &= \pi \vt \mp N_3 \vt^\dagger + \vt N_3 (\pi \vt \mp)^\dagger \\
		\tfrac12 [\beta, \vt N_2 \vt^\dagger] &= \beta \vt \mb N_3 \vt^\dagger
		+ \vt N_3 (\beta \vt \mb)^\dagger \\
	\end{split} \qquad
	\begin{split}
		\Re(\vt N_0\vt^\dagger) \beta &= \beta \vt \mb N_4 \vt^\dagger + \vt N_4 (\beta \vt \mb)^\dagger \\
		\tfrac12 [\pi, \vt N_2 \vt^\dagger] &= \pi \vt \mp N_4 \vt^\dagger
		+ s N_4 (\pi \vt \mp)^\dagger,	\\
	\end{split}
\end{equation}
we see that if $N_3$ or $N_4$ vanish, so must $N_0$ and $N_2$.  Equally, if $N_3$ vanishes
$N_4$ necessarily vanishes and vice versa due to
\begin{equation}
		- N_4 = \mh N_3 + N_3 \mh^\dagger \quad \text{and} \quad  N_3 = \mh N_4 + N_4 \mh^\dagger.
\end{equation}
Therefore, based on these dependencies, our investigation into this branch of possible super-extensions
of $\hat{\n}_+$ divides into the following sub-branches.
\begin{enumerate}
	\item $N_1 \neq 0$, and $N_0 = N_2 = N_3 = N_4 = 0$
	\item $N_1 \neq 0$, $N_3 \neq 0$, $N_4 \neq 0$, and $N_0 = N_2 = 0$
	\item $N_1 \neq 0$, $N_3 \neq 0$, $N_4 \neq 0$, $N_0 \neq 0$ and $N_2 = 0$
	\item $N_1 \neq 0$, $N_3 \neq 0$, $N_4 \neq 0$, $N_0 = 0$ and $N_2 \neq 0$
	\item $N_1 \neq 0$, $N_3 \neq 0$, $N_4 \neq 0$, $N_0 \neq 0$ and $N_2 \neq 0$
\end{enumerate}
Unlike the super-extensions of $\hat{\n}_+$ found in Branches 1, 2 and 3, the $[\Q, \Q, \Q]$ identity will not impose 
that either $N_0$ or $\mh$ must vanish.  In the first sub-branch above, we instantly see that $N_0=0$; therefore, the super-extensions found here are extensions satisfying $(\mathsf{i})$.  However, all other sub-branches have either non-vanishing $N_3$ or non-vanishing $N_4$.  Since $\mb\neq 0$ and $\mp \neq 0$, the $[\Q, \Q, \Q]$ identity will now form relationships between $N_0$, $N_3$ and $N_4$, with, in general, $\mh \neq 0$.  Therefore, these super-extensions, for which $N_0 \neq 0$ and $\mh \neq 0$, will be labelled $(\mathsf{iii})$ to distinguish them from the cases $(\mathsf{i})$ and $(\mathsf{ii})$.
\\
\paragraph{\textbf{Sub-Branch 4.1}} With only $N_1 \neq 0$, the conditions from Lemmas~\ref{lem:N2_011}
and \ref{lem:N2_111} reduce to 
\begin{equation}
	\begin{split}
		0 &= \mh N_1 +N_1 \mh^\dagger \\
		0 &= \mp N_1 - N_1 \mp^\dagger \\
		0 &= \mb N_1 - N_1 \mb^\dagger.
	\end{split}
\end{equation}
The latter two conditions tell us 
\begin{equation}
	0 = c, \quad 0 = \bb_3 \rr - \bar{\rr}\bar{\bb_3} \quad \text{and} \quad 0 = \bb_3 \uu \rr - \bar{\rr}\bar{\uu}\bar{\bb_3},
\end{equation}
which, when substituted into the first conditions, produce
\begin{equation}
	0 = \Re(\hh_1) \quad \text{and} \quad 0 = \Re(\hh_3 \rr).
\end{equation}
Therefore, the non-vanishing \hypertarget{N2_n+_41i}{matrices} for this super-extension are
\begin{equation}
	\mb = \begin{pmatrix}
		\zero & \zero \\ \bb_3 & \zero
	\end{pmatrix}, \quad 
	\mp = \begin{pmatrix}
		\zero & \zero \\ \bb_3\uu & \zero
	\end{pmatrix}, \quad 
	\mh = \begin{pmatrix}
		\hh_1 & \zero \\ \hh_3 & \bb_3(\hh_1 - \uu)\bb_3^{-1} 
	\end{pmatrix}, \quad
	N_1 = \begin{pmatrix}
		\zero & \rr \\ \bar{\rr} & d
	\end{pmatrix},
\end{equation}
subject to the constraints
\begin{equation}
	0 = [\uu, \hh_1], \quad 0 = \Re(\hh_1), \quad 0 = \Re(\hh_3 \rr), \quad 0 = \bb_3 \rr - \bar{\rr}\bar{\bb_3}, \quad 0 = \bb_3 \uu \rr - \bar{\rr}\bar{\uu}\bar{\bb_3}, \quad \uu^2 = -1.
\end{equation}
However, notice that the final three constraints listed above require one of $\bb_3$, $\uu$ or $\rr$ to vanish.  Since neither $\bb_3$ or $\uu$ can vanish in this sub-branch, it must be that $\rr = 0$.  Therefore, the final set of matrices is 
\begin{equation}
	\mathcal{M}_{\hat{\n}_+,\,\mathsf{4.1.i}} = \Big\{ \mb = \begin{pmatrix}
		\zero & \zero \\ \bb_3 & \zero
	\end{pmatrix}, \quad 
	\mp = \begin{pmatrix}
		\zero & \zero \\ \bb_3\uu & \zero
	\end{pmatrix}, \quad 
	\mh = \begin{pmatrix}
		\hh_1 & \zero \\ \hh_3 & \bb_3(\hh_1 - \uu)\bb_3^{-1} 
	\end{pmatrix}, \quad
	N_1 = \begin{pmatrix}
		\zero & \zero \\ \zero & d
	\end{pmatrix} \Big\},
\end{equation}
and the final set of constraints is
\begin{equation}
	\mathcal{C}_{\hat{\n}_+,\,\mathsf{4.1.i}} = \{ 0 = [\uu, \hh_1], \quad  0 = \Re(\hh_1), \quad \uu^2 = -1\}.
\end{equation}
To demonstrate that this sub-branch of $\cS$ is not empty, choose to set $\hh_1$ and $\hh_3$ to zero.  Using the endomorphisms of $\s_{\bar{1}}$ and $\Aut(\mathbb{H})$ on $\bb_3$ and $\uu$, respectively, we may write $\bb_3 = \ii$ and $\uu = \jj$.  Employing the scaling symmetry of $\sZ$, we arrive at the super-extension
\begin{equation}
	 \mb = \begin{pmatrix}
		\zero & \zero \\ \ii & \zero
	\end{pmatrix}, \quad 
	\mp = \begin{pmatrix}
		\zero & \zero \\ \kk & \zero
	\end{pmatrix}, \quad 
	\mh = \begin{pmatrix}
		\zero & \zero \\ \zero & \jj
	\end{pmatrix}, \quad
	N_1 = \begin{pmatrix}
		\zero & \zero \\ \zero & 1
	\end{pmatrix}.
\end{equation}  
Thus, this sub-branch is not empty.  We may then introduce $\hh_1$ while continuing to fix all the parameters of the super-extension; however, this cannot be achieved on introducing $\hh_3$.
\\
\paragraph{\textbf{Sub-Branch 4.2}} Now with $N_3 \neq 0$ and $N_4 \neq 0$ as well as $N_1\neq 0$,
we can use the conditions
\begin{equation}
	\begin{split}
		-N_3 &= \mp N_1 - N_1 \mp^\dagger \\
		N_4 &= \mb N_1 - N_1 \mb^\dagger
	\end{split} \qquad 
	\begin{split}
		0 &= \beta \vt \mb N_3 \vt^\dagger
		+ \vt N_3 (\beta \vt \mb)^\dagger \\
		0 &= \pi \vt \mp N_3 \vt^\dagger + \vt N_3 (\pi \vt \mp)^\dagger
	\end{split} \qquad
	\begin{split}
		0 &= \beta \vt \mb N_4 \vt^\dagger + \vt N_4 (\beta \vt \mb)^\dagger \\
		0 &= \pi \vt \mp N_4 \vt^\dagger
		+ \vt N_4 (\pi \vt \mp)^\dagger,
	\end{split}
\end{equation}
and the analysis of Branches 2 and 3 to write
\begin{equation}
	N_1 = \begin{pmatrix}
		\zero & \rr \\ \bar{\rr} & d
	\end{pmatrix} \quad 
	N_3 = \begin{pmatrix}
		\zero & \zero \\ \zero & \bar{\rr}\bar{\uu}\bar{\bb_3}-\bb_3\uu\rr
	\end{pmatrix} \quad 
	N_4 = \begin{pmatrix}
		\zero & \zero \\ \zero & \bb_3 \rr - \bar{\rr}\bar{\bb_3}
	\end{pmatrix}.
\end{equation}
This leaves only the $[\bH, \Q, \Q]$ conditions:
\begin{equation}
	\begin{split}
		0 &= \mh N_1 + N_1 \mh^\dagger \\
		- N_4 &= \mh N_3 + N_3 \mh^\dagger \\
		N_3 &= \mh N_4 + N_4 \mh^\dagger. \\	
	\end{split}
\end{equation}
We know from Sub-Branch \hyperlink{N2_n+_41i}{$\mathsf{4.1.i}$} that, since $c = 0$, the first of these produces
\begin{equation}
	0 = \Re(\hh_1) \quad \text{and} \quad 0 = \Re(\hh_3 \rr).
\end{equation}
Writing $\bar{\rr}\bar{\uu}\bar{\bb_3}-\bb_3\uu\rr=-2\Im(\bb_3\uu\rr)$ and $\bb_3 \rr - \bar{\rr}\bar{\bb_3}=2\Im(\bb_3\rr)$ to simplify our expressions, the second and third conditions give us
\begin{equation}
	\begin{split}
		\Im(\bb_3\rr) &= \hh_4 \Im(\bb_3\uu\rr) + \Im(\bb_3\uu\rr) \bar{\hh_4} \\
		- \Im(\bb_3\uu\rr) &= \hh_4 \Im(\bb_3\rr) + \Im(\bb_3\rr) \bar{\hh_4},
	\end{split}
\end{equation}
respectively, where $\hh_4 = \bb_3(\hh_1 - \uu)\bb_3^{-1}$.   Notice that since $\hh_1, \uu \in \Im(\mathbb{H})$, and $\hh_4$ is written
in terms of the adjoint action of $\bb_3 \in \mathbb{H}$, $\hh_4 \in \Im(\mathbb{H})$.
Therefore, using $\bar{\hh_4} = - \hh_4$, we find
\begin{equation}
	\Im(\bb_3\rr) = - [\hh_4, [\hh_4, \Im(\bb_3\rr)]] \quad \text{and} \quad 
	\Im(\bb_3\uu\rr) = - [\hh_4, [\hh_4, \Im(\bb_3\uu\rr)]].
\end{equation}
This imposes the constraint that $\Im(\bb_3\rr)$ and $\Im(\bb_3\uu\rr)$ must be perpendicular
to $\hh_4$ in $\Im(\mathbb{H})$.  The constraints for this sub-branch are summarised
as follows.
\begin{equation}
	\begin{split}
		\mathcal{C}_{\hat{\n}_+,\,\mathsf{4.2.iii}} = \{ 
			&0 = [\hh_1, \uu], \quad
			-1 = \uu^2 \quad
			0 = \Re(\hh_1), \quad
			0 = \Re(\hh_3 \rr), \\
		&\Im(\bb_3\rr) = - [\hh_4, [\hh_4, \Im(\bb_3\rr)]], \quad
		\Im(\bb_3\uu\rr) = - [\hh_4, [\hh_4, \Im(\bb_3\uu\rr)]] \}.
	\end{split}
\end{equation}
The non-vanishing \hypertarget{N2_n+_42i}{matrices} are then
\begin{equation}
	\begin{split} 
	\mathcal{M}_{\hat{\n}_+,\,\mathsf{4.2.iii}} = \Big\{ \mb &= \begin{pmatrix}
		\zero & \zero \\ \bb_3 & \zero
	\end{pmatrix}, \quad 
	\mp = \begin{pmatrix}
		\zero & \zero \\ \bb_3\uu & \zero
	\end{pmatrix}, \quad 
	\mh = \begin{pmatrix}
		\hh_1 & \zero \\ \hh_3 & \bb_3(\hh_1 - \uu)\bb_3^{-1} 
	\end{pmatrix}, \\
	N_1 &= \begin{pmatrix}
		\zero & \rr \\ \bar{\rr} & d
	\end{pmatrix}, \quad 
	N_3 = \begin{pmatrix}
		\zero & \zero \\ \zero & - 2 \Im(\bb_3\uu\rr)
	\end{pmatrix}, \quad
	N_4 = \begin{pmatrix}
		\zero & \zero \\ \zero & 2 \Im(\bb_3\rr)
	\end{pmatrix} \Big\}. \end{split}		
\end{equation}
To demonstrate the existence of super-extensions in this sub-branch, we will begin by simplifying our parameter set as much as possible.  In particular, we begin by setting both $\hh_3$ and $d$ to zero.  We then utilise $\Aut(\mathbb{H})$ and the endomorphisms of $\s_{\bar{1}}$ to set $\uu = \jj$ and impose that $\bb_3$ lies along $\ii$.  Notice that with $\uu$ along $\jj$, the first constraint in $\mathcal{C}_{\hat{\n}_+,\,\mathsf{4.2.iii}}$ tells us that $\hh_1$ must also lie along $\jj$, as must $\hh_4 = \bb_3(\hh_1 - \uu)\bb_3^{-1}$.  With these choices, the two constraints involving $\hh_4$ impose $\rr \in \mathbb{R} \langle 1, \jj \rangle$, and that $|\hh_1| = \tfrac12$ or $|\hh_1| = \tfrac32$.  Residual endomorphisms then allow us to scale $\rr$ such that it has unit norm.  Finally, we can scale the $\s_{\bar{0}}$ basis elements to arrive at
\begin{equation}
	\begin{split} 
	\mb &= \begin{pmatrix}
		\zero & \zero \\ \ii & \zero
	\end{pmatrix}, \quad 
	\mp = \begin{pmatrix}
		\zero & \zero \\ \kk & \zero
	\end{pmatrix}, \quad 
	\mh = \begin{pmatrix}
		\jj & \zero \\ \zero & \jj 
	\end{pmatrix}, \\
	N_1 &= \begin{pmatrix}
		\zero & 1 + \jj \\ 1 - \jj & \zero
	\end{pmatrix}, \quad 
	N_3 = \begin{pmatrix}
		\zero & \zero \\ \zero & \kk - \ii
	\end{pmatrix}, \quad
	N_4 = \begin{pmatrix}
		\zero & \zero \\ \zero & \kk + \ii
	\end{pmatrix}.
	\end{split}
\end{equation} 
\paragraph{\textbf{Sub-Branch 4.3}} The beginning of the investigation of this sub-branch is identical
to that of the previous sub-branch.  The $[\P, \Q, \Q]$ and $[\B, \Q, \Q]$ identities produce 
\begin{equation}
	\begin{split}
		-N_3 &= \mp N_1 - N_1 \mp^\dagger \\
		N_4 &= \mb N_1 - N_1 \mb^\dagger
	\end{split} \qquad
	\begin{split}
		0 &= \beta \vt \mb N_3 \vt^\dagger
		+ \vt N_3 (\beta \vt \mb)^\dagger \\
		0 &= \pi \vt \mp N_4 \vt^\dagger
		+ \vt N_4 (\pi \vt \mp)^\dagger,
	\end{split}
\end{equation}
where the first two conditions give $N_3$ and $N_4$ the form
\begin{equation}
	N_3 = \begin{pmatrix}
		\zero & c \bb_3\uu \\ - c \overbar{\bb_3\uu} & \bar{\rr}\bar{\uu}\bar{\bb_3}- \bb_3\uu\rr
	\end{pmatrix} \quad \text{and} \quad
	N_4 = \begin{pmatrix}
		\zero & - c\bar{\bb_3} \\ c\bb_3 & \bb_3\rr - \bar{\rr}\bar{\bb_3}
	\end{pmatrix}.
\end{equation}
Substituting this $N_3$ with $\vt = (0, 1)$ into
\begin{equation}
	0 = \beta \vt \mb N_3 \vt^\dagger
		+ \vt N_3 (\beta \vt \mb)^\dagger,
\end{equation}
we acquire
\begin{equation}
	0 = 2 c |\bb_3|^2 \Im(\beta \uu) \quad \forall \beta \in \Im(\mathbb{H}).
\end{equation}
As, by assumption, $\bb_3 \neq 0$ and $\uu \neq 0$, this imposes $c=0$, such that
\begin{equation}
	N_1 = \begin{pmatrix}
		\zero & \rr \\ \bar{\rr} & d
	\end{pmatrix} \quad 
	N_3 = \begin{pmatrix}
		\zero & \zero \\ \zero & \bar{\rr}\bar{\uu}\bar{\bb_3}-\bb_3\uu\rr
	\end{pmatrix} \quad 
	N_4 = \begin{pmatrix}
		\zero & \zero \\ \zero & \bb_3 \rr - \bar{\rr}\bar{\bb_3}
	\end{pmatrix}.
\end{equation}
With this form of $N_3$ and $N_4$,
\begin{equation}
	\begin{split}
		- \Re(\vt N_0\vt^\dagger) \pi &= \pi \vt \mp N_3 \vt^\dagger + \vt N_3 (\pi \vt \mp)^\dagger \\
		\Re(\vt N_0\vt^\dagger) \beta &= \beta \vt \mb N_4 \vt^\dagger + \vt N_4 (\beta \vt \mb)^\dagger, \\
	\end{split}
\end{equation}
have a vanishing R.H.S., showing that $N_0 = 0$.  This result contradicts our assumption
that $N_0 \neq 0$; therefore, this sub-branch does not contain any super-extensions.
\\
\paragraph{\textbf{Sub-Branch 4.4}} Letting $N_0 = 0$ and $N_2 \neq 0$, we can use 
\begin{equation}
	\begin{split}
		-N_3 &= \mp N_1 - N_1 \mp^\dagger \\
		N_4 &= \mb N_1 - N_1 \mb^\dagger
	\end{split} \qquad 
	\begin{split}
		0 &= \beta \vt \mb N_4 \vt^\dagger + \vt N_4 (\beta \vt \mb)^\dagger \\
		0 &= \pi \vt \mp N_3 \vt^\dagger + \vt N_3 (\pi \vt \mp)^\dagger,
	\end{split}
\end{equation}
to again write
\begin{equation}
	N_1 = \begin{pmatrix}
		\zero & \rr \\ \bar{\rr} & d
	\end{pmatrix} \quad 
	N_3 = \begin{pmatrix}
		\zero & \zero \\ \zero & \bar{\rr}\bar{\uu}\bar{\bb_3} -\bb_3\uu\rr
	\end{pmatrix} \quad 
	N_4 = \begin{pmatrix}
		\zero & \zero \\ \zero & \bb_3 \rr - \bar{\rr}\bar{\bb_3}
	\end{pmatrix}.
\end{equation}
Substituting these $N_i$ into 
\begin{equation}
	\begin{split}
	\tfrac12 [\beta, \vt N_2 \vt^\dagger] &= \beta \vt \mb N_3 \vt^\dagger
		+ \vt N_3 (\beta \vt \mb)^\dagger \\
	\tfrac12 [\pi, \vt N_2 \vt^\dagger] &= \pi \vt \mp N_4 \vt^\dagger
		+ \vt N_4 (\pi \vt \mp)^\dagger, 
	\end{split}
\end{equation}
the R.H.S. vanishes for both, showing $N_2 = 0$, contradicting our initial assumption in this 
sub-branch.
\\
\paragraph{\textbf{Sub-Branch 4.5}} With none of the $N_i$ vanishing, we start again by writing $N_3$ and
$N_4$ in terms of $N_1$ using conditions from the $[\P, \Q, \Q]$ and $[\B, \Q, \Q]$ identities:
\begin{equation}
	N_3 = \begin{pmatrix}
		\zero & c \overbar{\bb_3\uu} \\ - c \bb_3\uu & \bar{\rr}\bar{\uu}\bar{\bb_3}- \bb_3\uu\rr
	\end{pmatrix} \quad \text{and} \quad
	N_4 = \begin{pmatrix}
		\zero & - c\bar{\bb_3} \\ c\bb_3 & \bb_3\rr - \bar{\rr}\bar{\bb_3}	
	\end{pmatrix}.	
\end{equation}
Letting
\begin{equation}
	N_2 = \begin{pmatrix}
		\nn & \mm \\ -\bar{\mm} & \ll
	\end{pmatrix} \quad \text{where} \quad \nn, \ll \in \Im(\mathbb{H}), \quad \mm \in \mathbb{H},
\end{equation}
we can use
\begin{equation} \label{eq:N2_n+_branch4_N2_condition}
	\tfrac12 [\beta, \vt N_2 \vt^\dagger] = \beta \vt \mb N_3 \vt^\dagger
		+ \vt N_3 (\beta \vt \mb)^\dagger
\end{equation}
to write $N_2$ in terms of $\bb_3$ and $\uu$.  First let $\vt = (0,1)$ to find
\begin{equation}
	\tfrac12 [\beta, \ll] = -c [ \beta, \bb_3 \uu\bar{\bb_3}] \quad \forall 0 \neq \beta \in \Im(\mathbb{H}).
\end{equation}
Therefore, 
\begin{equation}
	\ll = -2c \bb_3\uu\bar{\bb_3}.
\end{equation}
Next, substitute in $\vt = (1, 1)$ to get
\begin{equation}
	[\beta, 2\Im(\mm)] + \tfrac12 [\beta, \ll] = -c [\beta, \bb_3\uu\bar{\bb_3}].
\end{equation}
Using the previous result, this tells us that $\Im(\mm) = 0$.  Analogous calculations
with $\vt = (0, \ii)$ and $\vt = (1, \ii)$ show that, in fact, all of $\mm$ must vanish.
Finally, substituting in $\vt = (1, 0)$ into \eqref{eq:N2_n+_branch4_N2_condition}, we find $\nn = 0$ since the R.H.S. vanishes.  
Therefore, we are left with
\begin{equation}
	N_2 = \begin{pmatrix}
		\zero & \zero \\ \zero & -2c\bb_3\uu\bar{\bb_3}
	\end{pmatrix}.
\end{equation}
We would have arrived at the same expression had we used $N_4$ and
\begin{equation}
	\tfrac12 [\pi, \vt N_2 \vt^\dagger] = \pi \vt \mp N_4 \vt^\dagger
		+ \vt N_4 (\pi \vt \mp)^\dagger.
\end{equation}
This form of $N_2$ automatically satisfies all other conditions it is involved in
from both the $[\B, \Q, \Q]$ and $[\P, \Q, \Q]$ identities.  Finally, we can
put this $N_2$ into 
\begin{equation}
	0 = \mh N_2 + N_2 \mh^\dagger
\end{equation}
to get
\begin{equation}
	0 = \hh_4 \ll + \ll \bar{\hh_4},
\end{equation}
where $\hh_4 = \bb_3(\hh_1 - \uu)\bb_3^{-1}$ and $\ll= -2c\bb_3\uu\bar{\bb_3}$.  Working through
some algebra, noting $\Re(\uu^2) = -1$ and the fact $c \neq 0$ for $N_2 \neq 0$, we arrive
at $\hh_1 \uu = \overbar{\hh_1\uu}$.  Since $\uu \in \Im(\mathbb{H})$, this forces $\hh_1 \in \Im(\mathbb{H})$
such that $\uu$ and $\hh_1$ are collinear. 
\\ \\
Now turn to $N_0$ and consider
\begin{equation}
	\begin{split}
		- \Re(\vt N_0\vt^\dagger) \pi &= \pi \vt \mp N_3 \vt^\dagger + \vt N_3 (\pi \vt \mp)^\dagger \\
		\Re(\vt N_0\vt^\dagger) \beta &= \beta \vt \mb N_4 \vt^\dagger + \vt N_4 (\beta \vt \mb)^\dagger.
	\end{split}
\end{equation}
Letting $\vt=(1, 0)$ in either of these conditions, we find that $a = 0$.  
Next, substituting $\vt = (0, 1)$ into the second condition produces
\begin{equation}
	- b = 2 c |\bb_3|^2.
\end{equation}
We would have arrived at the same result had we substituted into the first condition
and used the fact $|\uu|^2 = 1$.  Now substituting $\vt = (1, \ss)$ into the second condition, we find
\begin{equation}
	- 2 \Re(\ss\bar{\qq}) - b |\ss|^2 = 2c |\ss|^2 |\bb_3|^2.
\end{equation}
Therefore, using the previous result and letting $\ss = 1$, $\ss = \ii$, $\ss = \jj$
and $\ss = \kk$, we see that all components of $\qq$ must vanish.  All other conditions
on $N_0$ are now automatically satisfied, leaving
\begin{equation}
	N_0 = \begin{pmatrix}
		\zero & \zero \\ \zero & - 2c |\bb_3|^2
	\end{pmatrix}.
\end{equation}
Equipped with these $N_i$, we can now analyse the condition from Lemma~\ref{lem:N2_111}:
\begin{equation}
	\Re(\vt N_0\vt^\dagger) \vt \mh = \tfrac12 \vt N_2 \vt^\dagger \vt + \vt N_3 \vt^\dagger \vt \mb 
+ \vt N_4 s^\dagger \vt \mp.
\end{equation}
Letting $\vt = (0, 1)$:
\begin{equation} \label{eq:QQQ_branch4_n+}
	- 2c |\bb_3|^2 \begin{pmatrix}
		\hh_3 & \bb_3(\hh_1 - \uu)\bb_3^{-1}
	\end{pmatrix} = -c\bb_3\uu\bar{\bb_3} \begin{pmatrix}
	0 & 1
	\end{pmatrix} - \Im(\bb_3\uu\rr) \begin{pmatrix}
		\bb_3 & 0
	\end{pmatrix} + \Im(\bb_3\rr) \begin{pmatrix}
		\bb_3\uu & 0
	\end{pmatrix}.
\end{equation}
Concentrating on the second component, we have
\begin{equation}
	-2 c |\bb_3|^2 \bb_3 (\hh_1 - \uu) \bb_3^{-1} = -c \bb_3 \uu \bar{\bb_3}.
\end{equation}
Using the fact $|\bb_3|^2 \bb_3 = \bar{\bb_3}$ and cancelling relevant terms leaves 
\begin{equation}
	2 \bb_3\hh_1\bar{\bb_3} = 0.
\end{equation}
Since, by assumption $\bb_3 \neq 0$, we get $\hh_1 = 0$.  The first component of 
\eqref{eq:QQQ_branch4_n+} gives us a prescription for $\hh_3$,
\begin{equation}
	-2 c |\bb_3|^2 \hh_3 = - 2 \Im(\bb_3\uu\rr) \bb_3 + 2 \Im(\bb_3\rr) \bb_3\uu,
\end{equation}
therefore, we can fully describe $\mh$ in terms of $\mb$, $\mp$, and $N_1$.
\\ \\
The final conditions to consider are those from the $[\bH, \Q, \Q]$ super-Jacobi identity for $N_1$, $N_3$
and $N_4$.  First, the $N_1$ condition tell us
\begin{equation}
	0 = c\bar{\hh_3} + \rr \overbar{\bb_3\uu\bb_3^{-1}} \qquad 0 = \Re(\hh_3\rr).
\end{equation}
Notice that the second constraint here is automatically satisfied by the first, since $c \neq 0$ for a non-vanishing $N_0$.
Substituting this expression for $\hh_3$ into the previous prescription, we find
\begin{equation} \label{eq:h4}
	|\rr|^2 |\bb_3|^2 \bb_3 \uu \bb_3^{-1} = [\Im(\bb_3\rr), \Im(\bb_3\uu\rr)] + \Re(\bb_3\uu\rr) \Im(\bb_3\rr)
	- \Re(\bb_3\rr) \Im(\bb_3\uu\rr).
\end{equation}
Now, the constraints that the $N_3$ and $N_4$ conditions produce are the ones given in
Sub-Branch \hyperlink{N2_n+_42iii}{$\mathsf{4.2.iii}$}:
\begin{equation}
	\Im(\bb_3\rr) = - [\hh_4, [\hh_4, \Im(\bb_3\rr)]] \quad \text{and} \quad 
	\Im(\bb_3\uu\rr) = - [\hh_4, [\hh_4, \Im(\bb_3\uu\rr)]].
\end{equation}
These tell us that $\Im(\bb_3\rr)$ and $\Im(\bb_3\uu\rr)$ are perpendicular to $\hh_4$ in
$\Im(\mathbb{H})$.  Therefore, the expression in \eqref{eq:h4} becomes
\begin{equation}
	0 = \Re(\bb_3\rr), \quad 0 = \Re(\bb_3\uu\rr) \quad \text{and} \quad
	|\rr|^2 \bb_3 \uu \bar{\bb_3} = [\Im(\bb_3\rr), \Im(\bb_3\uu\rr)].
\end{equation} 
Putting all of these constraints together, we have
\begin{equation}
	\begin{split}
	\mathcal{C}_{\hat{\n}_+,\,\mathsf{4.5.iii}} = \{
		\uu^2 &= -1, \quad
		\Im(\bb_3\rr) = - [\hh_4, [\hh_4, \Im(\bb_3\rr)]], \quad
		\Im(\bb_3\uu\rr) = - [\hh_4, [\hh_4, \Im(\bb_3\uu\rr)]], \\
		0 &= \Re(\bb_3\rr), \quad
		0 = \Re(\bb_3\uu\rr), \quad
		|\rr|^2 \bb_3 \uu \bar{\bb_3} = [\Im(\bb_3\rr), \Im(\bb_3\uu\rr)] \},
	\end{split}
\end{equation}
for non-vanishing \hypertarget{N2_n+_45iii}{matrices}
\begin{equation}
	\begin{split}
	\mathcal{M}_{\hat{\n}_+,\,\mathsf{4.5.iii}} = \Big\{
		\mb &= \begin{pmatrix}
			\zero & \zero \\ \bb_3 & \zero
		\end{pmatrix}, \quad
		\mp = \begin{pmatrix}
			\zero & \zero \\ \bb_3\uu & \zero
		\end{pmatrix}, \quad
		\mh = \begin{pmatrix}
			\zero & \zero \\ -c^{-1} \bb_3\uu\bb_3^{-1} \bar{\rr} & \bb_3 \uu \bb_3^{-1}
		\end{pmatrix}, \\
		N_0 &= \begin{pmatrix}
		\zero & \zero \\ \zero & - 2c |\bb_3|^2
		\end{pmatrix}, \quad
		N_1 = \begin{pmatrix}
			c & \rr \\ \bar{\rr} & d
		\end{pmatrix}, \quad
		N_2 = \begin{pmatrix}
		\zero & \zero \\ \zero & -2c\bb_3\uu\bar{\bb_3}
		\end{pmatrix}, \\
		N_3 &= \begin{pmatrix}
		\zero & c \overbar{\bb_3\uu} \\ - c \bb_3\uu & - 2 \Im(\bb_3\uu\rr)
		\end{pmatrix}, \quad 
		N_4 = \begin{pmatrix}
		\zero & - c\bar{\bb_3} \\ c\bb_3 & 2 \Im(\bb_3\rr)	
		\end{pmatrix} \Big\}.
		\end{split}
\end{equation}
To demonstrate that there are super-extensions in this sub-branch, we will first simplify this system by letting parameters vanish where possible.  In particular, $\rr$ and $d$ in $N_1$ may be set to zero.  This reduces $\mathcal{C}_{\hat{\n}_+,\,\mathsf{4.5.iii}}$ to contain only $\uu^2 = -1$.  Now we can use the endomorphisms of $\s_{\bar{1}}$ to impose $\bb_3 = \ii$, $\uu = \jj$ and $c=1$.  With these choices, the matrices become
\begin{equation}
	\begin{split}
	\mb &= \begin{pmatrix}
			\zero & \zero \\ \ii & \zero
		\end{pmatrix}, \quad
		\mp = \begin{pmatrix}
			\zero & \zero \\ \kk & \zero
		\end{pmatrix}, \quad
		\mh = \begin{pmatrix}
			\zero & \zero \\ \zero & \jj
		\end{pmatrix}, \\
		N_0 &= \begin{pmatrix}
		\zero & \zero \\ \zero & - 2
		\end{pmatrix}, \quad
		N_1 = \begin{pmatrix}
			1 & \zero \\ \zero & \zero
		\end{pmatrix}, \quad
		N_2 = \begin{pmatrix}
		\zero & \zero \\ \zero & 2\jj
		\end{pmatrix}, \\ 
		N_3 &= \begin{pmatrix}
		\zero & -\kk \\ -\kk & \zero
		\end{pmatrix}, \quad 
		N_4 = \begin{pmatrix}
		\zero & \ii \\ \ii & \zero
		\end{pmatrix} .
		\end{split}
\end{equation}
As there are no restrictions on the parameter $d$, we may introduce it without affecting our other parameter choices; however, this is not the case for $\rr$.  There are several constraints in $\mathcal{C}_{\hat{\n}_+,\,\mathsf{4.5.iii}}$ involving $\rr$; therefore, we need to examine these constraints to determine whether new parameters must be chosen.  Interrogating 
\begin{equation}
		\Im(\bb_3\rr) = - [\hh_4, [\hh_4, \Im(\bb_3\rr)]] \quad \text{and} \quad 
		\Im(\bb_3\uu\rr) = - [\hh_4, [\hh_4, \Im(\bb_3\uu\rr)]]
\end{equation} 
with the parameter choices stated above, we find that $\rr$ must vanish.  In particular, due to $\hh_4 = \bb_3 \uu \bb_3^{-1}$ having unit length, we cannot replicate the analysis of Sub-Branch \hyperlink{N2_n+_42iii}{$\mathsf{4.2.iii}$}, where the magnitude of $\hh_4$ was necessarily either $+\tfrac12$ or $-\tfrac12$.  Thus, we cannot produce a super-extension in this sub-branch for which $\rr \neq 0$.  This simplifies the above $(\mathcal{M}, \mathcal{C})$, such that the remaining constraints are
\begin{equation}
	\mathcal{C}_{\hat{\n}_+,\,\mathsf{4.5.iii}} = \{ \uu^2 = -1 \},
\end{equation}
and the non-vanishing matrices are now 
\begin{equation}
	\begin{split}
	\mathcal{M}_{\hat{\n}_+,\,\mathsf{4.5.iii}} = \Big\{
		\mb &= \begin{pmatrix}
			\zero & \zero \\ \bb_3 & \zero
		\end{pmatrix}, \quad
		\mp = \begin{pmatrix}
			\zero & \zero \\ \bb_3\uu & \zero
		\end{pmatrix}, \quad
		\mh = \begin{pmatrix}
			\zero & \zero \\ \zero & \bb_3 \uu \bb_3^{-1}
		\end{pmatrix}, \\
		N_0 &= \begin{pmatrix}
		\zero & \zero \\ \zero & - 2c |\bb_3|^2
		\end{pmatrix}, \quad
		N_1 = \begin{pmatrix}
			c & \zero \\ \zero & d
		\end{pmatrix}, \quad
		N_2 = \begin{pmatrix}
		\zero & \zero \\ \zero & -2c\bb_3\uu\bar{\bb_3}
		\end{pmatrix}, \\
		N_3 &= \begin{pmatrix}
		\zero & c \overbar{\bb_3\uu} \\ - c \bb_3\uu & \zero
		\end{pmatrix}, \quad 
		N_4 = \begin{pmatrix}
		\zero & - c\bar{\bb_3} \\ c\bb_3 & \zero
		\end{pmatrix} \Big\}.
	\end{split}
\end{equation}
\paragraph{$\hat{\g}$}~\\ \\  Finally, substitute $\lambda = \eta = \varepsilon = 0$ and $\mu = -1$ into the second constraint in 
\eqref{eq:N2_branch4_condition} to investigate the $\hat{\g}$ case.  We
find
\begin{equation}
	[\vv, \hh_1] = -1,
\end{equation}
which, since $[\vv, \hh_1] \in \Im(\mathbb{H})$, is inconsistent.  Therefore, we cannot
get a super-extension of $\hat{\g}$ in this branch.
\subsection{Summary} \label{subsec:N2_summary}
Table~\ref{tab:N2_classification} lists all the sub-branches of $\cS$ we found that contain $\N=2$ generalised Bargmann superalgebras.  Each Lie superalgebra in one of these branches is an $\N=2$ super-extension of one of the generalised Bargmann algebras given in Table~\ref{tab:gb_algebras}, taken from \cite{Figueroa-OFarrill:2017ycu}.  It is interesting that $\bZ$ only appears in
\begin{equation}
	[\B, \P] = \bZ \quad \text{and} \quad [\Q, \Q] = \bZ.
\end{equation}
Therefore, in all instances, $\bZ$ remains central after the super-extension.  In particular, this means that we may always find a kinematical Lie superalgebra (without a central-extension) by taking the quotient of our generalised Bargmann superalgebra $\s$ by the $\mathbb{R}$-span of $\bZ$, $\s / \langle \bZ \rangle$.
\begin{table}[h!]
  \centering
  \caption{Sub-Branches of $\N=2$ Generalised Bargmann Superalgebras (with $[\Q,\Q]\neq 0$)}
  \label{tab:N2_classification}
  \setlength{\extrarowheight}{2pt}
  \rowcolors{2}{blue!10}{white}
    \begin{tabular}{l|l*{5}{|>{$}c<{$}}}\toprule
      \multicolumn{1}{c|}{(S)B} & \multicolumn{1}{c|}{$\k$}& \multicolumn{1}{c|}{$\mh$}& \multicolumn{1}{c|}{$\mz$}& \multicolumn{1}{c|}{$\mb$} & \multicolumn{1}{c|}{$\mp$} & \multicolumn{1}{c|}{$[\Q,\Q]$}\\
      \toprule
      \hyperlink{N2_a_1i}{$\mathsf{1.i}$} & \hyperlink{a}{$\hat{\a}$}  & \checkmark & & & &  \bZ  \\
      \hyperlink{N2_a_1ii}{$\mathsf{1.ii}$}  & \hyperlink{a}{$\hat{\a}$} & & & &  & \bH + \bZ \\
      \hyperlink{N2_a_21ii}{$\mathsf{2.1.ii}$} & \hyperlink{a}{$\hat{\a}$} & & & & \checkmark &  \bH  \\
      \hyperlink{N2_a_22i}{$\mathsf{2.2.i}$} & \hyperlink{a}{$\hat{\a}$} & \checkmark & & & \checkmark &  \bZ  \\
      \hyperlink{N2_a_23ii}{$\mathsf{2.3.i}$} & \hyperlink{a}{$\hat{\a}$}  & \checkmark & & & \checkmark &  \bZ + \B \\
      \hyperlink{N2_a_23i}{$\mathsf{2.3.ii}$} & \hyperlink{a}{$\hat{\a}$}  & & & & \checkmark & \bH + \bZ + \B \\
      \hyperlink{N2_a_45iii}{$\mathsf{4.5.iii}$} & \hyperlink{a}{$\hat{\a}$} & \checkmark & & \checkmark & \checkmark & \bH + \bZ + \B + \P \\
      \hyperlink{N2_ng_1i}{$\mathsf{1.i}$}  & \hyperlink{n-}{$\hat{\n}_-$} & \checkmark & & & &  \bZ  \\
      \hyperlink{N2_n-_22i}{$\mathsf{2.2.i}$} & \hyperlink{n-}{$\hat{\n}_-$}  & \checkmark & & & \checkmark &  \bZ  \\
      \hyperlink{N2_n-_23i}{$\mathsf{2.3.i}$} & \hyperlink{n-}{$\hat{\n}_-$}  & \checkmark & & & \checkmark &  \bZ + \P  \\
      \hyperlink{N2_n-_32i}{$\mathsf{3.2.i}$} & \hyperlink{n-}{$\hat{\n}_-$}  & \checkmark & & \checkmark & &  \bZ  \\ 
      \hyperlink{N2_n-_33i}{$\mathsf{3.3.i}$} & \hyperlink{n-}{$\hat{\n}_-$}  & \checkmark & & \checkmark & &  \bZ + \P \\
      \hyperlink{N2_ng_1i}{$\mathsf{1.i}$} & \hyperlink{n+}{$\hat{\n}_+$}  & \checkmark & & & &  \bZ  \\      
      \hyperlink{N2_ng_22i}{$\mathsf{2.2.i}$} & \hyperlink{n+}{$\hat{\n}_+$}  & \checkmark & & & \checkmark &  \bZ  \\ 
      \hyperlink{N2_n+_41i}{$\mathsf{4.1.i}$} & \hyperlink{n+}{$\hat{\n}_+$} & \checkmark & & \checkmark & \checkmark &  \bZ  \\      
       \hyperlink{N2_n+_42iii}{$\mathsf{4.2.iii}$} & \hyperlink{n+}{$\hat{\n}_+$} & \checkmark & & \checkmark & \checkmark &  \bZ + \B + \P  \\
      \hyperlink{N2_n+_45iii}{$\mathsf{4.5.iii}$} & \hyperlink{n+}{$\hat{\n}_+$}  & \checkmark & & \checkmark & \checkmark &  \bH + \bZ + \B + \P \\
      \hyperlink{N2_ng_1}{$\mathsf{1.i}$} & \hyperlink{g}{$\hat{\g}$} & \checkmark & & & &  \bZ  \\  
      \hyperlink{N2_ng_2i}{$\mathsf{2.2.i}$} & \hyperlink{g}{$\hat{\g}$}  & \checkmark & & & \checkmark &  \bZ  \\  
      \hyperlink{N2_g_32i}{$\mathsf{3.2.i}$} & \hyperlink{g}{$\hat{\g}$} & \checkmark & & \checkmark & &  \bZ  \\         
      \hyperlink{N2_g_33i}{$\mathsf{3.3.i}$} & \hyperlink{g}{$\hat{\g}$} & \checkmark & & \checkmark & &  \bZ + \P  \\  
       \hyperlink{N2_g_33ii}{$\mathsf{3.3.ii}$}  & \hyperlink{g}{$\hat{\g}$} & & & \checkmark & &  \bH + \bZ + \P  \\  
      \bottomrule
    \end{tabular}
    \caption*{The first column indicates the sub-branch of generalised Bargmann superalgebras,
    so that the reader may navigate back to find the conditions on the non-vanishing parameters
    of these superalgberas.  The second column then tells us the underlying generalised Bargmann algebra $\k$.
    The next four columns tells us which of the $\s_{\bar{0}}$ generators $\bH,\bZ, \B$, 
    and $\P$ act on $\Q$.  Recall, $\J$ necessarily acts on $\Q$, so we do not need to state this explicitly.
    The final column shows which $\s_{\bar{0}}$ generators appear in the $[\Q, \Q]$ bracket.
     }
\end{table}
\subsubsection{Unpacking the Notation}
Although the formalism employed in this classification was useful for our purposes, it may be unfamiliar to the reader.  Therefore, in this section, we will convert one of the $\N=2$ super-extensions of the Bargmann algebra in sub-branch \hyperlink{N2_g_33ii}{$\mathsf{3.3.ii}$} into a more standard notation.  The brackets for this algebra, excluding the $\s_{\bar{0}}$ brackets, which are shown in Table~\ref{tab:gb_algebras}, take the form
\begin{equation}
	[\sB(\beta), \sQ(\vt)] = \sQ(\beta \vt \mb) \quad \text{and} \quad [\sQ(\vt), \sQ(\vt)] = \Re(\vt N_0 \vt^\dagger) \sH + \Re(\vt N_1 \vt^\dagger) \sZ - \sP(\vt N_4 \vt^\dagger),
\end{equation}
where 
\begin{equation}
	\mb = \begin{pmatrix}
		\zero & \zero \\ \bb_3 & \zero
	\end{pmatrix} \quad
	N_0 = \begin{pmatrix}
		\zero & \zero \\ \zero & -2c |\bb_3|^2
	\end{pmatrix} \quad
	N_1 = \begin{pmatrix}
		c & \rr \\ \bar{\rr} & d
	\end{pmatrix} \quad
	N_4 = \begin{pmatrix}
		\zero & -c\bar{\bb_3} \\ c\bb_3 & \bb_3 \rr - \bar{\rr} \bar{\bb_3}
	\end{pmatrix}.
\end{equation}
Let $\{\sQ^1_a\}$ be a real basis for the first $\so(3)$ spinor module in $\s_{\bar{1}} = S^1 \oplus S^2$ where $a \in \{1, 2, 3, 4\}$ , and $\{\sQ^2_a\}$ be a basis for the second $\so(3)$ spinor module.  Letting $\vt = (\theta_1, \theta_2)$, and substituting the above matrices into the $[\s_{\bar{0}}, \s_{\bar{1}}]$ bracket, we get
\begin{equation}
		[\sB(\beta), \sQ^1(\theta_1)] = 0 \quad \text{and}  \quad [\sB(\beta), \sQ^2(\theta_2)] = \sQ^1(\beta \theta_2 \bb_3).
\end{equation}
Substituting $\vt = \vt'= (\theta_1, 0)$, $\vt = (\theta_1, 0)$ and $\vt'=(0,\theta_2)$, and $\vt=\vt'=(0,\theta_2)$ into the $[\s_{\bar{1}}, \s_{\bar{1}}]$ bracket we find
\begin{equation}
	\begin{split}
		[\sQ^1(\theta_1), \sQ^1(\theta_1)] &= c |\theta_1|^2 \sZ \\
		[\sQ^1(\theta_1), \sQ^2(\theta_2)] &= \Re(\theta_1\rr\bar{\theta_2})Z -\tfrac{c}{2} \sP(\theta_2\bb_3\bar{\theta_1}-\theta_1\bar{\bb_3}\bar{\theta_2}) \\ 
		[\sQ^2(\theta_2), \sQ^2(\theta_2)] &= -2c |\bb_3|^2 |\theta_2|^2 \sH + d |\theta_2|^2 \sZ - \sP(\theta_2 (\bb_3 \rr - \bar{\rr}\bar{\bb_3})\bar{\theta_2}).
	\end{split}
\end{equation} 
For the purposes of the present example, we will set the parameters of this super-extension as specified in \eqref{eq:notation_example}; therefore, we have $[\s_{\bar{0}}, \s_{\bar{1}}]$ brackets
\begin{equation}
		[\sB(\beta), \sQ^1(\theta_1)] = \sQ^2(\beta \theta_2 \ii) \quad \text{and}  \quad [\sB(\beta), \sQ^2(\theta_2)] = 0,
\end{equation}
and $[\s_{\bar{1}}, \s_{\bar{1}}]$ brackets
\begin{equation}
		[\sQ^1(\theta_1), \sQ^1(\theta_1)] = |\theta_1|^2 \sZ, \quad
		[\sQ^1(\theta_1), \sQ^2(\theta_2)] = -\tfrac{1}{2} \sP(\theta_2\ii\bar{\theta_1}+\theta_1\ii\bar{\theta_2}) \quad \text{and} \quad
		[\sQ^2(\theta_2), \sQ^2(\theta_2)] = |\theta_2|^2 \sH.
\end{equation}
Now, we can write
\begin{equation}
	[\bB_i, \bQ^2_a] = \sum_{b = 1}^4 \bQ^1_b \tensor{\beta}{_i^b_a} \quad [\bQ^1_a, \bQ^1_b] = \delta_{ab} \bZ
	\quad [\bQ^1_a, \bQ^2_b] = \sum_{i = 1}^3 \bP_i \Gamma_{ab}^i, \quad \quad [\bQ^2_a, \bQ^2_b] = \delta_{ab} \bH.
\end{equation}
Our brackets then produce the $\beta_i$ matrices
\begin{equation}
	\beta_1 = \begin{pmatrix}
		 -\mathbb{1} & \zero \\ \zero & \mathbb{1}
	\end{pmatrix} \quad
	\beta_2 = \begin{pmatrix}
		\zero & - \sigma_1 \\ - \sigma_1 & \zero
	\end{pmatrix} \quad
	\beta_3 = \begin{pmatrix}
		\zero & \sigma_3 \\ \sigma_3 & \zero
	\end{pmatrix},
\end{equation}
and the symmetric $\Gamma^i$ matrices
\begin{equation}
	\Gamma^1 = \begin{pmatrix}
		-\mathbb{1} & \zero  \\ \zero & \mathbb{1}
	\end{pmatrix} \quad
	\Gamma^2 = \begin{pmatrix}
		\zero & -\sigma_1 \\ -\sigma_1 & \zero
	\end{pmatrix} \quad
	\Gamma^3 = \begin{pmatrix}
		\zero & \sigma_3 \\ \sigma_3 & \zero
	\end{pmatrix},
\end{equation}
where $\sigma_1$ and $\sigma_3$ are the first and third Pauli matrix, respectively.  This $\N=2$ Bargmann superalgebra takes the same form as the $(2+1)$-dimensional Bargmann superalgebra utilised in \cite{Andringa:2013mma}.

\section{Conclusion} \label{sec:gb_superspace_conc}
In this chapter, we classified the $\N=1$ super-extensions of the generalised Bargmann algebras with three-dimensional spatial isotropy up to isomorphism.  We also presented the non-empty sub-branches of the variety $\cS$ describing the $\N=2$ super-extensions of the generalised Bargmann algebras.  To simplify this classification problem, we utilised a quaternionic formalism such that $\so(3)$ scalar modules were described by copies of $\mathbb{R}$, $\so(3)$ vector modules were described by copies of $\Im(\mathbb{H})$, and $\so(3)$ spinor modules were described by copies of $\mathbb{H}$.  We began by defining a universal generalised Bargmann algebra, which, under the appropriate setting of some parameters, may be reduced to the centrally-extended static kinematical Lie algebra $\hat{\a}$, the centrally-extended Newton-Hooke algebras $\hat{\n}_\pm$, or the Bargmann algebra $\hat{\g}$.  The most general form for the $[\s_{\bar{0}}, \s_{\bar{1}}]$ and $[\s_{\bar{1}}, \s_{\bar{1}}]$ bracket components were found before substituting them into the super-Jacobi identity and finding the constraints on the parameters for these maps.  Because of the formalism in use, solving these constraints amounted to some linear algebra over the quaternions, and paying attention to the allowed basis transformations $\G \subset \GL(\s_{\bar{0}}) \times \GL(\s_{\bar{1}})$.  Since we are only interested in supersymmetric extensions of these algebras, we limited ourselves to those branches which allow for non-vanishing $[\Q, \Q]$.  The results of the $\N=1$ and $\N=2$ analyses are in Tables \ref{tab:N1_classification} and \ref{tab:N2_classification}, respectively.  We found 9 isomorphism classes in the $\N=1$ case, and 22 non-empty sub-branches in the $\N=2$ case.  

\chapter{Conclusion} \label{chap:conc}
In this thesis, we presented a framework to explore kinematical symmetries beyond the standard Lorentzian case.  This framework consisted of an algebraic classification, a geometric classification, and a derivation of the geometric properties required to define physical theories on the classified spacetime geometries. We will now briefly run through the main results from each of the primary chapters.
\\ \\
In Chapter~\ref{chap:k_spaces}, each step in this framework was discussed in detail for the case of kinematical symmetries.  Section~\ref{sec:ks_klas} reviewed the classification of kinematical Lie algebras with spatial dimension $D=3$ (up to isomorphism).  These Lie algebras assumed spatial isotropy as well as homogeneity in both time and space.  Known and named Lie algebras were identified, and $18$ isomorphism classes were found.  In Section~\ref{sec:ks_kss}, the integration of these Lie algebras to spatially-isotropic simply-connected homogeneous spacetime was discussed, and the classification of these geometries was reviewed.  Five kinematical spacetime classes were identified; namely, Lorentzian, Riemannian, Galilean, Carrollian, and Aristotelian.  This classification was followed by identifying the characteristic Lie brackets of each spacetime class and identifying a procedure for connecting the spacetimes via geometric limits.  Section~\ref{sec:ks_gps} then derived various geometric properties for each kinematical spacetime including the fundamental vector fields, soldering forms, vielbeins, invariant structures, and the space of invariant affine connections.  
\\ \\
In Chapter~\ref{chap:k_superspaces}, the algebraic and geometric classifications of our framework were carried out in the super-kinematical case.  Section~\ref{sec:ks_superspace_ksa} gave the classification of the $\N=1$ kinematical and Aristotelian Lie superalgebras in three spatial dimensions, up to isomorphism.  We found $43$ isomorphism classes, some with parameters.  In Section~\ref{sec:ks_superspace_kss}, we then classified the corresponding simply-connected homogeneous $(4|4)$-dimensional kinematical superspaces, finding $27$ isomorphism classes.  It was then shown how these superspaces might be connected via geometric limits and the low-rank invariants of each superspace were determined.
\\ \\
Chapter~\ref{chap:gb_superspaces} presented the algebraic classification of our framework for the super-Bargmann case.   In Section~\ref{sec:gb_superspace_n1}, the classification of $\N=1$ generalised Bargmann superalgebras was presented, identifying $9$ isomorphism classes of Lie superalgebra.  Section~\ref{sec:gb_superspace_n2} then discussed the generalisation of this classification to the $\N=2$ case.  Here, owing to the increased complexity of the problem, we only identified non-empty branches in the real algebraic variety $\cS$ describing the possible generalised Bargmann superalgebra structures.  In particular, we found $22$ non-empty branches in $\cS$.
\\ \\
Outside the classifications and the derivations mentioned above, some other key results include the proof that boosts act with non-compact orbits in Lorentzian, Galilean, and Carrollian spacetimes.  The property of having non-compact boosts is an essential physical requirement, as discussed in~\cite{Bacry:1968zf}.  If this property does not hold, then a sufficiently large boost would no longer be a boost; we would arrive back at our starting point.  Thus, compact boosts are deemed unphysical.  
\\ \\
Another significant result presented in this thesis was the demonstration that kinematical and Aristotelian Lie algebras may admit several non-isomorphic super-extensions.  These non-isomorphic super-extensions may then integrate into inequivalent kinematical and Aristotelian superspaces.  Interestingly, due to geometric realisability and effectivity being, respectively, independent and dependent of the super-extension, Aristotelian Lie algebras may form effective Lie super pairs $(\s, \h)$ where the underlying Aristotelian Lie pair $(\k, \h)$ is not effective.  In these cases, the \emph{``boost''} generators act as R-symmetries, which transform the odd dimensions, but not the even dimensions. 
\\ \\
This thesis's results suggest several directions for future study, which we will present here in no particular order.  Given that the geometric classifications contain homogeneous (super)spaces, which may also be called Klein geometries, a natural generalisation would be to build Cartan geometries modelled on these spaces.  This construction is described at length in~\cite{sharpe2000differential}; however, for our purposes, the crucial point is that we can view our classifications as describing the possible local geometries for a spacetime manifold.  By modelling a Cartan geometry on these spaces, we allow for the introduction of curvature in the same way Riemannian geometry introduces curvature to Euclidean geometry.  This process was developed by Cartan in his rewriting of Newtonian gravity in~\cite{cartan1923varietes, cartan1924varietes}.  Additionally, for a discussion on this process in $2+1$-dimensions, see~\cite{Matulich:2019cdo}.  Note, the addition of Cartan geometries would constitute a new step in the presented framework, taking us closer to a complete set of kinematical spacetime theories. 
\\ \\
More immediate work, which is currently underway, is to complete the presented framework for all the given cases.  In particular, this would involve deriving the geometric properties for the kinematical superspaces, building the superspaces corresponding to the generalised Bargmann superalgebras, and calculating the necessary geometric properties in these instances.  Work on constructing the non-supersymmetric Bargmann spacetimes is also in progress and should appear soon. 
\\ \\
In addition to further classifications, we may look to build theories using some of the novel (super)spaces found here.  In particular, we may look to utilise some of these spaces in holographic theories, as in~\cite{Hartong:2014oma, Christensen:2013rfa, Hartong:2015wxa, Hartong:2015zia} or we may gauge some of the Bargmann superalgebras to arrive at novel non-relativistic supergravity theories.  Additionally, in a similar spirit to the Cartan generalisation mentioned above, we could look to generate interesting gravity theories by putting the classified Lie (super)algebras through the procedure which leads to MacDowell-Mansouri gravity~\cite{Wise:2006sm, DAuria:2020guc, Castellani:2018zey}. 

\bibliographystyle{utphys}
\bibliography{intro_refs.bib, math_prelims_refs.bib, KS_refs.bib, GBS_refs.bib, KSS_refs.bib, GBSS_refs.bib, conc_refs.bib}
\addcontentsline{toc}{chapter}{Bibliography}

\appendix


\end{document}